# Thermodynamic Design and Fouling of Membrane Distillation Systems

by

David Elan Martin Warsinger

M.Eng. Mechanical Engineering, Cornell University, 2011

B.S. Mechanical and Aerospace Engineering, Cornell University, 2010

Submitted to the Department of Mechanical Engineering
in partial fulfillment of the requirements for the degree of

Doctor of Philosophy in Mechanical Engineering

at the

MASSACHUSETTS INSTITUTE OF TECHNOLOGY

June 2015

Author . . . . . . . . . . . . . . . . . . . . . . . . . . . . . . . . . . . . . . . . . . . . . . . . . . . . .
Department of Mechanical Engineering
May 22, 2015

Certified by. . . . . . . . . . . . . . . . . . . . . . . . . . . . . . . . . . . . . . . . . . . . . . . . . . . . .
John H. Lienhard V
Abdul Latif Jameel Professor
Thesis Supervisor

Accepted by . . . . . . . . . . . . . . . . . . . . . . . . . . . . . . . . . . . . . . . . . . . . . . . . . . . .
David E. Hardt
Graduate Officer, Department of Mechanical Engineering

Thermodynamic Design and Fouling of
Membrane Distillation Systems

by

David Elan Martin Warsinger

Submitted to the Department of Mechanical Engineering
on 5/22/2015 in partial fulfillment of the
requirements for the degree of
Doctor of Philosophy in Mechanical Engineering

ABSTRACT

As water shortages intensify globally under the stresses of increasing demand, aquifer depletion, and climate change, the market for efficient desalination technologies has grown rapidly to fill the void. One such developing technology, membrane distillation (MD), has experienced keen academic interest and an increase in start-up businesses in the past decade. MD has expanded into a niche of small scale thermal desalination using solar and waste heat resources, due to its fouling resistance, scalability, and acceptable efficiency. Recent studies indicate that MD could attain the efficiencies of state-of-the-art mature thermal desalination technologies, although additional engineering and scientific challenges must first be overcome.

The aim of this research is to better understand and provide solutions for two major challenge areas for MD: efficiency and membrane fouling. Studies on improving MD efficiency included examining the effects of tilt angle on MD performance using numerical simulations paired with experiments, devising a novel MD system design for with superhydrophobic surfaces to improved efficiencies, and an entropy-generation comparison of MD to other desalination technologies. For fouling studies in MD, a review of MD fouling was undertaken to synthesize conclusions from the literature and to explore gaps in the literature. This review lead to studies of the effect of filtration and bulk nucleation on MD fouling, a study on heterogeneous nucleation of inorganic salts with a fouling regime map to avoid nucleation, and fouling prevention via induced air-layers.

Thesis Supervisor: John H. Lienhard V
Title: Abdul Latif Jameel Professor

## TABLE OF CONTENTS

# ACKNOWLEDGEMENT

I would like to thank my advisor, Professor Lienhard, for guiding my PhD.

This work was highly collaborative, and I give deep thanks to my collaborators Jaichander Swaminathan, Hyung Won Chung, Kishor Nayar, Karan Mistry, Emily Tow, Seongpil Jeong, Amelia Servi, Jehad Karraz, and Elena Guillen-Burrieza. I thank my other labmates too, who gave feedback and advice.

I would like to thank my committee members, Prof. Hassan Arafat, Prof. Karen Gleason, and Prof. Rohit Karnik for valuable feedback.

I would also like to thank all those who assisted in my work, including Prof. Mathias Kolle, Prof. Allan Myerson, and You Peng.

I give deep thanks to all the undergrads who worked with me, including Sarah Van Belleghem, Jocelyn Gonzales, Ann McCall Huston, Priyanka Chatterjee, Grace Connors, Joanna So, Sterling Watson, Aileen Gutmann, Laith Maswadeh, and Sarah Ritter.

I want to thank the Masdar Institute of Science and Technology for funding this work.

Finally, I would like to thank my friends, family, and my fiancée Andrea Carnie for their support.

# NOMENCLATURE

**Roman Symbols**

| | |
|---|---|
| $A$ | experimental pre-exponential factor |
| $A_m$ | membrane area [$cm^2$] |
| B | membrane distillation coefficient [$kg/m^2s$ Pa] |
| $c$ | concentration [$mol/m^3$] |
| $c_p$ | specific heat [kJ/kg-K] |
| $C_X$ | salt concentration [g/mL] |
| $d$ | channel depth [m] |
| $d_{gap}$ | air gap width [mm] |
| $D_{wa}$ | Diffusivity of water in air [$m^2/s$] |
| $f(\,)$ | function of |
| $g$ | specific Gibbs free energy [kJ/kg] |
| $\Delta G^*$ | Gibbs Free Energy barrier for the formation of a stable nucleus [J] |
| $h$ | specific enthalpy [kJ/kg] |
| $h_{fg}$ | latent heat of vaporization, kJ/kg |
| $J$ | mass flux [$kg/m^2s$] |
| $k$ | conductivity [W/m K] |
| $k_b$ | Boltzmann constant [$m^2$ $kg/s^2$ K] |
| $k_{gap}$ | effective conductivity of gap [W/m K] |
| $L$ | module effective length [m] |
| $\dot{m}$ | mass flow rate [kg/s] |
| $M$ | molecular weight [kg/kmol] |
| $\dot{m}_p$ | condensate flux [kg/hr] |

$n$ number of effects or stages

$N$ number of particles per unit volume [mols/L]

$p$ partial pressure [Pa]

$P$ pressure [kPa]

$q$ heat flux [W/m$^2$]

$\dot{Q}$ heat transfer [kW]

$\dot{Q}_{\text{least}}$ least heat of separation [kW]

$\dot{Q}_{\text{least}}^{\text{min}}$ minimum least heat of separation (zero recovery) [kW]

$\dot{Q}_H$ heat of separation added at $T_H$ [kW]

$\dot{Q}_0$ heat rejected to environment, kW, exiting at $T_0$

$R$ ideal gas constant [kJ/kg-K]

$r$ heat transfer [(kg/s product)/(kg/s feed)]

$\dot{S}_{\text{gen}}$ entropy generation rate [kW/K]

$s_{\text{gen}}$ specific entropy generation per unit flow [kJ/kg-K]

$S_{\text{gen}}$ specific entropy generation per unit water produced [kJ/kg-K]

$T$ temperature [°C]

$T_0$ ambient (dead state) temperature, K

$T_H$ temperature of heat reservoir, K

$t_{induction}$ induction time [s]

$v$ specific volume, m$^3$/kg

$v_p$ permeate velocity [m/s]

$\dot{W}_{\text{least}}$ least work of separation, kW

$\dot{W}_{\text{least}}$ minimum least work of separation, kW

$\dot{W}_{sep}$ work of separation, kW

$x$ mole fraction [-]

$z$ distance along module length [m]

$\alpha$ foulant sticking efficiency [-]

$\delta$ thickness of condensate film [m]

$\delta_c$ fouling layer average thickness [m]

$\theta$ contact angle [°]

$\rho$ density [kg/m$^3$]

ω stirrer rotation rate [rpm]

**Greek Symbols**

Δ change in a variable

$\rho$ density [kg/m$^3$]

$\eta_e$ isentropic efficiency of expander

$\eta_p$ isentropic efficiency of pump/compressor

$\eta_{II}$ Second Law/exergetic efficiency

$\eta_{reduced}$ efficiency reduction from decreasing

**Subscripts**

0 environment

a air

b brine

f feed

flash flashing

i air-liquid interface

m membrane

n stage

p product

sw feed seawater

**Superscripts**

IF incompressible fluid

IG ideal gas

$'$ stream before exiting CV

$\Delta p$ Pressure Change

**Superscripts**

AGMD Air Gap Membrane Distillation

CGMD Conductive Gap Membrane Distillation

DCMD Direct Contact Membrane Distillation

EES Engineering Equation Solver

PGMD Permeate Gap Membrane Distillation

MED Multi-Effect Distillation

MD Membrane Distillation

MSF Multi-Stage Flash Distillation

RO Reverse Osmosis

SI Saturation Index

PVDF Polyvinylidene Fluoride

PW Partial Wetting

VMD Vacuum Membrane Distillation

W Wetting

# Chapter 1 INTRODUCTION

## 1.1 EMERGING WATER CRISIS

Population growth, industrialization, global warming, dwindling aquifers, and dietary shifts are accelerating a global water crisis [1, 2]. The number of people experiencing water shortages is expected to multiply fourfold between 2000 and 2025 [2]. A global risk assessment from the World Economic Forum labelled a "water crisis" as the highest impact global risk, above infectious diseases and interstate conflict, and also labelled it among the most likely to occur [3]. Furthermore, lack of access to safe water accounts for roughly 3 million deaths annually [4], which is the second highest annual toll for preventable deaths not caused by lifestyle choices, exceeded only by deaths from hunger [5].

Already much of the world's populous areas use all available renewable water, as illustrated in red in figure 1, where renewable water is the water provided by the local rain cycle after deducting evaporation.

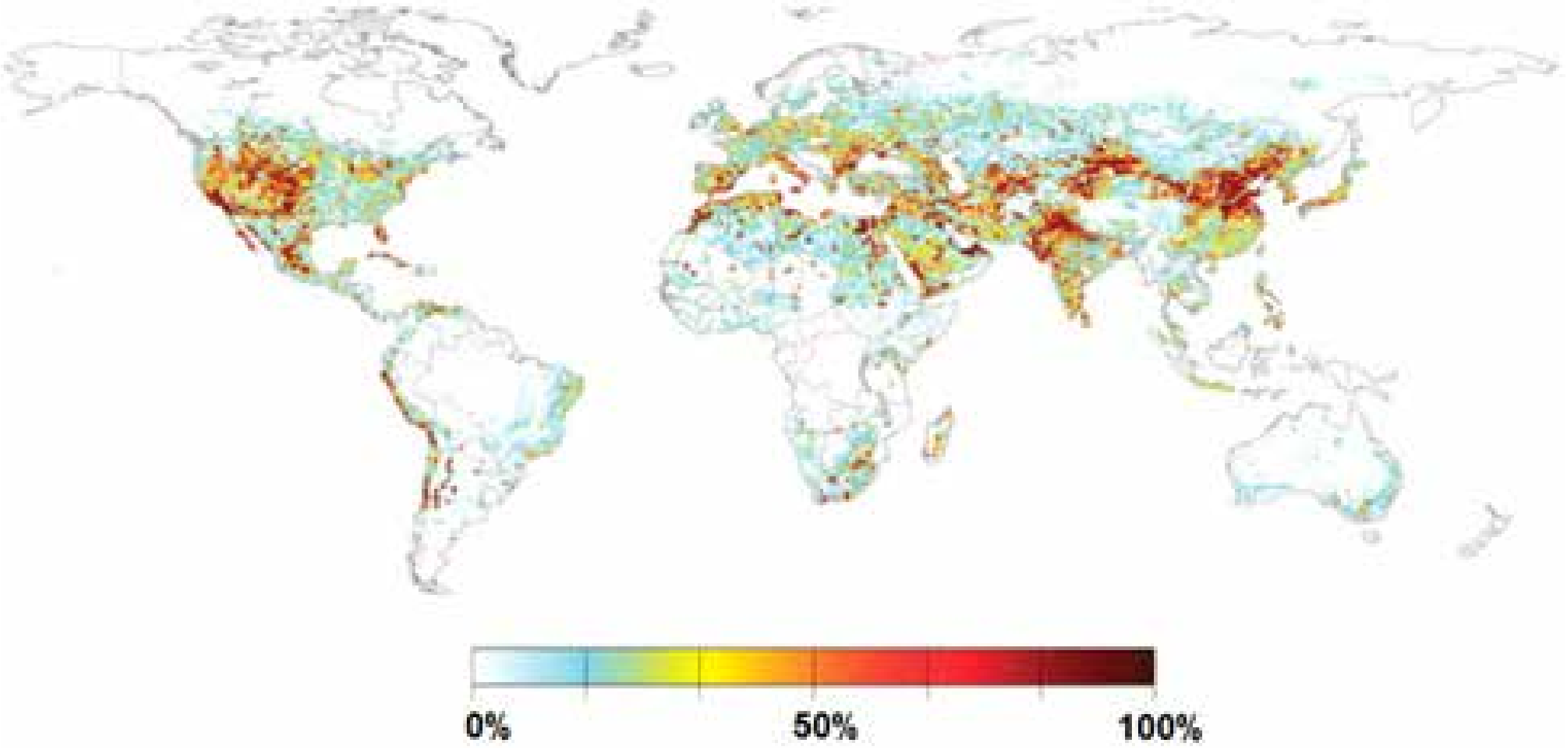

**Figure 1.1.** Water scarcity index: water withdrawn as a percent of net renewable water, modified from [6, 7]

To cope with the growing scarcities, people can use water more efficiently, shift from water intense foods and products, pipe water from distant areas that do not have shortages, improve water reuse, or desalinate water [1]. However, the potential gains from water efficiency cannot meet the new demands, and many developing societies are shifting to much more water intense foods like meat [1]. Adoption of reuse of sewage has been limited because of excessive fears of contamination and revulsion to the idea of consuming water that was recently faeces or other waste. Piping of water is expensive, contentious, and is limited to a few hundred miles, as distances farther than that have hefty pumping energy needs. Desalination, however, can supply nearly limitless water, but is not without trade-offs. Desalination can be energy intensive relative to renewable water [2], and if unregulated, may harm marine ecosystems.

Today, many nations, such as Israel and Saudi Arabia, get the majority of their water from desalination and water reuse [2], and soon Singapore will too. Arid nations, especially in the Middle East, adopted thermal desalination technologies such as Multistage Flash (MSF) as early as the 1970's [2]. The desalination technology reverse osmosis (RO) has expanded rapidly in the last two decades; its market share now dominates, and its costs have now approached near that of conventional water sources [8]. Other desalination technologies have grown as well, including the thermal technologies of membrane distillation (MD), humidification dehumidification (HDH), multi-effect distillation (MED), as well as other technologies such as ion exchange resins and mechanical vapour compression [2, 9]. The thermal technologies have advantages over RO, including superior fouling resistance, reduced pretreatment needs, lower energy use if paired with waste heat or renewable heat sources, and higher purity product water [10].

## 1.2 THESIS SCOPE

This thesis focuses on addressing the two biggest challenges and mysteries in the upcoming thermal desalination technology membrane distillation: efficiency and fouling [11]. This technology poses unique advantages over competitors, including scalability to small sizes, superior fouling resistance [10], and as this work shows, it can also have very high efficiencies [12]. This work aims to improve understanding of the challenges of MD and create inventions and enhancements to help propel this rapidly growing technology. Efficiency improvements on MD can help reduce impacts of humans on climate change, and improvements on fouling in MD can aid the adoption of this technology to play a role countering the global water crisis.

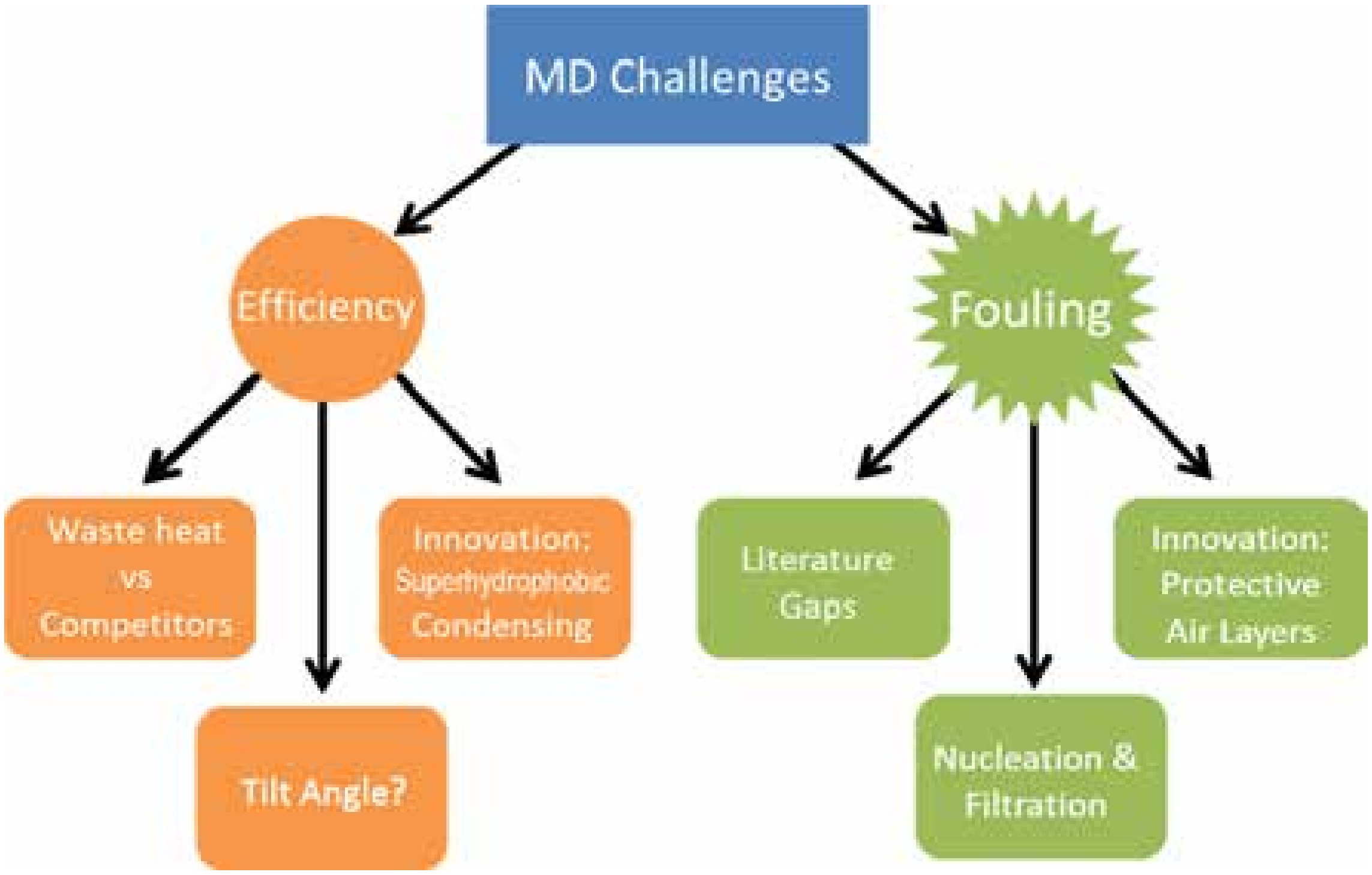

**Figure 1.2.** PhD Thesis Core Contributions

The waste heat work includes entropy generation analysis of each component of MD and five other desalination technologies, while using a waste heat source at varied temperature. The tilt angle and superhydrophobic work focus on understanding and improving air gap membrane distillation (AGMD). For fouling, the literature gaps includes a review paper on fouling and mitigation in MD. The nucleation & filtration work reveals the role of bulk heterogeneous nucleation in MD. Finally, injecting air to maintain a protective layer on a superhydrophobic membrane may act to reduce fouling.

## 1.3 Membrane Distillation Basics

Membrane distillation (MD) is an emerging technology for thermal desalination. MD desalinates by evaporating hot water at a membrane that will allow water vapour, but not liquid water, to pass [10]. MD offers unique advantages over other technologies by scaling down to smaller sizes, being fouling resistant, and as shown here, having first-rate efficiencies compared to other thermal desalination technologies. This thesis examines efficiency improvements and optimization in MD, and seeks to better understand fouling in MD so that it may be better implemented. The first part of this work, efficiency studies, includes an efficiency comparison with other technologies, studies on the effect of tilt angle on Air-Gap MD (AGMD), an efficiency enhancement on AGMD using superhydrophobic condensing, and contributions to other efficiency work, including multistage vacuum MD, conductive gap MD, and applications for MD powered by existing ocean temperature gradients. The second portion of this work focuses on fouling in MD, including a review of MD fouling, studies on filtration and bulk nucleation in MD, a temperature, saturation index, and time-based regime map for avoiding MD fouling, and the effect of introducing air to the system for fouling reduction.

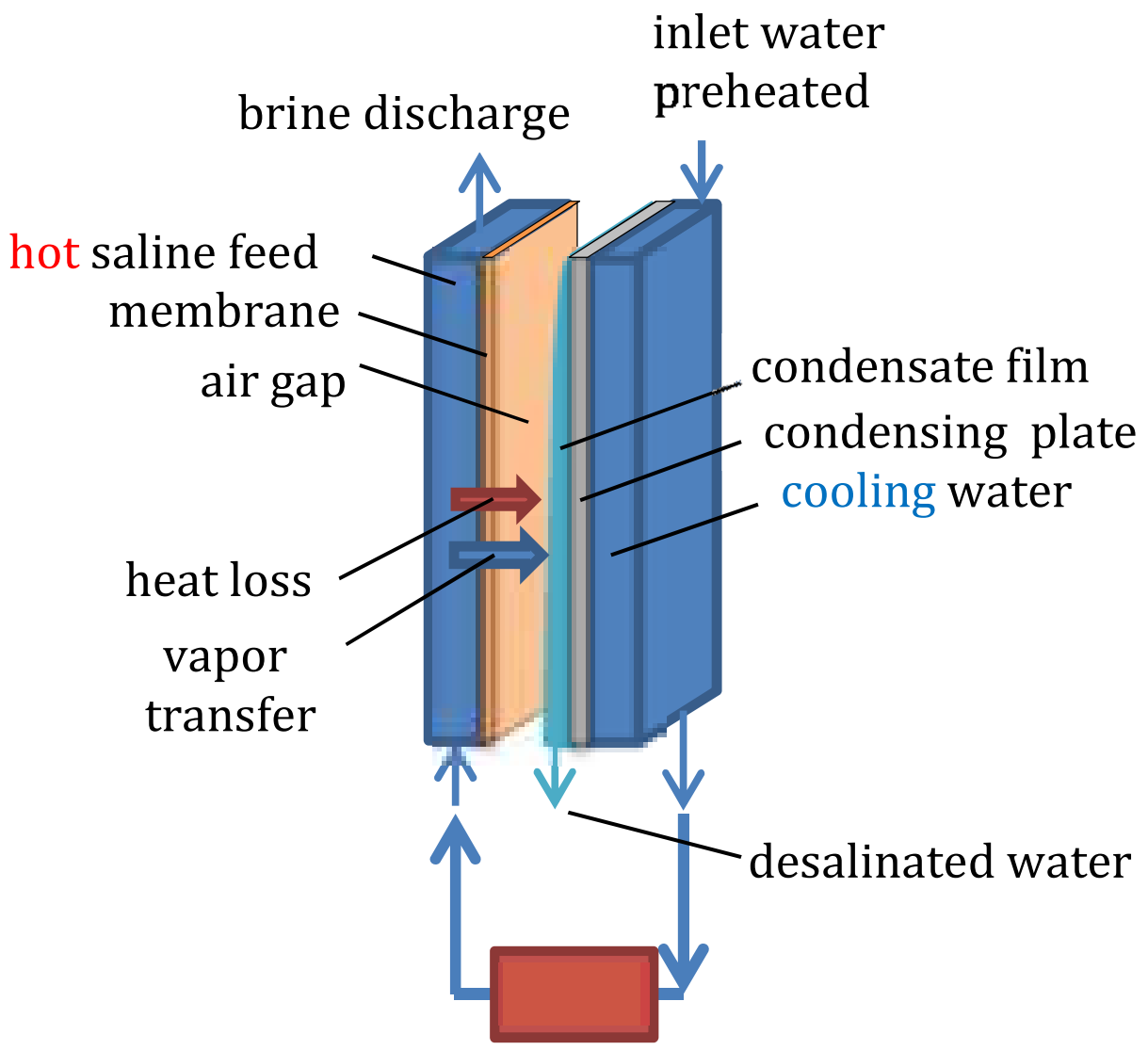

**Figure 1.3.** AGMD diagram

To provide the background for these contributions, a discussion is included of the MD heat and mass transfer basics, the types of MD configurations, the current existing MD systems and companies, a description of the MD state-of-the-art, an introduction to efficiency in MD, modeling in MD basics, a summary of fouling in MD, background for nucleation of salts, and implementations of air layers for fouling prevention.

## 1.4 HEAT AND MASS TRANSFER BASICS FOR MEMBRANE DISTILLATION

In an MD system, a hot saline solution with an associated high water vapor pressure flows across a hydrophobic membrane which selectively allows water vapor to pass through but not liquid water or dissolved salts. Pure water vapor diffuses through the membrane and is condensed and collected on the other side. Cool feed water enters the system in counter flow, becoming preheated as it absorbs the enthalpy of vaporization from the condensing vapor located on the opposite side of a condensing plate [10]. Several configurations exist; these vary in the region between the membrane and condensing surface, which is the dominant thermal and mass transfer resistance in the system. Module configurations may vary as well, including flat plate [13], spiral wound, tube, and hollow fiber setups [10], all of which were discussed in a recent review paper [11]. Multistage systems exist as well and are part of the work studied here.

## 1.5 CONFIGURATION TYPES OF MEMBRANE DISTILLATION

While all membrane distillation systems share several basic components, what differentiates them from one another is the configuration of the feed, cooling, and permeate channels. Firstly, all MD systems have a membrane, a top heater, and a saline feed, and they all produce liquid streams of pure distillate and saline brine. However, there may be variety in the methods or values for: conducting heat from the hot to cold side, the relative flow rate of the condensate, the phase of matter (liquid or vapor or vacuum etc.) on the side of the membrane opposite the feed, and the method for removing permeate from the system (e.g. removing as liquid or as vapor or with a gas) [10].

Six primary types of MD exist, as seen in the diagram: air gap membrane distillation (AGMD), permeate gap membrane distillation (PGMD), conductive gap membrane distillation (CGMD), direct contact membrane distillation (DCMD), vacuum membrane distillation (VMD), and sweeping gas membrane distillation (SGMD).

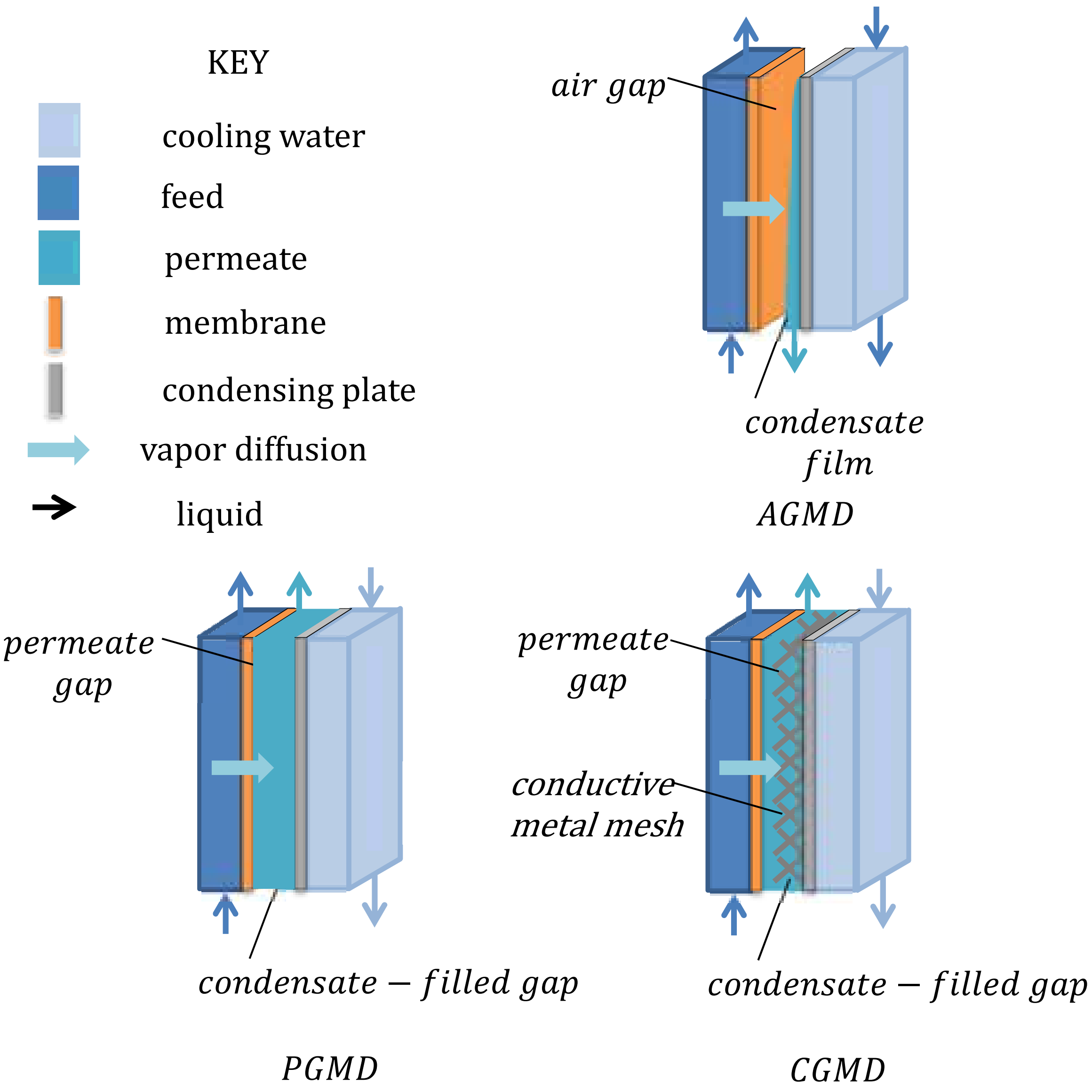

**Figure 1.4.** Configurations of MD that have been studied in the literature with feed preheating in the MD module. (3 channels of fluid)

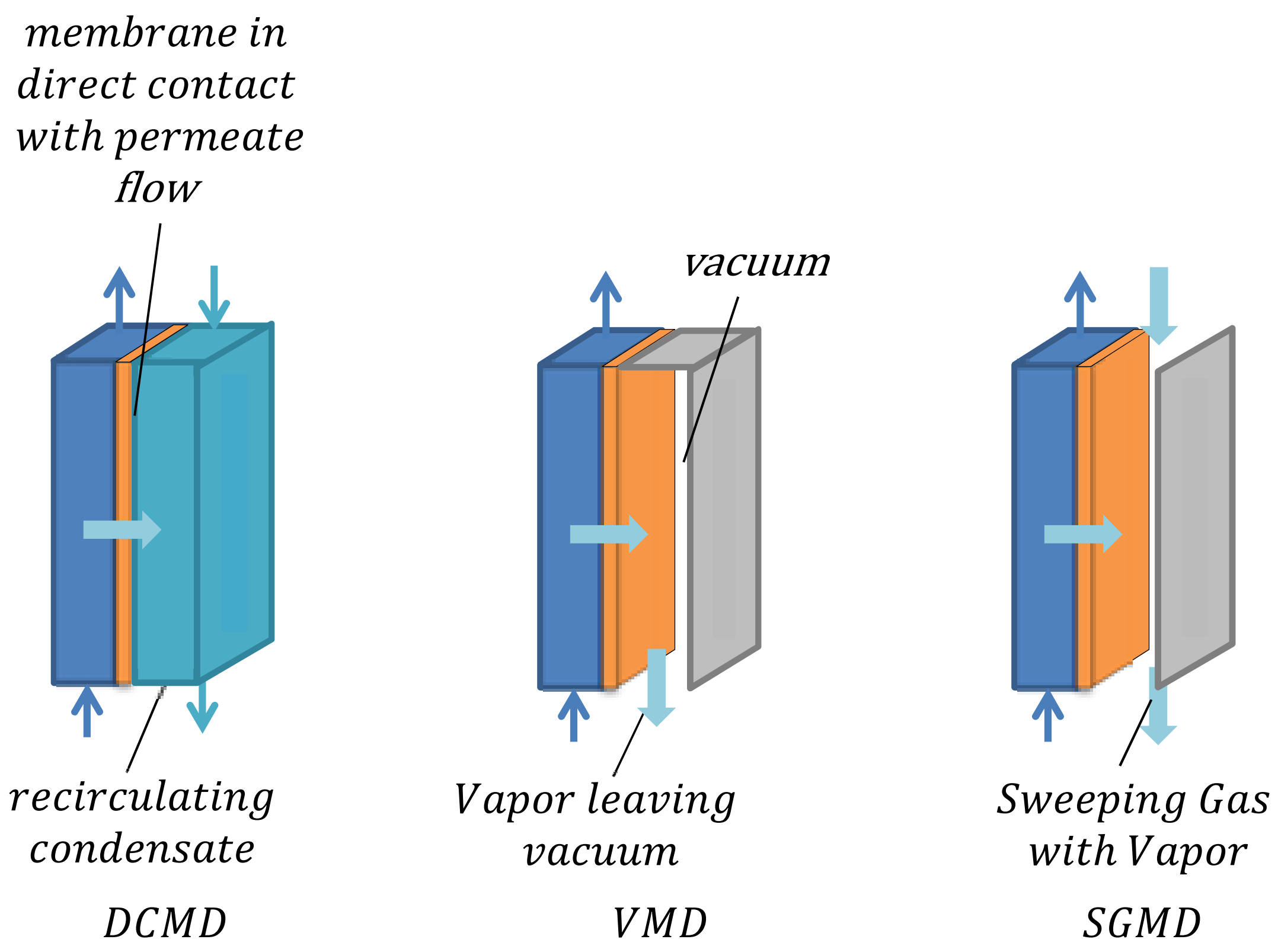

**Figure 1.5.** Configurations of MD that have been studied in the literature, with 2 channels of fluid

The first three types, AGMD, PGMD, and CGMD all share several similarities: 1. a condensing plate is cooling by cooling water, which is actually the incoming feed and 2. the condensate leaves the system as liquid water. What differs here is the content of the gap between the membrane and condensing plate, which will be referred to simply as "the gap." In AGMD, the gap is filled with air, in PGMD the gap is filled with permeate (like a version of AGMD that flooded), and CGMD resembles PGMD, but with a highly conductive structure within the gap that increases the gap's average thermal conductivity. These configurations have different trade-offs: AGMD has a conductivity resistance of the air gap, which helps prevent heat losses that don't contribute to desalination. However, this gap is also a mass transfer resistance, which impairs condensate production and thus overall system performance. PGMD and its enhancement, CGMD, have higher flux, but also higher thermal losses, because they lack a gap. The overall thermal and cost performance depends on system properties like membrane area:

AGMD will be more thermally efficient if very large membrane area is available, but will have higher capital costs relative to distillate production because of its lower flow rates.

Within these configurations, the condensation in the gap can also differ, as seen in figure 3. Past studies with AGMD have assumed film condensation, and innovations in this work have shown that jumping droplet and dropwise condensation may occur as well when superhydrophobic and hydrophobic surfaces are used. Air gap may turn into permeate gap (flooded) when condensate rates become too high for the device design. The tilt angle for AGMD also plays a role in the condensation in the gap as seen in figure 1.5, which this work also addresses.

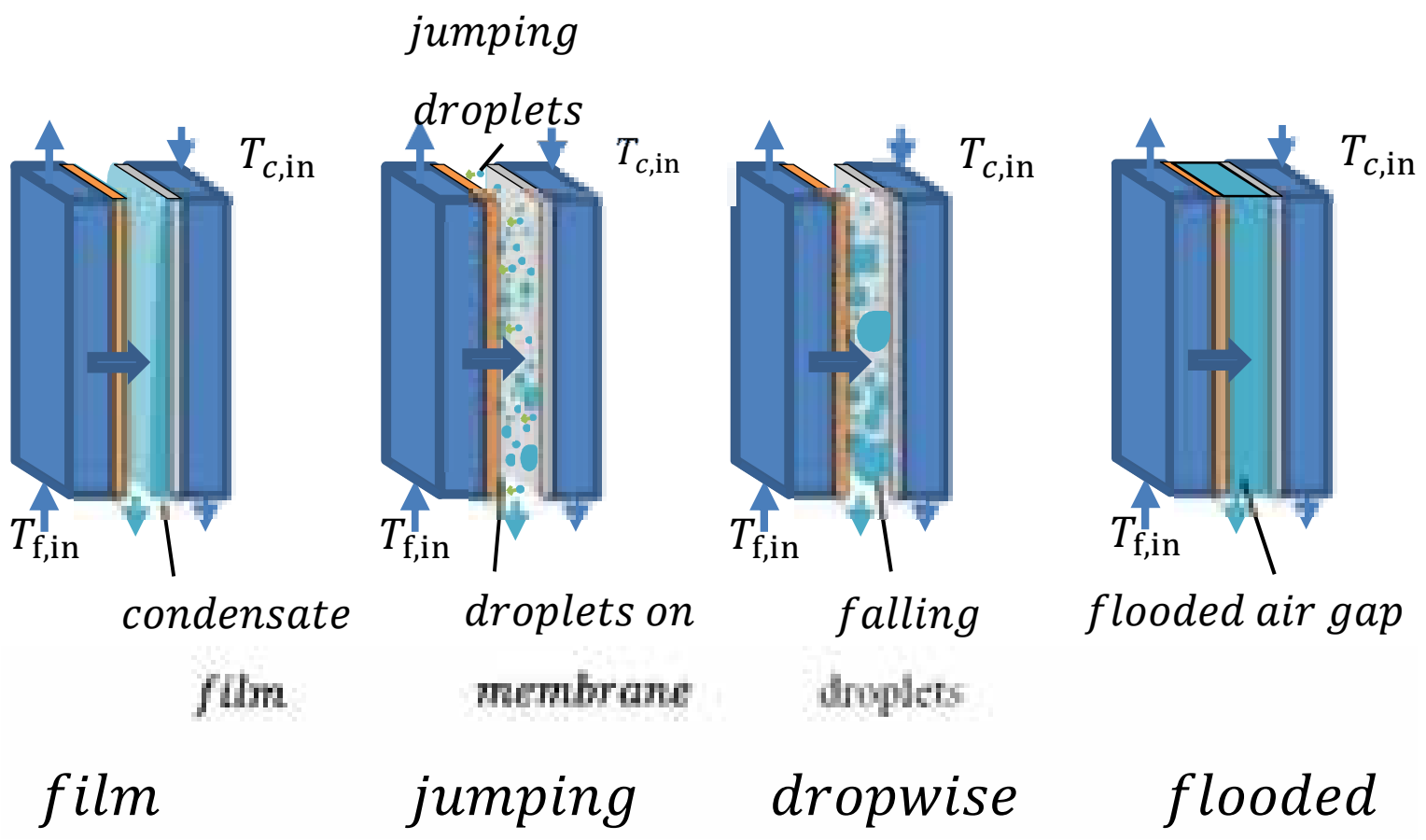

**Figure 1.6.** Diagram of condensation regimes which may occur in AGMD.

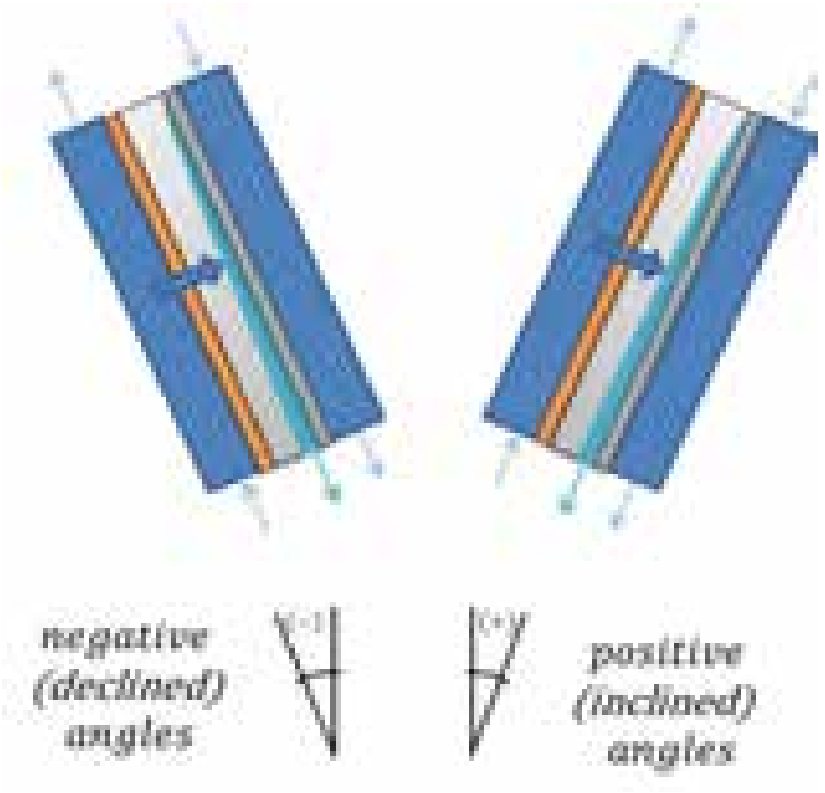

**Figure 1.7.** Tilting of AGMD flat plate module

The three other types of MD; DCMD, VMD, and SGMD, lack a condensing plate and preheating cooling channel [14]. The first, DCMD, circulates the distillate continuously at a similar flow rate of the feed, thus thermally acting much like a matched counter-current heat exchanger. This design, which is relatively efficient, is the only real option for hollow-fiber membranes, and is among the most common types of MD. In Vacuum MD (VMD), the vapor is continuously sucked out of the system and condensed outside the module. While the vacuum pumping adds capital costs and energy losses, this design allows MD to operate much like other thermal systems, such as desalination by multistage flash (MSF). Finally, SGMD, or sweeping gas MD, circulates a gas that continuously removes the water vapor. It is perhaps the least efficient and least-used configuration studied here.

In addition to different heat transfer designs for the MD module, the construction technique may also vary. Four types of MD modules have been used in the literature, as seen in figure 5 [15].

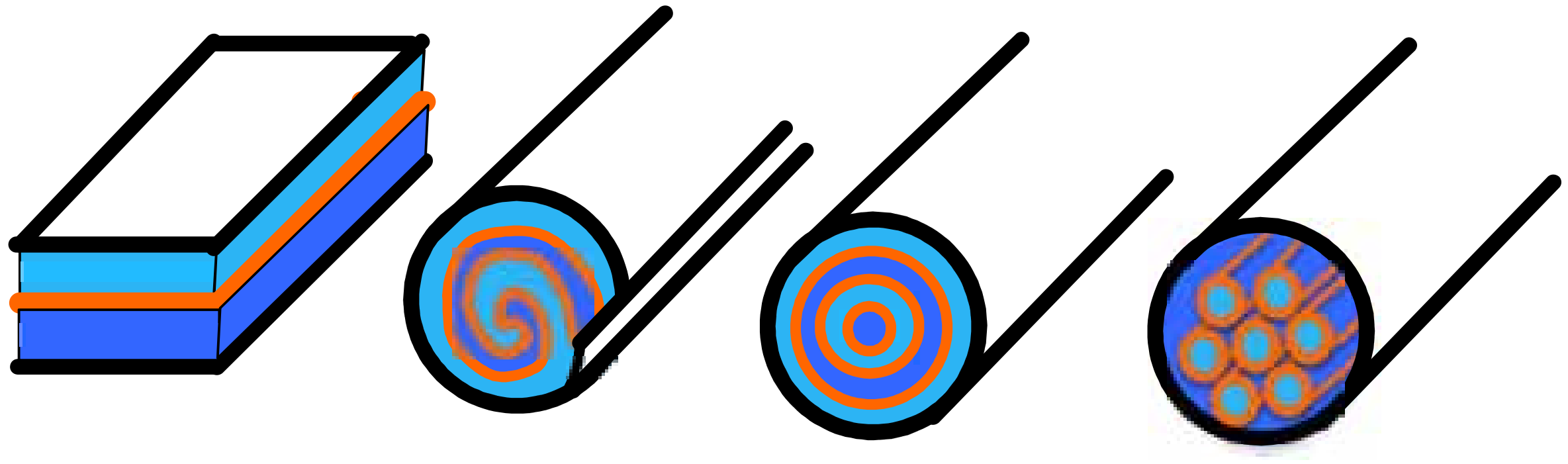

**Figure 1.8.** Types of MD Modules that have been studied in the literature

Flat plate systems are usually composed of sheets sandwiched together. Because they are so simple to construct, they are used very frequently in bench-scale systems, as well as in industry. Spiral wound functions very much like flat plate in terms of gap sizes, heat transfer, and modeling, and have the added benefit of reducing the amount of metal condenser surface and are more compact. Because of these benefits, other membrane technologies such as reverse osmosis very frequently use spiral wound membranes. However, because MD is a relatively immature technology, this potentially more cost-effective to mass produce design is rarely seen. MD systems can also be conical, with a series of tubes alternating between feed and condensate. In practice, these systems are still largely theoretical. All three of these types are similar enough in gap dimensions and flow regime that they can usually be modeled as flat plate systems, neglecting the curvature. The final type, hollow fiber membranes, is fundamentally different and must be modeled separately. These systems use small capillary tubes to transport permeate or feed, which are contained in a larger chamber. Because these tubes are very small (e.g. 1 mm) [16], these flows are laminar, while the larger flat plate systems tend to use turbulent flow to minimize temperature and concentration polarization. Additionally, because the flow channels are fully-filled cylinders, the curvature cannot be neglected, and they must be modeled in polar coordinates. These systems are usually DCMD, although VMD [16], and AGMD systems have been developed as well [17].

## 1.6 Implementation and State-Of-The-Art Membrane Distillation

Membrane distillation is a relatively new thermal desalination technology. While the ideas and first patents were in the late 1960's, suitable membranes weren't created for MD until the late 1980's, which were created for biological and other purposes [18]. Research took off in the late 1990's and early 2000s, and now MD has become fairly well understood. Meanwhile, companies began using MD [19]. The first applications were in creating ultra-pure water for computer manufacturing, as thermal technologies produce superior distillate quality to the market-dominating technology, reverse osmosis. Applications for juice concentration and other food uses soon followed. More recently, companies began using MD for seawater desalination. So far, they have found a niche in small to medium scale desalination applications, especially where waste heat is available [10].

At present, the primary competitors are MEMSTILL and MEMSYS. MEMSTILL relies on a modified AGMD process, while MEMSYS uses a multi-effect vacuum membrane distillation process. MEMSTILL is a venture by Keppel Seghers, and its projects include a waste-heat powered system on Jurong Island, Singapore with a capacity of 100 $m^3$/day. MEMSYS, meanwhile, is larger and has had many more installations. A particularly sustainable and clever application of MEMSYS has been on "energy positive" desalination in the Maldives and in Singapore, where the desalination systems recover the heat from air-cooled engines that produce electricity for the grid (e.g. diesel generators). Since water has superior heat transfer than air, using these desalination systems can lower the rejection temperature of these air-cooled engines, making them actually produce more electricity while also producing water.

MEMSYS systems in industry reach a GOR (thermal efficiency ratio) of about 4 [13], while MD systems in the literature have reached a GOR as high as 8.7[20]. For comparison, the most common thermal desalination technology, MSF, typically reaches a GOR of 7, and the second competing thermal desalination technology, MED, can reach a GOR approaching 10. However, MSF and MED have inferior GOR at smaller sizes, e.g. $<500$ $m^3$/day [21].

## 1.7 Numerical Modeling of Membrane Distillation

Mathematical modeling of the heat and mass transfer within MD systems is crucial for any study on MD energy efficiency or fouling. Modeling prior to experiment design allows for optimizing the components of the MD system, for purposes such as maximizing efficiency, condensate flux per unit area, or having well characterized conditions. Furthermore, not all characteristics of an MD system are readily measurable, such as the temperature and concentration at the membrane surface. To infer this information, modeling must be combined with measurements that are feasible.

To model the MD system, heat and mass transfer, fluids, and salinity-related equations were inputted into a simultaneous equation solver. The software used for this was EES (Engineering Equation Solver), expanding on a model originally developed by Dr. Edward K Summers [22]. This model uses the finite difference method to solve for several hundred successive cells of an MD system. One computational cell is shown in figure Figure 1.9.

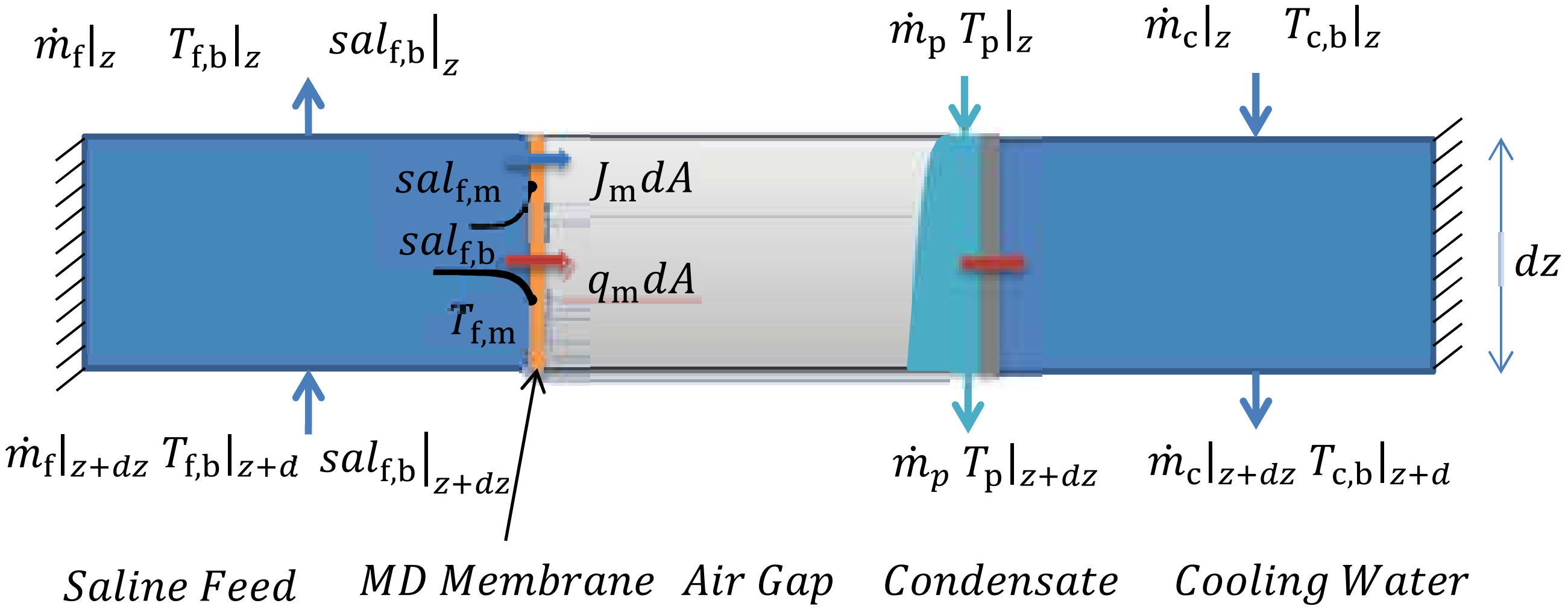

**Figure 1.9.** AGMD Computational Cell

This modeling was modified for AGMD, PGMD, CGMD, and the other systems modeled. As follows is a part by part explanation, which discusses the gap as an air gap, since most of the studies used an air gap.

### 1.7.1 Feed Modeling

Within the feed channel, hot saline feed gradually loses water mass as distillate evaporates and passes through the membrane. Meanwhile, convection with the colder membrane cools the bulk fluid. The loss of water near the membrane itself causes the salt concentration near the membrane to increase, an effect known as concentration polarization. The local reduction of temperature near the membrane surface is called temperature polarization. These effects reduce the driving force in MD: the vapor pressure difference across the membrane.

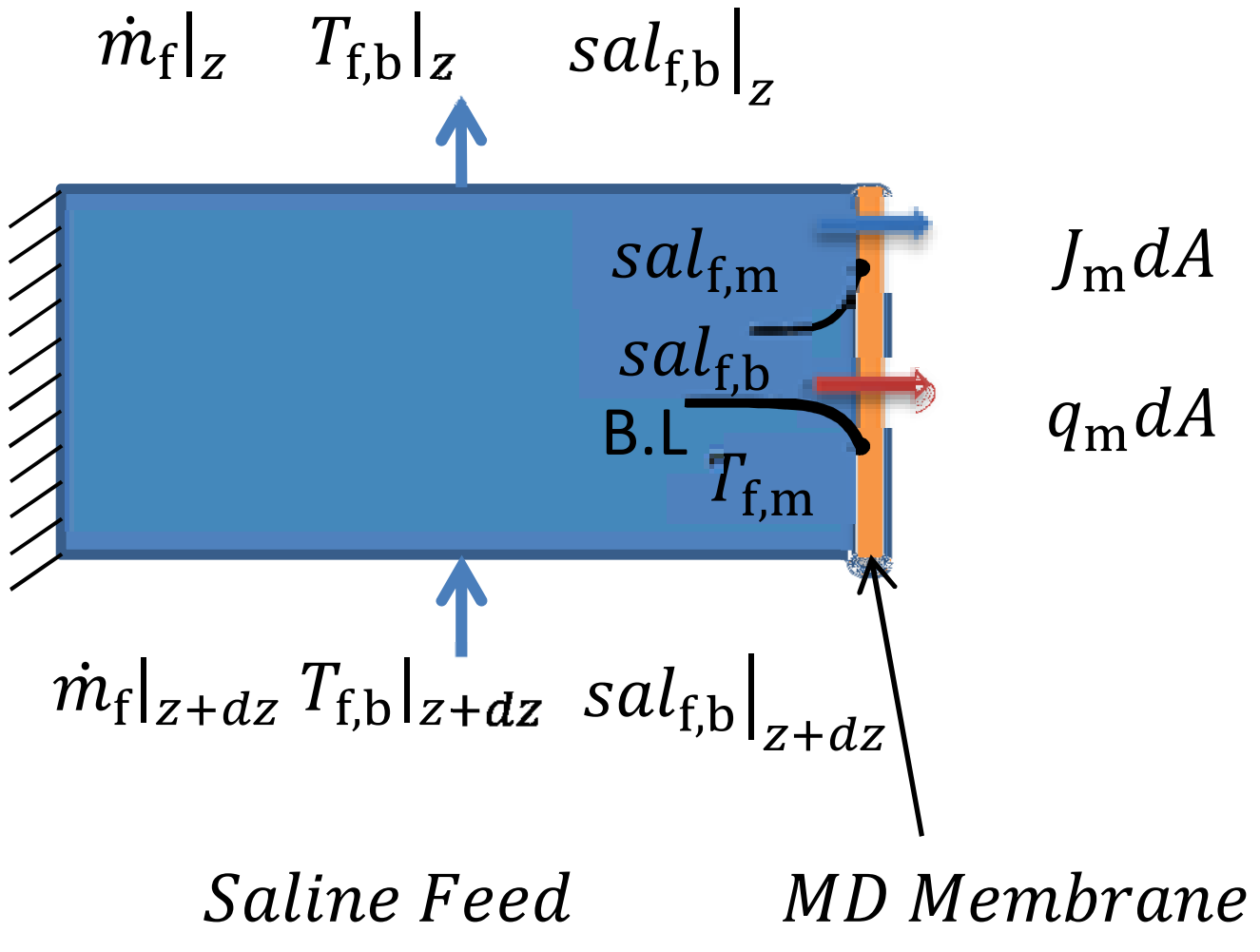

**Figure 1.10.** Numerical Modeling of the Feed and Membrane

The flux J of water vapor through an MD membrane is reliably approximated as linearly proportional to the vapor pressure difference across the membrane [23]. It is given by the following equation:

$$J = B \times (P_{vap,f,m} - P_{vap,a,m}) \tag{1.1}$$

where $P_{f,\ m,\ i}$ and $P_p$ are the vapor pressure of water on the feed and permeate side of the membrane. The parameter B, membrane permeability, is a membrane property, and is enhanced by more porous and thinner membranes. Improving it is often a focus of designing better MD membranes. The energy leaving the feed channel is a sum of the associated enthalpy of evaporation of exiting vapor, conduction losses, and the net losses from the difference in feed flow rates, as follows:

$$q_{out,i} = J_i \cdot (h_{fg,f,i}) + q_{m,i} - J_i \cdot (h_{f,b,i} - h_{f,m,i}) \tag{1.2}$$

Convection and mass transfer, which can be modeled via the mass transfer and heat transfer analogy, are simply given by turbulent channel flow as follows:

$$Nu = \frac{(f/8) \cdot (Re - 1000) \cdot Pr}{1 + 12.7 \cdot (f/8)^{1/2} \cdot (Pr^{2/3} - 1)} \tag{1.3}$$

For mass transfer, the heat and mass transfer analogy may be used, with the Sherwood number Sh being analagous to the Nusselt number Nu, and the Sc being analogous to the Prandtl number Pr.

$$Sc_f = \frac{\mu_f}{(\rho_{feed} \cdot D_{s,w})} \tag{1.4}$$

where $\mu_f$ is the dynamic viscosity, $\rho_{feed}$ is the feed density, and $D_{s,w}$ is the diffusivity of salt in the water. The bulk salt concentration increases as a simple ratio of the change in bulk mass, since only water leaves the system:

$$x_{f,b,i} = S_{in} \cdot \left(\frac{\dot{m}_{f,i}}{\dot{m}_{f,in}/n_{sheets}}\right) \tag{1.5}$$

where $S$ is the Salinity, $\dot{m}$ is the mass flow rate, $x_{f,b,i}$ is the mole fraction of salt in the feed bulk in cell $i$, and $n_{sheets}$ is the number of computational cells used in the numerical model. Concentration polarization near the membrane is an exponential function, and related to the competition between fluid movement toward the membrane (at flow rate J of the vapor exiting) and diffusion of salts back into the bulk.

$$x_{f,m,i} = x_{f,b,i} \cdot \exp\left(\frac{J_i}{k_{mass} \cdot \rho_{feed}}\right) \tag{1.6}$$

The vapor pressure on the feed side of the membrane is critical, as it is part of the previously mentioned distillate flux equation. It is a function of the saturation pressure of water at the membrane-feed interface, and a slight reduction due to the concentration of solutes in solution. It is given as follows:

$$P_{f,m,i} = \mathrm{P_{sat}}\,(Water,\ \mathrm{T} = T_{f,m,i}) \cdot \left(1 - \left(\frac{\frac{x_{f,m,i}}{MW_{solute}}}{\left(\frac{x_{f,m,i}}{MW_{solute}}\right) + \left(\frac{1000\ [\mathrm{g/kg}] - x_{f,m,i}}{MW_{water}}\right)}\right)\right) \tag{1.7}$$

where $P_{f,m,i}$ is the vapor pressure on the feed side of the membrane, $P_{sat}$ is the vapor pressure of saturated water at the temperature of the feed side of the membrane, and $MW$ is the molecular weight. The thermal conductivity of the membrane is important, since heat conduction within the membrane reduces the temperature and thus the vapor difference across it, reducing distillate flux. It is given as follows, which is a simple relation between the conductivity of air, the conductivity of the membrane material, the membrane thickness, and the porosity, which gives a ratio between them.

$$K_{cond} = \frac{k_m \cdot (1-\xi) + k_{air} \cdot \xi}{\delta_m} \tag{1.8}$$

where $K_{cond}$ is the effective thermal conductivity of the membrane, $k_m$ is the thermal conductivity of the membrane material, $\xi$ is the membrane porosity, $k_{air}$ is the thermal conductivity of air, and $\delta_m$ is the membrane thickness.

### 1.7.2 Air Gap Modeling

The heat transferred to the cold side is a sum of the enthalpy of evaporation and flow rate of distillate, and heat losses through the membrane itself.

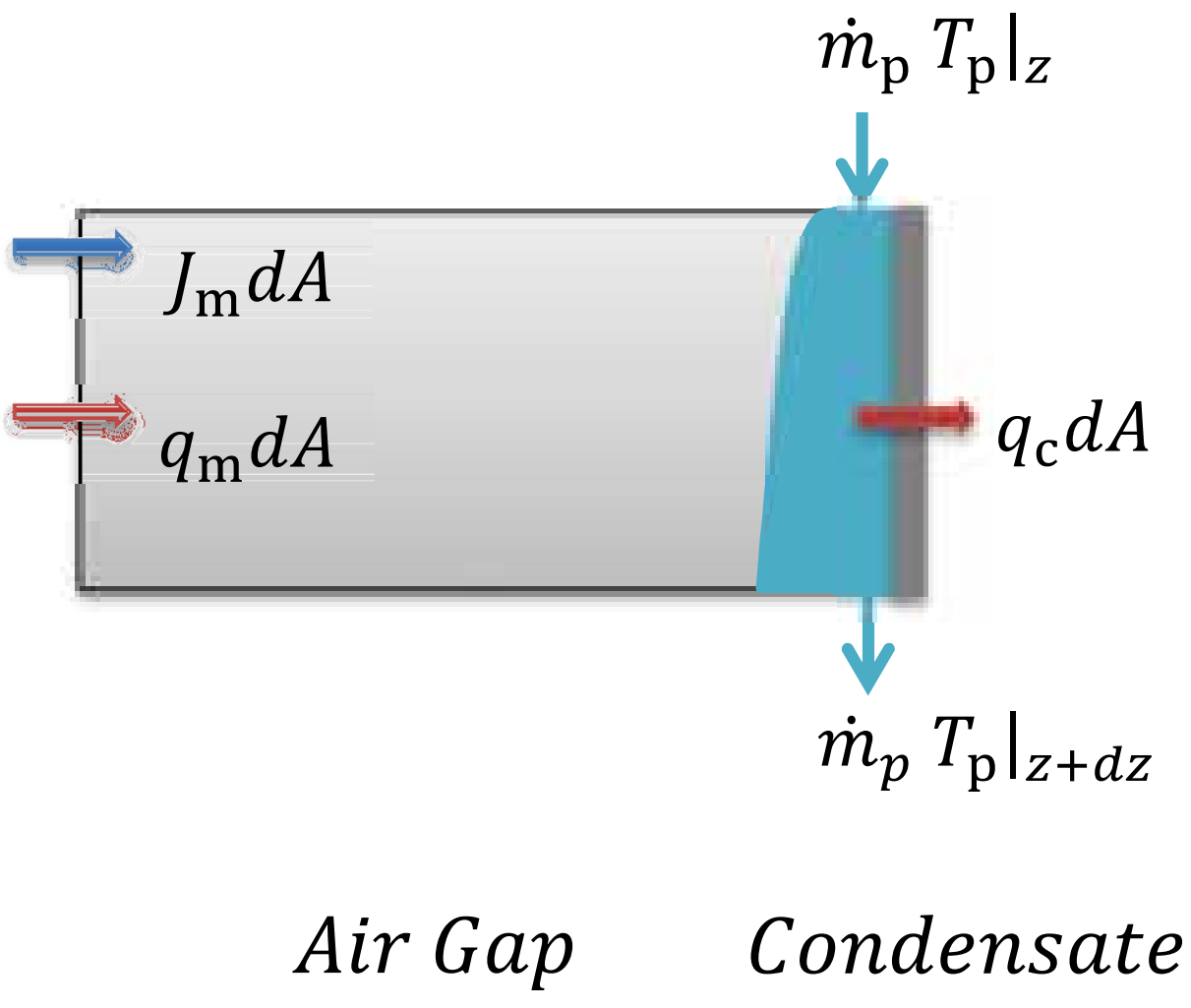

**Figure 1.11.** Numerical modeling of the air gap and associated condensation.

The heat absorbed by the cold channel $q_{c,i}$ is given by:

$$q_{c,i} = J_i \cdot h_{fg,c,i} + q_{m,i} \tag{1.9}$$

In standard AGMD with laminar film condensation, the film itself creates a conduction resistance, as given as follows. It is desirable to minimize the thickness of the film.

The film condensation resistance [24] is given by

$$q_{\mathrm{c,i}} = \frac{k_{\mathrm{film,i}}}{\delta_{\mathrm{i}}} \cdot \left(T_{\mathrm{i,i}} - T_{\mathrm{wall,i}}\right) \tag{1.10}$$

where $q_{\mathrm{c}}$ is the heat flux, $k_{\mathrm{film}}$ is the conductivity of the condensate film, $\delta$ is the local condensate film thickness, $T_{\mathrm{i}}$ is the local membrane temperature, and $T_{\mathrm{wall}}$ is the local wall temperature [25].

Laminar film condensation on a flat plate governs the condensation in traditional AGMD:

$$J_i \cdot dA = g \cdot \frac{\rho_f - \rho_g}{3 \cdot \nu_{f,i}} \cdot w \cdot \left(\delta_{i+1}^3 - \delta_i^3\right) \tag{1.11}$$

where $\nu_{f,i}$ is the viscosity of the fluid, and *w* us the width of the plate.

The air gap itself creates a significant diffusion resistance, as water vapor must diffuse through air to reach the condensing plate. This mass transfer resistance is the major drawback of AGMD. The diffusion of air through the gap is given by [26]:

$$\left(\frac{J_{\mathrm{i}}}{M_{\mathrm{H_2O}}}\right) = \frac{c_{\mathrm{a,i}} \cdot D_{wa}}{d_{\mathrm{gap}} - \delta_{\mathrm{i}}} \cdot ln\left(1 + \left(\frac{x_{\mathrm{i,i}} - x_{\mathrm{a,m,i}}}{x_{\mathrm{a,m,i}} - 1}\right)\right) \tag{1.12}$$

where $J_m$ is the flux through the membrane, $M_{H2O}$ is the molecular weight of water, $c_a$ is the local molar concentration of air, $D_{wa}$ is the diffusivity of water in air, $d_{gap}$ is the air gap depth, $\delta$ is the local condensation film thickness, $x_{\mathrm{i}}$ is the concentration of water vapor at the film-air interface, and $x_{a,m}$ is the local water mole fraction at the membrane interface [27]

The temperature between the membrane and condensation film is based on diffusion across the gap, and is given as follows:

$$T_{\mathrm{a,m}} - T_{\mathrm{i}} = \left(\frac{q_{\mathrm{gap}}}{k_{\mathrm{m}}}\right) \frac{\alpha\rho}{J} \left[\exp\left(\frac{J}{\alpha\rho}\left(d_{\mathrm{gap}} - \delta\right)\right) - 1\right] \tag{1.13}$$

where $q_{\mathrm{gap}}$ is the heat transfer across the gap, $k_{\mathrm{gap}}$ is the average thermal conductivity of the gap, $d_{\mathrm{gap}}$ is the air gap width, $\rho$ is the density of the gap mixture, $\alpha$ is the thermal diffusivity of the gap mixture, and $T_{\mathrm{m}}$ is the temperature of the gap side of the membrane[26].

### 1.7.3 COOLING CHANNEL MODELING

The cooling channel is simple, being merely flat plate turbulent internal flow with heating on one side.

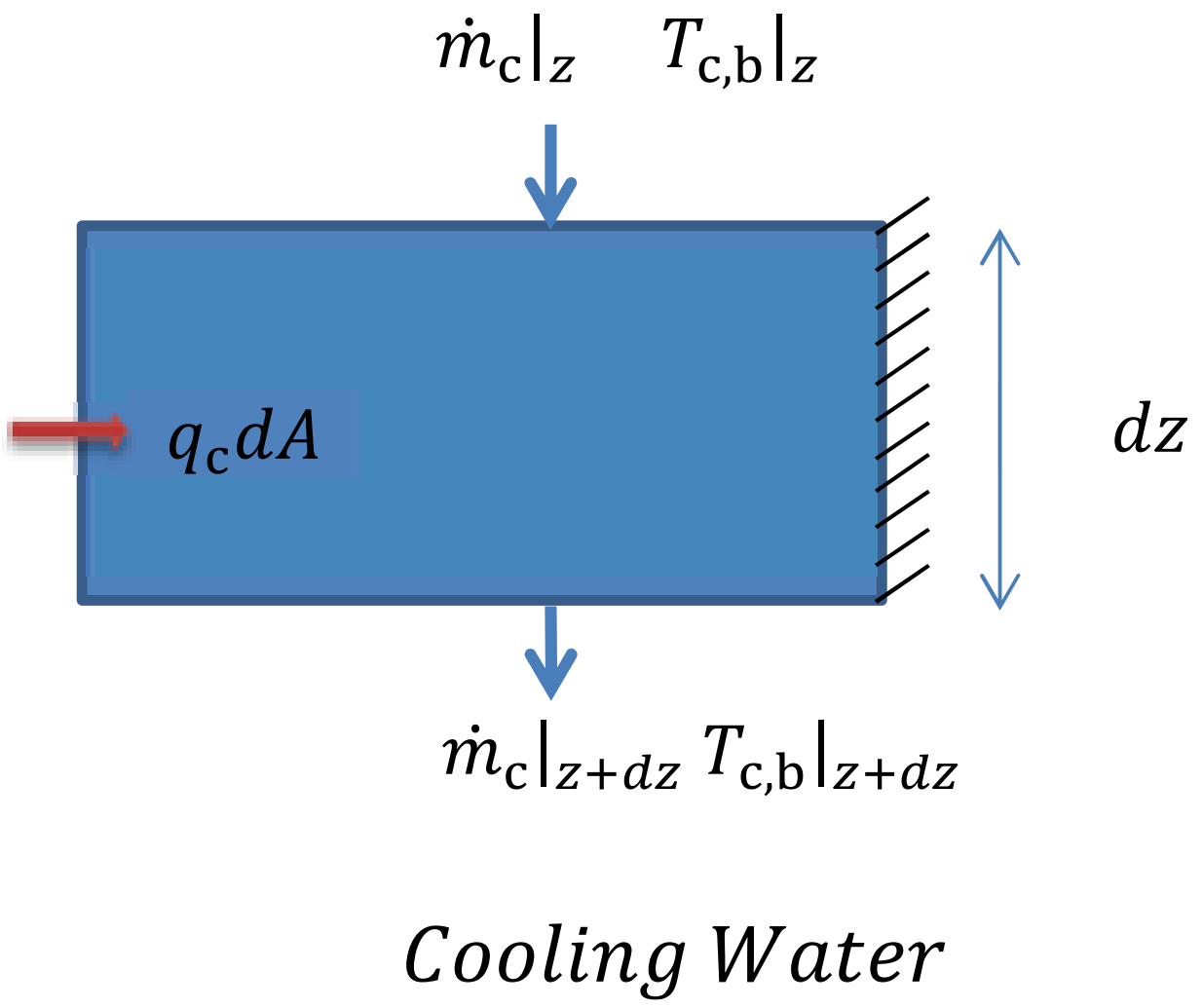

**Figure 1.12.** Numerical modeling of the cooling channel.

To calculate its properties, traditional internal flow with a hydraulic diameter may be used:

$$d_{h,c} = 4 \cdot \frac{w \cdot d_{cond}}{2 \cdot (w + d_{cond})} \tag{1.14}$$

where $d_{cond}$ is the gap depth. Temperature differences occur between the bulk flow and cold wall. The temperature change is a function of the enthalpy differences and flow rate, and the heat conducted to the cold side.

The bulk of the cooling side is at a lower temperature than the adjacent condenser wall because of the boundary layer and convective heat transfer resistance. The temperatures can be calculated as follows:

$$T_{wall,i} = T_{c,b,i} + \left( \frac{q_{c,i} + J_i \cdot (h_{c,i,i} - h_{c,wall,i})}{h_{t,c}} \right) \tag{1.15}$$

where $T_{wall}$ is the temperature of the cold plate on the side of the cooling channel, $T_{c,b}$ is the bulk temperature in the cooling channel, $h$ is the enthalpy, and $h_{t,c}$ is the total enthalpy of the cold stream.

The heat transfer to the cold stream is the enthalpy of the incoming cold stream plus the various heat transfers added through the rest of the system.

$$h_{c,b,i+1} = h_{c,b,i} - (q_{c,i} + J_i \cdot (h_{c,i,i} - h_{c,wall,i})) \cdot \frac{dA}{\dot{m}_{f,1} + (\dot{m}_{c,ex}/1)} \quad (1.16)$$

where $h$ is the enthalpy, the subscript $c$ represents the cold stream, and the subscript $b$ refers to the bulk.

## 1.8 Fouling in Membrane Distillation

The types of scaling and fouling that occur in MD can be divided into four categories: inorganic salt scaling, particulate fouling, biological fouling, and chemical degradation [28]. The appropriate mitigation methods vary dramatically for each of these [29, 30]. The causes also vary strikingly, although particulate fouling can be closely related to the others, as a result of coagulation. It is therefore most practical to analyze each of these four types separately. While MD has been implemented more broadly in the food and semiconductor industry [31, 32], its use for desalination has been minimal. As a result, expectations for the types of scaling in MD are often inferred from other desalination technologies, particularly reverse osmosis. While both technologies involve mass transfer through membranes, significant differences related to fouling exist, notably the significantly higher operating temperatures of MD, as well as the hydrophobic properties of MD membranes, the presence of temperature gradients in MD, and the larger pore sizes in MD [18, 33]. Also, MD lacks the high pressures of RO which are generally believed to aid the formation of compacted cake scales. Studies of fouling on heat exchangers of thermal desalination technologies are also relevant, since these systems have the high temperatures of MD. However, these are not concerned with mass transfer through the foulant layer, only heat transfer, and lack porous surfaces, limiting applicability of these studies.

Scaling and fouling in membrane distillation are found to be pervasive, but design and mitigation methods have proven effective at making MD technology resistant to scaling and fouling. Inorganic scaling risk, the primary focus of academic studies, varies greatly with the salts present. Alkaline salts such as $CaCO_3$, the most common scale by far, have proven to be readily prevented by decreasing feed pH or removed through acidic cleaning, while other scale has proven more tenacious and must be generally be limited by avoiding supersaturation. Biofouling has also been observed in MD, but can be largely mitigated through control of operating conditions. Particulate fouling in MD has proven difficult to remove, but it can largely be prevented by ultra- or microfiltration. Chemical degradation and damage to the membrane has proven to be a concern as well, but can be mitigated by selecting operating conditions that avoid fouling, extreme pH, and certain salts. The choice of membrane material and properties can also help to avoid chemical degradation; PTFE membranes, e.g., may be more susceptible to damage than PVDF. [34]

Fouling tendency had been perceived to be highly variable and perhaps unpredictable, but some consistent patterns are seen. Studies with extreme susceptibility to fouling have almost exclusively been performed with hollow fiber capillary membranes with the feed internal to the capillary tubes. These modules have fouled within hours to days in unsaturated conditions that would not cause fouling in other modules. Likewise, the studies showing high resistance to fouling tended to have highly hydrophobic membranes or coatings, and include hollow fiber studies with permeate in the capillaries [35]. Numerous studies have found substantial reduction in scale from superhydrophobic fluorosilicone coatings, and while the individual papers may question how large a role the coating played in the often complete lack of scale [36], the literature overall proves consistently that these coatings have a dramatic effect. Membrane materials that are somewhat more hydrophobic also show relative resistance to fouling.

Micro, nano, or ultrafiltration has proven effective in stopping particulate scale. Modifying pH in the feed or with cleaning may prevent or remove certain types of fouling very effectively as well. Keeping feed temperature above 60 °C has proven very effective in mitigating biofouling, with some exceptions. Rinsing with a basic solution such as with NaOH may resolve fouling for some substances, including humic acid. Mildly effective fouling prevention methods include boiling for removal of carbonate, ultrasonic cleaning, magnetic water treatment, flocculation, covering the membrane surface with a less porous smaller pore size layer, and for humic acid, oscillating the feed temperature. Antiscalant effectiveness studies in MD have been inconclusive; both strong reduction in scaling and actual decreases in permeate flux have been reported.

System design characteristics also influence fouling. Concentration polarization, closely related to feed Reynolds number and rate of permeate production, is critical in causing fouling, and can be mitigated by increasing the feed flow rate, or mixing technologies such as bubbling [37]. Temperature also remains important, as the most likely foulants have inverse solubility with temperature. Simple computational models were applied by the present authors to illustrate the effect of coupled heat and mass transfer on scaling. Finally, stagnation zones or high residence times in the module may contribute to fouling as well.

This information was synthesized in a review paper for fouling in MD [11]. The review process identified several promising areas for fouling mitigation, and consistent trends observed guided studies to improve understanding. The extreme effectiveness of filtration in preventing inorganic and organic scaling led to studies on the role of bulk nucleation in MD [38]. The impressive ability of superhydrophobic surfaces led to including membrane hydrophobicity in other work. The superhydrophobic insights, combined with successes in the literature of bubbling in air, led to the study on the possibility of intentionally adding air layers to mitigate fouling [39]. The results from these studies, and the literature as well, led to work developing a fouling regime map for inorganic scale in MD [40].

# Chapter 2. Entropy Generation of Desalination Powered by Variable Temperature Waste Heat

## 2.1 Abstract

Powering desalination by waste heat is often proposed to mitigate energy consumption and environmental impact of desalination. However, the literature lacks a thorough performance comparison of different desalination technologies and their components when they are powered by waste heat sources at various temperatures. This work estimates the energy efficiency of several technologies when driven by waste heat sources at temperatures of 50, 70, 90, and 120°C, where applicable. Entropy generation, Second Law efficiency analysis, and numerical modeling are applied. The technologies considered are thermal desalination by multistage flash (MSF), multiple effect distillation (MED), multistage vacuum membrane distillation (MSVMD), humidification-dehumidification (HDH), and organic Rankine cycles (ORC's) paired with mechanical technologies of reverse osmosis (RO) and mechanical vapor compression (MVC). For the model systems considered, the results indicate that RO is the most efficient waste heat-powered technology, followed by MED. Performances among MSF, MSVMD, and MVC were similar but the relative performance varied with waste heat temperature or system size. Entropy generation in thermal technologies increases at lower waste heat temperatures largely in the feed or brine portions of the various heat exchangers used. This increase occurs because lower top temperatures reduce recovery ratios, which increases the relative flow rate of feed to product water and also increases the temperature differences within heat exchangers. HDH (without thermodynamic balancing) showed a reverse trend of efficiency versus top temperature, making it competitive with other thermal technologies at low temperature. However, for the mechanical technologies, the energy efficiency only varies with temperature because of the significant losses from the ORC.  Karan Mistry, Hyung Won Chung, Kishor Nayar, and Professor John Lienhard V contributed to this work [12].

## 2.2 Introduction

Demand for water has been growing steadily due to growing population, industrialization, and consumer usage [42]. Additionally, supply to water is becoming increasingly scarce as climate change alters water availability [2]. In order to meet the demand for water, use of both thermal and electrical desalination technologies are increasing [43]. One of the greatest obstacles to more wide-spread usage of desalination is the large energy consumption associated with it [42]. Waste heat is often proposed as an inexpensive means of providing energy for desalination [12]; however, the discussion of waste heat pervasively lacks a practical assessment of energy efficiency and capital cost. Different sources of waste heat may include warm discharge streams from power plants, data centers, oil and gas refining, metal production, and other industrial processes [44].

The key variable in analyzing the performance of systems driven by waste heat is that low quality waste heat may be available at various temperatures, which substantially affects the exergetic input of the heat source [45]. Therefore, to understand the relative performance of technologies powered by waste heat, this work provides a comprehensive modeling analysis for a broad range of waste heat temperatures using shared approximations over a range of systems. The model values for parts of these systems were chosen from representative industrial installations. The modeled technologies include multistage flash (MSF) [46], multiple effect distillation (MED) [47], multistage vacuum membrane distillation (MSVMD) [41], humidification dehumidification (HDH) [48], and organic Rankine cycles [49] paired with reverse osmosis (RO) [50] and mechanical vapor compression (MVC) [51]. These technologies, where applicable, are examined at 50, 70, 90, and 110 °C.

Systems are compared to one another through the Second Law efficiency, which compares the least exergy required to produce a kilogram of fresh water to the actual exergy input required by the system at a given temperature [52]. To illuminate how heat source temperatures and system design affect the relative performance of each technology, the key sources of irreversibilities in key system components are analyzed at different waste heat temperatures using analytical models for several different technologies. The irreversiblities analyzed included entropy generation in throttling to produce vapor (flashing), fluid expansion without phase change, pumping, compression, heat transfer, and mixing of streams at different temperatures (thermal disequilibrium) and different salinities (chemical disequilibrium) [53]. These results are used to compare the relative performance of technologies at different heat source temperatures, and to analyze which components are responsible for the major inefficiencies at different temperatures.

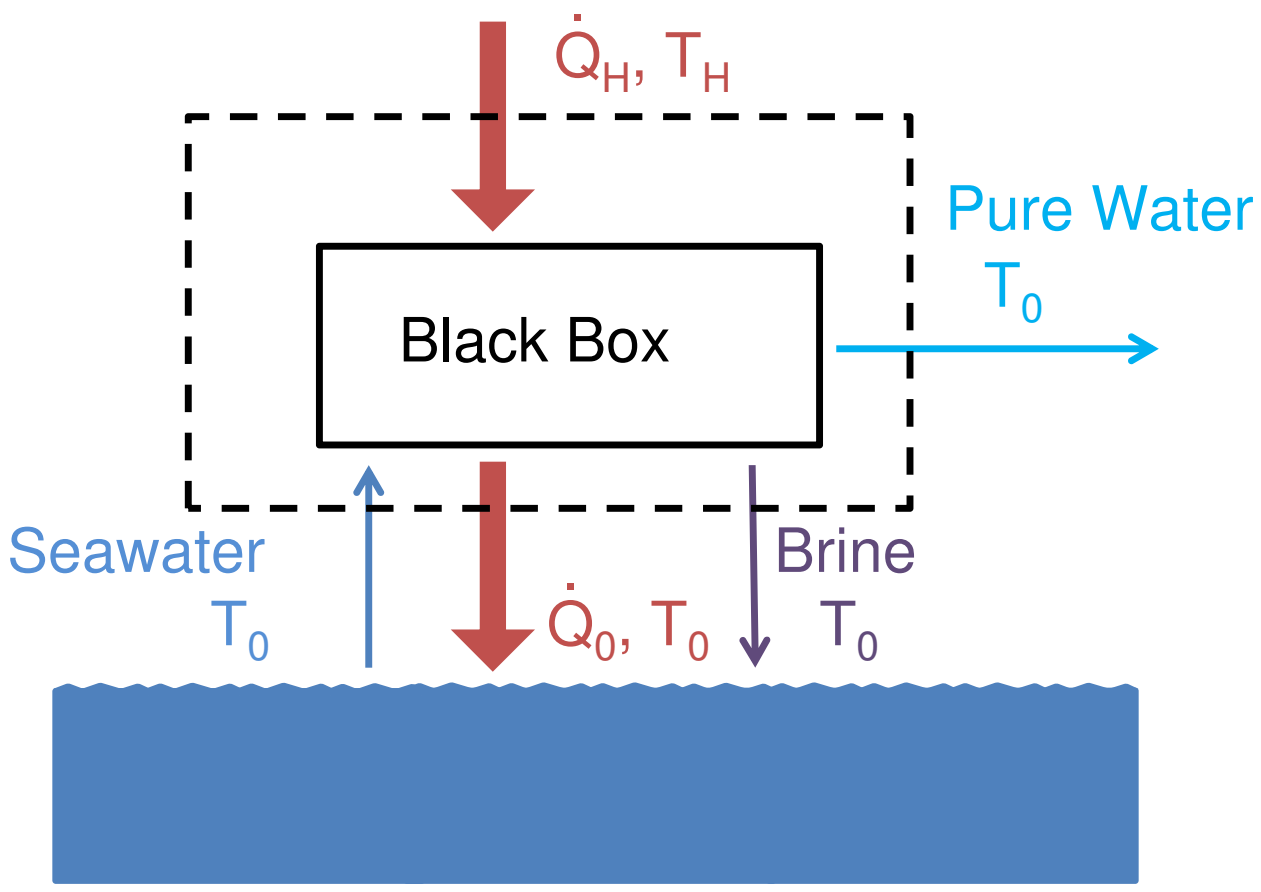

**Figure 2.1.** A control volume is selected around a black-box, waste heat driven desalination system such that the seawater (sw), product (p) and brine (b) streams are all at the environmental temperature, $T_0$, and pressure $p_0$.

## 2.3 Derivation of Performance Parameters for Desalination

A control volume analysis is used to derive the key performance parameters for waste heat driven desalination systems. Figure 2.1 shows a schematic diagram of a simplified black-box desalination system with heat transfer occurring from a waste heat source at temperature, $T_H$, and to the environment at temperature, $T_0$. The control volume is selected sufficiently far from the system boundary such that all material streams (seawater, product water, and brine) are in both thermal and mechanical equilibrium with the environment (restricted dead state, or RDS).

The First and Second Law of Thermodynamics applied to the control volume in Fig. 2.1

are as follows:

$$\dot{Q}_H - \dot{Q}_0 = (\dot{m}h)_{\mathrm{p}} + (\dot{m}h)_{\mathrm{b}} - (\dot{m}h)_{\mathrm{sw}} \tag{2.1}$$

$$\frac{\dot{Q}_H}{T_H} - \frac{\dot{Q}_0}{T_0} = (\dot{m}s)_{\mathrm{p}} + (\dot{m}s)_{\mathrm{b}} - (\dot{m}s)_{\mathrm{sw}} + \dot{S}_{\mathrm{gen}} \tag{2.2}$$

where $\dot{Q}_H$, $\dot{Q}_0$, $h$, $s$, and $\dot{m}$ are the heat transfer terms from the waste heat source, heat transfer to the environment, enthalpy, entropy and mass flow rate, respectively. Equations (2.1) and (2.2) can be combined by multiplying the Second Law by $T_0$ and subtracting from the First Law:

$$\left(1 - \frac{T_0}{T_H}\right)\frac{\dot{Q}_H}{\dot{m}_p} = (g_p - g_b) - \frac{1}{r}(g_{\mathrm{sw}} - g_b) + T_0\frac{\dot{S}_{\mathrm{gen}}}{\dot{m}_p} \equiv \frac{\dot{W}_{\mathrm{sep}}}{\dot{m}_p} \tag{2.3}$$

where $g$ is the Gibbs Free energy, defined as $h - Ts$, $r$ is the recovery ratio, defined as $\dot{m}_p/\dot{m}_{\mathrm{sw}}$, and $\dot{W}_{\mathrm{sep}}$ is referred to as the work of separation [52]. In the limit of reversible separation (*i.e.*, $\dot{S}_{\mathrm{gen}} = 0$), the work of separation reduces to the least work of separation, $\dot{W}_{\mathrm{least}} = \dot{W}^{\mathrm{rev}}_{\mathrm{sep}}$. Further, in the limit of infinitesimal recovery (*i.e.*, the recovery ratio approaches zero), the least work reduces to the minimum least work, $\dot{W}^{\mathrm{min}}_{\mathrm{least}}$ [52, 54]. Using seawater properties and assuming an inlet salinity of 35 g/kg, $T = 25\,^{\circ}\mathrm{C}$, the minimum least work is 2.71 kJ/kg [55].

For waste heat driven systems, such as those considered in this study, the least heat of separation is of more interest [52, 54]:

$$\frac{\dot{Q}_{\mathrm{least}}}{\dot{m}_p} = \frac{(g_p - g_b) - \frac{1}{r}(g_{\mathrm{sw}} - g_b)}{\left(1 - \frac{T_0}{T_H}\right)} \tag{2.4}$$

The least heat of separation is a function of both source temperature and recovery ratio, as shown in Fig. 2.2 [52, 54]. In the limit of infinite source temperature, the least heat approaches the least work.

To compare technologies to one another, a Second Law efficiency ($\eta_{II}$) comparison is performed; $\eta_{II}$ is the percent of an ideal Carnot efficiency that is achieved for each technology. This comparison is a ratio of the minimum least exergy over the actual exergy used, as seen in Eq. (2.5).

$$\eta_{II} = \frac{\dot{\Xi}^{\mathrm{min}}_{\mathrm{least}}}{\dot{\Xi}_H} \tag{2.5}$$

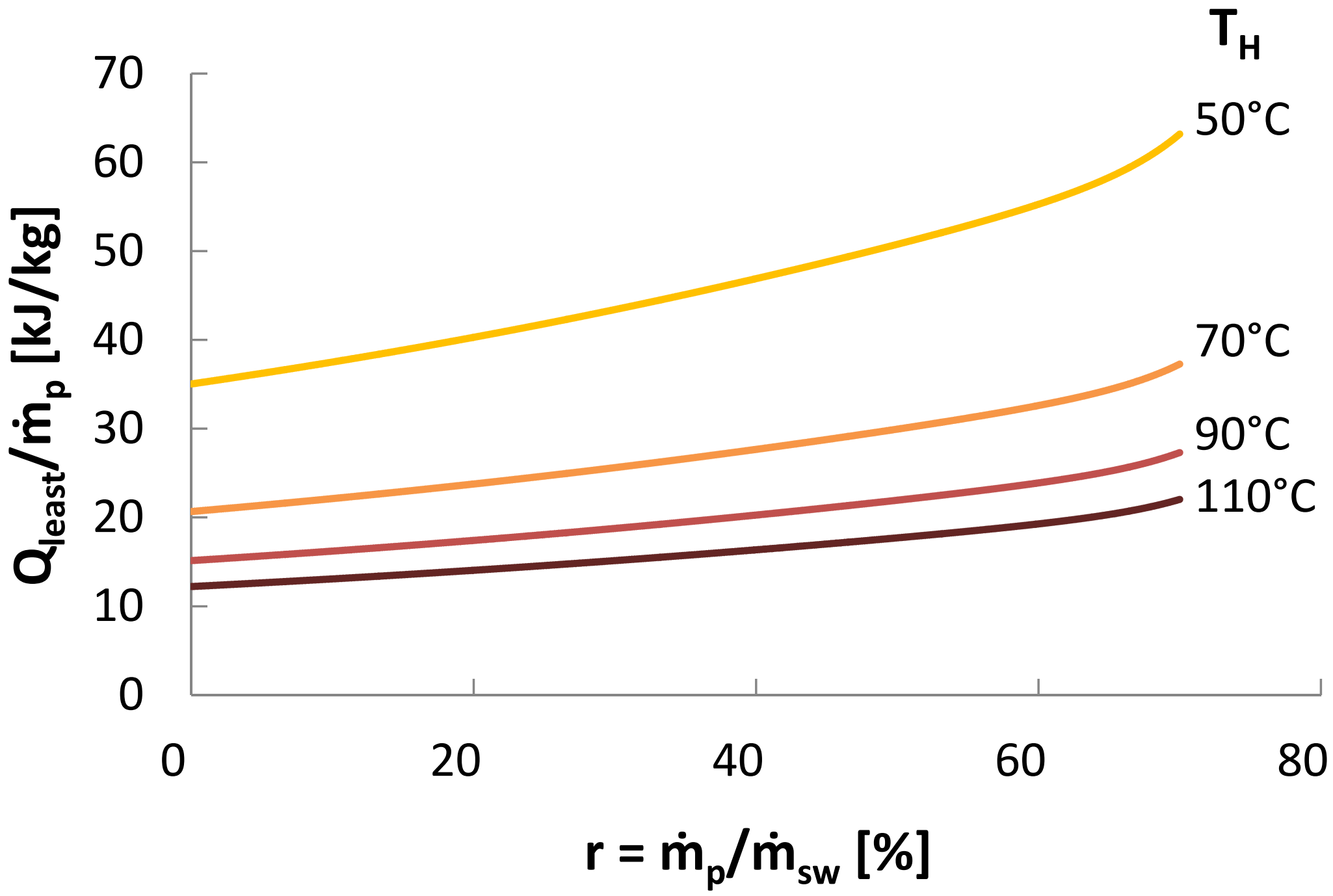

**Figure 2.2.** Least heat of separation, $\dot{Q}_{\text{least}}$ is a function of recovery ratio and source temperature. Results are shown for environmental temperature of 25 °C.

The minimum least heat is defined as the minimum heat necessary to reversibly desalinate the water with near-zero recovery, where the recovery ratio is the mass flow rate of the product water divided by that of the feed. Minimum least heat is equal to minimum least work (exergy change) dvided by the Carnot efficiency at the given temperatures.

Zero recovery refers to negligible concentration of the brine caused by removing an infinitesimal amount of pure water from the feed stream. For the purpose of the process considered here (fresh water production), only the product stream, not brine stream, is the desired output. Consequently, the minimum least heat required to produce a unit of fresh water—the value at zero recovery—is the appropriate benchmark against which to compare any actual process at finite recovery. Further, it may be noted that a reversible, finite recovery process will have a Second Law efficiency below unity by the present definition.

As shown in Eq. (2.5), Second Law efficiency for a heat-driven desalination system is simply the ratio of the minimum least heat of separation to the actual amount of heat required for the separation process, accounting for both finite recovery ratio and irreversible processes.

Substituting Eq. (2.4) into Eq. (2.3) and renaming $\dot{Q}_H$ as $\dot{Q}_{\text{sep}}$ yields:

$$\underbrace{\left(1-\frac{T_0}{T_H}\right)\frac{\dot{Q}_{\text{sep}}}{\dot{m}_p}}_{\dot{W}_{\text{sep}}/\dot{m}_p}=\underbrace{\left(1-\frac{T_0}{T_H}\right)\frac{\dot{Q}_{\text{least}}}{\dot{m}_p}}_{\dot{W}_{\text{least}}/\dot{m}_p}+T_0\frac{\dot{S}_{\text{gen}}}{\dot{m}_p} \tag{2.6}$$

The total entropy generation normalized to the product mass flow rate, defined here as "specific entropy generation" is represented as:

$$\mathcal{S}_{\text{gen}}=\dot{S}_{\text{gen}}/\dot{m}_p \tag{2.7}$$

The specific entropy generation term in Eq. (2.6) lowers the Second Law efficiency of any real system. By analyzing specific entropy generation, one can work to improve and adapt desalination systems to minimize exergetic losses. As specific entropy generation increases, the required heat of separation also increases. Equation (2.5) is plotted as a function of heat of separation for various source temperatures in Fig. 2.3.

Another commonly used efficiency parameter for thermal desalination technologies in the Gained Output Ratio (GOR), the ratio of the enthalpy of evaporation to the actual heat of

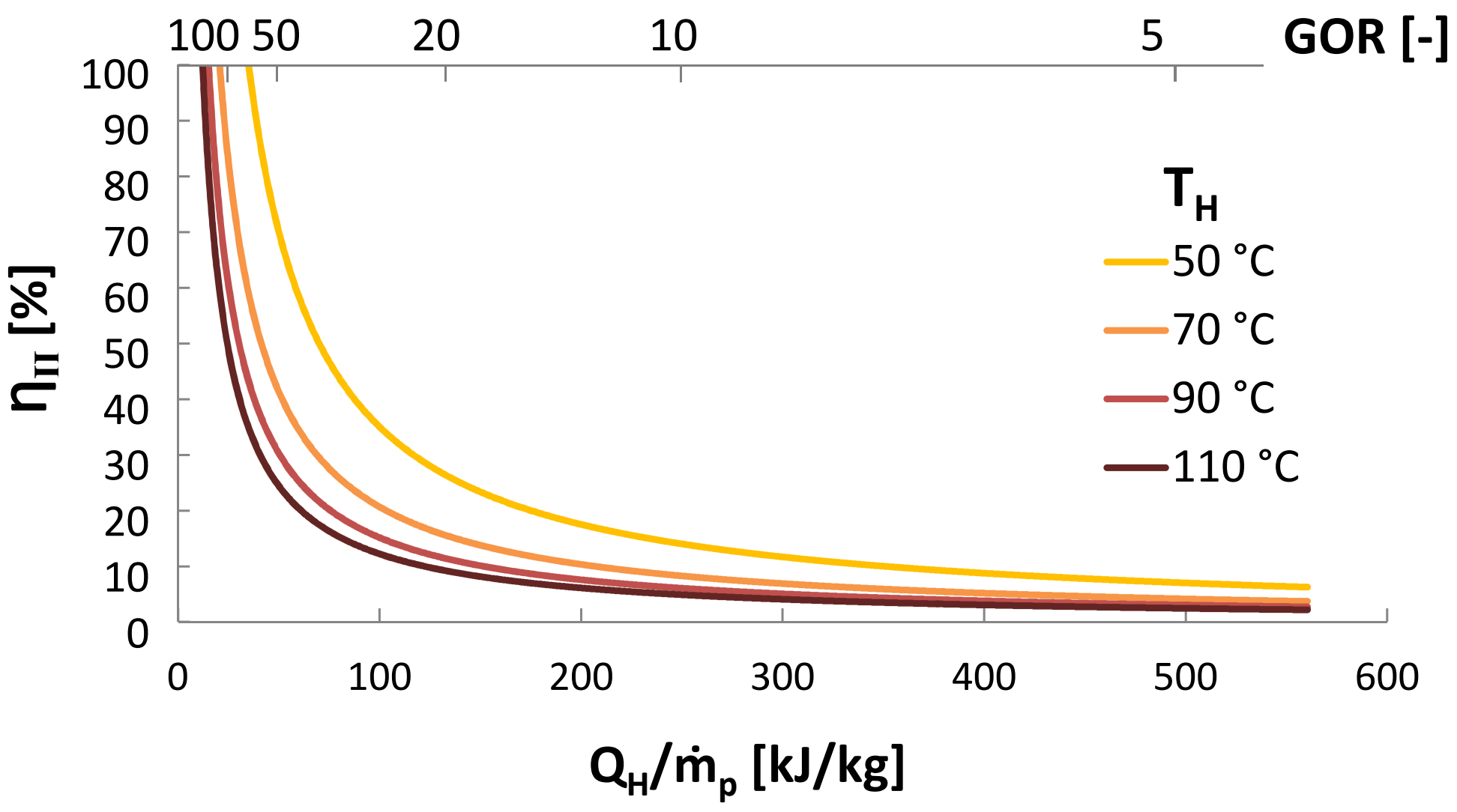

**Figure 2.3.** Second Law Efficiency $\eta_{II}$ for seawater desalination with an environmental temperature at 25 °C.

separation:

$$\text{GOR} \equiv \frac{\dot{m}_p h_{fg}(T_0)}{\dot{Q}_H} \tag{2.8}$$

Effectively, GOR measures how many times the enthalpy of evaporation is reused within a given system. Typical values for GOR range from 3–10, depending on the technology, operating temperatures, and other conditions. The relationship between GOR and Second Law efficiency is not simple, since the maximum GOR attainable depends on available temperatures and other conditions, while $\eta_{II}$ can only vary between 0 and 1.

## 2.4 Entropy Generation Mechanisms

Entropy generated through various physical processes can be calculated by using control volume analysis, where the control volume is selected such that it constrains the entirety of the process. Mistry et al. [52] analyzed the most common processes in desalination systems and a summary of their results is given in Table 2.1.

Temperature disequilibrium refers to the entropy generated by the exiting product and brine streams differing from ambient temperatures. Chemical disequilibrium, a result of salt concentration differences, refers to the entropy generated as the exiting saline brine mixes back in with the source seawater. The other variables and components are standard and well-known from thermodynamics. For a derivation of these equations, see Mistry et al. [52].

**Table 2.1.** Entropy Generation for Different Processes in Desalination, Partly Organized by Largest Total Contribution to the Technologies Modeled [52]

| Entropy Generation | Occurence | Equation |
|---|---|---|
| $S_{\text{gen}}^{\text{HeatExchanger}}$ | unbalanced heat exchangers | $[\dot{m}(s_2 - s_1)]_{\text{stream 1}} + [\dot{m}(s_2 - s_1)]_{\text{stream 2}}$ |
| $\dot{s}_{\text{gen}}^{\text{T gradient}}$ | heat transfer across $\Delta T$ | $\dot{Q}(\frac{1}{T_C} - \frac{1}{T_H})$ |
| $s_{\text{gen}}^{\text{R}ankine}$ | attainable Rankine Cycle losses | $\frac{\dot{W}_{\text{sep}}(1-\eta_{II})}{T_0}$ |
| $\dot{s}_{\text{gen}}^{\text{Q lost to environment}}$ | heat lost to environment | $\dot{m}c \ln \frac{T_{out}}{T_{in}}$ |
| $s_{\text{gen}}^{T \text{ disequilibrium}}$ | $\Delta T$ between output and environment | $c_i \left[\ln\left(\frac{T_0}{T_i}\right) + \frac{T_i}{T_0} - 1\right]$ |
| $s_{\text{gen}}^{\text{flashing}}$ | flashing: evaporation from rapid pressure drop (throttling) | $c \ln \frac{T_2}{T_1} + x\left\{(c_p - c)\ln T_2 - R \ln p_2\right\}$ $+x\left\{\left[s_{\text{ref}}^{\text{IG}} - s_{\text{ref}}^{\text{IF}} - (c_p - c)\ln T_{\text{ref}} + R\ln p_{\text{ref}}\right]\right\}$ |
| $s_{\text{gen}}^{\text{expansion,IF}}$ | reverse osmosis pressure recovery | $c \ln\left[1 + \frac{v}{cT_1}(p_1 - p_2)(1 - \eta_e)\right] \approx \frac{v}{T_1}(p_1 - p_2)(1 - \eta_e)$ |
| $s_{\text{gen}}^{\text{expansion,IG}}$ | turbines | $c_p \ln\left\{1 + \eta_e\left[\left(\frac{p_2}{p_1}\right)^{R/c_p} - 1\right]\right\} - R\ln\frac{p_2}{p_1}$ |
| $s_{\text{gen}}^{\Delta p,\text{IF}}$ | throttling (valves) for fluids | $c \ln\left[1 + \frac{v}{cT_1}(p_1 - p_2)\right] \approx \frac{v}{T_1}(p_1 - p_2)$ |
| $s_{\text{gen}}^{\Delta p,\text{IG}}$ | throttling (valves) for gases | $-R\ln\frac{p_2}{p_1}$ |
| $s_{\text{gen}}^{\text{compression}}$ | compressors, e.g. in MVC | $c_p \ln\left\{1 - \frac{1}{\eta_p}\left[1 - \left(\frac{p_2}{p_1}\right)^{R/c_p}\right]\right\} - R\ln\frac{p_2}{p_1}$ |
| $s_{\text{gen}}^{\text{pumping}}$ | pump efficiency for fluids | $c \ln\left[1 + \frac{v}{cT_1}(p_2 - p_1)\left(\frac{1}{\eta_p} - 1\right)\right] \approx \frac{v}{T_1}(p_2 - p_1)\left(\frac{1}{\eta_p} - 1\right)$ |
| $\dot{s}_{\text{gen}}^{\text{Chem disequilibrium}}$ | salinity difference between environment and output stream | $\frac{1}{T_0}\left[\dot{W}_{\text{least}} - \dot{W}_{\text{least}}^{\text{min}}\right]$ |

## 2.5 Unused Temperature Reduction of Waste Heat Sources

Some technologies have a maximum temperature threshold above which operation is not practical. In such systems, the source temperature may have to be reduced, resulting in significant entropy generation. It is therefore important to understand the thermodynamic losses caused by reducing the source temperature.

This becomes relevant when comparing technologies such as MED, which typically operate at 70 °C or below because of scaling issues, to technologies such as MSF, which may have a top temperature as high as 110–120 °C. To make a comparison where this temperature reduction is necessary for a fixed high temperature source, an efficiency $\eta_{\text{reduced}}$ may be used, which represents a ratio of the remaining available work divided by the original available work possible. This can be described in terms of waste heat temperatures by using $\eta_{\text{Carnot}}$, and assuming that the heat reduction occurs as a temperature gradient somewhere in the system, without losses to the environment.

$$\eta_{\text{reduced}} = \frac{\dot{W}_{\text{use}}}{\dot{W}_{\text{rev}}} = \frac{\dot{Q}_H \eta_{\text{Carnot}}^{\text{use}}}{\dot{Q}_H \eta_{\text{Carnot}}^{\text{rev}}} = \frac{(1 - \frac{T_0}{T_{\text{use}}})}{(1 - \frac{T_0}{T_H})} \tag{2.9}$$

The efficiency is graphed in Fig. 2.4. As expected, the lower the source temperature, the greater the impact on efficiency for a reduction in temperature prior to use. Degrading the heat source prior to use results in pure exergy destruction. The severity of the loss of efficiency resulting from lowering the temperature of the waste heat prior to use illustrates the importance of matching thermal technologies to the available energy sources. These losses can be improved by using a different technology for the higher temperatures that would otherwise be left unused. For example, thermal vapor compression (TVC) is often paired with MED, since TVC can handle the high temperatures MED cannot [56].

An example is reducing the top temperature from 90 °C to 70 °C for use in an MED system, which corresponds with an $\eta_{\text{reduced}}$ of 69%. For larger temperature differences, it is more efficient (and economical) to use another technology for the higher temperature region.

The downward curvature shows that efficiency losses accelerate as the temperature is decreased. This implies that for lower temperature thermal components such as heat ex-changers, it is desirable to focus on reducing entropy generation for the colder components of a system.

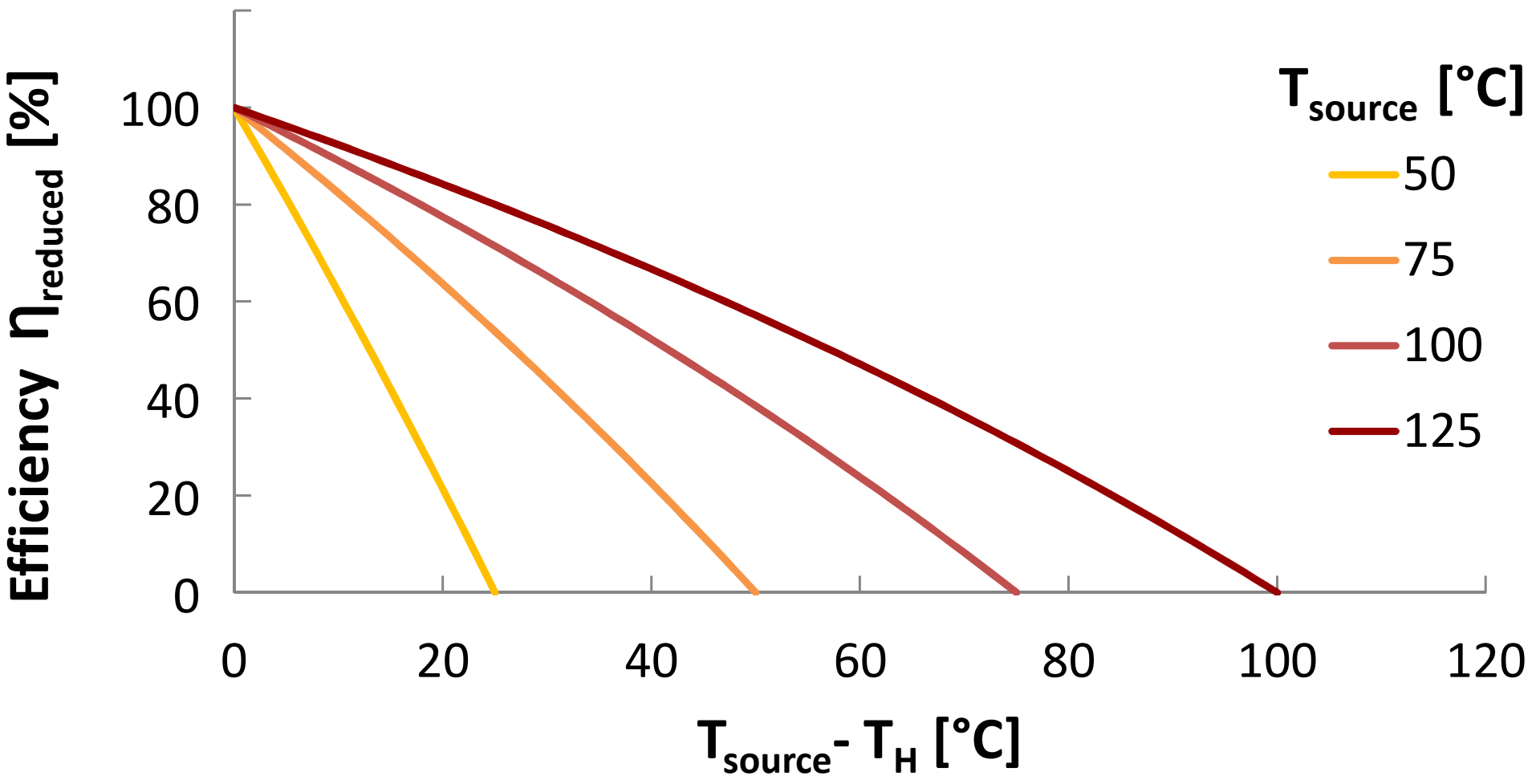

**Figure 2.4.** Efficiency factor $\eta_{\text{reduced}}$ versus temperature reduction $(T_{source} - T_H)$ for various source temperatures. All results shown for environmental temperature at 25 °C.

## 2.6 Entropy Generation Analysis of Seawater Desalination Technologies

Entropy generation analysis for the most common seawater desalination technologies is performed in the following sections. For each technology considered, specific entropy generation is evaluated at a component level. By doing a component level analysis, it is possible to identify the largest sources of inefficiency and determine how to best improve systems.

### 2.6.1 Modeling Approximations and Assumptions

In order to make a fair comparison of the various technologies considered, operating and input conditions must be consistent for all systems. These conditions include temperatures, pressures, and salinities of the input and output streams, as well as temperature pinches and efficiencies within individual components. Table 2.2 summarizes the temperatures and salinities of the input streams used in all modeling in subsequent sections. The temperatures were chosen from 50 °C to 110 °C as this represents the range from zero desalination possible ($T_H = T_0 = 25$ °C) to where the level at which scaling usual becomes prohibitive (110 °C).

**Table 2.2.** Standard input conditions used for desalination system models. Note, $T_{stage}^{last}$ is only applicable to thermal technologies.

| Input and output streams for all systems | | | |
|---|---|---|---|
| $T_{H1}$ | 50 °C | $T_{sw}$ | 25 °C |
| $T_{H2}$ | 70 °C | $T_{stage}^{last}$ | 35 °C |
| $T_{H3}$ | 90 °C | $S_p$ | $0g/kg$ |
| $T_{H4}$ | 110 °C | $S_{sw}$ | $35g/kg$ |

In addition to the standardized input conditions, the following general approximations are made:

1. All processes are modeled as steady state.
2. Heat transfer to the environment is negligible.
3. All streams are considered well mixed and bulk physical properties are used.
4. Heat transfer coefficients are constant within a given heat exchanger.
5. Seawater properties can be calculated using correlations from Sharqawy et al. [55].
6. Product water is pure (zero salinity).
7. In systems with multiple stages, the number of stages was proportionally reduced for lower waste heat temperature.
8. In systems with multiple stages, the recovery in each remaining stage stays roughly the same for lower temperatures.
9. Pumping power may be neglected in thermal systems.
10. Temperature drop across heat exchangers is between 2.5 and 3.3 °C

Approximation 7, assuming stages must be removed to compare to different temperature sources, is crucial to comparing thermal systems driven by waste heat. In this assumption, for lower top temperatures, higher temperature stages are eliminated and the remaining stages are not changed in all ways possible, retaining their temperatures and product production rates. The rational for keeping the number of stages constant for a given temperature difference is the assumption that the components and set-points of existing real-world systems are thermally and economically optimized. The lower stages, depending on the technology, are roughly independent from the higher temperature stages, so removing higher temperature stages should have minimal effect on the subsequent stages. This approximation neglects re-optimization for the change in brine salinity, which is acceptable since the brine salinity is minorally affected for thermal desalination systems. The approximation also neglects the different flow rate of the product stream, which is acceptable, because the entropy generation in the stages and product stream components varies little between different source temperatures.

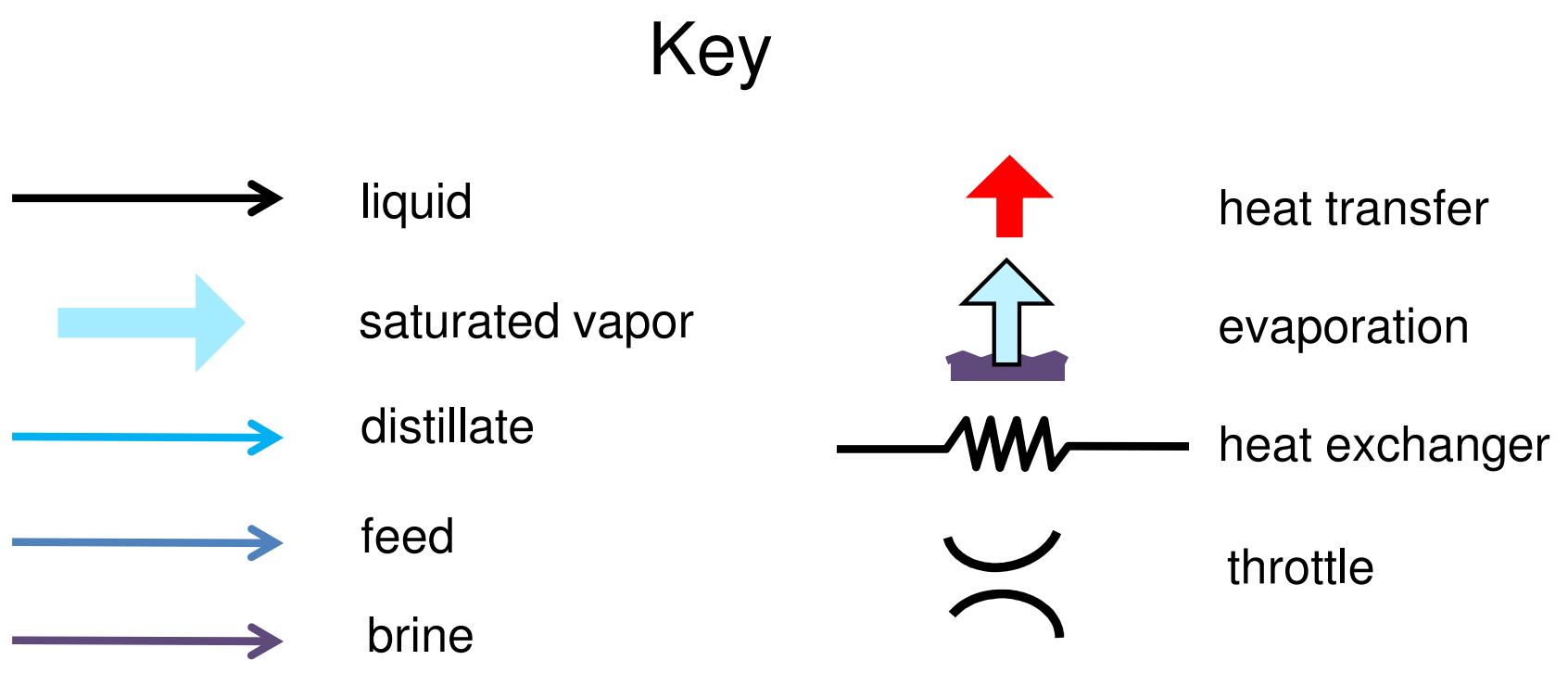

**Figure 2.5.** Symbolic key for all desalination system diagrams

Regarding approximation 10, the temperatures differences across heat exchangers were generally thermally consistent and most were set to 3 °C. This condition varied slightly because of optimization: for instance, the MED model optimized the distribution of heat exchanger area between stages, while keeping the average difference within all heat exchangers at about 3 °C. This temperature pinch is typical of desalination industry and many past models [57], and variation from this is small.

All the technologies and the specific modeling details are explained in diagrams using the symbols and colors shown in Fig. 2.5. In components where vapor coexists with liquid, the vapor arrows are used.

All of the modeling and calculations are done in Engineering Equation Solver (EES) [58]. EES is a simulation equation solver that automatically identifies and groups equations that must be solved and then solves the system iteratively. The default convergence values (maximum residual $< 10^{-6}$; change of variable $< 10^{-9}$) were used in this study. Additionally, EES has built-in property packages for seawater, vapor, and air.

### 2.6.2 Multistage Flash

A schematic diagram of a once-through MSF system is shown in Fig. 2.6. A numerical model for this system was built in EES with values based upon a representative system [46]. In MSF, seawater enters successive saturated stages, where the pressure is lower because of a throttle. This pressure drop, or flash, causes water to condense as vapor, or "flash." A counter-current heat exchanger acting between this flashed water vapor and the colder incoming seawater causes the vapor to condense, while preheating the incoming seawater [59]. A heater provides additional heat to the hot feed before the first and hottest stage, and after the coldest stage a regenerator exchanges heat between the exiting and entering streams.

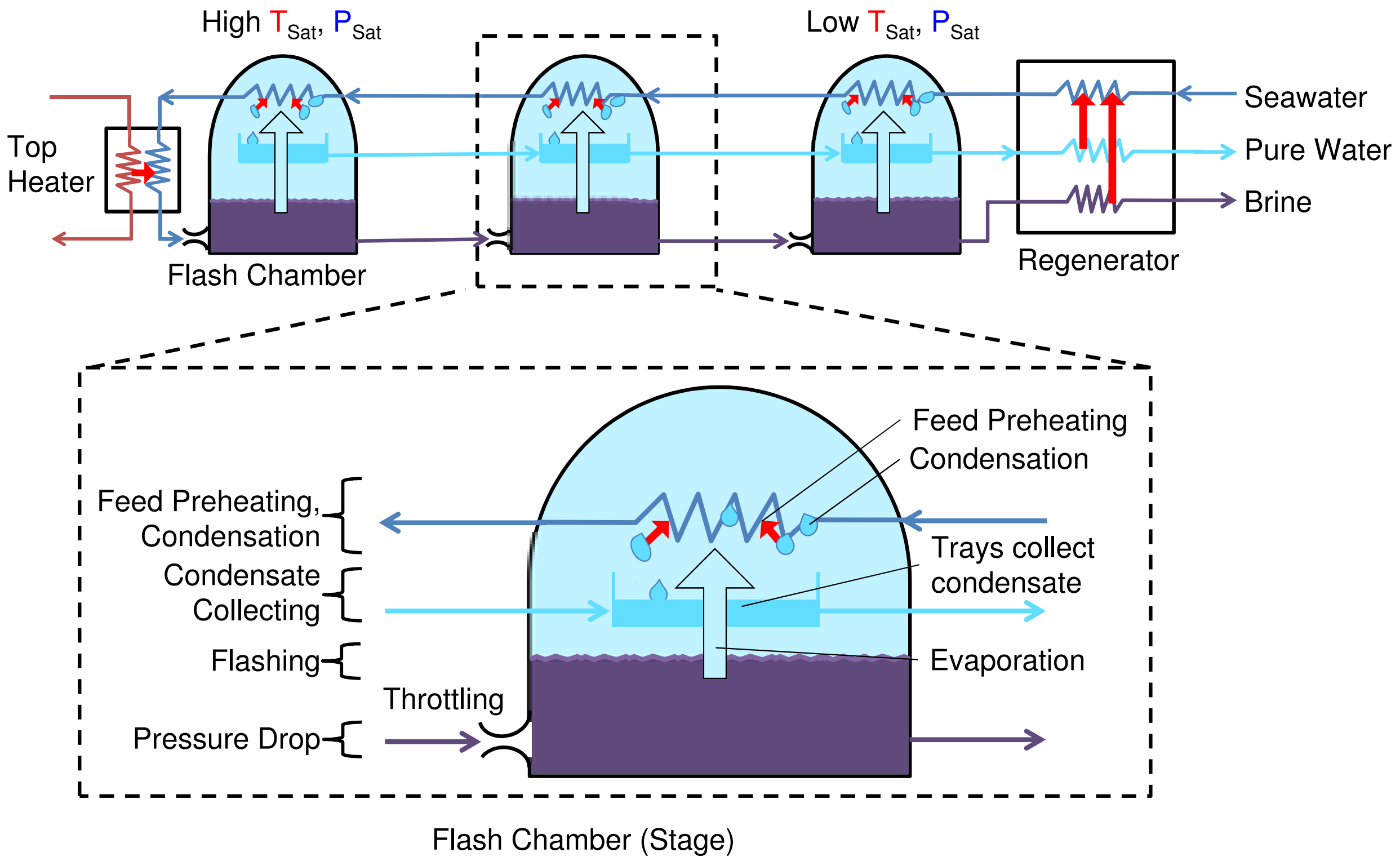

**Figure 2.6.** Flow diagram of once-through multistage flash (MSF) desalination.

A modification from typical MSF systems was the inclusion of a regenerator to exchange heat between entering and exiting streams, which was necessary to avoid large temperature gradients in the feed heaters for low top temperatures, which would have brought the GOR below 1.

**Table 2.3.** multistage flash modeling results

| Parameter | | | $T_H$ | | | |
|---|---|---|---|---|---|---|
| Output | | | 110 °C | 90 °C | 70 °C | 50 °C |
| Number of stages | $n$ | [-] | 24 | 17 | 11 | 4 |
| Performance ratio | PR | [-] | 10.2 | 7.3 | 4.5 | 1.5 |
| Gained output ratio | GOR | [-] | 10.3 | 7.3 | 4.4 | 1.4 |
| Recovery Ratio | RR | [-] | 11.1% | 7.7% | 4.3% | 1.3% |
| Steam flow rate | $\dot{m}_s$ | [kg/s] | 0.0983 | 0.137 | 0.222 | 0.674 |
| Brine salinity | $y_n$ | [g/kg] | 39.4 | 37.9 | 36.6 | 35.5 |
| Second Law Efficiency | $\eta_{II}$ | [%] | 6.28% | 5.67% | 4.98% | 3.25% |
| Entropy Generation | $\mathcal{S}_{\text{gen}}$ | [J/kgK] | 182.6 | 203.2 | 233.2 | 364.5 |

The model simultaneously solves a mass and energy balance for all system components (brine and feed heaters, flashing evaporators). The inputs include recovery ratio, number of stages, and top temperatures similar to [46], which was previously used to validate the model within 5% [52]. The solution gives the inlet and outlet temperature, phase, salinity, and entropy generation for each part.

Several common engineering approximations were made, including those from [52] and [46]. The following approximations were not included in the universal approximation list:

- In systems with multiple stages, the $\Delta T$ across each stage is constant.
- $\Delta T^{heat}_{\text{exchanger}} = 3\,°\text{C}$ and $\Delta T_{\text{stages}} = 2.85\,°\text{C}$, where the latter sets the number of stages [52]

Results from the present model are given in Table 2.3.

The results of entropy generation per stage show that the heating elements dominate, with the feed heater providing the largest source of entropy generation normalized by $\dot{m}_p$, or $\mathcal{S}_{\text{gen}}$, as seen in Fig. 2.7 and Fig. 2.8. Significant entropy is generated in the temperature gradients that occur when exchanging heat between streams, and is largest for the feed heaters as this is where most of the heat exchange for the largest stream, the brine, occurs.

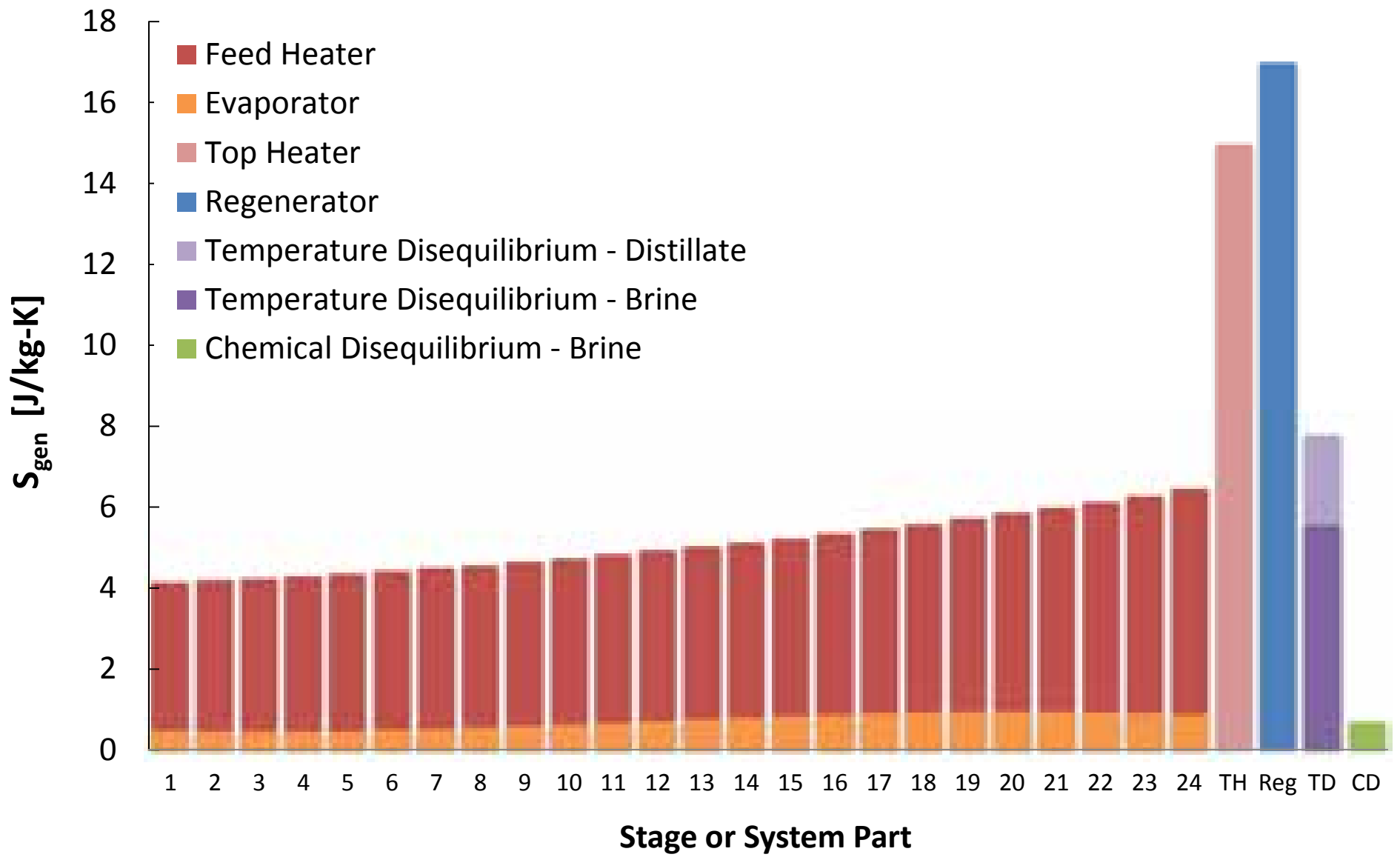

**Figure 2.7.** Entropy generation per kilogram product water produced in each multistage flash (MSF) component at a heat source temperature of 110 °C

For each stage, the entropy generation for flashing is very small: this occurs at constant temperature. At higher recovery ratios, the brine flow rate decreases, which reduces the $\mathcal{S}_{\text{gen}}$ of the feed heaters and chemical disequilibrium of the brine. Also for higher recovery ratios, more flashing will occur in each flashing chamber and thus more entropy will be generated there.

For multistage flash, the dominant entropy generation occurred in the feed heaters, generated by heat transfer across a temperature gradient. Similarly, there was significant generation in brine heater as well as the regenerator. Temperature disequilibrium entropy generation also played a role, especially in that of the brine. In contrast, generation from chemical disequilibrium was relatively negligible. At lower temperatures, these trends changed substantially. Because the stages are left unchanged (with the high temperature stages removed and the lower temperature stages operating at the same temperatures, distillate production rates, and other conditions) there was little difference in entropy generation within stage components of the feed heater and evaporator. More simply, the stages experience nearly identical conditions between different cases, and their $S_{gen}$ is associated with the distillate stream, which also changes little. Similarly, the entropy generation change in the regenerator is minimal due to minimal temperature change between cases. Since the seawater stream flow rate is much larger than the distillate stream, the seawater stream behaves like a heat reservoir at constant temperature with most of entropy generation in the distillate stream. The entropy generation from temperature disequilibrium in the distillate does not change between cases.

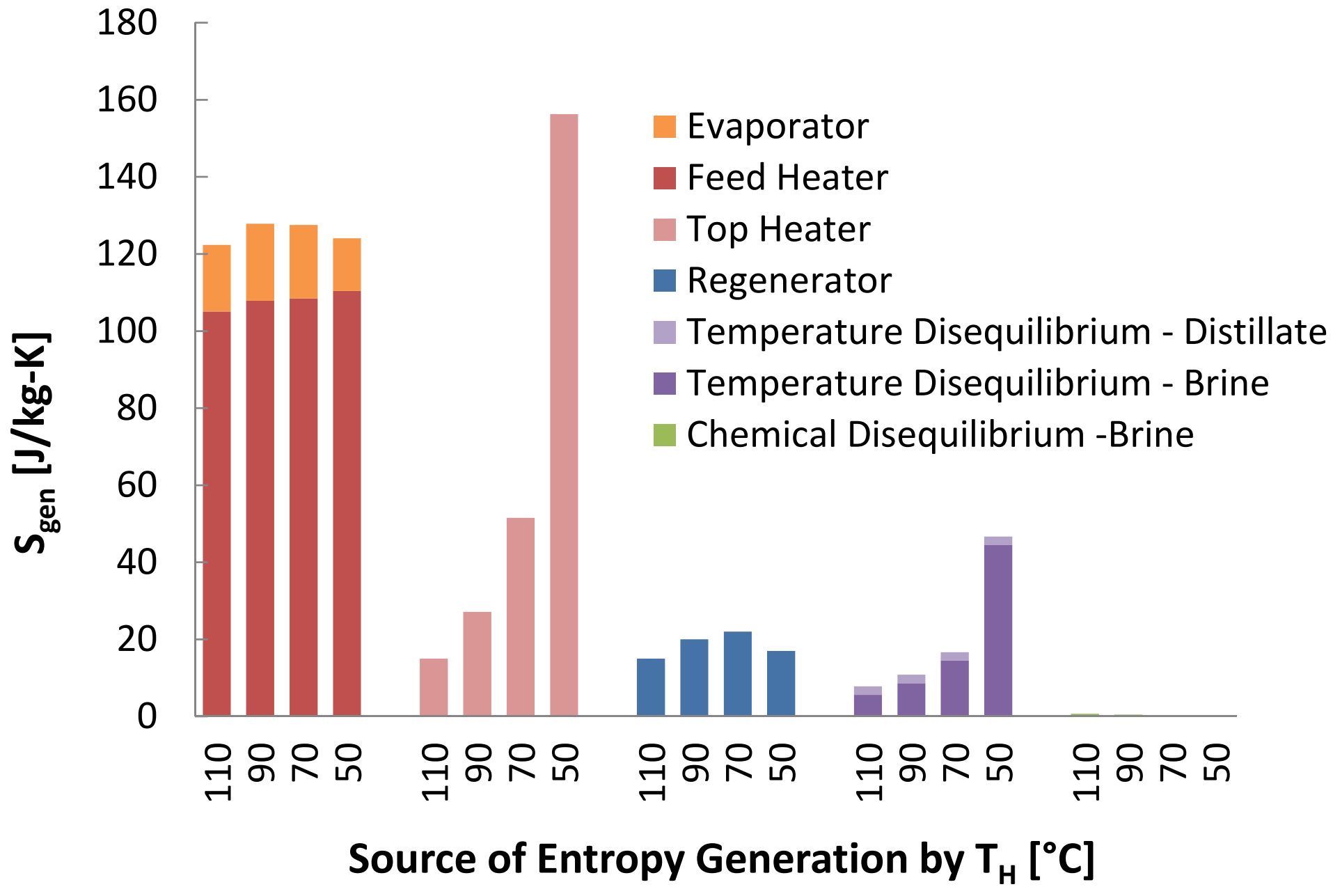

**Figure 2.8.** Entropy generation per kilogram product water produced in each multistage flash (MSF) component for all four temperatures modeled

For the lower temperature cases, removing the high temperature stages significantly increases entropy generation in the brine stream, since the product water lost from removing stages reduces recovery, making the feed stream much larger than that of the distillate. The most substantial increase in entropy generation occurs in the brine heater, where heat transfer occurs at the top temperature. This $S_{gen}$ increases substantially for lower temperature cases as well, for the above reason and because this heat transfer occurs at a lower temperature. The $S_{gen}$ increase in the brine heater is so substantial as it dominates at 50 °C; but it becomes minor at 110 °C. The $\mathcal{S}_{\mathrm{gen}}$ from brine temperature disequilibrium becomes more important because the flow rate of seawater becomes relatively large compared to that of the distillate.

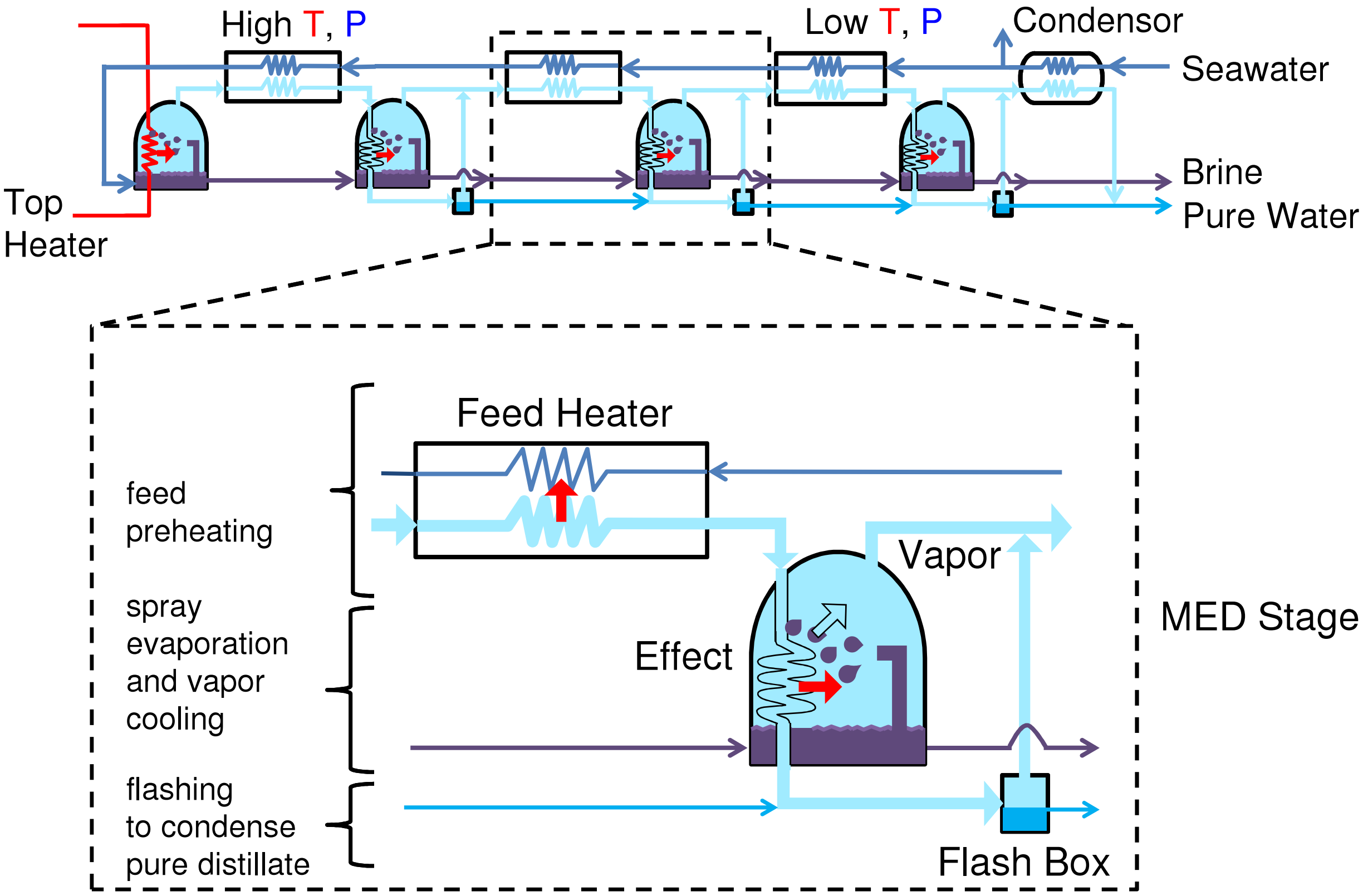

**Figure 2.9.** Flow diagram of multi-effect distillation (MED) desalination.

### 2.6.3 Multiple Effect Distillation

A detailed numerical model for multiple effect distillation (MED) was created by Mistry et al. [47], based upon more basic models in [60], [61], and [46]. A schematic diagram of a typical forward feed MED system is shown in Fig. 2.9.

MED shares features with MSF, including stages (called effects in MED) with generation of water vapor from seawater, and recovery of vapor latent heat to preheat the feed via a heat exchanger. However, in MED vapor passes through the subsequent (less hot) effect through a heat exchanger, and then goes through a flash box to condense pure distillate. For efficient evaporation, MED usually sprays seawater on the heat exchanger. The first effect has a steam heater to finish warming the feed, and no flash box.

The vapor produced in each effect is taken through a condenser where partial condensation helps preheat the incoming feed. The remaining vapor condenses while vaporizing feed from the subsequent effect. When the feed is passed from one effect to another, it is flashed. The vapor produced in this flashing process is also added to the vapor stream in the feed preheater.

In addition to the general approximations stated above, several additional standard engineering approximations are made in this analysis:

- Exchanger area in the effects is just large enough to condense vapor to saturated liquid (i.e., $x = 0$) at the previous effect's pressure.
- Seawater is an incompressible liquid and the properties are only a function of temperature and salinity.
- Non-equilibrium allowance (NEA) is negligible [46].
- Brine (liquid) and distillate (vapor) streams leave each effect at that effect's temperature. Distillate vapor is slightly superheated.
- The overall heat transfer coefficient in each effect, feed heater, and condenser is a function of temperature only [46].

The MED model is created by performing component level control volume analysis coupled with modeling the heat transfer within each effect. Specifically, mass conservation, First and Second Laws, as well as heat transfer rates are evaluated in each effect, flash box, and feed heater. Properties are evaluated using IAPWS 1995 Formulation [62]. Unlike many MED models in the literature, the model by Mistry et al. [47] relies on the simultaneous equation solver, EES.

For the following simulations, the temperature difference between the heat exchanger in the effects, which dominates entropy generation, varies between 2.65 and 3.3 °C, where the distribution of temperatures is optimized by the code, where the average is just under 3 °C. The terminal temperature difference between the feed heaters and condenser, which are both less important, was 5 °C, and was the only exception in this study to heat exchanger temperature differences being near 3 °C. These values were based upon a representative system [46]. The average heat exchanger temperature difference throughout the system is about 3 °C, and the variation does not effect the overall conclusions of this paper. The results for both the 50 and 70 °C cases are summarized in Table 2.4.

**Table 2.4.** Summary of results for a forward feed multi-effect distillation system operating at 50 and 70 °C.

| Parameter | | $T_H$ | | |
|---|---|---|---|---|
| Output | | | 70 °C | 50 °C |
| Number of stages | $n$ | [-] | 12 | 5 |
| Gained output ratio | GOR | [-] | 9.349 | 4.048 |
| Recovery Ratio | RR | [%] | 40% | 17.5% |
| Steam flow rate | $\dot{m}_s$ | [kg/s] | 0.1119 | 0.2526 |
| Brine salinity | $y_n$ | [g/kg] | 58.3 | 42.4 |
| Second Law Efficiency | $\eta_{II}$ | [%] | 10.59 | 6.58 |
| Entropy Generation | $\mathcal{S}_{\text{gen}}$ | [J/kgK] | 566.9 | 1044.7 |

Unsurprisingly, the MED system operating at 70 °C and 12 effects is significantly more efficient (GOR of 9.3) than the system operating at 50 °C with only 5 effects (GOR of 4.0). As is true of typical thermal systems, higher temperatures lead to higher thermodynamic efficiencies and therefore, higher temperature waste heat is desirable, when available. Given the large difference in performance ratios between the two conditions, it is clear that the total entropy generated in the 50 °C case is higher. Existing MED processes are generally limited to 70 °C, so higher temperature cases were excluded.

Detailed entropy generation results, by component, are illustrated in Fig. 2.10. From examining Fig. 2.10 and Fig. 2.11, it is clear that the condenser's performance has the greatest impact on the overall system performance. In particular, it is the most sensitive to varying input temperature. This substantial increase of entropy generation in the condenser is due to the relatively higher feed flow rate in the condenser, which increases entropy generation for two reasons. First, the streams have a larger average temperature difference beacause, the condenser streams larger heat capacity causes it to have a lower temperature drop. Second, the average absolute temperature at which heat transfer occurs is lower in the 50 °C system, resulting in a higher rate of entropy generation.

Between the different temperature cases, the results are similar to that in MSF: the entropy generated in the stages changes little, while it can increase elsewhere. $\mathcal{S}_{\text{gen}}$ for the feed heater and effects did not change much for the lower temperature cases, which is a result of the stage temperatures and product water production changing little, and occurring in the distillate. The total entropy generation in the effects increases slightly, partially due to the fact that the first effect has higher entropy generation, and this addition occurs at a lower temperature.

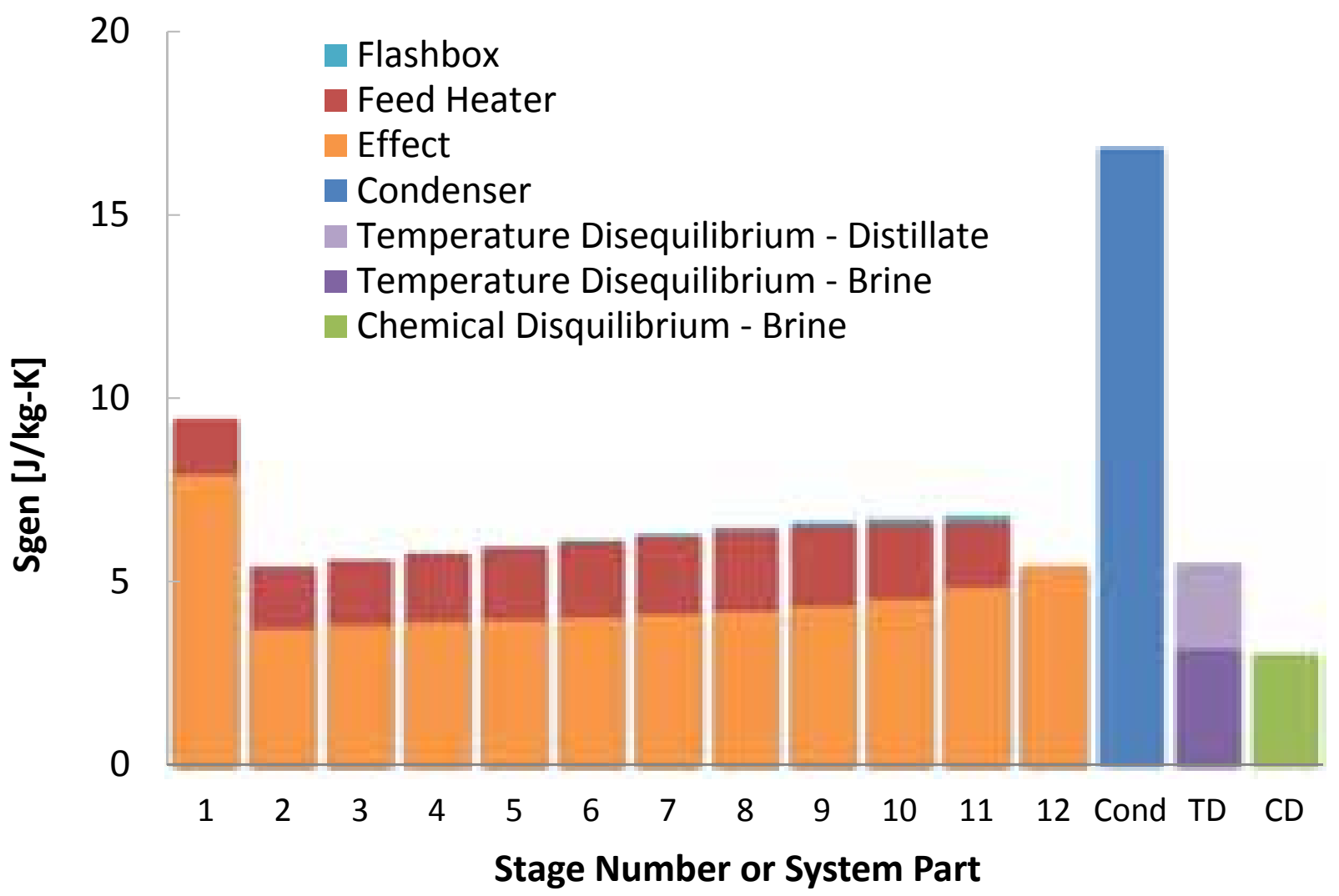

**Figure 2.10.** Entropy generation per kilogram product water produced in each multi-effect distillation component at a heat source temperature of 110 °C.

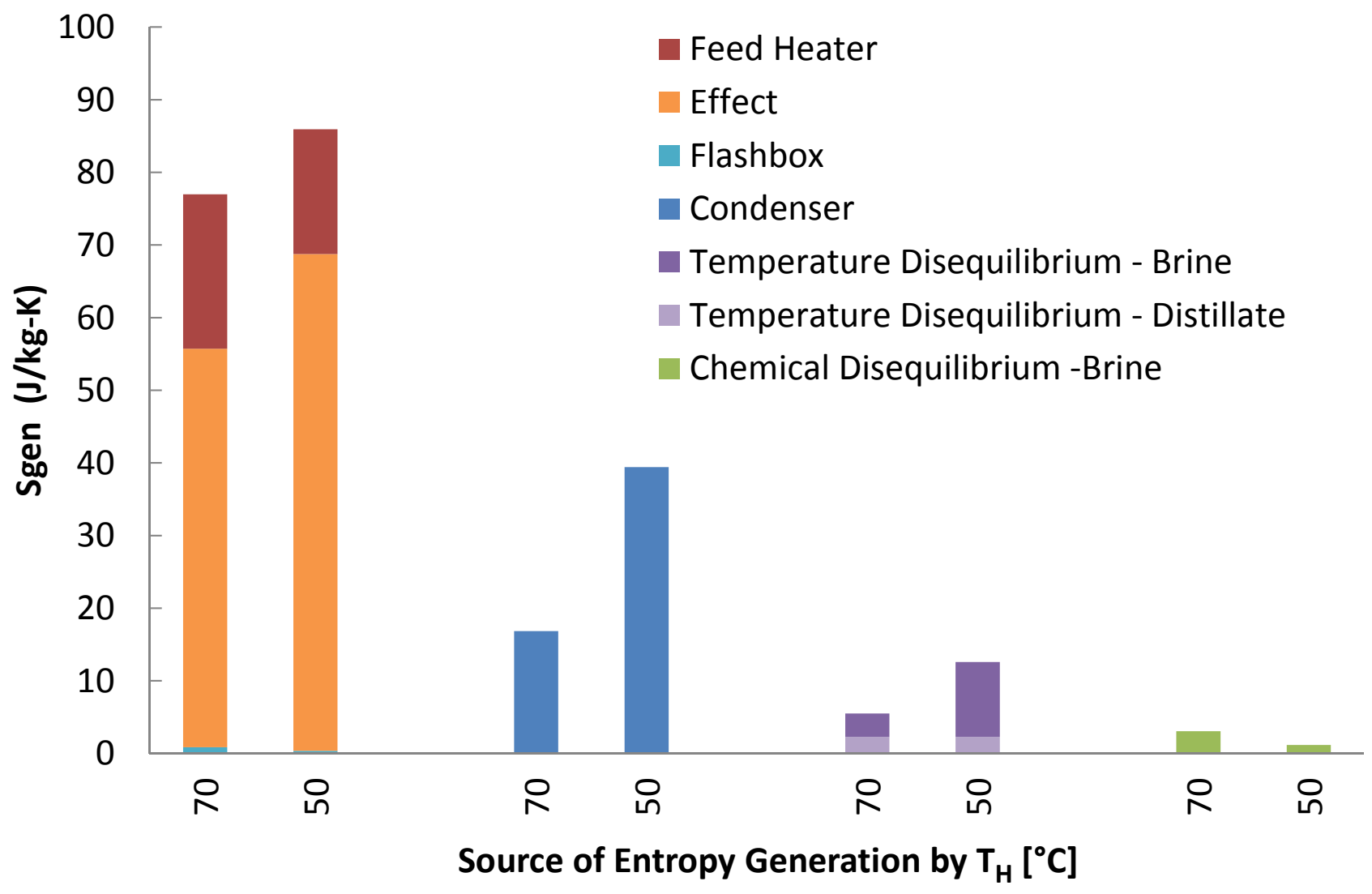

**Figure 2.11.** Entropy generation per kilogram product water produced in each multi-effect distillation component for all temperatures modeled.

Like the MSF case, the entropy generation in the system components that depend on the feed or brine flow rate increase significantly with the higher temperature states removed. As in the entirety of the paper, the entropy generation is normalized to distillate production, so the equations for entropy generation in the brine contain the ratio term $\dot{m}_{brine}/\dot{m}_{product}$. Since the volume of heated brine leaving the system has increased substantially relative to the distillate in the lower temperature case, the entropy generation from temperature disequilibrium of the brine increases. Again for the same reason of increased flow rate of the feed and brine streams, the entropy generation in the condenser increases as well. Finally, the lower recovery ratio results in lower chemical disequilibrium, although irreversibilities due to disquilibrium of the brine have a minimal effect on the overall performance of an MED system.

### 2.6.4 MSVMD

Vacuum MD (VMD) relies on a hydrophobic membrane that rejects liquid water but passes water vapor, where the water vapor passes into an evacuated chamber which is continuously emptied of vapor. The MSVMD system used here is similar to MSF; flashing chambers are replaced by VMD modules as shown in Fig. 2.12. One difference is that vapor produced in each stage is slightly superheated because feed temperature near the inlet of each stage is close to previous stage's saturation temperature. However, degree of superheat is small and extra enthalpy associated with superheat is negligible compared to latent heat of vaporization.

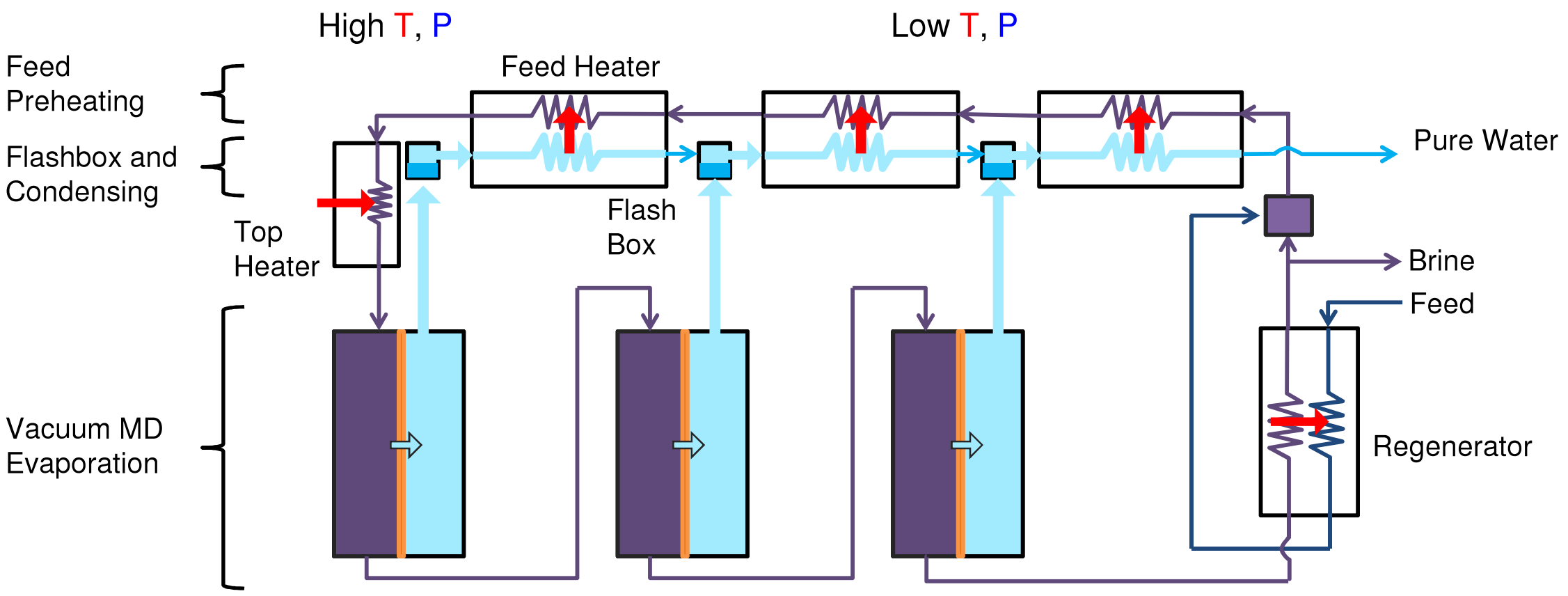

**Figure 2.12.** Flow diagram of multi-stage vacuum membrane distillation (MSVMD) system.

The feed stream is mixed with brine and enters a train of feed pre-heaters where it is heated by condensation energy released from water vapor. External heat input raises the feed temperature to a desired top brine temperature. The feed stream goes through stages of VMD modules which are maintained at progressively lower vacuum pressures. Some portion of brine leaving the last stage is recirculated and the rest is rejected to the environment. Pure water vapor produced from each VMD stage enters the flashing and mixing chamber where it is mixed with the product stream from previous stages. Detailed description of the system configuration and parameters can be found in [41]. There are several assumptions made for MSVMD models.

- Heat exchanger area is just enough to fully condense the vapor; saturated liquid leaves the exchanger.
- Smallest temperature difference between any heat exchanging streams are set to be 3 °C.
- Heat transfer coefficient and mass transfer coefficient are calculated using pure water properties.

Each stage of VMD module is discretized into computational cells in order to be solved by finite difference method developed by Summers et al. [16].

The energy, entropy and mass balance equations are solved in each computation cell.

**Table 2.5.** MSVMD Results

| Parameter | | | $T_H$ | | |
|---|---|---|---|---|---|
| Output | | | 90 °C | 70 °C | 50 °C |
| Number of stages | $n$ | [-] | 18 | 11 | 4 |
| Gained output ratio | GOR | [-] | 7.8 | 4.8 | 1.8 |
| Recovery Ratio | RR | [-] | 0.0848 | 0.0525 | 0.0197 |
| Second Law Efficiency | $\eta_{II}$ | [%] | 4.84 | 4.05 | 2.57 |
| Entropy Generation | $\mathcal{S}_{\text{gen}}$ | [J/kgK] | 178.7 | 215.5 | 344.1 |

Vapor flux, $J$, is calculated as:

$$J = B(P_{f,m} - P_p) \tag{2.10}$$

where $P_{f,m}$ is the saturation pressure of feed stream at the membrane surface, $P_p$ is the vacuum pressure on the product water side, and $B$ is membrane distillation coefficient that is calculated using Knudsen and viscous flow models. Interested readers are referred to Summers et al. [16] for more detail on the finite difference model for MD.

Simulation results are summarized in Table 2.5.

Just as other thermal systems have higher efficiency when the steam temperature is higher, MSVMD shows drastic improvement: GOR more than quadruples as the steam temperature increases from 50 °C to 90 °C. Like GOR, the Second Law efficiency increases with steam temperature. Component-wise analysis of entropy generation for the steam temperature of 90 °C is shown in Fig. 2.13. The more general component entropy generation analysis with all three temperatures tested is shown in Fig. 2.14.

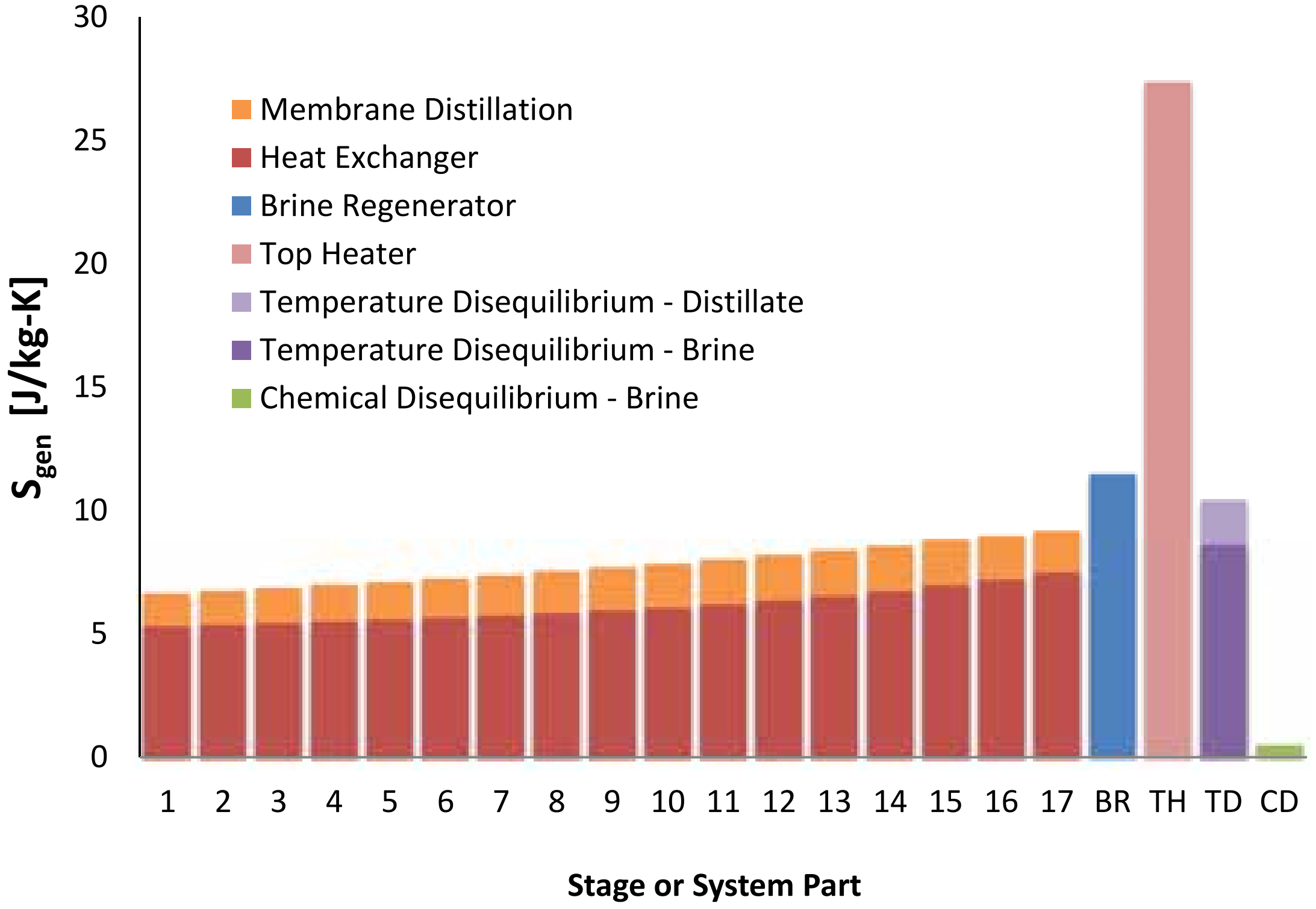

**Figure 2.13.** Entropy generation per kilogram product water produced for each component in multistage vaccuum membrane distillation (MSVMD) for a heat source temperature of 90 °C

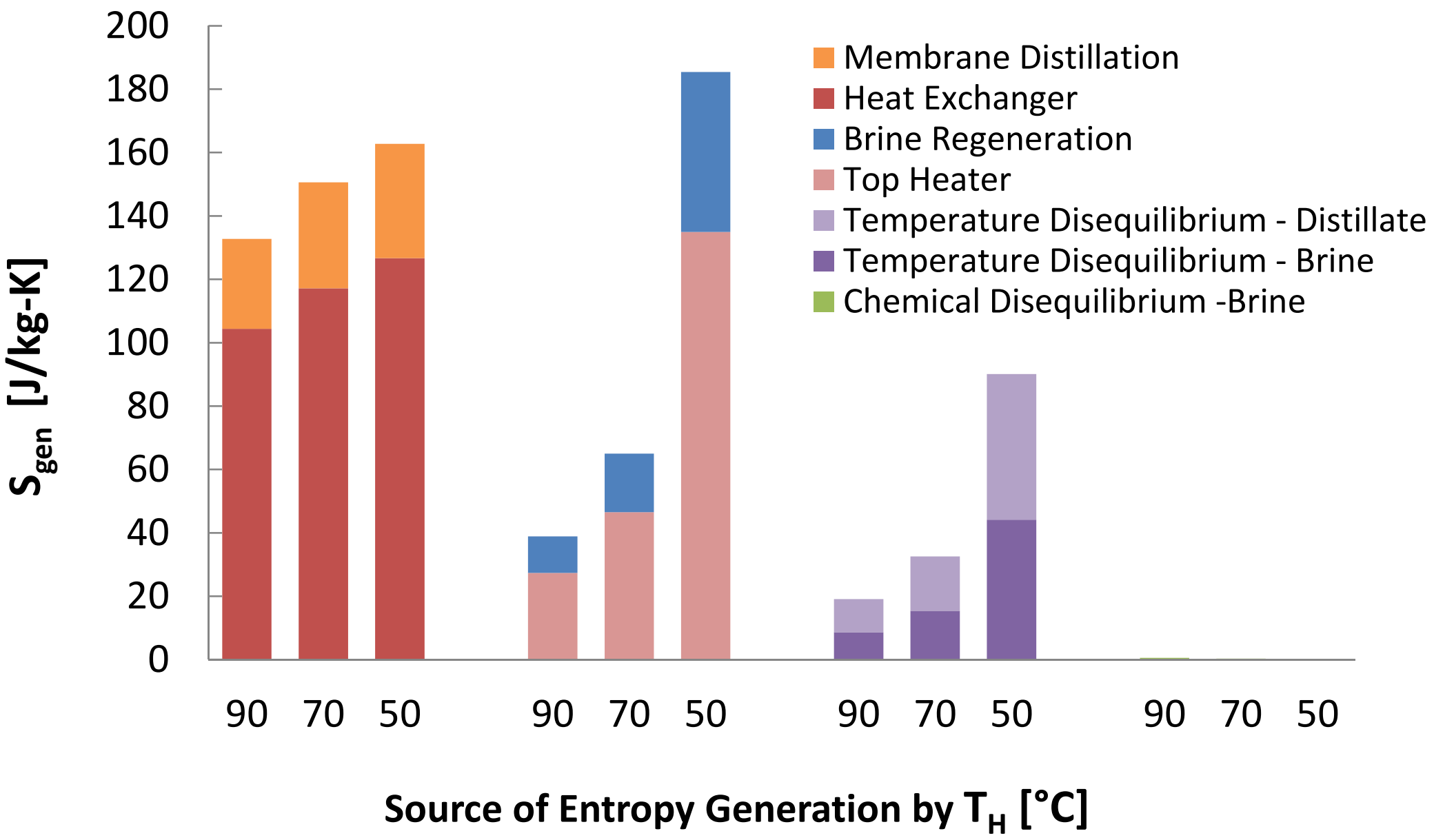

**Figure 2.14.** Entropy generation per kilogram product water produced for each component in multistage vaccuum membrane distillation (MSVMD) for all 3 temperatures modeled

The biggest contribution of specific entropy generation is due to heat exchangers in the MSVMD stages. The other heaters temperature disequilibrium is highest for 50 °C case for both brine and distillate. Total entropy generation is higher for the 50 °C for all components except chemical disequilibrium. Chemical disequilibrium is larger for higher steam temperatures because the recovery ratio is higher, which creates a larger difference between the least heat and minimum least heat of separation. But chemical disequilibrium is negligible compared to other terms. Another major contribution is the top heater, especially for the 50 °C case. This occurs because the energy regeneration is so low that significant energy has to be transferred to the feed stream from the heater, generating large amount of entropy, and because the heat transfer occurs at lower temperature. It should be noted that the absolute amount of entropy generation in each case do not differ by a lot. It is the low permeate production rate that significantly raises the specific entropy generation in each of the categories. In thermodynamic perspective, the main drawback of using low temperature waste steam is a low recovery ratio, resulting in high specific entropy generation.

### 2.6.5 Humidification-Dehumidification

Humidification-dehumidification desalination, inspired by the natural rain cycle, consists of a humidifier, heater, and dehumidifier. A closed air open water (CAOW) HDH configuration is depicted in Fig. 2.15. Feed water enters the system at ambient temperature and is pre-heated by hot moist air condensing in a bubble-column dehumidifier. The feed is then heated to a desired top temperature using an external heat source, in this case, a waste heat source. Subsequently, the feed water moves in to the humidifier where it is brought in direct contact with cold dry air, typically as a spray. The feed gets concentrated as it loses its water content to the cold dry air which in turn gets humidified and heated up. The hot moist air then condenses in the dehumidifier to give pure product water.

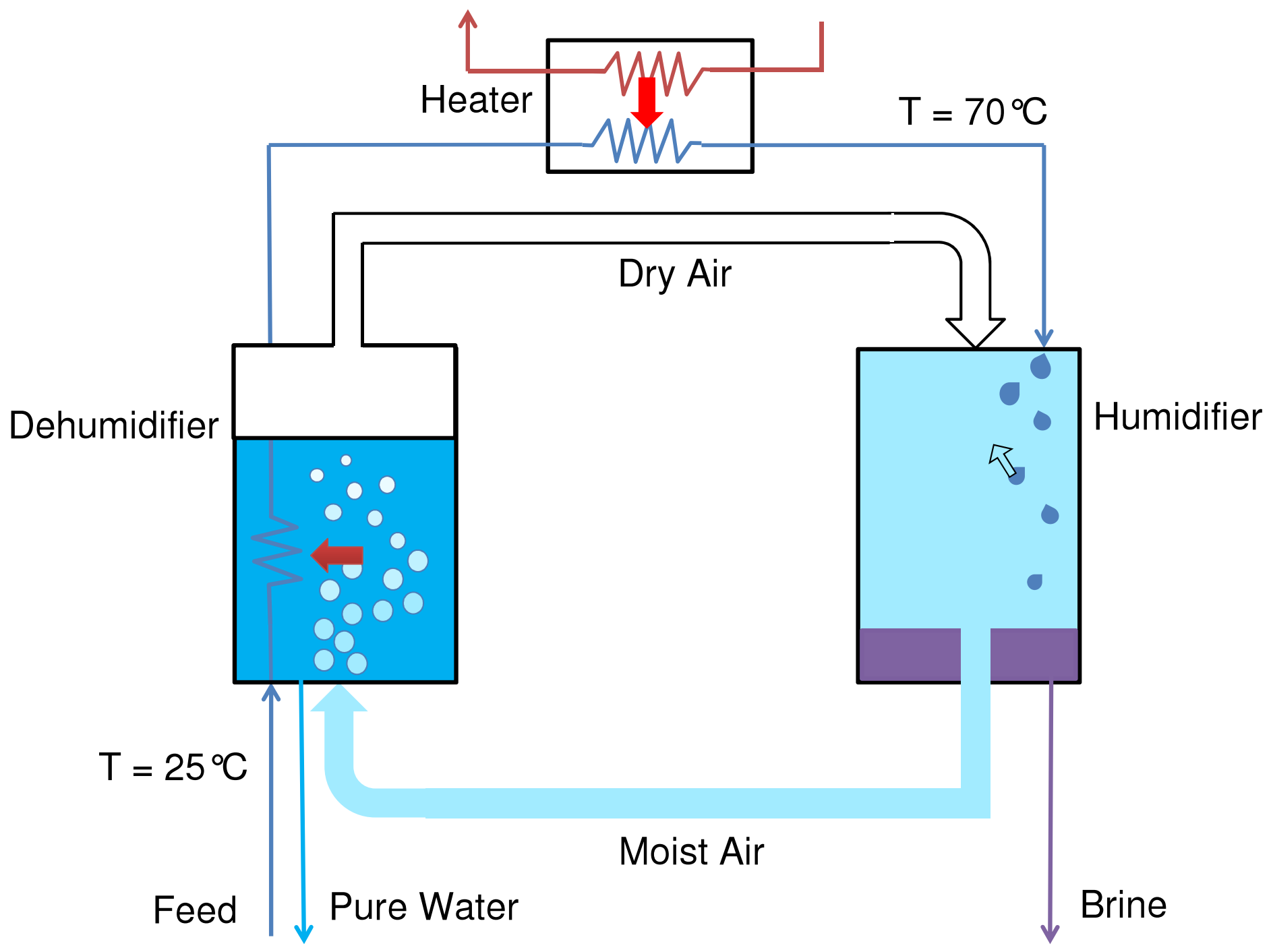

**Figure 2.15.** A flow path diagram of humidification-dehumidification (HDH) desalination

To simulate the HDH process, a numerical EES model developed by Mistry et al. [48] was used after modification to include a graphical temperature pinch analysis [63]. The inputs used in modeling are given in Table 2.6. A temperature pinch ($\Delta T_{\mathrm{pinch}}$) of 3 K between the moist air and water streams was maintained in both the dehumidifier and the humidifier. The Heat Capacity Ratio, defined as the ratio of the maximum possible enthalpy change of the cold stream to that of the hot stream, in the dehumidifier ($HCR_d$) was set to be 1 for balancing the dehumidifier, reducing entropy generation and optimizing the GOR of the system [64,65]. The moist air was assumed to be saturated. The HDH process was simulated for three different system top temperatures: $T_{\mathrm{H}}$ = 50 °C, 70 °C and 90 °C. The results are shown in Table 2.7.

**Table 2.6.** Input for a humidification-dehumidification desalination system

| Parameter | | | |
|---|---|---|---|
| Input | Symbol | Units | Value |
| Temperature pinch | $\Delta T_{\mathrm{pinch}}$ | [°C] | 3 |
| Dehumidifier Heat Capacity Ratio | $HCR_{\mathrm{d}}$ | [-] | 1 |
| Moist Air Relative Humidity | $RH_{\mathrm{ma}}$ | [%] | 100 |

**Table 2.7.** Summary of results for a Humidification-dehumidification system operating at 50, 70 and 90 °C

| Parameter | | | | $T_{\mathrm{H}}$ | |
|---|---|---|---|---|---|
| Output | Symbol | Units | 90 °C | 70 °C | 50 °C |
| Gained output ratio | GOR | [-] | 2.1 | 2.2 | 1.7 |
| Recovery Ratio | RR | [%] | 7.1 | 4.8 | 2.4 |
| Heat Input | $Q_H$ | [kJ/kg] | 1173 | 1118 | 1405 |
| Water-air mass flow rate ratio | MR | [-] | 4.1 | 2.5 | 1.6 |
| Brine salinity | $y_b$ | [g/kg] | 37.7 | 36.8 | 35.8 |
| Entropy Generation | $S_{gen}$ | [J/kgK] | 564 | 414 | 325 |
| Second Law Efficiency | $\eta_{II}$ | [%] | 1.29 | 1.85 | 2.49 |

With increasing top temperature, the recovery ratio steadily increased. However, unlike MED and MSF, in HDH, the entropy generated per product water increased with top temperature, an observation that was also noted previously by Narayan et al. [64]. GOR on the other hand increased with temperature and then decreased. This changing trend occurs because an increase in top temperature leads to both an increased permeate production for a given feed flow rate and increased $\mathcal{S}_{\text{gen}}$.

A breakdown of entropy generation with temperature is given in Fig. 2.16. Most of the entropy generation happened in the dehumidifier and in both the humidifier and the dehumidifier, the entropy generated increased with an increase in top temperature. This arises because the specific heat capacity of saturated air is nonlinear with temperature causing large temperature gradients in the heat exchangers in both components. The effect is clearly visible in the temperature-enthalpy profiles shown in Fig. 2.17. While the minimum temperature pinches are the same at 50 °C and 90 °C, the nonlinear specific heat capacity of moist air, causes the average temperature difference between moist and water to increase with increasing top temperature. This also causes the average temperature difference between moist air and water in the dehumidifier to be higher than that in the humidifier. The larger average temperature difference directly causes more entropy to be generated during heat transfer.

Performance at higher top temperatures can be however improved very substantially by using extractions of moist air from the humidifier to the dehumidifier to allow the water streams to better match the temperature of the moist air streams [66]. An extraction is simply removing some of the flow, typically the vapor and air mixture, to modify the $\dot{m}$ so that the enthalpies are better balanced, allowing for a smaller temperature difference when exchanging heat. For HDH systems which do not use extractions, the system is most efficient at lower top temperatures between 50 °C and 70 °C for inlet conditions considered here.

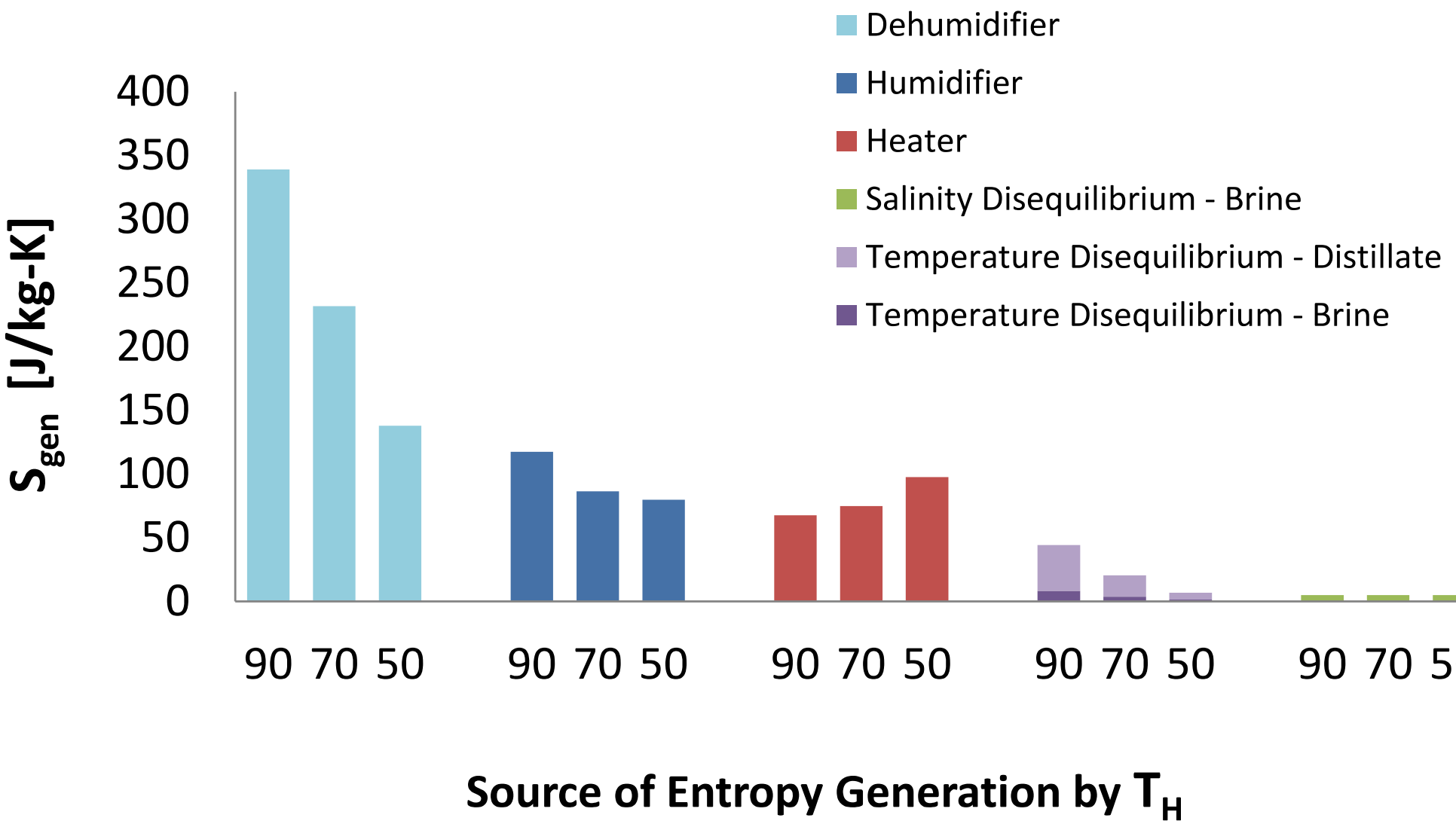

**Figure 2.16.** Entropy generation per kilogram product water produced in each humidification-dehumidification component for all 3 temperatures modeled

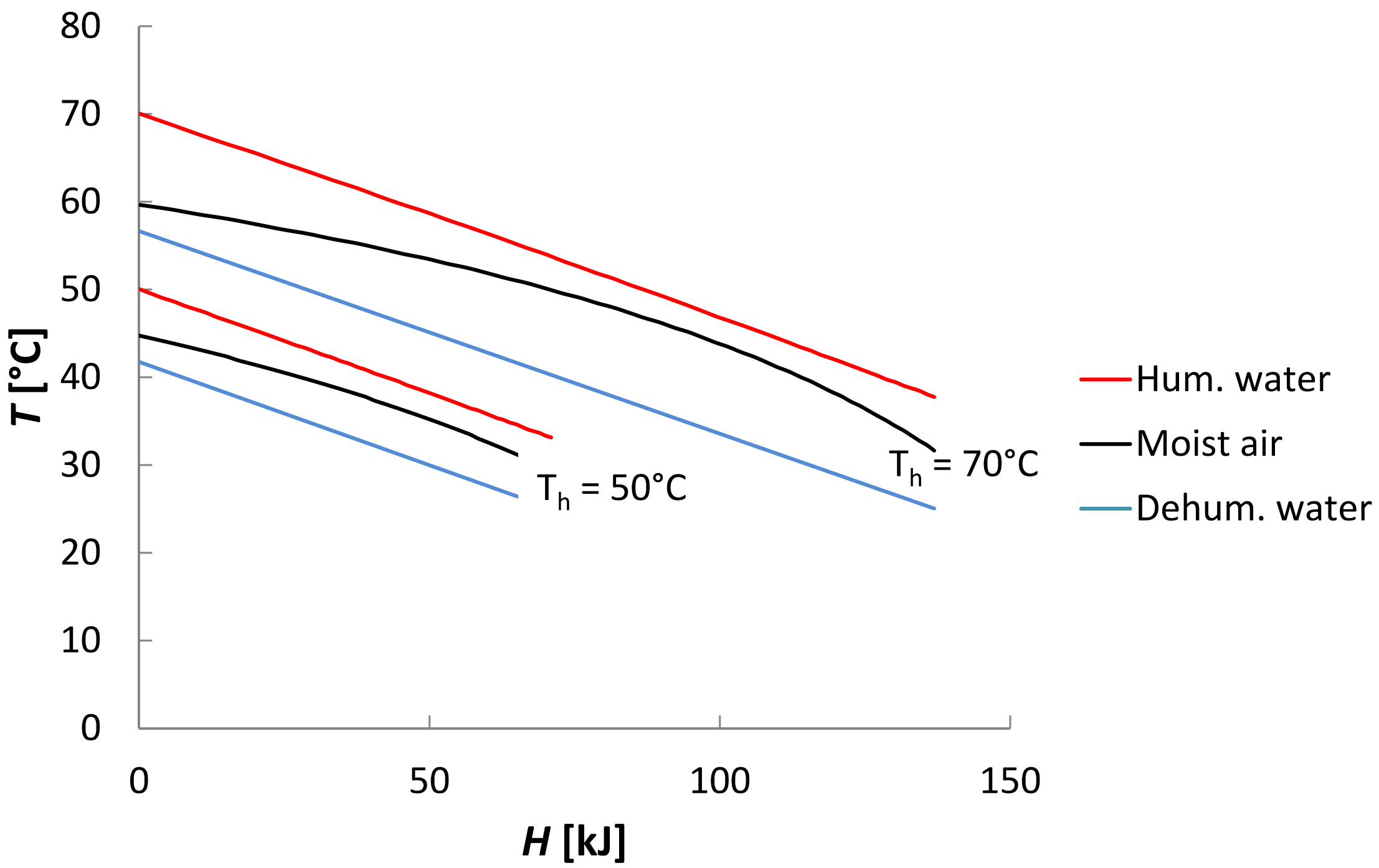

**Figure 2.17.** Temperature-enthalpy profiles for water and moist air streams in Humidification Dehumidification Desalination. The dehumidifier and humidifier T-H curves for $T_{\mathrm{H}} = 50\,^{\circ}\mathrm{C}$ and $90\ ^{\circ}\mathrm{C}$ are shown. The minimum temperature pinch is set between cases. The larger curvature in the $70\,^{\circ}\mathrm{C}$ case increased the average temperature gradient for heat exchange, and is the reason HDH efficiency decreases at higher temperature, unlike other technologies.

### 2.6.6 Organic Rankine Cycle

In terms of thermodynamics, electric energy has higher quality than thermal energy (*e.g.*, waste heat) because heat cannot be continuously converted into work with 100% efficiency. Therefore, for electrically driven desalination technologies to be compared in a fair way with waste heat, heat input that would have produced the electrical work should be considered. A holistic analysis and literature review was done on the available technologies that use low temperature waste heat as a heat source, including Organic Rankine Cycles (ORCs), thermoelectrics, Stirling engines, and other cycles. The literature consistently found that the most efficient and most technologically feasible technology for temperatures between 50 and 110 °C was ORC's [44, 67–70].

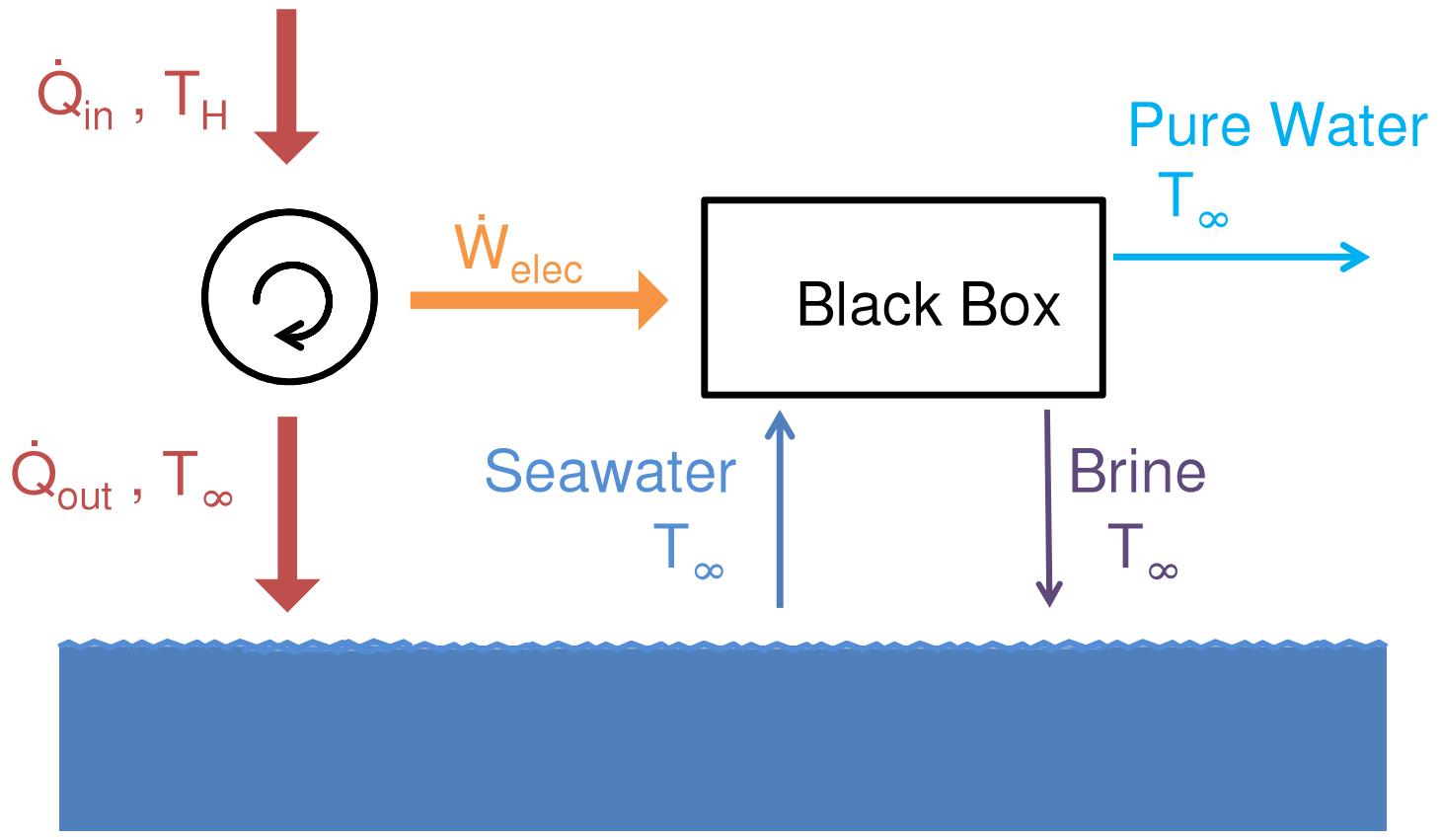

**Figure 2.18.** Blackbox diagram for powering electric desalination systems with waste heat

The Rankine cycle in some form is perhaps the most used thermodynamic cycle in the power industry, and are used with nuclear and fossil fuels alike. The Rankine cycle operates within the vapor dome of the working fluid, with both heating and heat rejection resulting in phase change at constant pressure and temperature. Work is produced by a turbine operating near the saturated vapor region, and work input, in the form of pumping, is required for pressurization in the liquid or saturated liquid region.

ORCs use an organic compound for the refrigerant, instead of water, due to a higher vapor pressure at lower temperature. A variety of cycle improvements may be implemented, including superheat, reheat, and regeneration, but the very low temperatures under study led to the conclusion that a standard Rankine cycle was suitable. The best performing working fluid, R123, was taken from a review on Rankine cycle compounds.

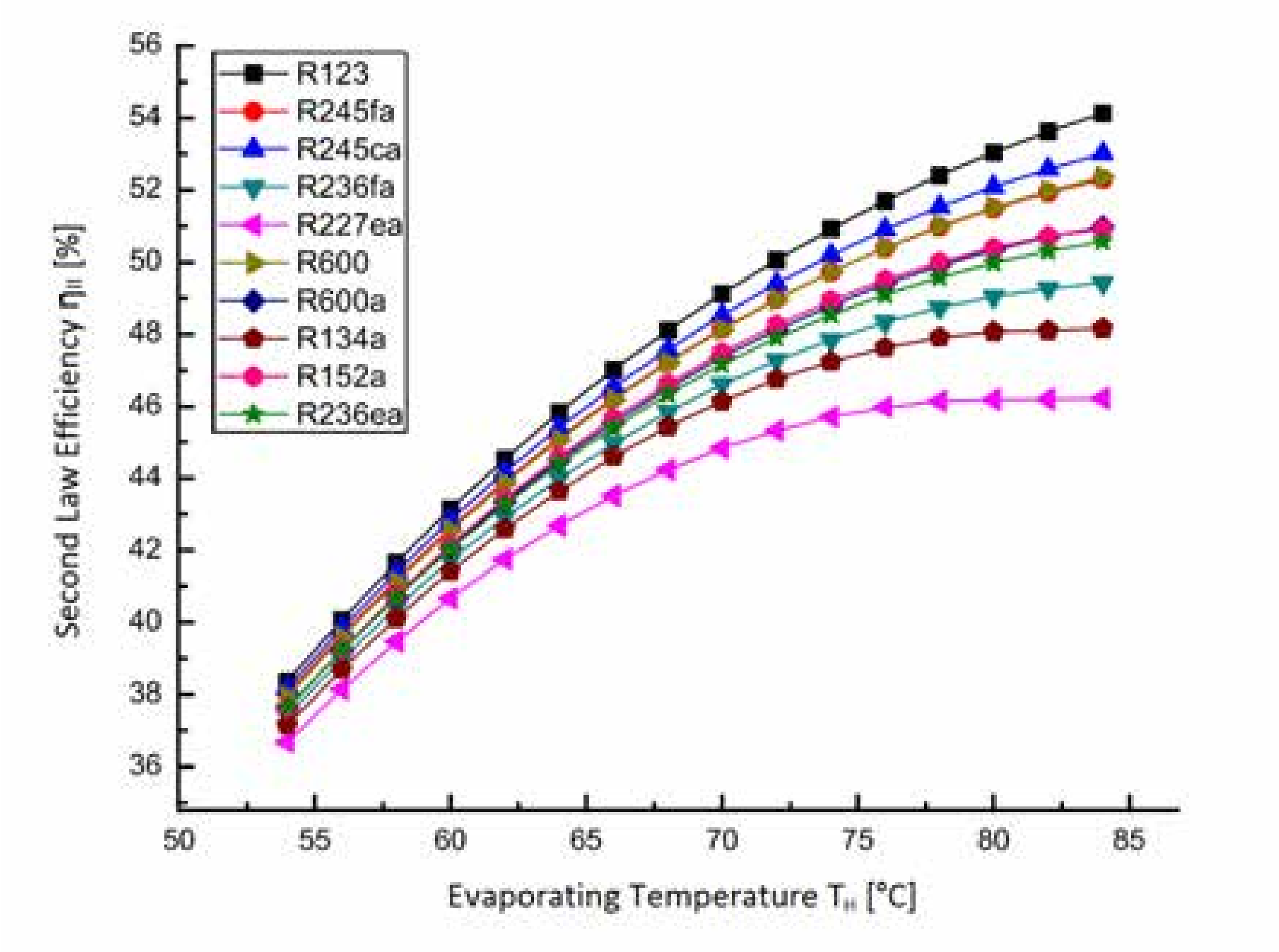

**Figure 2.19.** Organic Rankine cycle Second Law efficiency vs. evaporating temperature with $T_0 = 26\,°\mathrm{C}$ from Shengjun et al. [49].

The Second Law efficiency from this data was used in calculating the overall performance of the electrical desalination technologies operating on waste heat. The organic compounds have similar performance curves, so the refrigerant chosen does not make a large difference. This data was readily adaptable to the present study since the inlet temperature for this data roughly matches that of the thermal cycle modeling above (25 °C vs. 26 °C). A line of best fit was used for temperatures just outside this range (< 7 °C).

The entropy generation per unit product from using a Rankine cycle can be calculated from the definition of $\eta_{II}$ in terms of work, Eq. (2.5), and the relation between $\dot{W}_{\text{sep}}$ and $\dot{W}_{least}^{min}$, as seen in Eq. (2.6):

$$S_{gen,Rankine} = \frac{\dot{W}_{sep}(1-\eta_{II})}{T_0} \tag{2.11}$$

### 2.6.7 Mechanical Vapor Compression

Mechanical vapor compression (MVC) is a desalination technology that uses a compressor to pressurize and heat water vapor, then condenses this vapor into pure distillate using a heat exchanger with the incoming seawater. As seen in the schematic diagram in Fig. 2.20, seawater enters the system and is preheated by the exiting pure water and brine streams. The preheated seawater is then partially evaporated by spraying onto the hot evaporator heat exchanger. This water vapor exits into the compressor, where the adiabatic pressurization causes it to heat up. This vapor then goes inside the evaporator heat exchanger that helped evaporate it in the first place, but now gives off heat, causing it to cool and condense out pure distillate.

A numerical EES model was created which simultaneously solved the equations for energy and mass balances for a single stage mechanical vapor compressor desalination system. The results of this model previously appeared in prior work by Mistry et al. [51].

**Table 2.8.** MVC design inputs.

| Input | Symbol | Value |
|---|---|---|
| Top brine temperature | $T_b$ | 60 °C |
| Pinch: evaporator-condenser | $\Delta T_{evap}$ | 2.5 °C |
| Pinch: regenerator | $\Delta T_{regen}$ | 3 °C |
| Compressor inlet pressure | $P_{c,in}$ | 19.4 kPa |
| Recovery ratio | RR | 40% |
| Isentropic compressor efficiency | $\eta_c$ | 70% |

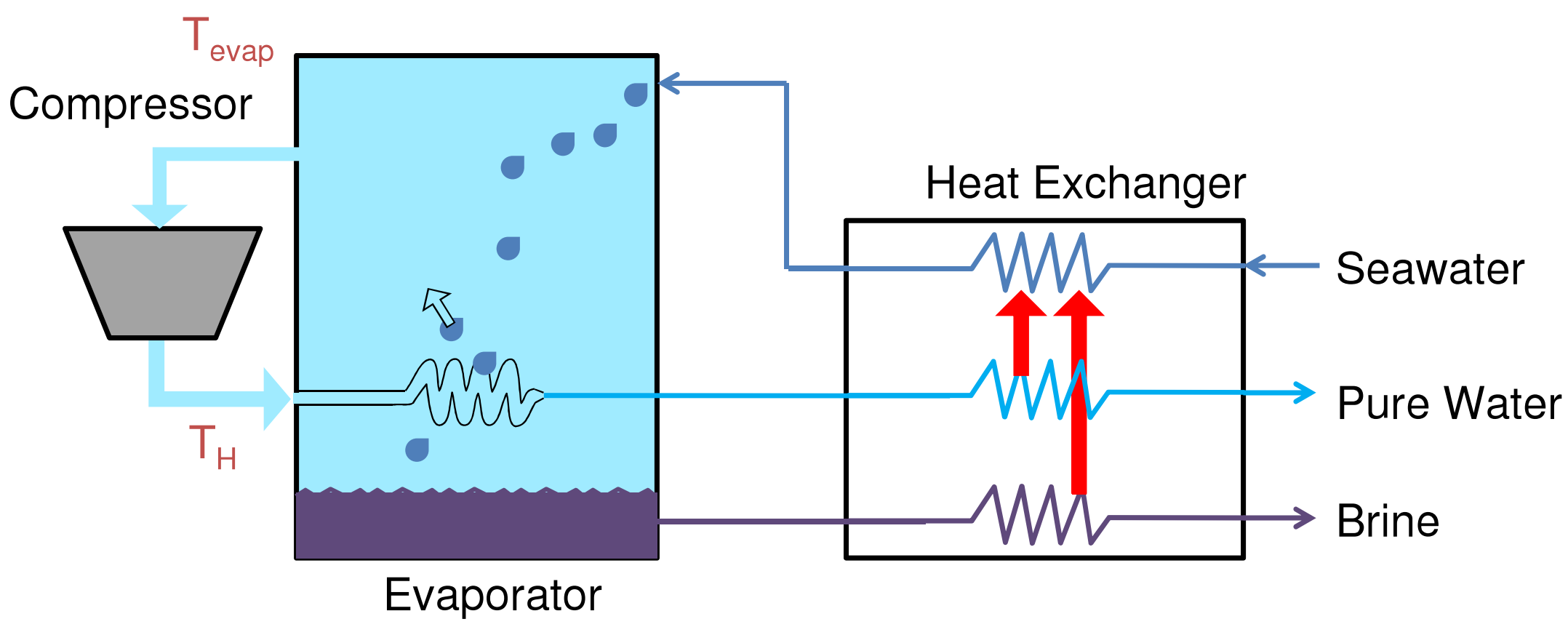

**Figure 2.20.** Single effect mechanical vapor compression process (MVC) with spray evaporation.

System operating conditions used in [71] and [72] were used in this study.

The smaller pinch in the evaporator-condenser is due to the high heat transfer coefficients found with phase change compared with conventional flow. The model outputs are presented in Table 2.9 and the entropy generation between different MVC components is presented in Fig. 2.21.

The Rankine cycle for converting a heat input to electrical work was the dominant source of entropy generation. The ORC was also the only component of the system whose $\mathcal{S}_{\text{gen}}$

**Table 2.9.** Mechanical vapor compression model outputs.

| Output | Symbol | Value |
|---|---|---|
| Specific electricity consumption | $\dot{W}_{elec}$ | 8.84 kWh/m$^3$ |
| Discharged brine temperature | $T_b$ | 27.2 °C |
| Product water temperature | $T_p$ | 29.7 °C |
| Compression ratio | $CR$ | 1.15 |
| Second Law efficiency | $\eta_{II}$ | 8.5% |

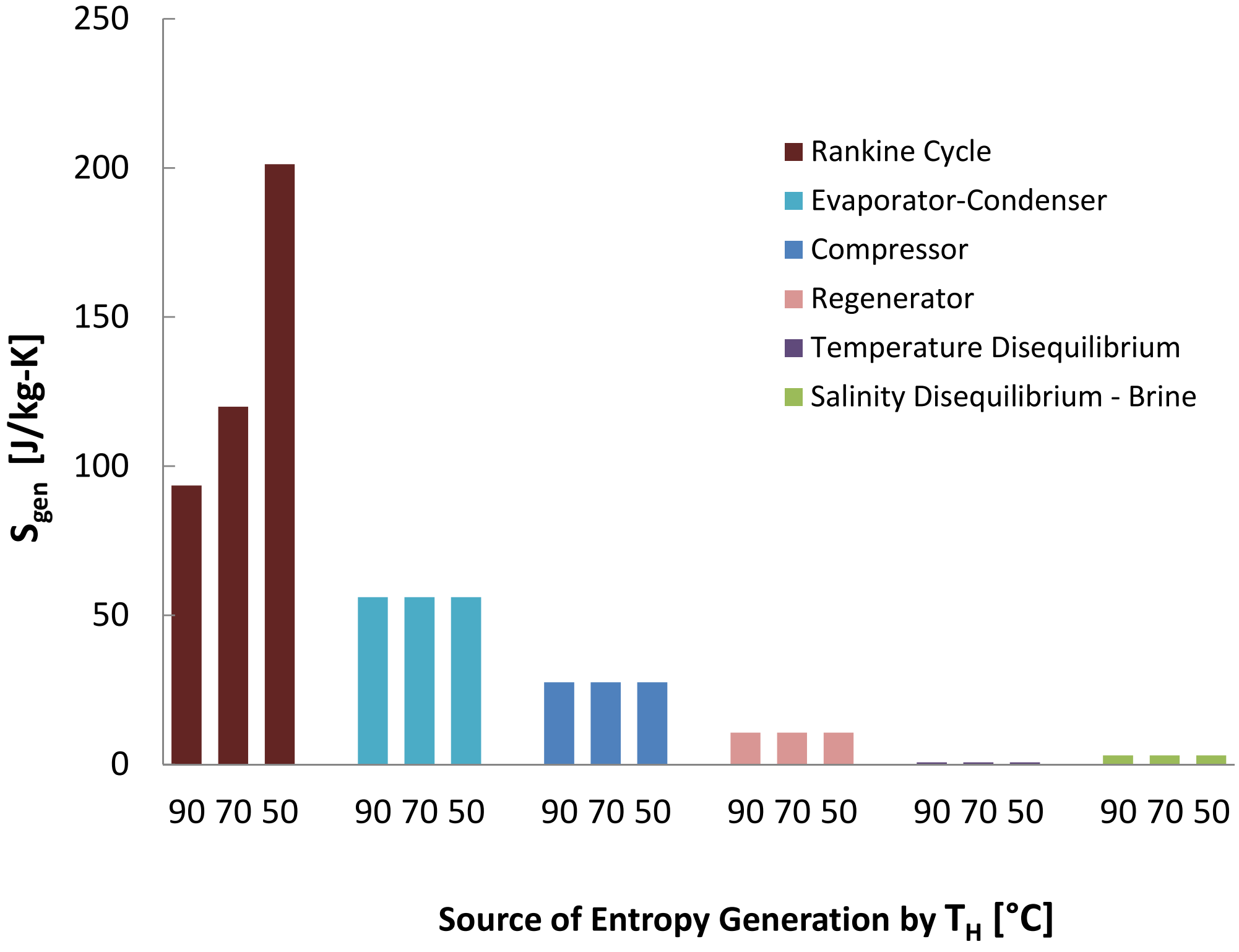

**Figure 2.21.** Entropy generation per kilogram product water produced in each mechanical vapor compression component for all three temperatures modeled

differed when operated at different temperature, with performance declining as the heat source temperature reduced. This performance decline is typical of any power cycle. This $\mathcal{S}_{\text{gen}}$ variation only with the ORC means that the design of MVC systems need not differ for waste heat applications, provided using a power cycle to convert heat input to electrical work is feasible for a particular project. The evaporator-condenser dominated entropy generation in MVC itself, as most heat transfer within the system occurs in this part. This inefficiency increases the required compression ratio and related work. It can be improved by reducing the temperature gradients in evaporation and condensation, e.g. by increasing condensation heat transfer coefficients [73] or increasing surface area [74]. The vapor compressor itself was the next largest source of entropy generation; more efficient air compression technologies may significantly improve MVC efficiency. Entropy generation from temperature gradients in the regenerator was the next largest source of entropy generation; the roughly equal incoming and outgoing heat capacity ( $\dot{m}_{in}h_{sw} = \dot{m}_p h_{H2O} + \dot{m}_b h_b$ ) means the system is balanced, reducing temperature divergence in the heat exchange. The higher recovery ratio and overall good efficiency led to a relatively large role of entropy generation from chemical disequilibrium. As this is a simple one stage MVC system, and since much of the entropy generation is in a mechanical component, the compressor, and in two phase heat transfer, significant gains in efficiency can be made with superior designs, such as multistage systems [75] and various technologies. It is worth noting that the assumptions for the MVC cycle here are less conservative than those of the other technologies, with smaller temperature differences in the heat exchangers.

Since the system is powered by electricity produced by the Rankine cycle, the operation of MVC itself is unaffected by the use of waste heat, and the component entropy generation remains identical between the different temperature cases. Like many other cycles, the Rankine cycle gets less efficient with lower temperature differences, causing worse performance at lower temperatures. The entropy generation in ORC dominates in MVC, so its performance worsens notably with lower temperature waste heat sources.

### 2.6.8 Reverse Osmosis

Reverse osmosis is globally the dominant seawater desalination technology, and relies on high pressures to overcome osmotic pressure to force water through a membrane that rejects the salts. Unlike the thermal technologies, reverse osmosis is so efficient and near the theoretical least work (typically 2-3 times $\dot{W}_{least}^{min}$) that temperature changes in the water are negligible, and the entropy generation study therefore focuses on pumping. This differs significantly from the previous systems, where thermal effects dominated and pumping effects were negligible, and can be observed with the equation for $\mathcal{S}_{\text{gen}}$ from temperature disequilibrium.

**Table 2.10.** RO System Input Parameters

| Input | Symbol | Value |
|---|---|---|
| Pump efficiency | $\eta_{pump}$ | 85% |
| Pressure exchanger efficiency | $\eta_{PX}$ | 96% |
| Feed pressure | $P_{feed}$ | 2 bar |
| RO pressure | $P_{RO}$ | 69 bar |
| Recovery ratio | RR | 40% |

A simple single stage RO system was modeled which incorporated pressure recovery from the brine. This standard system design was created with design and values from ERI, using their pressure exchanger [50], and was used previously by some of the authors [52].

The pressure recovery step requires equal mass flow rates between the brine and feed stream, so the feed stream is split and only a portion goes through the pressure exchanger. The feed pump pressurizes the rest of the feed. Two minor pumps are needed as well, one for circulating the feed, bringing it to a pressure of 2 bar, and another to finish pressurizing the portion of water that passes through the pressure exchanger. The system diagram is shown in Fig. 2.22.

The entropy generation is calculated for each component as explained in the component sections (Table 2.1), using entropy equations for pumping, chemical disequilibrium of brine, and for the pressure exchanger, compression and depressurizing. The entropy generation across the RO membrane can be calculated by summing the entropy generation caused by depressurizing the product stream as it passes from high to low pressure across the membrane with the compositional entropy change at the high pressure as evaluated from Eq. (2.12):

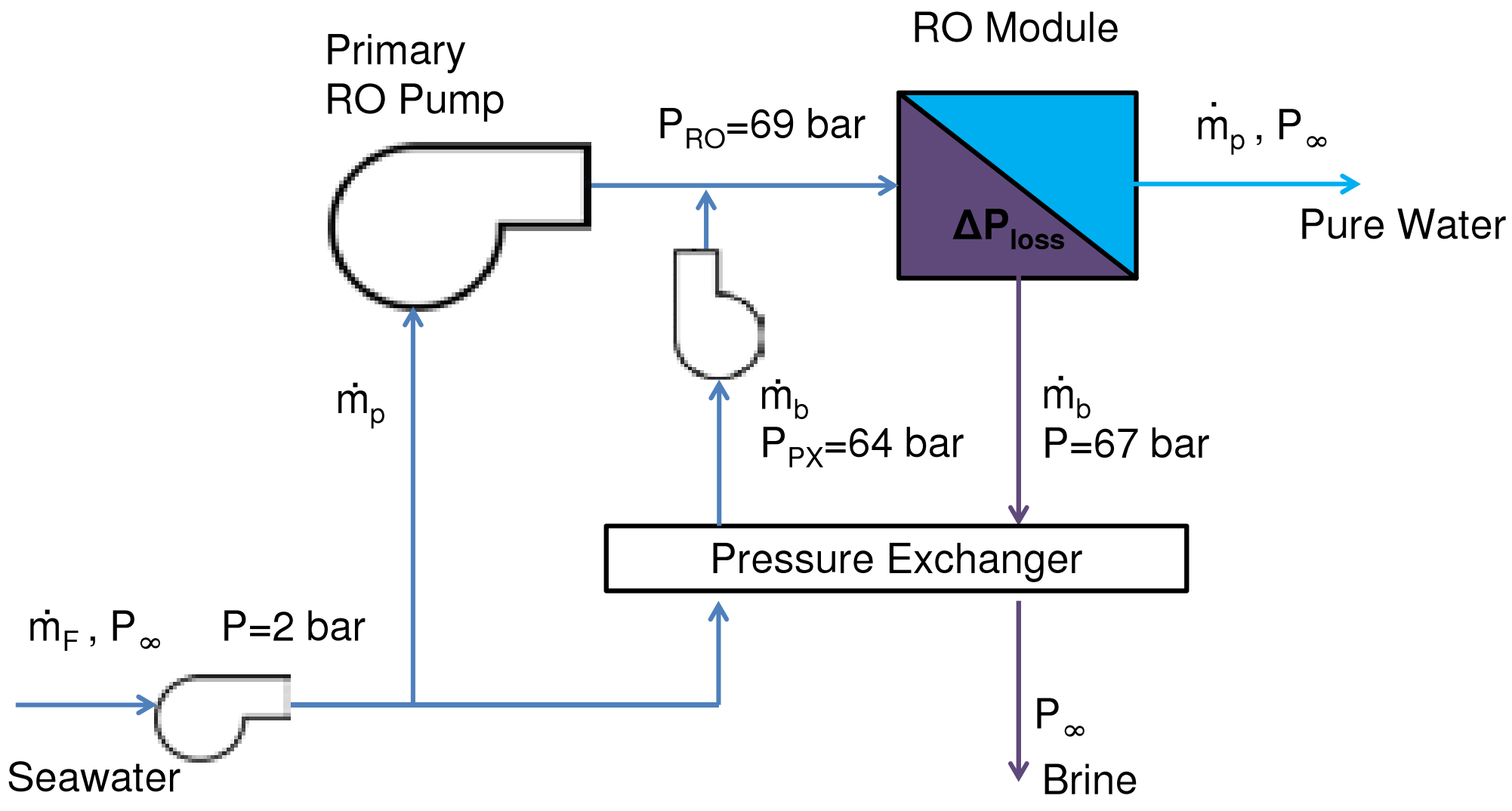

**Figure 2.22.** Flow path diagram of reverse osmosis (RO) desalination.

$$\Delta \dot{S}_{\text{composition}} = \dot{m}_p s_p + \dot{m}_b s_b - \dot{m}_F s_F \tag{2.12}$$

where the entropy, $s$, is a function of temperature, salinity, and pressure, but pressure dependence is neglected for this nearly incompressible situation. The entropy generation from the pressure exchanger is calculating assuming compressing and depressurizing steps, and is explained in [52].

As with MVC, the electricity for the RO unit first must be provided by converting the heat input at a given temperature to electrical work via the organic Rankine cycle. The results from the model are as follows.

The use of the Rankine cycle dominates the entropy generation for waste heat powered reverse osmosis. The next dominant source of entropy generation is across the RO module, and is significantly caused by the RO pressure (69 bar) being well above the osmotic pressure of seawater (27 bar), which is applied because the osmotic pressure grows as the salinity increases through the module, and because excess pressure is often used to increase the flow rate through the membrane, reducing needed membrane area. This can be improved by having multiple stages at different pressures [76], reducing recovery (and thus brine salinity) and top pressure, batch processes where pressure is increased over time [77], and other methods. The next dominant source of entropy generation is the high pressure pump, with the feed and booster pumps playing fairly insignificant roles. Chemical disequilibrium is the next major source of entropy generation, which is very high in RO due to the relatively good efficiency overall.

**Table 2.11.** Reverse osmosis modeling results

| Output | Symbol | Value |
|---|---|---|
| Specific electricity consumption | $\dot{W}_{elec}$ | 2.35 kWh/m$^3$ |
| RO unit Second Law efficiency | $\eta_{II,RO}$ | 31.9% |
| Total Second Law efficiency, $T_H = 90$ | $\eta_{II,90}$ | 17.3% |
| Total Second Law efficiency, $T_H = 70$ | $\eta_{II,70}$ | 15.3% |
| Total Second Law efficiency, $T_H = 50$ | $\eta_{II,50}$ | 10.8% |

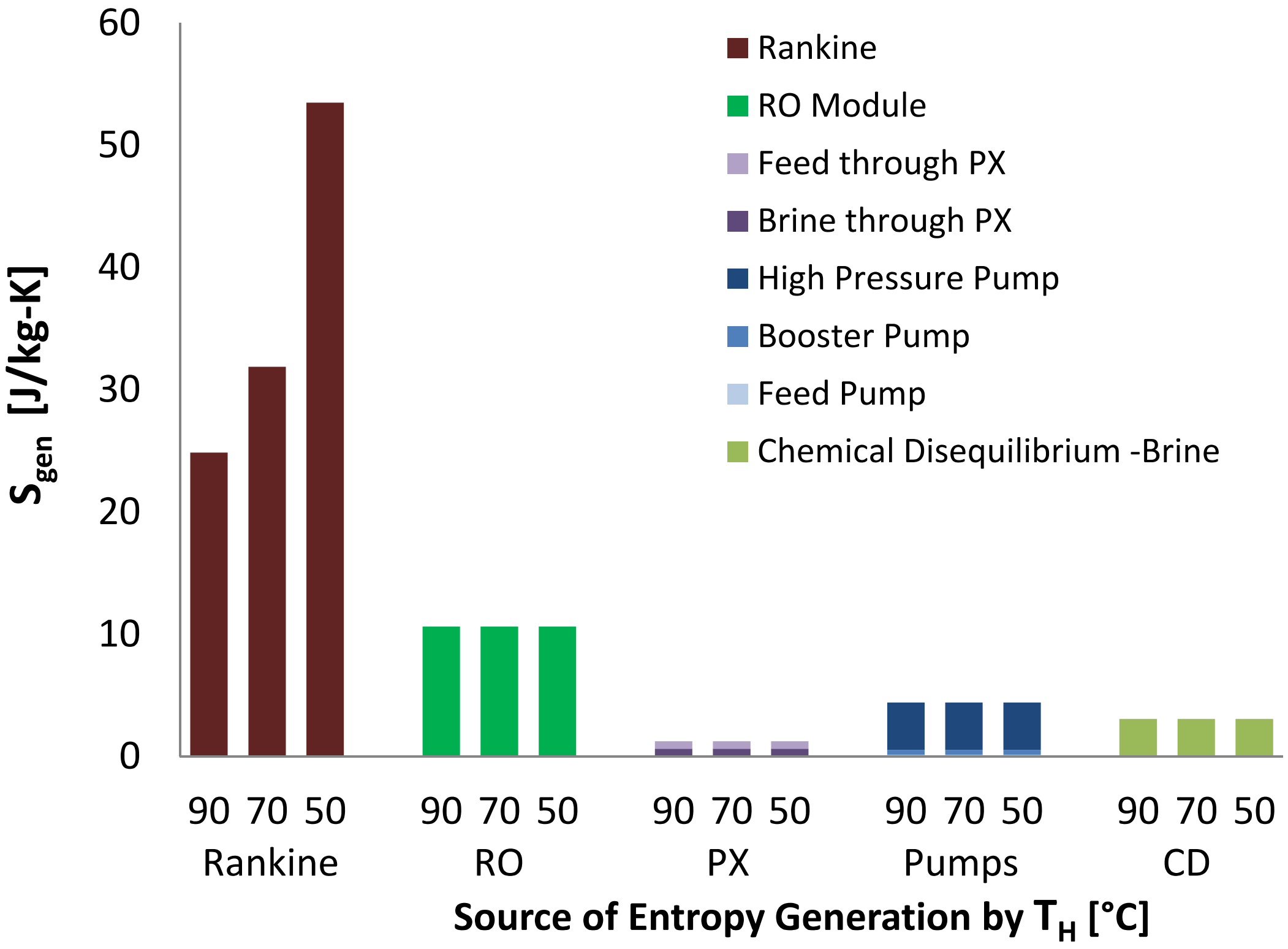

**Figure 2.23.** Entropy generation per kilogram product water produced in each reverse osmosis component for all four temperatures modeled

As seen in the case of MVC, the reduced performance of the Rankine cycle at lower temperatures causes the performance of RO to decrease.

## 2.7 Applicability of Analysis to Systems of Different Costs and Sizes

The effect of energy costs and system size on the efficiency of desalination plants must be understood to apply the results of this study to practical systems. The cost breakdown and data from existing plants can be examined to show the wide applicability of the thermodynamic analysis of this study.

Lower energy costs, which may occur in low temperature waste heat applications, reduce the need for energy efficient design. This affects the expenditure for heat exchanger area, which represents a trade off between large capital costs or large energy costs [78]. Heat exchangers are used in all heat transfers between feed, brine, and product water and from the top heater providing waste heat. As the analysis in this paper shows, the entropy production is largely due to heat transfer across temperature gradients. However, because the heat exchanger and efficiency trade off is universal for all thermally-powered desalination technologies, all exhibit the same trend when energy costs are reduced. Furthermore, a cost analysis shows that the heat exchanger area in representative systems is a small part of total cost of water ($< 1/4$) [53], which means the optimal area changes only little with large changes in the unit price of energy [53]. Comprehensive data on desalination plant sizes shows that for plants online now, MSF, MED, and RO systems of the same size but in regions of different typical energy cost vary very little in efficiency, almost always remaining withing $\pm 20\%$ of their individual average ranges [79, 80].

The efficiencies of online desalination plants varies only mildly across system sizes covering multiple orders of magnitude. For example, with few exemptions, the GOR of MSF plants varies within 7 and 10 for plants between 10,000 and 800,000 $m^3$/day [80], and modern RO unit energy use spans a range of $2.5 - 4$ kWh/$m^3$ for plants ranging from 600 $m^3$/day to 500,000 $m^3$/day [80]. Furthermore, the differences correlate strongly with size, with plants of similar sizes varying even less. Desalination systems do not get worse with increasing size, so the small MVC, MSVMD, and HDH systems modeled maintain or improve their efficiencies for large sizes. However, the comparison is limited for small systems, as MSF and MED systems of the types modeled change dramatically at small sizes, reducing the number of stages to 2-3 for the few installations below 1,000 $m^3$/day [15]. The RO model is not always applicable below 50 $m^3$/day as these plants may lack pressure recovery [80], and some instances of HDH and multistage membrane distillation systems can be extremely small, with real systems below even 1 $m^3$/day, while still providing GOR values similar to those in this paper [14, 81].

## 2.8 Technology Comparison

A summary of the Second Law efficiency from the previous modeling for each technology by heat source temperature is given in Fig. 2.24. The models used shared approximations and assumptions and shared real-world methods for modifying systems for lower temperatures.

The waste heat driven desalination technology with the highest Second Law efficiency, by a large margin, was RO, despite the entropy generated in the ORC. The next highest technology was MED. MVC, MSF, and MSVMD had similar performances. The similarity in performance and efficiency trends for MSF and MSVMD was expected due to the thermodynamic similarity between the MSVMD and MSF technologies. Notably, while a stand-alone MVC system is known to have a higher efficiency than other thermal technologies [52], when paired with an ORC, the efficiency of the MVC system decreases significantly, making MVC performance worse than that of MED but comparable to that of MSF and MSVMD. A simple closed cycle open water HDH system without moist air extractions, has lower efficiency than other thermal technologies at high waste heat temperatures. However, HDH efficiency at $T_H = 50\,^{\circ}\mathrm{C}$ was comparable to MSF, MSVMD and ORC-MVC [66], and HDH systems with extractions can demonstrate much higher thermodynamic efficiency [82, 83].

In considering applications for these technologies operating on seawater, a dominant consideration is the system size, since some technologies are more scalable than others. MSF and MED plants are generally complex and require very large capital costs, especially due to the large heat exchanger areas [15]. Overall, RO with a Rankine cycle is the most efficient for all system sizes: only other constraints such as relatively cheap energy or fouling issues could make the competitors a better choice [13].

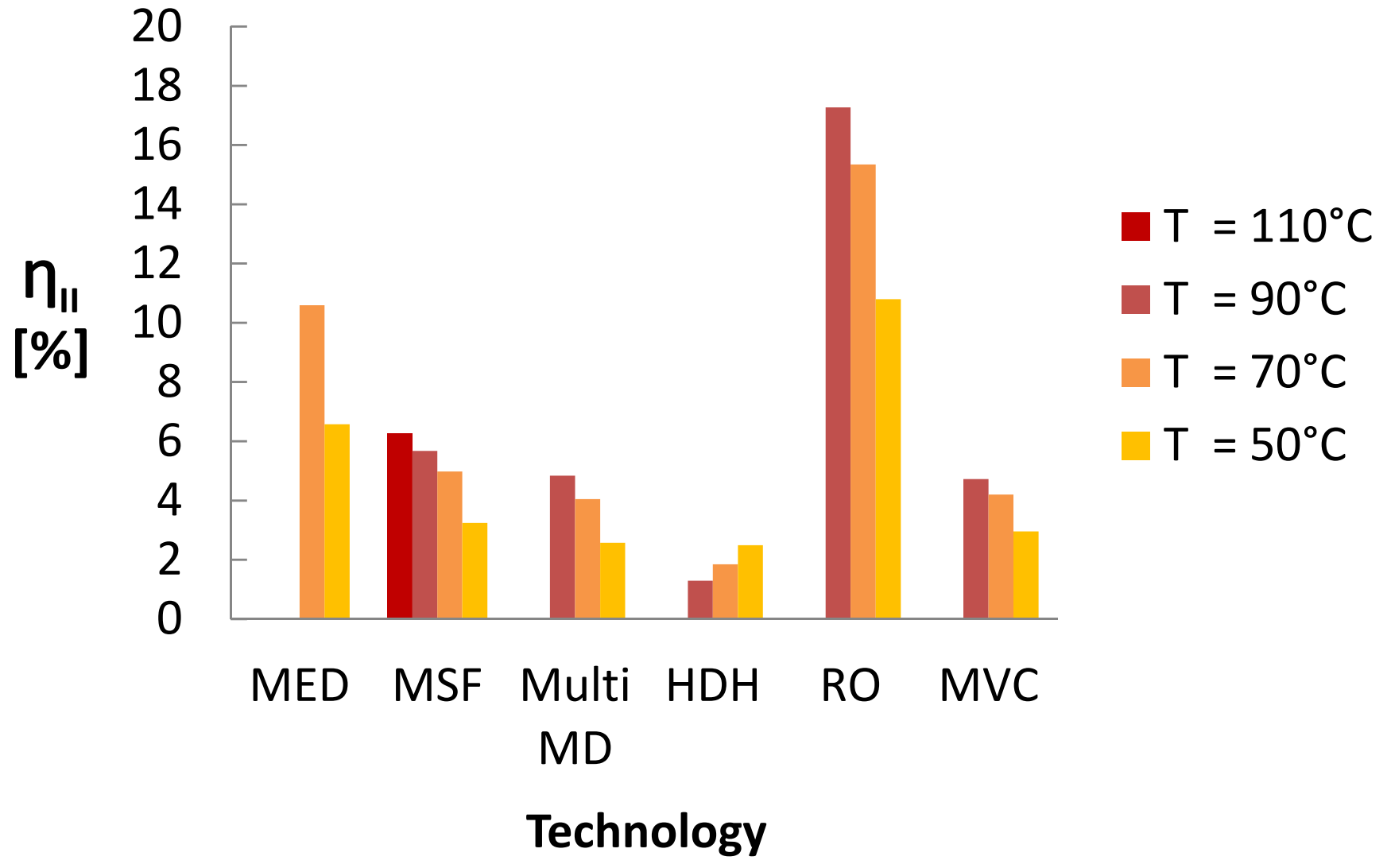

**Figure 2.24.** Second Law efficiency in each technology for all 4 temperatures modeled

## 2.9 Comparison of Entropy Generation in Desalination System Components at Different Temperatures

Sources of Entropy Generation for All Desalination Technologies, Sorted by Largest Maximum Contribution.

1. Temperature Gradients Across Heat Exchangers (In order from most to least $\mathcal{S}_{gen}$: Feed Heaters, Stages or Effects, Top Heaters, Condensers, Regenerators,) .
2. Rankine Cycle Losses.
3. Temperature Disequilibrium of Product and Brine
4. Compressors (MVC)
5. Phase Change (Flashing etc)

6. RO Module
7. Throttling/Expansion
8. Pumping Losses (RO)
9. Chemical Disequilibrium of Brine

Entropy generation was much greater in the thermal technologies, and was generally due to unbalanced heat exchange and temperature gradients in heat exchangers.

The variation in sources of entropy generation between different temperature cases was almost entirely localized to certain system components. The entropy generation for the stages of MSF, MED, and MSVMD changed little between the four temperature scenarios, including in stage sub-components such as the MD module, the MED effects, and the MSF feed heaters and evaporator. For lower top temperatures, the number of stages was also reduced to keep the temperature differences between stages approximately constant. The distillate production rates for these stages changed little as well. For components with heat exchangers involving the distillate, such as the feed heaters, the distillate flow rates were so much smaller than the feed flow rate that, in all cases $\mathcal{S}_{\text{gen}}$ changed little.

However, for components where the entropy generation occurred in the feed or brine streams, tremendous increases in specific entropy generation occurred at lower temperature for the thermal technologies. The reduction in top temperature reduces recovery, causing relatively very large brine and feed flow rates: in these streams entropy is generated from the conditions (e.g., temperature change from conduction across a temperature gradient), but since $\mathcal{S}_{\text{gen}}$ is normalized by $\dot{m}_p$, significantly more entropy is generated per kilogram product water produced. For example, in MSF the entropy generation in the brine heater went from minor to being the largest contribution. In all thermal technologies, the reduced recovery causes an increase in entropy generation for temperature disequilibrium of the brine, again since a larger volume of brine is mixing with the ocean water. This relative flow rate effect also increased the entropy generation in the condenser for MED. The reduced recovery is a natural feature of lower source temperature systems, and is difficult to reduce as that would require increased performance and investment in the stage components, increasing the stage number and reducing temperature differences in the stage heat exchangers. As the stage performance changes little at lower temperature, while other components perform much worse, it hardly makes sense to focus investment in the stages, further limiting the ability to increase recovery with lower temperature heat sources. The lower temperatures also performed worse because at lower temperatures, more entropy is generated by heat transfer across a temperature gradient, as made apparent by the equation for $\dot{s}_{\text{gen}}^{\text{T gradient}}$ in Table 2.1.

For electrically powered desalination technologies, entropy generation increased at lower temperatures due to the decreased efficiency of ORC's at lower driving temperature differences. The efficiency of the ORC significantly dominated their performance. Overall, the efficiency of RO and MVC did not vary quite as much with temperature as the efficiencies of the thermal desalination technologies, but the variation was proportionally very similar.

Entropy generation from chemical disequilibrium actually decreased for the lower temperature cases, since the recovery was reduced. Since this $\mathcal{S}_{\text{gen}}$ is a result of the concentration of brine, it was only proportionally significant for the most efficient systems, such as in RO.

Overall, most of the entropy generation in all these systems occurs due to heat transfer across temperature gradients in heat exchangers. This is seen in the feed heater, brine heater, and regenerator in MSF, the feed heaters and condenser in MED, and the humidifier and dehumidifier in HDH. Other methods experience lower losses for a few reasons. Where phase change occurs, temperature gradients can be small due to very high heat transfer coefficients. For mechanical systems such as pumps and compressors, technically it is easier to avoid losses with well-engineered components.

## 2.10 Conclusions

The impact of source temperature on six waste heat driven desalination technologies was studied. Through component level analysis, the impact on entropy generation within the components was identified which provides useful insight for how to adapt these technologies for lower temperature heat sources.

In comparing the technologies, reverse osmosis paired with an ORC had by far the highest Second Law efficiency.

Component-level entropy generation exhibited consistent trends. For the thermal technologies, entropy generation due to heat transfer across temperature gradients dominated; this occurred in the various heaters, regenerators, condensers, and other heat exchangers.

As top temperatures decreased, overall entropy generation increased because the reduced recoveries increased feed and brine flow rates, causing significantly more entropy generation in those heat exchangers. Meanwhile, $\mathcal{S}_{\text{gen}}$ in the product stream, including that in stages and effects of MSF, MED, and MSVMD. Increased temperature differences across heat exchangers also contributed substantially to the decreased performance at lower temperatures, as did heat transfer at lower temperatures.

For the electrically driven technologies, entropy generation in the ORC's dominated, and contributed to differences between different source temperatures. Entropy generation did not change for the components of the electrical technologies when the source temperature was varied.

When the effect of system size was analyzed with the literature, it was found that among the systems analyzed, HDH, MVC, RO, and multistage MD systems maintained similar efficiencies at small scale ($< 100\text{m}^3$/day). In general, efficiencies in the literature varied within a factor of two across many orders of magnitude, and much less for systems of similar size.

Generally, at lower source temperatures, all technologies performed poorly and, generated more entropy per kilogram of permeate produced than at higher source temperatures. The thermal technologies, excluding HDH, generally had their entropy generation affected more by a change in source temperature than electrical technologies, although all other technologies shared similar trends. HDH experienced the opposite trend, improving performance when the top and bottom temperatures are close together: this occurs because the enthalpy of moist air is nonlinear with temperature, inducing larger temperature differences in the dehumidifier when the temperature range is broader, which leads to higher entropy generation. For that reason, HDH systems with injection and extraction (not considered here) are preferred in practice.

## 2.11 Acknowledgments

This work was partially funded by the Cooperative Agreement Between the Masdar Institute of Science and Technology (Masdar University), Abu Dhabi, UAE and the Massachusetts Institute of Technology (MIT), Cambridge, MA, USA, Reference No. 02/MI/MI/CP/11/07633/GEN/G/00. The authors would also like to thank the King Fahd University of Petroleum and Minerals for partially funding the research reported in this paper through the Center for Clean Water and Clean Energy at MIT and KFUPM.

We would like to acknowledge Jaichander Swaminathan, Emily Tow, Greg Thiel, Sarah Van Belleghem, McCall Huston, Jocelyn Gonzalez, Priyanka Chatterjee, and Grace Connors for their contributions to this work.

## 2.12 Appendices

### 2.12.1 General First and Second Law of Thermodynamics for an open system

As follows is the First Law, conservation of energy, and the Second Law in their most general form for an open system. The First Law is comprised of the change in internal energy over time equaling a sum of the total work $\dot{W}$, the summation of all heat inputs, $\dot{Q}_i$, over all heat reservoirs, the enthalpy transfer into and out of the system from the incoming and outgoing streams, differences in the enthalpy of formation of the species present, and work from changes in volume of the system over time. The Second Law naturally includes the related entropy terms for these energy transfers, but naturally excludes the work term and volume changes over time since these don't create entropy, and also includes entropy generation, $\dot{S}_{gen}$.

$$\frac{\mathrm{d}U}{\mathrm{d}t} = \sum_{i=0}^{reserviors} \dot{Q}_i - \sum_{j=0} \dot{W}_j + \sum_{k=1}^{in} \dot{m}_k \; h_k - \sum_{l=1}^{out} \dot{m}_l \; h_l + \sum_{m=1}^{species} \dot{m}_{0,m} \; h_{0,m} - p_\infty \frac{\mathrm{d}V}{\mathrm{d}t} \tag{2.13}$$

$$\frac{\mathrm{d}S}{\mathrm{d}t} = \sum_{i=0}^{reservoirs} \frac{\dot{Q}_i}{T_i} + \sum_{k=1}^{in} \dot{m}_k \; s_k - \sum_{l=1}^{out} \dot{m}_l \; s_l + \sum_{m=1}^{species} \dot{m}_{0,m} \; s_{0,m} + \dot{S}_{\mathrm{gen}} \tag{2.14}$$

In the First Law, 2.13, and Second Law, 2.14, $\dot{m}_i$ is the mass flow rate, $h_i$ is the specific enthalpy, $h_0$ is the enthalpy of formation, $s_i$ is the specific entropy, $p_o$ is the atmospheric pressure, and $V$ is the volume.

Hypothetically, in addition to the heat input, $\dot{Q}$, additional inputs can be added to drive the system, including work, which is usually the case for RO, use of salinity differences (chemical potential) which is used in Reverse Electrodyalysis, or a pressure difference between the streams. While such terms are excluded above for the study of waste heat, they are included in the most general form of the First and Second Law of thermodynamics for this control volume, which this derivation will begin with as seen below in the following subsection.

For the purposes of a desalination power plant powered by waste heat, with the control volume defined above that allows for streams to reach equilibrium, the First and Second Law can be simplified significantly. The unsteady terms are eliminated, as is the term for the enthalpy of formation of species since no chemical reactions are performed (merely separation of chemicals). The summation of incoming terms becomes the seawater feed (subscript $sw$), and the outlet terms become the product (subscript $p$) and brine (subscript $b$). The First and Second Law for a thermally-driven desalination system simplify to 2.2

### 2.12.2 Other Technologies

While the main seawater desalination technologies were covered above, a few other technologies are worth mentioning and justifying their exclusion. These include electrodialysis (ED), thermal vapor compression (TVC), Forward Osmosis (FO), and Membrane Distillation (MD), as explained below.

First, electrodialysis (ED) is often used to desalinate brackish waters; it uses a voltage difference and membranes that only pass either positively or negatively charged ions to desalinate water. While cost effective and efficient at low salinities, ED is significantly inferior to RO at seawater salinities and is not used for seawater desalination at an industrial scale.

Instead of MVC, thermal vapor compression (TVC) may be suggested, since this would directly use the heat. However, this technology generally uses ejectors, which are very inefficient. The second law efficiency for these systems is generally similar to but inferior to MVC paired with an ORC.

Forward Osmosis (FO) is also excluded from this study. The technology differs from reverse osmosis since it causes osmotic flow through a membrane by using a highly saline draw solution instead of pressure.

Finally, other configurations of membrane distillation may be suggested, including direct contact (DCMD), air gap (AGMD), and other multistage configurations. The vapor compression multistage cycle chosen gives better GOR values.

nologies are worth mentioning and justifying their exclusion. These include electrodialysis (ED), thermal vapor compression (TVC), Forward Osmosis (FO), and other configurations of Membrane Distillation (MD), as explaiend below.

First, electrodialysis (ED) is often used to desalinate brakish waters; it uses a voltage difference and membranes that only pass either positively or negatively charged ions to desalinate water. While cost effective and efficient at low salinities, ED is significantly inferior to RO at seawater salinities and is not used for seawater desalination at an industrial scale.

Instead of MVC, thermal vapor compression (TVC) may be suggested, since this would directly use the heat. However, this technology generally use ejectors, which are very inefficient. The Second Law efficiency for these systems is generally similar to but inferior to MVC paired with an ORC

Forward Osmosis (FO) is also excluded from this study. The technology differs from reverse osmosis since it causes osmotic flow through a membrane by using a highly saline draw solution instead of pressure

Finally, other configurations of membrane distillation may be suggested, including direct contact (DCMD), Air gap (AGMD), and other multistage configurations.

### 2.12.3 Second Law Efficiency for Brine Reduction Applications

The previous analysis uses minimum least work as the comparison for ideal efficiency, which is the correct benchmark for seawater desalination systems where the only desirable result is product water. This can be referred to as the total dead state, or TDS, where the brine stream reaches chemical equilibrium with seawater. In desalination systems that work with brine concentration as an important result, such as in fracking, the least work should be used, which describes the work needed for a given recovery ratio of product water to seawater. Such analysis would resemble the above, with two major differences. First, all the $\dot{S}_{gen}$ would be replaced with $\dot{S}_{gen}^{\mathrm{RDS}}$, where RDS stands for reduced dead state, and gives the entropy generated without diluting the brine stream. Second, $\dot{W}_{\mathrm{least}}$ replaces $\dot{W}_{\mathrm{least}}^{\mathrm{min}}$, where the former is defined in 2.3.

### 2.12.4 Incompressible Fluid

For an incompressible fluid, the enthalpy only depends on $\Delta h = c\Delta T + v\Delta p$, and entropy depends on $\Delta S = ln(T_2/T_1)$ .

### 2.12.5 Ideal Gas

For an ideal gas, the enthalpy is the same as that for an incompressible fluid but with $c_p$ not $c$. The entropy is $\Delta s = c_p ln(T_2/T_1) - Rln(P_2/P_1)$.

### 2.12.6 Flashing

For a throttle, the enthalpy change and entropy generation can be calculated by experimental data table lookups for the saturated liquid and vapor values by using quality, $x$. These lookups can be combined with the incompressible and ideal gas relations for enthalpy and entropy to generate the equation in the table.

### 2.12.7 Heat Transfer

The entropy generation from heat transfer is the standard result from a heat exchanger control volume, totaling the flow of entropy in and out with the given temperatures, calculated assuming an energy balance and another condition, such as a set temperature difference.

### 2.12.8 Expanders

To calculate the entropy generation for expansion devices for gases or liquids with work output, the enthalpy change is set to the work plus losses, $\Delta h = \dot{W} + T_0 S_{gen}$. This is related to the device efficiency $\eta_{device} = \dot{W}_{out}/\Delta h$. This can be combined with the ideal gas or incompressible fluid enthalpy and entropy relations mentioned above to generate the entropy generation equation for expanders.

### 2.12.9 Compressors

Compressors may compress a gas, or if pressurizing a fluid, are simply pumps. The efficiency of a pump or compressor is simply an enthalpy change ratio of ideal over actual; $\Delta h_{rev}/\Delta h$. The entropy generation in such a device is simply the exergy destruction divided by the temperature difference, or the difference between the actual and ideal work divided by temperature: $s_{\text{gen}}^{\text{pumping}} = \Xi_d/T_0 = W - W_{rev}/T_0$. The entropy generation for compressors and pumps is derived by using this, along with the definition of pump efficiency and the entropy and enthalpy equations for ideal gases and incompressible fluids respectively.

### 2.12.10 Multistage Flash Modeling Results without Regeneration

**Table 2.12.** MSF-OT Results without a Regenerator

| Parameter | | | $T_H$ | | | |
|---|---|---|---|---|---|---|
| Output | | | 110 °C | 90 °C | 70 °C | 50 °C |
| Performance ratio | PR | [-] | 6.3 | 4.4 | 2.6 | .8 |
| Gained output ratio | GOR | [-] | 6.3 | 4.4 | 2.5 | .7 |
| Recovery Ratio | RR | [%] | 11.2 | 7.7 | 4.4 | 1.3 |
| Steam flow rate | $\dot{m}_s$ | [kg/s] | 60.6 | 59.2 | 57.9 | 56.7 |
| Brine salinity | $y_n$ | [g/kg] | 39.4 | 37.9 | 36.6 | 35.5 |
| Second Law Efficiency | $\eta_{II}$ | [%] | 4.21 | 3.64 | 2.98 | 1.45 |
| Entropy Generation | $\mathcal{S}_{\text{gen}}$ | [J/kgK] | 460.9 | 459.1 | 566.9 | 1044.7 |

# Chapter 3. Effect Of Module Inclination Angle On Air Gap Membrane Distillation

## 3.1 Abstract

Air gap membrane distillation (AGMD) experiments were performed with varied temperature and varied module inclination angles to characterize the effect of module angle on permeate production and thermal performance. While AGMD is potentially one of the most energy efficient membrane distillation configurations, transport resistances in the air gap typically dominate the thermal performance, resulting in degraded permeate production. Tilting the module away from vertical offers the opportunity to manipulate the condensate layer and its associated thermal resistance. In this study, we report experiments on varying module tilt angle performed with a flat plate AGMD module under fully characterized heat and mass transfer conditions. Numerical modeling is also performed to better understand the experimental results. The tests indicated that the AGMD system behaves as a "permeate gap", or flooded membrane distillation system for declined and extremely inclined positions. A key finding relevant to all AGMD systems is that at highly negative tilt angles (more than 30 degrees), condensate may fall onto the membrane causing thermal bridging and increased permeate production. Near vertical and positive tilt angles (<15 degrees from vertical) show no significant effect of module tilt on performance, in line with model predictions. Jaichander Swaminathan and Professor John Lienhard V contributed to this work [84].

## 3.2 Introduction

Membrane distillation (MD) is an emerging thermally-powered desalination technology with unique advantages at small scales, and potential to reach superior efficiencies compared to existing desalination technologies. In an MD system, a hot, saline solution with a high vapor pressure flows across a hydrophobic membrane which selectively allows water vapor to pass through but not liquid water or dissolved salts. Pure water diffuses through the membrane and is condensed and collected on the other side. Due to its scalability and low maintenance, most applications to date for this emerging technology have been for small installations, some driven by solar thermal power [85]. Recent papers have suggested that MD can theoretically provide superior efficiencies to all other thermal desalination technologies, including Multi-Stage Flash (MSF) and Multi-Effect Distillation (MED) [86, 87], although actual test results have shown more modest thermal performance [88]. Several configurations have been designed for MD, including the most commonly, Direct Contact Membrane Distillation (DCMD) and Air Gap Membrane Distillation (AGMD), as well as Vacuum Membrane Distillation (VCMD), and Sweeping Gas Membrane Distillation (SGMD) [10]. Of these configurations, it has been shown that AGMD has higher potential for superior thermal efficiencies [22].

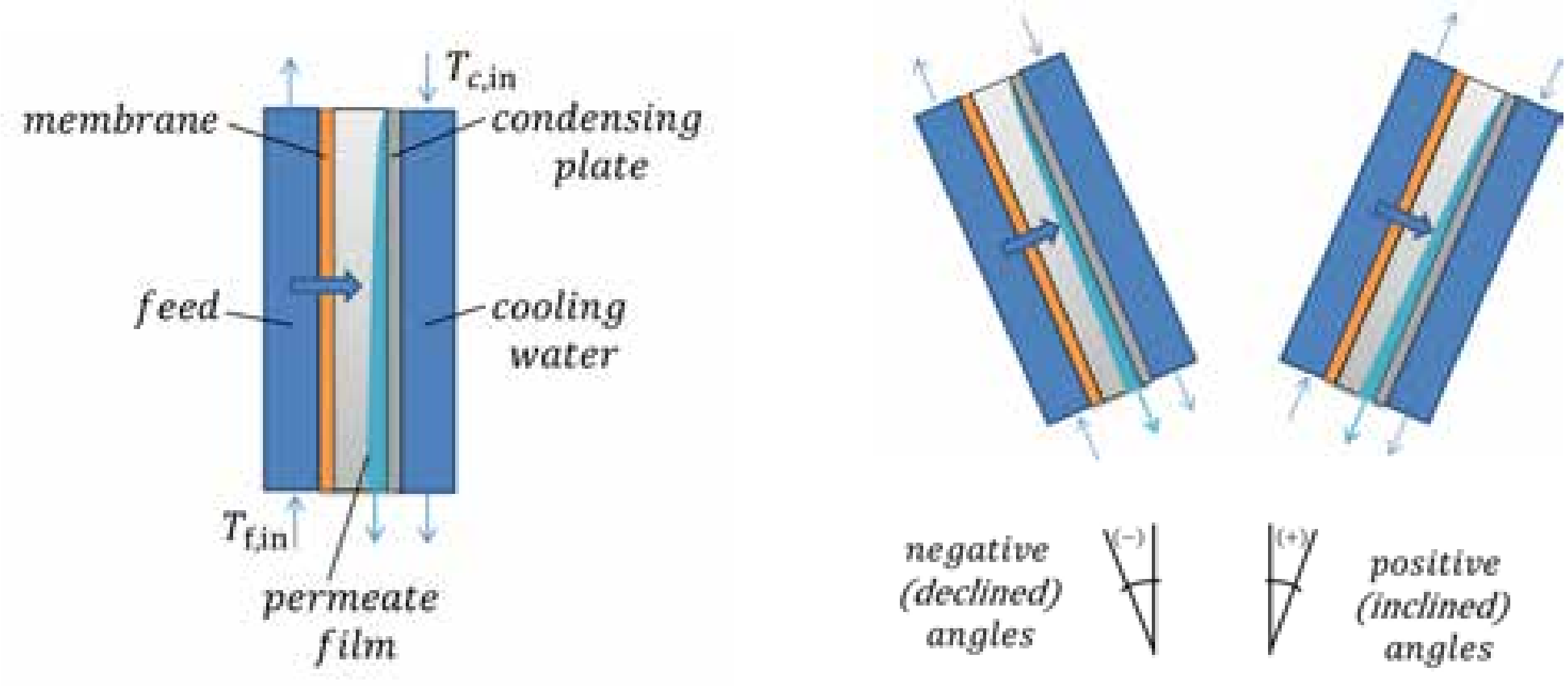

**Figure 3.1.** AGMD at varied angles

AGMD includes a saline feed, a hydrophobic membrane, an air gap behind the membrane, and a condensing surface beyond the air gap. Various configurations have been designed, including flat sheet designs, a tubular configuration with concentric cylinders, hollow fiber modules and spiral wound modules [10]. This study examines the effect of varying the tilt angle of AGMD modules, which is primarily applicable to flat plate configurations, but also to tubular and spiral wound systems. Flat plate systems are among the most common due to their ease of manufacturing and assembly, and are used in experimental systems as well as commercial modules such as those produced by MEMSYS [13].

In AGMD systems with small air gaps, an important phenomenon that can occur during operation is flooding and associated thermal bridging. Flooding occurs when the permeate production rate exceeds the rate of condensate removal from the air gap. This effect reduces mass transfer resistance, tending to increase permeate production rate, but it also increases heat loss from and temperature polarization in the feed channel as heat is conducted directly through the water from the membrane to the cooling plate. Thermal bridging may also occur at small, localized regions at which water falls from the condenser onto the membrane surface for declined tilt angles, if local bridging of the membrane and condensation plate occurs. As this phenomenon is unsteady and localized, it cannot be readily modelled in the generic model.

Module tilt angle has been studied in many condensation technologies. While the vertical module orientation has been the norm in AGMD studies [89], horizontal module orientation has also been used in some cases [90], which we define here as angles approaching positive 90° from vertical as seen in Figure 3.1 above. While relatively unimportant in other configurations such as DCMD, where tilt angle only has hydrostatic effects, the module tilt angle affects droplet flow and film thickness on the condenser surface in AGMD systems. The literature generally lacks experiments at other angles, although 45 degrees has been examined for DCMD, not AGMD, configurations in experiments using bubbles to encourage turbulence [37]. To the authors' knowledge, the present paper is the first detailed experimental and theoretical analysis of the effect and optimization of tilt angle on an AGMD process, and the first to examine declined angles for AGMD.

## 3.3 EXPERIMENT DESIGN

To analyze the effect of module tilt angle, experiments were performed on an AGMD test bed under fully characterized heat and mass transfer operating conditions. This system consists of two controlled fluid flow loops which serve a tilt-enabled central testing module.

### 3.3.1 APPARATUS DESIGN

As seen in Figure 3.2, a hot saline loop circulates feed water past one side of the membrane, and a separate cold side loop chills the condensing plate. Both loops use tanks for thermal storage to stabilize temperature, and include pumps, flow meters, and temperature controllers.

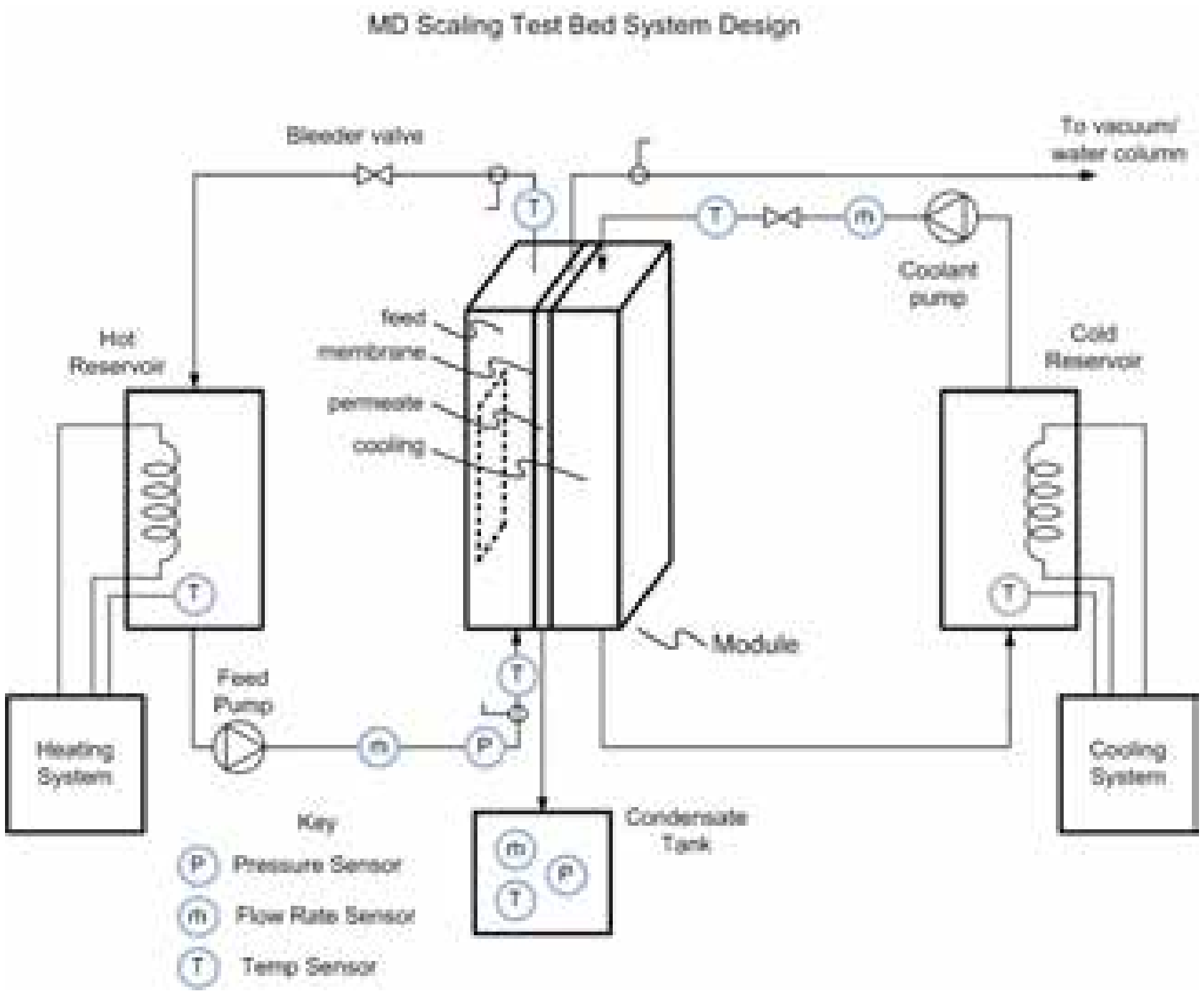

**Figure 3.2.** Experiment setup

As seen in Figure 3.3, the module consists of a series of plates for the various channels. The feed and cooling channels are made of polycarbonate plates with machined channels. A polycarbonate spacer separates the membrane and aluminum condensing plate.

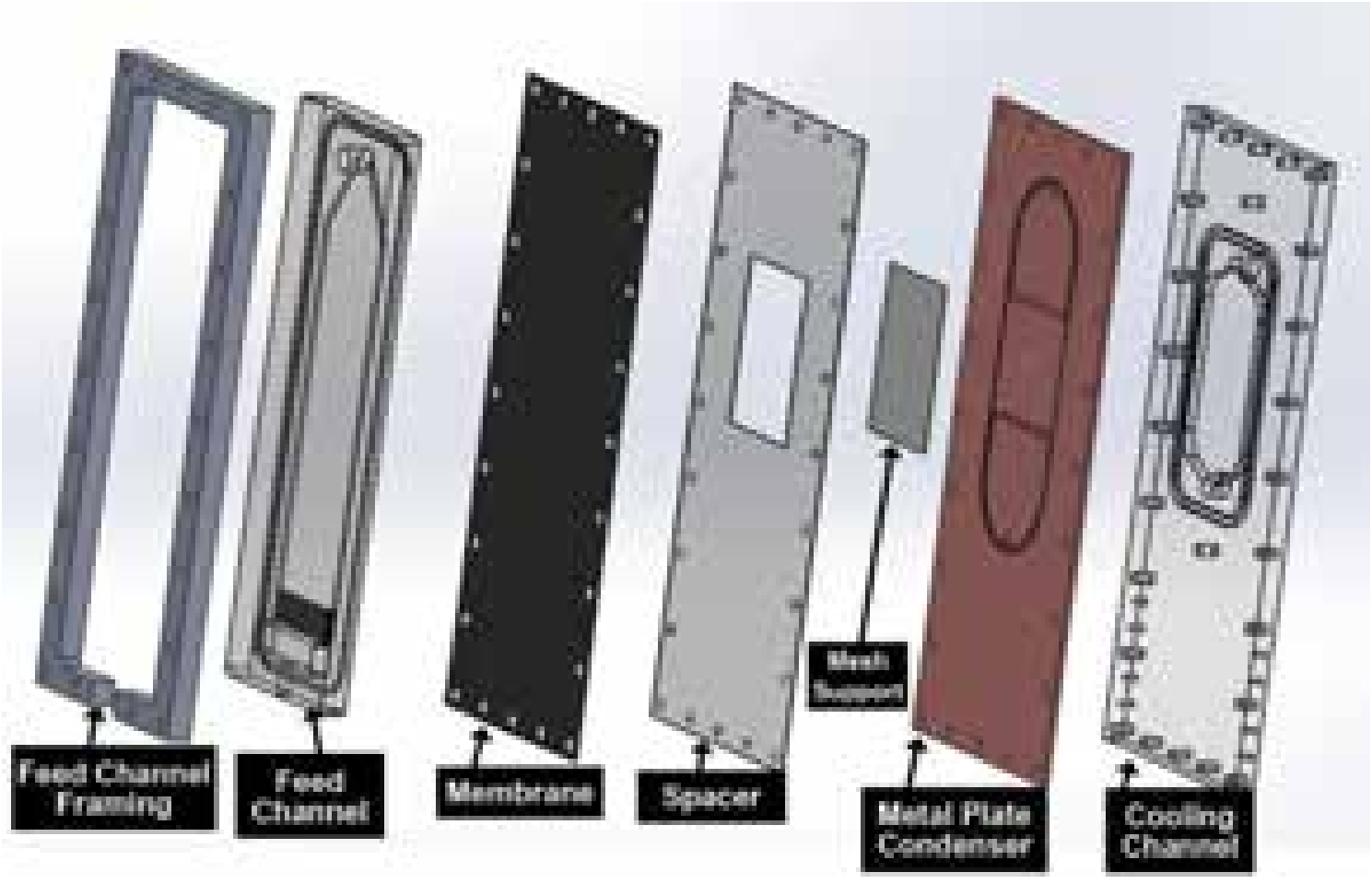

**Figure 3.3.** AGMD Module Plate design

The channels can be described by their internal dimensions, where the length is the stream-wise dimension from inlet to outlet and runs vertically in the above diagrams, the width runs horizontally, and the depth is the shallow dimension cut into each plate. While the feed plate is 35 cm long, only 16 cm is exposed to the membrane, with the region preceding it ensuring a fully developed turbulent flow beneath the active membrane region.

**Table 3.1.** Parameters for AGMD module experiments

| | **Feed Channel** | **Air Gap** | **Cooling Channel** |
|---|---|---|---|
| **Length** | 35 cm | 16 cm | 16 cm |
| **Width** | 12 cm | 12 cm | 12 cm |
| **Depth** | 4 mm | 1 mm | 10 mm |
| **Pressure** | 1.4 atm | 1 atm | NA |
| **Temperature** | 40°C - 70°C | 10°C - 40°C | 10°C - 50°C |

The membranes used are hydrophobic Immobilon-PSQ membranes. Although originally designed for protein binding, they have good characteristics for MD and have also been used in previous AGMD studies [22]. The membranes have an average pore size of 0.2 µm, a maximum pore size of 0.71 µm, and a porosity of 79.2%. A membrane coefficient of $B = 1.6\times10^{-7}$ s/m is used to characterize the permeability in the numerical models [85]. A fine mesh spacer is used to keep the membrane's flat shape; McMaster part number 9265T51. A course mesh spacer with most horizontal wires removed maintains the air gap width; McMaster part number 9275T65 [91].

## 3.4 NUMERICAL MODELING

### 3.4.1 MODELING METHODS AND FEED CHANNEL MODELING

The theoretical performance of the AGMD system at varied angles is estimated using numerical modeling techniques with Engineering Equation Solver (EES) [58], which is an iterative equation solver with several thermodynamic property functions built into it. A one-dimensional modeling approach is followed in which the temperatures and flow rates vary along the length of the module. The width of the module is assumed to be long so that the effect of the walls is negligible and properties are constant along this direction. In the depth direction, the boundary layers and associated resistances are taken into account, and the difference between the temperature and concentration for the bulk stream and at the membrane interface is evaluated using suitable heat and mass transfer coefficients.

The primary modeling calculations involve mass and energy conservation equations applied to each of the module sections: the feed channel, the air gap, and the coolant channel (Figure 4). Each section is coupled with suitable transport equations. A detailed description of the overall modeling methodology applied for the case of vertical module orientation is given by Summers et al. [85]. The heat and mass transfer coefficients are determined through the Nusselt and Sherwood numbers evaluated within each stream. The equations are solved using EES, and the number of computational cells was progressively increased to 120, by which point the results were seen to be grid independent.

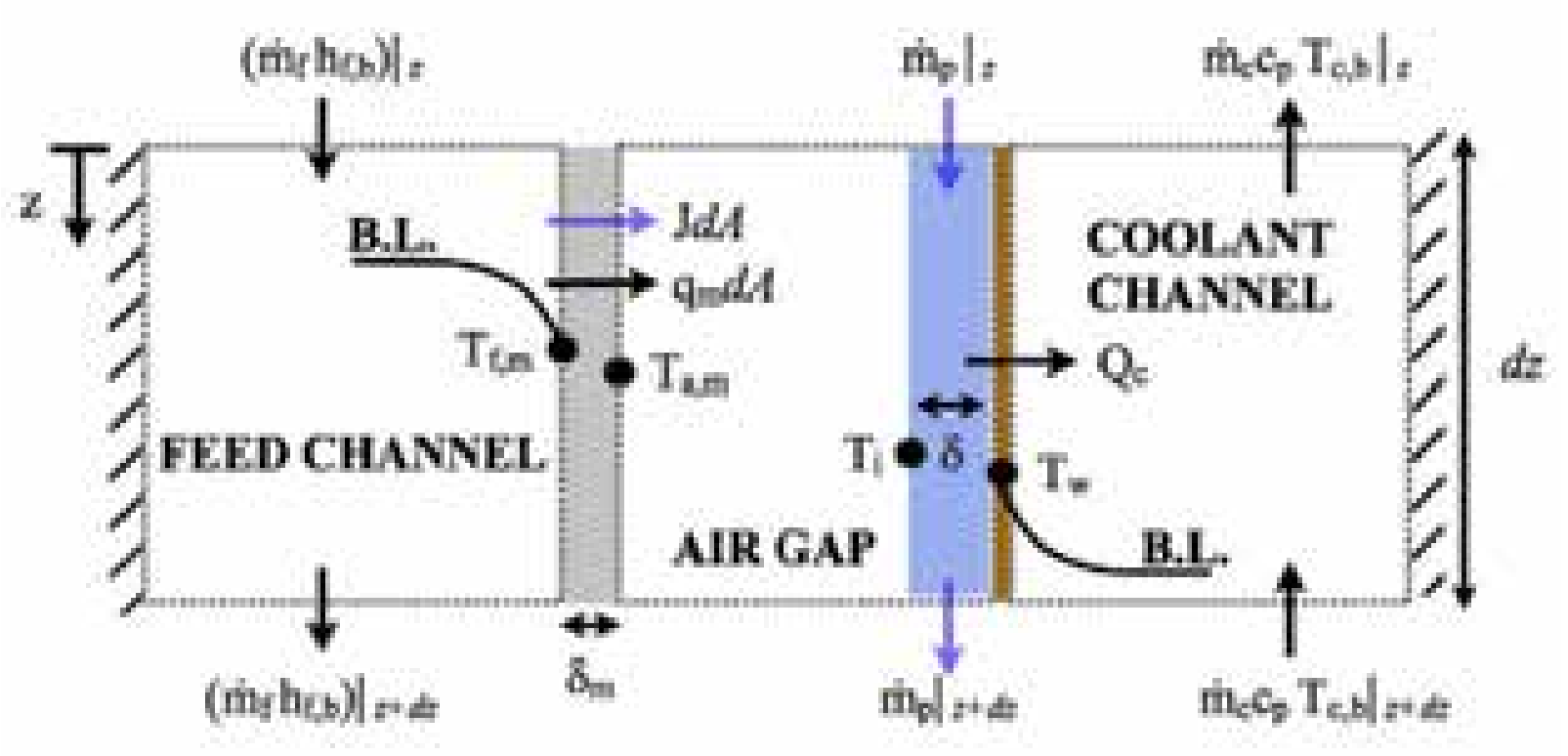

**Figure 3.4.** An integral control volume of a computational cell for the AGMD model from [22].

The mass flow of water vapor is given by the membrane characteristic equation (Eq. 1) and depends on the water vapor pressure on either side of the membrane. The mass flow rate in the feed stream reduces correspondingly.

$$J = B \times (P_{vap,f,m} - P_{vap,a,m}) \tag{3.1}$$

Here $J$ is the permeate flux, $B$ is the membrane permeability, $P_{\text{vap,f,m}}$ is the vapor pressure on the feed side surface of the membrane, and $P_{\text{vap,a,m}}$ is the vapor pressure on the air gap side surface of the membrane.

### 3.4.2 Air Gap and Condensing Channel Modeling

The air gap modeling is especially important since the transport resistances in this region significantly affect the AGMD module performance. Varying the tilt angle of the module can affect the process only through physical changes in this region. Typically, a spacer mesh is placed in this region to prevent the flexible membrane from collapsing onto the condensation surface. Like most other numerical models of the AGMD module, the spacer is not explicitly considered and a free air gap is modeled.

During operation, part of the membrane is pressed down into the gaps in the mesh and the effective air gap depth is reduced. For modeling purposes, the membrane is considered to be flat for simplicity. The effective depth of the air gap is therefore taken at a value lower than the design value in order to account for this effect.

No condensation or heat transfer through the polypropylene mesh spacer is considered, as the conductivity of the mesh is very low, limiting both conduction through it, and keeping it at a higher temperature than the condensation plate, limiting condensation on the mesh. The water vapor flux entering the air gap diffuses through the air layer to reach the film and condense. The air layer is less than 1 mm in depth and hence any convection effects are ignored.

The flux through the membrane is a function of the vapor fraction at the interface on the air gap side. The vapor diffuses to the water interface from this interface. The diffusion is governed by binary mass diffusion as described in Lienhard and Lienhard [27].

$$\frac{J_m}{M_w} = \frac{c_a D_{w-a}}{d_{gap} - \delta} \, ln\left(1 + \frac{x_i - x_{a,m}}{x_{a,m} - 1}\right) \tag{3.2}$$

Here, $J_m$ is the flux through the membrane, $M_w$ is the molecular weight of water, $c_a$ is the concentration of air, $D_{w-a}$ is the diffusivity of water in air, $d_{gap}$ is the air gap width, $x_i$ is the water mole fraction at the liquid-vapor interface, and $x_{a,m}$ is the water mole fraction at the membrane interface.

The thickness of the air gap increases as more liquid condenses into it. By mass conservation, assuming no shear at the liquid-air interface, the rate of growth of the film can be evaluated as

$$\delta_{i+1}^3 = \delta_i^3 + \frac{3\, J_i \, dA \, \nu_{f,i}}{g \cos\theta \left(\rho_f - \rho_g\right) w} \tag{3.3}$$

where $\delta_i$ is the condensation film thickness, $dA$ is the differential of area, $\nu_{f,i}$ is the fluid kinematic viscosity, g is the gravitational constant, $\rho_f$ is the fluid density, $\rho_g$ is the combined air and water vapor density, and $w$ is the width [25].

Note that the denominator here has $\cos\theta$ to account for the angle of inclination. When the module is vertical, $\theta = 0^{\circ}$. The formula is used for angles as high as $85^{\circ}$, since the evaluated film thickness is still smaller than the air gap thickness. The heat of condensation is conducted across the film, through the aluminum wall, and into the coolant liquid.

### 3.4.3 Modeling Inputs

The numerical model takes the following inputs: geometry of the experimental setup including the length, width, and depth of each of the channels; hot water flow rate and temperature as it enters the module; and cold water flow rate and temperature at module inlet.

### 3.4.4 Effect of Module Tilt Angle

Figure 3.5 shows the effect of tilt angle on flux in the model. The permeate production rate at a given module inclination is normalized with respect to the flux at vertical orientation and the ratio is plotted with respect to inclination angle. With all other conditions remaining the same, the thickness of the condensate film is affected by angle (Eq. 3). With a thicker film, the effective diffusion length ($d_{gap}$ minus $\delta$) for water vapor is reduced, and hence the flux increases (Eq. 2). This effect is independent of whether the module tilt angle is in the positive or negative direction.

An increase of around 4% is possible at very large module tilt angles as a result of the film becoming thicker. Figure 5 shows that this increase is higher for cases with lower cold side temperatures since viscosity of the film is higher and hence the film rolls off the surface slower. This small increase does not include the effects of thermal bridging or flooding.

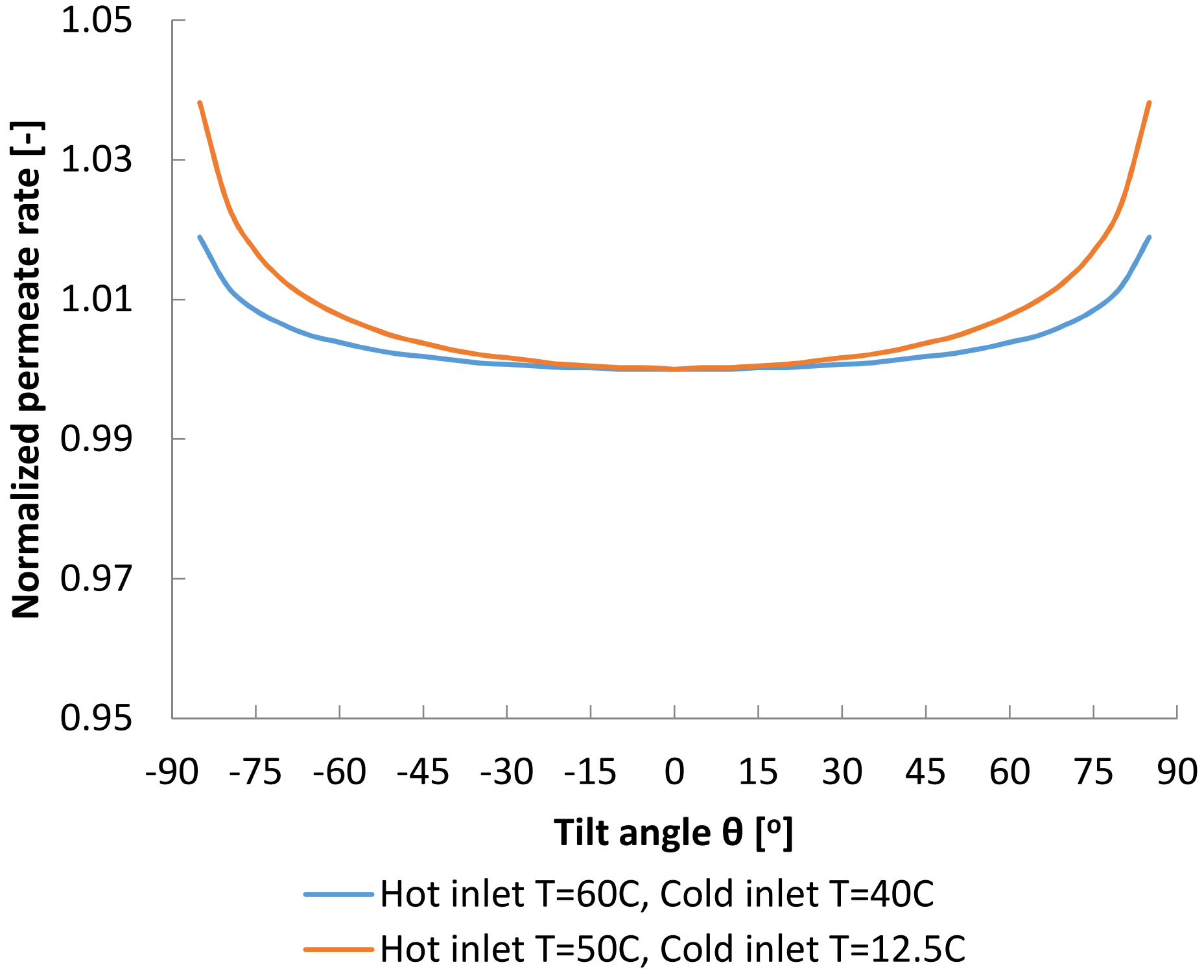

**Figure 3.5.** Effect of module tilt angle on flux predicted by model.

# 3.5 METHODOLOGY

## 3.5.1 EXPERIMENTAL METHODOLOGY

The measured permeate production rates, adjusted for temperature variability, and are presented. The hot side inlet and cold side inlet temperatures were set and maintained using temperature controllers during experiment operation. The hot side temperature in this experiment varies by ±0.2 °C over the course of an experiment; and on the cold side, the temperature controller, which actuates a coolant flow valve, achieves temperature control of ±0.5 °C around the set-point temperature. Between experiments at different angles, the average inlet temperatures of the streams can change. The difference in average temperature on the feed side is at most 0.4°C and on the cold side it is at most 1°C. While Figure 3.5 shows the relatively minute effect of tilt angle on flux, the stream temperatures have a much larger influence on flux. This is especially true because vapor pressure, which is the driving force for MD mass transfer, is an exponential function of temperature. In order to separate the effect of angle, the variability caused by small changes in average inlet temperatures must be identified and removed.

Using the corresponding average inlet temperature and flow rate conditions at each of the tilt angles, flux is estimated using the EES model for a vertical module operating under those conditions. The measured flux for the AGMD experimental system at vertical orientation was between 150-250 $l/m^2$day depending on the feed flow conditions and the type and thickness of the air gap spacer used. The measured permeate production rates are scaled by multiplying by vertical module flux obtained from EES at the same temperature conditions divided by the experimentally obtained flux at 0° tilt angle. The scaled experimental fluxes are then each divided by scaled flux at 0°. The ratio obtained is called the normalized permeate rate and will have a value of unity at 0° tilt angle. At other tilt angles, the deviation from unity would indicate the relative change in flux that results from tilting the apparatus if other conditions were to remain exactly the same.

### 3.5.2 Uncertainty Quantification

Uncertainty analysis was performed within the EES code to account for uncertainties in numerically predicted flux as a result of uncertainties in measured temperatures and flow rates of the hot and cold streams. The overall uncertainty in flux was generally dominated by temperature variations. While the hot side temperature has a larger impact on flux, the absolute uncertainty in cold side temperature is higher, so that both affect overall uncertainty. The uncertainty in actual temperature measurements was estimated conservatively as the standard deviation in temperature recorded during the experiment plus the maximum measurement uncertainty of ±0.2°C for the thermistors used.

Experimentally, the flux is determined over a period of 10-15 minutes by subtracting the final water mass from the initial water mass and dividing by the total elapsed time. Each measurement is carried out at time intervals of about 5 seconds by the data acquisition system. Hence a time measurement uncertainty of 10 seconds is considered. A conservative time error was used to account for variability in permeate drop collection from the apparatus. The measurement uncertainty on the mass scale is ±0.1g. In order to account for both the initial and final readings, an uncertainty of ±0.2g is considered for the total water mass collected.

## 3.6 EXPERIMENTAL RESULTS

### 3.6.1 EFFECT OF ANGLE ON PERMEATE FLUX

The experiments and modeling showed a minimal effect for small inclination angles on permeate flux. Very large angles, however, did produce a substantial increase in permeate production rate. This result is further supported by theoretical analysis of the relationship between the film thickness $\delta$ and the mass transfer resistance in the air gap and flux. The main parameter of interest is the permeate production rate; as the end product, permeate flux indicates the performance of the fixed size AGMD system with fixed top and bottom temperatures of the system. While energy efficiency of the system is also important, our apparatus does not incorporate energy recovery at the condenser, and is not intended for energy consumption studies.

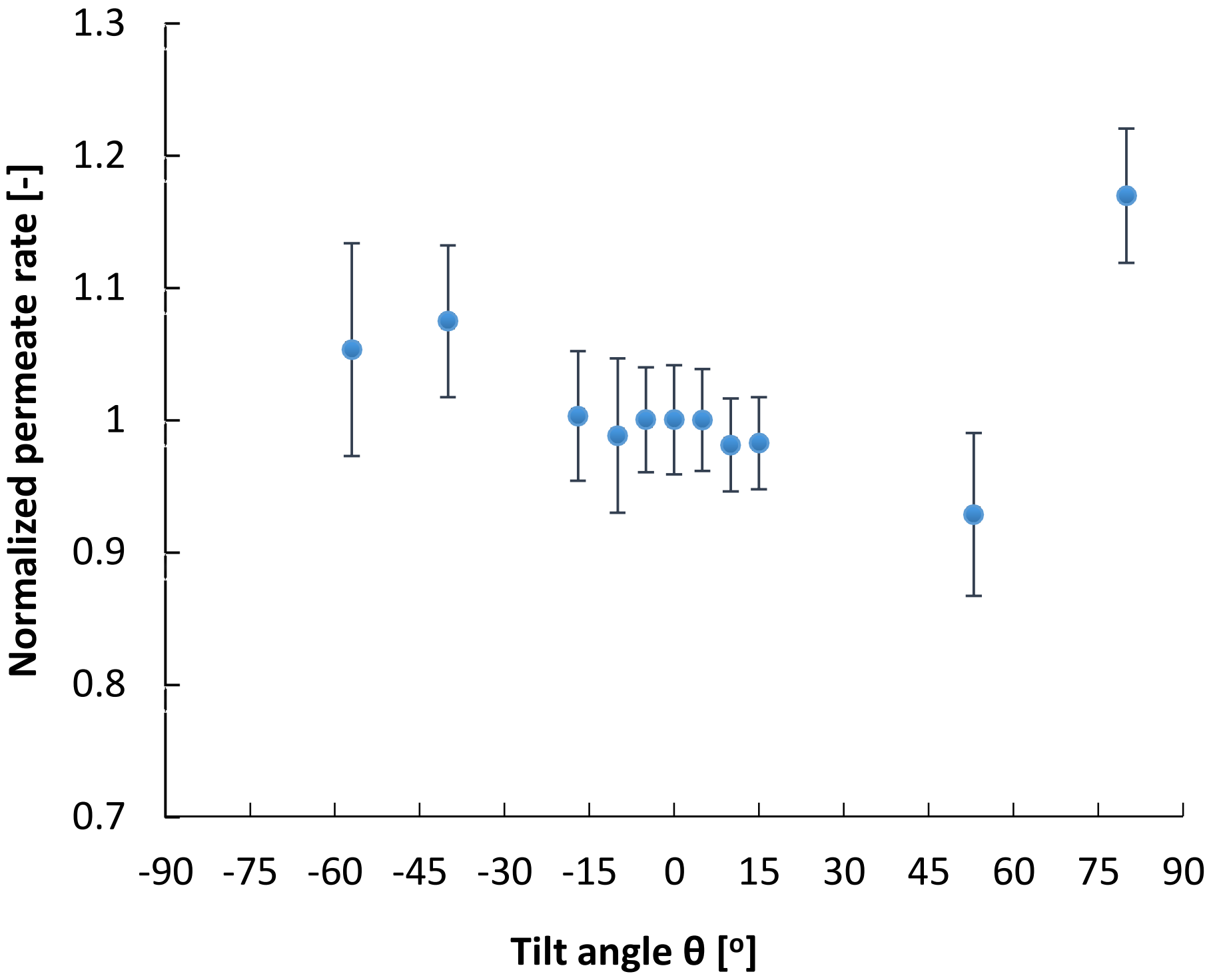

**Figure 3.6.** Effect of module tilt angle on permeate production: $T_{f,in} \approx 50°C$, $T_{c,in} \approx 12.5°C$

Figure 6 shows the normalized permeate flow rate vs. tilt angle. The normalized permeate flow rate is the permeate flow rate for the given angle normalized for temperature changes and divided by the permeate flow rate for vertical module orientation (0°). In Fig 6., MD was performed at moderate temperatures with a high temperature difference, ΔT, between condensate and feed streams and a very low condensate temperature. Permeate flux remained relatively stable but increased at extreme angles, both for negative and positive positions. The large ΔT and very low cold side temperature caused a relatively high permeate flux. The data show a substantial jump in permeate flux at 85°, which is attributed to air gap flooding. For this nearly horizontal case, the role of gravity in draining the film is substantially diminished (c.f., Eq. 3). Additionally, for the low condenser temperature of 12.5° C, the viscosity of liquid water is roughly twice the value in later trials at 30°C, also contributing somewhat to a thicker condensate film.

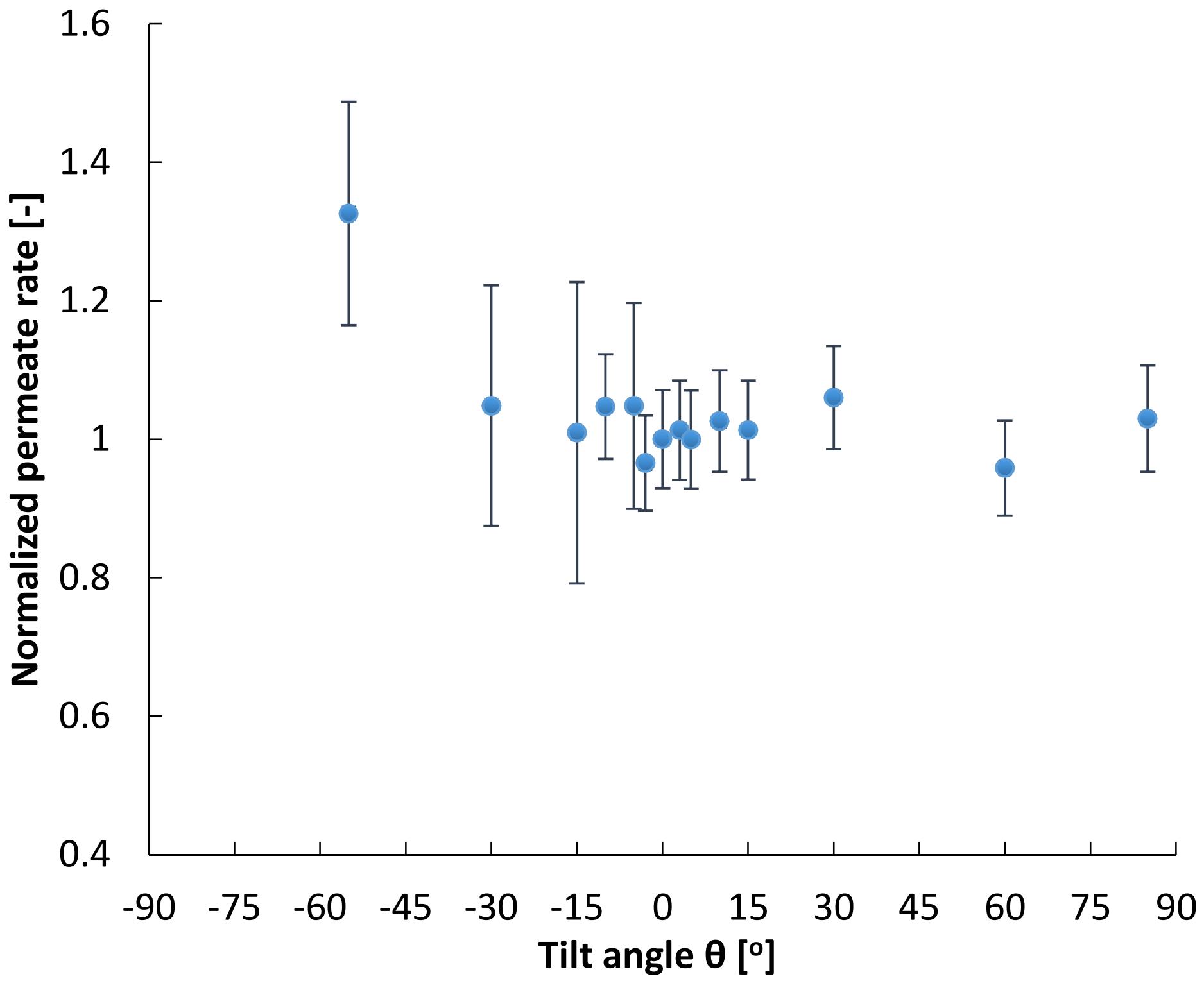

**Figure 3.7.** Effect of module tilt angle on permeate production: $T_{f,in} \approx 60°C$, $T_{c,in} \approx 40°C$.

Figure 7, for a trial at significantly higher temperature than Figure 3.6, shows that angle played a small role in positive and moderately negative angles, but that flux increased significantly at very large negative angles. This may be indicative of thermal bridging at relatively small negative angles such as -30$^{o}$, rather than flooding, and a tendency to flood as the module tilt angle furhter increases and approaches -60$^{o}$.

Since it is almost universally observed that thermal bridging or other effects of module tilt angle variation are effectively absent at low tilt angles, subsequent trials reduced the number of angles considered in the low angle range.

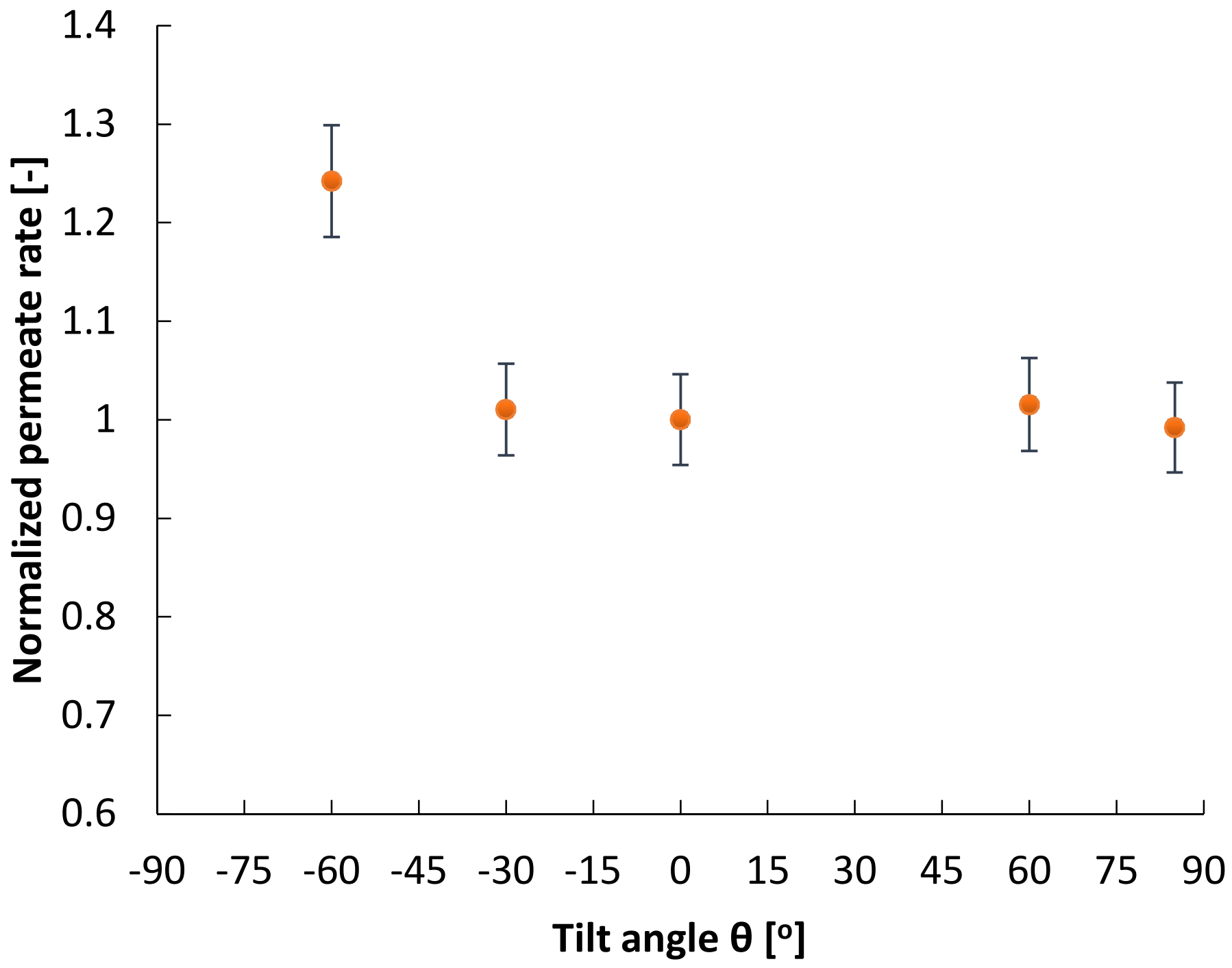

**Figure 3.8.** Effect of module tilt angle on permeate production: Smaller air gap. $T_{f,in} \approx 50°C$, $T_{c,in} \approx 20°C$.

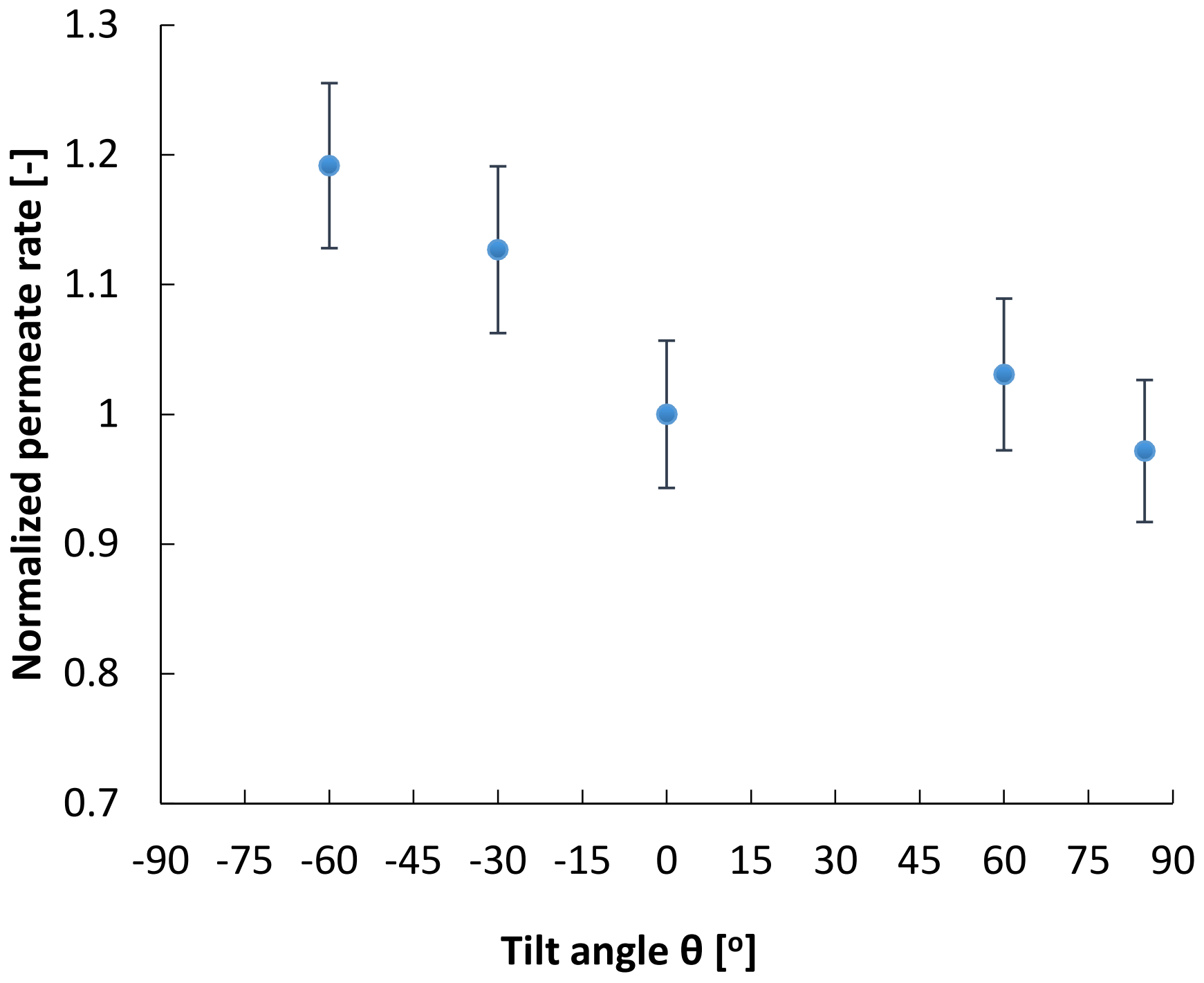

**Figure 3.9.** Effect of module tilt angle on permeate production: Smaller air gap. $T_{f,in} \approx 60°C$, $T_{c,in} \approx 40°C$.

To further investigate the hypotheses of thermal bridging and flooding, additional trials were performed for a smaller effective air gap thickness two thirds the original thickness, using only one spacer mesh instead of two (Figs. 8 and 9). While the absolute value of permeate production rate was higher for these cases due to the reduction in effective air gap size and a corresponding decrease in diffusion length for water vapor, the relative effect of tilt angle on flux remains similar. At small negative angle and even at large positive angles, the flux remains constant. At -60°, flux increases significantly. In some cases, the effect of thermal bridging can be discerned at angles as low as -30°. Thus, these observations suggest that changes in effective air gap thickness do not have a major impact on the induction of thermal bridging effects at various tilt angles in AGMD. Notably, the higher temperature trial experienced relatively significant thermal bridging starting at an angle of only -30°, indicating a temperature dependence on risk for thermal bridging.

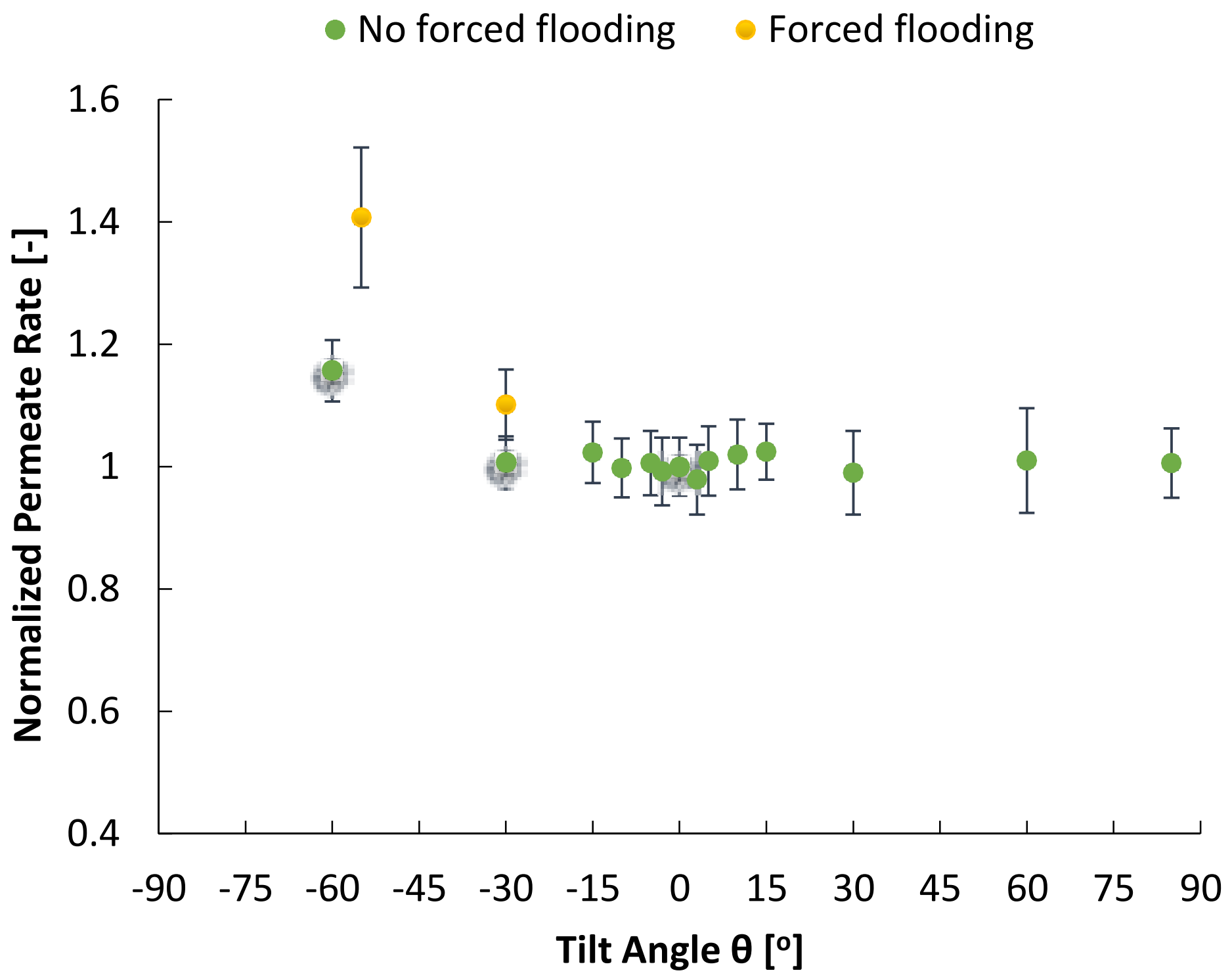

**Figure 3.10.** Effect of module tilt angle on permeate production: Comparison with modified experiment where hydrostatically forced flooding is avoided. $T_{f,in} \approx 50°C$, $T_{c,in} \approx 20°C$

An additional trial was performed with hydrostatically forced flooding, where the permeate outlet was at a larger height than some parts of the module, thereby partial filling the air gap with liquid water. At low negative angles (~ -30°C), partial flooding begins occuring; whereas when the angle is decreased further (~ -60°C), almost the entire active membrane area is beneath the outlet and hence filled with water. Identical experiments were also performed where care was taken to avoid hydrostatically forced flooding over the active membrane area even at extreme negative angles. The results showed that forced flooding, where the entire air gap is filled with liquid water produced significantly higher permeate flux (Figure 3.10). In the tests for which forced flooding was avoided, one can see that even at -30°, the permeate production rate does not change ralative to the vertical baseline. Only at higher negative angles we see an increase in relative flux, but the extent of increase is smaller than the case of forced flooding, pointing to the possibility of water falling back onto the membrane from the condensation plate and related local theral bridging. This further demonstrates that both flooding and thermal bridging are possibilites in AGMD at high inclinations and that the relative increase in flux is affected greatly by the extent of thermal bridging.

### 3.6.2 THERMAL BRIDGING HYPOTHESIS

The thermal bridging may occur if fluid falls from the condensing plate onto the membrane itself, forming a liquid bridge across the local air gap. This phenomenon is not captured by the numerical model. Thermal bridging is particularly likely in the case of negative tilt angles and hence the experimental flux results are asymmetric about the 0° orientation. As the mass transfer resistance is dominated by the air gap width, this phenomena should be easily observed by an increase in permeate flux at declined angles. Our initial hypothesis was that bridging could occur even at small negative angles. In the experimental data, however, we see no such change in permeate flux at small negative angles, indicating a lack of thermal bridging. The key insight is that thermal bridging is not a concern at vertical and near vertical angles for the air gap and spacer dimensions considered here.

An explanation for the why fluid does not fall onto the MD membrane even at negative inclinations may be obtained by the considering the hydrophobicity or hydrophilicity of the surface on which the fluid flows. The aluminum condensing surface is hydrophilic: typical aluminum condensing surfaces have contact angles around 5° [92]. Meanwhile, MD membranes are hydrophobic. Thin films of water can remain on the underside of inclined hydrophilic surfaces up to very high angles of inclination [93, 94, 95]. Additionally, in small gaps where droplets can easily exceed the size of the gap, droplets on the hydrophobic surface may touch and be reabsorbed into the film on the hydrophilic condensing plate. Therefore, large tilt angles, which favor thicker liquid films, may be required before thermal bridging effects occur.

## 3.7 CONCLUSIONS

The experimental and theoretical results indicate that moderate angles of inclination only had a minor effect on permeate flux, except at high angles where it was quite significant. Permeate flux varied by less than 5% within inclination angles of ±15°, but flux increased significantly as angles approached ±90°, in some cases rising by more than 40%.

The results indicate AGMD's susceptibility to two conditions in which liquid water spans the air gap, flooding and thermal bridging. Flooding, occurring when permeate production was rapid enough to fill the air gap with permeate, is a risk at large temperature differences across the air gap, at small gap sizes, and at very high inclined or declined angles. Flooding occurs primarily at very high inclined or declined angles.

The second condition, thermal bridging, occurs at high declined angles, for which condensate detaches from the laminar film on the condenser plate and makes contact with the hot membrane surface. This condition occurred more readily than flooding at declined angles, but was not observed for angles less than 30° off vertical. Results suggest that thermal bridging and flooding are likely to occur earlier, at smaller negative inclination angles, in systems that have lower air gap thickness.

While in many cases flooding and thermal bridging may reduce thermal efficiency, in single pass systems with limited size or lacking energy recovery, the increased permeate production from high incline angles with flooding or thermal bridging is often desirable, as more pure water is produced under the same cycle top and bottom temperature conditions.

## 3.8 ACKNOWLEDGMENT

We also thank Aileen Gutmann and Joanna K So for assistance in modifying and running experiments for the work in this chapter.

# Chapter 4. SUPERHYDROPHOBIC CONDENSER SURFACES FOR AIR GAP MEMBRANE DISTILLATION

## 4.1 ABSTRACT

Superhydrophobic surfaces for enhanced condensation in Air Gap Membrane Distillation (AGMD) may provide significantly improved distillate production rates and increased thermal efficiency. While AGMD is one of the most thermally efficient membrane distillation desalination configurations, large transport resistances in the air gap limit distillate production rates. AGMD experiments were performed with combinations of untreated, hydrophobic, and superhydrophobic condensation surfaces. A nanostructured copper oxide coated condensing surface produced durable 164°±4° contact angles and jumping droplet condensation. Tests were also performed on the air gap spacer, in this case a small diameter support mesh, to judge the effects of superhydrophobic treatment and conductivity on distillate production for AGMD. A novel visualization technique was implemented to see through PVDF membranes and confirm air gap behavior. The experiments were compared with numerical modeling of AGMD film-wise condensation and flooded-gap MD. The results indicate that the introduction of superhydrophobic surfaces can result in improvements in distillate production in excess of 60% over standard AGMD. However, for high distillate production condensation on the superhydrophobic plate transitions from a partially wetted droplet morphology to Wenzel flooded (wetting) conditions. Mildly hydrophobic condensing surfaces were found to provide moderate improvement in distillate production. Superhydrophobic support meshes made a negligible difference in distillate production, but high conductivity support meshes showed significant increases in flux at the expense of thermal efficiency. The results outline recommended superhydrophobic condensation conditions at varied feed and cold side temperatures for substantial improvement to distillate production rate for AGMD systems in a flat plate configuration. Jaichander Swaminathan, Laith Maswadeh, and Professor John Lienhard V contributed to this work [24].

## 4.2 INTRODUCTION

### 4.2.1 MEMBRANE DISTILLATION AND DROPLET CONDENSATION

Membrane distillation (MD) is a quickly advancing thermal desalination technology capable of providing low-maintenance water filtration using waste-grade or renewable heat. Membrane distillation units show promise for desalination on both the small and large scale and unlike reverse osmosis, are fouling resistant and have performance minimally affected by increases in salinity [11]. Recent work has suggested MD can potentially have similar or superior efficiencies to other state-of-the-art thermal desalination technologies as well [86, 96, 87]. Air Gap Membrane Distillation (AGMD) is an MD configuration with an air-filled cavity between the membrane and condensing surface. The thermal resistance of the air gap reduces conduction heat loss between the cooling surface and hot feed, and hence sustains the driving force for vapor transfer through the membrane. This comes at the cost of a large associated mass transfer resistance to vapor diffusion across the air gap [97]. Among implemented MD configurations, Summers et al. [22] has shown that AGMD is capable of the highest energy efficiency.

Two basic droplet morphologies exist for drop-wise condensation on microscopically rough surfaces: wetted and suspended. The wetted morphology, also known as a Wenzel droplet, adheres to the surface and grows rapidly. Water infiltrates the rough surface under the droplet, pinning it to the surface [98] and allowing for enhanced conduction through the surface and droplet [99]. Droplets can then grow to a large size before shedding due to gravity. The suspended morphology, also known as a Cassie-Baxter droplet, forms on top of surface features and leaves the surface non-wetted. This leads to lower conduction through the surface and slower growth, but also allows droplets to detach from the surface more easily and at smaller sizes [100, 101]. Jumping droplet condensing occurs in a partial wetting morphology, where a small wetted region develops on a nucleation site and grows into a droplet which is suspended over the surface features around that site. The behavior of partial wetting varies based on surface geometry, nucleation density, and local energy barriers [102]. Studies have shown that surfaces with higher contact angles tend to form smaller droplets that de-pin and jump from the surface more readily, leading to enhanced heat transfer [103]. It is the combination of rapid growth from partial wetting and the rapid ejection of droplets that enhances heat transfer: extremely pinned droplets cause flooding which often impedes overall heat transfer.

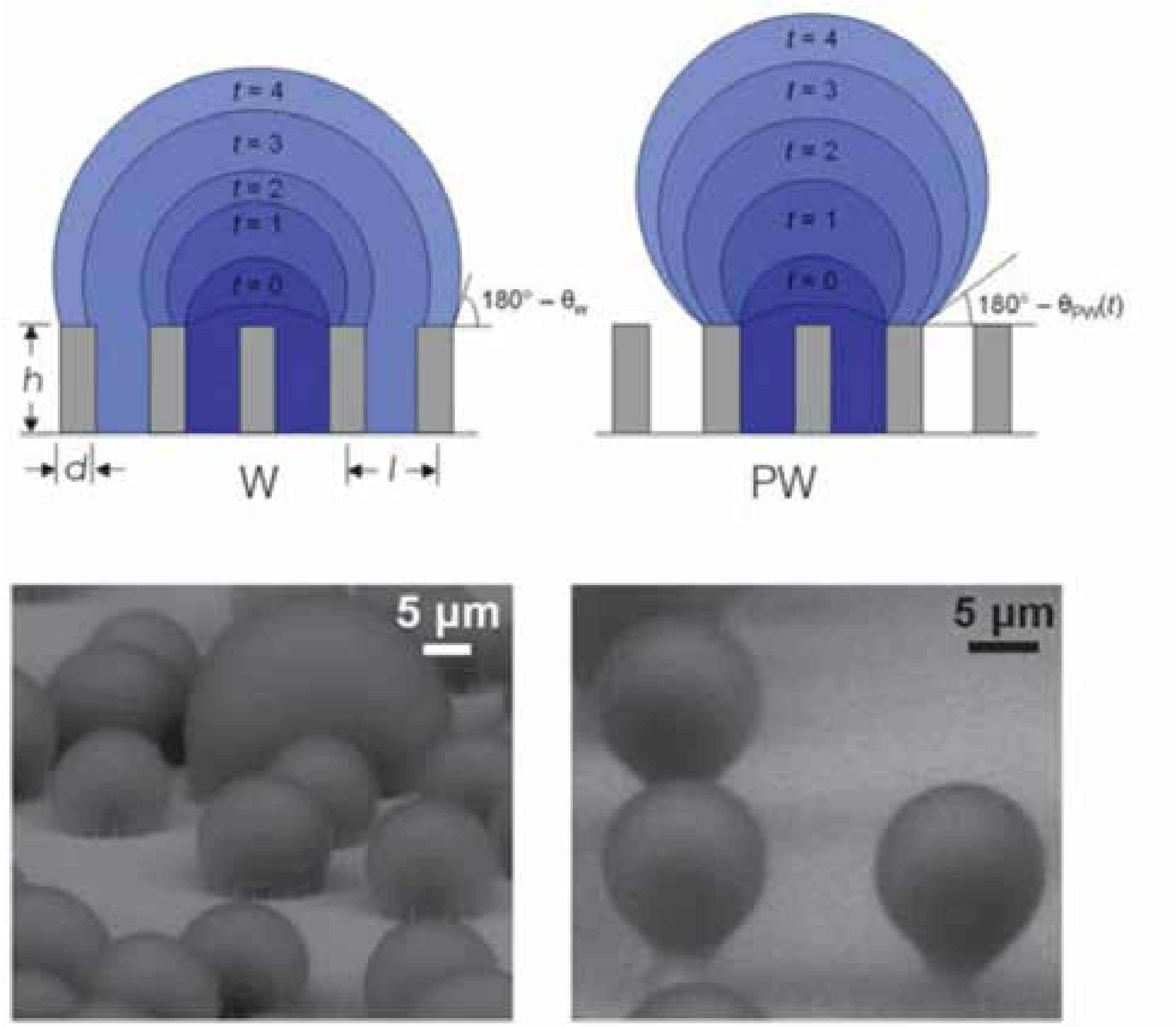

**Figure 4.1.** Left, top and bottom: diagram and ESEM image of Wenzel droplet (flooded) condensation, and right, top and bottom, partially wetting droplet diagram and ESEM image, modified from [102]. A suspended droplet regime, not pictured, resembles partially wetted, but with no liquid water between the surface features.

Partially wetted droplets as seen in Figure 4.1 have the advantage of improved thermal conductance between the droplet and surface, allowing for better heat transfer and faster nucleation and jumping. The copper oxide surfaces used here and in previous studies are designed specifically for that purpose [102].

### 4.2.2 CONDENSING IN AGMD

Several condensing regimes may occur in AGMD, depending on the condensation rate, air gap width, module height, and surface hydrophobicity among other parameters. Traditional AGMD simply condenses distillate in the laminar film regime, an understood and well-characterized process [104, 89]. With superhydrophobic surfaces, jumping droplets may occur, especially at low distillate flow rates and small air gap thicknesses. In this regime, small droplets combine and eject from the surface, with droplet sizes in the range of ≈10–100 μm [102].

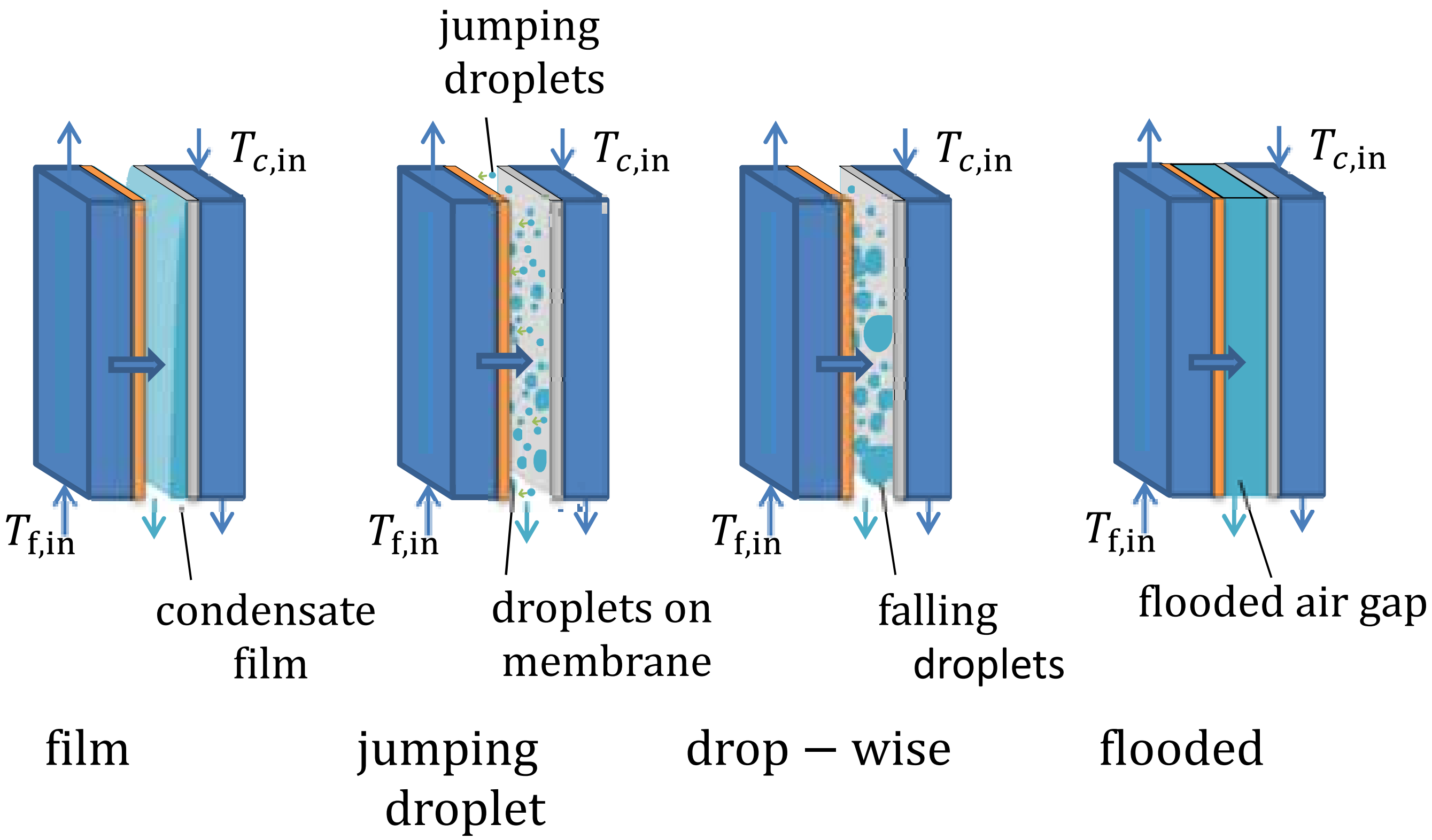

**Figure 4.2.** Diagram of condensation regimes that may occur in AGMD.

AGMD systems normally operate in a film condensation mode on the condenser surface. Recent research work in condensing surfaces has focused on drop-wise condensation on hydrophobic surfaces, which can have five to seven times the heat transfer coefficient of laminar film condensation [102]. Jumping droplet condensation on superhydrophobic surfaces provides further heat transfer coefficient improvement: for the coatings used in this study, previous work has shown a 25% increased heat flux and a 30% increased condensation heat transfer coefficient compared to state-of-the-art hydrophobic surfaces [105]. The jumping phenomenon is a result of the fusion of small droplets (10–100 μm) leading to a decrease in total surface area. The reduction in surface energy translates into a release of kinetic energy through a dynamic instability as the smaller droplets coalesce, which can launch the droplet from the surface [105]. The silanized copper oxide (CuO) surface used in this study produces a durable superhydrophobic surface which may provide a low-cost, scalable method for industrial use of drop-wise condensation.

The combination of superhydrophobic surfaces and membrane distillation, a novel implementation developed in this work, may provide significantly increased efficiency and condensate production rates for AGMD desalination. A flat plate AGMD module was designed for use with interchangeable condenser plates with different surface treatments. Two distinct behaviors were seen in superhydrophobic condenser testing. At lower rates of distillate flux, jumping droplet condensation was measured and observed and at higher rates of flux, flooding was seen to occur. The module also used a replaceable mesh air gap spacer and several mesh options were tested for effects on performance. It was found that mesh hydrophobicity had a small impact on distillate production but mesh conductivity could have a significant effect on distillate production. Experimental results were compared to a model of the system which used a finite difference analysis for film-wise condensation. The model accurately represented the film-wise AGMD tests and flooded conditions, with the jumping droplet results falling in between the two but closer to the flooded results.

## 4.3 METHODS

### 4.3.1 EXPERIMENTAL SET-UP

An air gap membrane distillation apparatus was constructed alongside a finite difference Engineering Equation Solver (EES) model [58]. The system was designed to minimize temperature change in the feed solution (<0.5°C) as it flows across the feed channel, allowing for fine control of conditions within the AGMD module.

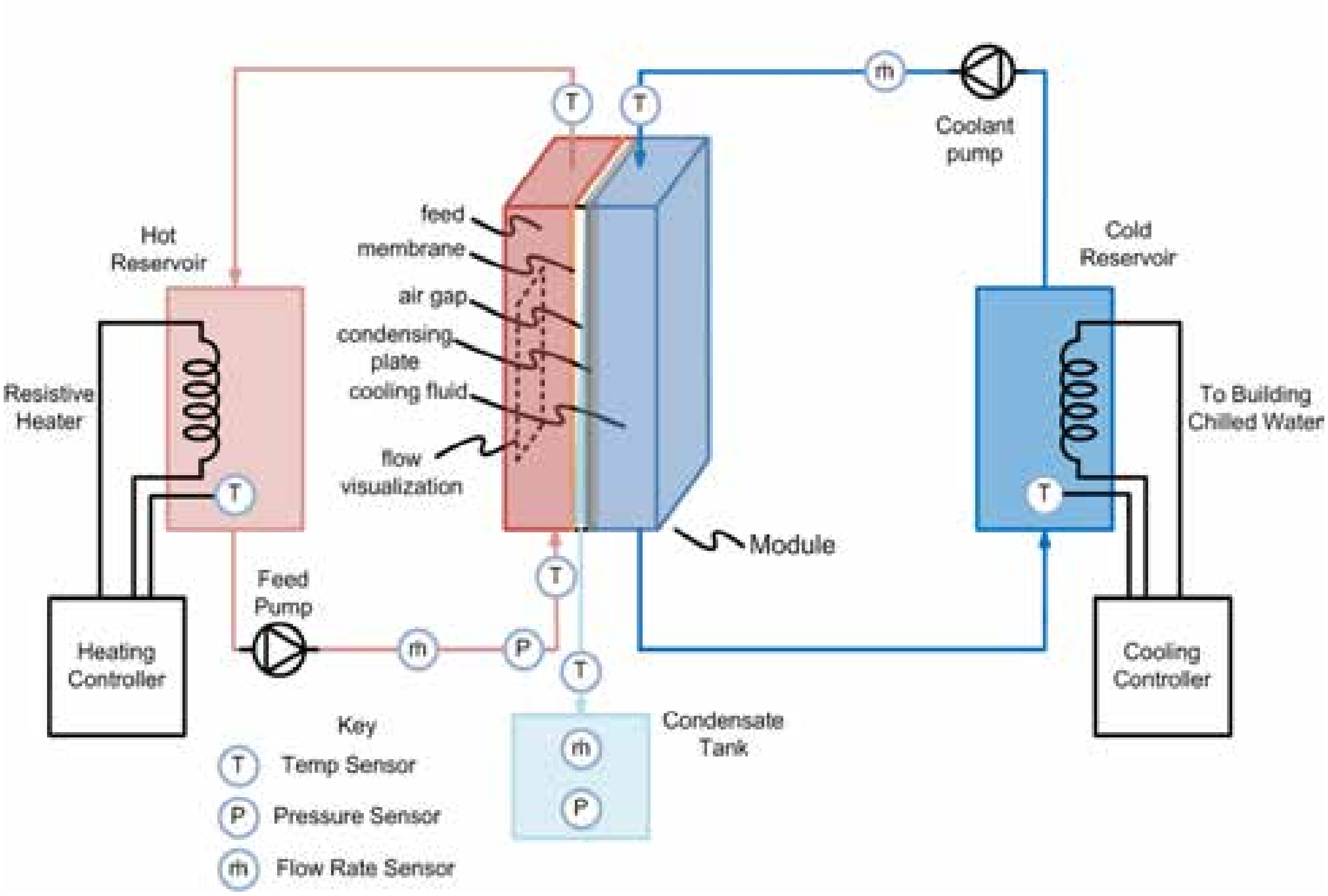

**Figure 4.3.** AGMD Apparatus diagram for superhydrophobic condensing and control tests

The system consists of a flat plate AGMD module connected to heating and cooling loops. The heating and cooling systems each contain a large tank with resistive heating elements connected to a temperature controller. The cold tank also connected to a cold water feed for testing at temperatures as low as 10 °C, and both tanks were sized to maintain consistent temperatures for the module feeds. Temperature and flow rate were measured at various points in the heating and cooling loops, including the entrances and exits of the module feed channels. Additionally, a small condensate collection tank was used to collect distillate and an electronic mass scale under this tank measured the mass flow rate of condensate. Components were chosen with temperature tolerances designed for a set range of operating conditions from 20°C to 90°C for the hot feed and from 10°C to 70°C for the cold side.

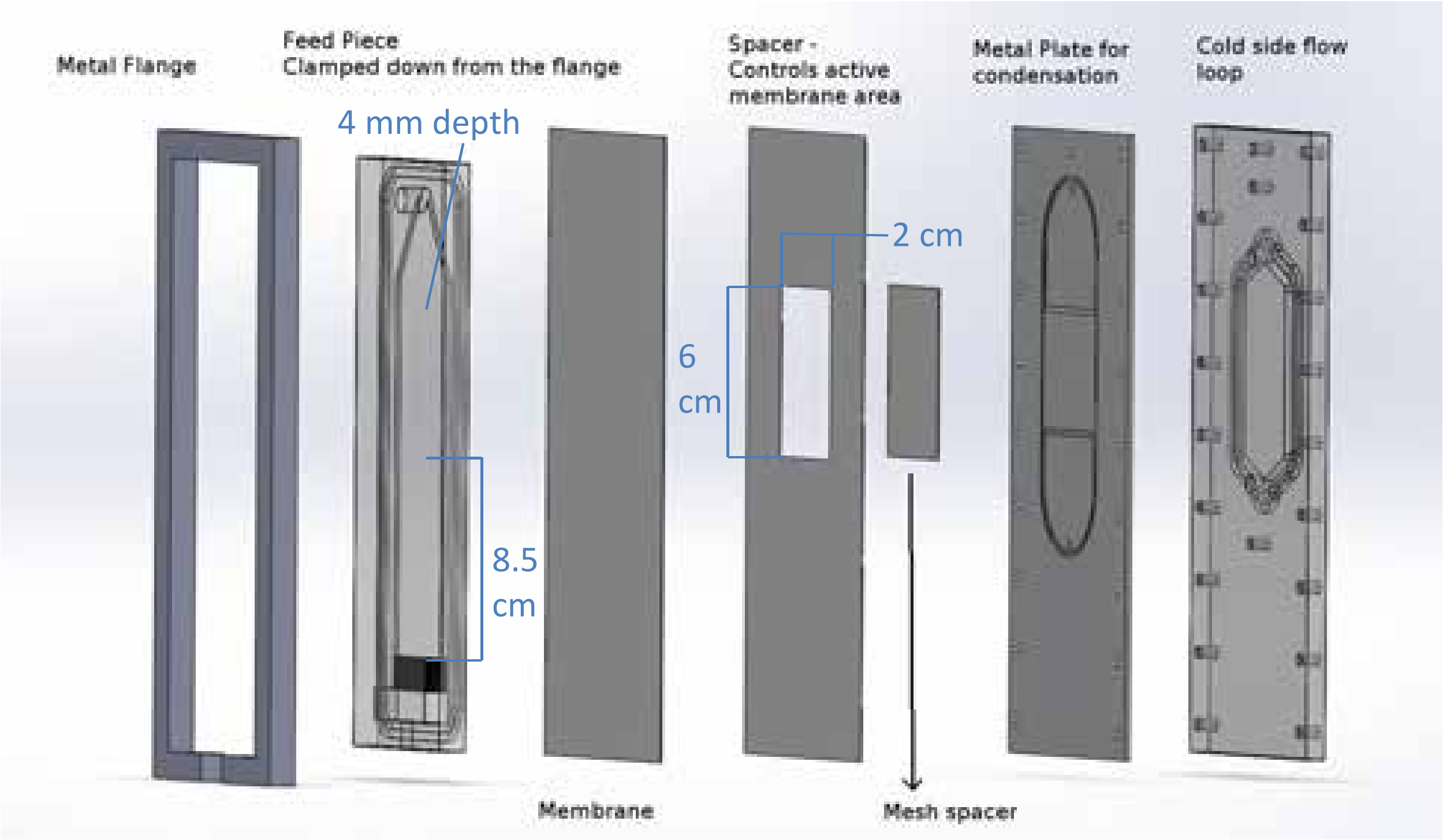

**Figure 4.4.** Membrane distillation apparatus module diagram. The flow channels are machined into polycarbonate or aluminum plates. Dimensions for the feed developing flow region (16 cm) and the active membrane area is shown.

The air gap module itself was constructed with a series of close-tolerance, CNC machined plates. The feed and cooling channels are milled into clear polycarbonate blocks, and the feed channel is longer than the active membrane area in order to ensure fully developed turbulent flow over the exposed membrane. An aluminum plate with a recessed collection region conducts heat from the condenser to the cold water feed and holds the copper condensing plate in the air gap. A small support mesh between the membrane and condenser plate maintains the air gap spacing.

The air gap size and system conditions were designed to maximize distillate flux. Repeatability experiments were performed for some of the trials to ensure reliable results, one of which is included here. More details are in the uncertainty analysis section. Influences from previous tests were eliminated by total dry out of the system (> 24 hours) between tests. To ensure stable conditions, each test was allowed to stabilize for at least 15 minutes before data was taken: this time frame was chosen because results showed stabilization of flux well within this. Tests were run starting at low temperature and continuing to high temperature, to avoid any effects from high temperature flooding. (The one exception is mentioned elsewhere in the paper. The air gap sizes used were on the smaller side of typical systems. Optimized systems that maximize distillate flux have smaller gap sizes in this range [106], often on the order of 1 mm.

| Variables | Symbol | Experiment Values | Control |
|---|---|---|---|
| Feed temperature | $T_{f,in}$ | 30-80 °C | ±0.1°C |
| Feed flow rate | $m_{f,in}$ | 0.25 kg/s | ±0.003 kg/s |
| Coolant temperature | $T_c$ | 10-50°C | ±0.5° |
| Coolant flow rate | $m_{c,in}$ | 0.2 kg/s | ±0.003 kg/s |

**Table 4.1.** Operating ranges and tolerances of variable parameters during testing

| System Parameters | |
|---|---|
| Air gap | |
| Thickness<br>Pressure<br>Active Membrane<br>Area | 0.45-2 mm<br><br>1 atm<br>6.3"x 4.72" |
| Feed Channel | |
| Turbulent, fully developed | |
| Length | 16 cm |
| Width | 12 cm |
| Height | 4 mm |

**Table 4.2.** Air gap and feed channel parameters

| Measurement | |
|---|---|
| Device | Uncertainty |
| Thermistors | ±0.1°C |
| scale (distillate) | ±0.1 g |

**Table 4.3.** Measurement uncertainty of instruments

### 4.3.2 UNCERTAINTY ANALYSIS

The uncertainty in experimental flux measurements is evaluated by considering the uncertainty in mass scale measurement (±0.1 g) and the total uncertainty in time (±10 seconds) and their effect on the effective uncertainty in flux measured.

Uncertainty evaluation for the numerical modeling results was calculated with the Engineering Equation Solver model. Uncertainties in flow rate, pressure, dimensions, membrane permeability, and temperature were all included in the analysis. The dominant sources of uncertainty were variations of the temperatures in the feed and cooling channels. The feed temperature on-off controller controls temperature within a range ±0.1 °C within the set point temperature, whereas the cold stream inlet temperature varies ±0.5 °C around the set point value. In addition to this variation, there is an uncertainty associated with differences between the measured temperature and the actual bulk fluid inlet temperature, especially in the case of the hot feed water input (1 °C). As a result this uncertainty (1 °C) was used as the estimated uncertainty in the feed inlet temperature. The B value (permeability) of the membrane is not likely to be constant over the entire range of temperatures, and so an uncertainty of 5% was assigned. The overall uncertainty in flux evaluated by the model is shown (Figure 4.11) at the highest and lowest temperature conditions in both air gap and water gap operating conditions and is found to be less than 6%. Repeatability experiments for select cases confirmed the accuracy of the system.

### 4.3.3 SUPERHYDROPHOBIC COATING

The superhydrophobic surface used for the condenser plate in this experiment is a silanized copper oxide (CuO) nanoscale surface which was found to provide a 25% higher heat flux and 30% higher condensation heat transfer coefficient than conventional copper at low supersaturation.

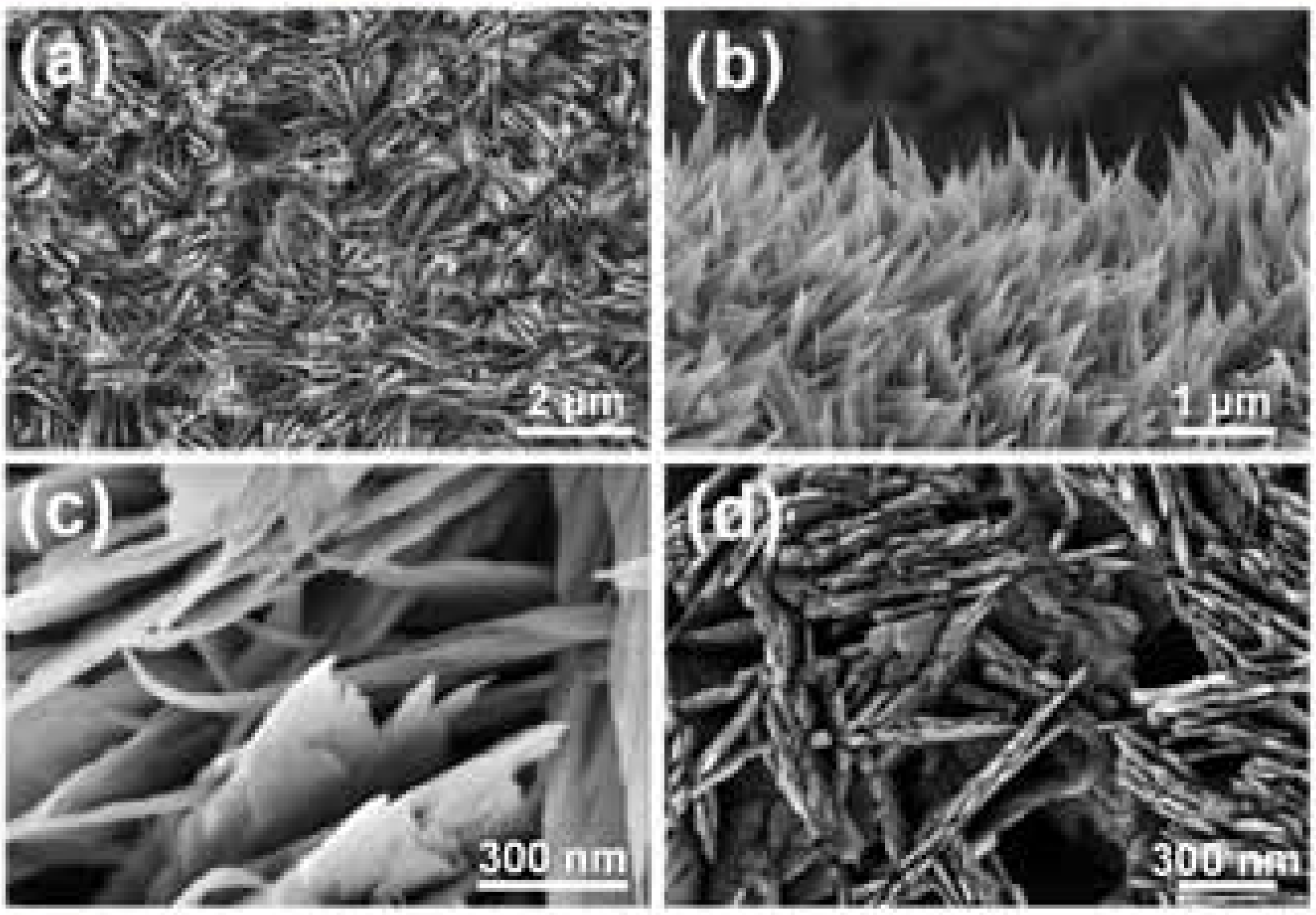

**Figure 4.5.** The field emission scanning electron microscope (FESEM) images above show the copper oxide nanostructure surface from the top view (a), side view (b), and close up on the oxide "blades" without the silane coating and the blades after they have been silanized (d). The resulting surface has an average height of ≈1 μm, a solid fraction of ≈2-3%, and rugosity (area ratio) of ≈10 [105].

As shown by Figure 4.5 above, the surface is covered in nanoscale copper oxide blades which allow for selective nucleation of partially wetted (PW) droplets and high nucleation densities.

The superhydrophobic surface was created through a low temperature, self-limiting process developed by the E.N. Wang group at MIT [105]. A polished, copper alloy plate (Alloy 110, 99.9% pure) was cleaned in an ultrasonic bath of acetone for 10 minutes and rinsed with deionized water, ethanol, and isopropyl alcohol. The plate was dipped in a 2.0 M HCl bath for 20 minutes to remove the surface oxide layer before being rinsed with deionized water and dried with pure nitrogen gas.

The plate was then immersed in a 96±3 °C solution of $NaClO_2$, NaOH, $Na_3PO_4 \cdot 12H_2O$, and deionized water (3.75:5:10:100 wt%) in order to create the copper oxide nanostructure. This process creates a thin layer of copper (I) oxide, $Cu_2O$, which then re-oxidizes into the sharp copper (II) oxide, CuO, nanostructure.

Fluorinated silane (trichloro (1H,1H,2H.2H-perflourooctyl)-silane) was then deposited from the vapor phase onto the CuO nanostructured surface to give the plate its hydrophobic character without changing the surface morphology [105].

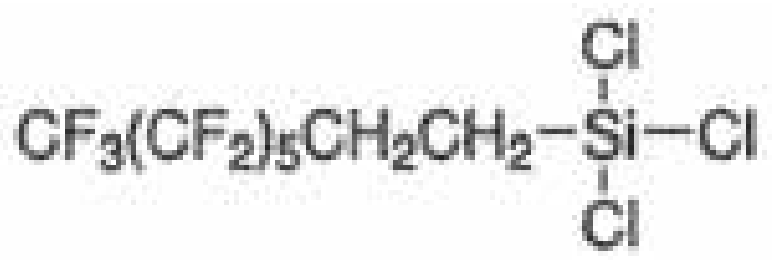

**Figure 4.6.** Coating molecule used for hydrophobicity: Fluorinated silane (trichloro (1H,1H,2H.2H-perflourooctyl)-silane).

For the experiments where the membrane support mesh was made hydrophobic, a commercial silicon-based superhydrophobic surface spray, Neverwet®, was used to coat the mesh. Neverwet is applied in two coatings of different roughness which results in a relatively robust superhydrophobic coating with a contact angle between 160 and 175° [107]. The mildly hydrophobic control surface included the silane coating, but not the rough copper oxide surface. The hydrophilic control surface was polished copper, which was also the substrate for the other surfaces. These surfaces represent realistic heat exchanger surfaces for thermal engineering systems.

### 4.3.4 MODELING

The experimental system's performance was predicted with numerical modeling using Engineering Equation Solver (EES). As modeling work has been previously published [22], only variations of the model will be examined here. The model calculates one dimensional transport of mass and energy across a unit cell, with about 400 unit cells used to describe the experimental system. The model takes input parameters including Reynolds number, bulk temperature and mass flow rate of the hot side feed, and condenser temperature and calculates a variety of parameters including Nusselt numbers, Sherwood Number, Schmidt number, effective conductivities, condensation film thickness, diffusion, thermal resistances, and MD membrane flux. The modeling includes concentration and temperature polarization effects in the feed channel near the membrane surface. The diagram below shows the temperature gradient of the hot side feed near the membrane surface.

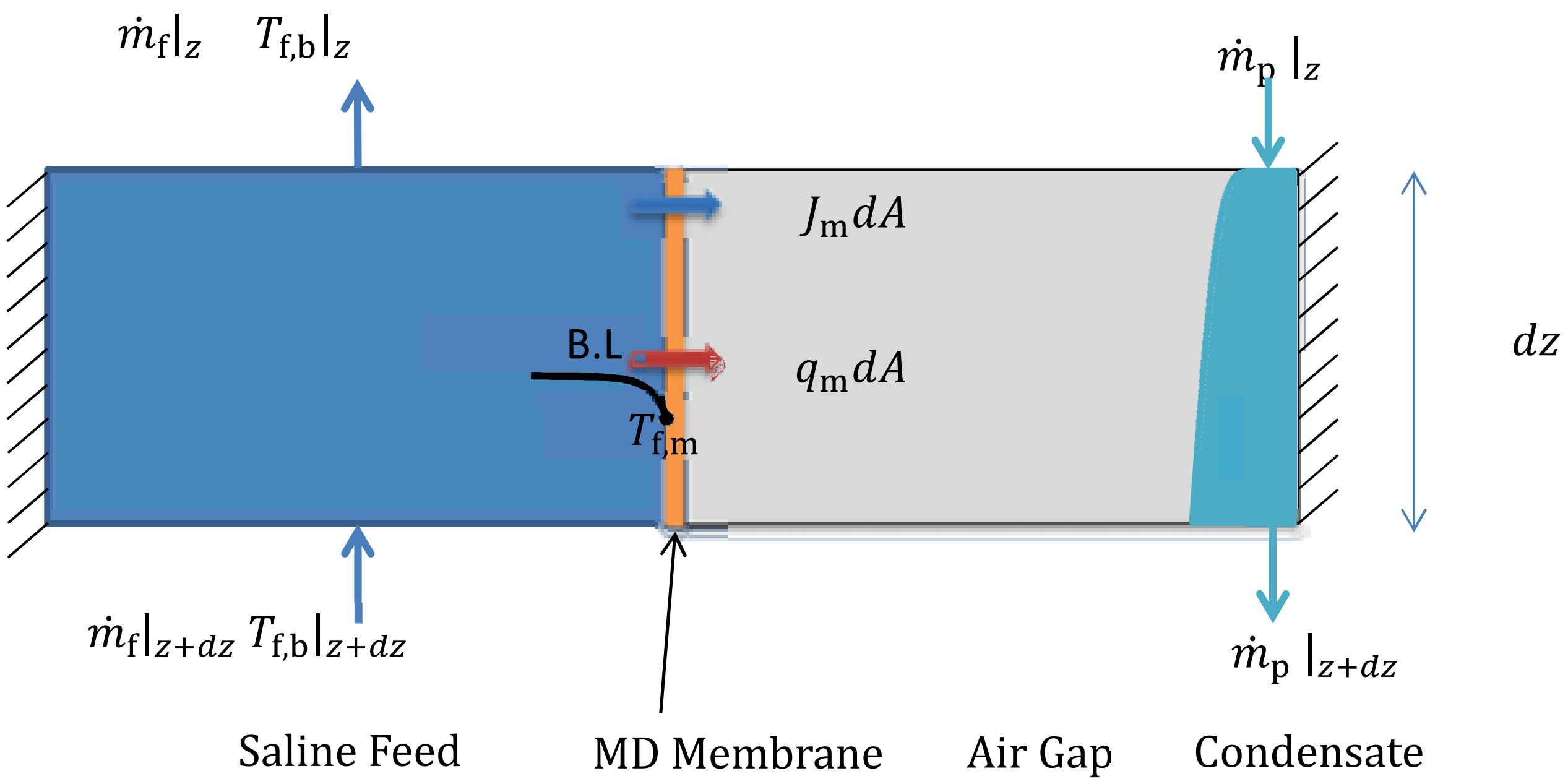

**Figure 4.7.** Unit Cell for AGMD EES Numerical modeling, depicting conservation of mass and energy for each of the several hundred cells

The model assumes incompressible, fully developed flow, and uses the finite difference method. Diffusion of water vapor through the membrane in MD is generally modelled as a linear function of the vapor pressure difference across the membrane, with the membrane coefficient B:

$$J = B \times \left(P_{vap,f,m} - P_{vap,a,m}\right) \tag{4.1}$$

In MD membranes, diffusion and Knudsen flow resistances can affect the *B* value [108]; however, the above flux equation has been supported by modeling and experiments.

The film condensation resistance is given by

$$q_\text{c}[n] = \frac{k_{\text{film}[n]}}{\delta[n]} \cdot (T_\text{i}[n] - T_\text{wall}[n]) \tag{4.2}$$

where $q_\text{c}$ is the heat flux, $k_\text{film}$ is the conductivity of the condensate film, $\delta$ is the local condensate film thickness, $T_\text{i}$ is the local membrane temperature, and $T_\text{wall}$ is the local wall temperature [25]. Heat transfer due to the enthalpy of evaporation also occurs as water vapor is advected through the gap, and is given by

$$T_\text{a,m} - T_\text{i} = \left(\frac{q_\text{gap}}{k_\text{gap}}\right)\frac{\alpha\rho}{J}\left[\exp\left(\frac{J}{\alpha\rho}\left(d_\text{gap} - \delta\right)\right) - 1\right] \tag{4.3}$$

where $q_{\text{gap}}$ is the heat transfer across the gap, $k_{\text{gap}}$ is the average thermal conductivity of the gap, $d_{\text{gap}}$ is the air gap width, $\rho$ is the density, $\alpha$ is the thermal diffusivity, and $T_{\text{m}}$ is the temperature of the gap side of the membrane.

This model has previously been validated with experiments [85], and further validation was performed by the authors as seen in previous publications [84].

While the model is well understood and was validated for film condensation, which occurs on typical hydrophilic surfaces, two-phase flows with droplets on hydrophobic surfaces are difficult to model due to significant variation and complexity in flow and regime. Therefore, the results of previous film condensation experiments and models, as well as the results of well-characterized flooded gap models, were compared to the present superhydrophobic condensation experiments. The modeling indicates the lower and upper limits for the flux of the system, as a function of the effective mass transfer resistance of the gap decided by the droplet configuration.

Several parameters of the system differ significantly in drop-wise condensation. The condensation resistance decreases dramatically, as the droplets have a greater surface area to volume ratio, shed at smaller diameters, and exit the system more quickly than films. Drop-wise condensation has heat transfer coefficients five to seven times higher than film conduction in pure vapor conditions [109]. This heat transfer in this case can be described as:

$$q_c[n] = h_{\text{eff}} \cdot \left(T_{a,m}[n] - T_{\text{wall}}[n]\right) \tag{4.4}$$

where $h_{\text{eff}}$ is the effective heat transfer coefficient between the membrane surface and the wall. This heat transfer coefficient is affected by both the transfer in the gap as well as across the water on the surface. If transfer across the gap is restricted to diffusion as in AGMD, a smaller diffusion distance would lead to a higher transfer coefficient. At the same time, a larger liquid thickness on the plate would result in a lower transfer coefficient for conduction across the liquid.

The water vapor diffusion equation through the air gap between the membrane and the condensing surface is modeled as follows in the AGMD model:

$$\left(\frac{J[n]}{M_{\text{H}_2\text{O}}}\right) = \frac{c_{\text{a}}[n] \cdot D_{wa}}{d_{\text{gap}} - \delta[n]} \cdot ln\left(1 + \left(\frac{x_{\text{i}}[n] - x_{\text{a,m}}[n]}{x_{\text{a,m}}[n] - 1}\right)\right) \tag{4.5}$$

where $J_m$ is the flux through the membrane, $M_{\text{H2O}}$ is the molecular weight of water, $c_{\text{a}}$ is the local molar concentration of air, $D_{\text{wa}}$ is the diffusivity of water in air, $d_{\text{gap}}$ is the air gap depth, $\delta$ is the local condensation film thickness, $x_{\text{i}}$ is the concentration of water vapor at the film-air interface, and $x_{a,m}$ is the local water mole fraction at the membrane interface [27]. The effective gap width $d_{gap}$-$\delta$ has a major effect on the heat and mass transfer resistances.

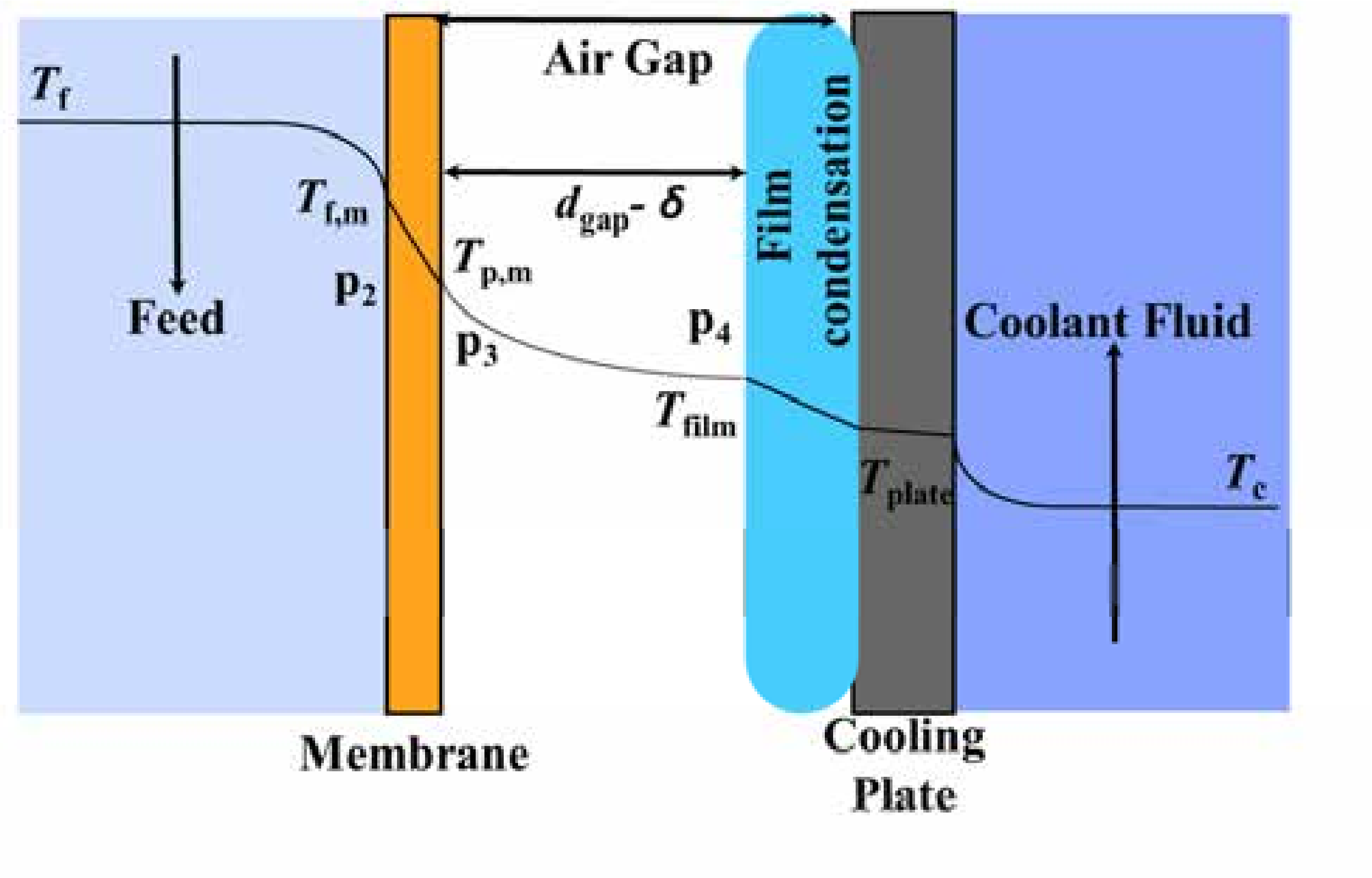

**Figure 4.8.** Temperature profile of film wise AGMD, roughly proportional to the model

As seen from the temperature profile in Figure 4.8, significance temperature gradients exist in the film condensation interface and the membrane. The main driving force for distillate production in MD is the temperature difference across the membrane. As heat transfer across a temperature gradient generates entropy, a significant temperature difference in any other region of the diagram, such as the condensate film, represent inefficiency. Drop-wise and jumping droplet condensation may reduce both the thermal resistance associated with conduction through a liquid film and the transport resistances in the air gap. These effects would each serve to increase distillate production.

### 4.3.5 NOVEL VISUALIZATION TECHNIQUE FOR VALIDATION

Visualization through the PVDF membrane was used to validate assumptions on the flow regime transitions occurring in superhydrophobic condensation experiments. In the module designed for this research, clear polycarbonate allowed the membrane to be seen clearly through the hot side feed channel, but the air gap was not visible behind the membrane. Though plastics such as PVDF are transparent at low thicknesses, MD membranes are generally opaque due to the presence of surface features which cause absorption and scattering of visible light. Regions of the membrane were made transparent by melting at low temperature in order to remove surface features and pores in the material.

A soldering iron with electronic temperature control was set to 3 °C above the melting point of the PVDF material used, and applied with moderate pressure (~50 kPa). Some trial and error was required to develop a technique which did not puncture the membrane. Seven visualization regions were successfully incorporated into one membrane without breakthrough.

The visualization technique was used to observe jumping droplet condensation and large pinned droplets (flooding) occurring in the flow rate regimes claimed in the paper. Through the transparent windows in the membrane, jumping droplet condensation was observed for lower distillate flow rates and validates the improvement seen over conventional AGMD in this regime. Larger pinned Wenzel droplets were visible at higher flow rates and partial flooding likely explains the improvement in flux over film-wise AGMD in these tests.

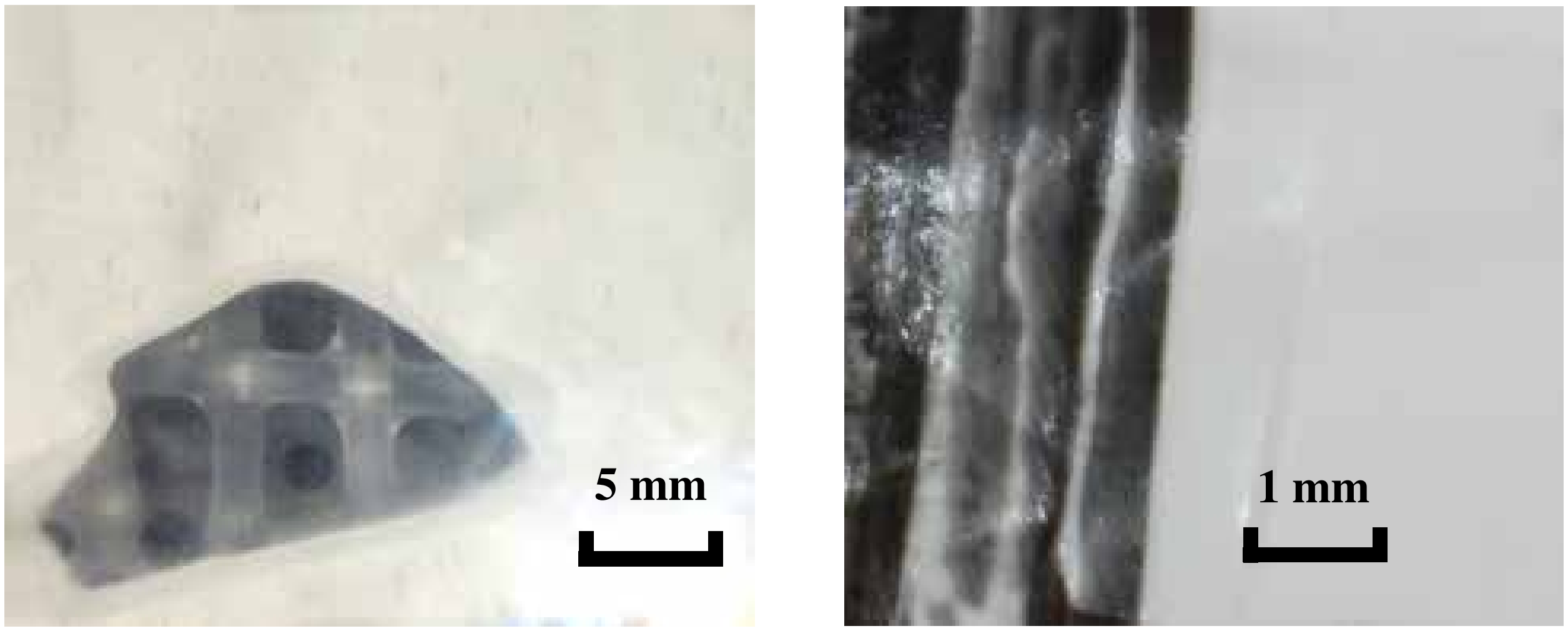

**Figure 4.9.** Images of a transparent region of membrane developed for visual confirmation of air gap behaviors. The image on the left shows partial flooding occurring at high temperatures while the image on the right shows jumping droplet surface at lower temperatures. An iPhone 5s camera (left) and an EOS Rebel T3i Canon digital camera (right) were used.

## 4.4 RESULTS AND ANALYSIS

### 4.4.1 EXPERIMENTAL RESULTS

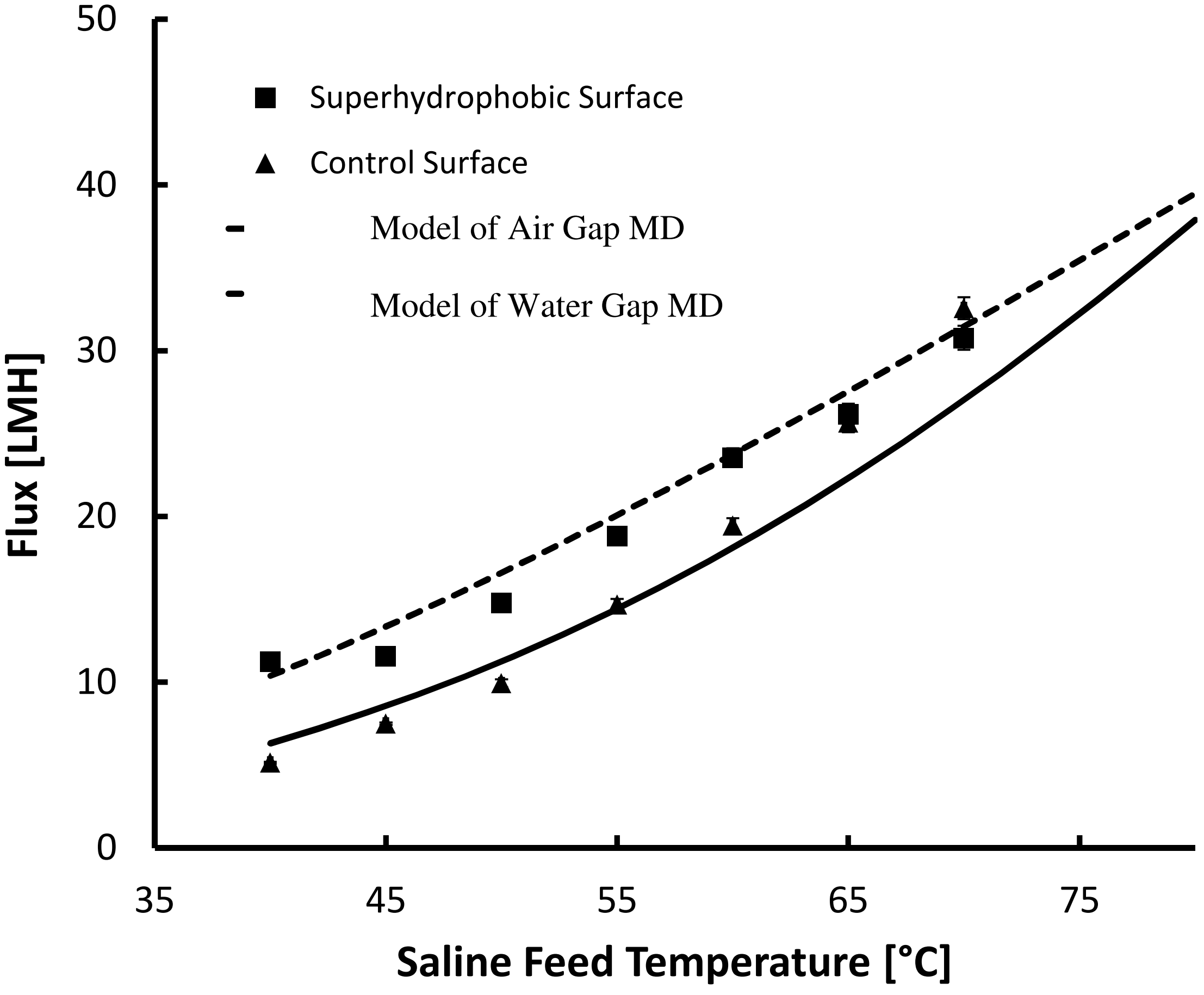

**Figure 4.10.** Superhydrophobic and control surface AGMD experiments with 0.45mm air gap and 13°C cold side temperature. At lower temperatures, flux improvement with the superhydrophobic surface was very significant; ~119%. A model of ordinary AGMD, as described in the modeling section, predicts the control surface tests accurately at lower temperatures, though at temperatures above 65°C flooding begins to occur within the module. The performance of the superhydrophobic surface is similar to that of a water-filled gap, with lower mass transfer resistance.

**Table 4.4.** The uncertainty evaluation from the model for air gap and water gap configurations. The relative contribution of the different measured parameters is also included. The hot feed inlet temperature uncertainty is the major contributor in all cases. In the case of water gap, the effects of uncertainties in B value are higher than in the case of water gap.

| **Variable ± Uncertainty** | | B | $m_{c,in}$ | $m_{f,in}$ | $T_{c,in}$ | $T_{f,in}$ | $J_{tot}$ |
|---|---|---|---|---|---|---|---|
| [Units] | | [kg/m$^2$Pa-s] | [kg/s] | [kg/s] | [°C] | [°C] | [kg/m$^2$hr] |
| Air Gap | $T_H$ = 40°C | 1.6E-6 ± 8.0E-8 | 0.2315 ± 0.005 | 0.2175 ± 0.005 | 13 ± 0.5 | 40 ± 1 | 6.313 ± 0.4182 |
| Water Gap | | | 0.1809 ± 0.005 | 0.2175 ± 0.005 | | | 10.39 ± 0.6333 |
| Air Gap | $T_H$ = 70°C | | 0.2315 ± 0.005 | 0.2175 ± 0.005 | | 70 ± 1 | 27.02 ± 1.048 |
| Water Gap | | | 0.1809 ± 0.005 | 0.2175 ± 0.005 | | | 31.53 ± 0.8886 |

The first AGMD hydrophobicity trial performed with a small air gap (~0.45 mm) and constant cold side temperature of 13°C showed a dramatic increase in flux with superhydrophobic condensation, especially at lower temperature differences. While the flux was 120% higher for a temperature difference of 27°C, at 53°C temperature difference, it was nearly identical to the control experiment. Figure 10 plots the results from the superhydrophobic as well as control surface experiments. The results are compared against numerical modeling predictions of flux for AGMD and water filled gap MD scenarios which are likely to be the lower and upper limits respectively for the flux. The $R^2$ value for the fit between the control surface and the AGMD model is 0.92. The value is reduced by the transition from air gap to flooded behavior observed for the control experiment at a saline feed temperature of around 65°C. The $R^2$ fit value for the superhydrophobic experimental data and water gap model predictions is 0.97. Observations with the visualization method showed that the initial, lower temperature regime involved superhydrophobic jumping droplet condensation, and the later, reduced flux regime had a mostly flooded water gap. Both the modelled water gap and experimental superhydrophobic condensing exhibit a relatively small mass transfer resistance in the gap, making their condensate flux production similar.

The CuO superhydrophobic surface was designed for partial wetting condensation, and has two observed regimes from this and previous experiments. First, superhydrophobic condensing can create small jumping droplets which eject from the surface after combining with nearby drops. This mode is known to have superior heat transfer coefficients and condensation rates, as was the case for the low temperature- difference experiments. The other mode is flooding, in which the gaps between the microstructured features of the CuO surface become filled with water. As a result, wetted droplet condensation occurs, where the droplets grow to larger sizes and only shed by gravity. These droplets become highly pinned to the microstructured surface, and do not de-pin until they reach a larger diameter (~2-3mm [110]) than for smooth hydrophobic surfaces. Thermodynamically, this is similar to a condensate flooded gap. A water-filled gap has higher heat transfer between the hot and cold streams than an air-filled gap.

The system compares favorably to literature reported values for flux. It exceeds the large majority of studies examined in recent reviews [11], and for similar conditions, is of similar magnitude to the optimized maximum flux reported by Khayet et al. for AGMD [106]. Their study reported AGMD flux of 47 LMH compared to 56 LMH here, where both studies had a 71°C hot side and close cold side temperatures (13.9°C vs 13°C here), a porosity of 80%, and similar membrane widths (178 μm vs 200 μm here). This LMH figure is seen in Fig 10, after dividing out the membrane area inactivated by the feed support spacer (43% of area). Notably, the model here predicts an LMH of 49.8 °C, which is in good agreement with their results. The results here are therefore both validated by the literature, and can sufficiently claim that superhydrophobic AGMD can be an improvement on the state-of-the-art.

The system was disassembled after the final 70 °C data point to examine the surface, which found mild flooding as seen in Fig 11**.**

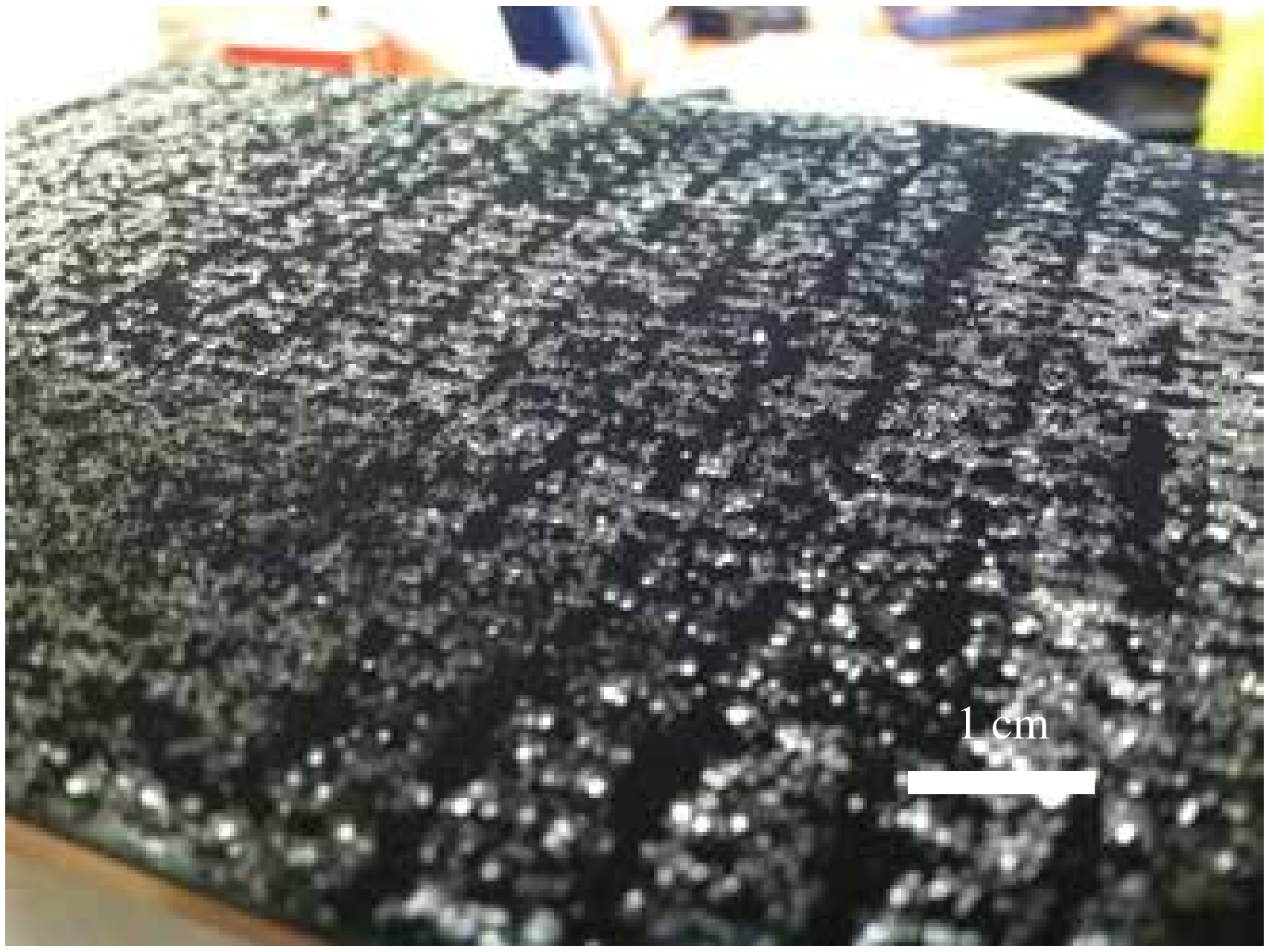

**Figure 4.11.** Partially wetted superhydrophobic surface after high temperature condensation

Because of a high degree of pinning, the wetted droplets remained on the surface even when the surface was held upside down. One important implication of this test is that flooded conditions are not easily reversible [98]: once a surface becomes flooded, it remains that way even at conditions where flooding would not initiate. Drying is often necessary to remove the wetted droplets.

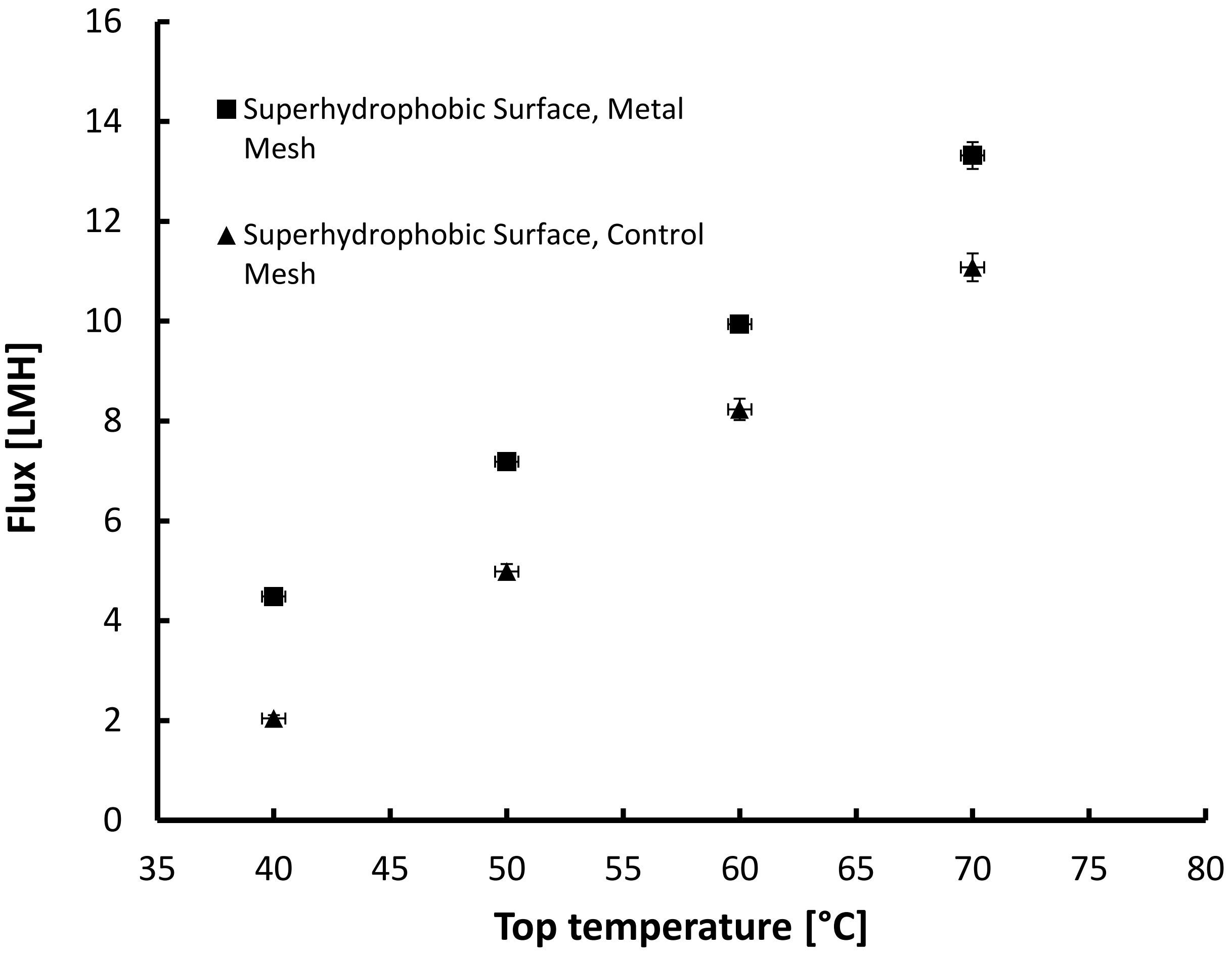

**Figure 4.12.** Superhydrophobic condenser with highly conductive metal support mesh and a control mesh, ΔT =20°C between hot and cold channels. Conductivity of the stainless steel mesh: 16 W/mK. Conductivity of the plastic support mesh: 0.2 W/mK

Air gap membrane distillation, and most other forms of membrane distillation, generally requires a support structure to hold the thin, hydrophobic MD membranes in place. In these experiments, woven meshes were used, where 2/3 of the horizontal weaves were removed to minimize interference with condensation phenomena. In addition to hydrophobic experiments, use of a highly conductive mesh was of interest as a mechanism to reduce heat transfer resistance in the air gap, which generally has a large temperature difference. The conductive mesh improves heat conduction across the air gap, and acts as an additional condensing surface with a smaller effective air gap.

This increase in flux comes with a significant trade off: the metal mesh allows for more direct heat transfer between the hot channel and the cold channel, independent of vapor transport. While it is difficult to model the complex two phase hydrophobic condensing heat transfer, calculating the adjusted effective conductivity of the gap, $k_{gap}$, can provide an estimate of thermal losses from the metal mesh.

In comparing experiments, it was found that the cold-side temperature did not have a significant impact on the distillate flux while the hot side temperature had a very significant effect: this agrees with previous studies [106].

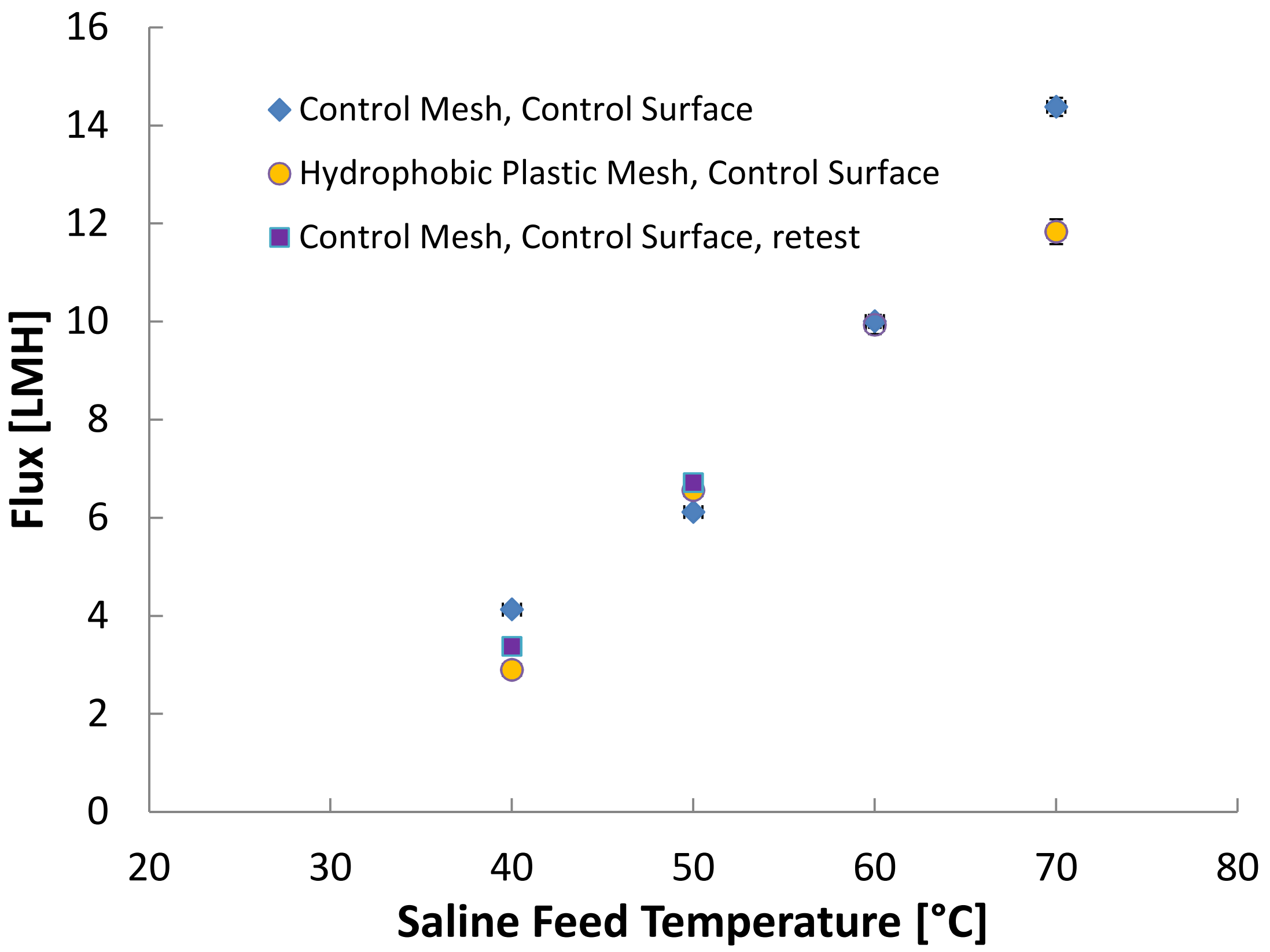

**Figure 4.13.** Ordinary (hydrophilic condenser) AGMD on a copper control plate with varied support mesh hydrophobicity, ΔT =20°C between hot and cold channels.

The effect of the hydrophobicity of the support mesh on condensate production was also examined as a possible method for improving AGMD systems. Most AGMD models assume the effects of the mesh are relatively small and model the system as a laminar film on a flat surface. The results from the hydrophobic mesh experiment supported this approach, as making the mesh superhydrophobic with Neverwet® had a negligible effect on condensate production. A slight effect appeared at the 70 °C test point, which may be related to effects on flooding in the air gap, which began occurring in the system in these conditions around 65°C.

The results show that modifying the mesh spacer's conductivity had a significant effect on performance, but that its hydrophobicity did not. This occurs because the mesh does not act as a significant condensing surface unless its conductivity is very high. In the case of a metal support mesh, the mesh's conductivity was almost three orders of magnitude larger than that of the air in the gap, and two orders of magnitude larger than the conductivity of a plastic mesh. This agrees with previous studies have found that copper fins in the gap increase distillate flux [111].

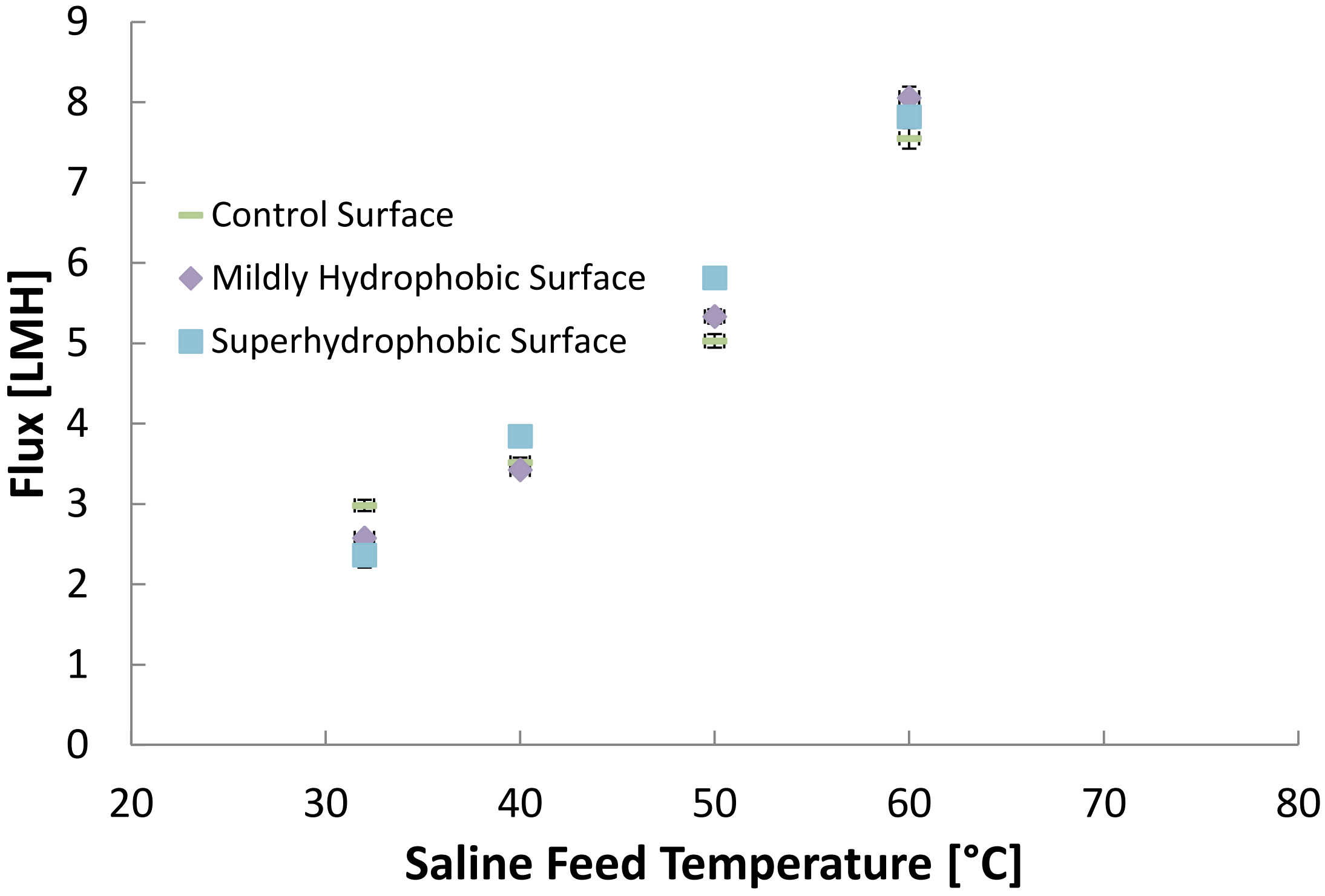

**Figure 4.14.** Superhydrophobic, mildly hydrophobic, regular AGMD with a large air gap (1.5 mm), ΔT =20°C between hot and cold channels.

An experiment including both superhydrophobic and mildly hydrophobic surfaces was performed with a larger air gap. With this roughly three times larger air gap (1.5mm), the effect of superhydrophobic surfaces appeared to be greatly reduced. This result was possibly caused by the increased vapor diffusion resistance of the larger air gap and resulting lower flow rate of droplets, but hysteresis effects may also have been involved. Additionally, with the larger gap size, it is expected that flooding occurs less readily, but may still occur at the highest temperatures studied, which may explain the similar flux at 60°C. At this temperature, the flux of the mildly hydrophobic surface slightly outperforms the others. Under flooding conditions, superhydrophobic surfaces have a lower heat transfer rate than a smooth hydrophobic surface. Both undergo Wenzel drop-wise condensation, with the superhydrophobic surface having larger, slower-moving drops due to a high degree of pinning which occurred as water infiltrated the rough copper oxide surface features.

However, unlike the others, this experiment was begun at a higher temperature for the superhydrophobic trial and then decreased, meaning that the detrimental effects of flooding may have continued down to lower temperatures. Such a result would indicate that superhydrophobic condensers might only be valuable in AGMD applications that operate within a certain range of conditions, and limit their usefulness in processes with a high peak flow rate and temperature difference.

Under most conditions the mildly hydrophobic plate performed similar to or better than the control plate, but did not outperform the superhydrophobic trial except under flooding conditions. The flux differences however remained relatively small. These results indicate that for larger air gaps, the beneficial effect of superhydrophobic surfaces on condensate production may be significantly reduced.

### 4.4.2 Comparison to Model

Simulations for a flooded gap, modeled as an elimination of air gap resistance and maximization of liquid film resistance, showed an increase in the rate of distillate flow. For the liquid gap case, the mass transfer coefficient and the distillate flux are higher compared to the air gap system, but at the expense of increased conductivity of the gap which reduces thermal efficiency. These results are seen in Fig 10.

In AGMD with jumping droplet condensation, the droplets may induce air circulation in the gap, leading to an improved heat and mass transfer coefficient and flux due to reduced air gap resistance. Due to rapid shedding of droplets at small diameters, the heat and mass transfer resistances of the condensate film are also reduced nearly to zero.

Due to the dynamics of a jumping droplet, the effective air gap thickness as modeled may be reduced. Droplets eject from the condenser surface and enter the air gap. For a horizontal air gap as used in our system, droplets will fall in towards the membrane and downwards. During this flight, the droplet's distance from the membrane is reduced which leads to a reduction in mass flow resistance. Because the droplet remains at a lower temperature than the air near the membrane, condensation will continue to occur on the droplet as it falls. Droplets may bounce off of the membrane and continue their flight with minimal heat transfer losses.

The jumping droplets may adhere to the hydrophobic membrane after ejection from the condenser plate. Initially when this occurs and the droplet temperature is still near the temperature of the condensing surface, it behaves as a locally flooded system, which increases distillate flux. Due to its small size, the temperature of the drop quickly approaches the temperature of the membrane and the rate of condensation is significantly reduced. When gravitational forces on the adhered droplet become significant compared to the surface-energy adhesion forces (Bond number >0.1-1 depending on surface hydrophobicity [112]), the droplet will fall down the membrane surface readily, as in standard drop-wise condensation.

In the case of flooded Wenzel droplets strongly adhering (pinning) to the superhydrophobic surface, which can occur in regimes of high condensate flux, the flooded droplets become large compared to the gap size (1 mm), touching the membrane. Thus the air gap width becomes locally zero in some places, improving distillate flux as seen in the flux equation and experiments. These larger flooded droplets are also responsible for thermal bridging between the condenser and membrane, which decreases resistance to heat flow and reduces the thermal efficiency of the system.

A comparison was made between the results of the jumping droplet experiment and the model for fully flooded air gap conditions, which is a simplification of the condensate flow model where the condensate film thickness becomes the entire width of the air gap. Jumping droplet condensation was found to have close-to but inferior condensate production when compared to the flooded gap and significantly higher flux than the ordinary AGMD model. Though this work did not include a direct model for jumping droplet condensation, which would involve an under-constrained two phase flow problem, the comparison to the standard and flooded AGMD models demonstrated the ability for superhydrophobic condensers in AGMD to minimize mass transfer resistance.

## 4.5 CONCLUSIONS

Superhydrophobic condensing surfaces have shown very significant distillate flux increases for AGMD, with more than a 100% improvement in some cases. However, at high heat transfer rates, the benefit of superhydrophobic surfaces becomes negligible. Visual validation confirmed that flooding and large droplet pinning occurred on the superhydrophobic surface at these high heat transfer rates, and confirmed that jumping droplet condensation occurred at lower heat transfer rates. This result is in line with past studies of superhydrophobic surfaces which showed a transition to flooding at high heat transfer rates. The temperature differences in realistic membrane distillation configurations are smaller (<10°C) than those studied here, and fit within the range of heat transfer rates where jumping droplet condensation provides significant benefit.

The hydrophobicity of the support mesh for the membrane was found to have minimal effect on the distillate flow rate, but high conductivity for the mesh showed notable improvement in distillate flux. Jumping droplet and wetted drop-wise condensation increase the mass transfer coefficient in the air gap, resulting in higher distillate flux and reduced temperature gradients within the condensate. Numerical modeling with EES accurately represented cases with film condensation and the EES model with no air gap resistance reasonably approximated the distillate production of superhydrophobic condensation, indicating that this jumping-droplet condensation significantly reduces the effective heat and mass transfer resistances of the air gap. Thus the mass transfer resistance behavior was similar to a flooded gap system, but without the conductive losses of a flooded gap.

AGMD is the membrane distillation configuration with the smallest sensible heat loss, and superhydrophobic condensation surfaces may be a valuable tool to further improve efficiency and the rate of distillate production in AGMD systems. Further work should be performed to accurately model the dynamics of jumping droplet condensation in AGMD and to compare the thermal efficiencies of AGMD using various condenser treatments, as well as flooded gap efficiency to quantify these values.

# Chapter 5. MISCELLANEOUS CONTRIBUTIONS TO MD EFFICIENCY

Collaborative work on MD efficiency has been briefly summarized in this chapter, please see the cited papers for details. This work includes using MD and flashing technologies (LTTD) for renewable desalinating using ocean temperature gradients, papers and a patent on new MD configuration we invented (CGMD), multistage vacuum MD studies, the effect of scale deposition in surface tension for MD, testing of membranes fabricated for MD, in-situ MD fouling, and a comparison between MD foulants.

The collaborative work of this chapter may appear in the Thesis of others, and has been or will be published as journal or conference papers. The author of this Thesis is not first author of the work in this chapter. Therefore, only brief summaries of the contributions, a selection of graphs and images by the author of this thesis, and verbal summaries of the results are included.

## 5.1 LOW TEMPERATURE THERMAL DESALINATION (LTTD) AND MEMBRANE DISTILLATION

In most oceans worldwide, significant temperature gradients exist between warm surface water and deep cold waters. This gradient represents a tremendous resource for sustainable energy, and has long been examined for electricity production, with numerous pilot plants built. While it has yet to prove economical for electricity production, the lower temperature gradients needed for desalination make it an ideal source to create renewable fresh water.

The first authors on this work were Jaichander Swaminathan and Kishor Nayar, and Professor Lienhard was a coauthor as well [113].

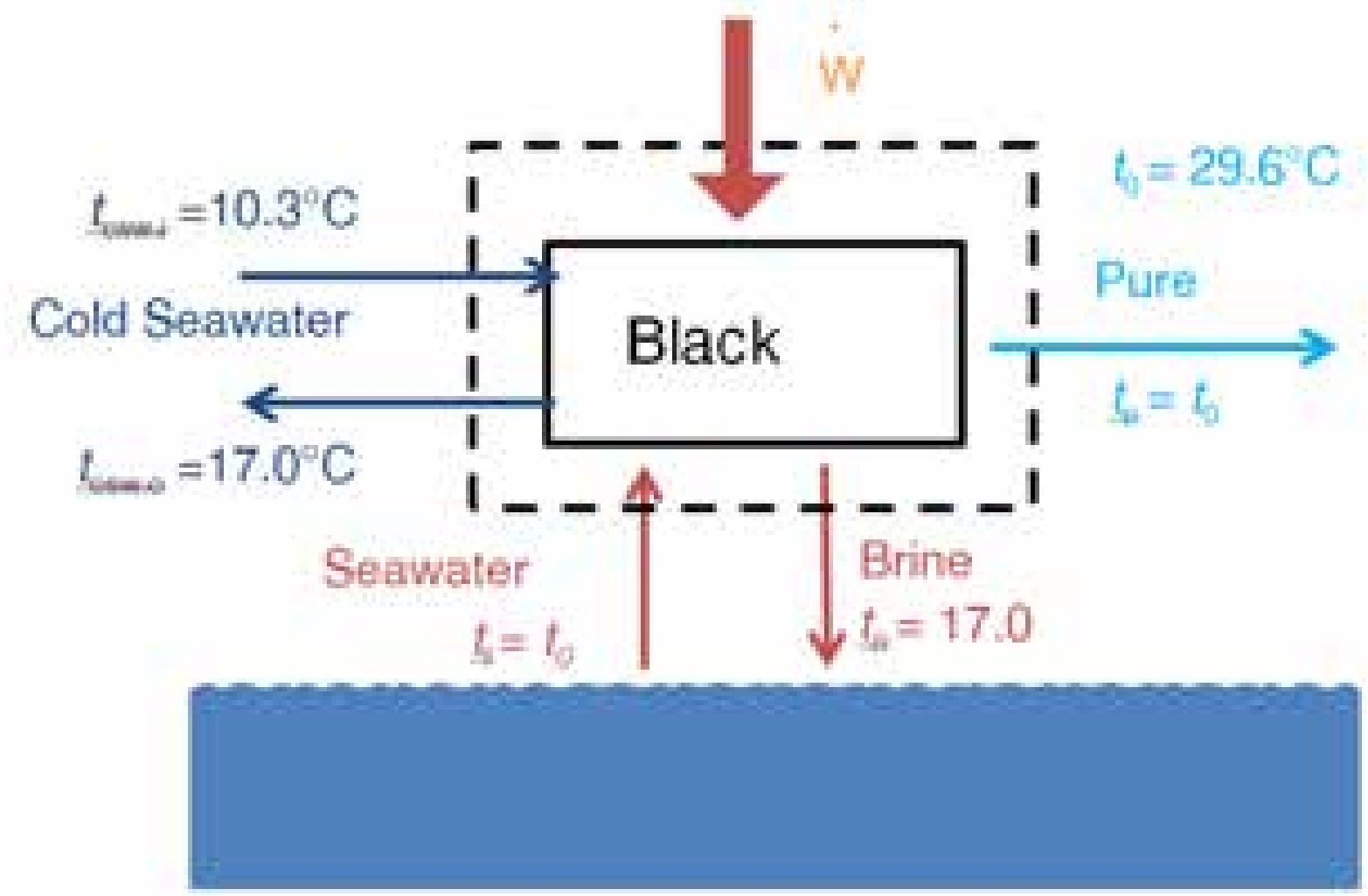

**Figure 5.1.** Pure water generation with ocean temperature gradients, showing typical temperatures. Ideally, the Work term $\dot{W}$ approaches zero, to only use the temperature gradient, but in practice, pumping power etc. is needed.

Previous papers have proposed using a pressure reduction of warm surface waters (flashing) and then recondensing pure water with a cooling coil from the cold ocean depths: this process is called Low Temperature Thermal Desalination, or LTTD. The desalination technology Membrane distillation may serve as a competitive replacement for LTTD, since MD operates effectively at low temperatures and requires fewer moving parts (e.g. no vacuum pump). The problem can be broken down into several questions:

1. What is the best theoretical performance from thermodynamics for desalinating using ocean temperature gradients
2. Can existing LTTD performance be improved?
3. Can MD compete with flashing LTTD?

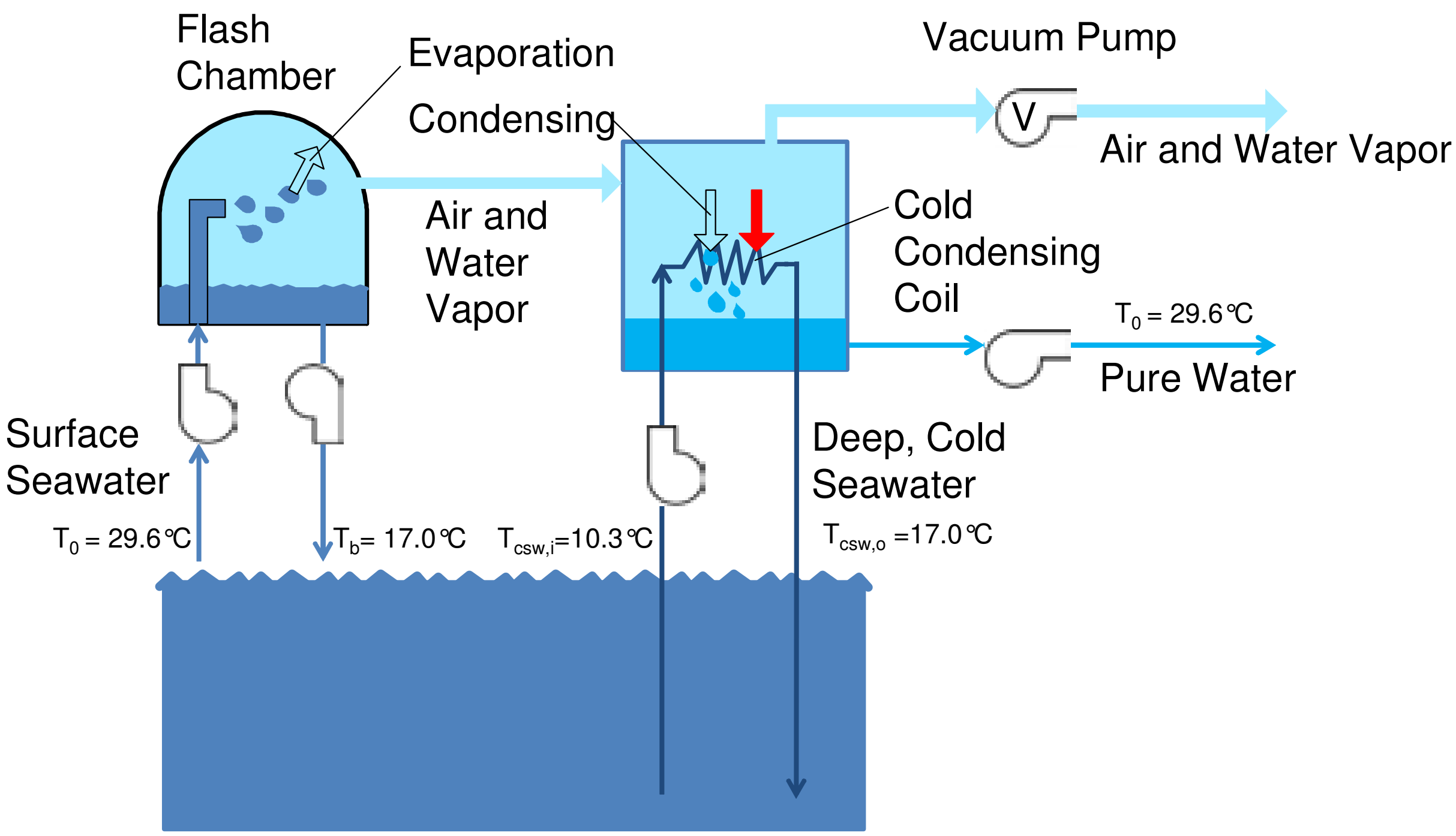

**Figure 5.2.** Low temperature thermal desalination process, modeled after literature

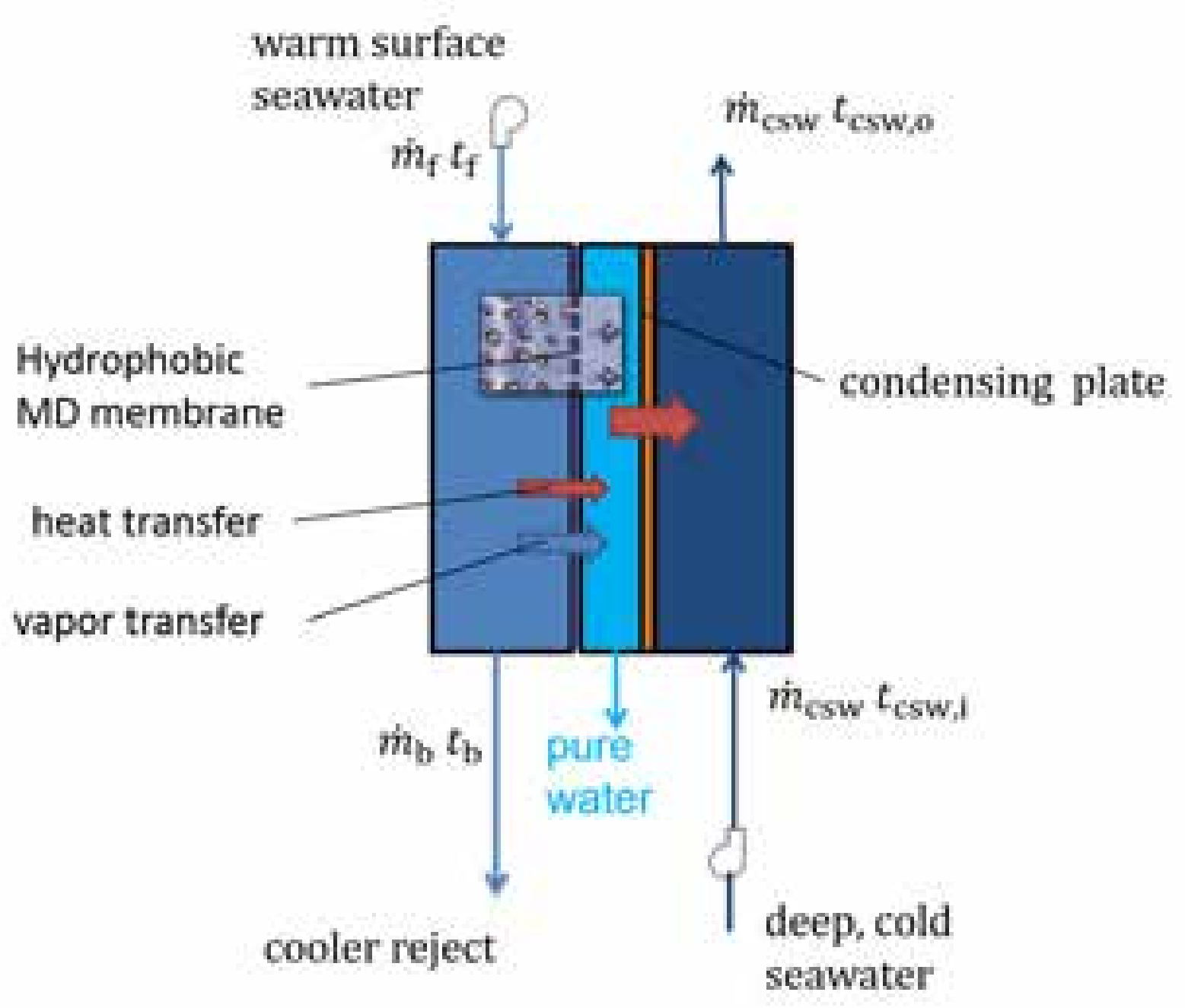

**Figure 5.3.** Permeate Gap Membrane Distillation (PGMD) system modeled for LTTD comparison.

The permeate gap membrane distillation system model was based on the other MD models used in this work: it was comprised of simultaneous heat and mass transfer across many computational cells, using the finite difference method.

The LTTD was modeled from experimental data from a study by Sistela et al for the NIOT LTTD plant in Kavaratti, India.

The entropy generation analysis analyzed the least work for separation, by comparing the entropy of the incoming and outgoing steams: it is process independent.

The results found that with reversible operation, the LTTD process could achieve 21.9 kWh/m$^3$, a performance significantly worse than RO, with losses largely due to pumping power and the vacuum pump. However, PGMD had much more promising results, being able to achieve about 2 kWh/m$^3$ including pumping losses from the MD module and sourcing the water: on par or superior to RO.

Only my contributions are shown here in any detail, for the full work, results, and analysis, see Nayar and Swaminathan et al. [113]. My contributions included idea generation, calculation choices for pump losses, modeling decision involvement, creating all diagrams, contributing to the PowerPoint and poster, and paper proofreading.

## 5.2 CONDUCTIVE GAP MEMBRANE DISTILLATION (CGMD)

Improvement of the configuration and module design for membrane distillation revolves around maximizing flux and minimizing heat transfer losses. The driving force for the MD desalination process is the vapor pressure difference across the membrane, which causes diffusion through it. It is therefore ideal to maximize the temperature across the membrane (to maximize the vapor pressure difference and thus permeate flux) and to minimize temperature gradients elsewhere in the system. It is also ideal to minimize heat transfer from the hot and cold streams in general.

Jaichander Swaminathan is first author on this work, and Hyung Won Chung and Professor Lienhard are coauthors [114].

While conventional modeling of MD has focused on minimizing heat transfer, our group has found the counter-intuitive approach of prioritizing flux with significant heat transfer to be more efficient, given the same membrane area. In contrast, traditional systems like Air-Gap Membrane Distillation (AGMD) employ the air-gap to minimize heat transfer from the cold and hot sides. But with limited system sizes, flux can more than double with improved reduction in the gap resistance, meanwhile keeping losses below 25%.

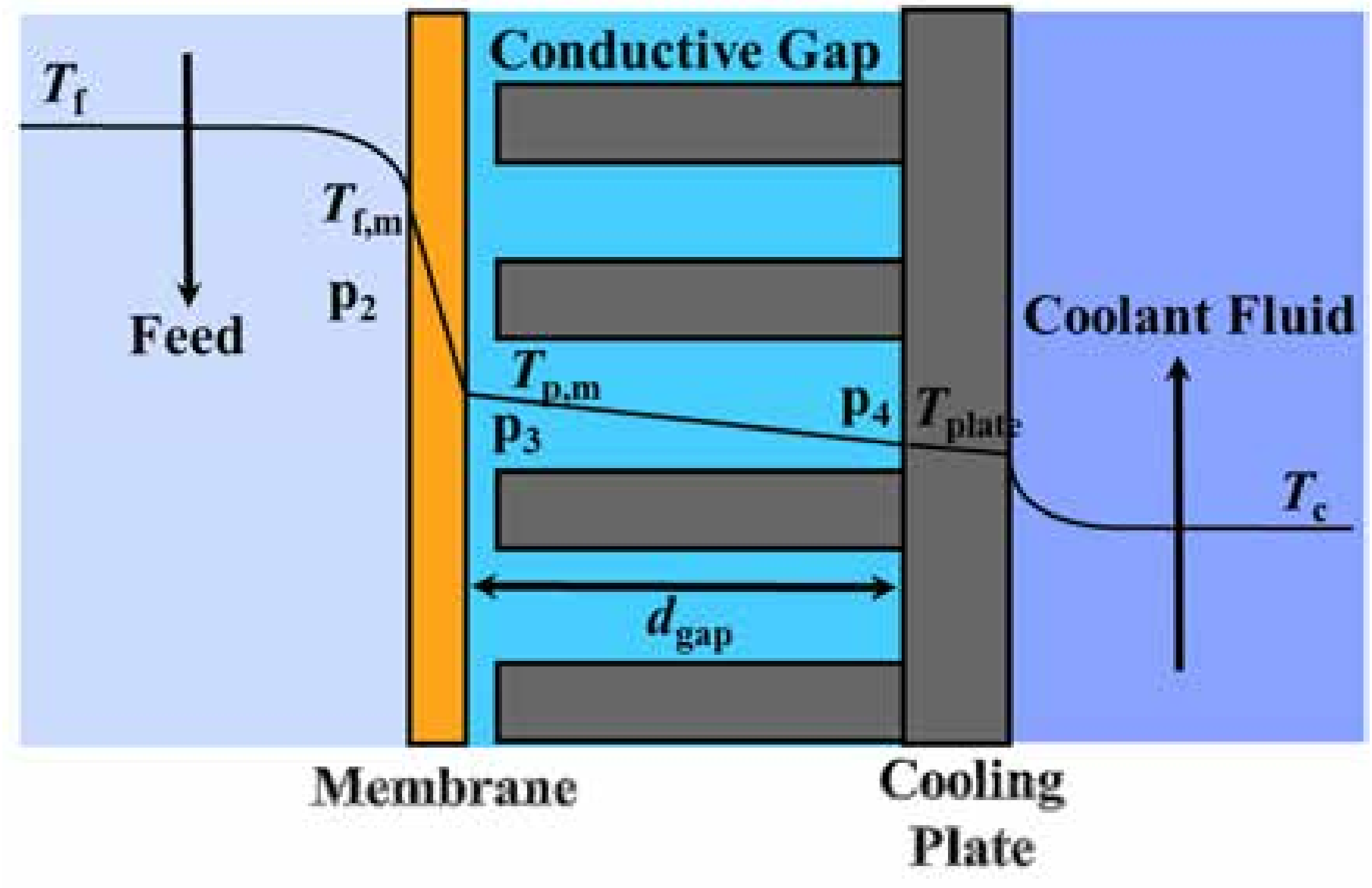

**Figure 5.4.** Approximate temperature gradients in CGMD. The conductive gap reduces the temperature difference across the gap, thus increasing it across the membrane to improve permeate flux.

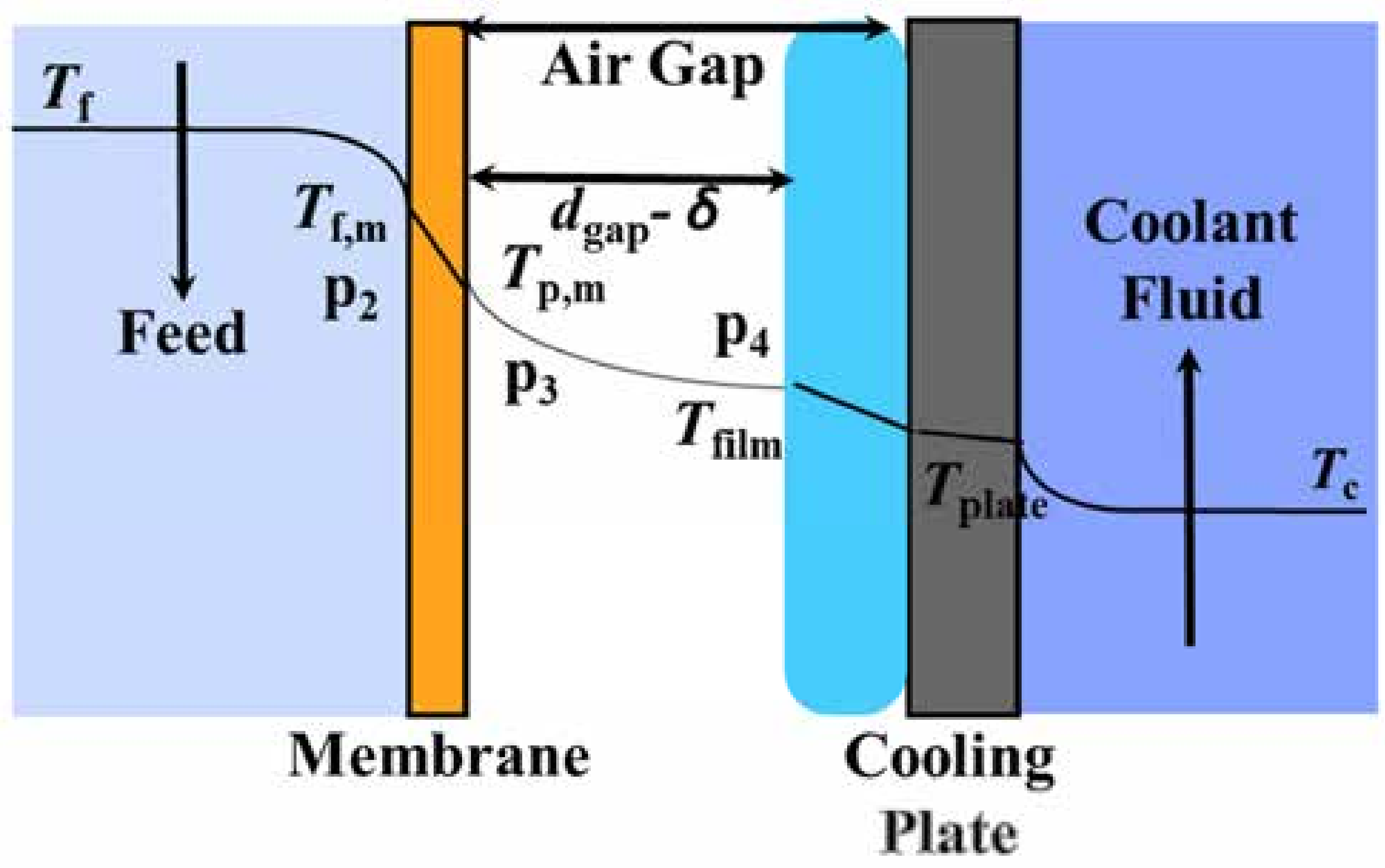

**Figure 5.5.** Approximate temperature gradients in AGMD.

However, reducing the gap size to reduce heat transfer resistance can only go so far, since the condensate itself requires a finite volume. To allow for low heat transfer resistance for larger real-world systems, metal meshes or fins can be placed in the gap to maintain a high conductivity. This is called Conductive Gap Membrane Distillation, or CGMD, an invention by our group.

Studies included creating several CGMD apparatuses for testing, numerical modeling for CGMD and other MD systems, testing of different conductive materials for the gap, and modeling and experiments to compare CGMD to other MD configurations. To date, two conference papers, a journal paper, and patent have been submitted or are nearing completion for this work.

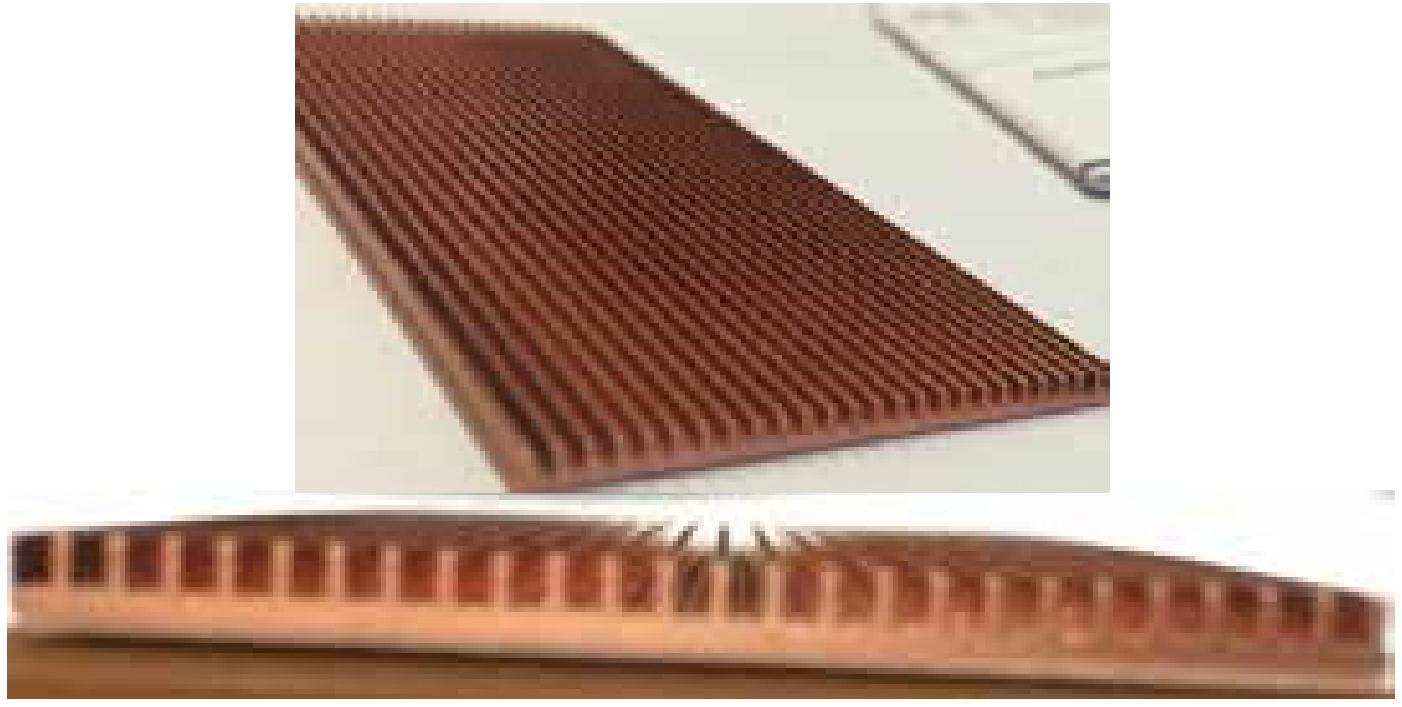

**Figure 5.6.** Fins for CGMD testing created for conductive gap comparison experiments at MASDAR. Fin was designed with SolidWorks and created with CNC machining.

**(a)** **(b)**

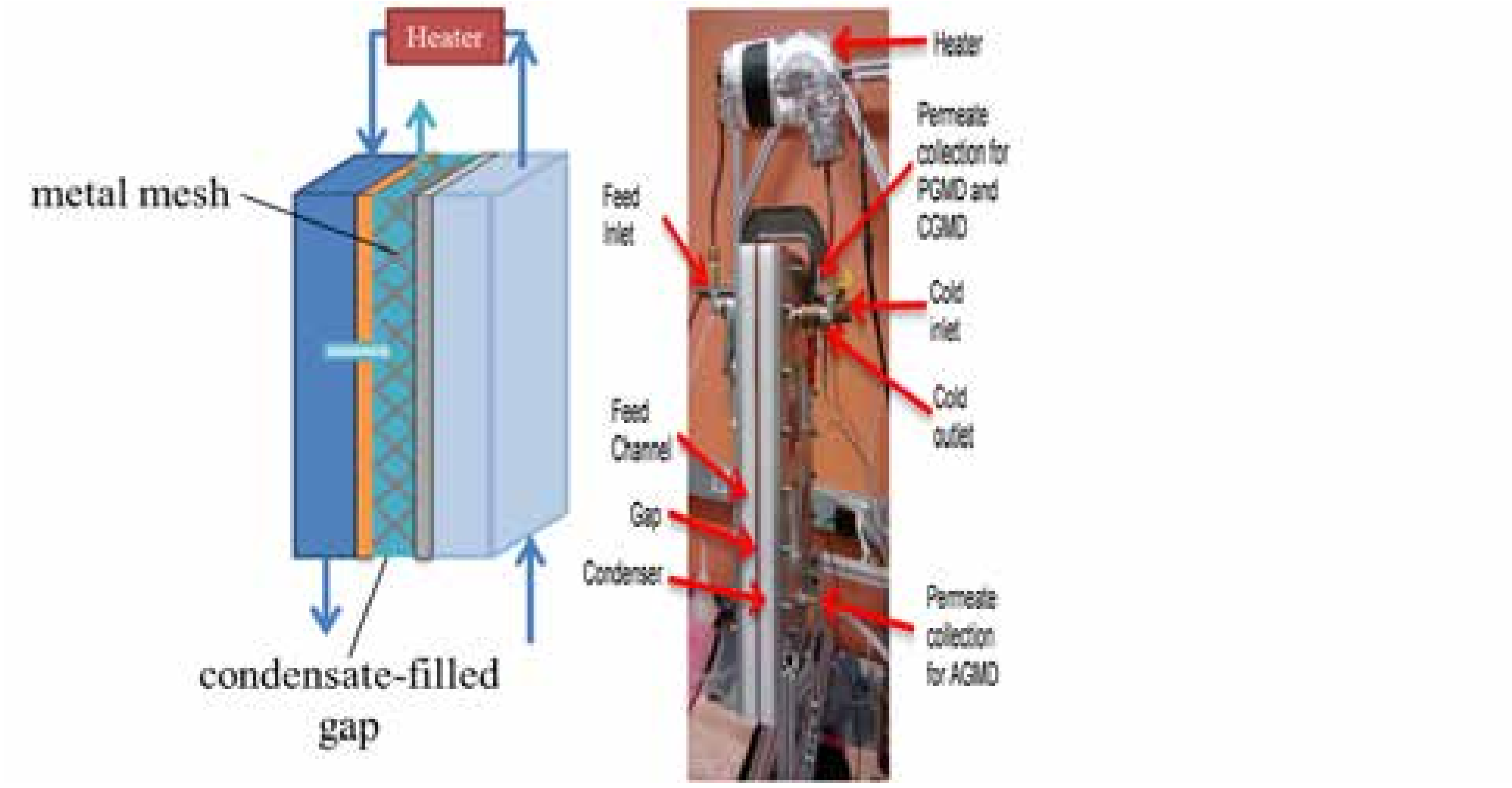

**Figure 5.7.** **a)** CGMD system diagram, showing top heater and preheat. **b)** CGMD test bed, created by modifying the apparatus by Ed Summers for CGMD.

The results of these studies have shown that for systems of the same membrane areas, the conductive gap can improve permeate flux so substantially that it can make MD competitive efficiency-wise with other mature thermal technologies (GOR of 6+), which has not previously been proven.

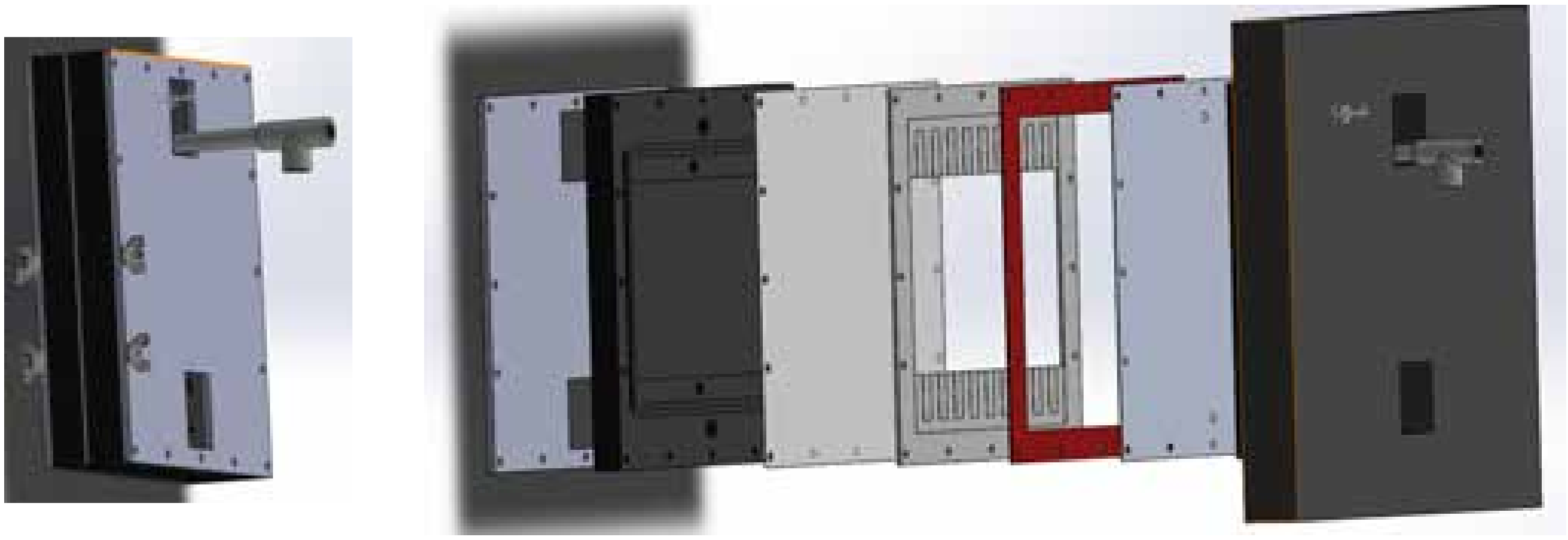

**Figure 5.8.** SolidWorks design for CGMD test bed intended to exceed state-of-the-art efficiencies. SolidWorks created by Jaichander Swaminathan.

My contributions to this work have included discussion of tradeoffs, idea generation, grant writing, proofreading, the idea of using fins, designing SolidWorks for fins, procuring parts, apparatus modification, construction, leak testing, patent content, flow channel design for the new CGMD system, and creating the CGMD diagrams for heat transfer, temperature gradients, apparatus description, and the general CGMD process.

## 5.3 MULTISTAGE VACUUM MEMBRANE DISTILLATION (MSVMD)

Multistage thermal systems are typically capable of superior efficiencies due to improvements in recovering and reusing energy. Multistage systems have the additional advantage of higher recoveries, allowing for systems that are designed for brine concentration, and that avoid problems with brine recirculation. For example, typical MD systems have recoveries of 5-10%, but the MSVMD systems studied here can exceed 40%.

Hyung Won Chung is first author on this work, and Jaichander Swaminathan and Professor Lienhard are coauthors [41].

To demonstrate the potential of multistage membrane distillation, our group created an MSVMD model thermodynamically similar to the dominant thermal technology by global market share, Multistage Flash.

This study implemented an EES numerical model of heat and mass transfer in MSVMD. It included the effects of Pitzer's equations for NaCl-solutions, boiling point elevation from salts present, and entropy generation. The performance was evaluated with energy efficiency (GOR), the second law efficiency, and specific membrane area.

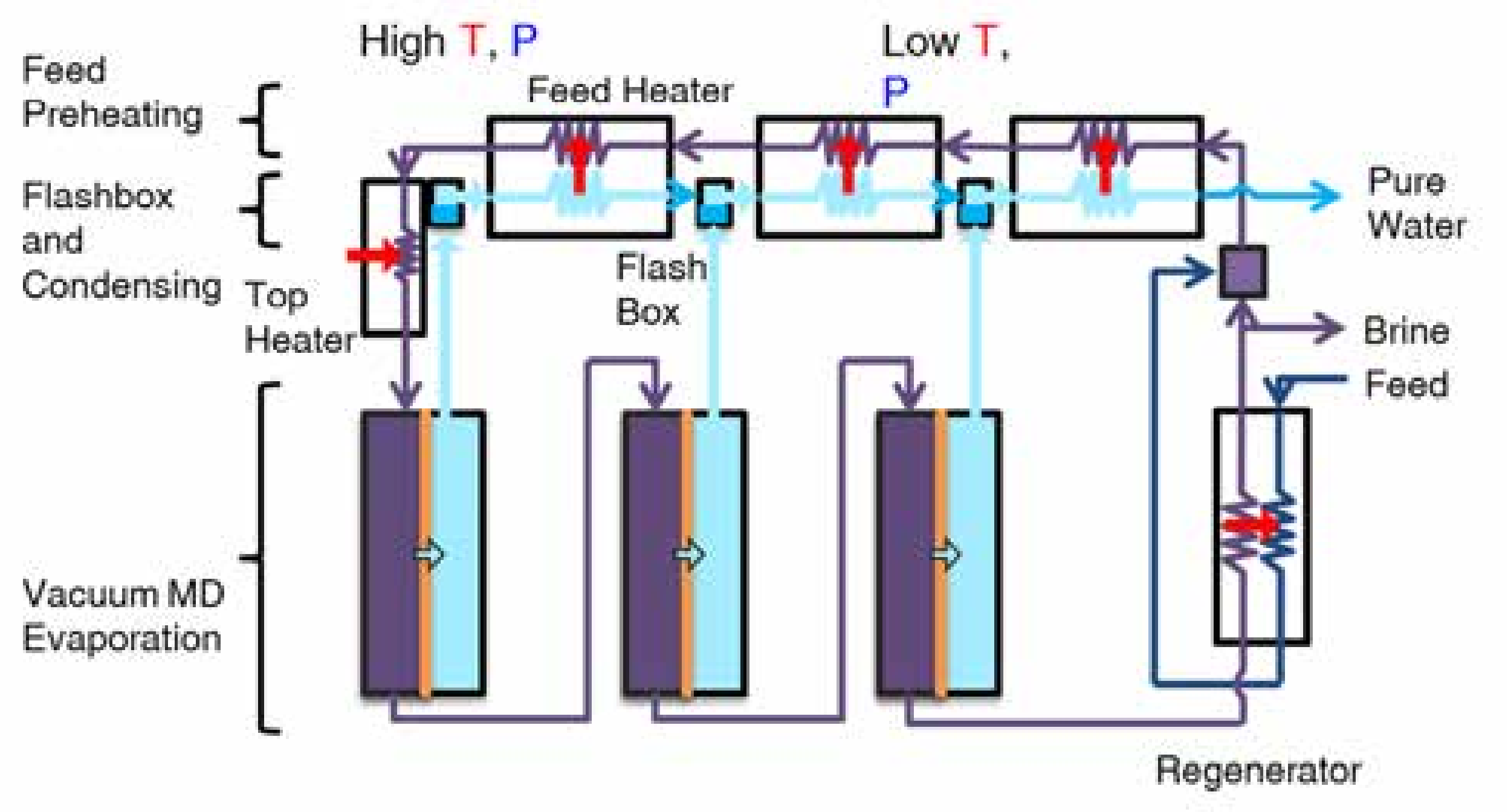

**Figure 5.9.** Multistage Vacuum Membrane Distillation (MSVMD) system diagram. The design is thermodynamically similar to MSF, but also includes a regenerator.

The results found that boiling point elevation decreased the overall GOR at higher concentrations, but that the second law efficiency increased. The second law efficiency improved because the energy needs of thermal systems is only slightly increased at high concentration, while the theoretical least work increases substantially at higher concentrations. Efficiencies were very similar to that of MSF while maintaining high recoveries, demonstrating that multistage MD can be competitive with the best thermal technologies. Notably, MD is more scalable and requires less metal heat exchanger area than MSF, making it suitable for ZLD or small scale applications.

A more notable result was that a non-uniform area for each stage is more optimal. Additionally, under certain conditions for non-uniform area systems, more permeable membranes can reduce efficiency by enhancing concentration polarization.

Hyung Won Chung was first author on this study [41]. My contributions to this work include help with the creating and troubleshooting of the numerical model, suggesting the inclusion of a regenerator to resolve temperature diverging issues within the heat exchangers, selection of stage spacing, making the system diagram, providing information for entropy generation calculations, proofreading, revisions, and writing parts of the journal paper.

## 5.4 Experimental Investigation Of High Efficiency Single-Stage Membrane Distillation Configurations

A new configuration of membrane distillation was invented by our group: conductive gap membrane distillation, or CGMD. To understand and validate the thermodynamics of this system, a comparison with both modeling and experiments was needed. An MD setup was adapted to change the gap configuration to compare the configurations under identical heat and mass transfer conditions. This was paired with EES numerical modeling.

Jaichander Swaminathan is first author on this work, and Hyung Won Chung and Professor Lienhard are coauthors [115].

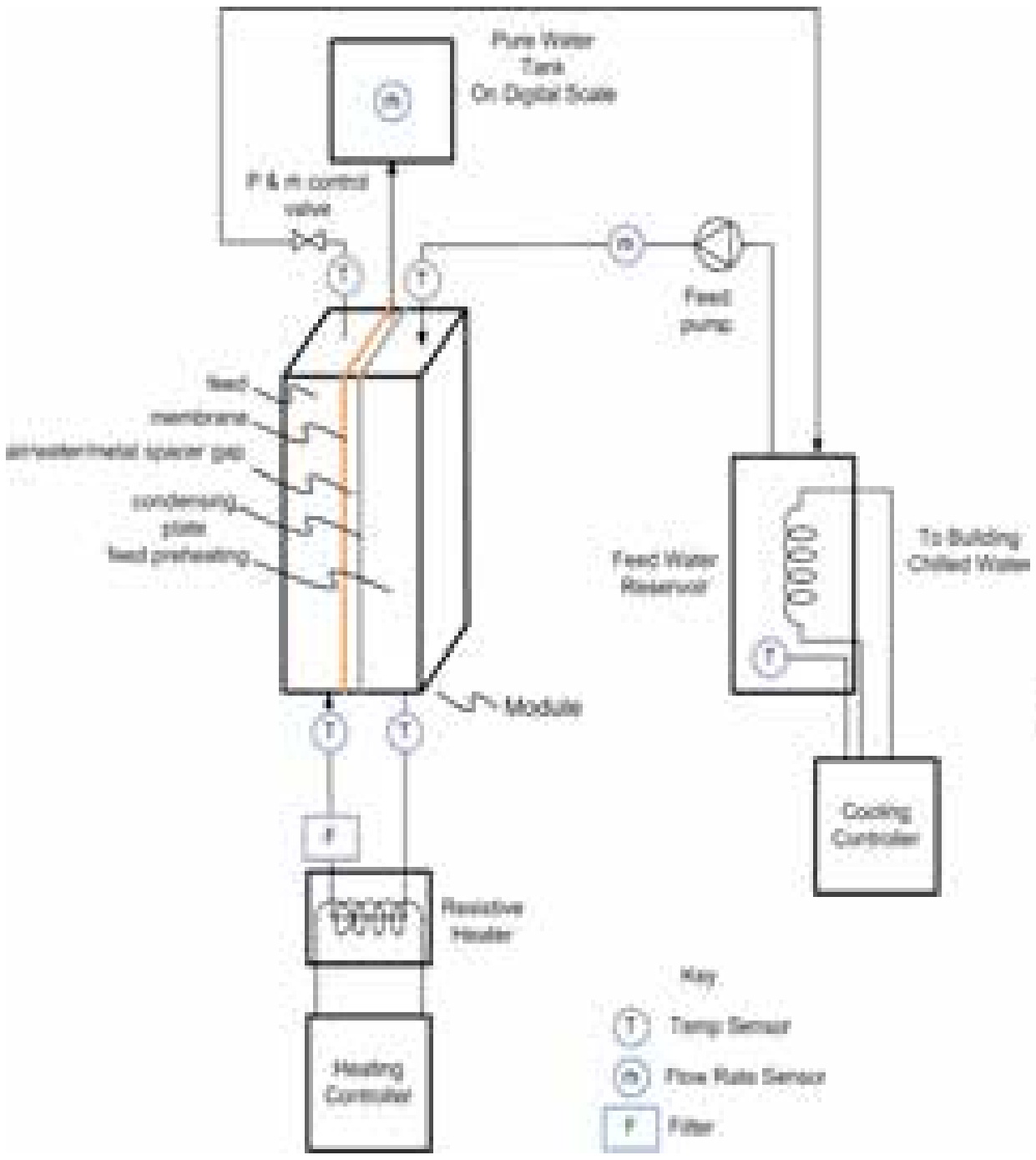

**Figure 5.10.** Schematic diagram of variable-module MD system

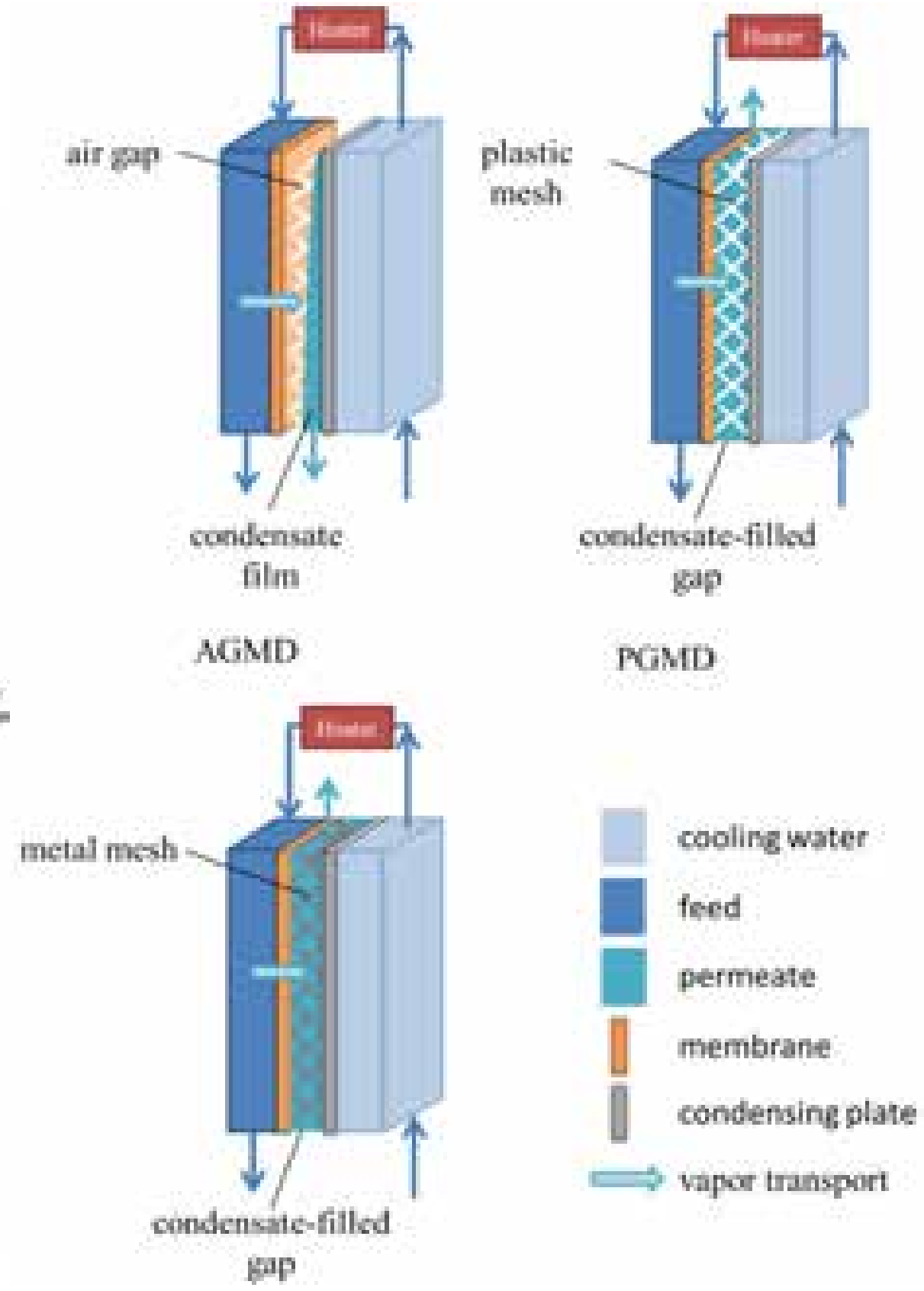

**Figure 5.11.** Membrane distillation configurations experimentally

The system was found to have a GOR 40-60% higher than that of AGMD, and 15-25% higher than PGMD. The modeling results matched these percent increases, and showed that further improvements on CGMD could realize even high efficiencies in better systems.

My contributions to this work included creating all diagrams, modifying the experimental setup, troubleshooting, idea generation, and paper revisions.

# Chapter 6. Scaling and fouling In membrane distillation for desalination applications: A review

## 6.1 Abstract

Membrane distillation (MD) has become an area of rapidly increasing research and development since the 1990s, providing a potentially cost effective thermally-driven desalination technology when paired with waste heat, solar thermal or geothermal heat sources. One principal challenge for MD is scaling and fouling contamination of the membrane, which has gained growing attention in the literature recently as well. The present paper surveys the published literature on MD membrane fouling. The goal of this work is to synthesize the key fouling conditions, fouling types, harmful effects, and mitigation techniques to provide a basis for future technology development. The investigation includes physical, thermal and flow conditions that affect fouling, types of fouling, mechanisms of fouling, fouling differences by sources of water, system design, effects of operating parameters, prevention, cleaning, membrane damage, and future trends. Finally, numerical modeling of the heat and mass transfer processes has been used to calculate the saturation index at the MD membrane interface and is used to better understand and explain some of trends reported in literature. Jaichander Swaminathan, Elena Guillen-Burrieza, Hassan A. Arafat, and John H. Lienhard V contributed to this work [11].

## 6.2 Introduction

Membrane distillation (MD) is a promising thermally driven desalination technology still in its infancy in terms of development and commercial deployment [10, 116]. The technology purifies water using a hydrophobic membrane, which is permeable to water vapor but which repels liquid water. In seawater desalination applications of MD, as hot saline feed solution flows over the membrane, the increased water vapor pressure from the higher temperature drives vapor through the pores ($d_p \approx 0.2 - 0.4\ \mu m$) of the hydrophobic membrane, where it is collected on the permeate side [18]. MD possesses unique advantages over other desalination technologies, including pressure-driven methods such as reverse osmosis (RO) and thermally-driven methods such flash distillation. MD is free of the specialized requirements of high-pressure RO systems, which includes heavy gauge piping, complex pumps, and maintenance demands [10]. Since MD is not a pressure driven process and only vapor is allowed to cross through the membrane, MD is more fouling resistant than RO [117] and has a potential 100% rejection of ions and macromolecules. MD can be run at lower temperatures than other thermal systems making untapped sources of waste heat usable, it requires significantly fewer parts, and can have a much smaller footprint as result of reduced vapor space [18]. Additionally, recent theoretical and computational work claims potential multistage DCMD configurations with efficiencies greater than that of other thermal technologies [86, 96, 87], assuming very large available heat exchanger areas. In practice, GOR values of practical state of the art MD systems with limited exchange areas are more modest [88]. Summers [118] has subsequently shown that multi-stage vacuum MD is thermodynamically identical to MSF, indicating that equivalent energy efficiencies can be achieved. The comparative simplicity makes MD more competitive for small-scale applications such as solar-driven systems for remote areas, especially in the developing world [18, 85, 119, 120]. However, significant advancements are needed in membrane technology for MD to reach the theoretical cost competitiveness and develop market share growth [121]. Fouling in MD is of particular importance, as fouling increases costs of energy consumption, downtime, cleaning, required membrane area, required membrane replacement, and creates problems with product water contamination from pore wetting [122, 123].

The first patents on MD were granted in the late 1960s, but it wasn't technologically feasible until ultrafiltration membranes in the 1980s enabled sufficiently high trans-membrane fluxes [18]. Currently, most MD work is done in the laboratory, although a number of test beds across the world for small-scale solar thermal MD have already been deployed, and a few other projects exist [18, 19, 120].

While increased research interest in MD is relatively recent [124], scaling under high temperature conditions has been a key problem in systems with water heating since the advent of the steam engine. Research in the area, especially for metal heat exchangers, originated well before 1900 [125]. However, with respect to thermal efficiency, these studies mainly focus on conductive resistance due to scale formation, and often do not address the type of transport phenomena that are important in the context of fluid-membrane systems [125]. A somewhat more relevant area of scaling research is that for RO. However, RO membranes are not specifically hydrophobic, are virtually non-porous, are comprised of different materials, and operate at much lower temperatures but much higher pressures. Hence, RO membranes exhibit significantly different fouling characteristics than MD membranes [18, 122, 125, 126].

Studies focused on scaling in MD largely originated in the 1990s, and since then have become more numerous [121, 124]. Between 1991 and 2011, sixteen solar-driven MD systems were tested at the pilot or semi-pilot scale [127]. Limited fouling data from those plants constitute most of what we know about the fouling potential of MD membranes and the damage they may sustain under actual field operation conditions. Parallel to those pilot studies, a number of dedicated lab-scale studies were also conducted to understand fouling in MD. For many years, it was believed that the hydrophobic nature of the membrane, the maximum pore size and the low feed pressure in the MD process are sufficient to prevent the feed solution from penetrating the membrane pores (often referred to as pore wetting), and from causing significant scaling on its surface. For example, in 2003, Koschikowski et al. [128] stated that "*the membranes used in MD are tested against fouling and scaling. Chemical feed water pre-treatment is not necessary. Intermittent operation of the module is possible. Contrary to RO, there is no danger of membrane damage if the membrane falls dry.*" Indeed, for years it was widely accepted that MD has this described ability to withstand dry out from intermittent operation. In fact, this is how most solar-powered MD plants operated, intermittently (shutting down overnight) and allowing the membranes to fall dry for hours every day [128, 129, 130, 34]. Intermittent operation can also result from unstable solar conditions or an uneven distribution of flux [34]. In contrast, the present review shows that while MD membranes are relatively resistant to fouling, they remains vulnerable to it and often require well engineered designs and operating methods to avoid and mitigate damage or destruction of the membranes by fouling. These design choices, especially in the case of inorganic scaling, are often related to maintaining the concentration of ions and the temperature at the membrane interface within limits where crystallization is not favored. Understanding temperature and concentration polarization effects (relative reduction in temperature and increase in solute concentration at the membrane interface compared to the feed bulk, due to the removal of energy and water mass through the membrane) therefore becomes key. Section 6.7 considers these factors in further detail while interpreting scaling data available in the literature.

Importantly, current MD membranes are adapted from microfiltration and similar markets, as yet there are no commercially available membranes specifically made for MD desalination [124]. An aim of this paper is to summarize differences in membrane properties for desalination from the literature so as to provide a background for the development of future, specialized membranes. The paper also aims to better understand fouling mitigation methods, and the effects and risks of different foulants.

## 6.3 TYPES OF FOULING IN MD

Fouling is commonly defined as the accumulation of unwanted material on solid surfaces with an associated detriment of function. The types of fouling that can occur in membrane systems and therefore potentially found in MD systems can be divided into four categories: inorganic salt scaling or precipitation fouling, particulate fouling, biological fouling, and chemical membrane degradation [131, 132, 133, 134]. The appropriate mitigation methods vary dramatically for each of these [29, 30]. The causes also vary strikingly, although particulate fouling can be closely related to the others, as a result of coagulation. It is therefore most practical to analyze each of these four types separately. After MD was introduced in the late 60s, the first commercial applications for MD were in the food and semiconductor industries, not desalination [30, 31]. Since 1985, the number of publications dealing with MD desalination has increased [31, 32]. As a result, expectations for the types of scaling in MD are often inferred from other desalination technologies, particularly RO. While both technologies involve mass transfer through membranes, significant differences related to fouling exist, notably the significantly higher operating temperatures of MD, as well as the hydrophobic properties of MD membranes, the presence of temperature gradients in MD, and the larger pore sizes in MD [18, 33]. Also, MD lacks the high pressures of RO which are generally believed to aid the formation of compacted cake scales.

### 6.3.1 INORGANIC SCALING IN MD

Study of inorganic scaling dominates the fouling literature for MD. Inorganic scaling, or simply scaling, in RO and MD generally falls into one of three categories: alkaline, non-alkaline, and uncharged molecule scale [122, 135]. Alkaline salts, or basic salts, have a tendency to make a solution more basic when added through hydrolysis, forming hydroxide ions. Generally, acidifying solutions below pH 7 decreases the tendency for many alkaline salts like calcium carbonate to scale or precipitate [136]. Non-alkaline salts include most other charged ions that dissolve in water without pH-raising tendencies [137]. Uncharged molecules that may scale, such as silica, are generally less soluble than salts because the charge on salts allows highly polar water molecules to break up and dissolve the salts. Uncharged scale can be considered with particulate scale [138], and is done so in this work.

#### 6.3.1.1 Alkaline scale in MD

Calcium carbonate is perhaps the most common scale in thermal desalination systems, often limiting operating conditions in brackish, groundwater, and seawater desalination [126, 139]. In thermal and RO desalination, $CaCO_3$ scale is regarded as a pervasive scale and among the first to reach supersaturated conditions in many feed solutions [139]. Calcium carbonate scale often forms after the breakdown of bicarbonate, $HCO_3^-$, as shown in the equation below:

$$Ca^{2+} + 2HCO_3^- \rightarrow CaCO_3 + CO_2 + H_2O$$

For typical MD conditions, the breakdown of bicarbonate plays the dominant role in calcium carbonate precipitation [140]. However, the carbonate equilibrium and scaling is more complex than this simple equation implies. Increased pH and higher carbonate concentration are strongly associated with calcium carbonate scale [141]. The solubility of CaCO3 changes dramatically with the concentration of CO2, and may decrease at higher temperatures as CO2 comes out of solution, which raises the pH [126, 142]. Adding to this effect, CaCO3 has inverse solubility, so high temperatures will decrease its solubility irrespective of CO2 concentration [140]. According to Shams El Din, in typical thermal desalination systems such as MSF, a temperature of 37 oC can be considered as the minimum temperature for the formation of CaCO3 from mildly concentrated ocean water [143], which is well below the typical operation temperatures of MD (60-80oC). However, the alkaline scaling process is strongly dependent on many factors such as heat transfer rate, brine concentration, residence time, flow conditions, etc. [126]. In general, bicarbonate dissociates more readily with increased temperatures. Notably, at higher temperatures carbonate has a tendency to hydrolyze into carbon dioxide [144], as follows:

$$CO_3^{2-} + H_2O \rightarrow 2\ OH^- + CO_2$$

This reaction makes the solution more basic, which influences the solubility of other scales, notably making $Mg(OH)_2$ more likely to precipitate. Carbon dioxide gas may come out of solution; this process is related to thermal water softening that can be used as a pretreatment strategy and has been discussed in the fouling mitigation (Section 5).

Calcium carbonate precipitates can take six different forms. Three anhydrous crystalline polymorphic forms may occur, known as calcite, aragonite and vaterite. These forms, all $CaCO_3$, differ in crystal morphology, color, hardness, and refractive index [28]. Three hydrated forms occur as well: amorphous calcium carbonate (ACC), calcium carbonate monohydrate (MCC), and calcium carbonate hexahydrate (CCH) [126, 145, 146, 147, 148]. Calcite is the most stable form found in MD system operation, but vaterite is common as well [149, 36]. In MD experiments, it was found that calcite formation was promoted by laminar flow [149]. Aragonite has been observed for MD as well [150], but it is relatively rare. The form of calcium carbonate is highly dependent on temperature, as seen in Figure 1 [140].

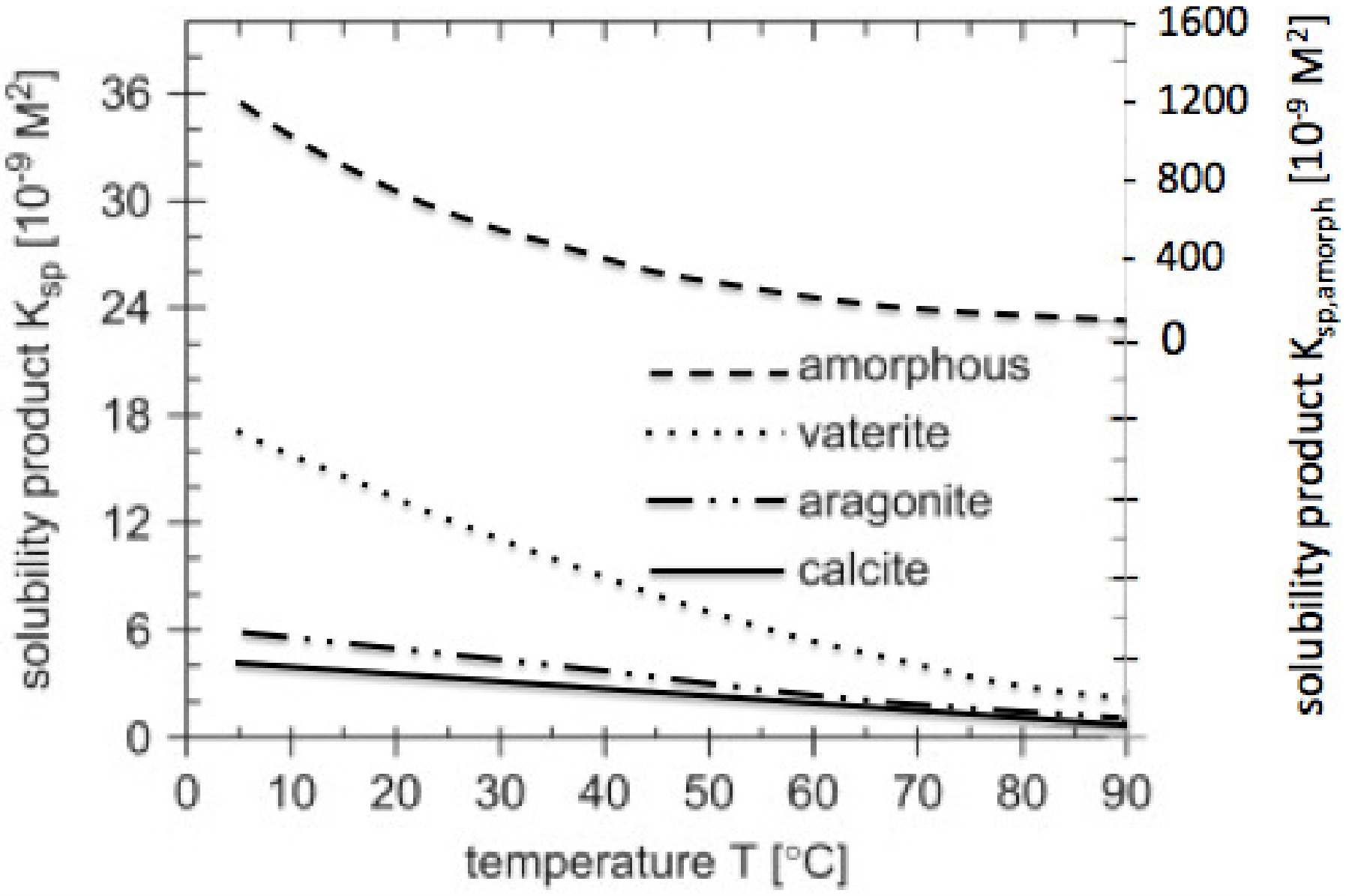

**Figure 6.1.** Solubility of several forms of calcium carbonate in water as a function of temperature [151].

Importantly, calcium carbonate scale in MD is usually only observed at relatively high saturation indices (SI). However, the presence of a microporous membrane substantially reduces induction times in $CaCO_3$ precipitation, causing an increase in nucleation rate regardless of the SI index [152, 153]. With hollow fiber membranes, Gryta observed that for saturation indexes between 5 and 20 (supersaturated), which he considered low levels, the induction period for $CaCO_3$ scale exceeded 30 min for tests ranging from 20°C to 100°C [140, 154]. Fei et al. [36] found that the system required very high SI values for precipitation of $CaCO_3$ in MD, recording concentration 32 times higher than saturation concentration to initiate scaling for calcite [36]. Fouling of $CaCO_3$ may be highly variable. As solubility decreases significantly with temperature, Gryta recommended a feed temperature below 80°C to avoid calcium carbonate scale for experiments with lake water [140]. Although more common at higher temperatures due to reduced solubility, calcium carbonate scale has been an issue for MD even at low temperatures, such as 40°C in a study on untreated tap water with stacked membrane modules [155]. Calcium carbonate fouling can even occur at ambient temperatures such as in reverse osmosis at sufficiently high concentrations [139].

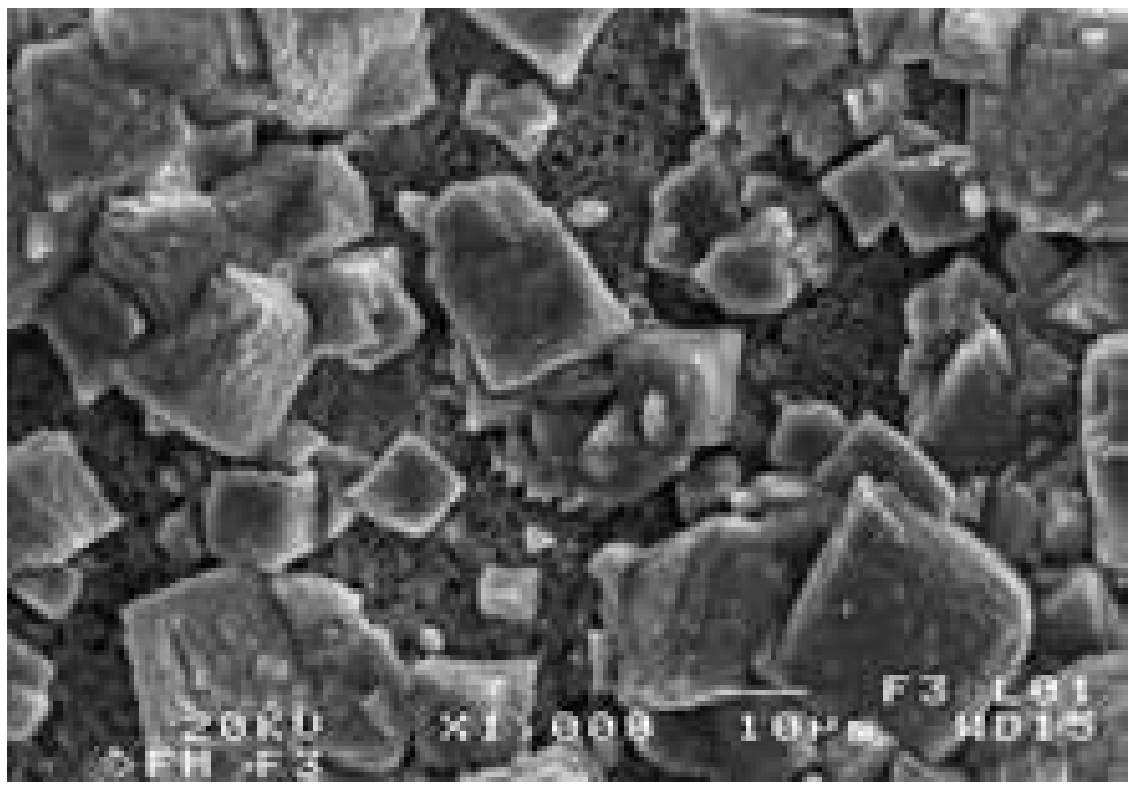

**Figure 6.2.** SEM image of calcium carbonate scale on a polypropylene Accurel PP S6/2 membrane surface with tap water feed for long duration DCMD experiments. Experiment performed with 86°C feed and 20°C distillate [156].

Studies of calcium carbonate scale in MD have consistently found that pure calcium carbonate scale causes significant flux decline, and that it is generally nonporous in nature [140, 157]. However, it has also been found that the feed flow rate can modify the morphology of the carbonate deposits. Figure 6.2 illustrates $CaCO_3$ crystal scale formation under small Reynolds number flow in hollow fiber membranes. In general, lower flow rates promote the growth of bigger crystals and more compact scaling layers, while higher flow rates reduce the size of the crystals and create comparatively more porous layers [140]. The flux decline varies widely, from zero or near zero to overall declines as high as 66% [158]. Furthermore, many experiments have found that calcium carbonate penetrates and scales in pores [157]. The scaling of calcium carbonate frequently causes wetting and thus results in contamination of the permeate by the feed [116, 149]. Long term performance studies, on the order of thousands of hours, found that calcium carbonate was consistently associated with wetted membranes, while deionized water experienced no such wetting [156]. Discussion of membrane damage from calcium carbonate scale can be found in the following sections, and results of calcium carbonate experiments, including flux decline or fouling rate, are detailed in Table 5.1.

Notably, some studies have pointed out that the content of impurities in the $CaCO_3$ solutions used as feed can definitely play a role in flux decay. For example, He et al. [36] reported almost negligible impact on permeate flux using a very pure $CaCO_3$ solution. Fast homogeneous precipitation in the bulk solution and transport of $CO_2$ across the membrane explained the negligible membrane scaling, which is in good agreement with the results presented by Nghiem et al. [159].

Co-precipitation of foulants is common in desalination systems, and complicates the prediction of scaling behavior. Calcium carbonate co-precipitation has been frequently observed in MD systems, especially with calcium sulfate [157]. However, systematic studies of co-precipitation are lacking [131, 160]. Gryta found that $CaCO_3$ co-precipitation with $CaSO_4$ resulted in bimodal crystal size distribution, using scanning electron microscopy with energy dispersive X-ray spectrometers (SEM-EDS) post mortem analysis. Importantly, co-precipitation with $CaSO_4$ was found to weaken the negative effects of $CaCO_3$, including reducing wetting and reducing membrane damage [157]. However, in a different paper, Fei found that co-precipitation of $CaCO_3$ and $CaSO_4$ caused an increased flux decline relative to $CaCO_3$ alone [36]. Studies on $CaCO_3$ and $CaSO_4$ precipitation kinetics have reported that an increased carbonate level may make $CaSO_4$ scale more tenacious and fine, but slows the rate of $CaSO_4$ precipitation [160]; this may explain the seemingly conflicting results in literature on the co-precipitation of these two compounds.

Curcio et al. [131] analyzed the fouling of $CaCO_3$ in the presence of humic acid (HA), with synthetic seawater concentrated 4 to 6 times. They found that the presence of other ions, including magnesium, sodium, sulfate, and HA, all inhibited the precipitation of $CaCO_3$ [131], which agrees with other studies [131, 161]. With 2 mg/L of HA present in hollow fiber membrane MD, the induction time of $CaCO_3$ was increased from 16 to 30 seconds [131]. It was found that HA increases the interfacial energy of vaterite by 7%, from 45 to 48 $mJ/m^2$ [131, 161]. Other authors have shown that different humic substances, such as humin or fulvic acid, have dissimilar degrees of inhibition on calcite growth [131, 162].

Scaling literature predicts that the interactive effects of mixed salt solutions are significant and can alter the thermodynamics of precipitation [163]. Gryta found that when $CaCO_3$ co-precipitated with iron oxides, the scale was porous, and the flux reduction was not very high [132].

Several magnesium scales may also be a concern in MD for feed solutions with high levels of $Mg^{2+}$. Magnesium hydroxide is another commonly observed alkaline scale in desalination applications, especially in groundwater, albeit not nearly as pervasive as calcium carbonate scale. Like $CaCO_3$, it exhibits inverse solubility with temperature, increasing its scaling propensity in MD [140]. Gryta tested lake-derived tap water with an Mg concentration of 15 mg/L in a Direct Contact Membrane Distillation (DCMD) system, and found that $Mg(OH)_2$ scaling occurred at above 348 K.

#### 6.3.1.2 Non-alkaline scale in MD

Calcium sulfate is one of the most common non-alkaline scales that occur in membrane systems [126]. In thermal desalination, $CaSO_4$ scale is regarded as a tenacious and very adherent scale [157], and it has behaved as such in MD processes as well [157]. Cleaning calcium sulfate is relatively difficult compared to alkaline scales, so modifying operating conditions to avoid such scale is the most common method of mitigation [126]. Calcium sulfate scale may occur in one of two hydrate forms, the dihydrate $CaSO_4 \cdot 2H_2O$ (gypsum) and the hemihydrate $CaSO_4 \cdot 0.5H_2O$ (Plaster of Paris), or as an anhydrite, $CaSO_4$. The form precipitated depends strongly on temperature, with gypsum common around 20°C [126, 164], and the anhydrite form more common at higher temperatures. Calcium sulfate solubility peaks around 40 °C [157], but does not vary dramatically across typical MD operating conditions. A study by Gryta on calcium sulfate in MD found that the concentration of sulfate ions should not exceed 600 mg/L, but up to 800 mg/L can be tolerated if bulk removal is available [157]. The study, specifically focused on $CaSO_4$ scaling in MD, found that membrane damage caused wetting and leaking and that it prevented further use of the membrane [157]. Gypsum scale was found to scale and penetrate the membrane pores. SEM images revealed needle-like gypsum crystals in typical orthorhombic and hexagonal prismatic needle. The crystals were tightly packed and tended to grow outward from initiated sites. As a consequence, exponentially worsening flux decline was observed, roughly experiencing a 29% decline over 13 hours [157]. Studies suggest that a supersaturated condition alone is not enough to start the crystallization of $CaSO_4$ on the membrane surface. Sufficiently long induction times (i.e., 53, 43, and 30 h for feed concentrations of 500, 1000, and 2000 mg/L of $CaSO_4$, respectively) [159] are also a requisite [159]. This long induction time suggests a strategy to control the $CaSO_4$ membrane scaling.

Calcium phosphate, another potential scale, is a non-alkaline scale that has frequently occurred in wastewater treatment and in RO membranes [126, 165, 166, 167]. Phosphate often exists in water supplies as phosphoric acid, which is relatively weak and which dissociates through several stages; significant concentrations of phosphate ion do not occur until the pH becomes relatively basic. Therefore, maintaining a low pH is an effective method to avoid phosphate scale [126]. It is often treated with use of dispersants in nanoparticle form in the feed as well [126]. However, calcium phosphate scale has not been found in the MD desalination literature. A most likely potential risk of calcium phosphate scaling arises when phosphates additives are used as antiscalants. These additives prevent calcium carbonate precipitation by sequestering $Ca^{2+}$. However, under relatively high temperatures and neutral pH (MD conditions), the rate of polyphosphate hydrolysis increases [168] decreasing the scale inhibition efficiency and creating a potential for calcium phosphate scaling [169].

MD experiments on non-alkaline magnesium scale have been performed as well. Tung-Weng et al. [170] found that $MgCl_2$ and $MgSO_4$ scale significantly more on polytetrafluoroethylene (PTFE) membranes compared to polyvinylidene fluoride (PVDF) membranes in a flat sheet module. They further report flux rate reduction of about an 86% of the initial value with the addition of 0.1% of either $MgCl_2$ or $MgSO_4$ to a 4.4% NaCl solution. In contrast, increasing the NaCl concentration to a 10% reduced the flux rate only to a 96% of the initial value, suggesting that the former registered decrease is a result of the Mg salts precipitating at the membrane surface rather than just a result of increased concentration and concentration polarization effect.

While not a common scale for most MD installations, sodium chloride, as the principal constituent of most desalination feeds, has been very widely used in MD literature including scaling studies. Sodium chloride, a non-alkaline scale, is characterized by a very high solubility and lengthy induction times. In scaling experiments by Tung-Wen Cheng et al. [170] under DCMD conditions at 50°C, increasing NaCl concentrations from 4.5% to 10% by weight only resulted in a 3-4% flux reduction, an expected level because of the decreased mole fraction of water at the membrane surface and not indicative of scaling-induced flux reduction, confirmed by the SEM micrographs which showed small levels of crystallization [170]. A MD paper by Fei He operating at 10% wt. NaCl reported similar results [87]. By contrast, extreme concentrations of NaCl, roughly 26-27.5 wt%, resulted in significant fouling [171]. After about 26% NaCl, the feed concentration seemed to asymptote while the flux dropped dramatically, indicating the onset of significant scaling. In the experiment, for roughly the first 250 min., 26% NaCl gave very good agreement to theory, indicating that even at this high concentration, the flux only dropped due to reduced vapor pressure, not scaling [172]. The extremely high solubility of NaCl relative to other salts and consideration of available sources of water indicates that virtually no natural source of water for desalination would have NaCl fouling as a concern. However, in cases of drying out membranes, NaCl will be among the salts to easily form on the membrane surface, as it will be discussed later. Other studies have shown that high concentrations of salts, exceeding saturation, can also cause wetting [157, 173].

### 6.3.2 PARTICULATE AND COLLOIDAL FOULING IN MD

Particulate and colloidal fouling risk is common in many feed water solutions. Larger particles can often be addressed with modern filtration technology (i.e., UF, MF, NF), but smaller particles can be an issue in fouling. Notably, particles vary greatly by water source, and, in the case of surface water, by season. However, in many MD desalination pilots, the use of cartridge filters or screens is widespread and proves to be effective for particulate matter [174, 175, 176]. Compared to ocean water; lake, ground, and especially river water are more likely to have particulate fouling concerns. Particles and colloids include clay, silt, particulate humic substances, debris, and silica [177].

Silica is particularly notable because its small size makes it harder to remove with pretreatment stems such as microfiltration. Silica is generally found in water supplies in three forms: colloidal silica, particulate silica, and dissolved silica (or monosilicic acid). The latter can cause severe fouling in RO and FO systems when supersaturation is reached and the silica starts polymerizing on the membranes [126]. PH can also play a role in the ionization and polymerization of silica; nevertheless at most natural pH levels (including that of the SW, around 8.5) silica is relatively unionized, lowering the risk of scaling [178]. However, silica solubility increases with temperature and should be much less of a problem in thermal systems such as MD. In an MD experiment with hollow fiber membranes and tap water, Karakulski et al. [158] found precipitation of silica compounds on the membrane [158]. The silica clogged capillary membrane inlets, causing a gradual decline of the module efficiency. The flux declined by 30% after 1100 hours of operation, mostly during the first 200-300 hours. Removal of the foulant with acidification combined with drying the membranes restored the initial flow rates only briefly [158]. This happened despite nanofiltration of the feed upstream of the MD membranes [158]. SEM-EDS analysis indicated that the deposit consisted of silicon, with small amounts of iron, calcium, zinc, and chlorine. Unlike conventional fouling, which coats the surface and blocks the pores, the SEM analysis indicates that the decline was not from a deposit layer, but from clogging membrane capillaries. The clogging reduced the feed flow rate, thus increasing temperature and concentration polarization, which reduced the module flux. The fibrous structure of the deposits blocked the foulants from further entry into the membranes capillary, stabilizing the flux [158]. This indicates that silica scale may be of significantly more concern in hollow fiber capillary MD systems than in flat sheet membranes. Silica fouling, while not causing a flux decline as fast as calcium carbonate, is a concern because it is difficult to clean. Acids that are commonly used to break down crystalline scale are not very effective on silica, which is uncharged [158]. When the feed has significant silica present, the authors recommend avoiding hollow fiber membranes with feed flow inside the capillaries.

An important and typically particulate foulant investigated in MD is iron oxide. Iron oxide fouling may consist of a variety of compounds, including iron oxides, iron hydroxides, and iron oxide-hydroxides [132]. These compounds are usually crystalline, and also may consist of hydrated forms. Iron oxide scale is not anticipated to be present in typical feed waters, but it is a risk of scaling due to the high propensity to rust on steel and even stainless steel parts in distillation systems. Corrosion fouling cannot only cause clogging problems, but also cause membrane damage by surface erosion (corroded flakes and chunks in motion through the narrow flow passages). Gryta found significant iron oxide fouling unexpectedly in a study on MD for treating effluents from regeneration of an ion exchange system in a water treatment plant [132]. However, the study found that iron oxide deposits did not significantly affect flux, undergoing only an 8% decline in permeate flux over 20 hours of operation. It was inferred that the iron scale was relatively porous. The composition of the iron oxides foulants was determined by x-ray diffraction, including maghemite, lepidocrocite, akaganéite, and hematite. It was found that the "iron oxides, hydroxides, and oxide-hydroxide" scales were crystalline [132]. These oxides exhibited high tendency to accumulate both on the membrane surface and within membrane pores.

The expected wetted corrosion reaction that causes iron oxide scale, also known as electrochemical corrosion, is as follows [132, 179]:

$$4\mathrm{Fe} + 2\,\mathrm{H_2O} + 3\,\mathrm{O_2} = 2\,\mathrm{Fe_2O_3}\cdot\mathrm{H_2O}$$

Corrosion reactions vary, but consist of oxidized forms of the metal and particles of positive, negative, or zero charge. Reactions also vary by oxygen content present [132]. In the various corrosion reactions, at least one of the products will be an oxidized metal, a metal cation, metal anion, or uncharged solid compound [132]. Oxidized metals may be $Fe_2O_3$ , FeO, or $Fe_3O_4$; iron metal cations are $Fe^{2+}$ or $Fe^{3+}$; metal anions include $HFeO_2^-$ and $FeO_4^{2-}$; and uncharged solid compounds include $Fe(OH)_3$, $Fe_3O_4$, and $Fe_2O_3$*$H_2O$ [132]. The oxidation reactions are complex, and are affected by conditions in the feed, including salinity, feed composition, and oxygen content. Under oxygen limited conditions, black magnetite $Fe_3O_4$ is often formed. When other salts are present, such as $Cl^-$ or $SO_4^-$, they may be incorporated into iron oxides or hydroxides. Hydrolysis of $Fe^{3+}$ ions may occur in basic conditions, from heating, or dilution of a salt with the ion. Hydrolysis may form hexaaquocation $Fe_3(H_2O)_6^{3+}$, and its $H_2O$ ligands again experience hydrolysis, creating FeOH or $Fe_2O_3$. The oxides present and their crystalline structure generally vary by conditions of formation, including temperature, other ions present, and pH. The presence of water, high or low pH, and other dissolved ions are conditions existing in MD that encourage corrosion of steel elements [132].

Additional Iron Oxide fouling in Gryta's study occurred as a consequence of acid cleaning (HCl) of the feed side. The volatile acid was capable of getting through the membrane to a small degree as gaseous HCl, acidizing the permeate and causing oxidation of the stainless steel elements on the permeate side of the system. Even concentrations of less than 50 g/L HCl can lead to significant reaction with the stainless steel elements [132]. Therefore, Gryta recommends using acid resistant high-grade steel or plastic as materials for MD systems [132, 180]. However, it is important to note that Gryta used very acidic conditions to clean the module, 18% and 36% HCl, which are relatively high compared to 3% or 5% HCl used in other experiments [132]. Therefore, acidic permeate may not be guaranteed to be a concern. The acid wash trials from the study are discussed in Section 5.7.7. The fouling layer was observed with SEM-EDS, as seen in Figure 6.3, and small amounts of Cu, Zn, Ca, P, Al, Mg, Na, S, Cl and Si were also observed in the membrane.

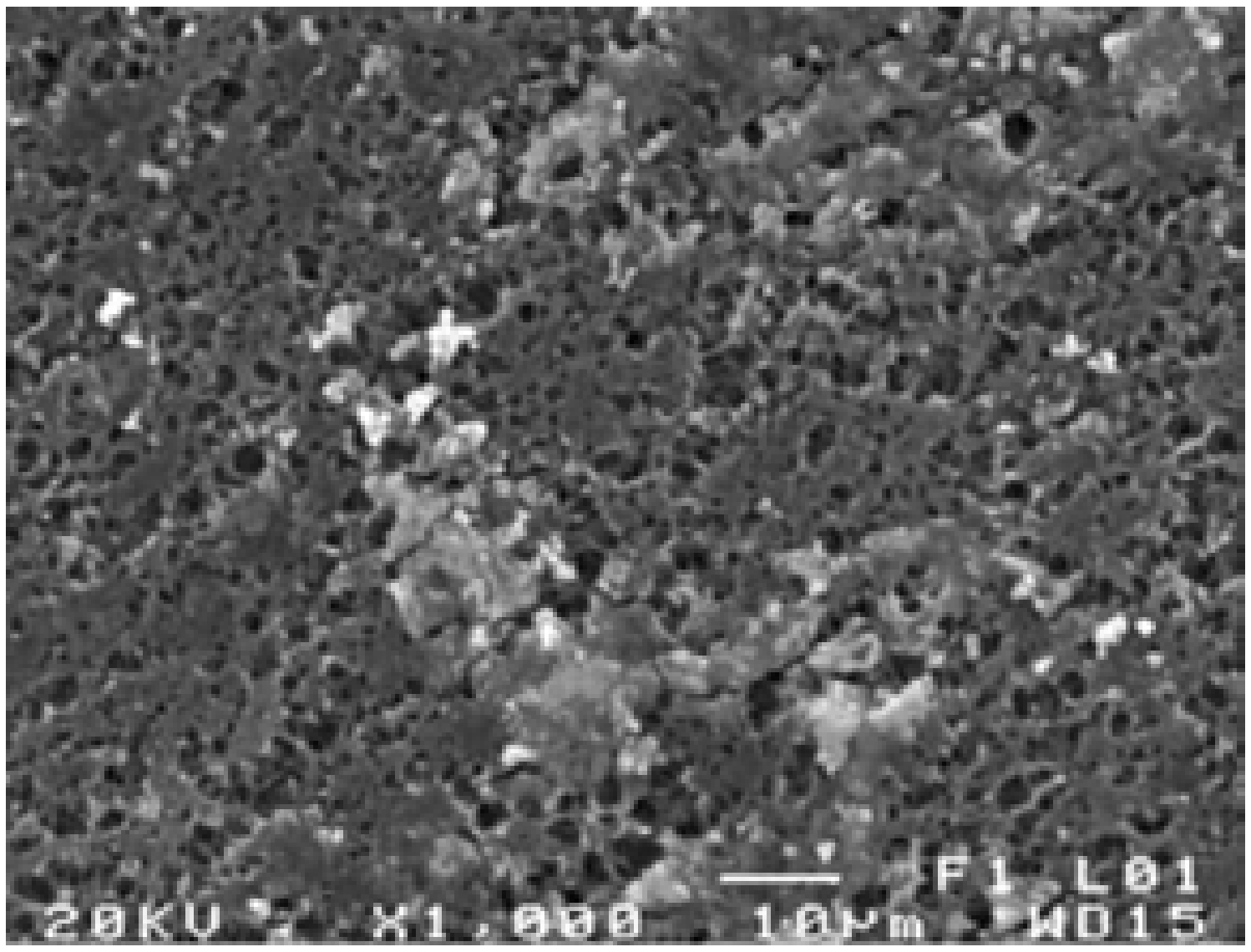

**Figure 6.3.** SEM micrograph of the polypropylene membrane surface covered with iron oxides after cleaning with 18 wt.% HCl failed to remove them, with a 353 K feed and a 293 K distillate temperature [132].

Compared to other scale forms in MD, iron oxide can be judged as unlikely to occur with proper system design. It is relatively less harmful to permeate flux but still a major cause of wettability and very difficult to remove. The lower operating pressure typical of the MD process makes it possible to use plastic components, which can potentially eliminate most iron from the system.

### 6.3.3 BIOFOULING IN MD

Biofouling of hydrophobic membranes applicable to MD has been a key research interest in the food, beverage, and wastewater industries [133, 181]. Many of the membranes used in MD originated in these industries. The majority of these studies focus on very high chemical oxygen demand (COD) effluents, although recently MD biofouling studies have extended to MD for desalination [133]. The high COD content comes from using MD or similar systems for animal products processing, fermentation, and other processes [133]. For the sake of brevity, this review focuses only on biofouling relevant to clean water production.

Biofouling relevant to MD desalination includes bacteria, fungi, and biofilm studies. Biofouling is pervasive in most waters [177] and has been a critical issue for RO membranes [182] and it is likely to be a concern in practical MD systems. However, the operating conditions of the MD process, especially the high temperatures and salinity, can restrict to a great extent the microbial growth in MD installations [133]. As a consequence, the problems caused by biofouling in membrane processes including NF, UF or RO should not occur in such a high degree in MD systems. [183]. However, organic fouling can play a more important role in membrane wetting in MD.

#### 6.3.3.1 Bacteria and biofilms in MD

Bacteria and microorganisms are pervasive in water systems. While chlorination is effective in killing bacteria, it can be damaging to many common MD membrane materials [177]. Bacteria can be very difficult to remove from membranes, as they excrete an extracellular polymer substance (EPS) to adhere to the surface [177]. In typical biofilm formation, bacteria colonize and excrete EPS, and then organic compounds accumulate in the film. These compounds are typically composed of polysaccharides, proteins, lipids, humic substances, nucleic acids and aromatic amino acids, and often contain trapped particles and absorbed substances [133]. These biofilms are typically 75-95% water and are relatively porous compared to alkaline scale [177].

Biofouling impairs MD process through wetting and pore blocking. Additionally, the relatively porous biofouling layers reduce diffusion and create a hydrodynamically stagnant layer of water at the feed side [133]. Biofilms are mostly constituted by a hydrated EPS matrix which makes diffusion the main mass transport mechanism. Water diffusion coefficient in biofilms has been estimated to be 15% lower than that in bulk water [184], conferring biofilms an extra mass transfer resistance and increasing temperature and concentration polarization effect. As a consequence, they can hinder convective heat transfer to the membrane while favoring diffusion and conduction. The thermal conductivity of biofilms has been estimated to be around 0.57–0.71 W/mK (close to that of water), almost 75% less than that of inorganic scale (i.e. $CaCO_3$ and $CaSO_4$) [185].

In Krivorot et al.'s experiments on hollow fiber MD with ocean water and a high biological load ($1\times10^8$ CFU/ml) at 40°C, permeate flux declined by 34% over 19 days [133]. However, minimal flux decline was detected in samples with normal biological loads. The high biological load was attributed to local wastewater spills to the sea. In the high biological load sample, a conditioning biofilm was formed in as little as 4 hours, and a recognizable biofilm was apparent after 28 hours. Over the 19 day experiment, all samples showed a biofilm. Temperature cycling to 70°C was found to reduce the biofouling behavior. In general, processes with a top brine temperature of 70°C minimized any biological presence [133].

In a study done by Gryta [183], MD was performed on a bioreactor with saline wastewater that contained yeast, *Pseudomonas* and *Streptococcus Faccalis* bacterium, and the fungi *Penicillium* and *Aspergillus*. The DCMD hollow fiber membranes failed to prevent *Streptococcus* bacteria from entering the distillate. With a temperature of 90°C and salts concentrated up to 300,000 ppm, no bacteria was detected at the membrane surface, indicating these conditions prevented bacterial growth. However, when the temperature was decreased to 80°C, bacteria and fungi were detected at the membrane (Figure 6.4).

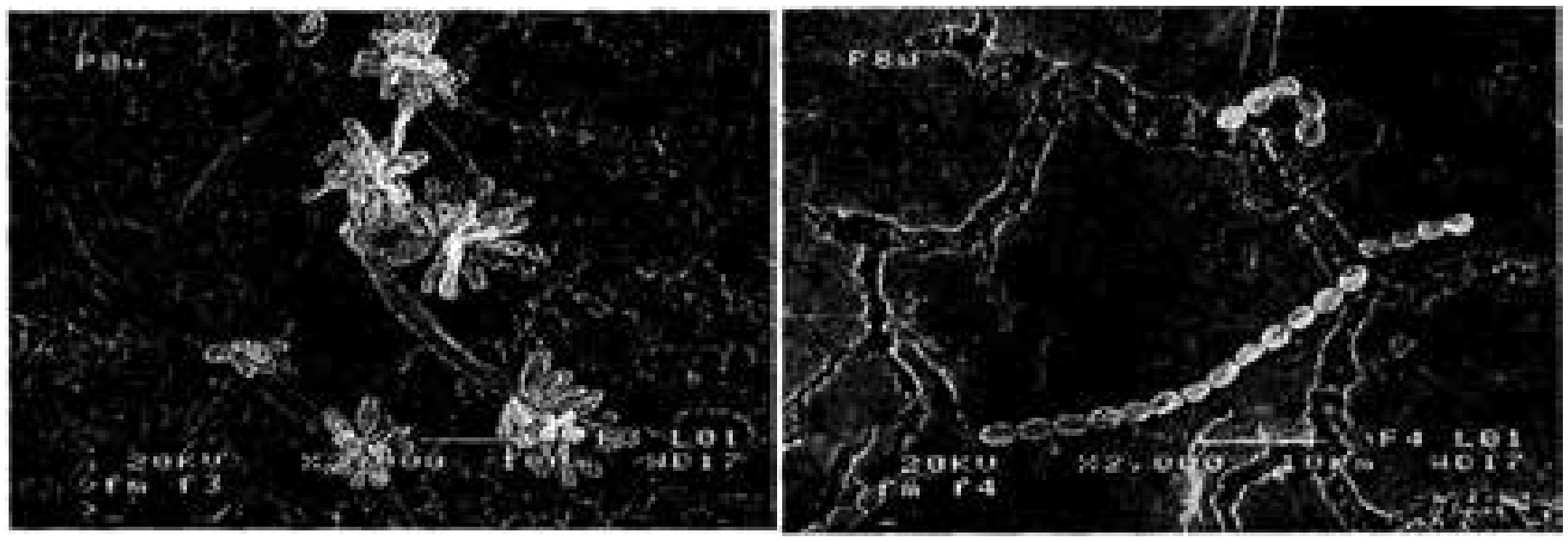

**Figure 6.4.** Left: SEM image of *Aspergillus* fungi on MD membrane surface. Right: SEM image of *Streptococcus faecalis* bacteria, 65°C feed solution, at roughly 5 times the concentration of seawater [183].

Notably, while the anaerobic bacteria *Streptococcus* grew on the membrane, no aerobic bacteria were observed on the membrane, despite being present in the original feed. The same feed caused significant fouling in MSF and RO, but tolerable and unproblematic fouling in MD. Elevated temperature, significant salt concentrations, and low pH values all can hinder bacterial growth. Similar MD studies by Meindersma et al. using an AGMD system and pond water as feed, reported a flux decline after 800 h from biofouling. No organism break through was noted however [186]. In this same study, original flux was almost restored by reversing the direction of the flow.

Biofilms, especially in seawater settings, often contain more microorganisms in addition to bacteria. Although direct MD experiments with marine microorganisms couldn't be found in the literature, highly relevant superhydrophobic materials similar to MD membrane materials have been examined in seawater conditions. A study by Zhang [187] compared ocean fouling in submerged hydrophobic and superhydrophobic surfaces over 6 months. Polysiloxane and PTFE surfaces were examined. The results showed that the hydrophobic surface exhibited fouling within a day, but the superhydrophobic surface (contact angle 169°) resisted fouling for about three weeks. However, after 2 months, both surfaces were heavily fouled and wetted with 10-20% macroalgae, 5-10% barnacles, and 50-60% bryozoans, although they resisted the tubeworms, ascidians, and to some extent algae that covered the control plate. The loss of biofouling resistance was attributed largely to air bubble loss through the membrane, as the air dissolves into the surrounding water [187]. These results can be extended to MD surfaces, where seawater is passed on the feed side over a hydrophobic membrane.

Specific conditions for avoiding biofouling cannot be determined presently from the literature because biofouling depends on many factors such as: salt concentration, feed composition, residence time, pre-treatment, bacterium present, operating temperatures, membrane type and cleaning frequency [133].

### 6.3.3.2 Natural organic matter (NOM) in MD

NOM compounds at risk for MD are especially prevalent in wastewater and certain lake and ocean water samples. NOM includes various constituents such as proteins, amino sugars, polysaccharides, polyhydroxy-aromatics, and humic substances [188]. These compounds are often present where traditional biofouling is a concern, as they often provide sustenance to bacteria and may originate from similar processes. The fouling caused by NOM can affect both the permeability and the dissolved solids rejection of the membranes. Membrane fouling in the presence of organic compounds is affected by ionic strength, pH, ions present, membrane surface structure and chemistry, molecular weight, polarity, permeate flux, and hydrodynamic and operating conditions [135]. In general, hydrophilic surfaces are less susceptible to organic fouling [189, 190] but MD employs hydrophobic surfaces, making organic fouling a concern.

The principal NOM foulant is HA, a general term for complex mixtures of organic acids with carboxyl and phenolate groups [191]. HA is produced by biodegradation of organic matter, and it gives many swamps and rivers a characteristic yellow brown color [137]. HA are complex and vary greatly; molecular weights typically occur between 700 and 200,000 Daltons [192], and may even be above or below this range. HA systems are thus typically measured with averages. For instance, the average HA particle in a solution may act as an acid with two and sometimes three free hydrogen atoms [191]. HA production in rivers and lakes is often seasonal, with large quantities produced annually from deciduous tree leaf decomposition. Importantly, HA may readily nourish bacteria, and thus may instigate significant bacterial fouling [133]. HA fouling in MD may vary based on feed composition, membrane hydrophobicity, temperature, membrane pore characteristics, and pH [193, 194]. For example, the addition of multivalent cations increases the electrolyte or ionic strength of the feed water and can favor the aggregation of the HA and therefore the fouling [195]. Divalent ions, including $Ca^{2+}$, act as binding agents to the carboxyl functional groups reducing the charge and the electrostatic repulsion between macromolecules and encouraging particulate precipitation [131, 13]. HA at 100mg/L concentration in a MD study with added $CaCl_2$ at 3.775 mM caused significant flux reduction of 40% on flat-sheet membranes [29]. This solution treated with MD produced a thick fouling layer, blocked pores, and increased heat transfer resistance [29]. In more acidic conditions, HA dissociates less because fewer $Ca^{2+}$ ions are available [29, 196]. However, HA can also affect other types of scaling and has significant scaling inhibition effects on calcium carbonate (i.e. inhibit heterogeneous nucleation and increases induction time) as explained earlier [131].

Other metal ions (i.e., $Fe^{3+}$, $Mg^{2+}$, $Al^{3+}$, etc.) can also contribute to NOM fouling in a similar way: increasing the ionic strength and causing metal ion-induced aggregation [197]. In a study with UF hollow fiber polyethersulfone (PES) membranes, the presence of $Fe^{3+}$ in the HA feed solution reduced the flux to one fifth of the original rate in one hour. The effects of $Ca^{2+}$ and $Mg^{2+}$ were similar but not as significant [198]. In this same study, the authors found that the use of EDTA as a chelating agent inhibited the crosslinking of HA induced by the presence of the metal ions and reduced the fouling.

HA deposits are typically loosely packed and porous and traditionally in UF and MF systems they are effectively eliminated through backwashing. In MD systems, they can be cleaned rather effectively with basic solutions. Srisurichan [29] found that HA fouling was easily removed with a 0.1 M NaOH solution, while still achieving full recovery of permeate flux. Alternating temperature changes, in one case between 25 °C and 35 °C, was found to clean HA as well, resulting in a flux recovery of 98.2% with proven repeatability [13].

## 6.4 FACTORS THAT INFLUENCE SCALING IN MD

### 6.4.1 TEMPERATURE

Temperature is among the most dominant factors related to scaling and fouling of MD membranes. In particular, the solubility and crystal formation of salts vary widely over the temperature range relevant to the MD systems. Importantly, the solubility of individual salts may be positively or negatively correlated with temperature. For example, the solubility of sodium chloride increases with temperature, whereas those of calcium carbonate, magnesium hydroxide, and calcium phosphate decrease with temperature. This negative correlation of solubility with temperature is typical for alkaline salts, which depend on the breakdown of water into hydrogen and hydroxide in order to form scale; such dissociation increases at higher temperatures [137].

Salts such as calcium sulfate and calcium carbonate that exhibit inverse solubility are also often the closest to being saturated in desalination feed solutions (calcium sulfate concentration is higher in the case of seawater as a feed while calcium carbonate concentration is higher in ground water sources). Generally, for common feed solutions, increased temperature causes increased risk of scaling. Higher temperatures also reduce induction periods for some salts [199].

Temperature can have a significant effect on biofouling due to microorganisms' lack of tolerance for high temperature and also because of thermal effects on organic compounds. According to M. Krivorot et al.'s experiments with hollow fiber membranes, at temperatures above 60 °C, most environmental organisms will not function and hence not grow on MD membranes [133]. Temperature increase causes the decomposition of HA and other biological compounds. In fact, for RO membrane systems, temperature increase may be used as an effective cleaning method for HA [13]. However, for MD at higher temperatures, permeate flux increases: this may lead to higher concentration of organic compounds at the membrane interface due to the concentration polarization effect. Srisurichan [29] found that at higher temperatures, flux decline was greater for solutions containing HA, observing a 16% decline at 50°C and a 43% decline at 70°C. Severe protein fouling was observed at temperatures higher than 20–38°C for aqueous solutions containing organic compounds at representative concentrations (i.e., wastewater, NOM, bovine serum albumin, etc.) [30, 183, 200] but it was practically absent at lower temperatures [201]. Notably, hydrophobic surfaces show an especially high tendency to get fouled by proteins [202], making MD membranes problematic for waters containing proteins, amino sugars or polysaccharides [203].

### 6.4.2 DISSOLVED GASES

Dissolved gases are present in almost all feed waters of interest in desalination [137], and these gases may have some limited effects on scaling and fouling in MD. Gases dissolved in feed water, as well as those resulting from chemical processes such as the breakdown of bicarbonates, travel into the membrane along with the water vapor, providing an additional diffusive resistance for the water vapor [140, 171].

The effect of dissolved gases on fouling is indirect and small; dissolved gases impede the permeate flow process, reducing concentration polarization and scaling. This occurs because dissolved gases in the feed stream may flow into the membrane pores, providing mass transfer resistance to water vapor, and may also contribute to mass transfer resistance in the air gap after the membrane (depending on the system configuration). The effect is to reduce the condensation heat transfer rate, possibly making the system mass transfer limited on the air-side, thus reducing the overall vapor flux [19]. On the other hand, the absence of dissolved gases can increase membrane wettability by removing the air trapped in the membrane pores, which was experimentally confirmed by Schofield et al. [204]. The presence of these gases in the membrane can act as a barrier to fouling. Therefore, reducing dissolved gases by deaeration or other means may be expected to increase fouling potential [19].

Dissolved gases, especially carbon dioxide, may alter the pH of the solution, affecting the solubility of various salts as described previously. Dissolved $CO_2$, common in many feed waters and often produced by breakdown of calcium carbonate in ground water, may acidify the water by the creation and dissolution of carbonic acid as follows [134]:

$$CO_2(aq) + H_2O \leftrightarrow H_2CO_3 \leftrightarrow HCO_3^- + H^+ \leftrightarrow CO_3^{2-} + 2H^+$$

So despite increasing carbonate concentrations, typically increased dissolution of $CO_2$ reduces scaling by the associated pH decrease. This result may differ depending on temperatures, concentrations, and pH, as discussed in Section 5.1 (on thermal softening). Dissolved gases or lack of them may significantly affect biofouling, as the presence of dissolved oxygen supports aerobic bacteria and microorganism fouling. Therefore, deaeration of oxygen may be used to inhibit microbial growth.

### 6.4.3 WATER SOURCE

As seen in previous sections, fouling likelihood and type of fouling in MD depend on the salts and other foulants present in the feed water and thus are highly dependent on the water source. Generally, specific sources have fairly consistent conditions and thus consistent expectations for fouling, although surface waters' quality and algae blooms may be seasonal.

Possible water sources for desalination include lake, river, ocean, ground waters as well as industrial waste water. Generally calcium carbonate saturation is a significant concern in most water sources relative to other salts. However, water sources are variable enough that other insights on types of fouling susceptibility may not be comprehensive. Compared to ocean water, lake and river water are typically characterized by high silica content, biological compounds, suspended solids, and calcium concentrations, but vary widely between different rivers and lakes [205]. Due to the low salinity of these waters, MD use is unlikely, but electrodialysis desalination may be used if slightly reducing salt concentrations is desirable. Ocean waters often have relatively high scaling potential for calcium sulfate compared to other surface waters, as well as calcium carbonate, possible biological compounds and organisms, and a significant dry-out concern from very high levels of sodium chloride [137]. Groundwater sources are perhaps the most variable. Groundwater often has high levels of salts compared to non-ocean surface water [206], and may be especially rich in calcium, bicarbonates, magnesium and sulfate. Groundwater is also commonly rich in iron (reduced $Fe^{2+}$) that can oxidize in contact with air and form iron hydroxides (nearly insoluble in water), producing heavy fouling. Due to its variability, groundwater may cause some of the worst scaling and flux reduction seen in desalination systems. Finally, the composition of industrial wastewater varies depending on the source, but can be extremely saline with a variety of dissolved metals (as in, e.g., produced water from hydraulic fracturing operations). The compositional variability of these waters makes it difficult to single out problem-causing compounds in a general fashion.

Gryta investigated MD fouling for water from river, lake, and groundwater sources for hollow fiber MD [171]. In that study, within 20 hours of MD testing, river tap water experienced the largest flux decline, lake tap water experienced the smallest flux decline, and groundwater had a flux decline in between. The flux decline was largely caused by calcium carbonate, with bicarbonate ion concentrations being the limiting factor for causing scale.

In contrast, seawater has very consistent constituents, generally has greatly more sodium chloride than other bodies of water, very high salt content overall, and lower concentrations of many other ions than some lake or river waters. Due to typically low concentrations of magnesium and organic HA, seawater may be less prone to fouling by these components than other sources of water. Like many tap water sources, calcium carbonate is a significant component of expected scale for ocean water, but is even more of an issue in ground and lake water. Calcium sulfate is also a concern for scaling in seawater [207]. Ocean water is susceptible to a variety of biological fouling types, including algae and microorganisms that may differ significantly from inland waters [133]. Curcio et al. found scaling of $CaCO_3$ in DCMD of seawater at concentration factors of 4-6 and 40°C [131]. The consistency of seawater salts and literature studies indicate that calcium sulfate, calcium carbonate, particulate fouling, and biofouling are the expected fouling concerns in seawater.

Wastewater treatment has also been attempted through MD, and fouling studies can even be found in literature. Wastewater contains numerous fouling compounds that may affect the flow significantly, especially biological compounds. In a study with hollow fiber MD membranes, a rapid decline in permeate flux was observed in solutions containing $HCO_3^-$, expected from $CaCO_3$ fouling [208]. The presence of bacterium *S. faccalis* was detected as well, but temperatures of 85°C prevented bacterial growth. Silica was detected as a deposit as well. Wastewater constituents vary dramatically by the source, so inorganic scale types may be hard to predict, although organic fouling of various kinds is common in wastewater.

## 6.5 Scaling and fouling effects on MD operating parameters

### 6.5.1 Wetting and permeate water quality change

An important requirement for the MD process to perform well is that the membranes have to remain hydrophobic, thus allowing only vapor and not liquid water to pass through. Wetting refers to the process whereby the membrane starts allowing liquid water to flow into the membrane pores. While wetting can be caused by the pressure in the feed channel exceeding the liquid entry pressure (LEP), fouling induced wetting is the concern for real MD systems. Hydrophobicity of the membrane material is the reason why the interior of the pores are not normally wetted. Scaling along the pores with salt crystals growing into the pore tends to reduce the net hydrophobicity and non-wetting character of the pores. Wetting caused by scaling is an important long-term performance issue for MD since the maximum concentration of the salts is expected to occur close to the pore openings where water evaporates; therefore the potential for precipitation is highest in this region.

Once wetting occurs, the MD process is affected in several ways. In several studies, water has been observed to more easily penetrate adjacent pores [25, [158, 156, 132, 183]. Further crystallization can also lead to accelerated wetting. Once the membrane is wetted, MD is no longer selective and hence doesn't achieve its goal of desalination or other types of separation. Figure 5 shows that scaling can occur within the membrane pores following wetting. The presence of this layer of salt deposits renders the top surface more hydrophilic, making it more prone to wetting [34, 209, 210]. In some cases, such propensity to wetting was shown to affect only the top-most portion of the pores of a polypropylene (PP) membrane, leaving the pores beneath un-wetted [210]. In other cases, wetting across the whole membrane thickness was reported, which has been seen in PTFE and PVDF membranes, evidenced by the presence of salt crystals at various depths of the membrane's cross section [34, 200].

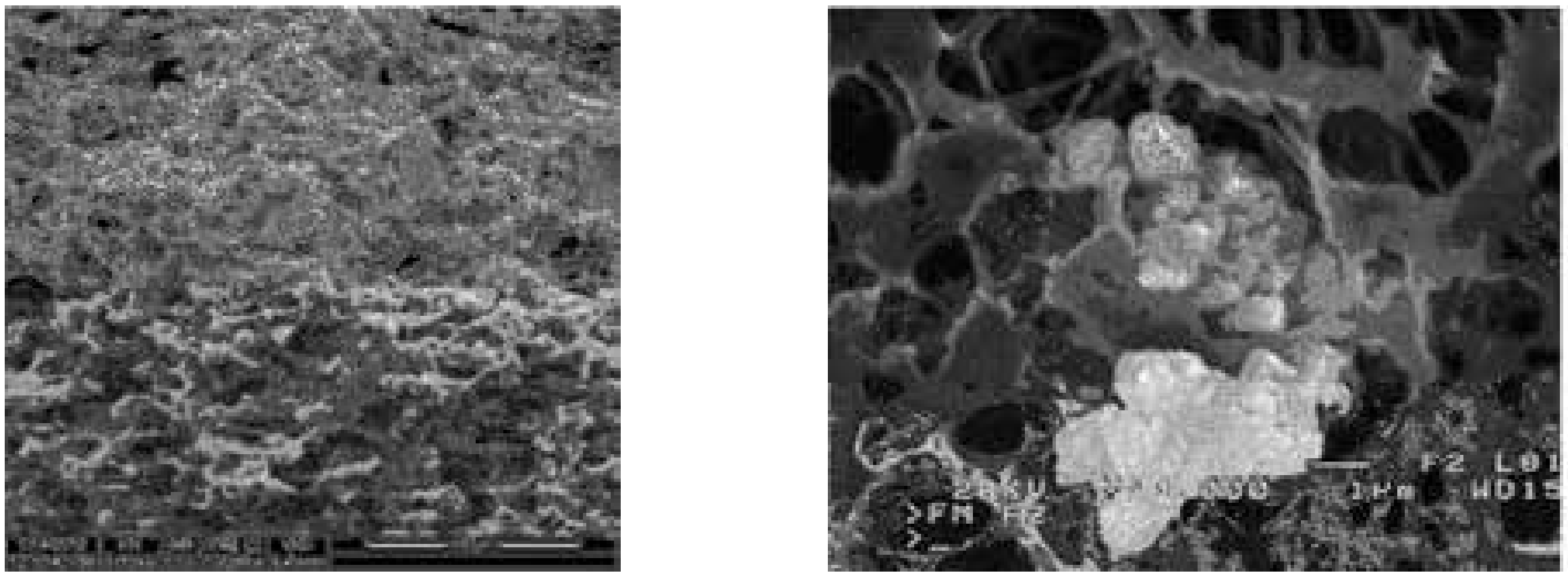

**Figure 6.5.** Left: Cross sectional SEM image of salt deposits inside a PVDF membrane after the 4$^{th}$ week of real seawater exposure [51]. Right: Cross sectional SEM image of deposits located inside a PP MD membrane after rinsing with HCl solution [35].

Pore wetting may degrade the performance of the MD process either because it reduces the interface for evaporation and therefore the production of vapor, or because, once a pore is wetted, saline water may flow through and contaminate the distillate [34, 158, 156, 200, 210, 211].

Finally, one very interesting impact of scaling on MD membranes is the occurrence of "negative flux," reported by Guillen-Burrieza et al. [34] in one of their PVDF membrane tests. According to Guillen-Burrieza et al. [34], one of their fouled PVDF membranes showed a flux from the permeate side to the feed side, until a minimum trans-membrane temperature gradient of 10°C was reached, after which a positive flux (from feed to permeate side) was obtained. They attributed this negative flux to the osmotic pressure created by the localized high salinity at the membrane surface on the feed side due to the deposited salt layer. Franken et al. [212] observed a 30% decay in DCMD flux over a period of one month and postulated that this must be a result of membrane wetting and possibly "back flow." Laganá et al. [213] has also reported a similar phenomenon.

### 6.5.2 PERMEATE FLOW RATE REDUCTION

Relatively few studies report data on fouling and its effect on the MD process performance. Different experiments report fouling differently, so a comparative metric is desirable. The average percentage flux reduction is defined as the reduction in flux as a percentage of the initial flux. The rate of change of the flux reduction can be used to examine the rate at which fouling lowers flux, as shown:

$$fouling\ rate\ \left[\frac{\%}{hr}\right] = \frac{flux_{initial} - flux_{final}}{flux_{initial}} \times \frac{100}{t} \tag{6.1}$$

The fouling rate may eventually level off at some system steady state. In a modeling context, flux decline is associated with a decrease in the MD coefficient B. The MD coefficient, or B coefficient (kg $m^{-2}$ $Pa^{-1}$ $s^{-1}$), characterizes the permeability of a membrane under the MD process. The coefficient B is a function of the membrane material properties (i.e. pore size, thickness, etc.) and the operating temperature [22].

The degree of permeate decline varies widely by experiment. In some processes, no decay occurred after months of operation [18, 214], while in others flux decline was as high as 66% in less than two days [158]. Permeate decline is dependent on the porosity of the scale that occurs and the thickness of scale layer [132], thus it is largely dependent on the salt or type of scale, the local concentration, and the scale's solubility under the given operating conditions. Decline may be slow and gradual [215], or it can occur rapidly from rapid crystal growth after exceeding critical levels of supersaturation [172]. As discussed previously, uncharged fouling including biofouling and iron oxide fouling were found to be relatively porous, while inorganic scale such as calcium carbonate were relatively nonporous.

Theory on fouling in membrane systems suggests different profiles for decline in flux over time based on different models of fouling [216]. Srisurichan et al. [216] applied the cake model of scale deposition to DCMD experiments using HA as a foulant with NaCl and $CaCO_3$ in the feed and found that it was capable of explaining the decline in flux. This paper illustrates that previous research on the use of additional transport resistances to model a fouling layer can be adapted to MD systems, accounting for the fact that the driving force in MD is a vapor pressure difference rather than a pressure difference.

Table 5.1 shows data gathered from publications that report fouling induced flux reduction. The program WebPlotDigitizer [217] was used to analyze the graphs and estimate initial and final flux values and the elapsed time between them. This data is used in subsequent analysis to draw conclusions about the effect of system parameters on rate of fouling.

**Table 6.1.** Fouling induced flux changes reported in the literature.

| | Study | MD configuration | Feed | Feed, Permeate Inlet Temperatures [°C]. Permeate Pressure ($P_p$) [kPa] for VMD | Initial Flux [$L/m^2$ day] | Final Flux [$L/m^2$ day] | Time of experiment [hours] | Additional Information | Average Fouling Rate [% flux decre / hr] |
|---|---|---|---|---|---|---|---|---|---|
| 1 | [218] | VMD | 7 ppt NaCl solution | 60, $P_p$ = 1.5 | 680.2 | 597.4 | 150.0 | $T_{ind}$ = 100.3 hrs | 0.08 |
| 2 | | | 7 ppt NaCl solution | 40, $P_p$ = 3 | 391.0 | 332.2 | 150.0 | $T_{ind}$ = 90.4 hrs | 0.10 |
| 3 | | | Ground Water | 60, $P_p$ = 1.5 | 529.7 | 376.1 | 75.0 | $T_{ind}$ = 40.9 hrs | 0.39 |
| 4 | | | Ground Water + 0.1 mol/l HCl | 60 , $P_p$ = 1.5 | 529.8 | 506.9 | 70.0 | | 0.06 |
| 5 | [171] | DCMD-hollow | Ground water GW | 85, 20 | 400.0 | 350.0 | 23.0 | $v_f$ = 0.4m/s | 0.54 |
| 6 | | | Boiled GW | 85, 20 | 395.0 | 383.0 | 23.0 | | 0.13 |
| 7 | | | Tap Water TW1 | 85, 20 | 418.0 | 247.0 | 44.0 | | 0.93 |
| 8 | | | Boiled TW1 | 85, 20 | 412.0 | 320.0 | 38.0 | | 0.59 |
| 9 | | | Tap water TW2 | 85, 20 | 795.0 | 705.0 | 65.0 | | 0.17 |
| 10 | | | Boiled TW2 | 85, 20 | 720.0 | 590.0 | 36.0 | | 0.50 |
| 12 | | | Tap Water TW1 + HCl pH = 4 | 85, 20 | 407.0 | 404.0 | 27.0 | | 0.03 |
| 13 | [140] | DCMD-hollow | Tap Water ($CaCO_3$) | 90, 20 | 729.7 | 653.1 | 24.7 | | 0.42 |
| 14 | | | Tap Water ($CaCO_3$) | 85, 20 | 612.0 | 577.0 | 25.0 | | 0.23 |
| 15 | | | Tap Water ($CaCO_3$) | 80, 20 | 548.0 | 534.0 | 24.4 | | 0.10 |
| 16 | | | Tap Water ($CaCO_3$) | 80, 20 | 414.4 | 242.5 | 43.0 | $v_f = 0.3$ m/s | 0.96 |
| 17 | | | Tap Water ($CaCO_3$) | 80, 20 | 476.0 | 416.1 | 66.0 | $v_f = 0.6$ m/s | 0.19 |
| 18 | | | Tap Water ($CaCO_3$) | 80, 20 | 537.7 | 506.4 | 89.8 | $v_f = 1.0$ m/s | 0.06 |
| 19 | | | Tap Water ($CaCO_3$) | 80, 20 | 599.3 | 479.5 | 82.9 | $v_f = 1.4$ m/s | 0.24 |
| 21 | [200] | DCMD-hollow | Waste water | 85, 20 | 321.0 | 288.0 | 47.5 | | 0.22 |
| 22 | | | Waste water boiled | 85, 20 | 364.4 | 348.8 | 45.0 | | 0.10 |
| 23 | [199] | DCMD | $CaSO_4$ (batch mode) | 60, 20 | 265.3 | 32.9 | 37.3 | $T_{ind}$ = 28.6 hrs | 2.35 |
| 24 | | | $Na_2SiO_3$ (batch mode) | 60, 20 | 265.3 | 172.6 | 59.3 | $T_{ind}$ = 21.6 hrs | 0.59 |
| 25 | | | $CaSO_4$ (batch mode) | 60, 20 | 739.2 | 38.4 | 15.5 | $T_{ind}$ = 4.4 hrs | 6.12 |
| 26 | | | $CaSO_4$ (batch mode) | 50, 20 | 496.8 | 100.8 | 18.1 | $T_{ind}$ = 8.5 hrs | 4.40 |
| 27 | | | $CaSO_4$ (batch mode) | 40, 20 | 259.2 | 28.8 | 37.0 | $T_{ind}$ = 30.1 hrs | 2.40 |
| 28 | [215] | DCMD | Seawater | 60, 20 | 570.2 | 344.6 | 720.0 | $v_f = 0.14$ m/s | 0.05 |
| 29 | [211] | DCMD | MF treated Seawater | 45, 20 | 350.0 | 276.1 | 168.4 | | 0.13 |
| 30 | | | Raw Seawater | 45, 20 | 230.6 | 185.0 | 167.3 | | 0.12 |

### 6.5.3 INCREASED TEMPERATURE AND CONCENTRATION POLARIZATION

Scaling may increase temperature and concentration polarization by creating a hydrodynamically stagnant or slow moving layer of water at the surface of the membrane [133, 209]. If scaling impedes flow velocity, which occurs particularly in hollow fiber membranes, the slower velocity will increase temperature polarization as the water residence time is lengthened [219]. The deposited salt layer creates an additional thermal resistance, reducing the heat transfer coefficient from the feed bulk to the evaporation and condensation interfaces [203], and in some cases possibly accelerating the degradation of polymeric materials [209]. As explained in Section 5.4.3.1 Bacteria and Biofilms, a fouling layer impedes convective heat transfer in the bulk. Although fouling layers have greater thermal conductivities than the polymers composing the membrane, they do not increase temperature polarization directly. Instead, they impart mass transfer resistance, altering the heat transfer associated with evaporation and consequently increasing the temperature polarization [220].

The mass transfer coefficient close to the membrane surface is also often reduced due to the presence of biofilms or other foulants. This leads to an increase in the concentration of dissolved ions close to the membrane interface, reducing the local vapor pressure and thereby reducing flux, in addition to increasing tendency to precipitate [221].

Recent studies by Goh et al. [220] found that hydrophilic biofouling on membrane surfaces that reduce the average pore size of the evaporating surfaces can reduce the MD driving force by causing vapor pressure reduction, as described by the Kelvin equation.

### 6.5.4 MEMBRANE DAMAGE AND CHEMICAL DEGRADATION

Many studies, especially at the lab-scale, have reported physical damage to the membrane as a result of scaling [34, 136]. The damage to the MD membrane was observed to take several forms. These include: a reduction in hydrophobicity of the membrane surface by altering its chemistry [34, 210], 119]; alteration of the membrane's pore structure and pore size distribution [34]; reduction in the membrane's mechanical strength [34]; reduction in the membrane's permeability via surface blockage [34, 210]; and the formation of defects (e.g. cracks) within the membrane structure [34]. In most cases, the damage was associated with a deterioration of the distillate quality (lower salt rejection) [129, 130, 136, 156, 176]. The most frequently reported membrane damage when scaling occurs during MD is the formation of a scale layer on the membrane's top surface in contact with the feed (Figure 6). This layer is composed of insoluble salts, such as $CaSO_4$, $MgCO_3$ and $CaCO_3$ [34], in addition to NaCl [34, 210].

Interestingly, membrane damage due to scaling can lead to either higher or lower flux than that of an intact membrane. While lower flux can be attributed to pore blockage by scale deposits on the surface [136, 210], higher flux is primarily due to pore wetting, usually accompanied by a lower salt rejection [34, 176]. For example, Hsu et al. [211] reported severe fouling and flux decline, but without permeate quality deterioration, when PTFE membranes were used to desalinate pre-treated sub-tropical seawater, with high NOM and biofouling potential. Ultrasonic cleaning was applied periodically in this study, which restored most of the flux. However, the data suggest a small degree of irreversible fouling, as evidenced by a slow reduction in the flux with time [211].

Guillen-Burrieza et al. [34] reported that the scale layer formed on the membrane surface during MD operation reduced the gas permeability of the membrane [34]. However, MD experiments showed apparently higher permeate fluxes for the fouled membranes which were attributed to heavy pore wetting processes caused by the inorganic scaling and membrane damage. Post-deposition washing with de-ionized water did very little to remedy that. Gryta et al. [210] reported similar findings for PP MD membranes exposed to NaCl solution as feed. A similar behavior was also reported for fouled membranes [203] when a biofouling protein layer was formed after concentrating saline wastewater.

Changes in membrane morphology upon fouling are the second observed damage in MD membranes. Gryta et al. [210] noticed only a minor change in membrane porous structure in PP membranes and concluded that the polymer material used for PP membranes production exhibited good thermal stability [210]. This was not the case for the PVDF and PTFE membranes studied by Guillen-Burrieza et al. [34], who conducted a study on the effect of cyclic wet-dry MD operation using seawater. The latter reported a noticeable variation in membranes' porosities and a shift in their pore size distributions (PSD) upon fouling after two weeks of seawater exposure. The shift in PSD was more pronounced in the PVDF membranes (which had a broader PSD to begin with) than in PTFE membranes. They attributed this change to: i) buildup of a relatively thick (4-7 μm) salt deposit layer on the membranes' surface; and, ii) damage to the membranes, especially PTFE ones, in the form of cracked fibrils (Figure 6.6) and altered structure. The cracks in PTFE membranes, described as being similar to those observed during shrinkage of dried clay, were attributed to the dry out periods. In PVDF membranes, on the other hand, a buildup of salt crystals within the membrane was observed. Guillen-Burrieza and co-workers concluded that PVDF and PTFE behaved differently in their reaction to a fouling medium under MD [34]. A series of AFM studies were also conducted by this group, which revealed that the two membrane materials behaved differently in terms of their attraction forces to $CaCO_3$ salt crystals, as well as their surface roughness [34]. Collectively, this strongly suggests that the nature of membrane material, in addition to its surface morphology, has an important role to play in resisting fouling in MD. The mechanisms behind this role are yet to be explored.

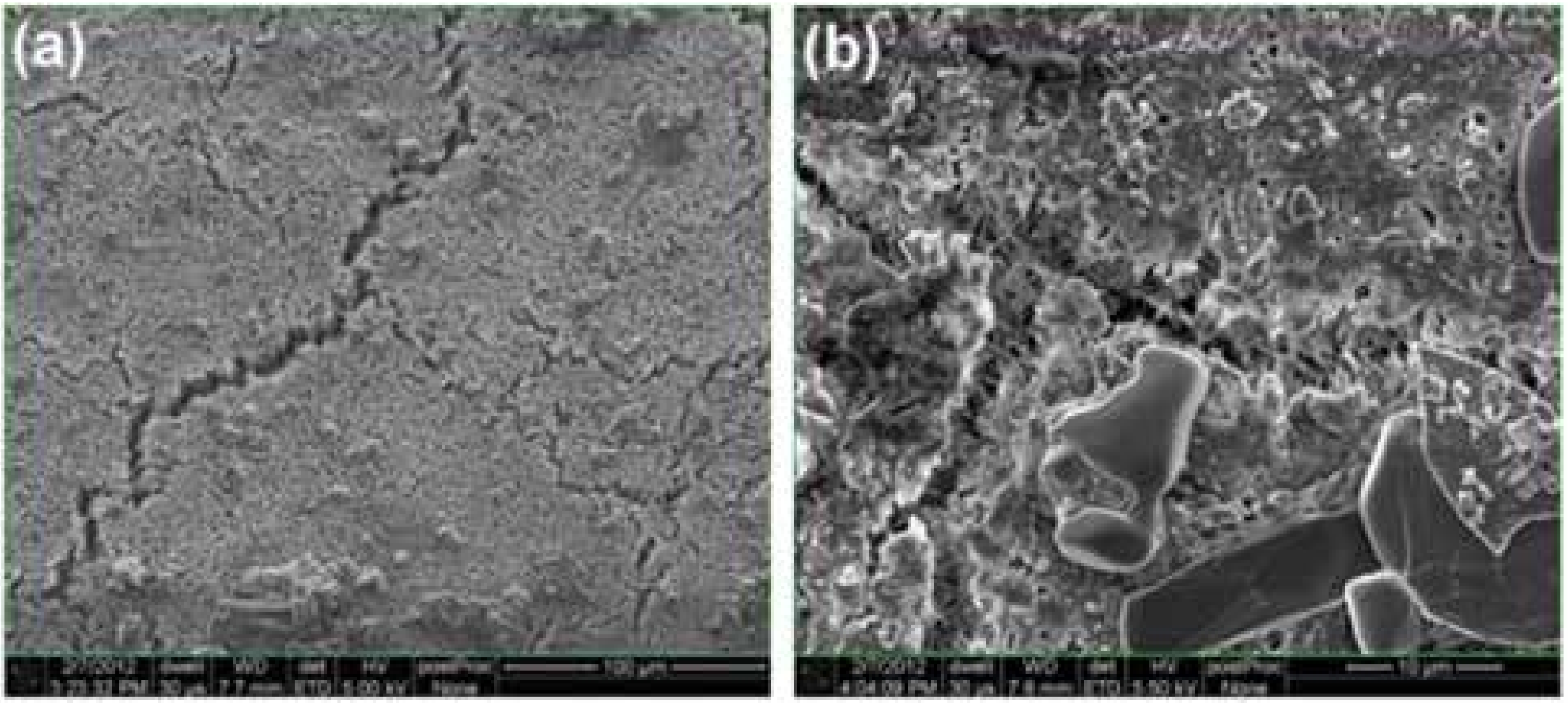

**Figure 6.6.** (a) and (b): Cracking and thick salt layer deposit on two different PTFE membranes under intermittent drying conditions after 2 weeks of exposure to real seawater [34].

Using Mullen burst test, Guillen-Burrieza et al. [34] were able to demonstrate a third impact of fouling on MD membranes: the loss of their mechanical strength. They showed a difference between the Mullen burst pressure of PTFE and PVDF membranes before and after fouling. While PVDF membranes remained unaffected after seawater exposure and maintained their integrity, fouled PTFE membranes showed a decreasing trend in the pressure required to initiate the burst (even with the presence of the PP backer material). This strength deterioration was attributed to the fibril damage and crack formation seen in PTFE during the dry out processes.

Another reported form of membrane damage during MD operation is that which is due to chemical degradation. Using Accurel PP membranes, Gryta et al. [210] reported significant hydrophilization of the membrane surface, leading to increased rate of wetting. They found that the operating conditions of their MD process, including elevated temperature and the presence of oxygen, enabled the degradation of the PP polymer by forming hydrophilic groups (hydroxyl and carbonyl) on its surface [210]. These surface groups reacted with the concentrated NaCl solutions and consequently sodium carboxylate was formed. This was supported by FTIR analysis [210]. Attempting to regenerate the wetted membranes by rinsing and drying was found ineffective, since the presence of those hydrophilic groups caused a rapid membrane rewetting during the consecutive MD operation [210]. The presence of salt, according to Gryta et al. [210], has also stiffened polymer molecules (i.e. PP), preventing their chain disentanglement, and leading to chain scission. However, the quality of the distillate in their study was found to have remained unaffected [210].

## 6.6 FOULING MITIGATION IN MD

The main scaling prevention tools employed in MD are feed pretreatment and chemical cleaning [10, 157]. Other fouling prevention methods attempted include increasing the feed flow rate, hydraulic cleaning, reducing surface roughness, changing the hydrophobicity of the membrane, magnetic water treatment, and changing surface charges on the membrane [19, 222]. The effects of the filtration and antiscalants have been studied in MD, as well as less commonly used technologies like feed heating or boiling, pH changes, flocculation, and magnetic water treatment [149, 158, 171, 181].

Pretreatment of the feed is standard practice in most desalination systems, and pretreatment needs vary significantly by technology and feed water quality. Common pretreatment methods include filtration, antiscalants, flocculation, and chlorination [205]. The market dominant desalination technology, RO, has intense pretreatment demands to protect the membranes, and some reports anticipate that MD will not need this level of pretreatment, as liquid water does not pass through the membrane and no cake compaction takes place. [205]

### 6.6.1 THERMAL WATER SOFTENING

Certain water conditions, such as groundwater with high hardness, may benefit from the intentional breakdown of bicarbonate ions through high temperatures or boiling. This process is called thermal softening, and helps reduce scale by causing $CaCO_3$ and other salts to precipitate out in a heating step, and also reducing $CO_2$, which is linked to bicarbonate ions as explained previously. For an MD system, M. Gryta boiled feed water for 15 min. and paper filtered it prior to undergoing hollow fiber MD, which lowered the bicarbonate ion content 2 to 3 times, to a concentration of 1.5 mmol $HCO_3^-$/L [171]. While boiling is an expensive pretreatment option in terms of energy use, since MD needs a hot feed stream, the boiled water is used in the MD setup after some salts precipitate out of the solution. As a result, flux declined only 3% over 23 hours, instead of 12% without boiling. However, for tap water with a concentration of 0.4-0.95 mmol $HCO_3^-$/L, $CaCO_3$ fouling was worse than in the untreated case [171]. A comparison of the rate of fouling is shown in Figure 6.7. Although not addressed in the paper, boiling likely reduced the concentration of dissolved $CO_2$, which in turn made calcium carbonate more likely to precipitate, and that may have been the cause of this non-intuitive result that contradicts other typical results. Boiling may therefore only be advisable with certain feed water solutions for certain durations.

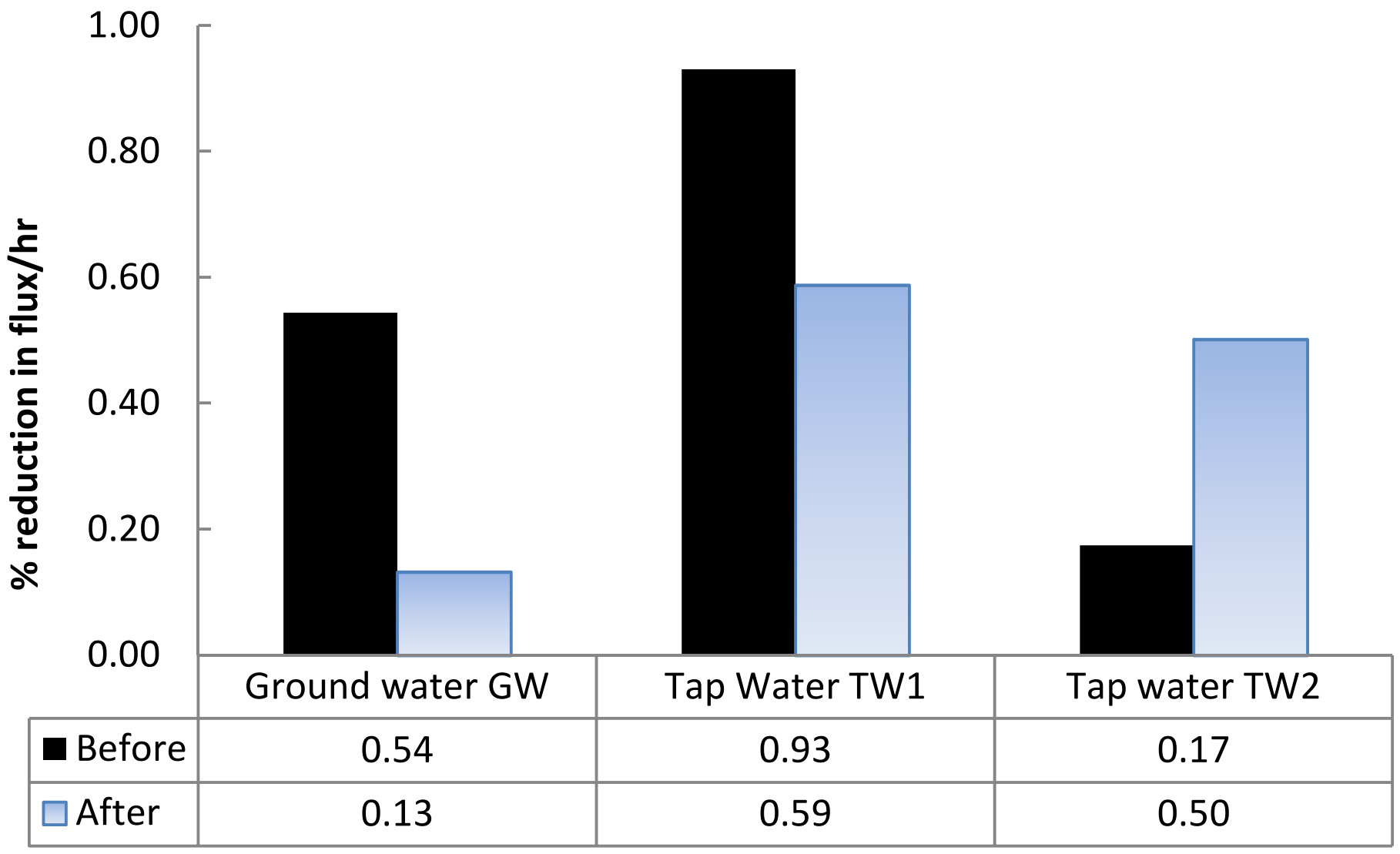

**Figure 6.7.** Literature data. The effect of boiling as pretreatment on the fouling rate (represented as the % of flux reduction per hour) for different source waters (see Table 5.1 for references).

### 6.6.2 MICRO/NANO FILTRATION PRE-TREATMENT

Microfiltration and nanofiltration (NF) is sometimes used before the MD process to remove particulate material and large molecules in the feed [186, 223]. Filtration may also be used to remove scaling the bulk, such as in the study by Kesieme et al where a cartridge filter captured calcium scale, allowing high recovery of groundwater RO concentrate with MD [224]. While ultrafiltration is good for removing suspended solids and colloids, NF can be an effective pretreatment for rejecting organic matter and removing hardness from feed water. As commonly practiced in RO, some researchers have also suggested using chemical coagulation followed by sand filtration or microfiltration to decrease fouling potential in MD [171]. Lawson and Lloyd [18] note that "several investigators … reported a reduction in the degree of flux decay for pre-filtered (≈1 μm) process liquids" [18, 214, 225, 226]. Alkalaibi and Lior [227] have observed that following pretreatment with nanofiltration and acid addition to a pH of 5, the microorganism count in the water was close to zero even after 1400 hours.

Flocculation has been examined in conjunction with microfiltration for wastewater MD treatment, which found microfiltration to be very effective in reducing suspended solids, with relatively mild to weak improvement from $FeCl_3$ coagulation/flocculation [181].

### 6.6.3 USE OF ANTISCALANTS

Antiscalants are a common tool to prevent inorganic scaling, and are potent for carbonate scales, as well as phosphate, sulfate, and fluoride, disperse colloids, and metal oxides [140, 228, 229]. Antiscalants act through a variety of mechanisms, including delaying nucleation, reducing the precipitation rate, distorting crystal structure, and altering CO2 concentration [230]. They are generally the most common technique for scale control because of low costs, and usually require dosing of less than 10 ppm [230]. However, antiscalants molecules, typically organic, often reduce the surface tension of the water, which can promote membrane wetting [231].

A notable antiscalant study was performed for hollow fiber MD using a polyphosphate-based antiscalant designed for RO, and compared with laboratory grade sodium polyphosphate [134]. The particular antiscalant works by sequestering calcium, thus inhibiting precipitation. The name and composition of the commercial Polyphosphate based anti-scalant were not given. With the antiscalant, the formation of $CaCO_3$ crystals was virtually eliminated, but a thin amorphous non-porous layer deposited on the MD membrane. Surprisingly, the flux decline was worse when the antiscalant was present. The higher the antiscalant concentration, the lower the permeate flux. However, the associated scale was mostly on the membrane surface, so periodic HCl cleaning was effective. Because of orthophosphate deposit risk from the breakdown of this antiscalant, residence times in MD were recommended not to exceed one hour [134]. Gryta concluded that regular HCl cleaning with an antiscalant system could make it useful in MD [134]. A different study by He et al. [232] found more positive results using polyacrylic acid antiscalants in MD, which were particularly effective in reducing calcium sulfate scale. Organo-phosphorus antiscalant compounds were also very effective in reducing calcium carbonate scale and mildly effective in reducing calcium sulfate scale [232]. An antiscalant blend of carboxylic and phosphoric acids was moderately effective in reducing both calcium carbonate and calcium sulfate [232]. Surface tension of the proposed antiscalant solutions (concentrations varying from 0.6 to 70 mg/L,) was very close to that of tap water (71.5 mN/m against 71.8 mN/m for tap water), and no wetting phenomena were detected under the experimental conditions.

### 6.6.4 pH Control of the MD feed

The pH control of the feel has been a common method to reduce or eliminate MD scaling. In almost all cases, this means acidifying the feed, as alkaline salts, the main component of scale, become drastically more soluble at acidic pH. However, according to Karakulski's findings, acidification of the feed failed to prevent silica scale [158, 208]. In one DCMD hollow fiber study with solutions of $CaCO_3$ at SI of 49 and $CaSO_4$ at SI 1.12, acidification to a pH of 4.1 fully prevented scale under conditions that experienced rapid scaling of MD membranes at neutral pH [36, 158, 208]. Numerous researchers found that bringing the feed to a pH of 4 with acid addition was sufficient to virtually eliminate calcium carbonate scale even at extreme SI indices [36, 171, 208]. Figure 8 shows the relative reduction in fouling rate (calculated as the rate of % flux decrease per hour, see Section 4.2) achieved by controlling the pH of the feed solution. However, adding solution to modify the pH can quickly become prohibitively expensive, depending on the pH [233].

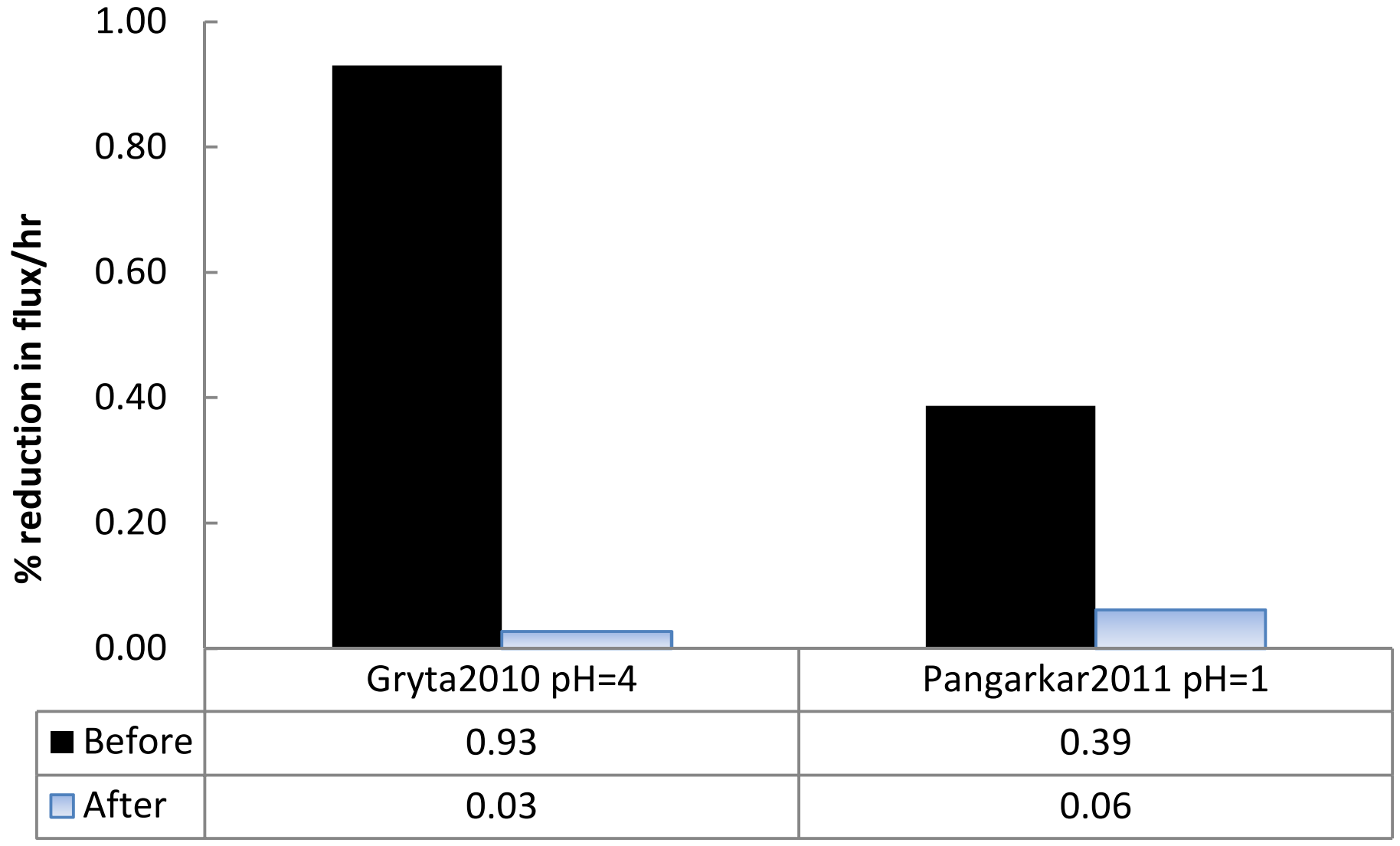

**Figure 6.8.** Literature data. The effect of adding HCl to the feed solution and reducing pH as pretreatment on the fouling rate (represented as the % of flux reduction per hour) [171, 182].

### 6.6.5 Magnetic water treatment

Magnetic water treatment (MWT) is a technology developed for scale reduction in water treatment [149, 234, 235] and power plant heat exchangers [149]. The magnetic field slows nucleation while increasing crystal growth rate [210, 211, 236, 237], and it can alter the precipitate morphology and properties. Gryta performed a study with a commercially available MWT device called Magnetizer RWE-S on hollow fiber MD with a tap water feed [149]. The experiment ran with a 0.1 T magnetic field, feed temperature of 85°C, and alkalinities of concentrations of 2.72 and 3.61 mmol/dm$^{-3}$ $HCO_3^-$. The magnetic field caused significant changes in the morphology of crystal deposits on the membrane, causing crystals to be larger, but the deposits to be more porous. MWT caused the deposit layer thickness to be 10-25% smaller, and significantly mitigated flux decline. It was expected from previous studies that MWT would shift $CaCO_3$ crystal type towards vaterite and aragonite forms, but under the temperature and slow laminar conditions, predominately calcite deposits occurred [149]. No effect on membrane wettability was found.

### 6.6.6 TAILORING MD MEMBRANE PROPERTIES

Increased hydrophobicity of MD membranes has shown to have a dramatic effect in reducing scale formation. Hydrophobicity differences between materials may also account for different propensities to foul. For example, according to Gryta, polypropylene membranes do not have an optimal hydrophobicity, and undergo wettability readily, often after a few days in his experiments [209].

Superhydrophobic coatings provide an additional hydrophobic layer, which acts as a buffer layer from scale [238] by reducing the surface nucleation and the particulate attachment. Superhydrophobicity can reduce membrane wetting and improve membrane recovery from acid cleaning [222, 239]. One method of creating superhydrophobic MD membranes is depositing $TiO_2$ nanoparticles using LTH (low temperature hydrothermal synthesis). The superhydrophobic membranes generally have fluorosilicone coatings, and may achieve contact angles of 166° [222]. Coating the membranes with fluorosilicone for hollow fiber MD was found to create very long induction times of 194 minutes for gypsum [232], compared to no induction period otherwise. Other sources found significant $CaCO_3$ scale reduction [133, 36] and $CaSO_4$ scale [36] using fluorosilicone coatings [35]. These coatings can also reduce pore sizes, increasing resistance to scale [36].

As mentioned previously, superhydrophobicity also has shown to prevent microorganism fouling. However, more hydrophobic membranes are known to preferentially absorb HA [131]. Studies by Meng et al. found that superhydrophobic PVDF membranes fouled similarly to humic acid as ordinary PVDF, despite significantly reduced inorganic scaling [194]. According to Meng et al., humic acid fouling on the superhydrophobic membrane occurred via an adsorption-desorption mechanism. A commonly used method to prevent biofouling is based on the hydrophilization of the membranes. Using sodium alginate hydrogel as coating may reduce the adsorption of organic compounds such as citrus oil on PTFE membranes [202]. Additionally, UV-induced grafting of zwitterionic polymers on PP membranes has been used to prevent protein fouling with very good results [240] as well as interfacial surface crosslinking of PP membranes [241] without compromising the rejection factor.

Increased surface porosity of membranes seems to be a factor in increased tendency to scale as well. Gryta performed a study of the effect of porosity on membrane performance using polypropylene capillary membranes and tap water. He found that the presence of larger pores on the surface allowed for the deposition of $CaCO_3$ crystals in the membrane interior, causing wettability, especially wetting during HCl cleaning [209]. The surface with higher porosity was found to have significantly increased wettability. However, no difference in flux decline was observed between the highly porous and non-porous surfaces of otherwise identical membranes, indicating that high porosity may not increase clogging or membrane blocking [209]. Gryta found that membranes with a low porosity coating 1 μm thick has significantly less tendency to wet, while exhibiting similar average properties and permeate production to the uncoated membrane [209]. Gryta inferred that crystal growth inside the membranes can be restricted by pore diameters much smaller than the crystal size. Such low porosity coatings may therefore be recommended to avoid wettability with only minor reduction in permeate production.

Finally, some studies suggest that membrane material type may have a significant impact on scaling. Curcio et al. found that in the presence of polypropylene hollow fiber membranes, the induction time for $CaCO_3$ was 18 seconds, compared with over 80 seconds for no membrane present [131, 152, 153]. Tung-Weng et al. [170] note that PVDF membranes wet more than PTFE. Non-polymer membranes such as glass membranes may have superior thermal and chemical robustness [239].

### 6.6.7 MD MODULE DESIGN AND OPERATION

The conditions within the MD module may also significantly affect fouling. Temperature and concentration polarization may cause scaling preferentially at the membrane. Residence time in the module may have an effect as well, as stagnation areas can promote scalant precipitation from the bulk. Scaling is caused by conditions on the saline feed side rather than pure permeate, so system configuration (e.g., AGMD, DCMD, VCMD, or SGMD) plays a role only because they have different flux magnitudes [14]. Technologies with more conduction losses and with higher permeate flux, such as DCMD [22, 242] or VMD respectively, may be expected to have somewhat worse scaling issues from temperature and concentration polarization effects, as explained below. However, based on a numerical model developed by the authors elsewhere [243], the evolution of the Saturation Index (SI) as a methodology to predict extent of scaling, has been calculated and detailed. Understanding the theory associated with inorganic salt precipitation can help design safe operating conditions where the extent of precipitation can be controlled.

#### 6.6.7.1 Temperature polarization

Temperature polarization is one of the most important secondary phenomena that affect the MD process. Water that evaporates into the pores of the MD membrane removes the corresponding latent heat of evaporation from the liquid feed stream. This cools down the fluid close to the membrane and results in a thermal boundary layer. Since the vapor pressure of water rises exponentially with temperature, any reduction in temperature of the water at the membrane interface leads to a significant drop in MD driving force. As a result, care is often taken in MD design to reduce and limit this effect as much as possible. At a given flux, the most important factor that determines the extent of temperature polarization is the effective heat transfer coefficient in the feed stream. At a higher heat transfer coefficient, a smaller ΔT is required across the thermal boundary layer to transfer the same amount of heat of vaporization. The most common design strategies are to increase the flow rate of feed, operate in the turbulent regime or have turbulence promoters in the form of a spacer to increase the heat transfer coefficient on the feed side.

The simplest model for temperature polarization is given by Eq. 1, where $h$ is the heat transfer coefficient in the feed channel and $\dot{q}_{out}$ is the total heat loss across the membrane including sensible heat transfer and latent heat of evaporation of the vapor [85]:

$$T_{f,b} - T_{f,m} = \frac{\dot{q}_{out}}{h} \qquad (6.2)$$

In general, temperature polarization reduces vapor flux, but it may also reduce scaling tendency since many critical salts' solubilities vary inversely with temperature. Conditions giving rise to a large temperature polarization may also create a large concentration polarization, which has the opposing effect of raising concentration near the membrane and potentially promoting scale formation.

#### 6.6.7.2 Concentration polarization

The MD process allows only water vapor to pass through while retaining non-volatile salts, causing an increase in salt concentration near the membrane which is referred to as concentration polarization. A simple model of the concentration polarization process is the film model [124] described by Eq. 2:

$$\frac{x_m}{x_c} = exp\left(\frac{J}{\rho k}\right) \tag{6.3}$$

where $J$ is the water flux through the membrane, $k$ is the mass transfer coefficient for the salt in solution, and $x_m$ and $x_c$ are the molar concentrations at the membrane interface and the bulk of the fluid streams, respectively.

Concentration polarization affects the MD driving force through its effect on vapor pressure, but this influence is minor relative to the effect of temperature polarization on vapor pressure. The more important effect of concentration polarization is that the elevated concentration may promote scaling formation on the membrane. As can be seen from Eq. 2, the effect rises with an increase in flux and falls with an increase in the mass transfer coefficient. It should be noted that temperature and concentration polarization are coupled phenomena since both depend on and influence the water vapor flux through the membrane.

#### 6.6.7.3 Saturation index, numerical modeling

Thermodynamically, membrane fouling is expected to be a function of the salt supersaturation at the membrane surface. SI is a commonly used measure of supersaturation [244]:

$$SI = \log\left(\frac{\text{activity product}}{K_{sp}}\right) \tag{6.4}$$

$$K_{sp} = f(T, P) \tag{6.5}$$

According to these equations, temperature and concentration have an effect on the thermodynamic tendency for precipitation. These two parameters also influence the permeate flux and are in turn influenced by the flux and heat transfer rates (Eq. 1, 2). Engineering Equation Solver (EES) was used to numerically solve the implicit set of equations that connect these parameters along with the mass and energy balance equations, in order to predict the value of the SI on the membrane interface and feed bulk. The modeling methodology is described in detail in [85, 243]. In the following sections, results from the model are used to better understand the trends reported in literature. The numerical modeling provides a quantitative understanding of the competing effects and the reason for the overall trend observed. Data on changes in the extent of membrane fouling with modifications in feed flow velocity and feed inlet temperature are available in literature. In the following sections, data from reference [140] is used to illustrate the effect of feed flow rate and feed inlet temperature on the extent of $CaCO_3$ fouling induced flux decline. Modeling results for an AGMD setup on the effect of system variables on SI at the membrane interface are also presented.

For the numerical model the following assumptions have been made: an inlet salinity of 13 mg/kg is assumed which is approximately equal to the solubility of $CaCO_3$ at room temperature; diffusivity of $CaCO_3$ is assumed to be $1.469 \times 10^{-9}$ $m^2/s$ and the dependence of solubility product of aragonite on temperature is modeled as in [140] (T is temperature in K):

$$K_{sp} = -171.9773 - 0.077993T + \frac{2903.293}{T} + 71.595 \log T \quad (6.6)$$

It is important to note that only the trends in SI variation with changing system parameters are being compared with the reported experimental fouling data. The actual value of SI in the experiment is likely to be different due to differences in concentration of the feed water, presence of other ions in the feed and differences in geometry and MD configuration.

#### 6.6.7.4 Effect of temperature

As discussed previously, temperature has a dominating effect on scale formation due to salt solubility variation with temperature. Similarly, temperature polarization may have a significant effect on scaling and fouling since the solubilities of common foulants are highly temperature dependent, and since biological fouling as discussed previously can be limited by temperature. Also, the effect varies greatly depending on the salt. Reduced membrane temperature due to temperature polarization might prevent precipitation of $CaCO_3$ and $CaSO_4$ which are less soluble at higher temperatures, but may make non-alkaline salts scale more readily. Higher temperatures as stated previously may limit bacteria growth, so high degrees of temperature polarization may enable biofouling. Figure 9 shows data on the effect of temperature on fouling rate in an experiment where the major scalant was $CaCO_3$ [140].

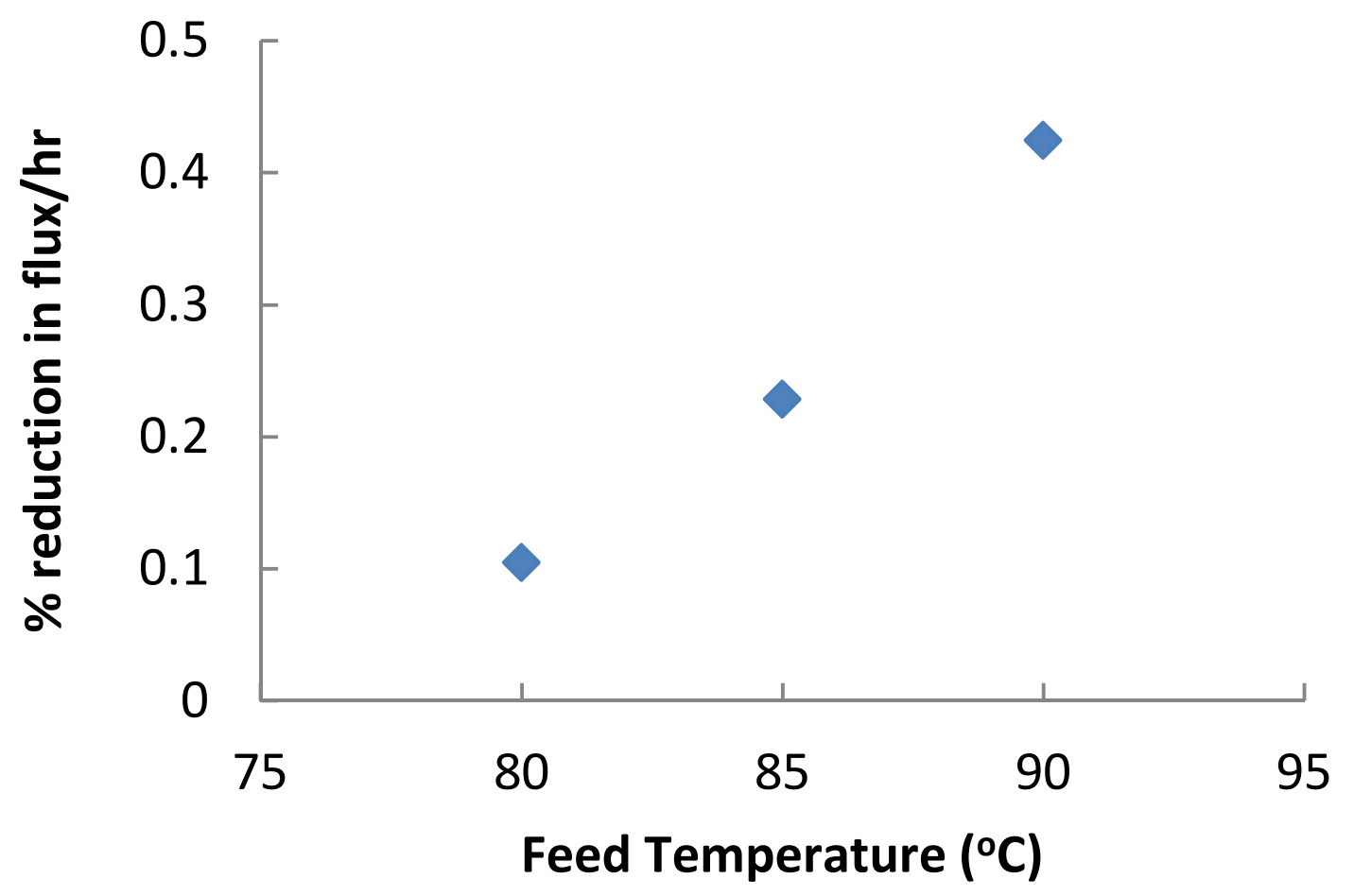

**Figure 6.9.** Literature data by Gryta. The effect of the feed temperature on the fouling rate (represented as the % of flux reduction per hour) for Lake water with bicarbonates for hollow fiber DCMD [140].

Due to the inverse solubility of the salt and increased flux contributing to higher concentration polarization in the feed stream, there is a significant increase in fouling rate with change in feed temperature. The rate of fouling induced flux reduction is four times higher at 90°C than at 80°C.

To numerically explain this behaviour, the previously described EES model was used to estimate the effect of increasing feed temperature on the SI at the membrane interface and feed bulk. The simulations were carried out for a system with a feed flow rate of 0.1575 kg/s in a feed channel with a depth of 4 mm and a width of 12 cm. Figure 10 shows the predicted variation in SI at the membrane interface and in the bulk of the feed stream.

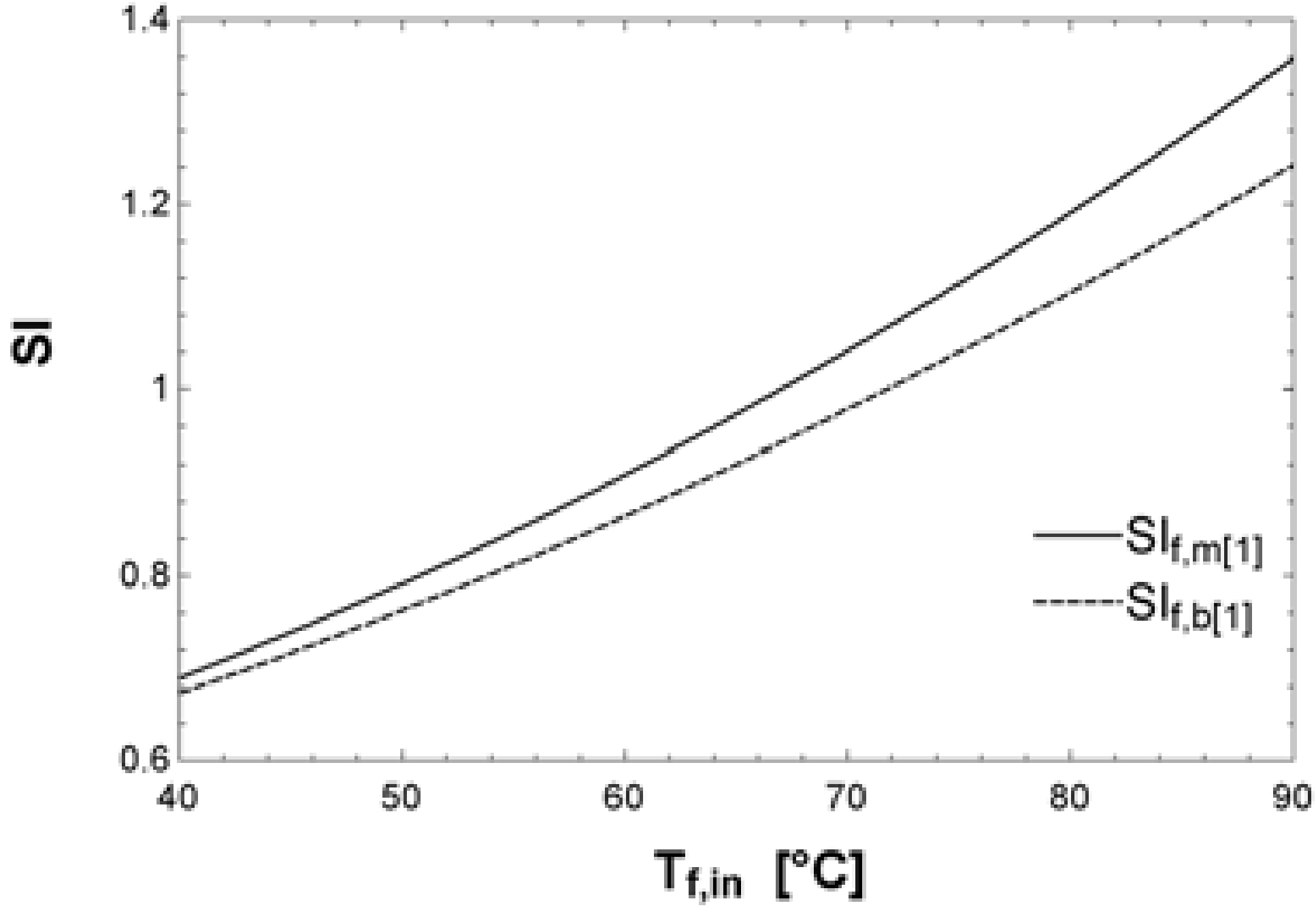

**Figure 6.10.** Simulation results at fixed $\dot{m}_{f,in} = 0.1575$ kg/s. Saturation Index of $CaCO_3$ in the feed stream at the membrane interface ($SI_{f,m}$) and in the bulk ($SI_{f,b}$) as a function of feed inlet temperature. Increase in $SI_{f,m}$ correlates with higher scaling observed. $Si_{f,m}$ increases faster than $SI_{f,b}$ illustrating the effect of higher concentration polarization.

With an increase in feed inlet temperature, the SI of the bulk fluid increases associated with a reduction in the $K_{sp}$ for $CaCO_3$ (Aragonite). At the membrane surface, the increase in SI is higher, since there is a larger water flux and an associated increase in membrane concentration of ions compared to the bulk. The increase in predicted SI at the membrane surface correlates well with the increased fouling rate at high temperatures as illustrated in Figure 6.9.

### 6.6.7.5 Polarization abatement: feed flow rate and bubbling

Feed flow velocity has a direct influence on the heat and mass transfer coefficients in the feed channel. With an increase in feed velocity, both heat and mass transfer coefficients increase, and transition from laminar to turbulent flow may cause this change to be discontinuous. As a result, with an increase in flow rate, flux increases as a result of reduced temperature and concentration polarization.

For inverse solubility salts such as $CaSO_4$, $K_{sp}$ decreases with an increase in temperature (Figure 6.1). Since the heat transfer coefficient (h) increases with flow rate, the temperature at the membrane surface increases. At the same time, an increase in mass transfer coefficient will lead to a decrease in salt ion concentration at the membrane surface. Since both the numerator (activity product) and the denominator ($K_{sp}$) of Eq. 4 decrease, the relative change in SI is determined by the rate at which the two quantities change.

Numerical modeling was again used to understand the overall effect of increasing the feed velocity on the SI at membrane interface. The parameters used for the study were the same as mentioned earlier, with the feed inlet temperature fixed at 80°C. Results are shown in figures 11 and 12.

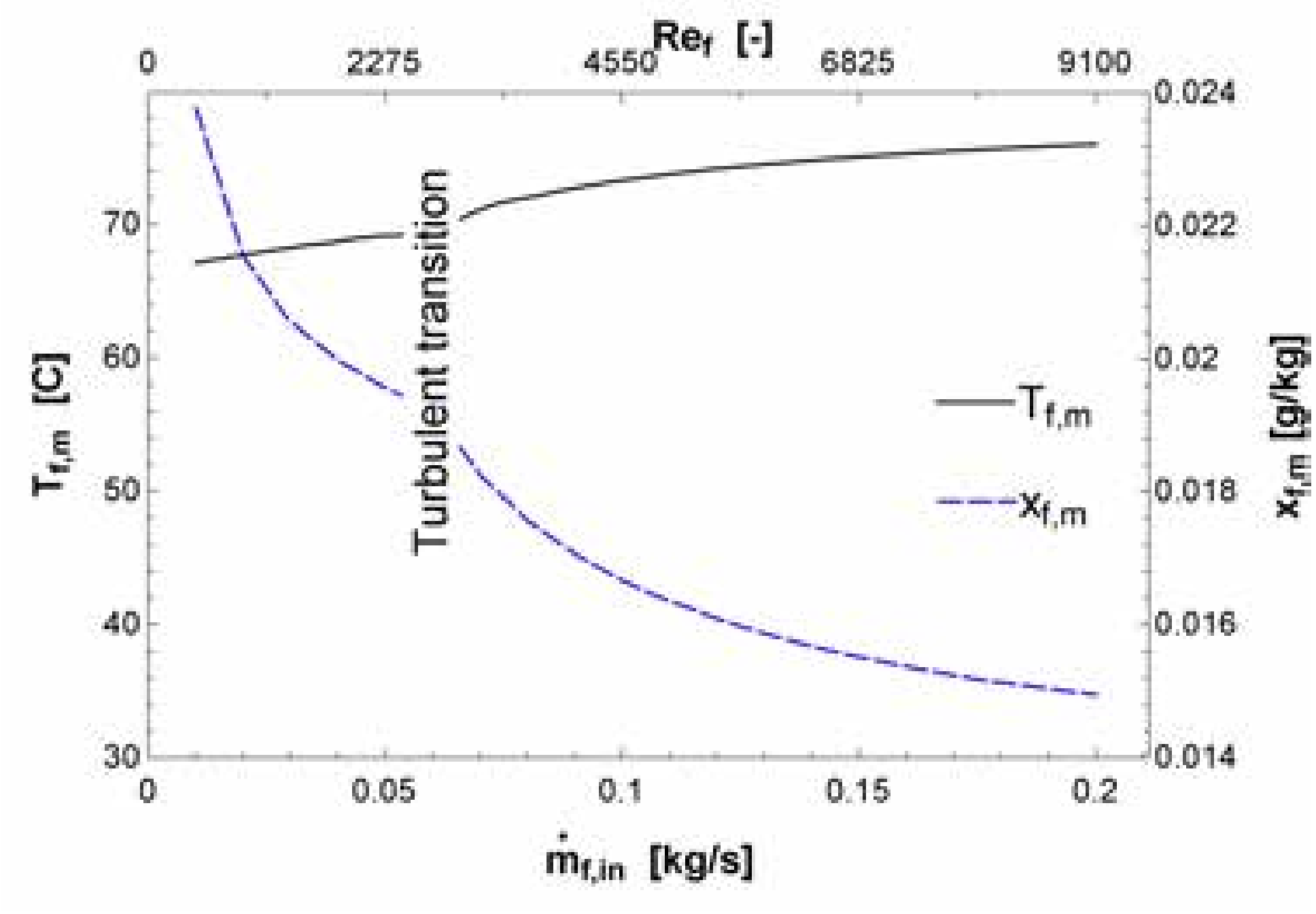

**Figure 6.11.** Simulation results at fixed $T_{f,in} = 80$ °C. Effect of feed mass flow rate ($m_{f,in}$) and Reynolds number ($Re_f$) on temperature ($T_{f,m}$) and $CaCO_3$ concentration ($X_{f,m}$) at the membrane interface. Increase in temperature would increase tendency for precipitation of $CaCO_3$, whereas decrease in concentration would reduce tendency for scaling.

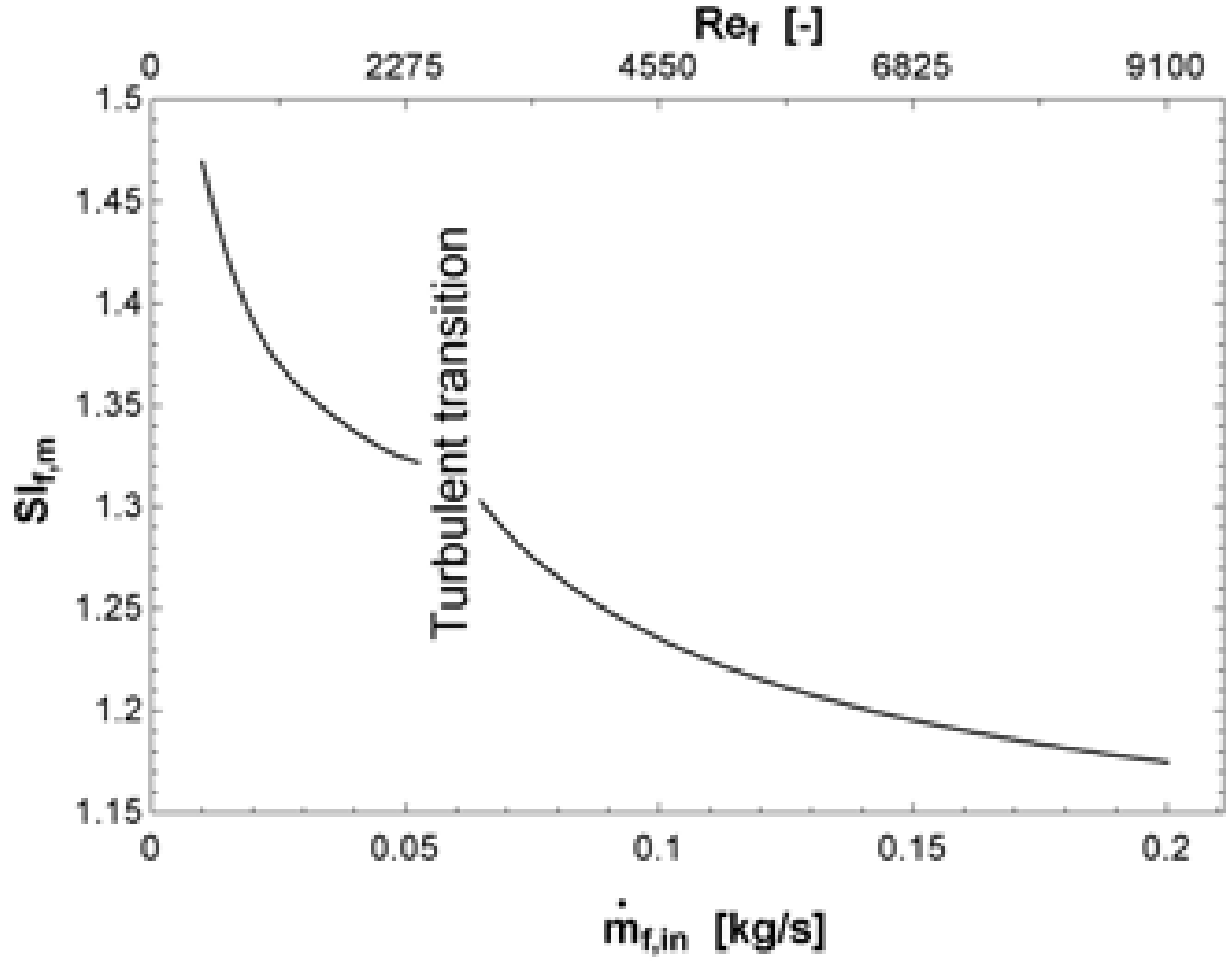

**Figure 6.12.** Simulation results at constant $T_{f,in} = 80$ °C. Overall effect of the feed mass flow rate (m f,in) and Reynolds number (Ref) on the SI at the membrane interface ($SI_{f,m}$). Decrease in SI correlates with experimentally observed decreased rate of fouling (Figure 6.13).

Salt ion concentration decreases and temperature at membrane surface increases with increase in feed flow rate (Figure 6.11). The overall effect of these two opposing effects was a decrease in SI at the membrane surface with increased feed flow rate, as shown in Figure 6.12, since the reduced concentration polarization leads to a larger decrease in ion activity product as compared to the decrease in $K_{sp}$ associated with lower temperature polarization. This result is consistent with data from Gryta [140] (Table 5.1, Figure 6.13) which shows that an increase in feed velocity from 0.3 m/s to 1 m/s leads to a reduction in the rate of flux decline.

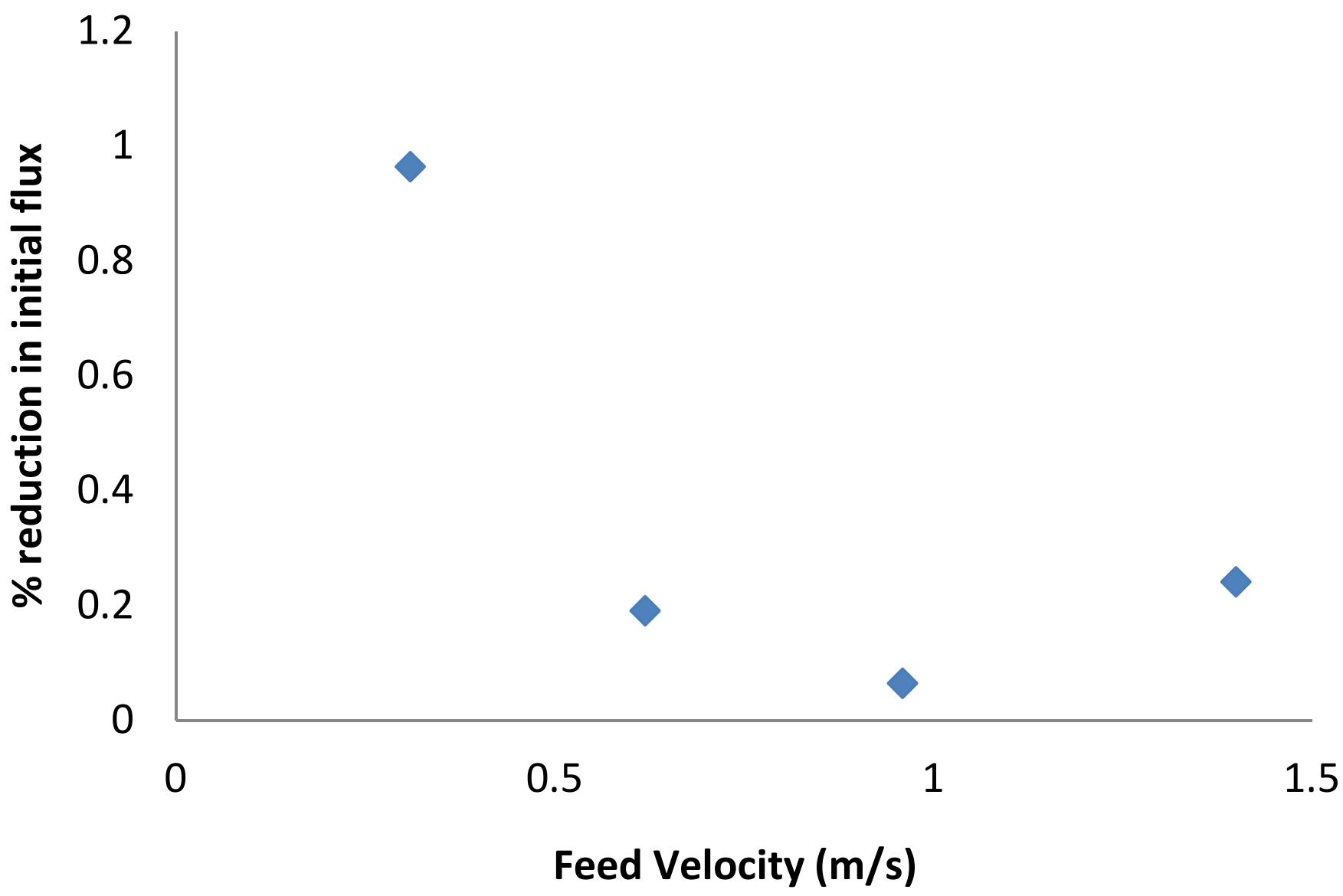

**Figure 6.13.** Literature data. The effect of the feed flow velocity on the fouling rate at constant feed inlet temperature of 80 °C (represented as the % of flux reduction per hour). Decrease in fouling rate with increase in feed velocity is shown [140].

The marginal increase in fouling rate between 1 m/s and 1.5 m/s (feed flow velocity), is inconsistent with the theory presented above and could be associated with transition to turbulence that is expected to happen at these velocities in the experiment.

In addition to the thermodynamic effects discussed above, at higher flow rates, the shearing action of the water in removing deposited precipitates is higher. This can help reduce the overall membrane scale deposition and fouling rate. Gryta [140] also reported that the size of salt crystals observed was much larger at the lower flow rates (0.31 m/s) as compared to higher flow rates (0.62 m/s), which can be explained by considering the kinetics of the fouling process as explained above.

Other methods to reduce concentration and temperature polarization have also been explored. Recently, Chen et al. [37] have demonstrated the effectiveness of gas bubbling through the feed as a means of promoting mixing and thereby improving flux and reducing scale deposition. Traditional methods to promote better mixing include increased flow rate and the use of spacers to induce turbulence in the feed channel and to improve mixing. Chen et al. compare three different configurations of hollow fiber membrane MD with feed outside the fibers: no-spacer, spacer and air bubbling without spacer. Figure 14 shows the type of spacer used in that study.

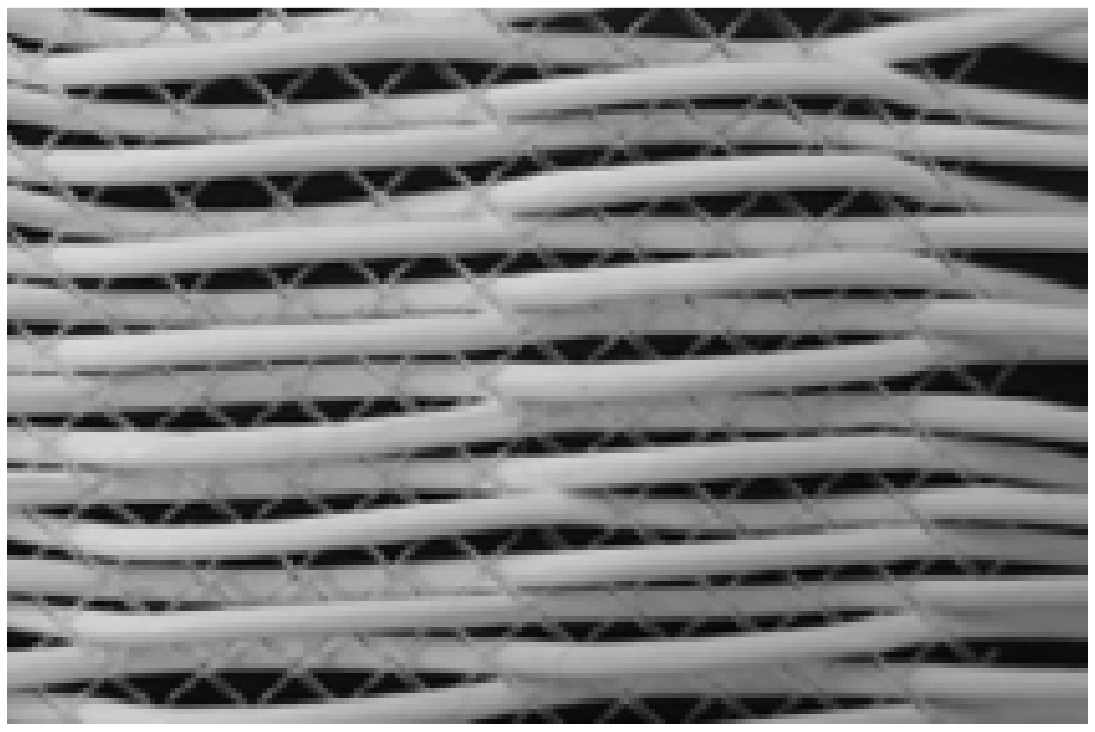

**Figure 6.14.** Example of mesh spacers used in hollow fiber MD experiments [37].

The flux is lower for the no spacer case as compared to the other two. With time (about 7 hours of experiment), when the bulk NaCl concentration by weight is about 23%, a significant fouling layer and associated decline in flux rate is observed in all three cases. While the air bubbling case showed less scale formation, the spacer actually leads to more scaling than the no-spacer case, since it enables local trapping of NaCl close to the membrane. Air bubbling achieves the goal of mixing the feed stream and improving performance, while at the same time retards local scale formation. Amongst the air bubbling cases, performance improvement is higher for cases where significant temperature or concentration polarization is expected (at higher feed water temperature and laminar flow conditions). Air bubbling will also increase the dissolved gas concentrations in the feed which has potential impact on gas content in the membrane pores, permeate-side mass transfer resistance (for AGMD or SGMD), and on the formation of salts depending on the gas chemistry (e.g., $CO_2$ content) [187].

#### 6.6.7.6 Types of module

Several MD module configurations have been designed and studied for fouling, with significant differences. Although direct comparisons are lacking, evidence from similar feed solutions indicates dramatic differences in fouling risk. The types of MD modules include: hollow fiber membranes which consist of small tubular capillaries; flat plate modules; tubular modules which consist of concentric cylinders that resemble flat plate membranes; and spiral wound membrane modules, which consist of a membrane and spacer wrapped around a collection tube [10]. The largest differences can be found between hollow fiber membranes, which have been studied for MD fouling relatively extensively, and all the rest. Hollow fiber membranes are known to have higher fouling potential [36], but the present review has found that their fouling potential is often extremely high, occurring under many conditions for which other units have not been known to foul, and they have even experienced fouling with bulk concentrations well below saturation conditions, which is highly unusual [18, 245]. Crucially, this high degree of fouling has generally been found in the experimentally more common setup in which feed is *inside* the capillary tubes [134]; in other experiments with distillate inside the capillary tubes, fouling was in some cases regarded as unlikely [35], although it still can occur, as shown in Figure 6.15.

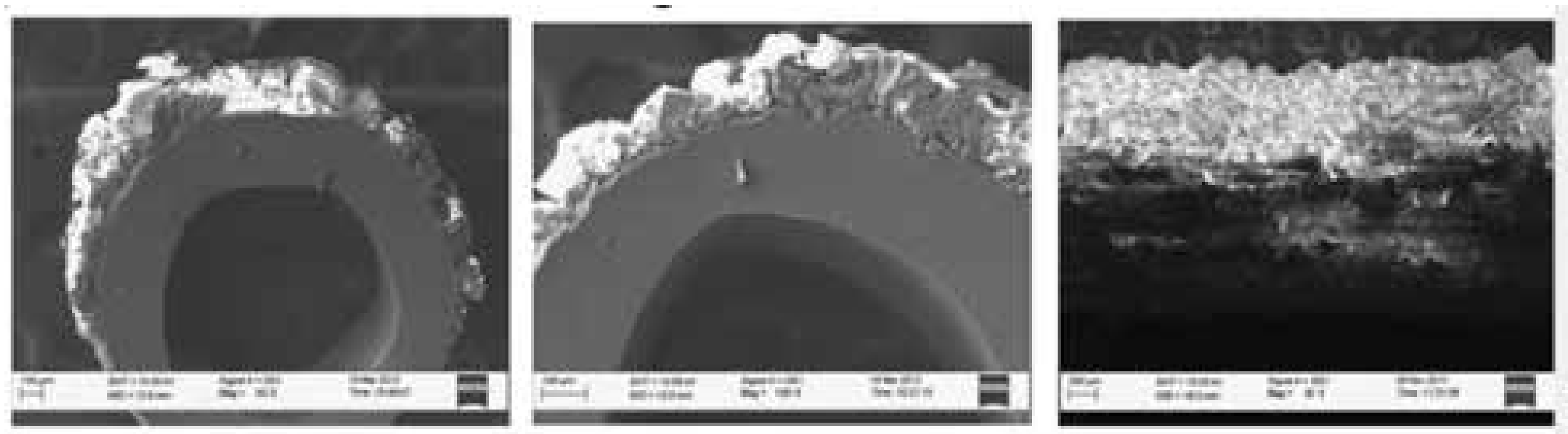

**Figure 6.15.** SEM images of PVDF Hollow Fiber DCMD fouled modules in cross-section with NaCl taken from 18 wt.% to saturation after 7 hours of operation with feed external to the capillary tubes [37].

In the case of hollow fiber membranes, the fine fibers cause slow internal laminar flow and often extreme concentration polarization. They are also known to clog readily by colloidal particles in the feed. Additionally, hollow fiber MD modules are difficult to clean and maintain [10], and if the feed penetrates through the membrane pores, the whole module should be replaced.

Flat plate modules have been examined thoroughly as well in this review, and while tubular and spiral wound modules have been investigated less, they are essentially flat plate systems with mild curvature, and should not behave much differently. Flat plate modules are relatively easy to clean [10]. However, since hollow fiber and spiral wound modules can hold greater membrane areas they are regarded as more cost effective MD modules. Therefore, cleaning protocols should be adapted to the most commercially successful module option.

The various types of MD operation, including DCMD, AGMD, SGMD, and VCMD are not thought to significantly affect MD scaling, but their differences on permeate flux and heat conduction alters concentration and temperature polarization effects. These configurations differ on the permeate side, not the saline feed side, so effects on fouling are secondary feed design [10].

#### 6.6.7.7 Membrane cleaning

Acid cleaning is one of the most common methods used for fouled membrane reclamation. Several studies have used different types of acids (both strong and weak) to remove scale components from the membrane surface. HCl is by far the most commonly used acid in MD experiments for membrane cleaning. It is particularly effective in removing basic salts such as $CaCO_3$ by dissolving them [140, 156]. Acid cleaning is carried out as a batch process with the feed water being replaced by an acidic solution. During this period, the permeate should be discarded, as noted by Gryta [132, 246] since HCl is volatile and can be carried into the permeate side along with water vapor. This was observed in a DCMD experiment with hollow fiber membranes [132].

Gryta [134, 212] reported the efficacy of HCl rinsing in restoring flux back to original value in the case of membranes with CaCO3 scale layer. However, after repeating the washing procedure with 2-5% HCl, Gryta found that maximum restored flux declined [209]. Yang et al. [155] were able to restore flux completely by cleaning with a 5% HCl solution following the use of untreated 'tap water' in their experiments. Curcio et al. [131] used a two-step cleaning strategy using a citric acid solution followed by a NaOH solution, allowing each of them to act for 20 minutes. They reported complete recovery of both flux and hydrophobicity in an MD experiment using synthetic seawater. Some authors [29] have also reported cleaning using bases such as 0.1 M NaOH for 20 minutes followed by pure water run for 2 hours. While pure water achieved 87.5% flux recovery, the NaOH solution was able to restore the flux to the initial value. Gryta et al. found that pretreatment using $Ca(OH)_2$ helped mitigate fouling from silicates and sulfates [200].

Gryta [132] used HCl also to clean a membrane with iron oxide scale layer. It was found that 18% HCl solution cleaning brought the flux close to the initial value and did better than 5% or 36% HCl. The experiment showed that iron oxides were very difficult to remove from the MD hollow fiber membrane, requiring high levels of acidification (36 wt. %) for 1 hour cleaning duration to obtain full removal, which damaged the membrane [132]. Complete removal of the iron oxide scale did not eliminate all the negative effects of fouling, which included decreased membrane flux after cleaning and increased membrane wettability. Moreover, drying was needed to actually recover the flux and restore the distillate quality because both were hindered (i.e., flux reduction from 800 to 650 $dm^3/m^2$-day and distillate electrical conductivity of 20 μS/cm) after the acid cleaning. Therefore, Gryta recommends only partial removal of iron oxide scale.

Table 5.2 shows data gathered from literature on the effectiveness of various cleaning protocols for restoring permeate flux.

**Table 6.2.** Common cleaning methods used in MD and reported flux recovery.

| S No | Ref | Membrane Material | Solution type | Cleaning method | Recovery % |
|---|---|---|---|---|---|
| 1 | [218] | PTFE | 7 g/L NaCl | Water wash | 98.48 |
| 2 | | | 7 g/L NaCl | Water wash | 97.48 |
| 3 | | | Ground Water | Water wash | 94.43 |
| 3 | [171] | PP-hollow fiber | Ground Water (CaCO3) | 3 wt.% HCl | 98.75 |
| 4 | | | Tap water (CaCO3) | 3 wt.% HCl | 98.56 |
| 5 | | | Boiled TW (CaCO3) | 3 wt.% HCl | 98.79 |
| 6 | [131] | PP-hollow fiber | Seawater (HA, CaCO3) | 20 min citric acid followed by 20 min NaOH two stage cleaning | 100 |
| 7 | [211] | PTFE | Seawater | Piezoelectric transducer at 35kHz used to induce cavitation and cleaning of membrane. Initially flux restored. | 91.20 |
| 8 | [133] | PP-hollow fiber | Microbial Biofilm | NaOH at pH=12, 40C distilled water, 70% ethanol for disinfection, distilled water | 100 |
| 9 | [140] | PP-hollow fiber | Ground Water (primarily CaCO3, CaSO4) | 2-5 wt.%HCl | 100 |
| 10 | [157] | PP-hollow fiber | Tap Water (CaCO3) | 5 wt.% HCl | Recovery reduces with number of cleaning cycles: Cycle 1 - 99.9 Cycle 6 - 96.5 |

The effectiveness of acid washing is known to vary dramatically by the type of fouling. Alkaline solutions that are more soluble at lower pH have been found to be very effectively removed by acid cleaning, including the most common scale, $CaCO_3$. Other crystalline scale, like iron oxide, has required very strong acid to remove, [132] as described above. Silica scale has proven to be similarly difficult to remove. Acid cleaning has caused mild but incomplete removal in cleaning organic matter on MD membranes [30].

Sometimes cleaning has been achieved by simply running de-ionized (DI) water through the system. DI water readily absorbs salts, and is even known to leach salts from surroundings [218]. Mericq et al. [207] for example, completely restored membrane flux using RO permeate water to remove and re-dissolve scale deposits from synthetic seawater in their VMD system.

In terms of other cleaning methods, a common and simple approach is reversing the flow direction, which successfully restored flux from biofouling in AGMD Memstill experiments with pond water performed by Meindersma et al. [186].

## 6.7 TRENDS IN SCALING IN MD

Future developments in MD technology and industrial applications will create new issues and areas needing investigation for MD fouling. Multi-stage designs of MD promise much better efficiency, with some theoretical studies claiming lower energy consumption that the existing state-of-the-art thermal technologies, such as MSF and MED [87]. Fouling in staged systems and for energy recovery devices that recirculate fresh feed into later stages needs further research [132, 183]. As MD is developed industrially, field experience and optimization of pretreatment requirements will be developed. Additionally, as membranes designed specifically for MD are created, studies fouling with those materials, porosities, B values, and other properties will be critical. MD use will continue to grow as well in non-desalination areas such as the food, chemical, and dye industries [247], which were not considered in this review. One final area of for further study is monitoring of scale development in plants operations and long-term experiences.

## 6.8 CONCLUSION

Scaling and fouling in MD are found to be pervasive, but design and mitigation methods have proven effective at making MD technology resistant to scaling and fouling. Four principal types of fouling and membrane damage have been found in MD: inorganic salt scaling or precipitation fouling, biofouling, particulate fouling, and chemical degradation. Inorganic scaling risk, the primary focus of academic studies, varies greatly with the salts present. Alkaline salts such as $CaCO_3$, the most common scale by far, have proven to be readily prevented by decreasing feed pH or removed through acidic cleaning, while other scale has proven more tenacious and must be generally be limited by avoiding supersaturation. Biofouling has also been observed in MD, but can be largely mitigated through control of operating conditions. Particulate fouling in MD has proven difficult to remove, but it can largely be prevented by ultra- or microfiltration. Chemical degradation and damage to the membrane has proven to be a concern as well, but can be mitigated by selecting operating conditions that avoid fouling, extreme pH, and certain salts. The choice of membrane material and properties can also help to avoid chemical degradation; PTFE and PVDF membranes exhibited different fouling characteristics, such as in wetting, internal crystal growth, cracking, and mechanical strength [34]. A study on glass membranes claimed glass membranes had superior thermal and chemical robustness than polymer membranes [239]. However, there is a general lack of information about the effect of the polymer type on the prevention of fouling/scaling in MD processes.

Fouling tendency had been perceived to be highly variable and perhaps unpredictable, but some consistent patterns are seen. Studies with extreme susceptibility to fouling have almost exclusively been performed with hollow fiber capillary membranes with the feed internal to the capillary tubes. These modules have fouled within hours to days in unsaturated conditions that would not cause fouling in other modules. Likewise, the studies showing high resistance to fouling tended to have highly hydrophobic membranes or coatings, and include hollow fiber studies with permeate in the capillaries [35]. Numerous studies have found substantial reduction in scale from superhydrophobic fluorosilicone coatings, and while the individual papers may question how large a role the coating played in the often complete lack of scale [36], the literature overall proves consistently that these coatings have a dramatic effect. More hydrophobic membranes show higher resistance to wetting and associated internal crystal precipitation. Since internal fouling often leads to further wetting, this too is reduced. Studies have shown that increased hydrophobicity increases LEP and, in experiments, is associated with reduced wetting, fewer fouling deposits, and purer condensate [222].

Micro, nano, or ultrafiltration has proven effective in stopping particulate scale. Modifying pH in the feed or with cleaning may prevent or remove certain types of fouling very effectively as well. Keeping feed temperature above 60 °C has proven very effective in mitigating biofouling, with some exceptions. Removal of oxygen via deaeration may be expected to reduce biofouling and the authors recommend further investigation into the effectiveness of this technique. Rinsing with a basic solution such as with NaOH may resolve fouling for some substances, including HA. Mildly effective fouling prevention methods include boiling for removal of carbonate, ultrasonic cleaning, magnetic water treatment, flocculation, covering the membrane surface with a less porous smaller pore size layer, and for HA, oscillating the feed temperature. Antiscalant effectiveness studies in MD have been inconclusive; both strong reduction in scaling and actual decreases in permeate flux have been reported.

System design characteristics also influence fouling. Concentration polarization, closely related to feed Reynolds number and rate of permeate production, is critical in causing fouling, and can be mitigated by increasing the feed flow rate, or mixing technologies such as bubbling [37]. Temperature also remains important, as the most likely foulants have inverse solubility with temperature. Simple computational models were applied by the present authors to illustrate the effect of coupled heat and mass transfer on scaling. A correlation between theoretical prediction of higher SI at the membrane interface and increased rate of fouling induced flux decline was observed. Choice of safe operating conditions should therefore consider temperature and concentration polarization effects and solubility characteristics of the salts. Finally, stagnation zones or high residence times in the module may contribute to fouling as well.

# Chapter 7. THE EFFECT OF FILTRATION AND PARTICULATE FOULING IN MEMBRANE DISTILLATION

## 7.1 ABSTRACT

Fouling and scaling in membrane distillation (MD) is one of the significant barriers to its continued growth. Fouling in MD blocks the pores, causing a decline in permeate flux, and may eventually lead to wetting of a hydrophobic membrane by the saline feed, contaminating the permeate. Many previous studies on MD have observed that while MD is more fouling resistant than reverse osmosis (RO), inorganic salt precipitation on the membrane surface under supersaturated conditions may readily foul and cause wetting of the membranes. While most studies have assumed that crystal growth occurs directly on the membrane surface, precipitation of particles in the bulk which then migrate to the surface may play a significant or dominant role. In this study, membrane distillation experiments are run at varied supersaturated salinities with and without filtration in equivalent operating conditions to examine the effectiveness of particulate fouling in MD, and thus also the role of bulk precipitation in inorganic fouling. Conditions were designed to favor surface crystal growth over bulk nucleation, to examine the limiting case. The experiment is paired with heat and mass transfer and solubility numerical modeling to analyze the effects of concentration polarization to accurately calculate the saturation index (SI) in the bulk and at the membrane surface. The results show that the removal of particles precipitating in the bulk makes a tremendous difference in reducing flux decline, visible crystals on the membrane surface, and wetting. This implies that bulk nucleation followed by deposition, rather than surface crystal growth, dominates the scaling process in MD when the bulk is supersaturated. Jaichander Swaminathan and John H. Lienhard V [38].

## 7.2 INTRODUCTION

Membrane distillation is a thermal desalination technology which relies on porous hydrophobic membranes that repel liquid water but allow water vapor to pass. MD possesses relative fouling resistance compared to other membrane-based technologies such as reverse osmosis [10]. The process is driven by a difference in vapor pressure across the pores. Numerous studies have shown that MD membranes can withstand concentrations several fold higher than saturation without the onset of fouling [140, 154]. Indeed, materials used for MD are naturally scaling resistant, and are often used to coat heat exchangers to prevent surface scaling [248]. This advantage enables the use of MD in further desalinating brine from other desalination systems or to concentrate industrial waters. Understanding and control of fouling prevention is therefore imperative to the technology's competitive advantage. Fouling in MD is generally categorized as scaling, which is the process of inorganic salts growing on the membrane surface, and particulate fouling, which generally refers to insoluble particles like silt, silica, and organic matter depositing on the membrane surface. While scaling is often described as largely crystal growth on the surface, this study investigates whether a particulate mechanism is substantially important in conditions thought of as scaling. Many studies have shown that inorganic fouling at high SI may block membrane pores, thus decreasing flux and wetting the hydrophobic membrane, causing saline feed water to contaminate the permeate [157, 149]. Therefore, it is highly desirable to determine the limiting supersaturation conditions that can allow MD to operate without fouling degrading performance. However, the literature lacks systematic studies determining which saturation conditions cause the fouling to originate at the membrane surface or in the bulk solution. The concentration of salts can be meaningfully compared across experiments with the saturation index, or SI, which is a log-based scale for the concentration compared to the saturation, where an SI = 0 indicates a saturated solution and SI = 1 is 10 times the saturated concentration.

Pretreatment with filtration can help address this question as it removes incoming particles from the bulk. Previous MD studies have shown that systems with filtration had low levels of fouling [223, 186], and a literature review study showed that filtration was among the most effective methods in reducing inorganic fouling [11]. Furthermore, additional studies found that incorporating filters in membrane distillation reduced flux decline [226, 225, 18, 214]. These studies lack a systematic examination of how the bulk SI and the SI at the membrane surface relate to both flux decline and wetting of the membrane, which the present study addresses. In this study, a range of experiments of supersaturated SI at the membrane surface are run in an Air Gap Membrane Distillation (AGMD) system, with and without a filter which removes particles that nucleate in the bulk before recirculation. The effects of nucleation and scaling are examined by measuring flux decline, increases in permeate conductivity caused by wetting, and polarized-light microscopy.

## 7.3 METHODOLOGY

### 7.3.1 EXPERIMENT DESIGN

The experimental setup for the study is a flat plate, air-gap MD (AGMD) module with turbulent flow conditions in the feed. The system has a small temperature drop over the feed and cooling channels (<0.2°C) to keep the conditions finely controlled. Numerous flow rate, temperature, and pressure sensors are arranged in the system as shown in Figure 7.1. The system uses large 40 L tanks for the feed solution and cooling liquid to act as thermal masses, keeping the temperature constant. The module consists of a series of plates machined out of polycarbonate and aluminum. The module and its numerical modeling have been previously described and validated [84]. The system was designed with no metal parts exposed to the flow, including the heater, since even resistant metal parts such as stainless steel may corrode when in contact with warm, saline feed [132], thus creating nuclei for crystallization. DI water and reagent grade chemicals were used to minimize contamination. This abated the availability of crystal growth nuclei in the bulk, making conditions adverse for bulk nucleation to play a significant role. When these opposing conditions of filtration and highly pure solutions still occur in tandem with significant bulk nucleation, it indicates that bulk nucleation plays a substantial role in MD fouling.

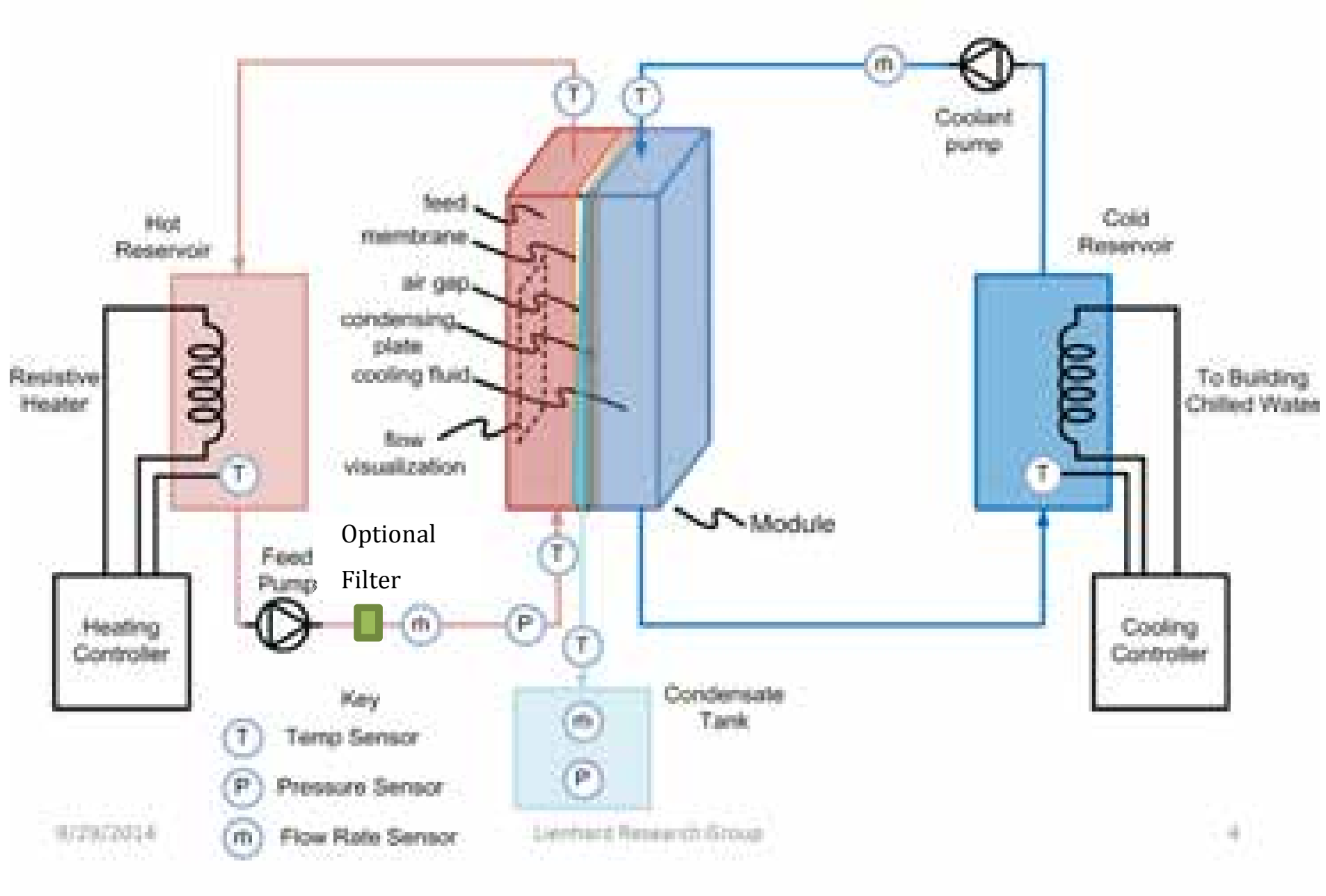

**Figure 7.1.** AGMD scaling test bed.

The AGMD system operated in a continuous mode with the SI and temperatures held constant throughout the experiment. The air gap size was 1 mm. The SI concentrations were determined with PHREEQC chemistry modeling software.

Error analysis was performed, and has been discussed in detail for this apparatus in the previous work by the authors [84]. The uncertainty in temperature for the thermistors was ±0.2 °C, and the temperature controllers cycled on and off to maintain a temperature range of ±0.1 °C. However, occasional temperature downward spikes caused by maintenance on the system was as large as 4 °C for a few minutes, causing slight variation in permeate flow rate, but not increasing or significantly effecting fouling. The permeate mass scale, measuring grams with intervals of 30 minutes each containing ~200 g thus had an uncertainty of around 0.5%. The uncertainty in the conductivity readings was <2%. The flow rate varied within ±0.6%. Feed static pressure was set to 1.5 bar, varying by ±3%, but static pressure has minimal effect in MD permeate flux and fouling. Two of the four experiments were performed twice (SI = 0.4, no filter; and SI = 0.2, with filter), and each produced the same results regarding whether notable flux decline or any wetting occurred, confirming repeatability of the system.

### 7.3.2 Mathematical Modeling

An Engineering Equation Solve (EES) discretized model based on finite difference method was used for determining the saturation index at the membrane surface, and estimating permeate flux [58]. The model balanced simultaneous heat and mass transfer through an AGMD system with approximately 300 computational cells over the length of 16 cm, as shown in Figure 7.2. This balancing includes flow of the feed and permeate in and out of each cell, diffusion of the water vapor through the membrane and air gap, and convective heat transfer in the feed and permeate, advection of water vapor from the feed to the air gap, and conduction across the membrane. The model included the vapor flux equation through an MD membrane as a function of the membrane permeability B and the vapor pressure difference across it, and modeled the condensation film of the permeate as a classic laminar condensation film on a vertical flat plate. The details and equations of this model have been previously published by the authors [84, 24].

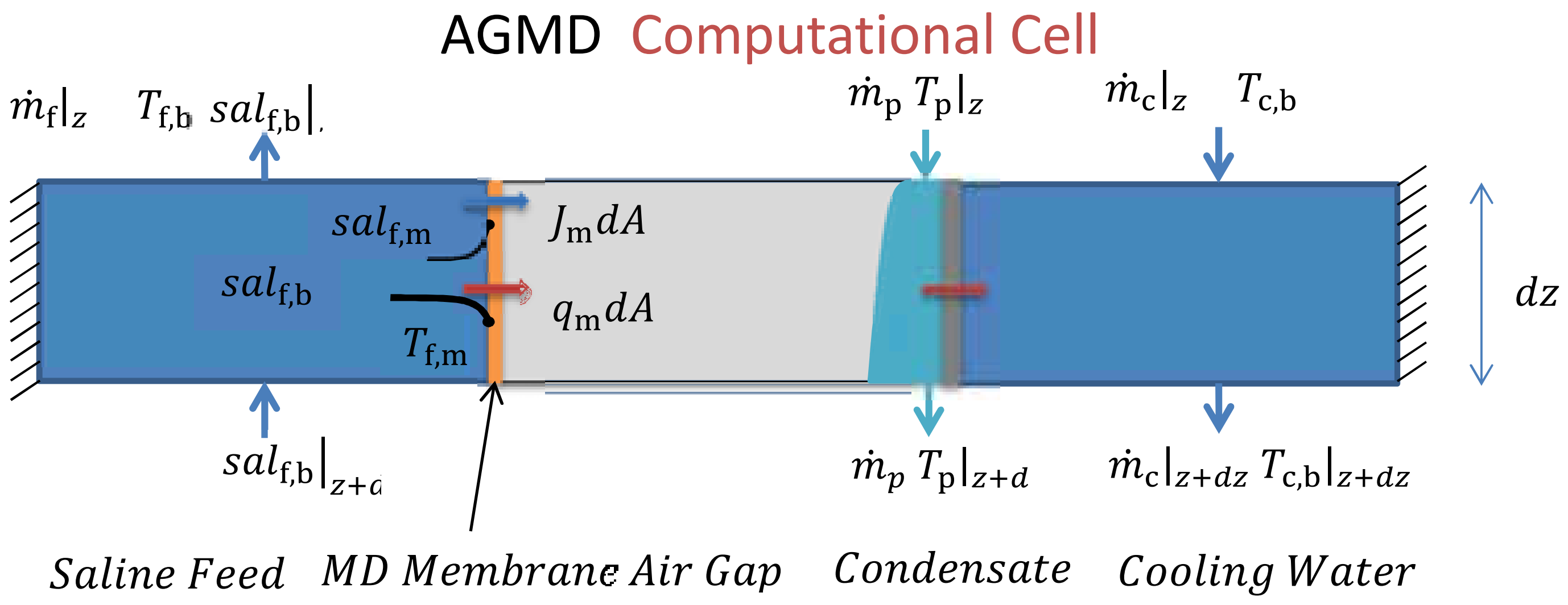

**Figure 7.2.** Computational Cell for AGMD Analysis and Concentration Polarization Calculation

### 7.3.3 Membrane and Filter

A commercial PVDF membrane was used with a nominal pore size of 0.2 μm and a thickness of 200 μm. The effective area of the membrane was 192 $cm^2$. The static contact angle of the membrane with water was 125°. The membrane was uncoated: improvement above the typical contact angle of PVDF (89°) can be attributed to surface features and roughness.

A filter was used for the feed to remove small nucleating particles, which could deposit on the membrane and serve as seeds for further growth within the AGMD module. The filter used was a Pentair pleated cellulose cartridge filter, with a 20 μm nominal pore size, allowing it to block large nucleated particles of $CaSO_4$.

### 7.3.4 Scaling conditions

The experiments focused on calcium sulfate in a SI range from 0.2 to 0.4 at the membrane surface with an inlet bulk temperature of 70 °C. Calcium sulfate is among the most significant scales that may occur in concentrated seawater brine and other desalination applications [157, 159]. Unlike the common calcium carbonate scale, calcium sulfate does not have a strong pH dependence and has less complex chemistry, making the experimental study more tractable.

The setup was designed so that concentration polarization near the membrane surface caused further supersaturated conditions, increasing the concentration by 16%. Due to its inverse solubility behavior, the salt is supersaturated near the heater element in the feed tank as well, leading to nucleation and precipitation before the feed reaches the module. This effect was minimized by separating the heater from the feed solution with a plastic bag that acted as a thermal barrier. Real world MD systems also experience the precipitation near the heater, as salts such as $CaCO_3$ and $CaSO_4$ would be most supersaturated at higher temperatures.

Avoiding particles in the bulk was important to prevent alternative means of fouling the membrane. To achieve this, the water was created with DI water, and the feed was created by mixing two fully dissolved solutions: one with calcium chloride and another containing sodium sulfate, both of high purity from Sigma-Aldrich. Between experiments, system cleaning was performed by recirculating DI water. To avoid excessive supersaturation conditions near the heater, good mixing and heat transfer ensured that the heater temperature was close to that of the feed tank.

## 7.4 RESULTS

Experiments of $CaSO_4$ membrane fouling were performed at varied super-saturated SI at the membrane surface. These experiments, ranging from 24-60 hours in duration, identified fouling by permeate flux decline and an increase in permeate conductivity caused by wetting. By studying the effect of using or excluding a filter, the experiments can determine how significant a role bulk nucleation plays in MD membrane fouling. These experiments are detailed in the order of more adverse fouling conditions to less susceptible conditions. All experiments were run with an AGMD module with a hot feed temperature of 70°C and a coolant temperature of 20°C.

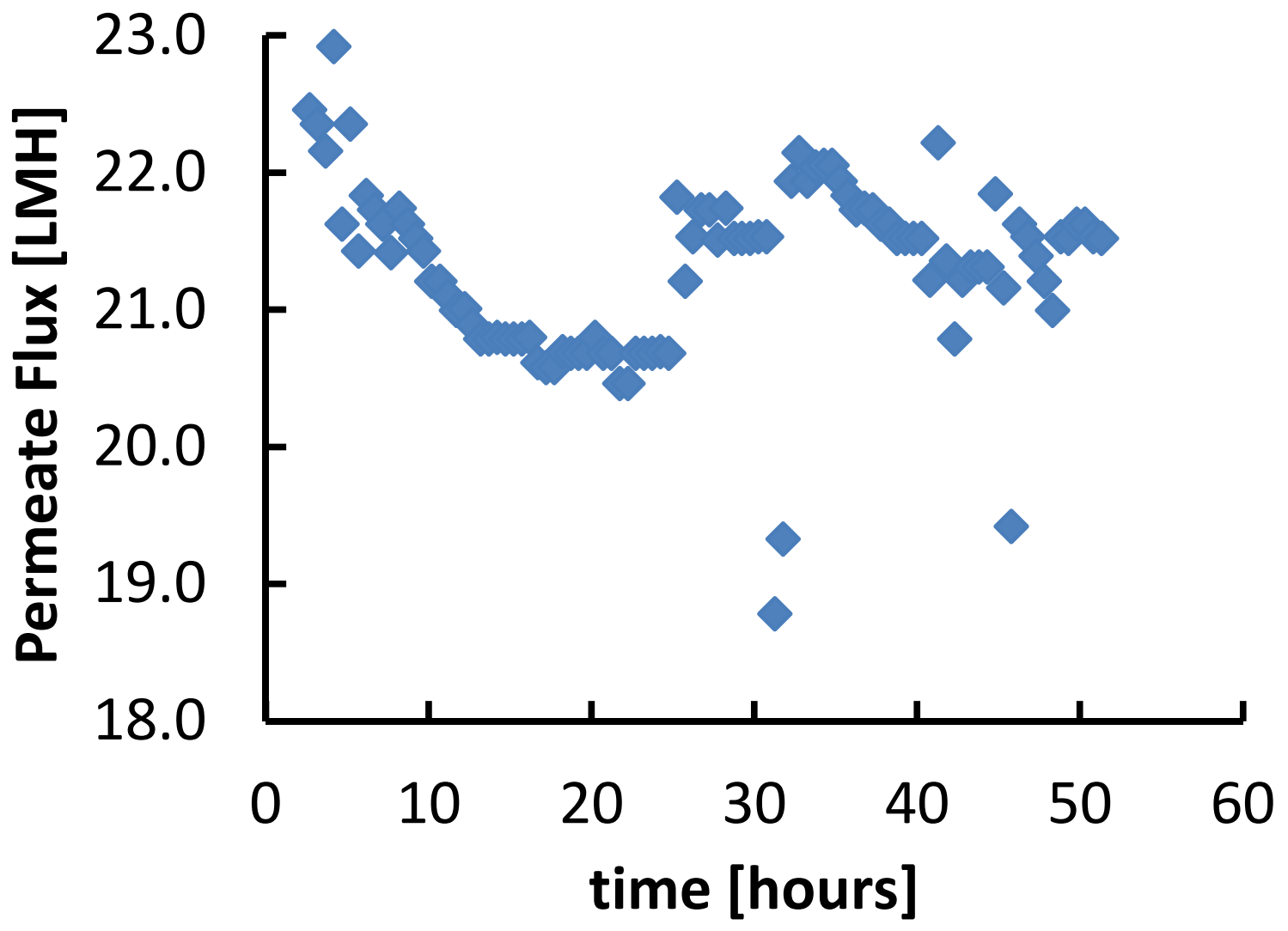

**Figure 7.3.** Permeate flux decline averaged over 0.5 hours for $CaSO_4$, SI = 0.4, no filter. The increase in flux between 25 and 35 hours corresponds with wetting through the membrane, as seen in Figure 7.4.

Figure 3 shows that in the more highly saturated experiments without a filter, at SI = 0.4, permeate flux declines significantly by about 11%, as a result of foulant deposition clogging the membrane pores.

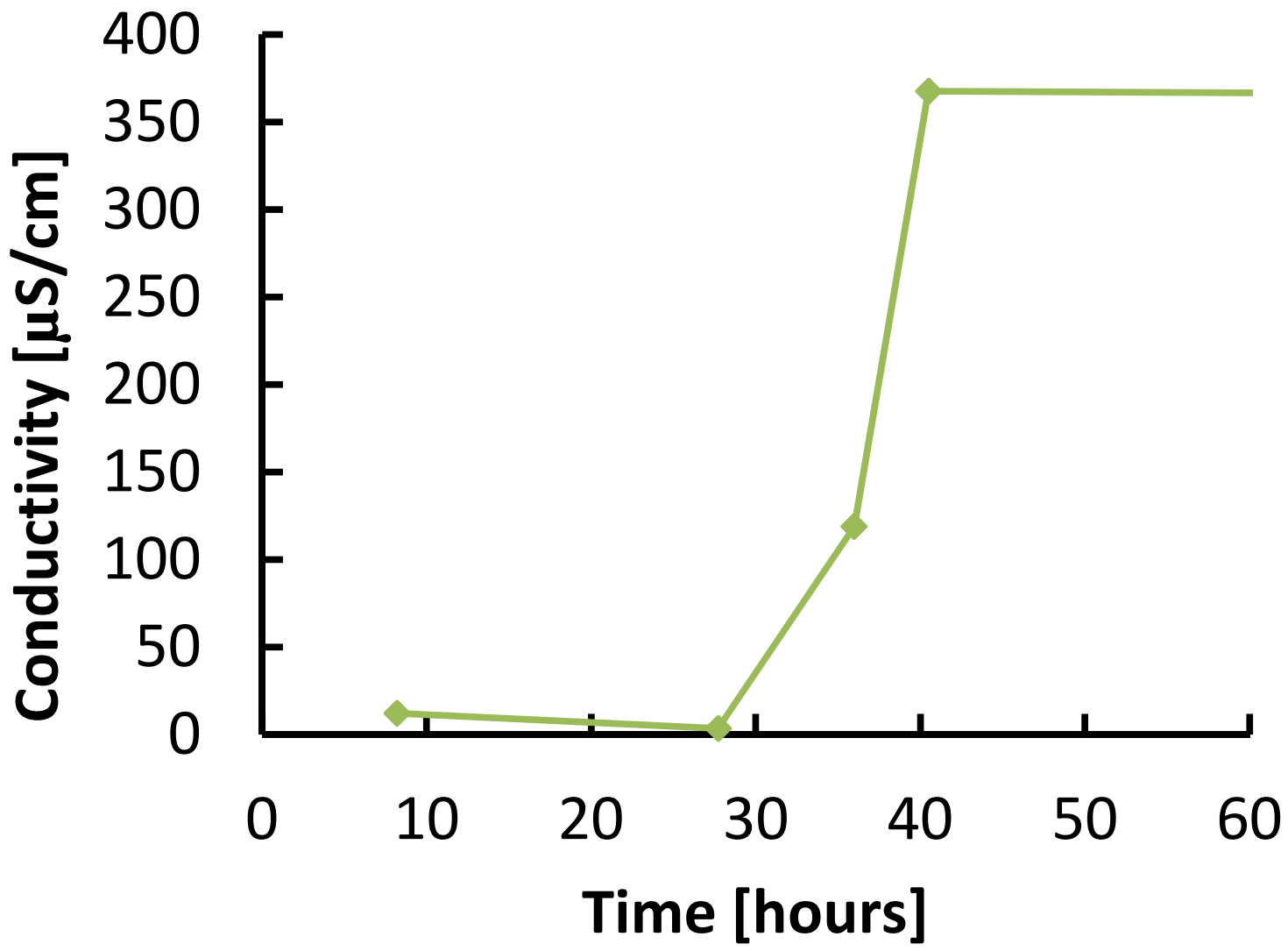

**Figure 7.4.** Permeate conductivity for CaSO4, SI = 0.4, no filter. Flux increase between 28 and 35 hours corresponds to wetting of the feed through the MD membrane.

The increase in permeate flux in Figure 7.3 corresponds to the amount of feed water that would leak through to cause the conductivity (and thus salinity) results seen in Figure 7.4. There is a slight time delay between flux increase and conductivity rise (~3 hours), which is due to the piping after the module containing a significant volume of water, thus causing a time gap between water being flushed out and measured and water at the membrane surface.

This saline experiment without the filter experienced significant wetting of the membrane and contamination of the permeate. By the end of the experiment, the rate of leaking feed into the permeate accounted for ~8% of the total permeate flux leaving the module. This represents a failure of the MD process, unless the standards for permeate quality are not high.

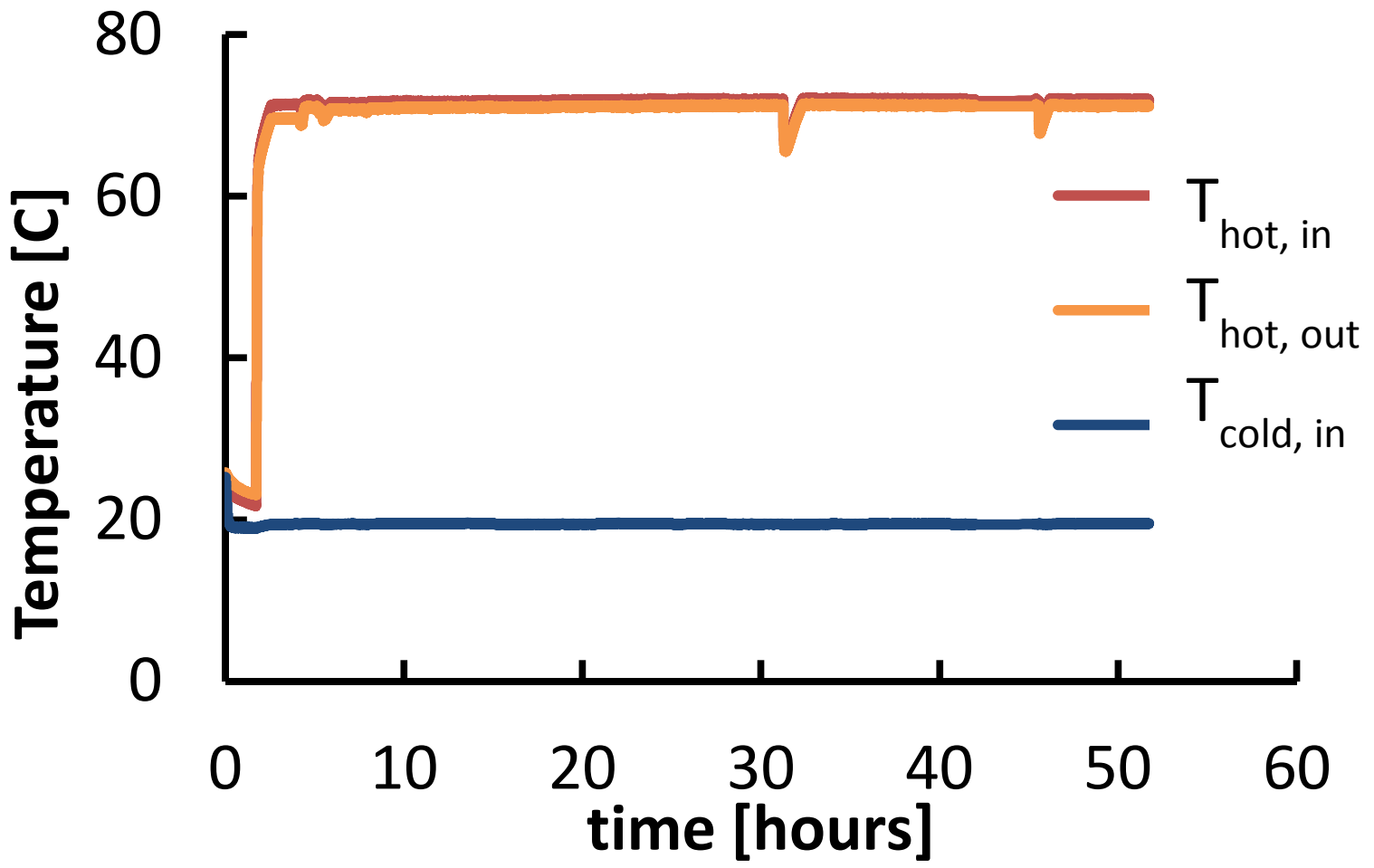

**Figure 7.5.** Cold side and hot side inlet and outlet temperature readings.

The temperature results shown in Figure 7.5 indicate minimal variation in system temperatures, and represent those found in all the other trials. The dips in temperature in Figure 7.5 at t = 32 and 46 hours and result from water addition to make up for evaporation losses and maintain inlet feed salinity constant. This process resulted in minor and brief flux declines observed throughout the permeate flux data for all experiments.

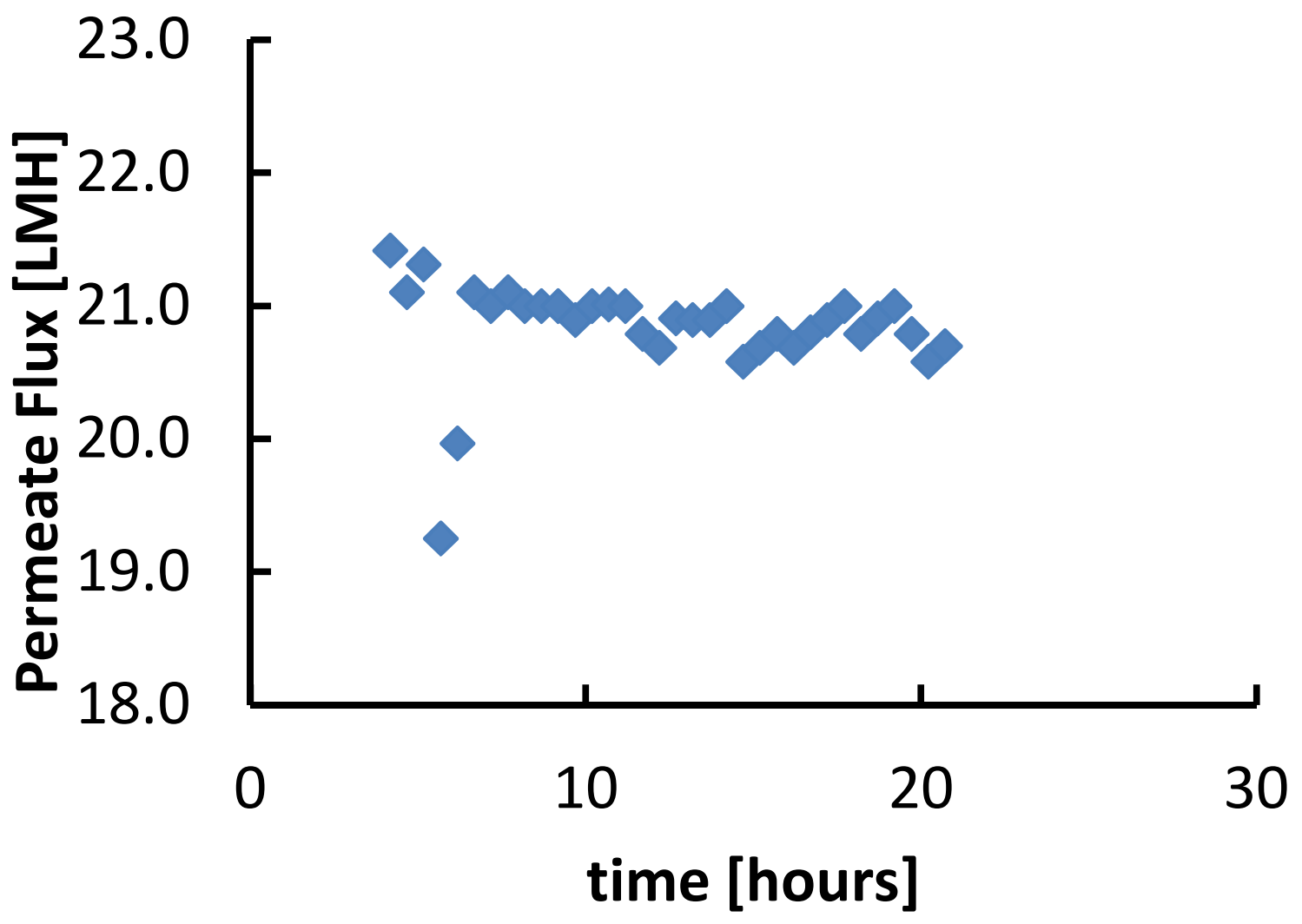

**Figure 7.6.** Permeate flux decline averaged over 0.5 hours for $CaSO_4$, SI = 0.4, with a filter.

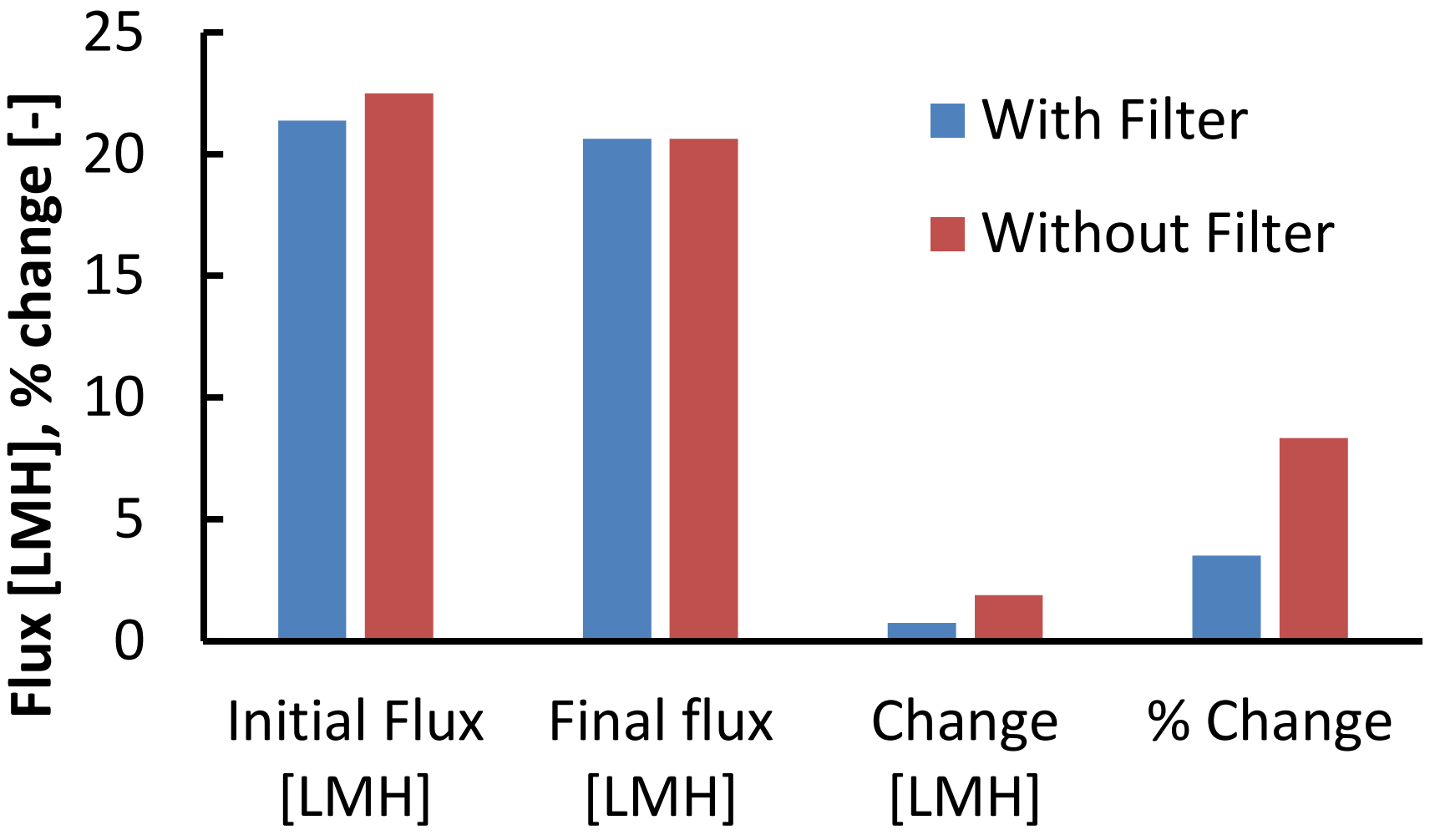

**Figure 7.7.** Flux comparison with and without filter, $CaSO_4$, SI = 0.4

As seen in Figs. 6 and 7, in the high SI case (0.4), the flux declined significantly less when a filter was in place. Furthermore, no wetting was observed with a filter. The results indicate that even filters with fairly large pore sizes (20 micrometers) significantly reduce scaling in MD.

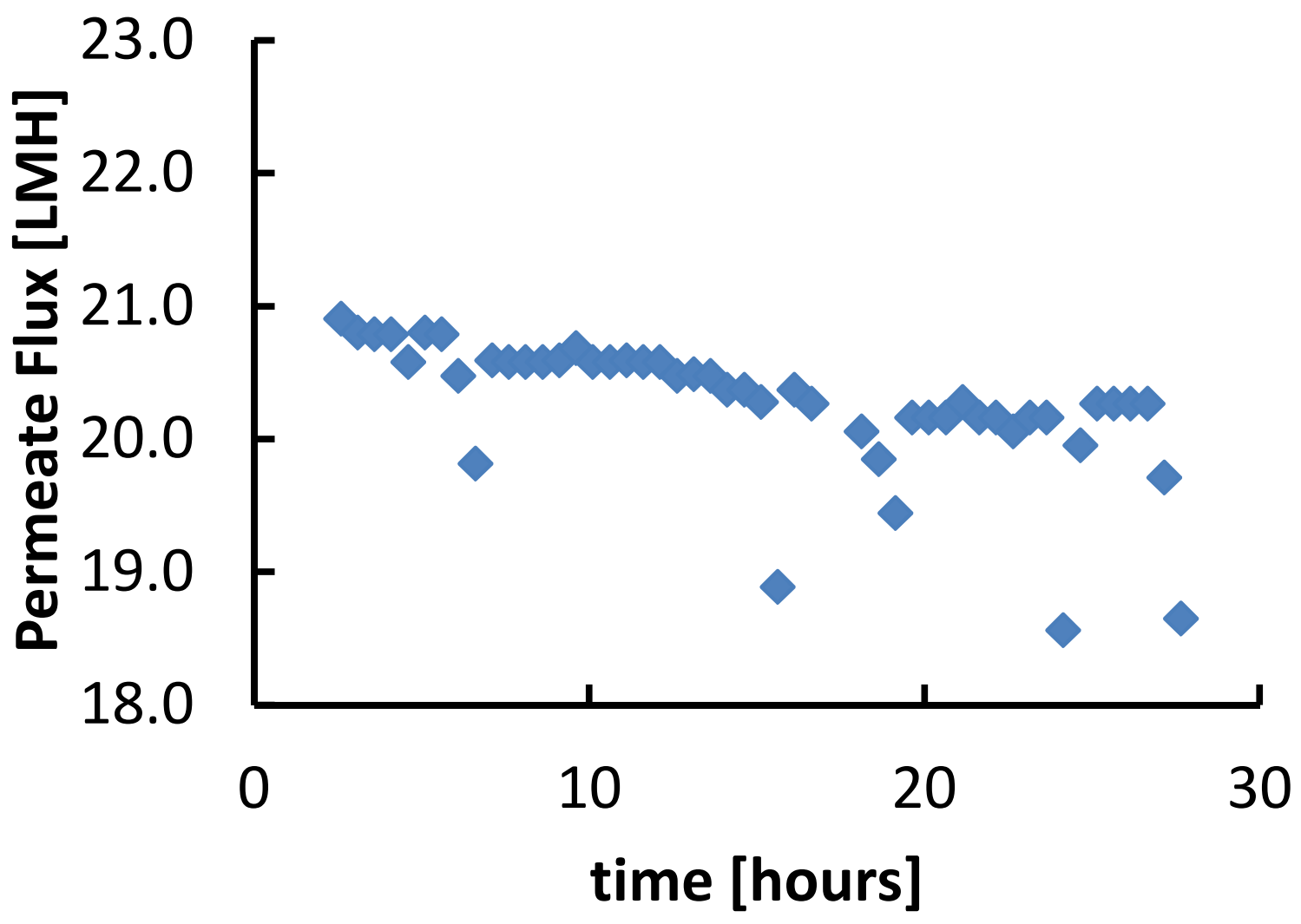

**Figure 7.8.** Permeate flux decline averaged over 0.5 hours for $CaSO_4$, SI = 0.2, without a filter.

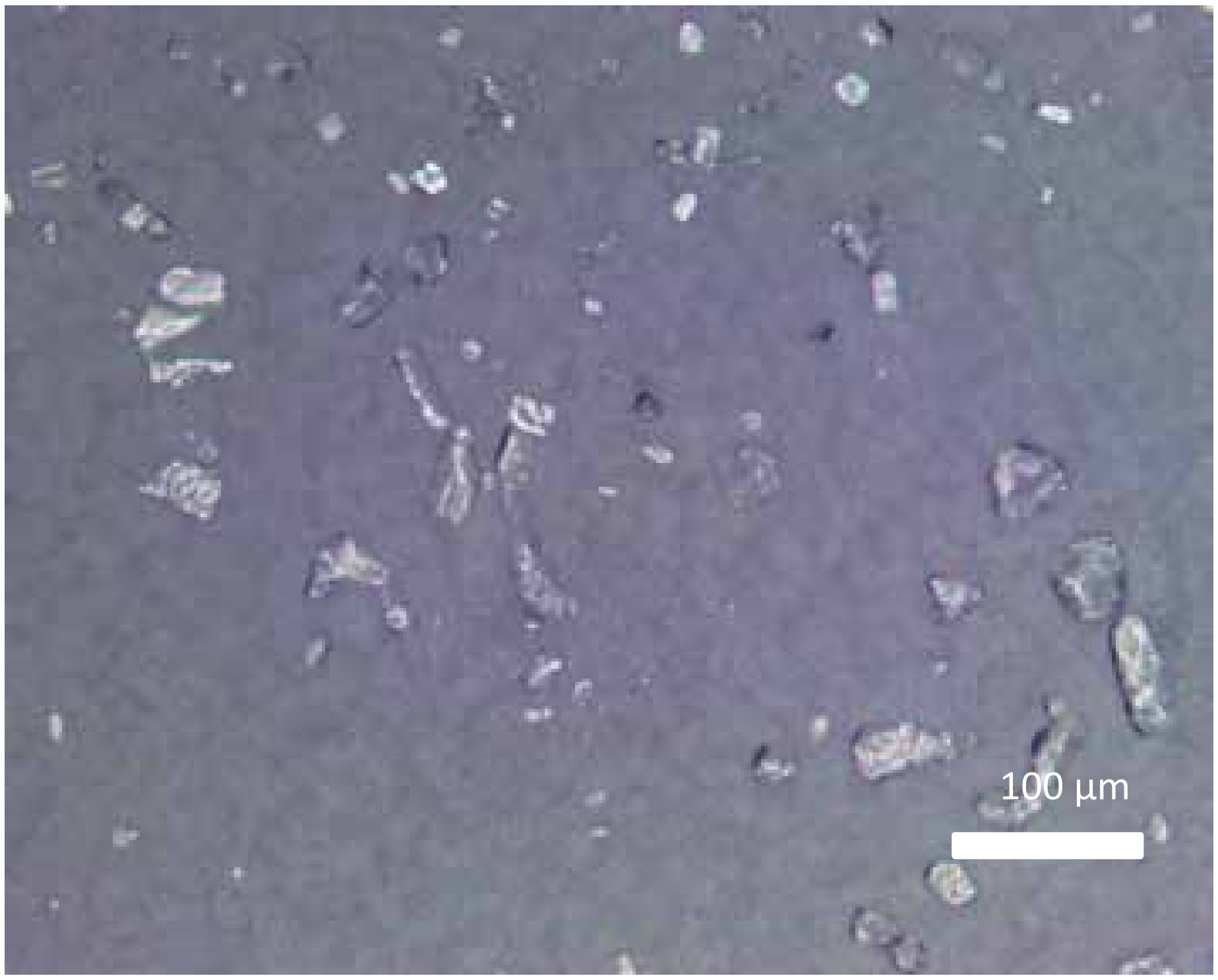

**Figure 7.9.** Polarizing microscopy image of MD membrane surface after SI = 0.2 experiment without a filter. The crystals on the surface clog the membrane pores, reducing the permeate flux

Microscopy with polarizing light helps make the white $CaSO_4$ scale visible on the white PVDF MD membrane. The scale occurs in discrete crystals, with most of the membrane remaining unscaled. This indicates that heterogeneous nucleation on crystals or colloids in the bulk and pipe surfaces is much more favorable than growth on the PVDF surface itself. In cases with the filter, relatively few crystals were observed.

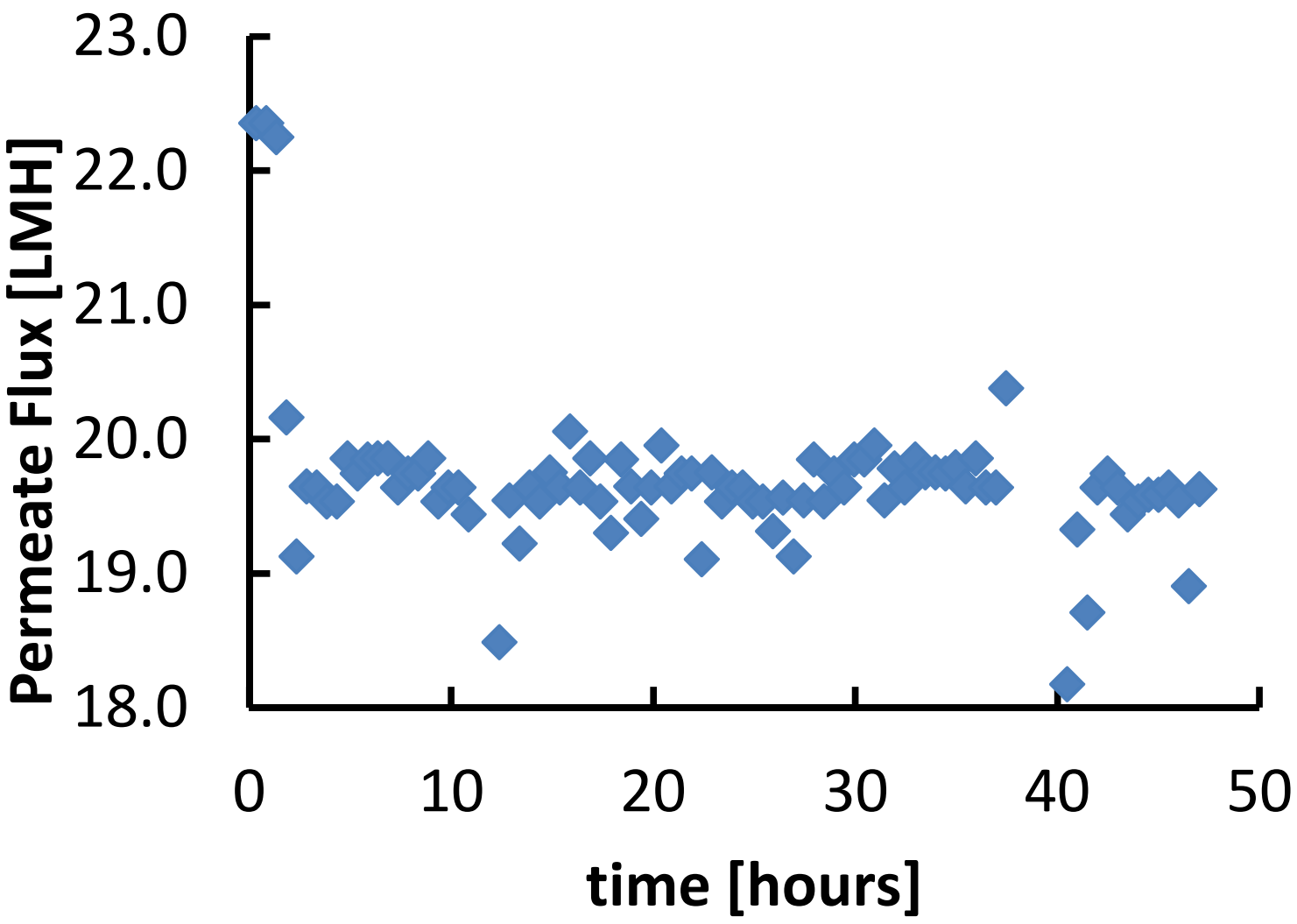

**Figure 7.10.** Permeate flux decline averaged over 0.5 hours for $CaSO_4$, SI = 0.2, with a filter.

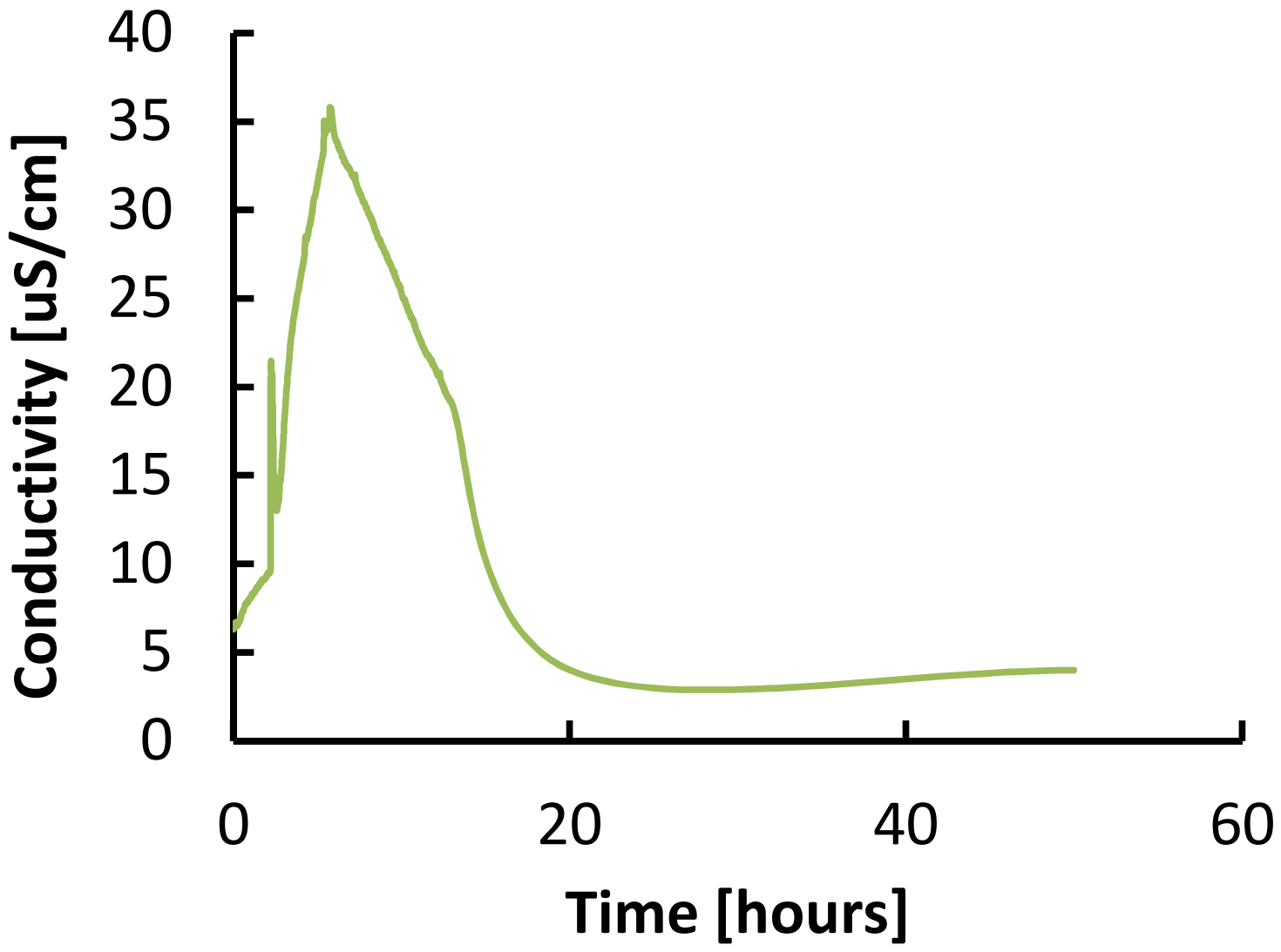

**Figure 7.11.** Permeate conductivity for $CaSO_4$, SI = 0.2, with a filter. No significant flux declined occurred.

The permeate and conductivity graphs for the lower SI (0.2) experiment with a filter showed neither significant flux decline nor membrane wetting. Under these conditions, even for a long duration experiment, the scaling is minimal enough to not affect the performance of the MD process. The flux is slightly lower than the other experiments because the hot side temperature was maintained at 1.2 °C lower than those in the hottest experiments, where variability between experiments was caused by changes in heat losses. The first two hours were at 71.4°C, which caused higher initial flux.

In testing and commissioning the system, experiments were tried with subsaturated (SI<0) and saturated (SI = 0) solutions under the same conditions with $CaSO_4$ without a filter, and no effects of fouling were observed, which included no flux decline, no wetting, and no visually observable crystals on the membrane surface. It should be noted that all the SI values reported are those at the membrane surface, which are calculated via the numerical modeling explained previously.

**Table 7.1.** Summary of the effects of filtration and salinity on fouling in membrane distillation.

| SI (membrane surface) | Filter | Flux Decline | Wetting | Duration [hr] | Final Permeate Conductivity [μS/cm] | Permeate Flux Decline Percent | $T_H$ [°C] |
|---|---|---|---|---|---|---|---|
| 0.2 | no | yes | no | 27 | 4.9 | 3.5% | 69.8 |
| | yes | no | no | 47 | 4.0 | 0.6% | 69.5 |
| 0.4 | no | yes | yes | 29 | 367 | 12.6% | 71.8 |
| | yes | yes | no | 23 | [not recorded] | 3.3% | 70.9 |

**Table 6.1** summarizes the experimental results for all trials performed. Filtration was extremely effective in reducing membrane fouling, indicating bulk nucleation of crystals plays a very significant role in these experiments.

The pore size of the filters was still quite large. The theoretical critical size for stability of a $CaSO_4$ crystal in the bulk solution is much smaller than the 20 μm pore size of the filter, so smaller crystals can presumably still pass through. Nanofiltration would be required to entirely remove smaller but stable crystals. Nevertheless, the use of a filter made a more significant difference in reducing flux decline than did changes in the saturation index. This indicates that bulk nucleation is likely dominant in the present MD system, and that it may be important in other system configurations. System design to reduce that process, including filtration and control of residence times, may therefore be effective in mitigating MD fouling.

The results also have important implications for MD systems that use brine recirculation. Since MD typically has low recovery (e.g., 6-10%), recirculation is common, especially for systems trying to reach high water recovery or zero-liquid-discharge. Such systems will have substantial residence time for potentially supersaturated brine, which would allow significant bulk nucleation to occur. The effective residence time for long term operation could be an order of magnitude or larger than for the present experiments.

## 7.5 CONCLUSIONS

Experiments were conducted by separately mixing $Na_2SO_4$ and $CaCl_2$ in DI water and then combining these solutions to form the feed. With the concentrations, the measured temperature and flow rates, and an Engineering Equation Solver (EES) numerical model, the concentration and temperature polarization effects were estimated, and the saturation index of $CaSO_4$ was evaluated at the feed-membrane interface, $SI_{f,m}$.

**Table 7.2.** $CaSO_4$ fouling effects summary for the filtration experiments

| SI | Filter | No Filter |
|---|---|---|
| 0.2 | none | Flux Decline |
| 0.4 | Flux Decline | Wetting and Flux Decline |

The experiments have shown that filtration can significantly reduce flux decline at high saturation index as seen in Table 6.2, which indicates that precipitation in the bulk may play a dominant role in the flux decline and wetting of MD membranes by inorganic fouling. This occurred despite conditions that were favorable to surface scaling over nucleation, including a weakly hydrophobic membrane material (PVDF), notable concentration polarization at the membrane surface, the minimize of available crystallization nuclei provided by lab-grade purity solutions and avoiding of metal parts, and a minimization of temperature elevation at the heater. Therefore, bulk heterogeneous nucleation, not surface crystallization, is indicated as the dominant fouling mechanism in the present system.

For the conditions studied, the presence of the filter reduced flux decline by an amount similar to decreasing the saturation index to SI =0.2 from 0.4. The present experiments have shown that bulk precipitation dominates at both low and high SI (0.2-0.4). Significantly, the study implies that real MD systems using brine recirculation without filtration may suffer a significant penalty in performance under supersaturated conditions, as brine residence times are effectively endless. Additionally, benchtop experiments using recirculation may not properly represent real systems without recirculation.

## 7.6 ACKNOWLEDGEMENTS

We would like to thank Mathias Kolle, Jocelyn Gonzales, Sarah Van Belleghem, Ann McCall Huston, and Priyanka Chatterjee for their contributions to this work.

# Chapter 8. The Combined Effect of Air Recharging and Membrane Superhydrophobicity on Fouling in Membrane Distillation

In previous studies of the desalination technology membrane distillation (MD), superhydrophobicity of the membrane has been shown to dramatically decrease fouling in adverse conditions, but the mechanism for this is not well understood. Air present on the membrane surface may also play an important role in certain MD technologies and perhaps reduces certain types of fouling. Previous MD studies have used air-bubbling to reduce concentration polarization, and other surface studies have combined superhydrophobicity and air layers to dramatically reduce biofouling. Air layers can reduce the fraction of the membrane in contact with water, and reduce the shear force of flowing water needed to remove small particles. The present work studies the effect of air layers on the membrane surface and superhydrophobicity on fouling of MD membranes by salts, particulates, and organic particles. Superhydrophobic MD membranes were prepared using initiated chemical vapor deposition (iCVD) of perfluorodecyl acrylate (PFDA) on poly(vinyldene fluoride) PVDF membranes and used to study the effects of hydrophobicity on fouling. A static MD setup with evaporation through an MD membrane but no condensing of permeate was used to examine the effect of air exposure on fouling, by measuring the increase in weight of the membrane caused by scale deposition. The method of applying air and salinity were analyzed. The study shows that the presence of air on the membrane surface significantly reduces biological fouling, but has mildly exacerbating effects on fouling of salts. Air recharging combined with superhydrophobicity reduced fouling in several cases where hydrophobic membranes did little. Amelia Servi, Sarah Van Belleghem, Jocelyn Gonzalez, Jaichander Swaminathan, Jehad Kharraz, Hyung Won Chung, Hassan A. Arafat, and John H. Lienhard V also contributed to this work [249] .

## 8.1 INTRODUCTION

### 8.1.1 SCALING IN MEMBRANE DISTILLATION

Membrane distillation is an emerging thermal desalination technology that relies on a porous hydrophobic membrane that passes water vapor but rejects liquid water [24]. MD is known to be relatively more resistant to scaling of salts compared to other membrane-based desalination processes such as reverse osmosis (RO) [11, 250]. However, the mechanism for this resistance is poorly understood. [34]. Fouling of the membrane surface impairs MD performance by blocking the surface, which reduces permeate flux, and may cause wetting of the saline feed through the membrane, contaminating the permeate [40].

Past studies on superhydrophobic MD membranes have shown extreme resistance to scaling [222, 239], which included a reduction in surface nucleation and particulate attachment [238]. Additionally, past studies on MD have also found that nucleation in the bulk feed fluid contributes significantly to MD fouling [38]. Studies on submerged superhydrophobic surfaces with visible air layers have shown extreme resistance to biofouling [187].

In desalination systems including membrane distillation, several types of scale dominate. Calcium scale, including calcium sulfate and calcium carbonate, are among the least soluble and most problematic inorganic scale in seawater and various groundwater sources [122]. For systems that experience regular dry out such as remote solar thermal desalination, significant sodium chloride is often left behind after evaporation [34], since it is present in such high levels in most waters. Finally, in seawater applications, the remains of algae often cause biological fouling, and in fact the polysaccharide alginate is often used to study algae fouling [251]. Alginate can form a gel layer on membrane surfaces that causes significant diffusion resistance [252].

For inorganic scaling to occur, two steps are involved: first nucleation of crystals from the solution, followed by crystal growth [253]. Crystal growth on stable crystals is spontaneous in saturated solutions [254], so the key to avoiding crystallization is to extend the induction time before nucleation occurs. The degree of saturation is measured by the saturation index (SI), which is a log scale of saturation, where 0 is saturated and 1 is 10 times the saturation concentration.

$$\mathrm{SI} = \log_{10}\left(\frac{C_x}{C_{sat}}\right) \tag{8.1}$$

where $C_x$ is the local concentration and $C_{sat}$ is the saturated concentration.

Previous studies have shown bubbling of air in the MD feed could reduce fouling, which was largely attributed to reduced concentration polarization by increased mixing [37]. Because of the effectiveness of air layers and superhydrophobicity for fouling prevention of MD membranes and other surfaces, as well as the desire to reduce wetting, the it is hypothesized in the present work that deliberately introducing air into the MD feed stream periodically could reduce fouling. Under this hypothesis, the air layer formed on the membrane surface may act as a barrier to particulate fouling, reducing the adhesion rate and reducing particle advection to the surface by physically blocking them. The periodic introduction of air may allow wetted membrane sections to recover lost hydrophobicity, reducing the risk of the feed contaminating the permeate.

In the present study, NaCl, $CaSO_4$, Silica, and Alginate foulants were each tested in a beaker-based MD setup with different methods of air exposure. Lifting horizontally and vertically, as well as the introduction of water-vapor saturated air bubbles, were techniques applied to either regular hydrophobic or coated superhydrophobic MD membranes. The rate of scale deposition by mass was used to examine the effect of air recharging on reducing fouling on different types of MD membranes.

### 8.1.2 Fouling and Nucleation Kinetics

The physical behavior of the types of foulants that may be present in the feed water must be understood to predict and explain fouling phenomena.

If particles deposit on the membrane, either from biological fouling, particulate fouling, or bulk nucleation, then the flux of particles to the surface can be modeled as follows, which was previously applied for particulate deposition on RO membranes [255]:

$$\frac{\mathrm{d}\delta_c}{\mathrm{d}t} = \alpha \delta_c v_p C_x \tag{8.2}$$

Where $C_X$ is the salt mass concentration, $v_p$ is the permeate velocity, $\delta_c$ is the fouling layer average thickness, $\alpha$ is called the "foulant sticking efficiency," and $t$ is the nucleation induction time.

An air layer on the surface of the membrane may alter deposition by reducing the depth of wetting into the surface, which may reduce the likelihood that particles convected to the surface will stick to it, thus reducing the foulant sticking efficiency, $\alpha$.

Previous studies have examined the Gibbs energy of formation ($\Delta G^*$) on the microporous hydrophobic membrane distillation surfaces as a function of the membrane-water-crystal static contact angle, $\theta$, and the porosity, $\varepsilon$. This formation energy is given as follows [256, 131]:

$$\frac{\Delta G^*_{heterogeneous}}{\Delta G^*_{homogeneous}} = \frac{1}{4}(2+\cos\theta)(1-\cos\theta)^2\left(1-\varepsilon\frac{[1+\cos\theta]^2}{[1-\cos\theta]^2}\right)^3 \tag{8.3}$$

Homogeneous nucleation refers to classical nucleation theory in the bulk, which can be derived from the surface energy (proportional to area) and the energy of phase transformation (volumetric). Meanwhile, heterogeneous nucleation refers to nucleation at interfaces, including on surfaces and particles in the bulk. The equation above shows that heterogeneous nucleation is favored, as its energy barrier is much smaller. Figure 12 shows the approximate Gibbs free energy barrier for homogeneous nucleation on the superhydrophobic membranes from this study using the equation above, and heterogeneous nucleation on the same membranes.

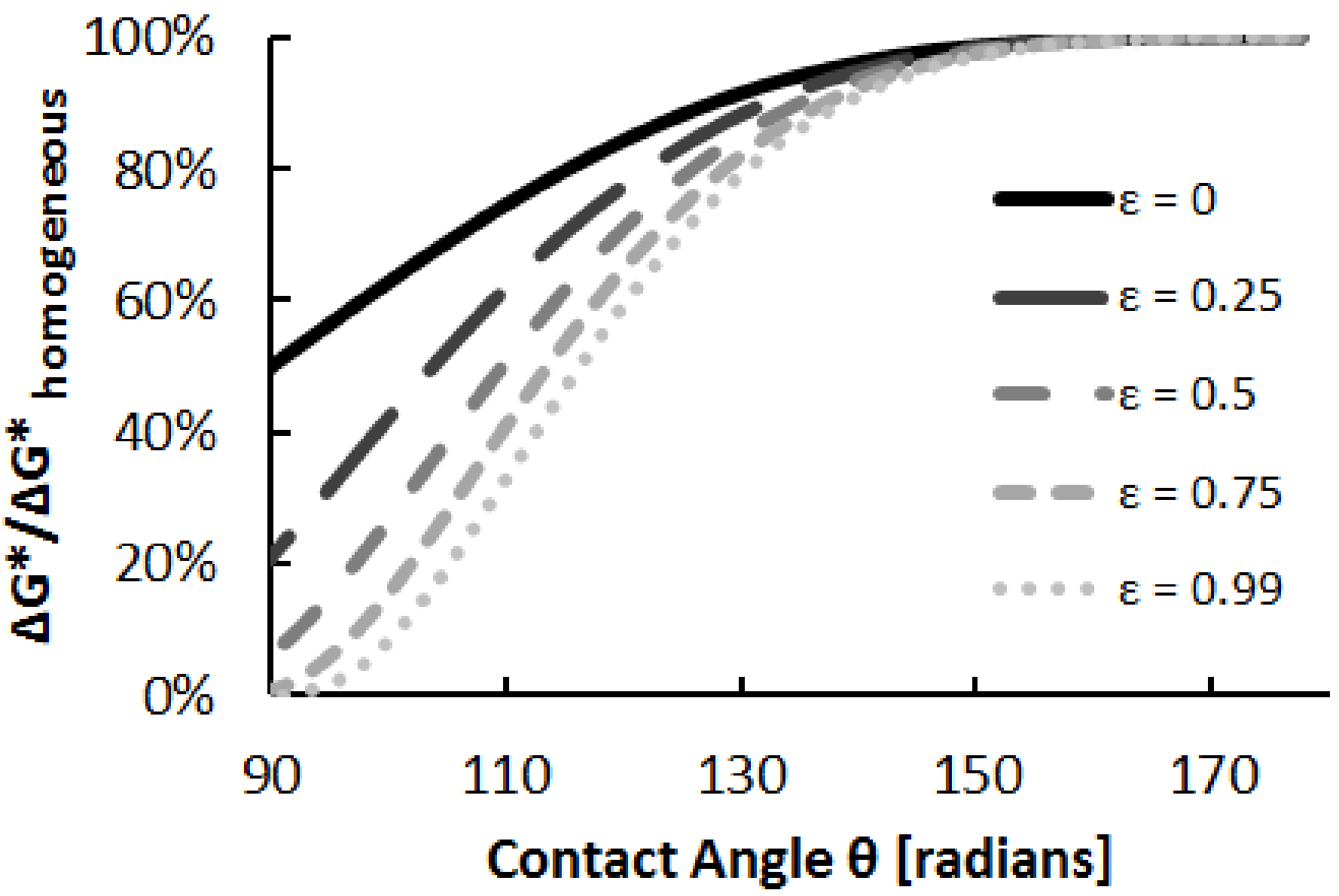

**Figure 12.** Plot of the energy of formation of a critical nucleus versus substrate-crystal-liquid contact angle (equation 8.3) with curves for variable porosity ε.

This equation and figure show that for larger contact angle, the Gibbs free energy required to nucleate on the membrane surface is increases, approaching that required for homogeneous nucleation in the bulk liquid. Physically, this means that MD membranes with large contact angles are not really favored as a location for nucleation. Heterogeneous nucleation, however, can still occur on any small foreign particles in the bulk liquid, and it would be favored over either homogeneous nucleation in the bulk or heterogeneous nucleation on the membrane surface. While this contact angle of substrate-crystal-liquid is hard to measure, it is closely linked with the substrate-liquid-air contact angle that defines hydrophobicity. Both angles depend on the magnitude of the surface free energy term of the substrate: for a lower surface free energy, the higher both the nucleation energy barrier $\Delta G^*_{heterogeneous}$ and hydrophobicity are. [257]. An explanation for this link can be found using the Van Oss–Chaudhury–Good approach, acid–base theory [257]. This means that the superhydrophobic surface is expected to reduce surface nucleation.

Nucleation induction time is exponentially dependent on the Gibbs free energy barrier; small differences in this barrier can make nucleation much more rapid. The induction time for homogeneous nucleation is

$$t_{induction} = \frac{N}{A} exp\left(\frac{\Delta G^*}{2kT}\right) \quad (8.4)$$

where $N$ is the number of particles per unit volume, $k$ is the Boltzmann constant, $T$ is temperature in kelvin, and $A$ is a pre-exponential factor usually found through experiments [131]. Crucially, the air-water interface can induce heterogeneous nucleation [258, 259, 260, 261], which is expected because of the lower energy barrier. As a result, the presence of air, other surfaces, or colloids may cause heterogeneous nucleation. A fraction of the particles nucleating elsewhere may then deposit on the membrane surface, as seen in equation 8.2. This may foul MD systems in conditions where the induction time of the membrane and homogeneous nucleation would otherwise be hours or days, long enough to avoid fouling. Therefore, since air layers may block the membrane surface but encourage nucleation, they may either help or hinder fouling: the present study is the first to study both these competing effects.

## 8.2 METHODOLOGY

### 8.2.1 STATIC MEMBRANE DISTILLATION SETUP

A static MD setup was created to analyze the effect of introducing air layers to reduce fouling. This simple set up lacks water recovery like a full MD system; instead, it simply analyzes the effect of air layers on an MD membrane that separates a hot well-mixed saline solution from turbulent dry air. This design allows for typical MD conditions on the feed side, while allowing for rapid results and more precise weight measurement of foulants.

In a static MD setup (Figure 13), the MD membrane rests on the surface of a beaker, with various foulants in the water. A stirrer keeps the solution well mixed, and a hot plate with temperature controls keeps it at a constant temperature. A fan situated one meter away from the apparatus blows arid air over the system. Fouling was observed by weighing the membrane before and after running the experiment. After the experiment, the weight was measured twice to account for water remaining on the membrane: first, a few seconds after the experiment to weigh water left clinging to the membrane, and again a few hours later to measure the dry weight. The three weights were then compared, using the salt concentration of the water on the membrane surface, to determine how much weight of salt precipitated onto the membrane during the MD process itself and how much was added by foulants in water remaining on the surface.

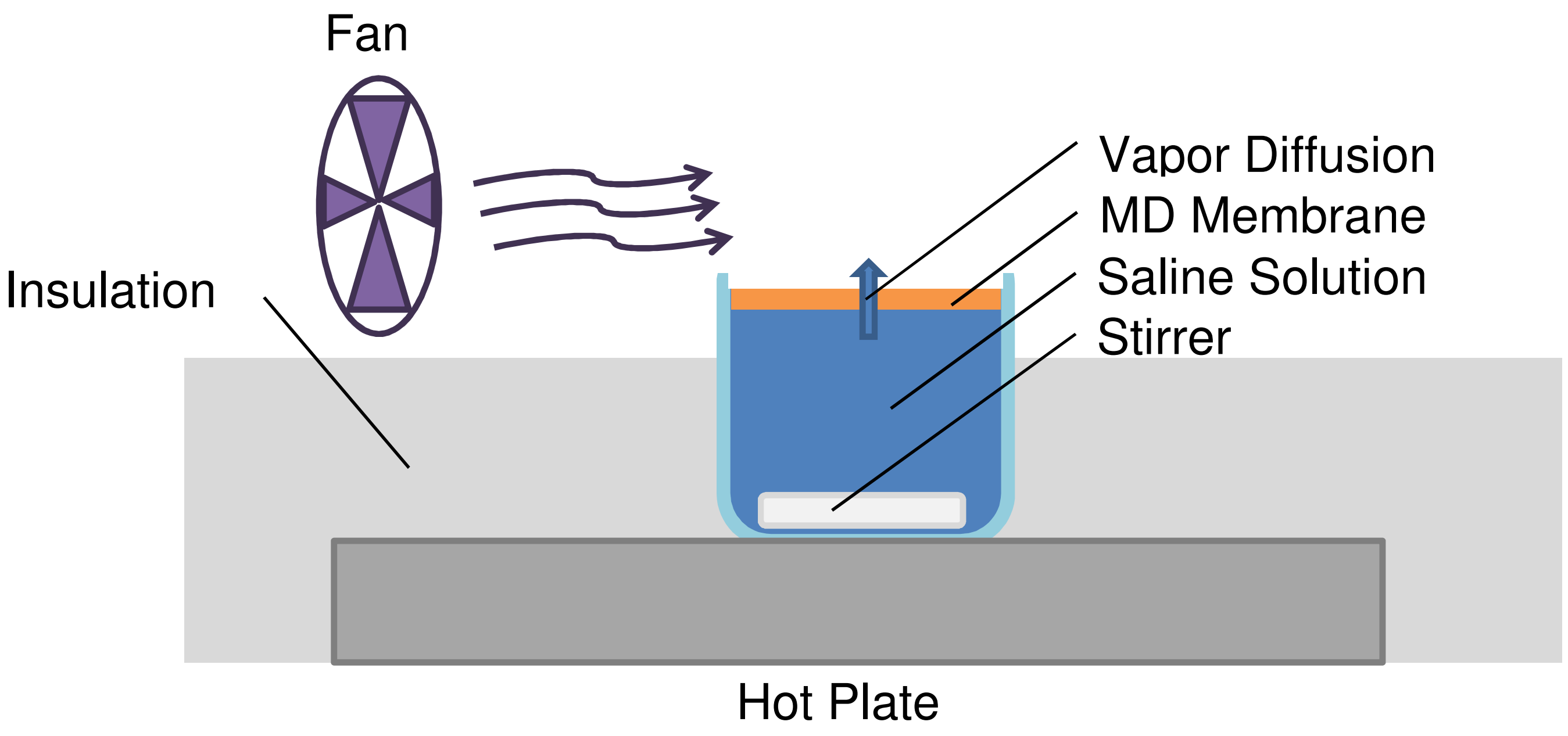

**Figure 13.** The hot Experimental setup of evaporation and scaling through a MD membrane

The experiment took place under a fume hood with consistent conditions between trials. A humidity meter was placed inside plate was well insulated with a 2.5 cm thick foam insulation sheet so that it would not significantly warm the air circulated by the fan. For temperature readings, measurement was done with a handheld Omega Microprocessor Thermometer, Model HH23, with a type J-K-T thermocouple. Humidity was measured with an Avianweb Digital Mini Instant-Read Temperature & Humidity Gauge, part number B00U2S6JSC.

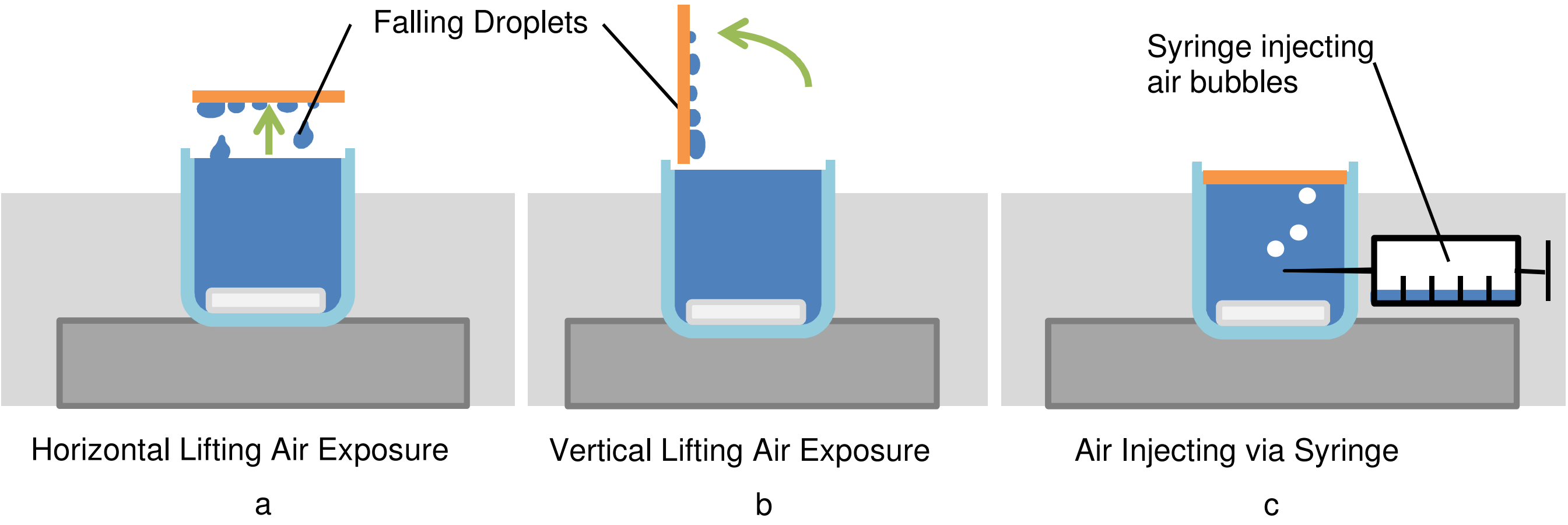

**Figure 14.** Methods for air recharging examined in the experiment

Several different methods of air recharging were tried. This ensured that variables outside the scope of study, such as the evaporation of water on the membrane surface during air recharging, did not affect the results. In the lifting methods (Figure 14 a and b), the membrane is gently lifted vertically, and the saline water rolls off the hydrophobic surface back into the solution. The vertical lifting case allowed more water droplets to roll off the surface than did the horizontal lifting. For the syringe method (Figure 14c), saturated air bubbles were injected periodically near the center of the device, 5 mL per injection. The syringe contained water to ensure saturation. The added air to the small 80 mL beaker caused the membrane to briefly lose touch with the surface.

The air recharging was done by periodically lifting the membrane off of the surface for a specific duration and frequency. A lift time of 10 seconds was used as this was sufficient for most of the water to drain off due to gravity, but not so long such that significant dry out by evaporation would occur on the membrane surface. A variety of configurations and lifting frequencies was examined to determine which was the most effective. A horizontal lift position (Fig. 3a) means that the membrane was lifted without any tilting, whereas a vertical lifting configuration (Fig. 3b) rotated the membrane 90°. Lifting was done with tweezers, with two ~0.5 mm small lift tabs cut into the edges of the original membrane, ~0.5 mm x 0.5 mm. The experimental cases considered are summarized in Table 8.1, and values of related variables are given in Table 8.2.

**Table 8.1.** Experimental conditions for air recharging of MD membrane for all trials performed

| Trial | Feed water conditions [% weight] | Lifting Frequency [min] | Air recharging | Test A | Test B | Test C | Test D |
|---|---|---|---|---|---|---|---|
| 1 | 20% NaCl | 10 | horizontal lifting | hydrophobic, control | hydrophobic, lifting | super-hydrophobic, control | super-hydrophobic, lifting |
| 2 | 20% NaCl | 5 | vertical lifting | hydrophobic, control | hydrophobic, lifting | super-hydrophobic, control | super-hydrophobic, lifting |
| 3 | 25% NaCl | 10 | vertical lifting | hydrophobic, control | hydrophobic, lifting | | |
| 4 | 30% NaCl | 10 | vertical lifting | hydrophobic, control | hydrophobic, lifting | super-hydrophobic, control | super-hydrophobic, lifting |
| 5 | 0.58% $CaSO_4$ | 10 | syringe | hydrophobic, control | hydrophobic, lifting | super-hydrophobic, control | super-hydrophobic, lifting |
| 6 | 15% Silica [262] | 10 | syringe | hydrophobic, control | hydrophobic, lifting | super-hydrophobic, control | super-hydrophobic, lifting |
| 7 | 0.04 % Alginate 0.029% $CaCl_2$ | 10 | syringe | hydrophobic, control | hydrophobic, lifting | super-hydrophobic, control | super-hydrophobic, lifting |

Several foulants were examined. NaCl and $CaSO_4$ are common salts that scale in desalination systems, silica is a common inorganic foulant, and alginate is a component of biofouling from algae, which forms a fouling gel layer. The concentrations of the salts were chosen to be supersaturated at the membrane surface, except for some of the NaCl cases. For the colloids Silica and Alginate, concentrations known to produce fouling in MD systems were chosen. All substances were reagent grade from Sigma-Aldrich, and prepared with DI water. All hydrophobic cases (tests A and B) used the PVDF membrane, and all superhydrophboci cases (tests C and D) used the same PVDF membrane coated with PFDA via iCVD. All superhydrophobic samples used were created during one iCVD experiment run, ensuring nearly identical properties.

**Table** 8.2. Experimental Variables.

| Variables | Symbol | Values | Uncertainty |
|---|---|---|---|
| temperature | $T_{f,in}$ | 60°C | ±3°C |
| humidity | $m_{f,in}$ | $30\%^{-1}$ | ±5% |
| condensate flux | $\dot{m}_p$ | 5 LMH | ±0.5 LMH |
| stirrer rotation | ω | 60 rpm | ±1 rpm |
| membrane area | A | 19.63 $cm^2$ | – |

### 8.2.2 Superhydrophobic Membrane Preparation and Testing

The hydrophobic membranes were commercial polyvinylidene fluoride (PVDF) membranes (Millipore Immobilon-PSQ, 0.2 μm pore size, part # ISEQ 000 10). The superhydrophobic membranes were prepared using the same PVDF membranes treated with a conformal coating of poly-(1H,1H,2H,2H-perfluorodecyl acrylate) (PPFDA). The coating was produced using initiated chemical vapor deposition (iCVD). iCVD of the PPFDA was conducted using a custom-built reactor using a process described previously [263]. iCVD of PPFDA has been previously used to create hydrophobic, conformal coating on membranes [264, 265].

PFDA monomer (97% Sigma-Aldrich) and t-butyl peroxide initiator (TBPO) (98% Sigma-Aldrich) were used without further purification. The monomer was heated to 80 °C and fed into the chamber at a rate of 0.03 sccm(standard cubic centimeter per minute). The initiator was kept at room temperature and was fed into the chamber at a rate of 1.0 sccm. The total pressure in the chamber was maintained at 45 mTorr throughout the deposition using a mechanical pump (45 CFM pumping speed, Alcatel). The reactor was equipped with an array of 14 parallel filaments (80% Ni, 20% Cr) resistively heated to 210 °C. The membranes were placed on a stage that was maintained at 30 °C using a recirculating chiller/heater (NESLAB). A 200 nm thick PFDA film was deposited after which the filaments were turned off and deposition was halted. The deposition rate was 1.8 nm/min.

Contact angles of water on the membrane surfaces were measured using a goniometer equipped with an automatic dispenser (model 590, Ramé-Hart). DropImage software was used to acquire images for measurement. A 3 μL drop of room-temperature DI water was first placed onto the membrane surface. The contact angle of the drop on the surface was measured at this time to determine the static contact angle. Water was then added to this drop in increments of 2 μL, and the angle between the advancing drop and the membrane surface was measured 1 second after each addition. The maximum of these measured angles was considered the advancing contact angle. Receding contact angle was measured by removing 2 μL of water at a time from the drop and measuring the angle between the receding drop and the surface 1 second after each removal. The lowest value observed was the receding contact angle. Measurements on at least five locations on each membrane were taken and averaged.

Air permeability was measured using a custom setup. A syringe-pump (PHD 22/2000, Harvard Apparatus) was used to push room-temperature air through a membrane held in a membrane holder (GE healthcare biosciences) at a rate of 210 mL/min. While the air was being pushed through the membrane, the pressure difference across the membrane was monitored using a USB pressure transducer with a precision of +/- 0.03 kPa (PX409, Omega). This pressure difference was used to calculate the permeability to air of the membranes.

Scanning electron microscope (SEM) images (JEOL 6010a) and a porosity test were used to verify the conformity of the coating, and SEM was also used to study scaling after the tests.

## 8.3 Results and Discussion

### 8.3.1 Superhydrophobic membranes and their properties

The effect of the PPFDA coating on the PVDF membranes was determined by measuring contact angles and air permeability and taking SEM images of the membranes before and after coating. As expected, the PPFDA coating increased the hydrophobicity of the membranes (Table 8.3), Advancing and static contact angles increased from hydrophobic to superhydrophobic (increases of 11° and 22° respectively). Receding contact angle had the most significant increase, transforming from less than to greater than 90° (an increase of 78°). These results were more dramatic than expected considering that both PVDF and PPFDA are fluorinated polymers. However, PPFDA has a higher concentration of fluorine than PVDF which likely increases its static and advancing contact angles. Its sidechains also form a semi-crystalline structure which prevents the fluorine atoms from orienting away from water after contact. This feature may explain why the PPFDA-coated membranes have hydrophobic receding contact angles unlike the uncoated PVDF membranes. For both membranes, the contact angles measured on the membranes are affected by the roughness of the surface and are significantly higher than the same chemistry would be if measured on a flat surface.

Air permeability was measured (Table 8.3) and SEM images of the membranes were taken before and after coating (Figure 15) to verify the conformity of the coatings. The minor change in air permeability (< 7.5%) suggests that the coating did not significantly alter the total porosity or pore structure of the membranes. The hydrophobicity of the membranes after iCVD coating was also further illustrated by observation of an air layer during submersion in water (Figure 16).

**Table 8.3.** Summary of MD membrane properties used in this study. The membranes are commercial PVDF membranes used as received. The membranes are the PVDF membranes coated with PPFDA using iCVD.

| Membrane | Air permeability (kg/m2-Pa-s) | Advancing contact angle (°) | Static contact angle (°) | Receding contact angle (°) |
|---|---|---|---|---|
| Superhydrophobic | 2.96E-06 | 156 | 157 | 134 |
| Hydrophobic | 3.19E-06 | 145 | 125 | 56 |

For each angle measurement, 4 or 5 trials were performed, with a standard deviation between 2.1° and 11.9°. However, it is clear that the coated membranes have significantly higher receding contact angles than the uncoated membrane. Both membranes had a porosity of 80%.

**Table 8.4.** Selected trials for the measurement of MD membrane contact angles

| | | Standard deviation | | |
|---|---|---|---|---|
| Membrane | # Trials | Static contact angle (°) | Advancing contact angle (°) | Receding** contact angle (°) |
| Superhydrophobic | 4 | 4.39801 | 2.894823 | 2.124461 |
| Hydrophobic | 5 | 8.10728 | 11.862209 | 8.193107 |

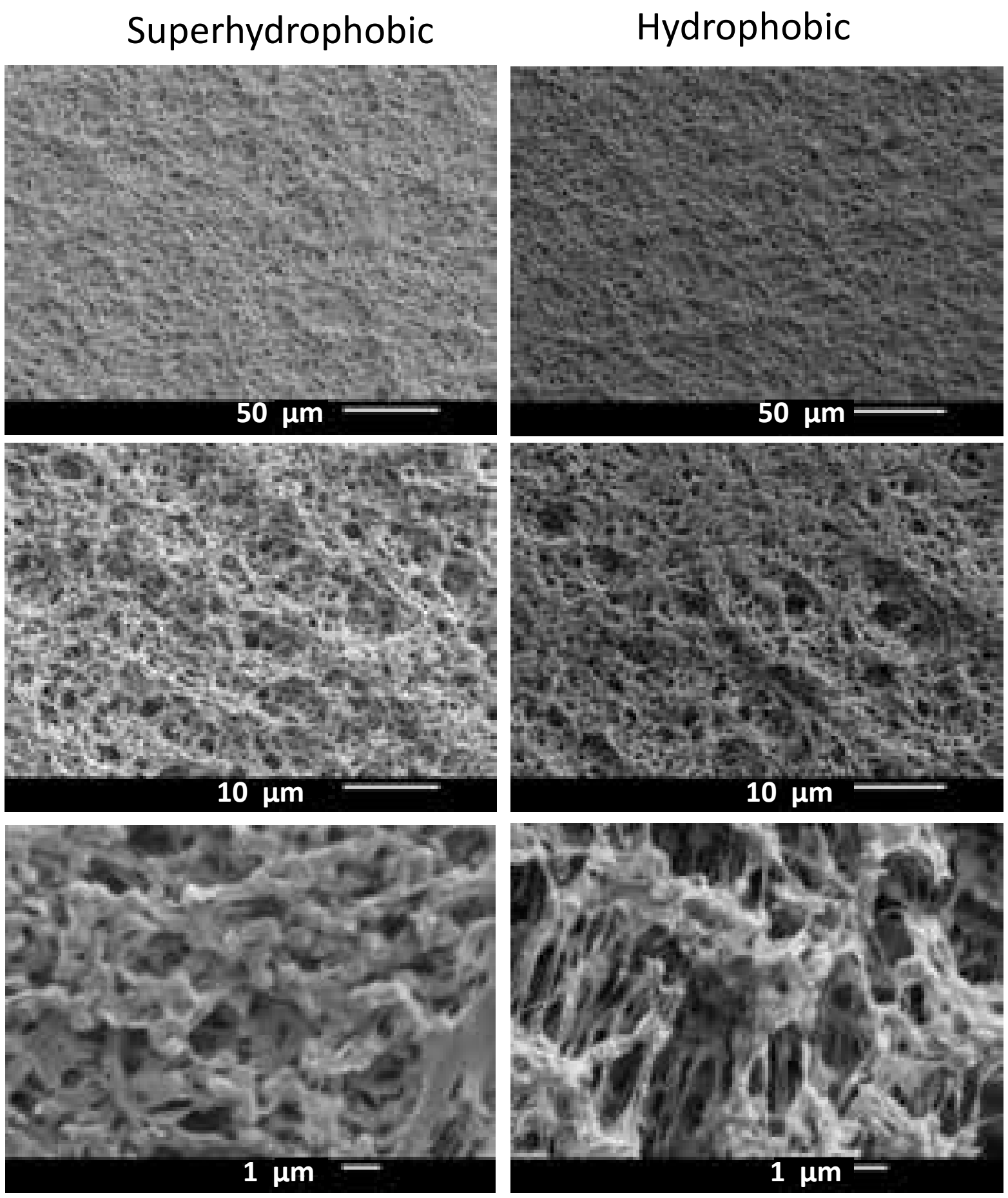

**Figure 15.** SEM images of MD membrane surface. The coating layer of PPFDA deposited by iCVD on the membranes does not significantly decrease porosity, and does not drastically change the membrane surface structure.

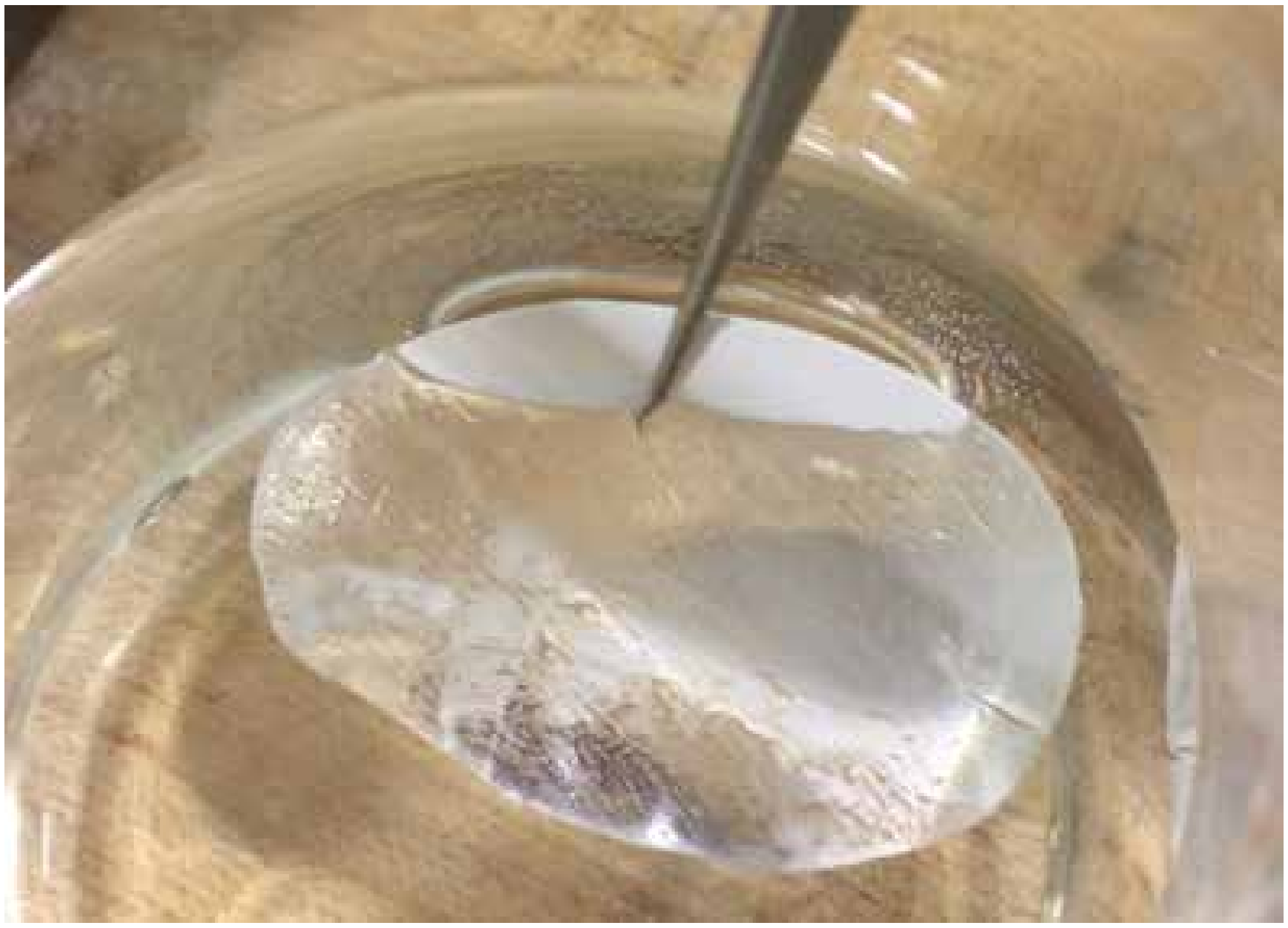

**Figure 16.** Photograph of submerged superhydrophobic MD membrane. The membrane is visibly shiny due to the thin air layer on its surface.

### 8.3.2 Scaling Results

The effectiveness of the air recharging and membrane superhydrophobicity were measured by the mass of salt adhered to the membrane after the experiment. SEM was performed as well for select cases. The salinity is characterized by the Saturation Index (SI), a log scale with 0 being saturated. The concentration was determined by measuring salt added with a mass scale, and the saturation concentration was calculated using the software PHREEQC (version 3) by USGS.

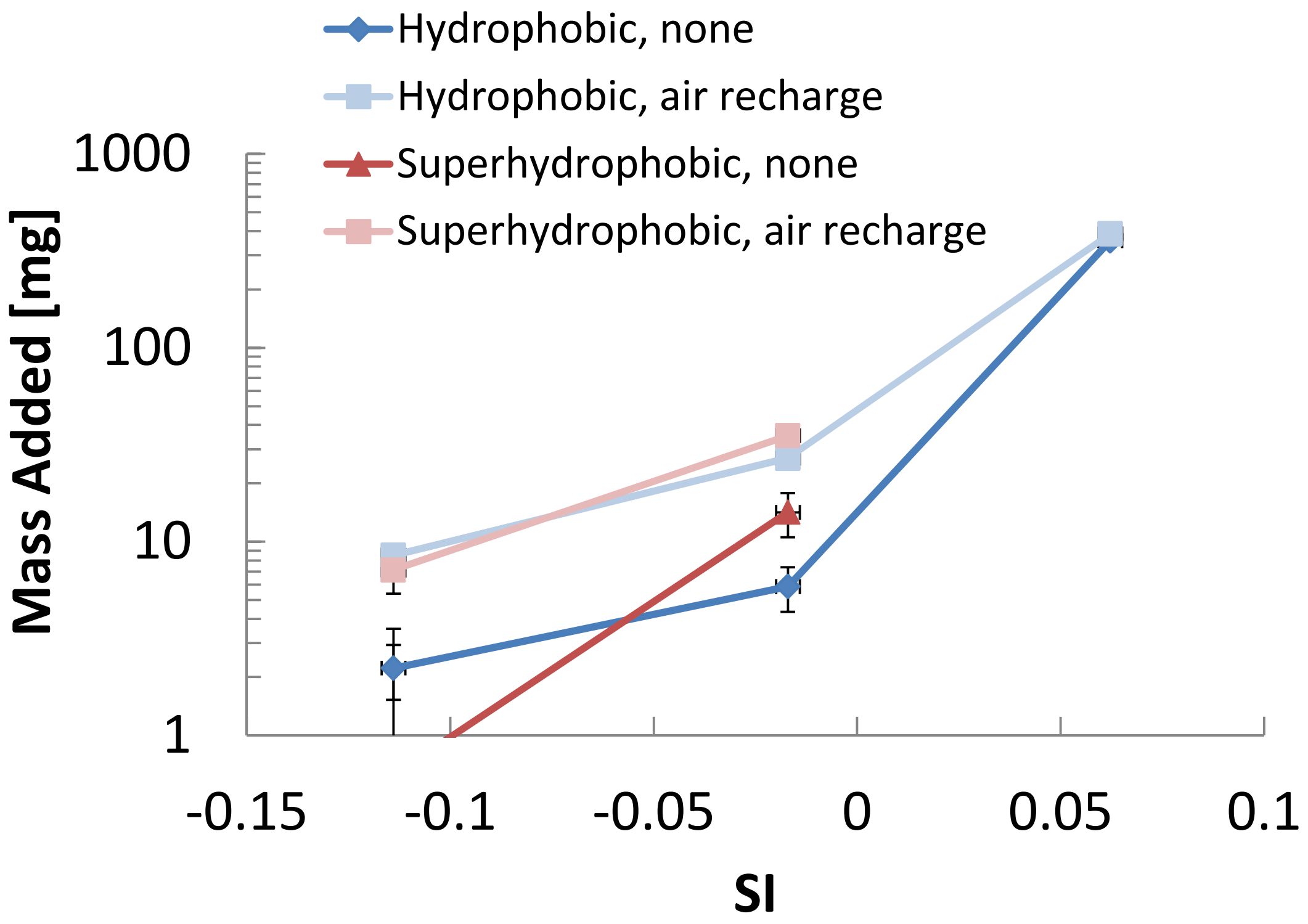

**Fig. 17.** SI vs Mass Added, NaCl, horizontal lifting. (Trials 1, 3 and 4)

In the case of NaCl scaling, shown in Fig. 17, the air recharging trials consistently had more salt adhered to the membrane. As the water contains a high fraction of NaCl, water evaporating off the membrane after the experiment was over (dry out) left significant amounts of NaCl on the membrane, increasing the overall error.

Notably, the total salt adherence for the trials with added air had similar masses added, regardless of the membrane used. This may suggest that nucleation in the bulk, not on the membrane itself, dominated over any nucleation or crystal growth at the membrane. Bulk conditions do not depend on the membrane, and particles in the bulk may then precipitate on the membrane, increasing its mass. This aligns with the expected thermodynamics of MD systems with added air, where scaling on the hydrophobic membrane is relatively unfavorable but heterogeneous scaling at the air-liquid interface may be significant.

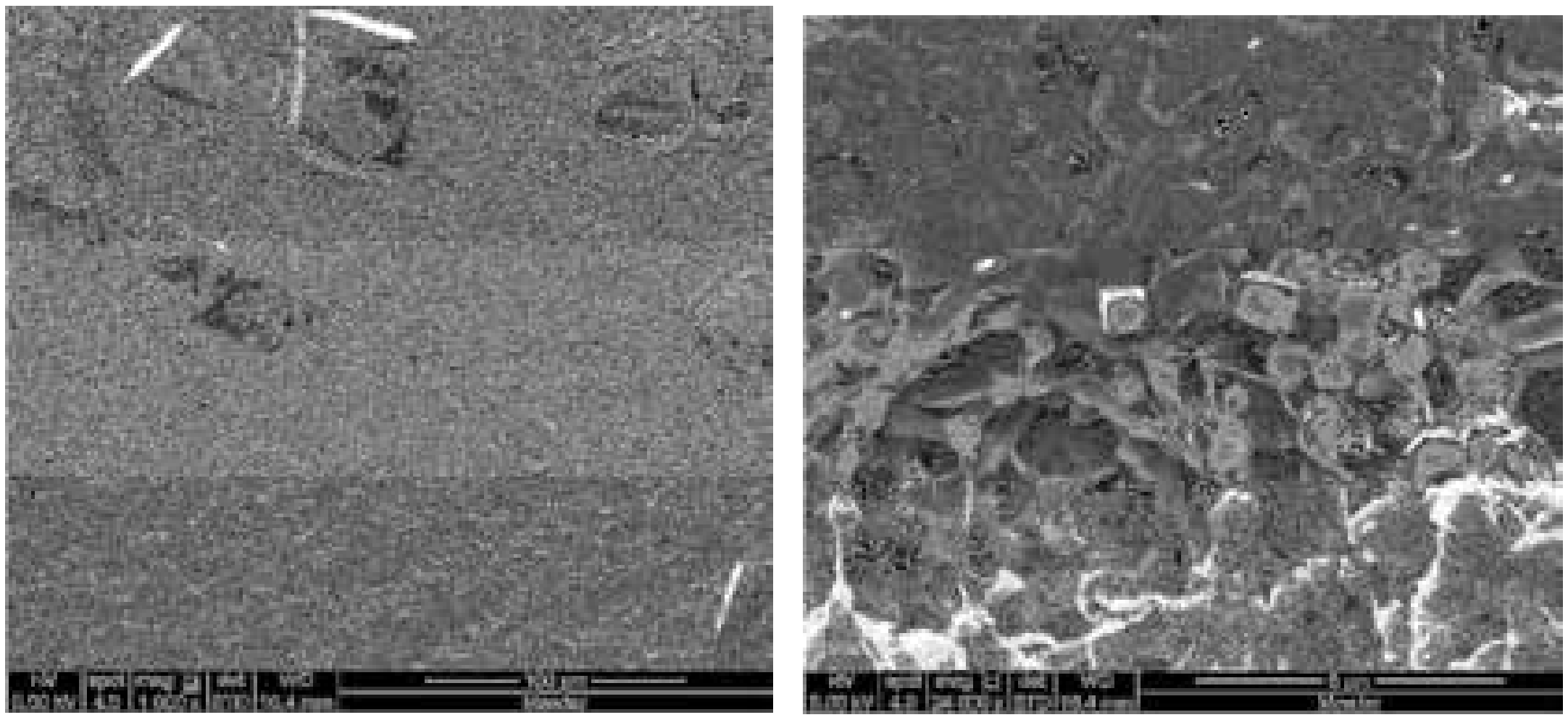

**Fig. 18.** SEM Images of the superhydrophobic PVDF membrane after air recharging, in a solution with 30% NaCl, with vertical lifting**.** (Trial 4)

NaCl salt crystals can be seen in Fig. 18. These crystals span multiple orders of magnitude. Notably, there were few crystals visible embedded deep into the pores, likely because of minimal wetting of the water into the superhydrophobic surface. This is very desirable, as crystals forming in the pores have been shown to cause wetting, contaminating the permeate [34]. Since crystal growth is much more thermodynamically favorable on the crystals themselves rather than the membrane, most of the membrane surface is not blocked by any crystal.

No differences were seen between the vertical and horizontal lifting methods of adding air (trials 1-4). Evaporation into the air during lifting may leave salt crystals behind, but this can be avoided if the air is supersaturated and is at least at the temperature of the feed solution. The salt deposition was compared between the brief 10 second lifting periods, and with the syringe containing hot saturated air. Both trials showed that the air addition to NaCl exacerbated scale deposition, and repeated experiments repeated experiments confirmed that result. The salt deposition was reduced with the hot saturated air, but did not appear dominant as the mass added changed little. While having similar trends, the graph for the NaCl syringe experiment is not shown here, since that experiment's temperature varied significantly due to an improper heater setting.

The frequency of air recharging was varied as well (trials 1 and 2), but this did not show significant differences over the range of frequencies examined.

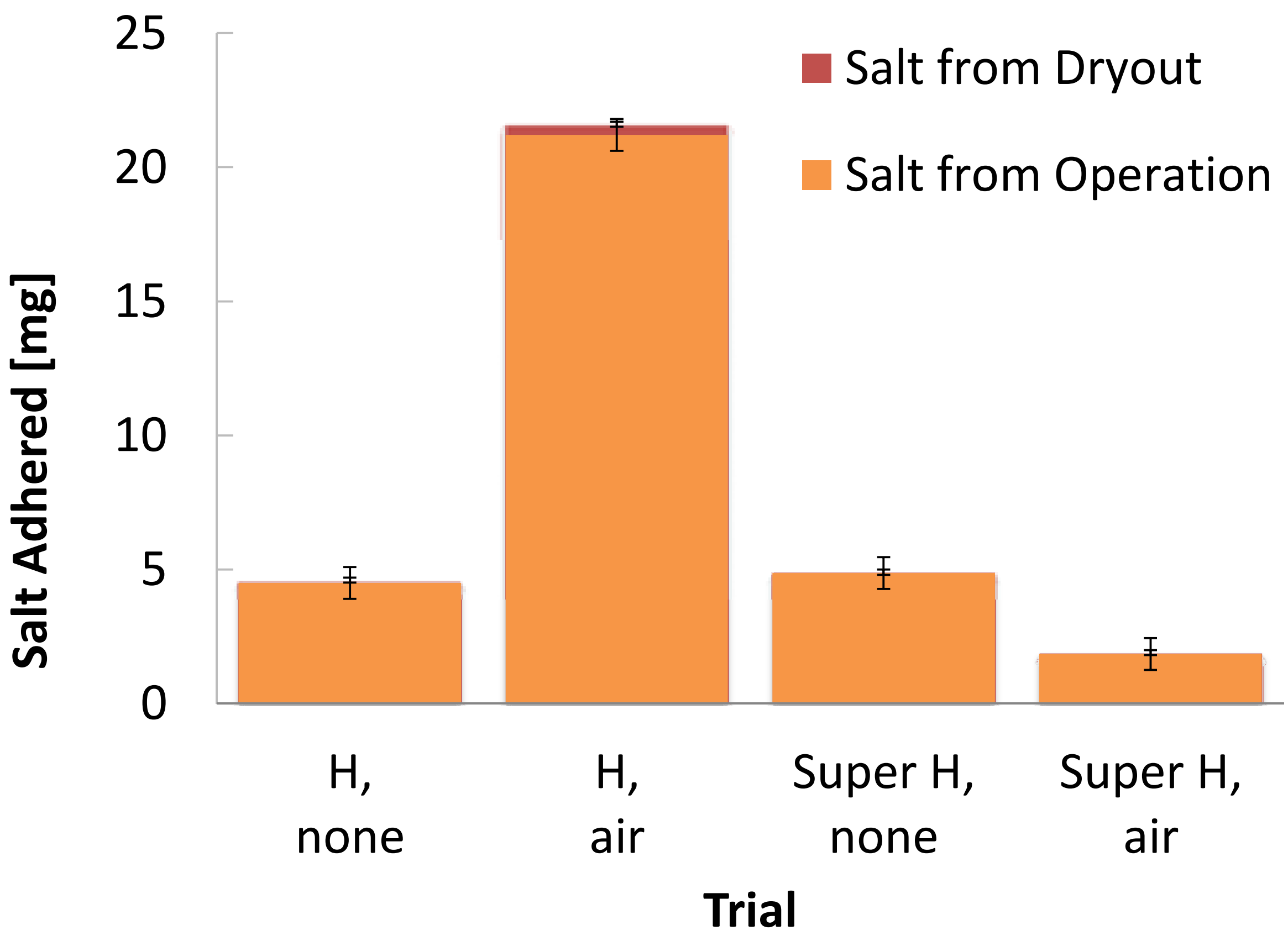

**Figure 19.** Mass of Salt Adhering to MD Membrane, $CaSO_4$ (Trial 5)

For the $CaSO_4$ trials, the introduction of air had a varied effect, as seen in Figure 19. The trials without air recharging did not vary substantially from one another, while the trials with air recharging did. The superhydrophobic membrane show a vast reduction in salt mass relative to the hydrophobic case. The superhydrophobic membranes were able to sustain a substantially thicker air layer, and the buffering effect of this layer may be the reason for the difference. Meanwhile, as in the NaCl experiments, the introduction of air may have caused heterogeneous nucleation at the interface, increasing the salt adherence in the hydrophobic case. Because of the low solubility of $CaSO_4$ in water and thus the small concentration, the calcium sulfate salt deposited while drying the membrane was minimal, reducing the uncertainties. The opposing fouling effects for superhydrophobic air recharging of the salts NaCl and $CaSO_4$ is striking, and it may be related to several factors. First, the induction times of the salts may differ: calcium sulfate is known for its particularly long induction times [126]. Calcium sulfate also tends to form long needle-like structures, compared to the more squat structures of NaCl. Other factors may also have an influence: less salt was deposited in the $CaSO_4$ trials, the crystal growth rates of the salt vary, and the salinity for the NaCl experiments is large enough to affect water properties such as surface tension, which the presence of crystals may affect as well [266]. Since the hydrophobic case of $CaSO_4$ with air layers also performed poorly, it is likely that the difference is not related to nucleation in the induced from the air, as that would have been the same. However, there is insufficient data to explain this trend adequately, so no speculations are made here.

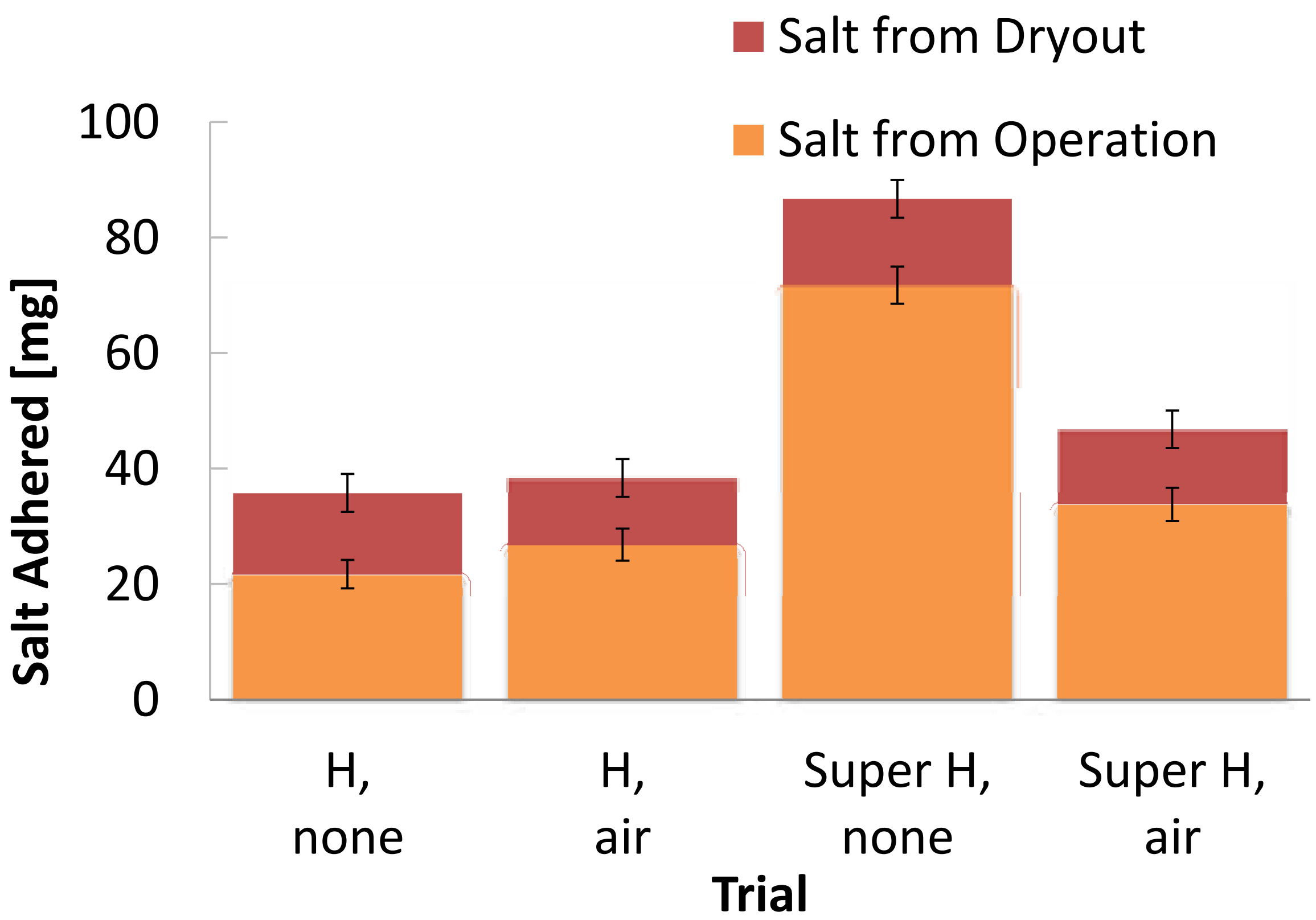

**Fig. 20.** Weight of Salt Adhered to MD Membrane, Silica (Trial 6)

For the case of Silica, the superhydrophobic membrane faired worse, but improved with air layers. However, this improvement still had similar and slightly worse performance than the control.

For the deposition of Silica, generally, polymerization by dehydration occurs, including cross-linking and aggregation by Van Der Waals forces, creating negatively charged colloids. The aggregation leads to gels on the membrane surface [267]. Only in this trial did the superhydrophobic surface perform worse that the control surface. Perhaps the surface plays a role in the coalescing steps. Alternatively, superhydrophobic surfaces are known to have a lower charge density [268], and thus may repel these negatively charged colloids less, causing a relative increase in colloidal adherence to the membrane.

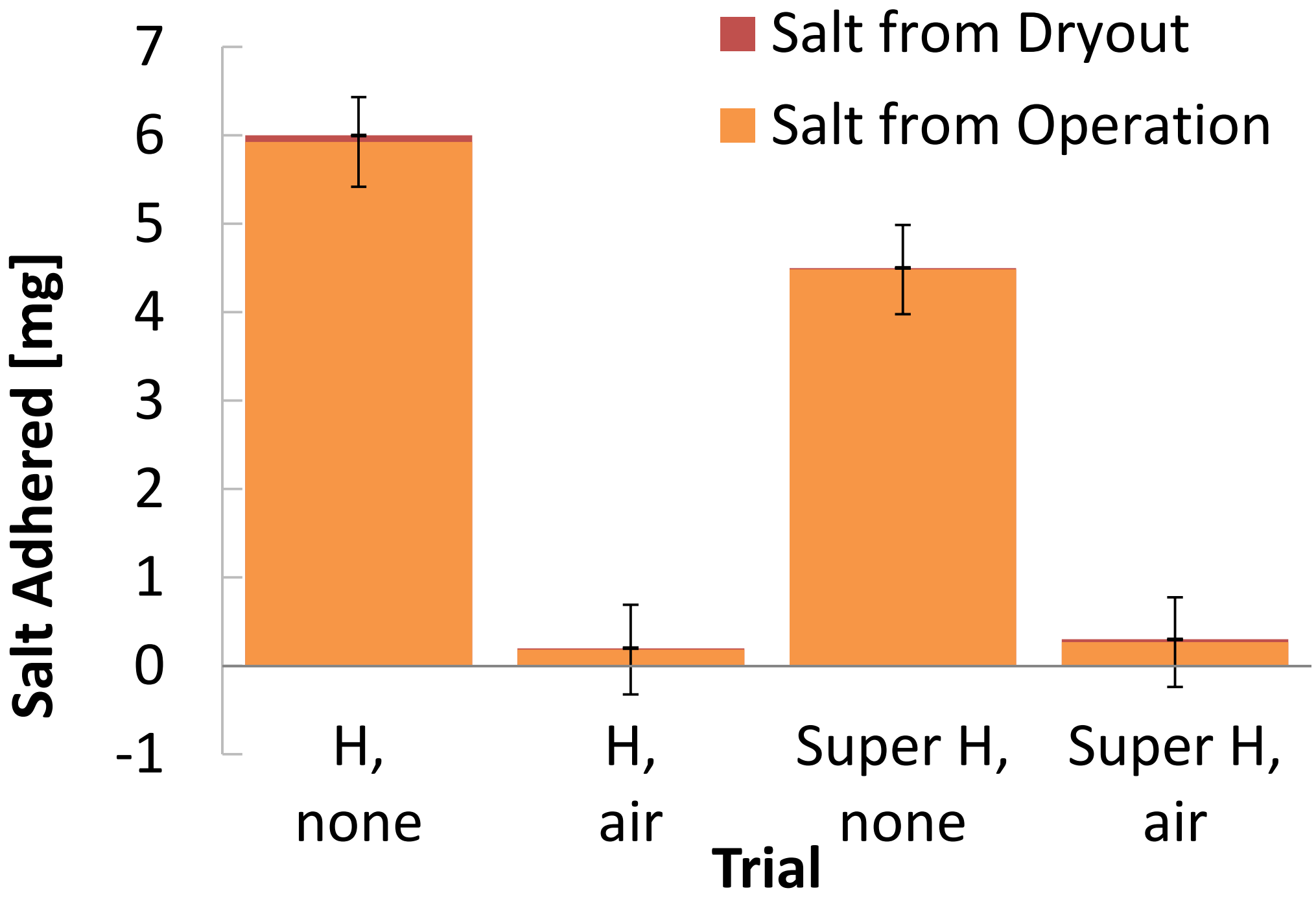

**Fig. 21.** Weight of Salt Adhered to MD Membrane, 0.04% Alginate, $CaCl_2$ (Trial 7)

The introduction and maintenance of air layers on the membrane surface caused a profound and consistent reduction in biofouling of alginate. Alginate, which forms a gel in the presence of calcium [269], does not follow classical crystalline nucleation. These large molecules are already colloids, and so do not have a prolonged induction time like nucleating salts [270]; given sufficient minimum concentration, gels form in a matter of seconds. Therefore, the induction of nucleation by air layers was not a factor. The air layers were thus able to reduce the contact area and adherence of the gel, substantially reducing fouling.

Notably, while $Ca^{2+}$ is part of the gel, no scaling of $CaCl_2$ occurred here, as it is extremely soluble and was orders of magnitude below saturation concentrations. This result shows that air layers may be helpful in reducing biofouling in MD, which is particularly important for MD because the hydrophobicity of MD membranes tends to make them oleophilic [11].

As discussed in the introduction, nucleation thermodynamics indicate that the air-water interface encourages heterogeneous nucleation more than hydrophobic microporous membranes do [256, 131]. Furthermore, several experiments in the literature have shown the presence of air interfaces to encourage nucleation [258, 259, 260], and [261]. Nucleation is the initial step, and typically the limiting step for salt scaling, although does not describe gel formation of alginate biofilms [251].

Some other notable trends were observed as well. Solutions with a larger mass fraction of foulant had larger masses adhered to the membrane. In fact, the total average mass added, when ordered, has the same ordering as that of the mass fraction of foulant: NaCl, Silica, $CaSO_4$, and finally Alginate. Solutions with a smaller mass fraction of foulant also benefitted more from the presence of air layers. A related trend: the foulants showing the biggest reduction in the present of air layers were the ones that were the least soluble.

A summary of the results is shown given in Table 8.5.

Table 8.5. Summary of the effect of air layers and superhydrophobicity on mass of foulant left on the membrane, compared to the hydrophobic control

| | Effect of Air Layers | Effect of Super-hydrophobicity | Effect of Both |
|---|---|---|---|
| NaCl | +282-359% | -10 to +141.% | +22-500% |
| CaSO4 | +371% | +8.2% | -59% |
| Silica | 23% | 230% | 55% |
| Alginate | -96% | -24% | -95% |

color key

strongly reduced fouling

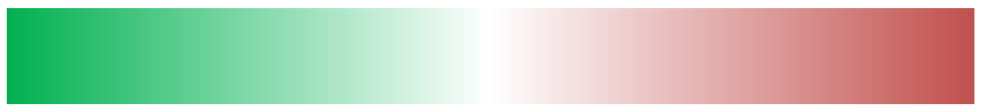

strongly increased fouling

## 8.4 CONCLUSION

The introduction of air layers had significant but varied effects, depending on the foulant studied. In the case of NaCl in all cases, and for $CaSO_4$ with a less hydrophobic membrane, air layers worsened fouling, increasing the amount of mass left on the membrane. However, biofouling, studied with Alginate and calcium ions, had the opposite effect. The air layers reduced fouling by as much as 96%. The introduction of air layers also reduced fouling for superhydrophobic membranes paired with calcium sulfate and colloidal silica. However, the improved value for silica was still worse than the control membrane. The anti-fouling effects and exacerbating-fouling effects can be explained by two separate mechanisms:

The presence of air layers reduces the membrane area in contact with the solution, preventing fouling on hydrophobic and especially superhydrophobic surfaces. These results indicate that maintaining air layers on MD membranes can dramatically reduce biofouling, but with varied and often detrimental results to preventing inorganic scale. This may suggests that studies from the literature encouraging air bubbling to reduce fouling by concentration polarization reduction should be viewed with caution.

A second mechanism explains the observed increased fouling for salts but decreased biofouling: nucleation of salts at the air-water interface. The thermodynamics of nucleation support this idea: the Gibbs free energy barrier for heterogeneous nucleation at interfaces is much lower than that of the membrane. Furthermore, numerous studies in the literature have observed heterogeneous nucleation preferentially occurring at air-water interfaces. Another mechanism may worsen fouling as well: evaporation of salts during the introduction of air layers, which can leave crystals behind. Comparison tests between air sub-saturated and saturated with vapor yielded the same results, indicating that nucleation dominated evaporation in worsening fouling.

## 8.5 ACKNOWLEDGEMENT

This work was funded by the Cooperative Agreement between the Masdar Institute of Science and Technology (Masdar University), Abu Dhabi, UAE and the Massachusetts Institute of Technology (MIT), Cambridge, MA, USA, Reference No.

02/MI/MI/CP/11/07633/GEN/G/00. This work made use of the Cornell Center for Materials Research Shared Facilities which are supported through the NSF MRSEC program (DMR-1120296).

The authors would like to thank Allen Myerson, You Peng, Emily Tow, McCall Huston, and Grace Connors for their contributions to this work.

# Chapter 9. SUPPLEMENTAL MATERIAL

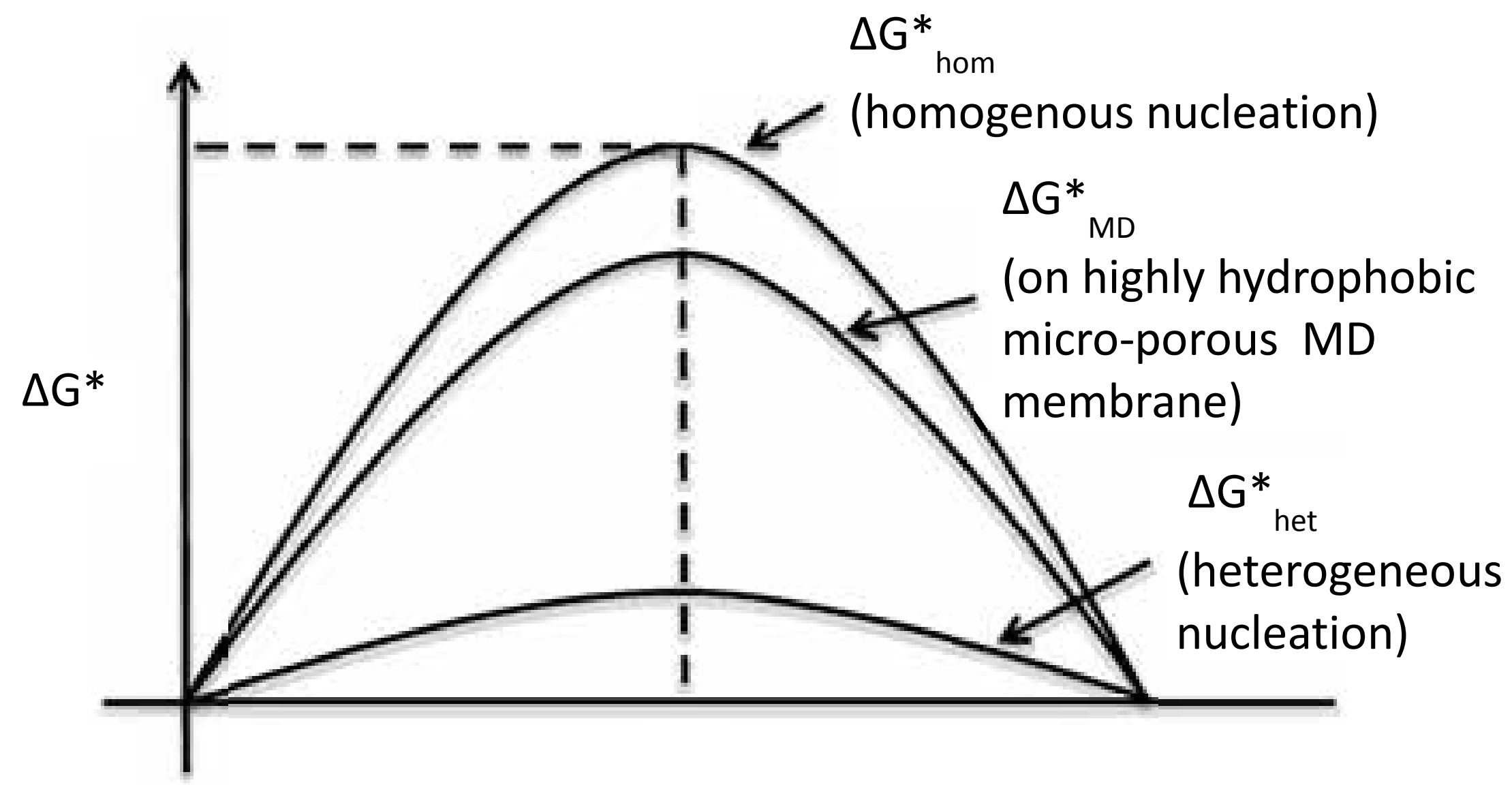

**Fig. 22.** Plot of the energy barrier of formation ΔG* versus r, the radius of minimum stable crystal nuclei, with examples of membrane and surface heterogeneous nucleation, adapted from (Ragone, 1994)

As seen in Fig. 22, the energy barrier for heterogeneous nucleation is much smaller than that of homogeneous nucleation. As the induction time is exponentially related to the Gibbs free energy to form a minimum stable nucleus, it can be expected that most nucleation will be heterogeneous, if surfaces, particles, or interfaces are present to cause it.

**Table 9.1.** Selected trials for the measurement of MD membrane contact angles

| | | Standard deviation | | |
|---|---|---|---|---|
| Membrane | # Trials | Static contact angle (°) | Advancing contact angle (°) | Receding** contact angle (°) |
| Superhydrophobic | 4 | 4.39801 | 2.894823 | 2.124461 |
| Hydrophobic | 5 | 8.10728 | 11.862209 | 8.193107 |

Three mass measurements are made of the membrane, and are numbered in chronological order. The mass of the membrane before the experiment is $M_1$. The mass of the membrane just moments after the experiment is ended (~30 seconds) is $M_2$, and includes mass added from water droplets adhered to the membrane. The third mass, $M_3$, is measured after all water has evaporated from the membrane surface (4+ hours after removal). The mass of foulant left by evaporation of adhered water droplets can be calculated from the first two masses, and the concentration of salt, as follows.

$$m_{evap,after} = \frac{M_2 - M_1}{1 - (M_2 - M_1)} C_x$$

The mass of foulant that sticks to the membrane during MD operation, $m_{stick}$, is calculated as follows:

$$m_{stick} = (M_3 - M_1) - m_{evap,after}$$

In the graphs presented, $m_{stick}$, is shown in orange, and $m_{evap,after}$ is shown in red.

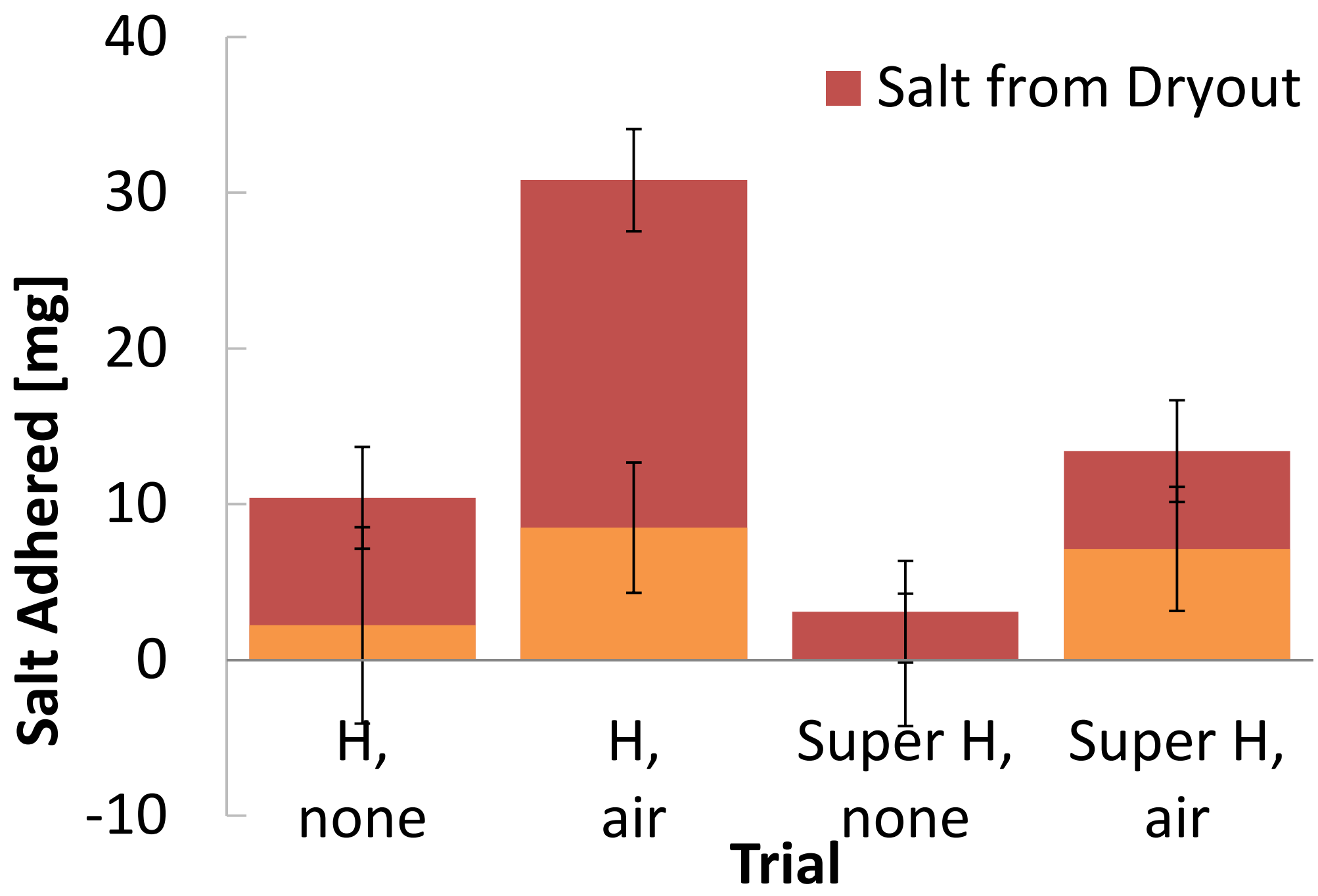

**Fig. 23.** Weight of Salt Adhered to Static MD Membrane, 20% NaCl (Trial 1)

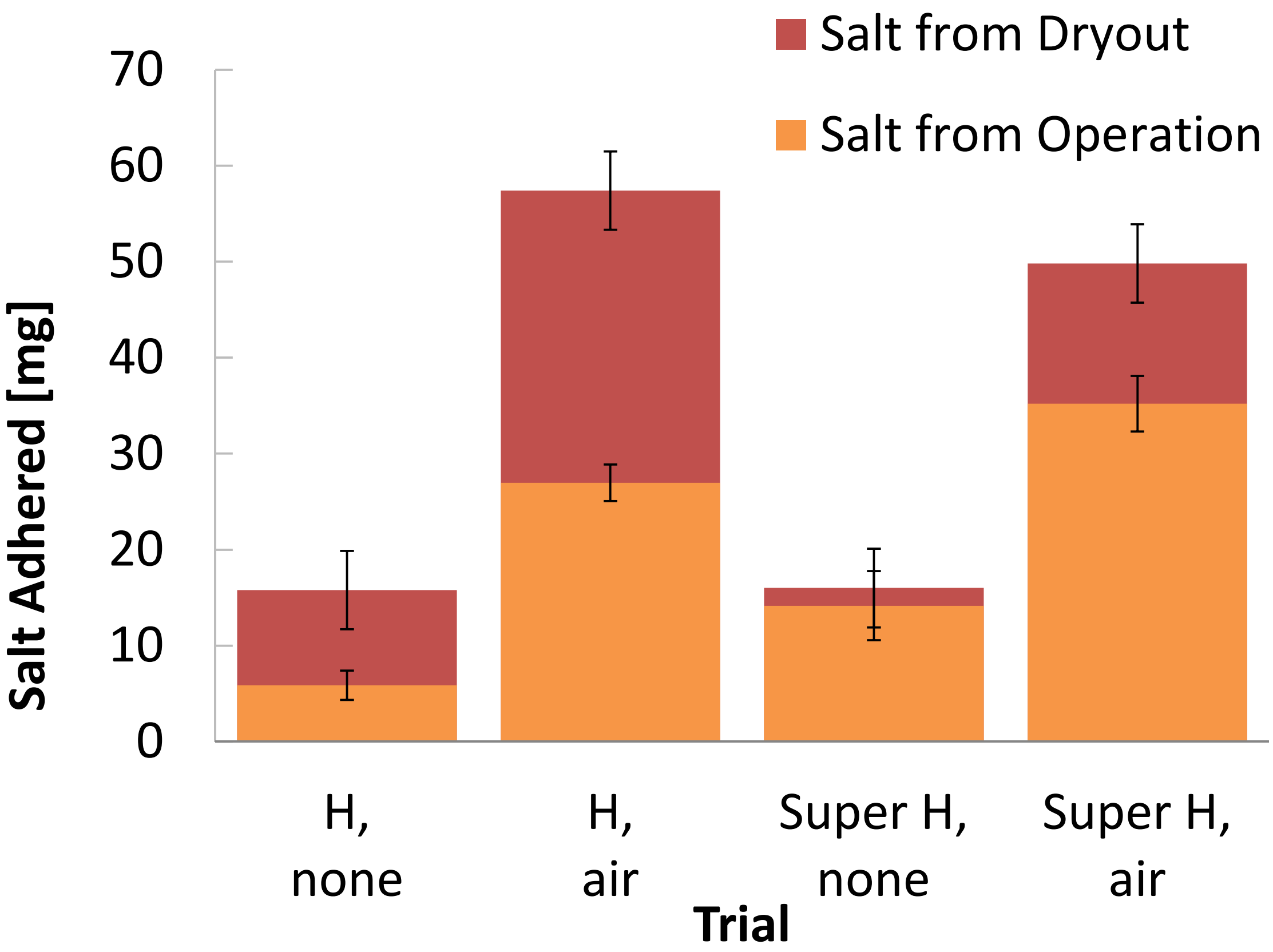

**Fig. 24.** Weight of Salt Adhered to MD Membrane, 25% NaCl (Trial 3)

The salt from operation refers to precipitation while membrane distillation is ongoing, while dryout refers to salt left from adhered water, which dries after the experiment is over. As seen in Fig. 23 and Fig. 24, significant salt was left by dryout (shown in red). This is significant due to the very high solubility of NaCl in water, making the mass fraction of the water very high (e.g. 25%). As a result, water evaporating off the membrane after the experiment contains significant salt, leaving a large mass behind in dry out.

# Chapter 10. MD Fouling Contributions Not As First Author

## 10.1 Effect Of Scale Deposition On Surface Tension Of Seawater And Membrane Distillation

Membrane distillation desalinates water with the principle that only water vapor, not liquid water, can pass through the microporous hydrophobic membrane. A membrane's ability to prevent saline water contaminating the feed is described by the liquid entry pressure (LEP), which is the pressure difference needed across the membrane to force saline feed through. Highly concentrated solutions are known to foul MD membranes, which can lead to wetting of the feed through the membrane into the permeate, especially with low LEP. Previous studies have not examined the effect of surface tension on LEP, which can be significant, and may vary since the surface tension of seawater decreases at high temperature and with high concentration of salts.

Kishor Nayar is first author on this work, and Jaichander Swaminathan, Divya Panchanathan, and Professor John Lienhard V are coauthors [266].

LEP is a function of the membrane hydrophobicity, feed liquid surface tension and the size of pore. LEP can also be defined locally at different regions of the membrane with the overall LEP being the least among these local values. Using an idealized model of the MD membrane consisting of circular pores, the LEP of the membrane can be written in terms of the liquid-vapor surface tension, the contact angle between the membrane and the liquid, and the membrane pore diameter as:

$$LEP = \frac{-4B_g \gamma \cos\theta}{D_{\max}} \tag{10.1}$$

where $B_g$ is a geometric factor for pores ($B_g$ = 1 for cylindrical pores), $\gamma$ is the solution's surface tension, $\theta$ is the contact angle between the membrane and solution (typically $\theta \geq \pi/2$), and $D_{max}$ is the membrane's maximum pore size.

The decrease in LEP leading to breakthrough has been related to changes in contact angle $\theta$ and pore structure resulting from scaling, whereas the effect of surface tension variation during operation has not been investigated.

To understand the effect of surface tension on MD, Wilhelmy plate surface tension measurement experiments were performed, and were paired with AGMD tests with fouled and unfouled membranes. It was hypothesized that the dramatic drop in surface tension under hot supersaturated conditions was related to precipitation of salts on the Wilhelmy plate surface, and therefore the presence of salts on an MD membrane would similarly reduce surface tension and thus LEP.

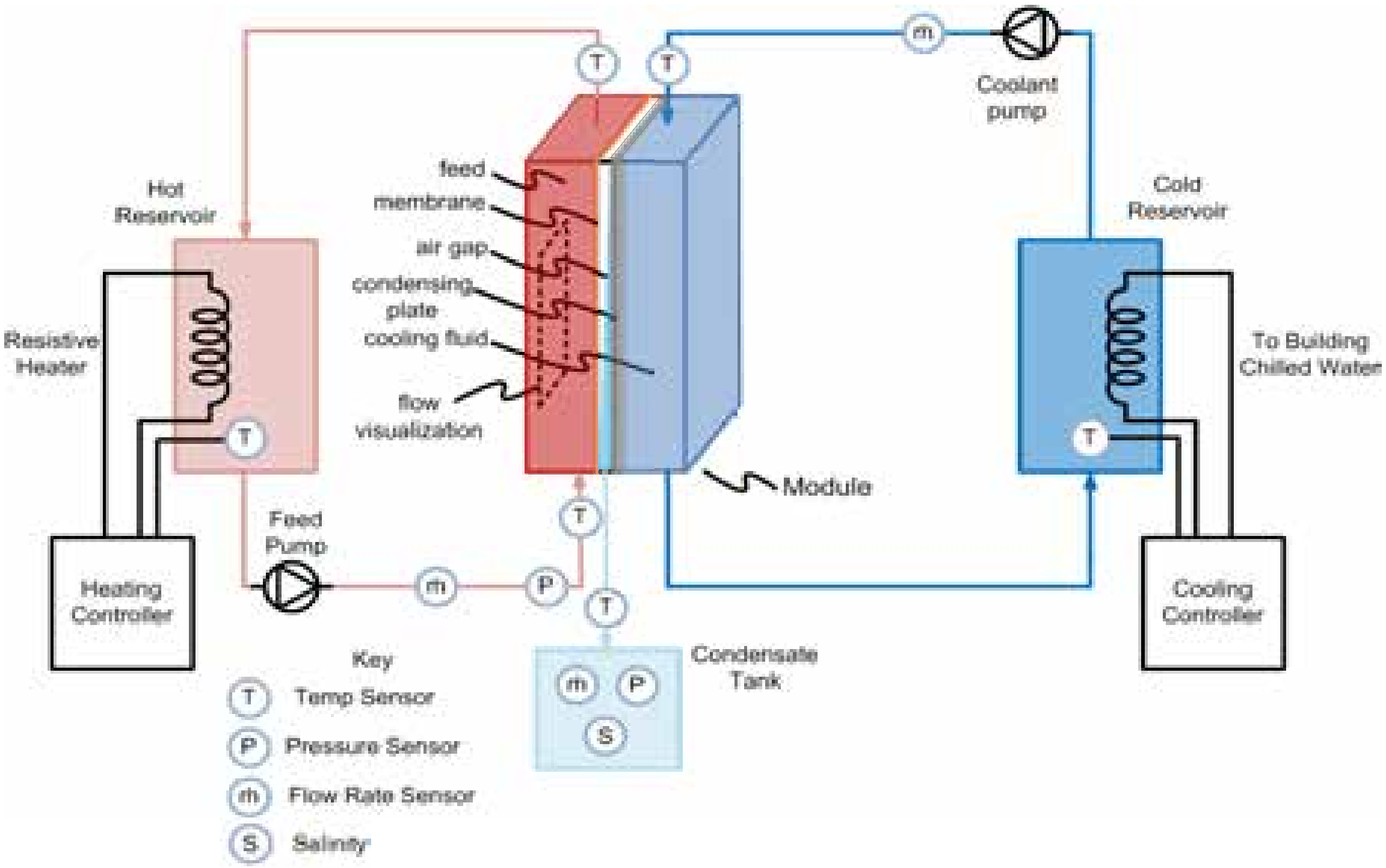

**Figure 10.1.** Membrane distillation fouling system diagram

The Wilhelmy plate seawater tests were done for 1°C to ≤ 92 °C and salinities between 0 and 131 g/kg. The AGMD tests were performed with a feed salinity of 120 g/kg and temperature of 70 °C.

(a) (b)

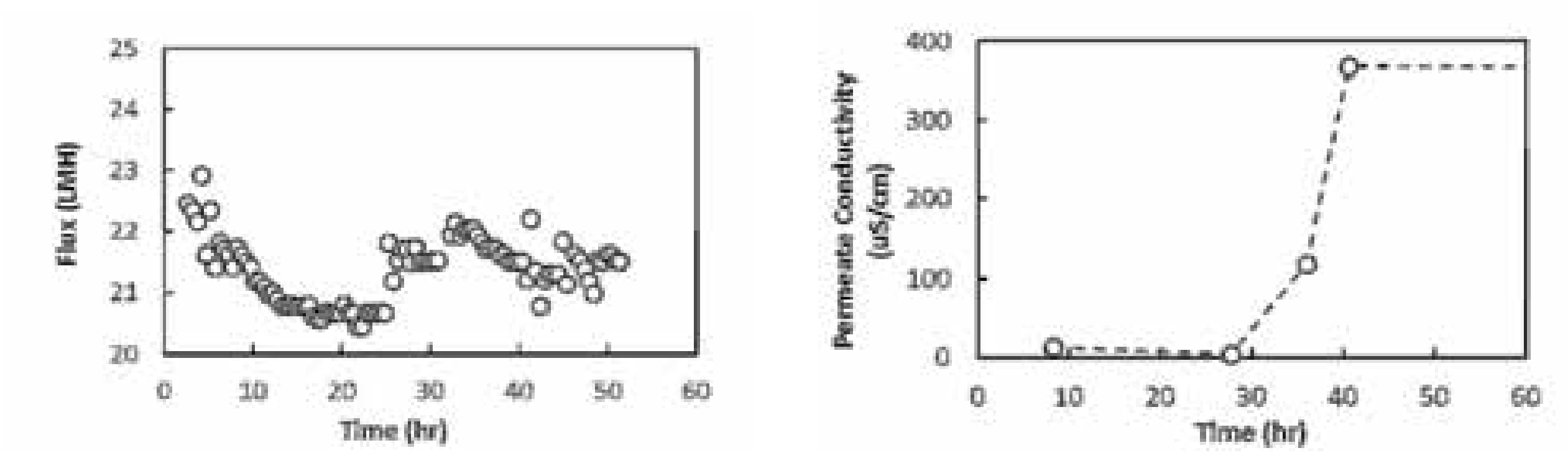

**Figure 10.2.** Results from AGMD fouling test with supersaturated $CaSO_4$ solution. Permeate conductivity increase observed following flux increase

(a) (b)

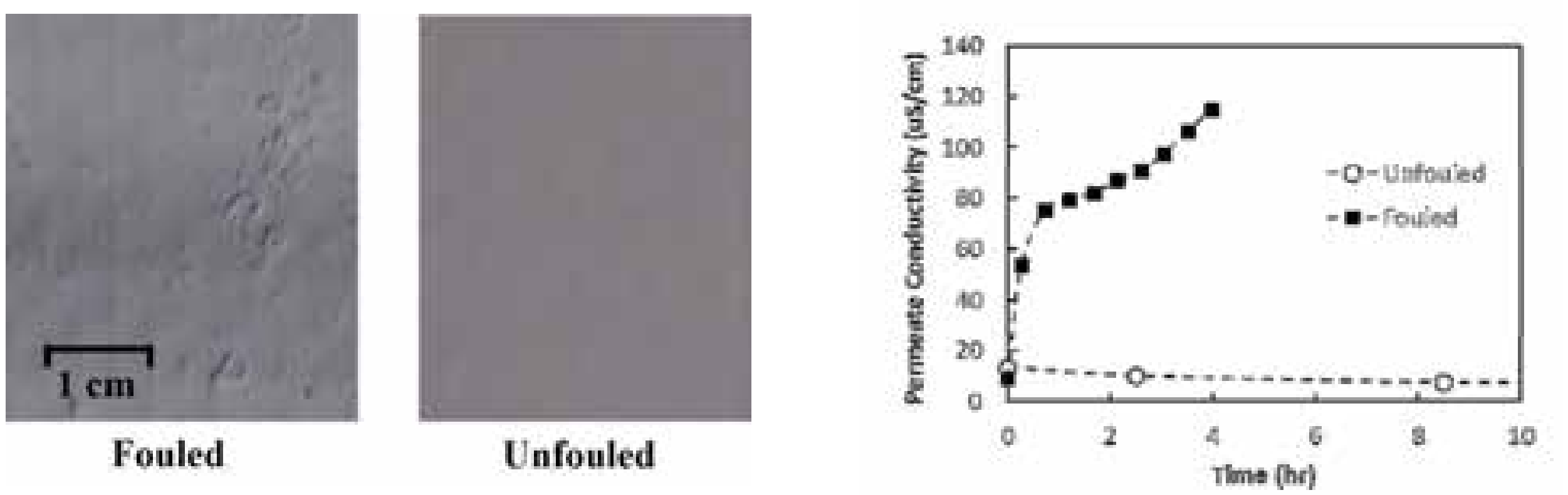

**Figure 10.3. .** Comparison of salt rejection by unfouled membrane and membrane using *S* = 120 g/kg seawater brine as feed at *t* = 70 °C with $CaSO_4$, $CaCO_3$ and NaCl cake deposits

The study showed that dramatically decreased salt rejection occurred when salt layers were left on the membrane surface. This occurred under conditions where the temperature and concentration reduced the surface tension by nearly 30%. Therefore, the decline in surface tension may play a significant role in LEP decrease under realistic MD operating conditions.

My contributions to this work include running the MD experiments, discussing and writing the theory on how surface tension affects fouling and LEP, and paper writing and proofreading.

## 10.2 TESTING OF ELECTROSPUN MD MEMBRANES

This in-progress work is focusing on the creation of electrospun MD membranes via iCVD. Amelia Servi is first author on this work, and Professors Hassan Arafat, and Karen Gleason are coauthors [271].

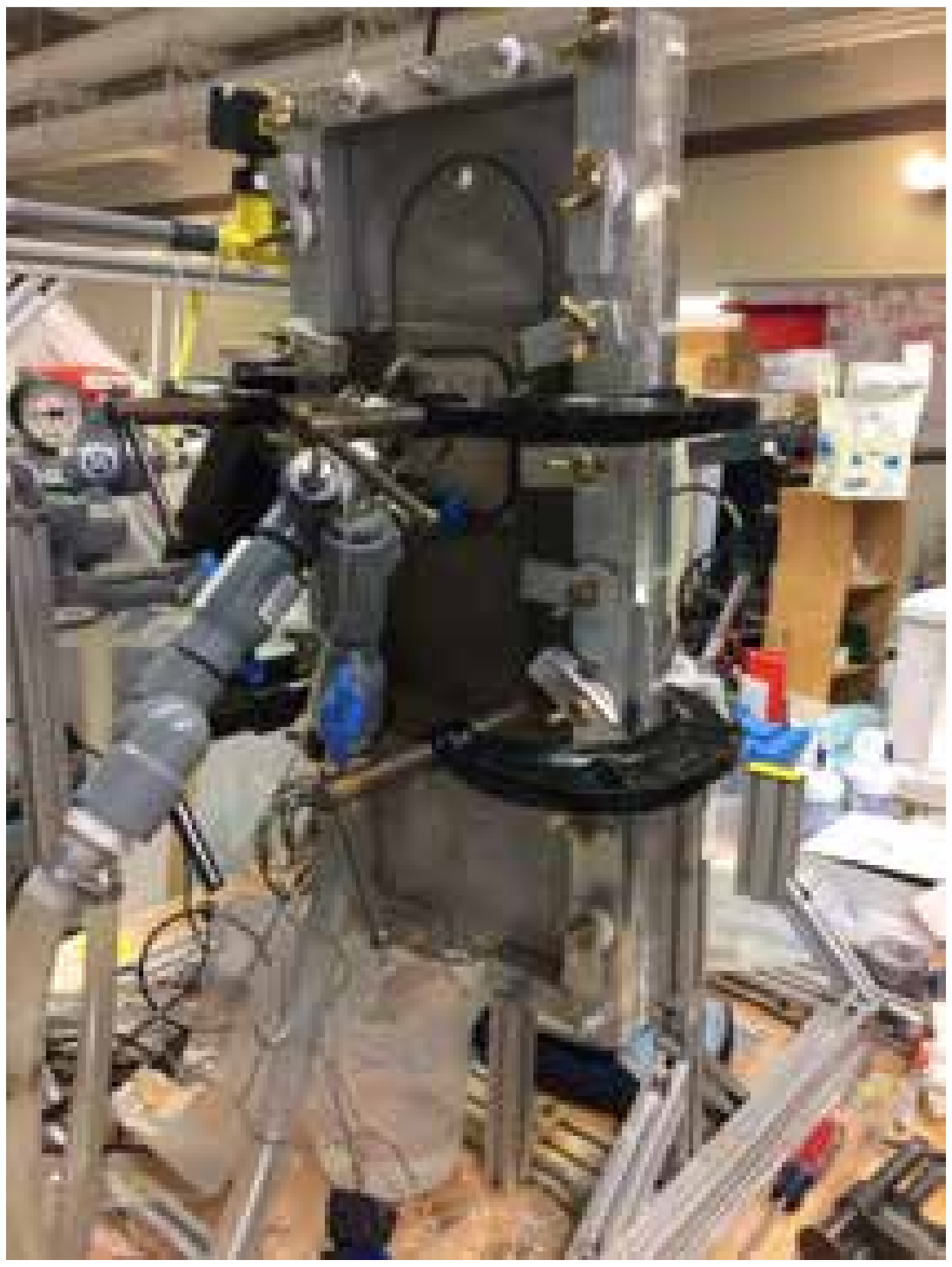

**Figure 10.4.** MD apparatus modified for testing electrospun membranes

I aided with membrane characterization, including testing them in my apparatus, evaluating the B value, and testing for leaks. To do so, I had to design a smaller feed plate, get it into SolidWorks, have it CNC machined, and then modify the setup to accommodate it.

## 10.3 IN-SITU VISUALIZATION OF MD FOULING WITH FLUORESCENCE

The MD fouling literature is lacking in-situ visualization of fouling occurring. Such analysis can help better understand the full process of fouling, and avoids effects such as dry out that occur because an experiment has ended. A challenge to this is that the most common scale, such as $CaSO_4$ and $CaCO_3$, are white, and many of the polymer membranes are also white. To deal with this issue, membranes can be made to be fluorescent, making it obvious when scale occurs.

Seongpil Jeong is first author on this work, and Jaichander Swaminathan, Hyung Won Chunng, and Professor John Lienhard are also coauthors [272].

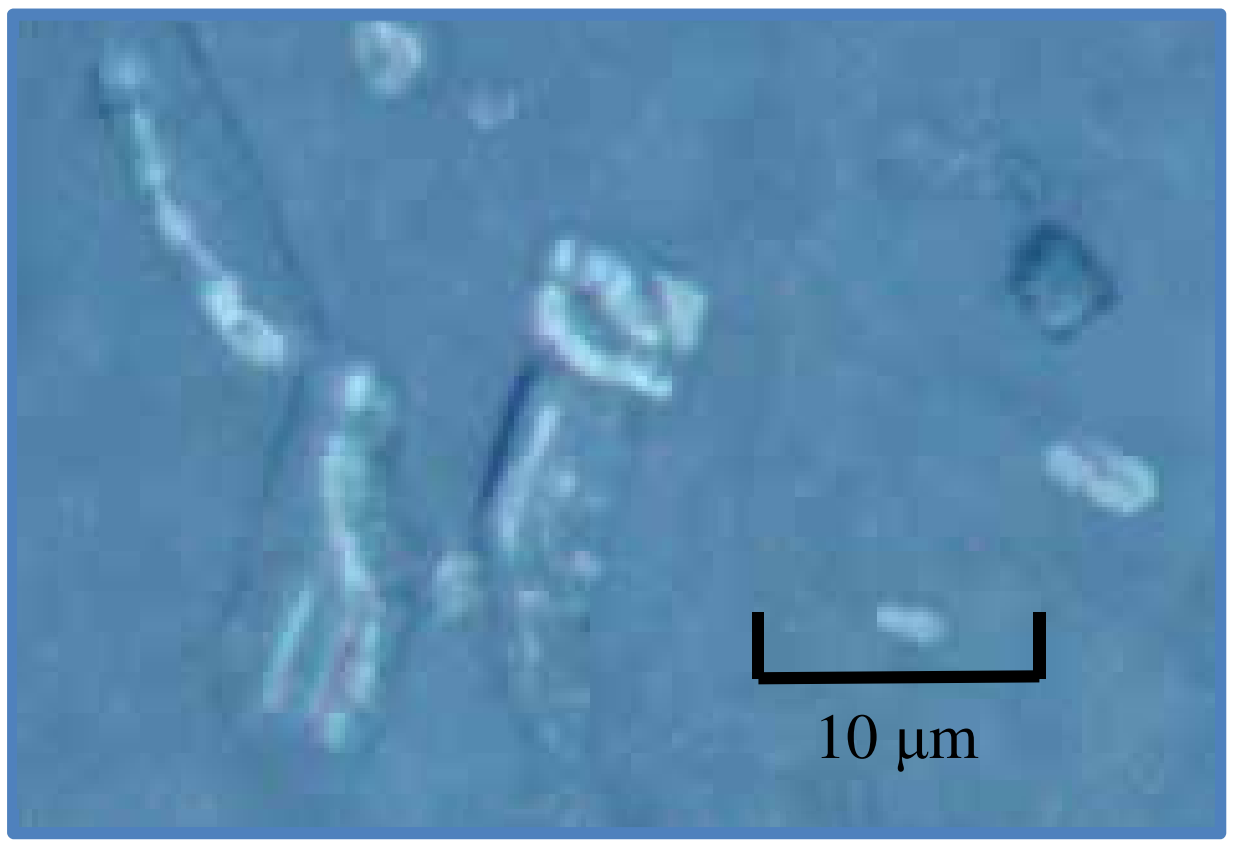

**Figure 10.5.** Microscope image of MD membrane after seawater experiments

This work was led by PostDoc Seongpil Jeong and run on my experimental apparatus [272].

## 10.4 Foulant Comparison

Studies in the literature and these experiments have shown that different foulants vary in their effect to reduce permeate flux and to cause wetting. Calcium sulfate generally creates fairly porous scale, which does not significantly reduce permeate flux. Iron oxide scale however tends to be tenacious, hard to clean [132], and readily causes wetting in this MD system.

Jaichander Swaminathan is first author on this work, and Amelia Servi and Professor John Lienhard V are also coauthors [273].

To better understand the effect of foulant type on MD performance degradation, we are performing experiments with foulants placed on an MD membrane and allowed to dry out, followed by LEP tests. This method allows for a wide variety of salts to be tested rapidly.

# Chapter 11. CONCLUSION

This work on membrane distillation thermodynamics and fouling improved understanding in several significant ways and provided several innovations to improve the technology. Thermodynamic studies included waste heat modeling on MD compared with other technologies, studies on the effect of MD tilt angle, and an innovation of enhancing AGMD with superhydrophobic surfaces. Fouling was analyzed with a review paper of MD fouling literature, studies on the effect of filtration and bulk nucleation, and examining the effect of air layers on MD fouling.

Numerical modeling analyzed the efficiency of multistage membrane distillation and five other representative desalination technologies using variable temperature waste heat. The study found that multistage vacuum membrane distillation can perform with similar efficiencies and recoveries to all other thermal desalination technologies excluding MED, performed similarly to ORC-MVC too. Another crucial contribution from the study was that the increased entropy generation of thermal technologies operating on lower temperature waste heat occurs almost exclusively in feed and brine portions of heat exchangers. Additionally, this work provided detailed component-level entropy generation analysis with shared assumptions for all technologies, the first to do so with variable-temperature waste heat.

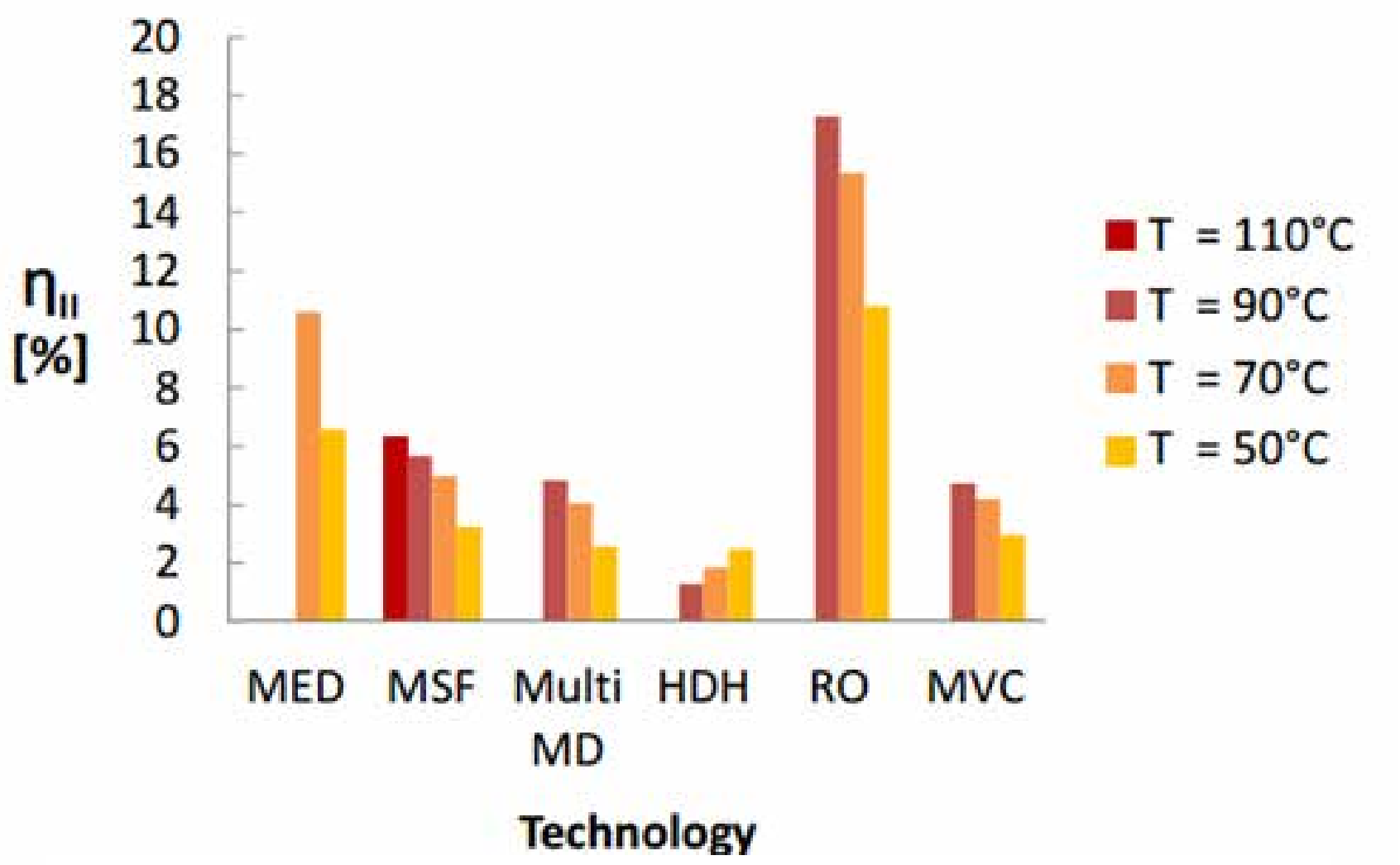

**Figure 11.1.** Second Law efficiency of desalination technologies using variable-temperature waste heat.

Some of the early work addressed an important but previously unaddressed question: how does the effect of module tilt angle effect the performance of Air Gap MD? Experiments and numerical modeling were performed to test the condensate flux that occurred at variable tilt angle. The study found that the performance varied with angle little except at extreme angles. Declined angles caused thermal bridging of water falling from the condensing plate to membrane surface, while extreme angles in either direction (80°<) caused flooding.  A trade-off exists between the desire to maximize the thermal resistance but minimize the mass transfer resistance in the gap. Therefore, flooded or thermal bridging conditions enhanced permeate flux but at a cost of heat transfer losses. Systems that desire AGMD instead of PGMD should use near vertical angles.

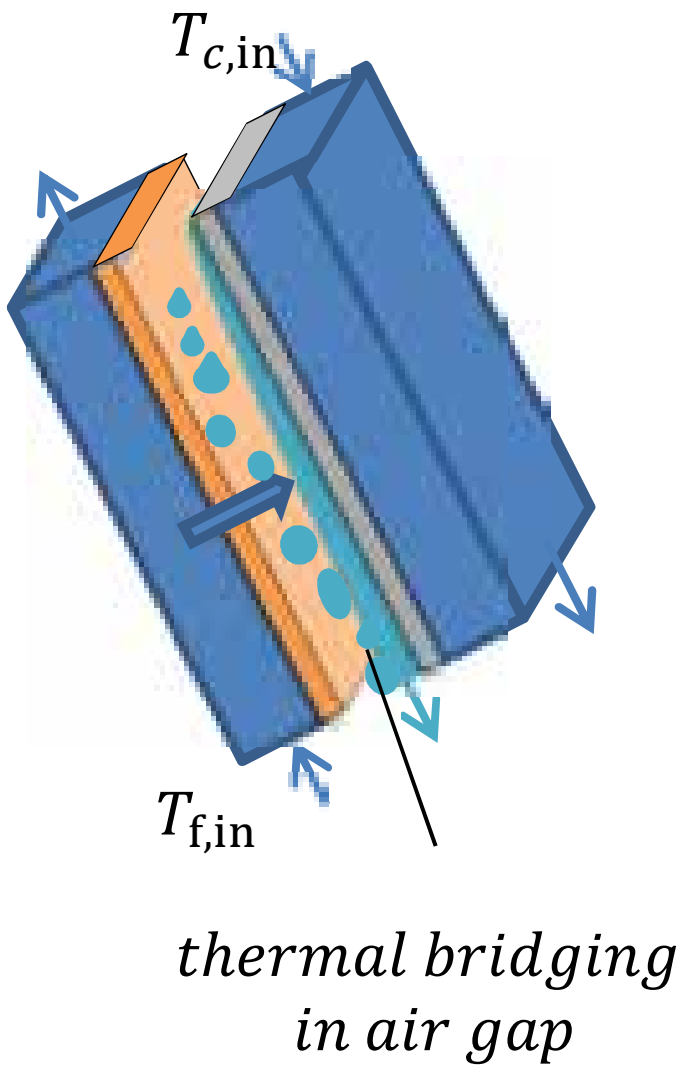

**Figure 11.2.** AGMD thermal bridging at declined

Superhydrophobic condensing has long been known to have superior heat transfer, but only recently have superhydrophobic coatings been sufficiently robust and thermally conductive for practical use. Air Gap MD experiments were performed with a fabricated superhydrophobic (164° contact angle) jumping droplet surface and compared with regular and mildly hydrophobic surfaces.  The hydrophobicity of the support spacer was examined as well, and the results were compared to numerical modeling of flooded gap and regular AGMD. The results showed that superhydrophobic condensing can enhance condensate flux more than 100%. As AGMD is the most efficient MD module configuration when sufficiently high surface areas are used, this innovation can expand on the maximum performance of MD. A journal paper and full patent were submitted on this work.

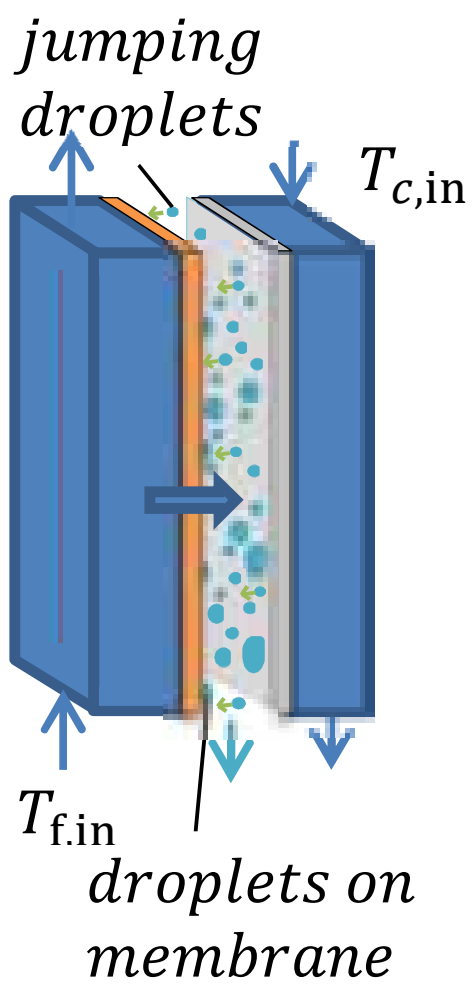

**Figure 11.3.** Superhydrophobic AGMD

To begin fouling studies of MD, a comprehensive review paper was completed on MD fouling and scaling for seawater applications. This 85-page review integrated results from about

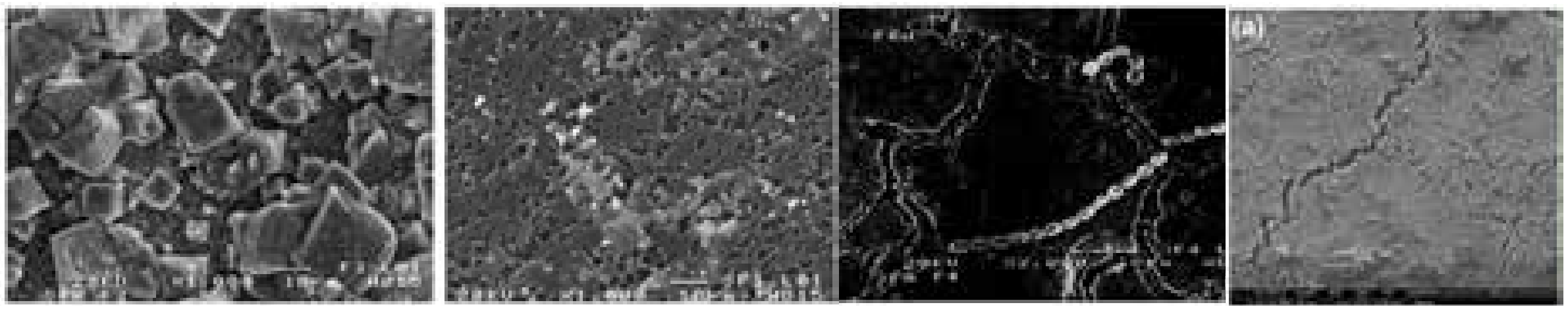

inorganic [156] particulate [132] biofouling [183] chemical degradation [34]

**Figure 11.4.** Types of Fouling in MD

200 papers, and was the first of its kind for MD. Four types of fouling occur in the MD literature: inorganic salts, biofouling, particulate fouling, and chemical degradation of the membrane. The susceptibility to fouling depends on the concentration of foulants, flow conditions, membrane, temperature, effects of foulants on one another, and other factors. The review studied fouling prevention technologies and found two extremely effective: superhydrophobic membranes and particle filtration. This inspired the subsequent studies. Other fouling prevention technologies were also reviewed, with mildly effective results for ultrasonic cleaning, magnetic water treatment, flocculation, boiling to remove carbonate, bubbling of air, oscillating feed temperature, and low-porosity surface coatings.

The review paper on MD fouling led to work on MD filtration and bulk precipitation. It was observed in the literature that filtration was extremely effective at preventing fouling, and that the Gibbs free energy barrier for nucleation on hydrophobic membranes was similar to that of the bulk. Therefore, bulk nucleation may dominate for inorganic scale. To test this, experiments were performed with the common scale CaSO4 with and without a filter, where the permeate was examined for the damaging symptoms of fouling: condensate flux decline from blocked pores, salinity increase of the permeate from membrane wetting, and SEM images of crystals to examine deposition and growth. The results found that filtering particulate matter out of the bulk was extremely effective in reducing fouling, even under conditions adverse to bulk nucleation. This supports the idea that focusing on bulk conditions can substantially reduce fouling in MD.

| SI | Filter | No Filter |
|---|---|---|
| 0.2 | none | Flux Decline |
| 0.4 | Flux Decline | Wetting and Flux Decline |

**Figure 11.5.** Effect of filtration on MD fouling of $CaSO_4$

In the literature, superhydrophobic membranes prevent MD fouling extremely effectively, and air bubbling was also effective. Visible air layers on solid superhydrophobic surfaces have also shown robust biofouling resistance, until the air layer dissipates. These observations led to a study on a potential innovation: intentionally introducing air into the feed of MD systems to maintain an air layer on the membrane surface to reduce fouling. MD experiments were performed on a small desktop beaker MD setup with NaCl, $CaSO_4$, Silica, and Alginate to test scaling, particulate fouling, and biofouling. The results found two competing effects from maintaining air layers. The air layers could prevent fouling, likely by reducing the surface area of membrane in contact with solution, acting to protect it. However, for salts especially, the introduction of air exacerbated fouling. One possible cause was evaporation of vapor into the introduced air, leaving behind salt crystals. However, revised tests with hot saturated air showed the same results, so there is a more likely cause: the air-water interface can induce nucleation of salts in the bulk, as the energy barrier for formation is small. The air layers were extremely effective at reducing biofouling, with reductions up to 97%. As biofouling is more of an issue on hydrophobic surfaces such as MD membranes, this innovation could prove invaluable for using MD despite algae from seawater or biological compounds from wastewater.

# Chapter 12. APPENDIX

## 12.1 BIOGRAPHY

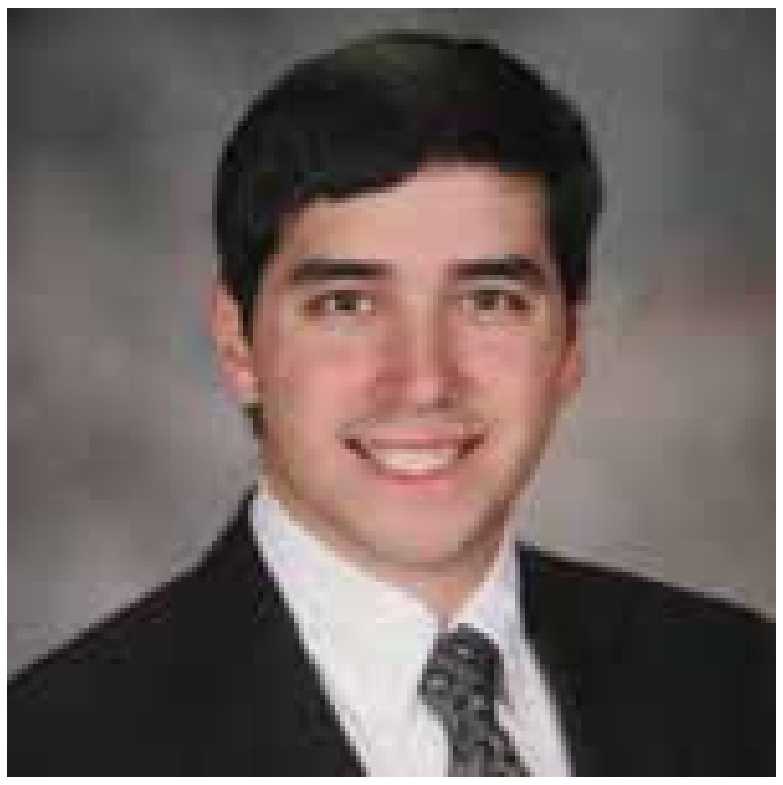

David Warsinger is a scientist, teacher, inventor, and entrepreneur. At MIT, his work has focused on heat transfer and fouling in the desalination technology membrane distillation. David has also done other heat transfer research, and work on a water-borne disease prevention technology he invented. Previously as an undergraduate and Masters student at Cornell, David's research included a scaled test bed for building technology, gas turbine blade failure, oxycombustion for carbon sequestration, and high temperature solar thermal collectors.

David has issued or submitted patents in membrane distillation desalination, geothermal cooling, hybrid vehicles, and wind turbines. Before MIT, David worked as an engineer at Arup creating energy models, designing heating and cooling systems, and performing sustainability analysis. David has worked for or consulted for several startups, including EcoVent, and Pareto Energy, and is currently CTO of Coolify.

Some of David's past awards include an award for highest GPA in his Masters program, first place in a national entrepreneurship competition ($100k Agriculture Innovation Prize), and best mentor of undergraduates at MIT (2015 Outstanding UROP Graduate Mentor). Some of David's interesting achievements include drawing against a chess grandmaster (Palatnik), and completing his Masters and PhD in a combined 3 years.

## 12.2 PUBLICATIONS

### 12.2.1 FIRST AUTHOR PAPERS

### 12.2.2 COLLABORATIVE AUTHOR PAPERS

### 12.2.3 Patents

### 12.2.4 Presentations

## 12.3 PhD Committee Members

Professor John Lienhard, PhD Advisor

Massachusetts Institute of Technology,

Room 3-166 , 77 Massachusetts Ave, Cambridge, MA 02139

617-253-3790

lienhard@mit.edu

Professor Hassan Arafat, PhD Committee Member, MASDAR Collaborator

P.O. Box 54224, Masdar City,

Abu Dhabi, United Arab Emirates

971-2810-9119

harafat@masdar.ac.ae

Professor Rohit Karnik, PhD Committee Member

Room 3-461A, Massachusetts Institute of Technology,

Room , 77 Massachusetts Ave, Cambridge, MA 02139

617-324-1155

karnik@mit.edu

Professor Karen Gleason, PhD Committee Member

Room 66-466, Massachusetts Institute of Technology,

Room , 77 Massachusetts Ave, Cambridge, MA 02139

617.253.5066

kkg@mit.edu

## 12.4 Additional Experiments

Most of the experiments performed were not included in the papers that comprise the Thesis. This work was performed for other papers, was omitted because conditions weren't consistent for comparison to others in a particular paper, were left out because it was less novel than others, or was left out because it was decided certain system modifications were needed, such as the inclusion of a filter. Some were excluded simply because the scope of the papers was narrowed, such as the full system tests for Silica and $CaCO_3$. Despite the lack of inclusion of this work, it provided significant insights which led to the fouling papers.

These experiments in the full MD apparatus usually lasted between 2 and 7 days, with a few outside that range. The salts tested include Synehetic RO brine, tap water, $CaSO_4$, $CaCO_3$, $FeCl_3$, NaCl, and combinations thereof.

**Table 6.1** MD Fouling Experiments Performed

| # | Salt | Details | SI | Th [°C] | Tc [°C] | Duration [hours] | Filter | Date |
|---|---|---|---|---|---|---|---|---|
| 1 | CaSO4 | | 1.4 | | | | no | |
| 2 | CaSO4 | | 2 | | | | | |
| 3 | CaSO4 | | 1 | | | | | |
| 4 | Synthetic SW | | 2x sw | 70 | 20 | | yes | |
| 5 | Synthetic RO brine (60% of Ca2+) | < 10 uS/cm | 2x sw | 70 | 20 | 70 hr | yes | |
| 5 | Synthetic RO brine (60% of Ca2+) +0.5 ppm FeCl3 | 2.5 uS/cm | 2x sw | 70 | 20 | 50 hr | yes | |
| 5 | Synthetic RO brine | 30 uS/cm | 2x sw | 70 | 20 | 50 hr | yes | |
| 5 | Synthetic RO brine +0.5 ppm FeCl3 | 40 uS/cm | 2x sw | 70 | 20 | 30 hr | yes | |
| 4 | Synthetic RO brine (9-10 times of SW) +2 g of FeCl3 | | 2x sw | 70 | 20 | 18 hr | yes | |
| 7 | Tap | | <0 | 70 | 20 | 18 hr | yes | |
| 1 | NaCl | | <0 | 70 | 20 | | no | |
| 2 | NaCl | Concentration Polarization | <0 | 70 | 20 | | no | |
| 3 | NaCl | DI water | <0 | 70 | 20 | | no | 21-Feb-14 |
| 4 | NaCl | | <0 | 70 | 20 | | no | 3-Apr-14 |
| 5 | DI water | metal mesh | <0 | 70 | 20 | | no | Jul-14 |
| 6 | DI water | sautered, see-through | <0 | 70 | 20 | | no | Jul-14 |
| 7 | CaCl2 | 52g, RO-brine | 2x sw | 70 | 20 | | | Jul-14 |
| 8 | FeCl3 | RO-brine | 2x sw | 70 | 20 | | | Jul-14 |
| 9 | FeCl3 | | 2x sw | 70 | 20 | | | Jul-14 |
| 10 | RO-brine | RO-brine | 2x sw | 70 | 20 | | | Jul-14 |
| 11 | | flitering one light, "wavelength" | | 70 | 20 | | | Jul-14 |
| 12 | Silica | | colloidal | 70 | 20 | | no | Aug-14 |
| 13 | CaSO4 | | 0.2 | 70 | 20 | | | Aug-14 |
| 14 | CaSO4 | long | | 70 | 20 | | | Aug-14 |
| 15 | CaCO3 | | | 70 | 20 | | | 4-Aug-14 |
| 16 | CaSO4 | 31.9g | | 70 | 20 | | | 11-Aug-14 |
| 17 | CaSO4 | 68.84g | | 70 | 20 | | | 12-Aug-14 |
| 18 | | | | 70 | 20 | | | 18-Aug-14 |
| 19 | CaSO3/NaCl | | | 70 | 20 | | | 20-Aug-14 |
| 20 | CaSO4/NaCl | w/ DI water, repeat experiment | 0.4 | 70 | 20 | | | Sep-14 |
| 21 | DI water | cleaning membrane | 0.251 | 70 | 20 | | | 14-Sep-14 |
| 22 | CaCl2/Na2SO4 | | 0.2 | 70 | 20 | | | 15-Sep-14 |
| 23 | CaSO4/NaCl | no filter, 70+ hours. Flux decline 0.12 to 0.11 in 20 hours. Permeate conductivity Increases to 350 uS/cm and stablizes. Starts increasing after 20 hours of operation | 0.4 | 70 | 20 | 52 | no | 28-Sep-14 |
| 24 | CaSO4/NaCl | | 0 | | | | | 6-Oct-14 |
| 25 | | | | | | | | 17-Oct-14 |
| 26 | CaSO4/NaCl | filter, 23 hours | 0.4 | 70 | 20 | 20.8 | yes | 5-Nov-14 |
| 27 | NaCl | Yes, flux decline only in 260 o | 260 ppt | | | | no | |
| 28 | NaCl | Petri Dish ordinary (tiny) - no air recharg | 200 ppt | 60 | 25 | 1 | no | Dec-14 |
| 29 | NaCl | Petri Dish ordinary (tiny) -air recharge | 200 ppt | 60 | 25 | 1 | no | Dec-14 |
| 30 | NaCl | Petri Dish Superhydrophobic (tiny) -no recha | 200 ppt | 60 | 25 | 1 | no | Dec-14 |
| 31 | NaCl | Petri Dish Superhydrophobic (tiny) -air recha | 200 ppt | 60 | 25 | 1 | no | Dec-14 |

These studies helped reveal several trends, which are elaborated on by other experiments, and elsewhere in the thesis. These trends include:

- Nucleation is largely occurring in the bulk, made apparent by the other filtration experiments, including those with synthetic seawater.
- Certain foulants such as iron are much more likely to cause wetting than others. This was shown by problems with exposed stainless steel heater elements rusting, as well as $FeCl_3$ experiments. This has led to ongoing collaborative work.

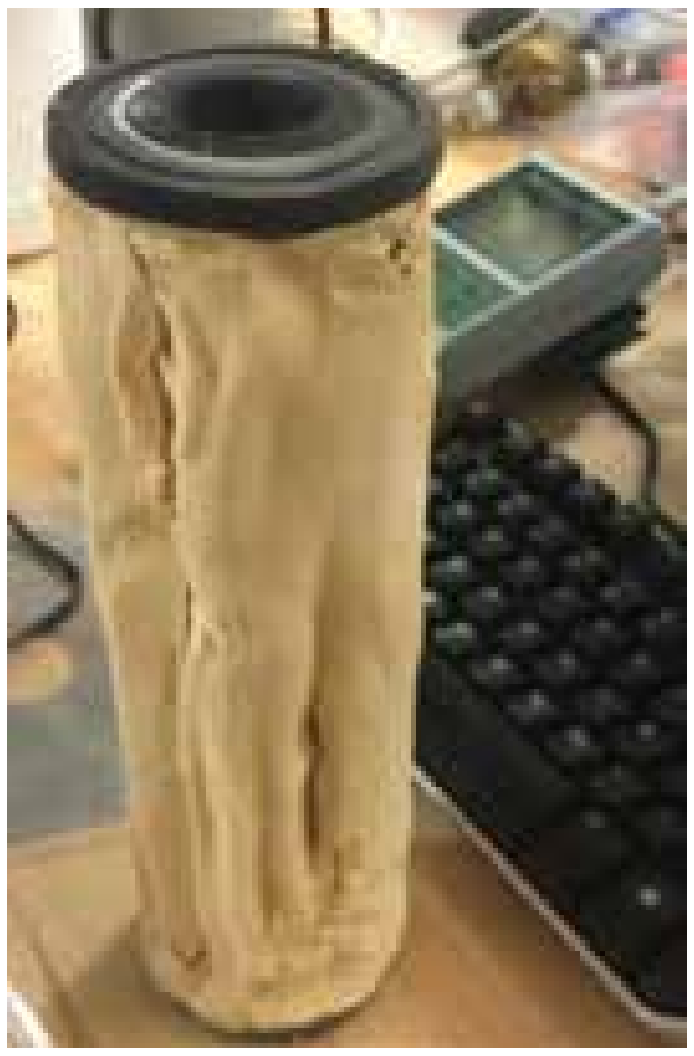

**Figure 12.1.** Completed fouled filter from synthetic RO brine

- If the air gap is wetted, and there are two metals within the gap, very bad corrosion occurs on the otherwise resistant surfaces. This occurred with copper meshes and plates which rested on the Aluminum plate. Likely, the two metals with a wetted electrolyte solution are acting like a battery.
- $CaSO_4$ is a very porous scale, and doesn't cause flux decline readily.
- Silica in the MD system may cause flux decline and wetting, even in small concentrations.
- Significant fouling may occur on heaters, due to higher temperatures, and likely because the surface of metals is typically hydrophilic, providing a small energy barrier for nucleation.

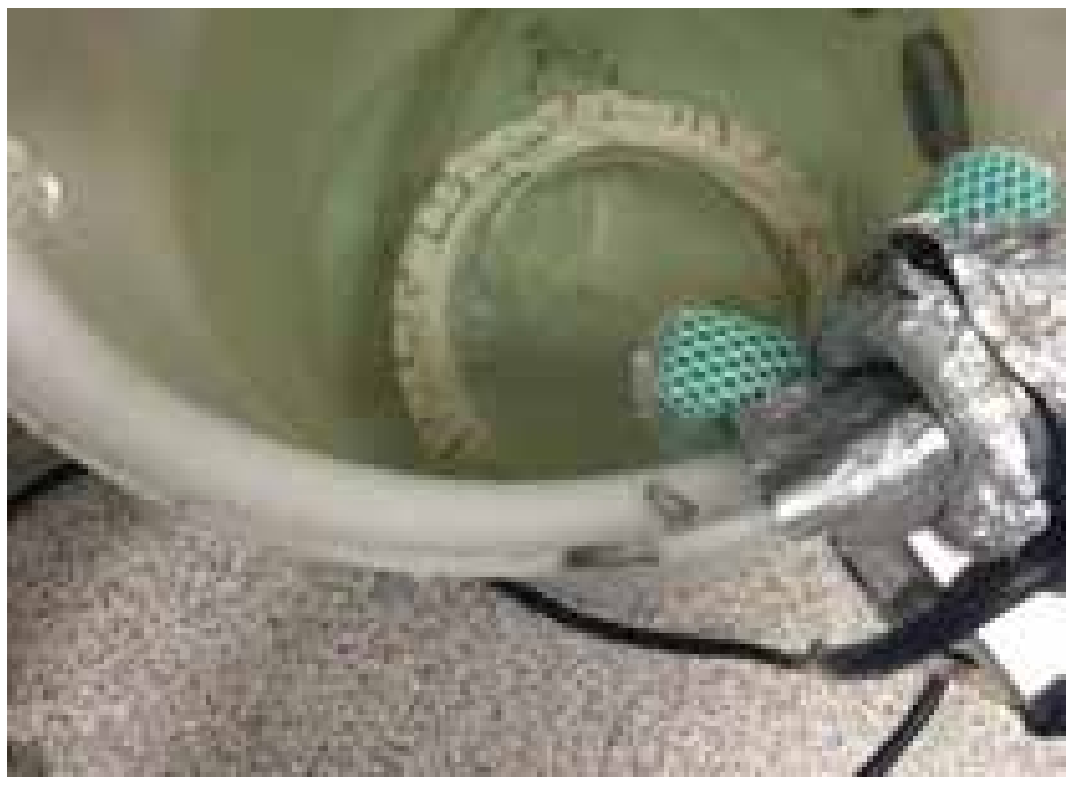

**Figure 12.2.** Substantial $CaSO_4$

- Recirculation systems can have nearly infinite residence times if left super saturated.
- Both $CaSO_4$ and $CaCO_3$ were observed when using RO brine (concentrated seawater)
- $CaCO_3$ is very pH sensitive, which makes modelling more complex, and requirements for consistent results more stringent. It is very temperature sensitive too

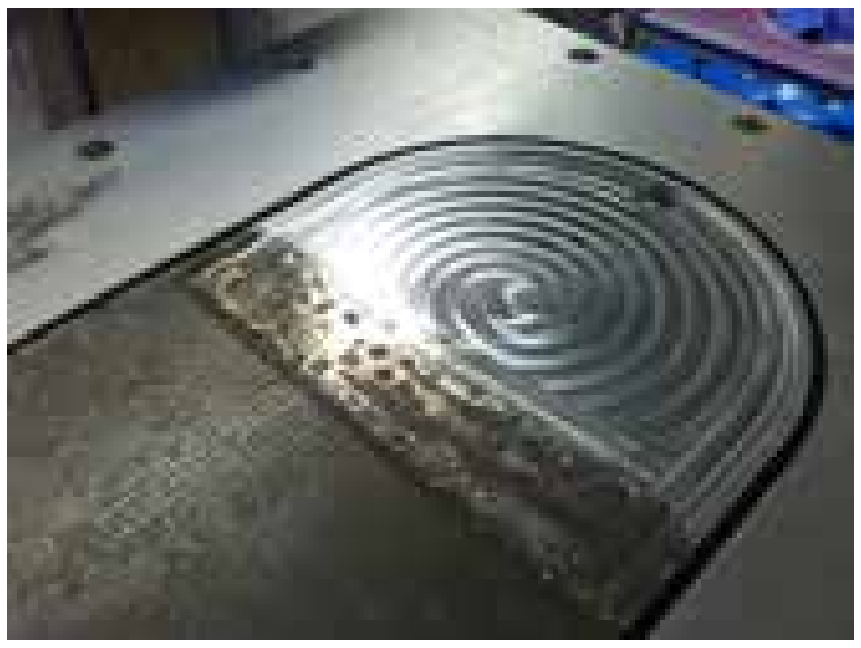

**Figure 12.3.** Corrosion of Aluminum condensing plate in the presence of salts (from wetting) and copper

Note that while the work in this section is unpublished, that is not the case for the other sections. MIT copyright allows for the reproduction of journal and conference papers authored by the student in student Theses.

## 12.5 Apparatus Design

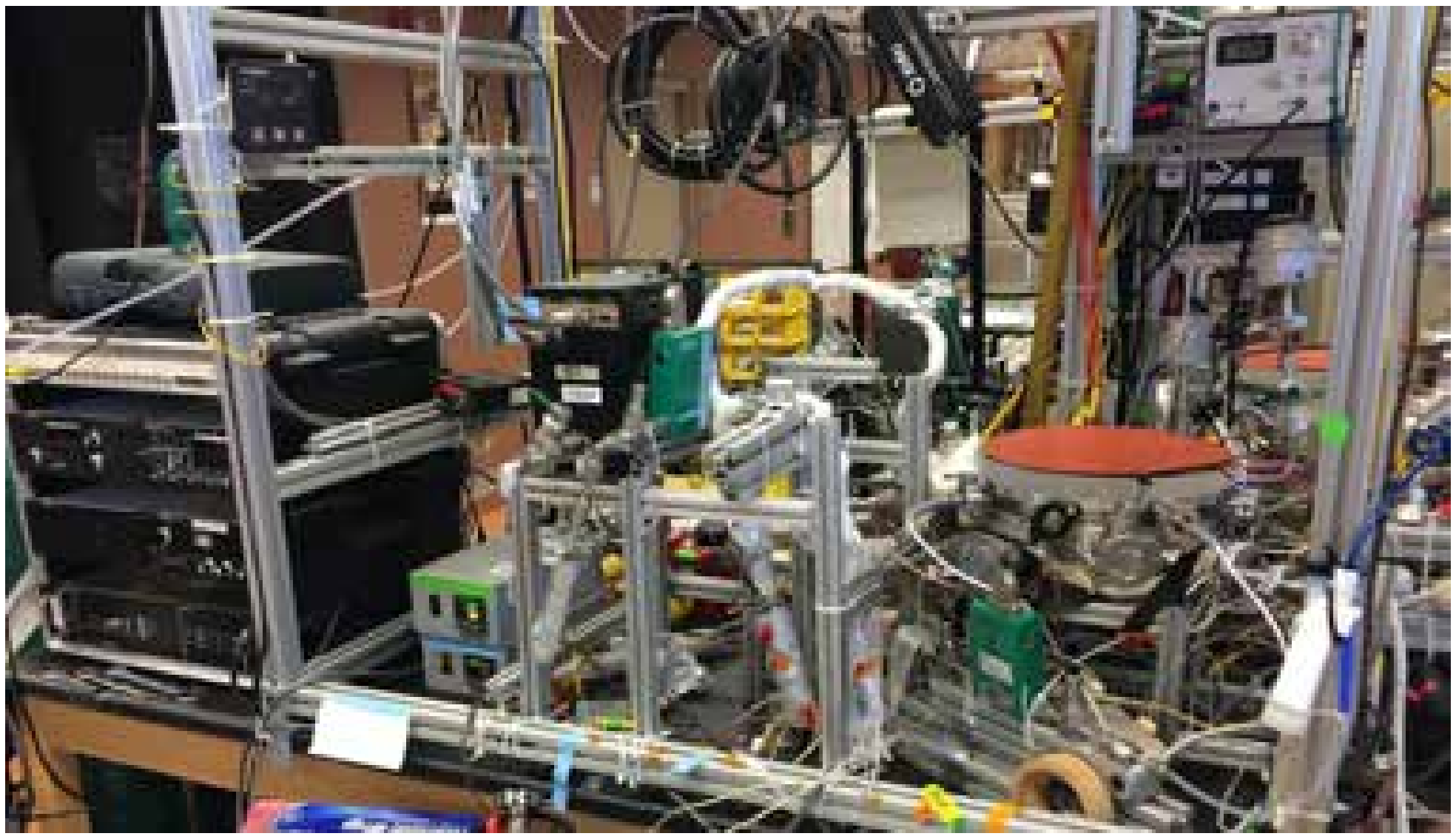

**Figure 12.4.** iCVD apparatus for coating membranes with PFDA for superhydrophobicity. Used for collaborative work as well.

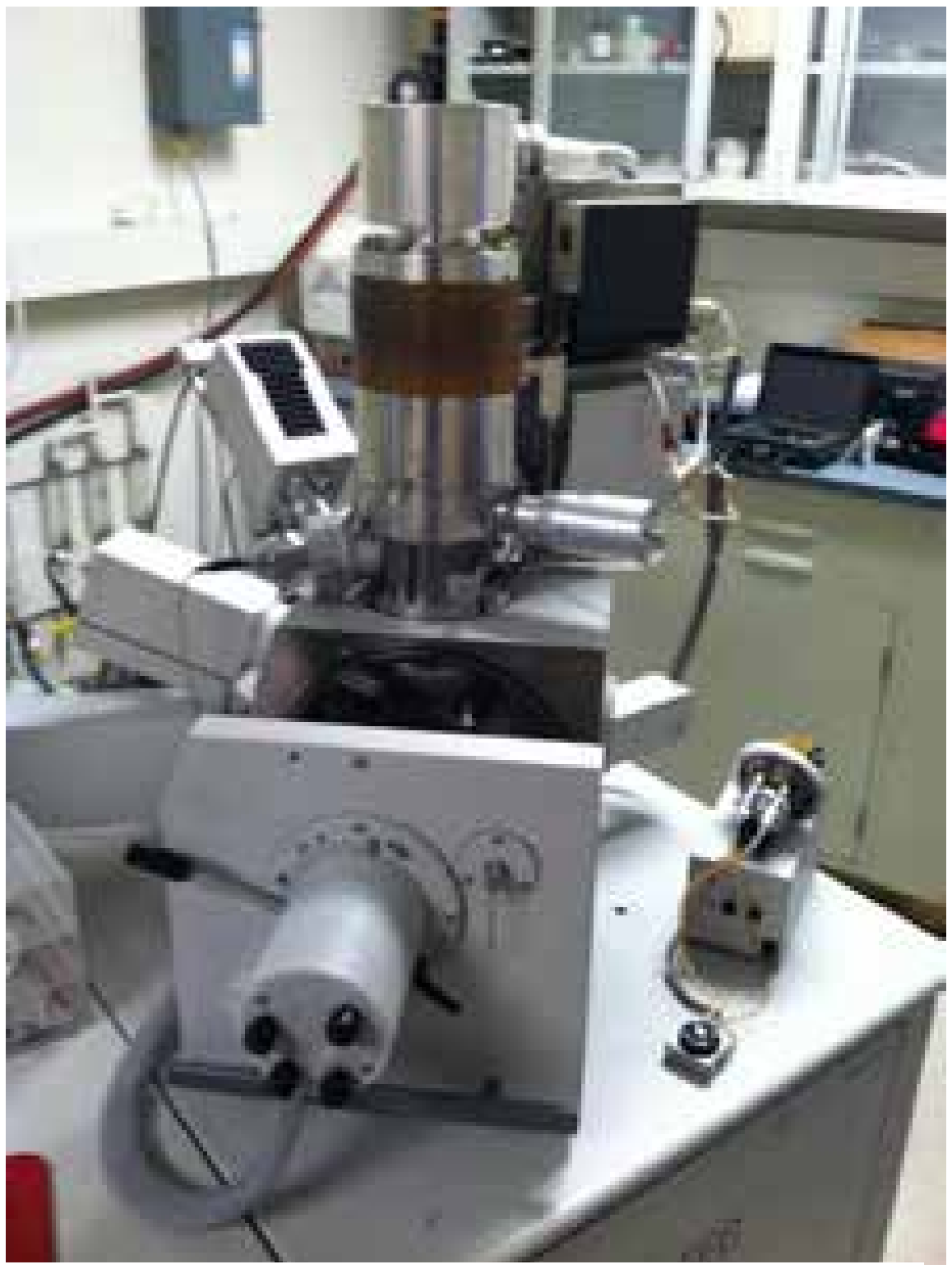

**Figure 12.5.** Environmental Scanning Electron Microscope (ESEM) at MIT used for imaging membrane fouling

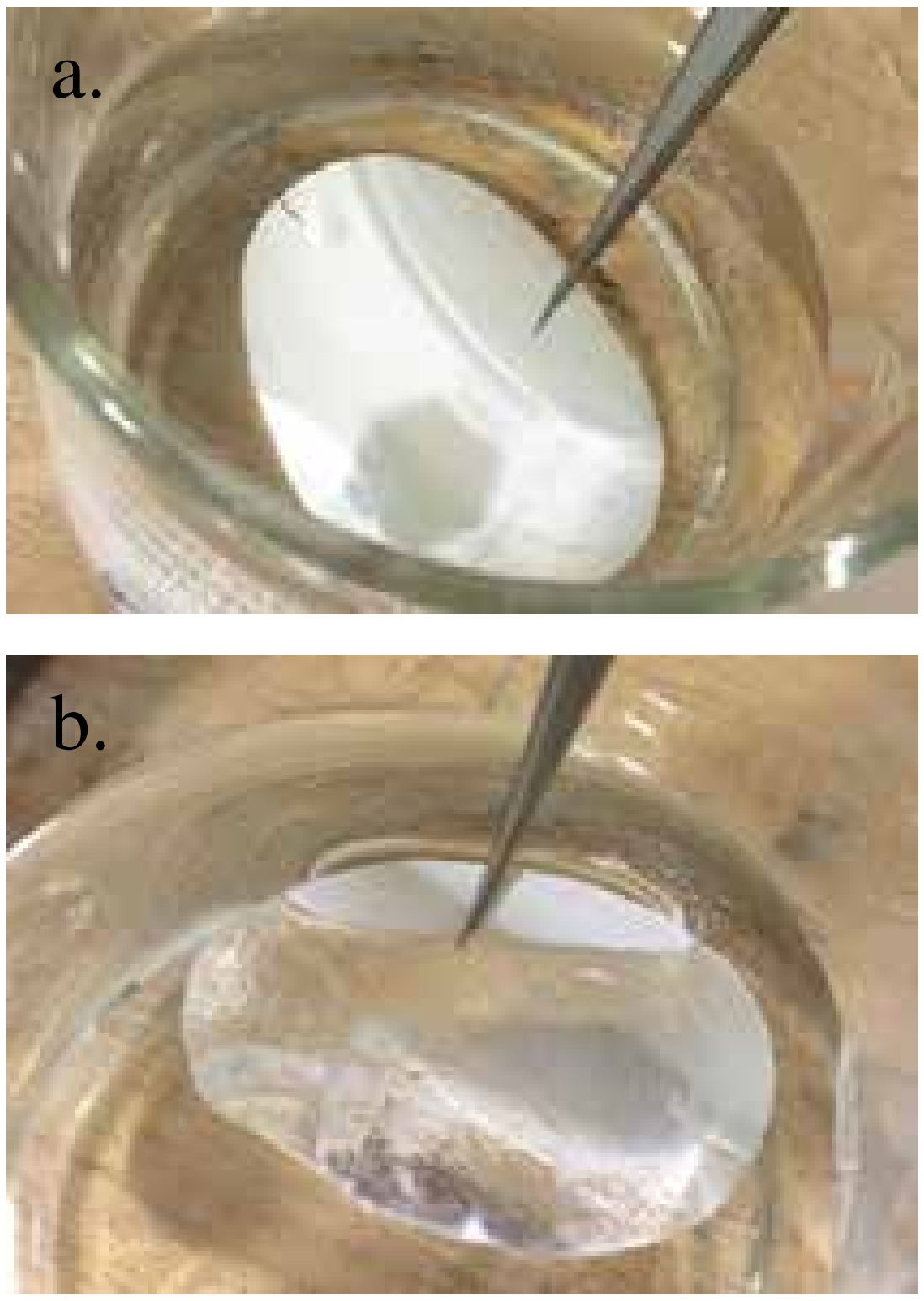

**Figure 12.6.** Comparison of uncoated (a) and superhydrophobic membranes (b)

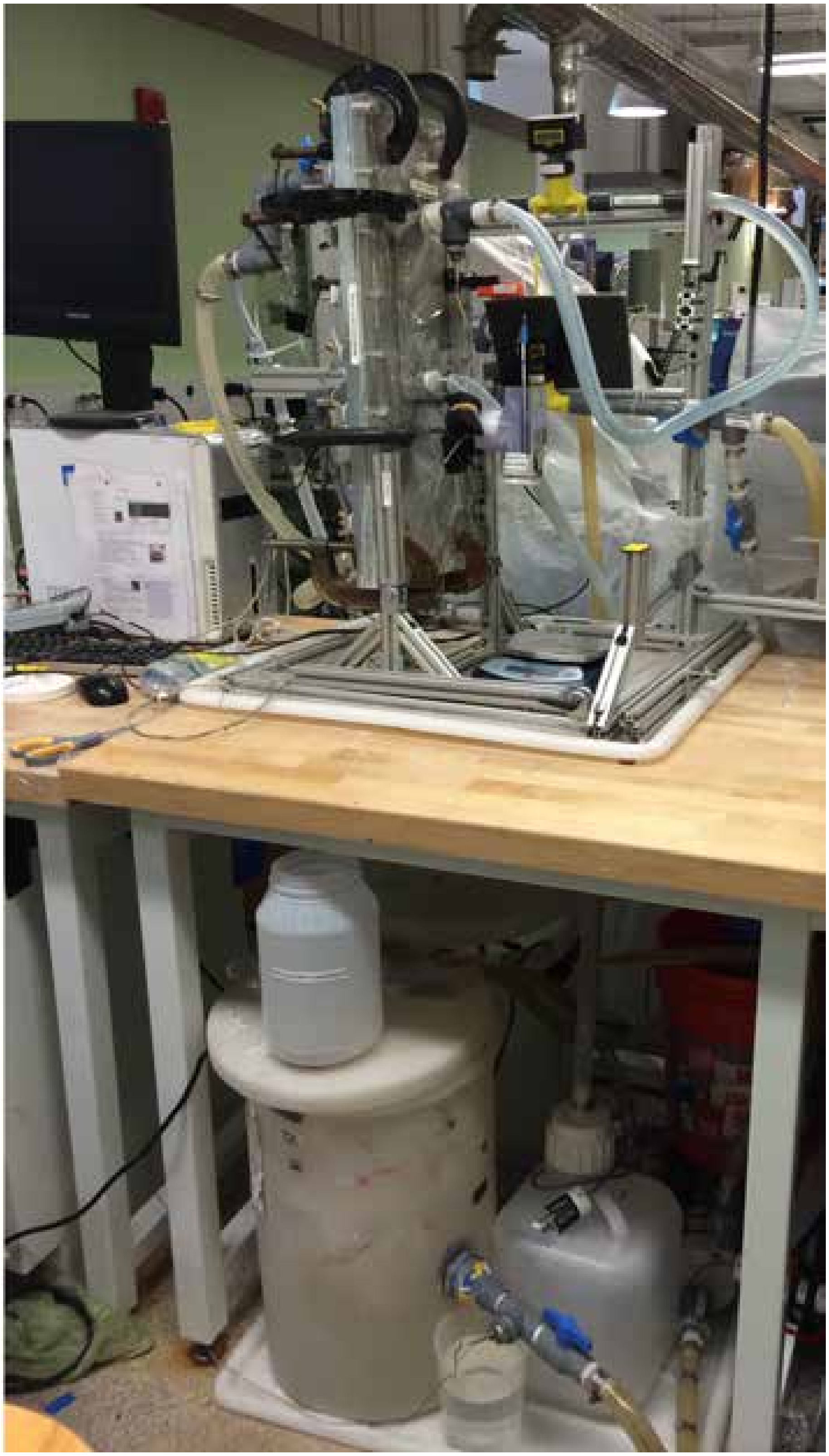

**Figure 12.8.** Membrane distillation apparatus for fouling and efficiency studies

**Table 6.1** Design Needs and Approach for MD fouling system

| Design Need | Approach |
|---|---|
| Avoid Leaks | Use of hose clamps, filling leaks with caulk & Sugru, threaded fittings, placing components in containers. O-rings used instead of gaskets |
| Easy to visualize | Module composed of clear polycarbonate, clear tygon tubing, melting method developed for seeing through membranes, scaling tests in clear beaker |
| Avoid corrosion | No metal parts exposed to the feed: plastic piping, magnetic drive centrifugal pumps, heater enclosed in thin plastic bag |
| Waterproof Components | Control components neat the apparatus were protected with thin plastic. Other control equipment was placed under the table, away from the piping components of the system |
| Record Temperature | Thermistors with input into computer, processed by the program TracerDAQ. Exported data compatible with MS Excel |
| Record Permeate Mass Flow Rate | Drops fall onto large container placed on a scale. The scale data reads to the computer using the program RealTerm |
| Record Permeate Salinity | Use a small reservoir after the module that collects enough water to hold the conductivity probe |
| Make system quick to take apart | The feed plate is clamped onto the module instead of using bolts which most systems do. Reduces disassembly time to ~2 minutes, compared to ~15 for previous MD systems in the group |
| Component temperature tolerance | Plastics with a higher temperature tolerance (CPVC, polyethelene, polycarbonate, etc) were selected |
| Cleaning system between tests | The feed water was fully emptied, replaced with DI or tap water, and then a cleaning cycle was run. After cleaning, the water was replaced again with DI |
| Pressure Drops and Pressure Control | Excessive pressure drops were reduced by replacing small opening fittings with much larger ones. Pressure was controlled by nearly closed valves before and after the MD module, which had a pressure meter in between |
| Providing Cooling | Piping the building's chilled water through a copper-coil heat exchanger in the cold water tank. Notably, due to the high pressure of the cold line (7-9 bar), high pressure tubing was used. Solenoid valve turns the flow on and off |
| Temperature Control | Temperature controller paired with thermistor. |
| Return permeate without temperature dip | A tank with a small hole in the bottom was used to slowly drip permeate back into the feed tank |
| Electrical Safety | All plugged-in components used an outlet with a built-in circuit breaker, or a surge protector with a circuit-breaker |

### 12.5.1 VISUALIZATION

Visualizing the fluid inside the apparatus is invaluable for diagnosing fouling, and related studies including examining air layers. Several techniques were used to take images, and see inside the membrane. For image taking of fouled membrane, in-situ studies were done with a web camera and an EOS Rebel T3i Canon digital camera. For autopsy analysis of membranes, SEM, ESEM, XRD, EDS, and polarizing light microscopy were used.

Where possible, apparatus components were made of clear materials. The feed channel was composed of clear polycarbonate, which is sturdy, has a high melting point, and is easy to machine. For beaker studies, clear glass beakers were used. To observe air bubbles in pipes, all flexible pipe components were made of clear plastics, including Tygon.

To visualize jumping droplet condensation, membranes were melted to consolidate the pores, as seen in Figure 11.9.

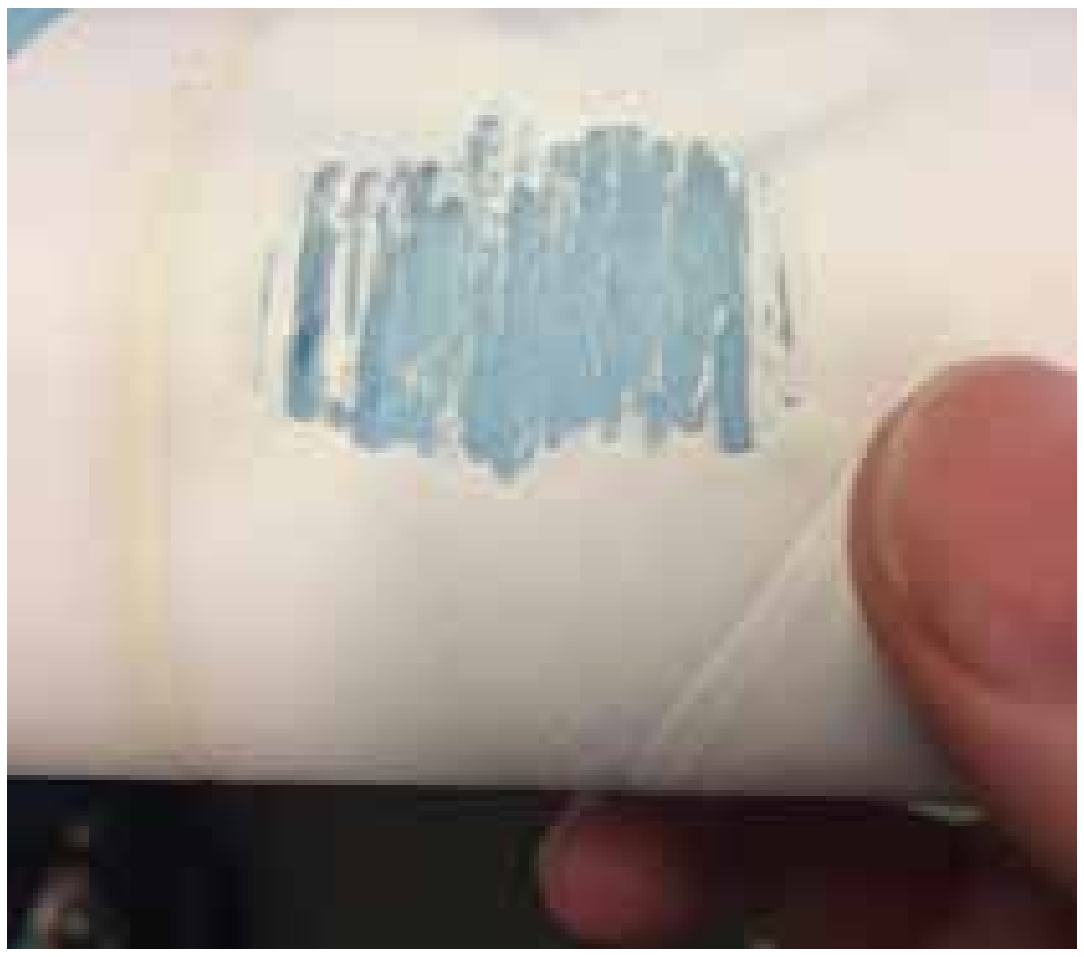

**Figure 12.9.** Melting of membrane for visualization of condensation

## 12.5.2 Experiment Parts

**Table 12.1.** Key Parts Ordered

| Item | Part # | Vendor | Quantity | Cost | Total Price |
|---|---|---|---|---|---|
| | | | | | |
| **Equipment -Pumps, Tanks, etc** | | | | | |
| Cold Water Tank, 10 gal | 569098 | All Glass Aquarium Co. | 1 | 56 | $56.00 |
| Hot Water Tank, 10 gal | 4085K35 | McMaster | 1 | 152.7 | $152.70 |
| Permeate Tank | SI-04884106 | HoldnStorage | 1 | 6.5 | $6.50 |
| Extra Heater Bags | 8852K63 | McMaster | 20 | $2.91 | $58.20 |
| Mag Drive PP Centrifugal Pump w/Enclosed Motor; 13.7 GPM/21.3 ft, 115V | EW-72010-50 | Cole-Parmer | 1 | $443.00 | $443.00 |
| | | | | | |
| **Instrumentation** | | | | | |
| Weight Scale | 1760T86 | McMaster | 1 | 361.32 | $361.32 |
| USB interface for weight scale | 1760T42 | McMaster | 1 | 111.6 | $111.60 |
| Immersion Thermistor Sensors with Threaded Mounting Fitting | ON-910-44004 | Omega | 3 | $55.00 | $165.00 |
| Pipe Plug Thermistor Probe | TH-44004-1/4NPT-80 | Omega | 4 | $68.00 | $272.00 |
| Pressure sensor | 4147K42 | McMaster | 1 | $125.69 | $125.69 |
| Diverting 3-Port CPVC Ball Valve | 4697K42 | McMaster | 4 | $134.80 | $539.20 |
| 1⁄8 DIN Ramp and Soak Controller | CN1501-TH | Omega | 1 | $245.00 | $245.00 |
| Solid State Relays | SSRL240DC50 | Omega | 1 | $37.00 | $37.00 |
| Finned Heat Sink | FHS-2 | Omega | 1 | $19.00 | $19.00 |
| Fast Blow 25A Fuse | KAX-25 | Omega | 1 | $30.00 | $30.00 |
| Fuse Holder | FB-1 | Omega | 1 | $20.00 | $20.00 |
| | | | | | |
| **Wire** | | | | | |
| Wire 22 AWG black | 7587K931 | McMaster | 1x100 ft | $9.31 | $9.31 |
| Wire 22 AWG red | 7587K932 | McMaster | 1x100 ft | $9.31 | $9.31 |
| Wire 18 AWG black | 7587K951 | McMaster | 1x25 ft | $9.31 | $4.32 |
| | | | | | |
| **Apparatus Support Structure** | | | | | |
| Locking Pivots | 47065T221 | McMaster | 2 | $18.98 | $37.96 |
| Connectors Adjustable 90° | 47065T153 | McMaster | 12 | $3.39 | $40.68 |
| Heavy Duty 90° Braces | 47065T186 | McMaster | 7 | $15.67 | $109.69 |
| Extended 90° Brackets | 47065T175 | McMaster | 12 | $4.56 | $54.72 |
| Extended 90° Brackets Fasteners | 47065T139 | McMaster | 12 | $1.85 | $22.20 |
| 3 Way Corner Connector | 47065T244 | McMaster | 1 | $9.86 | $9.86 |
| Al T-Slot Extrusion | 47065T101 | McMaster | 1x6" | $14.2/ 4 ft | $19.20 |
| Double Al T-Slot Extrusion | 47065T107 | McMaster | 2 x 2", 1x6" | 22.98/ 4 ft | |
| Double Al T-Slot End Cap | 47065T92 | McMaster | 6 | $1.50 | $9.00 |
| Al T-Slot End Cap | 47065T91 | McMaster | 6 | $1.20 | $7.20 |
| Al T-Slot Extrusion | 47065T121 | McMaster | 1x4' | $14.20 | $14.20 |
| Double Al T-Slot End Cap | 47065T124 | McMaster | 1x4' | $22.98 | $22.98 |
| 90° 4-hole bracket | 47065T169 | McMaster | 4 | $5.58 | $22.32 |
| Fasteners | 47065T139 | McMaster | 8 | $1.85 | $14.80 |
| Extended 90° Brackets | 47065T175 | McMaster | 12 | $4.56 | $54.72 |
| Double Al T-Slot Extrusion | 47065T213 | McMaster | 2ft | $12.85 | |
| Fasteners | 47065T139 | McMaster | 12 | $1.85 | $22.20 |
| Locking Pivots | 47065T221 | McMaster | 1 | $18.98 | $18.98 |
| Al T-Slot Extrusion | 47065T209 | McMaster | 4x2' | $8.35 | $33.40 |
| | | | | | |
| **Clamps** | | | | | |
| Hook Clamp | 8954A11 | McMaster | 1 | | $25.60 |
| Swivel Heel Clamps | | Carrlane | 1 | | |
| SWING CLAMP ASSEMBLIES | | Carrlane | 1 | | |
| Grizzly 6 Piece 3" Clamp Set | G8093 | Amazon | 1 | $17.95/pack | $17.95 |
| 2" Industrial C Clamp | 37844 | Harbor Freight Tools | | $1.99 | |
| JET 2 1/2" C Clamp, 3 way side | 53430 | Walmart | | $7.99 | |
| IRWIN 2" Locking C Clamp | 43344 | | | $8.47 | |
| Smalles toggle clamp | | Carrlane | | | |
| Compact Hold down toggle clamps | 5004A12 | McMaster | | | |
| Jergens Hook Clamp Assembly | 41905 | Amazon | | $8.91 | |
| MSC Hook Clamp | 6971808 | MSC | | $36.79 | |
| Standard Arm Hook Clamps | | Carr-Lane | | | |
| | | | | | |
| **Heating and Cooling Components** | | | | | |
| Side-Mounted Immersion Heaters | 35605K69 | McMaster | 1 | 288 | $288.00 |
| Incoloy Element for Water | 3583K91 | $68.82 | 1 | $68.82 | |
| Cole-Parmer Stirring Hot Plate Ceramic top, 12"x12" | WU-04803-15 | Cole-Parmer | 1 | $740.00 | $740.00 |
| Side-Mounted Immersion Heaters | 35605K69 | McMaster | 1 | $288.00 | $288.00 |
| **Sealing & Other** | | | | | |
| O-ring cord stock | 9616K12 | McMaster | 10ft | $0.40/ft | $4.00 |
| O-ring cord stock for feed side | 9679K16 | McMaster | 12ft | $0.51/ft | $6.12 |
| O-ring cord stock | 9616K16 | McMaster | 10ft | $0.72/ft | $7.20 |
| Filter With Easy-View Bowl, 3/4", | 4422K3 | McMaster | 1 | 35.28 | 35.28 |
| O-ring cord stock | 9616K15 | McMaster | 10ft | $0.66/ft | $6.60 |
| O-Ring (package) | 1KEV3 | Grainger | 1 | $14.05 | $14.05 |
| O-Ring (package) | 2KAN4 | Grainger | 1 | $8.85 | $8.85 |
| O-ring joining glue | 7569A22 | McMaster | 1 | $24.57 | $24.57 |
| | | | | | |
| **Flexible Piping** | | | | | |
| sanitary Clear PVC tubing, 3/4" ID, 1" OD, 25FT | 5231K385 | McMaster | 25 | $1.44 | $36.00 |
| Masterkleer PVC Tubing 0.5" ID, 5/8" OD | 5233K66 | McMaster | 25 | 0.40/ft | $10.00 |
| Vacum Rated Clear 3A Sanitary Tubing | 5393K43 | McMaster | 10 | 2.24/ft | $22.40 |
| sanitary Clear PVC tubing, 5/8" ID, 7/8" OD, 10FT | 5231K381 | McMaster | 10ft | 1.25 | $12.50 |
| | | | | | |
| **Instrumentation** | | | | | |
| Pipe plug thermistors | TH-44004-1/4NPT-80 | Omega | 6 | 68 | $408.00 |
| Temperature 8-channel USB Data acquisition | OM-USB-TEMP | Omega | 1 | 549 | $549.00 |
| Splash Resistant Digital Water Flowmeter 0.8-8 GPM | 3562K11 | McMaster | 1 | 313.86 | $313.86 |
| Splash Resistant Digital Water Flowmeter 2-20 GPM | 3562K12 | McMaster | 1 | 313.86 | $313.86 |
| OEM Style Acrylic Rotameters | FL7604 | Omega | 1 | 145 | $145.00 |
| Conductivity meter | CDH-45 | Omega | 1 | 99 | $99.00 |
| ph, Conductivity | PHH-60BMS | Omega | 1 | 360 | $360.00 |
| Pressure sensor | 4147K42 | McMaster | 1 | 125.69 | $125.69 |
| Conductivity probe | VWR | 89174-028 | 1 | $326.12 | $326.12 |
| Pipe Plug Thermistor Probe | TH-44004-1/4NPT-80 | Omega | 4 | $68.00 | $272.00 |
| Immersion Thermistor Sensors with Threaded Mounting Fitting | ON-910-44004 | Omega | 3 | $55.00 | $165.00 |
| | | | | | |
| **Cold Side** | | | | | |
| Heat exchanger for cold piping | | | | | |
| | | | | | |
| **Solid Piping, Fittings, Hot Side** | | | | | |
| CPVC Fittings - 3/4 inch -Tee | 4589K43 | McMaster | 3 | $12.37 | $37.11 |
| CPVC hex reducing bushing | 4589K91 | McMaster | 2 | $8.25 | $16.50 |
| CPVC Fittings - 3/4 inch – Coupling | 6826K92 | McMaster | 2 | $3.43 | $6.86 |
| CPVC Cement | 24285K31 | McMaster | 1 (8oz) | $5.40 | $5.40 |
| CPVC Primer | 18815K71 | McMaster | 1 (8oz) | $5.16 | $5.16 |
| 2" threaded CPVC piping | 6810K33 | McMaster | 2 | $1.98 | $3.96 |
| 6" threaded CPVC piping | 6810K73 | McMaster | 2 | $3.55 | $7.10 |
| 5' unthreaded CPVC piping | 6803K53 | McMaster | 1 | $11.78 | $11.78 |
| fittings - connect to pump and flowmeter | | | | | |
| CPVC hex reducing bushing | 4589K154 | McMaster | 3 | $8.25 | $24.75 |
| CPVC hex reducing bushing | 4589K155 | McMaster | 1 | $3.11 | $3.11 |
| CPVC Fittings - 3/4 inch -Elbow | 4589K23 | McMaster | 4 | $4.79 | $19.16 |
| CPVC Coupling | 4589K53 | McMaster | 4 | $7.47 | $29.88 |
| 12" threaded CPVC piping | 6810K43 | McMaster | 2 | $8.05 | $16.10 |
| 12" 3/8"dia threaded piping | 6810K41 | McMaster | 2 | $7.52 | $15.04 |
| 12" 1/2"dia threaded piping | 6810K42 | McMaster | 2 | $6.04 | $12.08 |
| 3/4" male/male threaded coupling | 6810K13 | McMaster | 4 | $1.62 | $6.48 |
| 1/4" male/male threaded coupling | 6810K111 | McMaster | 1 | $1.30 | $1.30 |
| hex reducing bushing | 4589K154 | McMaster | 1 | $8.25 | $8.25 |
| hex reducing bushing | 4589K109 | McMaster | 1 | $7.48 | $7.48 |
| 1/4" barbed fitting | 5116K87 | McMaster | 1 | $3.17 | $3.17 |
| 1/2" barbed fitting | 5372K132 | McMaster | 1 | $5.33 | $5.33 |
| 1/2" female coupling | 4589K52 | McMaster | 1 | $5.79 | $5.79 |
| 3/4" barbed fitting | 5047K22 | McMaster | 16 | $3.00 | $48.00 |
| 1/2" nipple | 6810K12 | McMaster | 1 | $1.35 | $1.35 |
| 3/4" Tee | 4589K43 | McMaster | 5 | $12.37 | $61.85 |
| hex reducing bushing | 4589K154 | McMaster | 2 | $3.11 | $6.22 |
| shutoff valve: disconnect 3/4" hot side | 4724K72 | McMaster | 2 | $11.55 | $23.10 |
| 1" coupling | 4589K54 | McMaster | 1 | $8.50 | $8.50 |
| hex reducing bushing | 4589K154 | McMaster | 1 | $8.25 | $8.25 |
| 3/4" barbed fitting | 5047K22 | McMaster | 4 | $3.00 | $12.00 |
| hex reducing bushing | 4589K91 | McMaster | 4 | $5.25 | $21.00 |
| hex reducing bushing | 4589K78 | McMaster | 2 | $3.11 | $6.22 |
| 90deg barbed elbows | 5047K86 | McMaster | 10 | $3.13 | $31.30 |
| 3/8" female coupling | 4596K69 | McMaster | 2 | $3.10 | $6.20 |
| 3/8" pipe x4"long | 4882K52 | McMaster | 1 | $2.81 | $2.81 |
| 1/2" pipe x4" long | 4882K53 | McMaster | 1 | 1.07 | $1.07 |
| 1/2" female coupling | 4589K52 | McMaster | 1 | $5.79 | $5.79 |
| Low-Pressure CPVC Ball Valve | 4724K72 | McMaster | 4 | $11.55 | $46.20 |
| 3/4" | | | | | |
| | | | | | |
| **Fittings, Cold Side** | | | | | |
| Thogus Adapter 3/4" barb nylon | TA101212/N | Grainger | 1 pack of 10 | $7.61/pack fo 10 | $7.61 |
| fittings - connect to pump | | | | | |
| fittings - connect to flow meter | | | | | |
| Piping adaptor for pump (1/2" to 3/4") | 6826K141 | McMaster | 2 | 5.69 | $11.38 |
| 3/8" barbed fitting | 5372K149 | McMaster | 1 | $5.65 | $5.65 |
| pipe adaptor | 5372K523 | McMaster | 1 | $3.86 | $3.86 |
| 3/8" coupling | 4596K69 | McMaster | 2 | $3.10 | $6.20 |
| 3/4" barbed fitting | 5372K133 | McMaster | 1 | $6.24 | $6.24 |
| 3/4" pipe nipple | 4882K14 | McMaster | 1 | $0.74 | $0.74 |
| hex reducer | 4880K345 | McMaster | 2 | $0.87 | $1.74 |
| PVC tees | 4596K33 | McMaster | 2 | $4.09 | $8.18 |
| pipe adaptor | 2974K277 | McMaster | 1 | $7.89 | $7.89 |
| | | | | | |
| **Pumps** | | | | | |
| Feed Pump | 72010-60 | Cole-Parmer | 1 | 484 | $484.00 |
| Condensate Pump | 72010-30 | Cole-Parmer | 1 | 261 | $261.00 |
| | | | | | |
| **Valves** | | | | | |
| Bypass Valve | 4017T14 | McMaster | | 33.58 | $33.58 |
| Needle Valves | 4995K19 | McMaster | 2 | 81.85 | $163.70 |
| Other Valves | | | | 200 | $200.00 |
| Cold Side Valve | 9762K32 | McMaster | 1 | 9.58 | $9.58 |
| Brass Check Valve | 7768K41 | McMaster | | 35.01 | |
| | | | | | |
| **Misc** | | | | | |
| Stuff | | McMaster | | | $500.00 |
| Other Misc | | McMaster | | | $500.00 |
| 3 prong plug | 7196K31 | McMaster | 1 | 9.1 | $9.10 |
| | | | | | |
| **Connections** | | | | | |
| Bolts | 90911A637 | McMaster | 2 bags | 6.10/per bag of 10 | $12.20 |
| Square Head Bolts | 91465A176 | McMaster | 4bags | 4.48/per bag of 5 | $17.92 |
| More Bolts | | MSC | | | |
| Wing Nuts | 92771A535 | McMaster | 2 | $5.96/pack of 5 | $11.92 |
| Nuts | 93827A225 | McMaster | 1 bag | 8.05/per bag of 100 | $13.11 |
| Washers | 93852A104 | McMaster | 2 bags | 3.31/per bag of 10 | $6.62 |
| O-ring joining glue | 7569A22 | McMaster | 1 | 24.57 | $24.57 |
| | | | | | |
| **Module Components** | | | | | |
| Feed Channel: Polycarbonate 12x24" Impact Resistant, 1" | 8574K5 | McMaster | 1 | 396.76 | $700.00 |
| Condensate Channel Polycarbonate 12x24", 1/2" | 8574K5 | McMaster | 1 | 52.63 | $200.00 |
| Cooling Channel | 8560K382 | McMaster | 1 | 121.76 | $300.00 |
| .04" thick polycarbonate solid sheet. 12x24" | | Amazon | 4 | 10 | $40.00 |
| Wire mesh spacer | 9275T65 | McMaster | | | |
| | | | | | |
| | | | | | |
| **Visulization** | | | | | |
| GE Panametrics contact transducer 10 MHz - V111-RB | U8403032 | Olympus | 1 | 307.8 | $307.80 |
| 2-channel PC oscilloscope, 100 MHz bandwith, AWG | OMSP-3205 | Omega | 1 | 1153 | $1,153.00 |
| Pulser/Reciever - Panametrics 5057PR-15-U | U8120080 | Olympus | 1 | 3213 | $3,213.00 |
| Camera | | | 1 | 300 | $300.00 |
| | | | | | |
| **Membrane** | | | | | |
| Immobilon-PSQ Membrane (0.2 µm pore size) PR02533 | ISEQ 000 10 | Millipore | 2 | $289.54 | $579.08 |
| | | | | | |
| **Salts** | | | | | |
| Na2CO3, 1 kg | 791768-1KG | Sigma-Aldrich | 1 | $106.50 | $106.50 |
| NaCl, 12 kg | 746398-12KG | Sigma-Aldrich | 1 | $185.00 | $185.00 |
| CaCl2, 1kg | C1016-500G | Sigma-Aldrich | 2 | $51.80 | $103.60 |
| Na2SO4, 1kg | 746363-1KG | Sigma-Aldrich | 1 | $92.60 | $92.60 |
| | | | | | |
| **Other to add** | | | | | |
| support mesh for membrane | | | | | |
| Al T-Slot Extrusion | 47065T209 | McMaster | 4x2' | $8.35 | $33.40 |
| UN-Compliant Plastic Shipping Pail | 40015t41 | $13.73 | 4 | | $54.92 |
| Magnetic Stirring Bar | 5678K129 | McMaster | 1 | $20.35 | $20.35 |
| Replacement filters | 7191K11 | McMaster | 5 | $4.78 | $23.90 |

### 12.5.3 SOLIDWORKS DESIGN

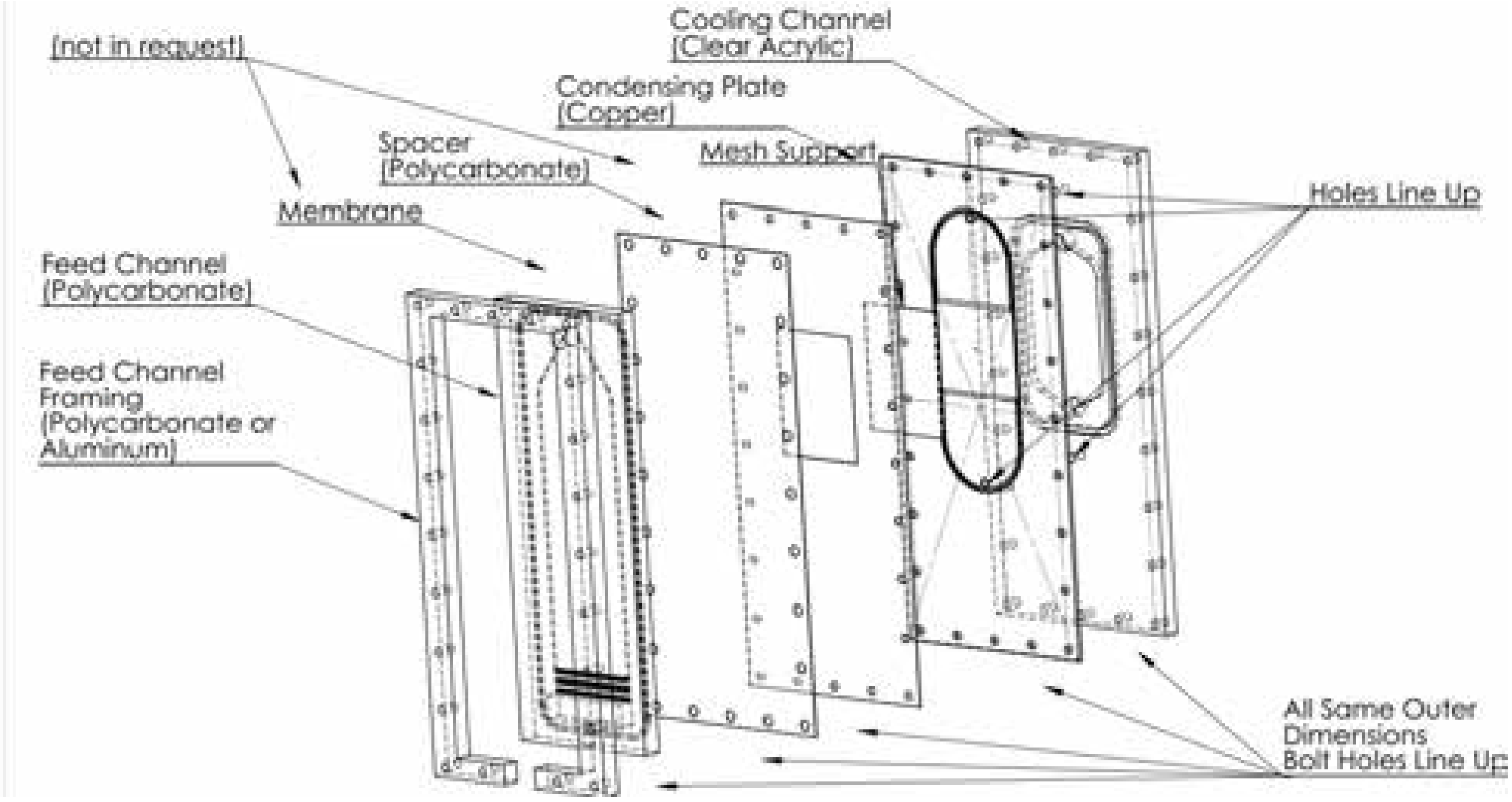

**Figure 12.10.** Module Assembly

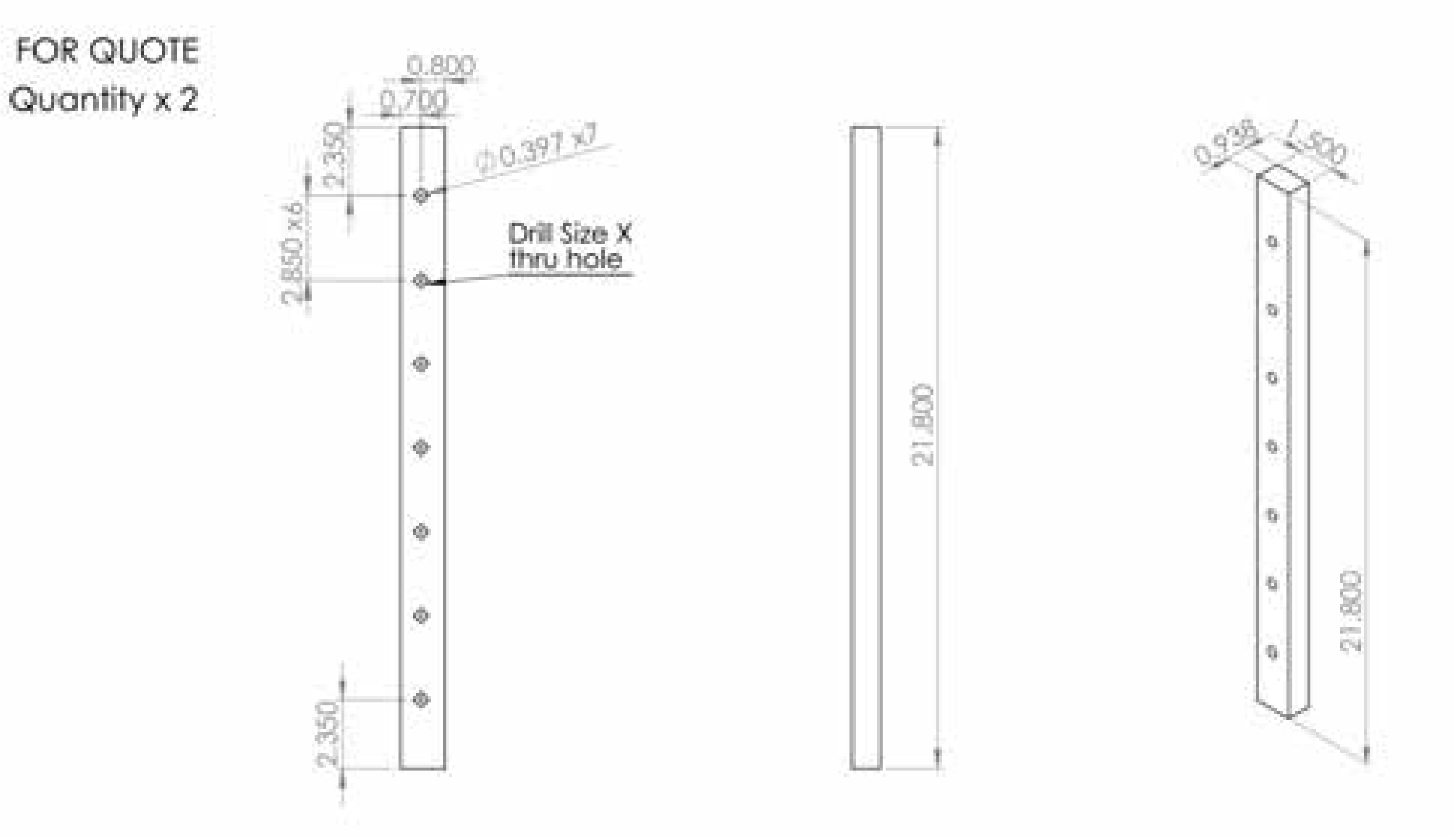

**Figure 12.11.** Feed Plate Frame 1

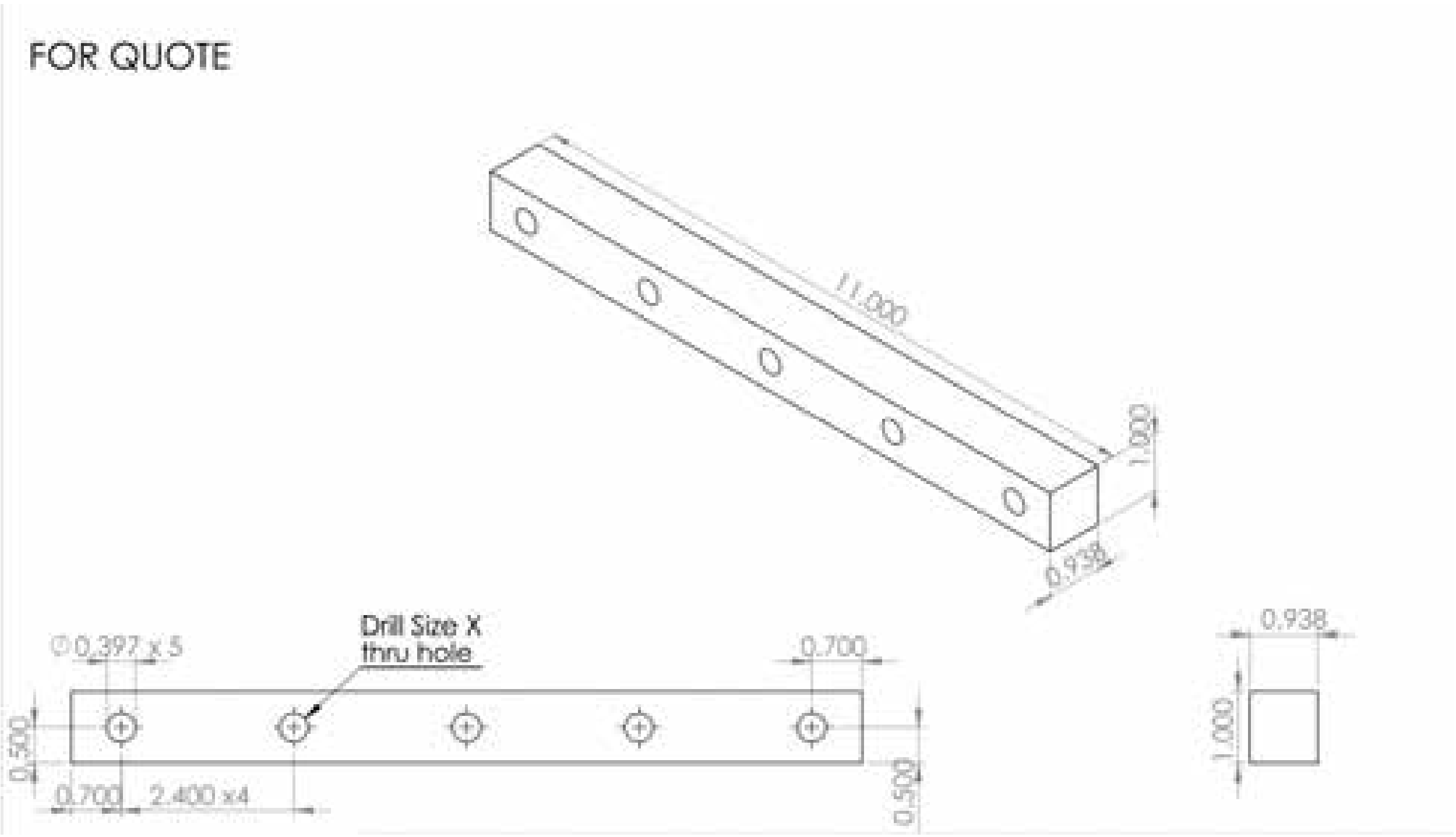

**Figure 12.12.** Feed Plate Frame 2

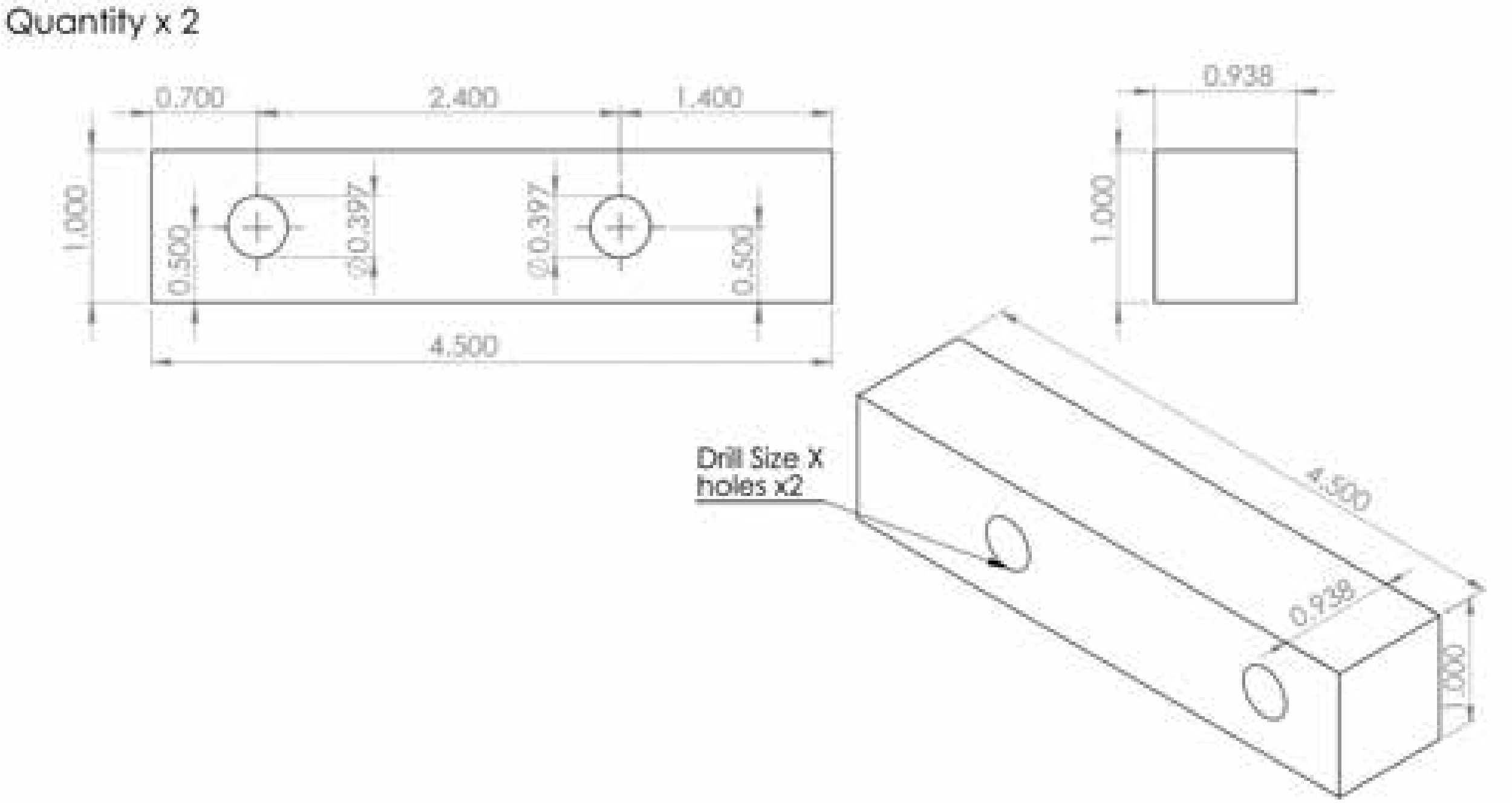

**Figure 12.13.** Feed Plate Frame 3

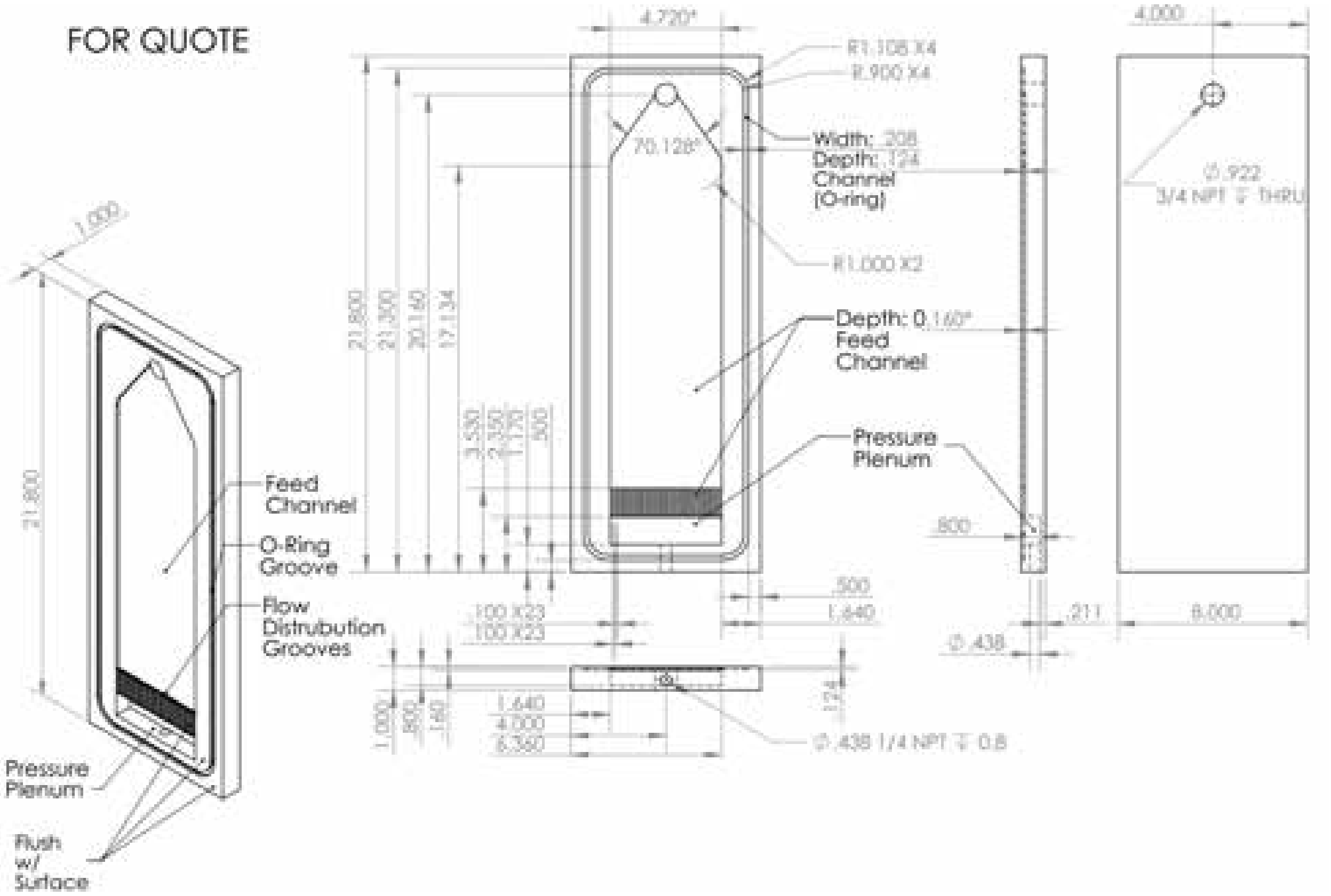

**Figure 12.14.** Feed Plate

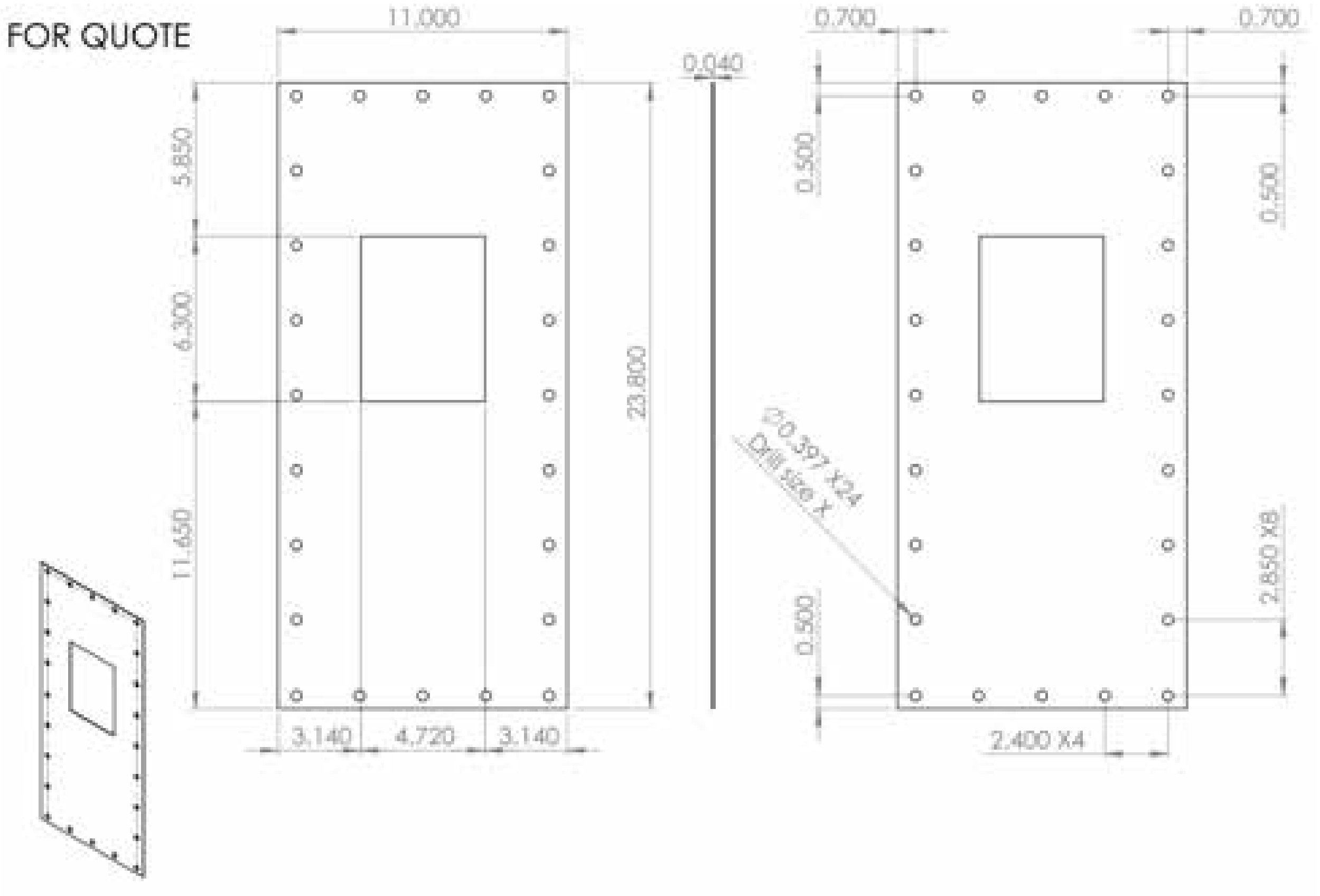

**Figure 12.15.** Air Gap Spacer

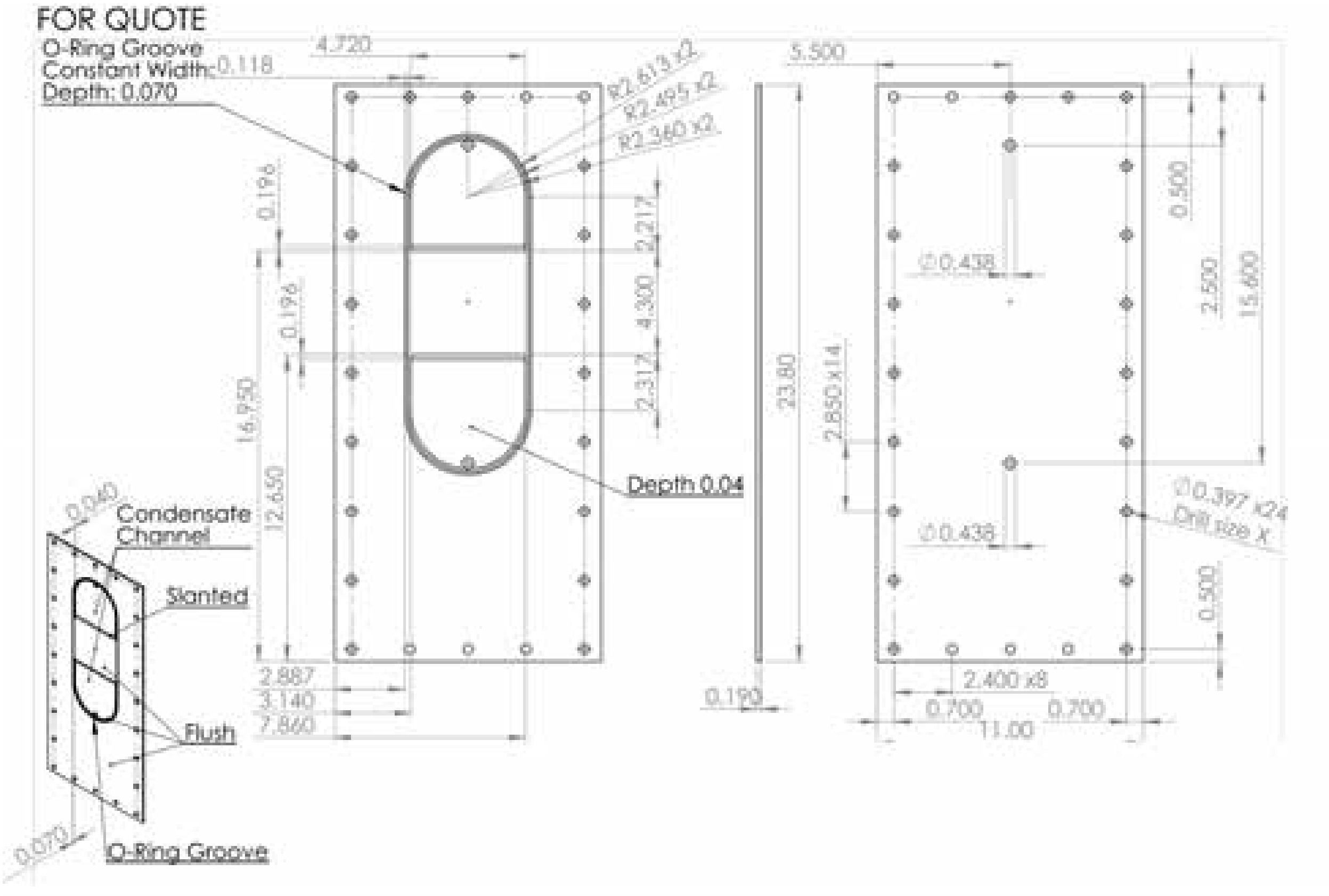

**Figure 12.16.** Condensing Plate

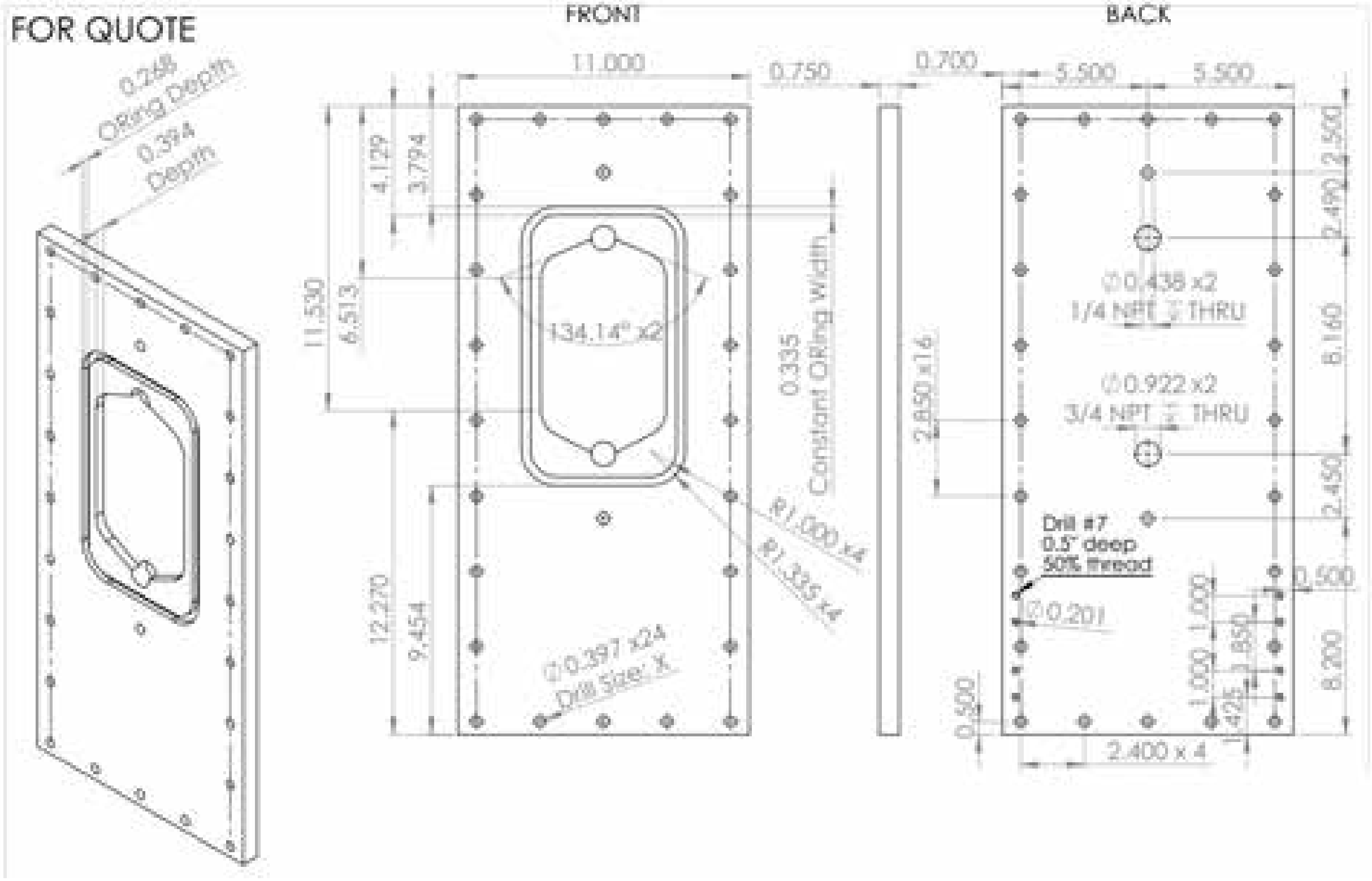

**Figure 12.17.** Cold Channel Plate

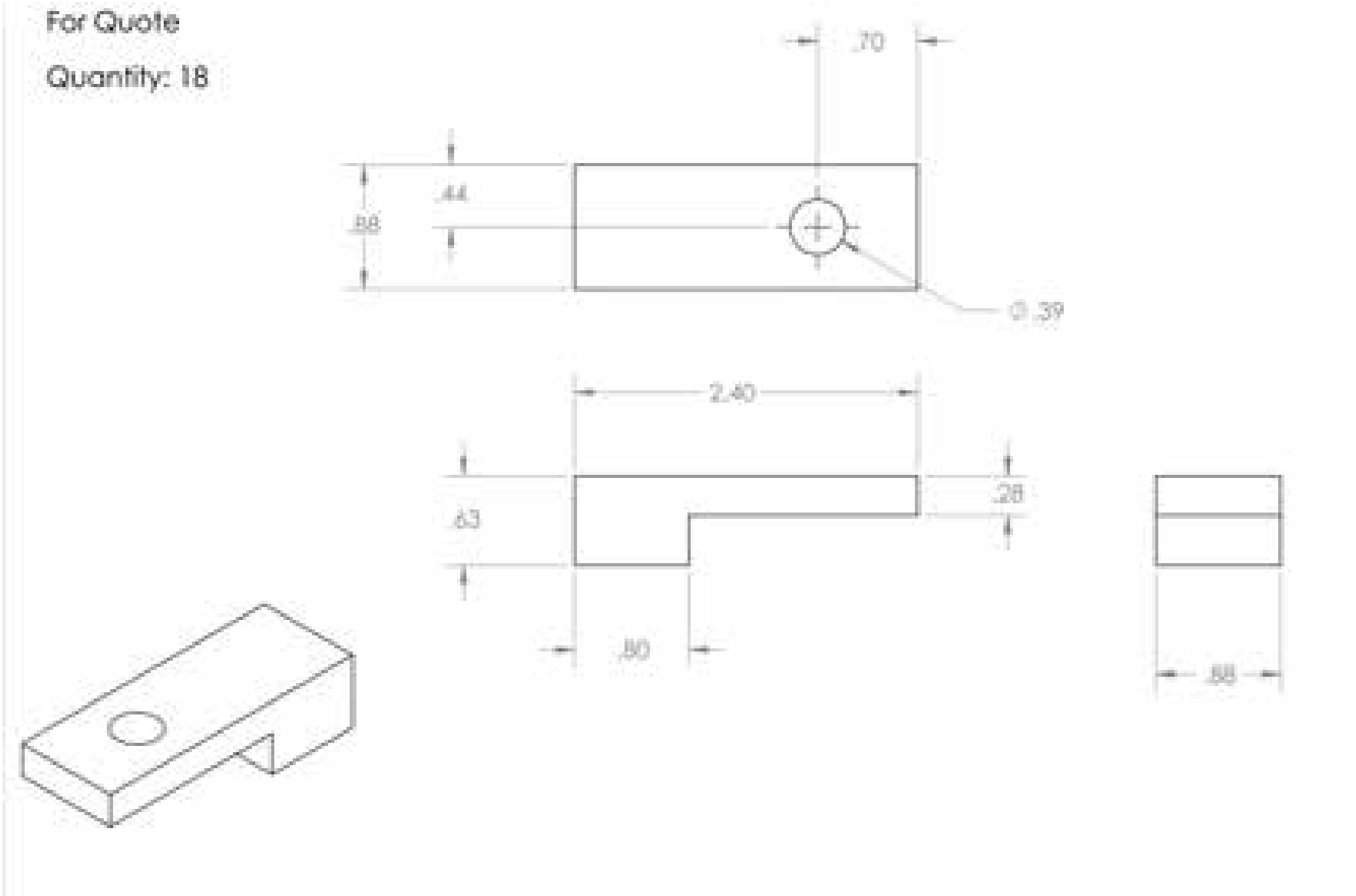

**Figure 12.18.** Clamps for Feed Module

## 12.6 MD Experiment Procedure

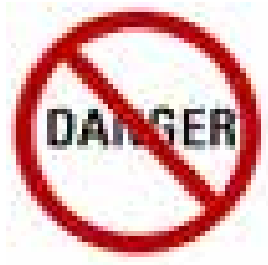

1. **Safety Check**

   Tank is sufficiently full of water, no spills/exposed salts, no disconnected pipes or wires.

   Clean feed loop (empty, refill with DI or tap, run for ~30 min, replace w/ DI)

-----------------------------------------------Phase 1 --Prep System---------------------------------------

2. **Turn on heater**

   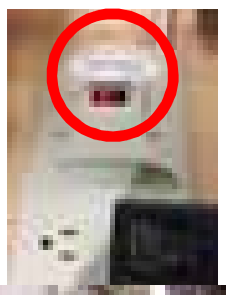
   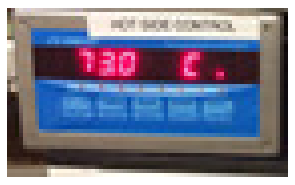

   (since it takes time to heat up). Requires turning on heating controller and setting a temperature. Heater power plug to be connected directly to supply.

3. **Check Pipes/Valves/Module**

   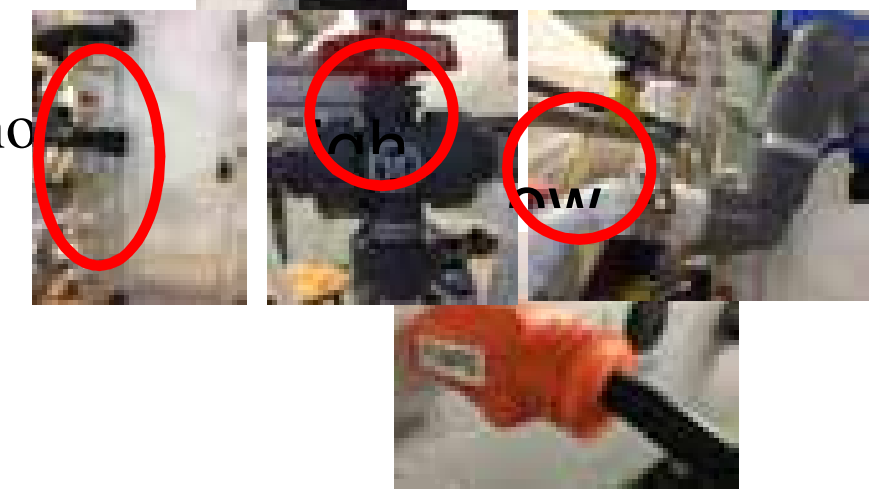

   Ensure the module is ready (clamped, membrane present) and the correct flow paths (valves) are cho

4. **Turn on Pumps**

   Plug in pumps. (Done for even circulation.)

   Send through bypass only typically

-----------------------------------------Phase 2 -- Start Experiment -----------------

5. **Set experiment conditions**

   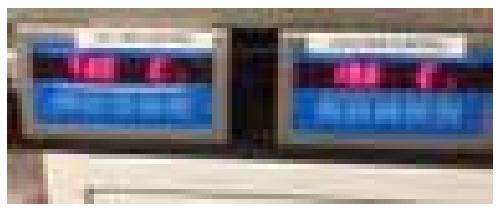

   Set temperatures, flow rate

6. **Prepare Salts**

   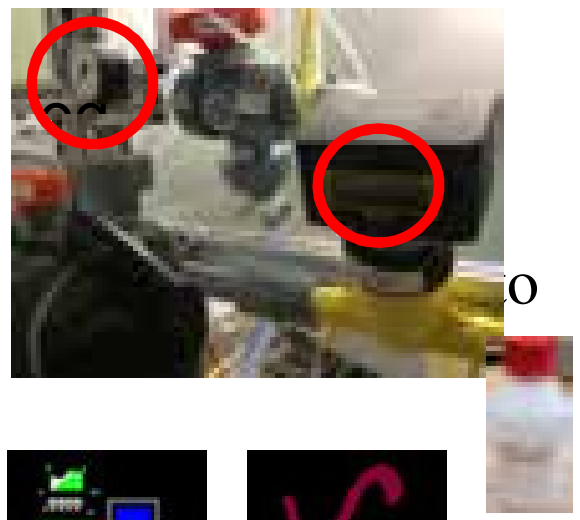

   Measure, pre-dissolve in DI water as o feed tank

7. **Start data recording**

   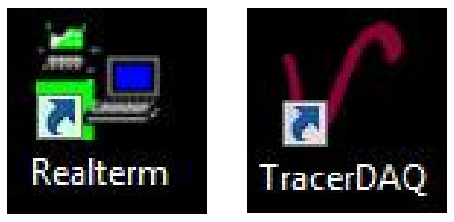
   

   Turn on measurement for permeate mass, temperature, and salinity.

   (see other page)

---------------------------------Phase 3 –Experiment Operation-------------------------------------

8. **Periodically Check up**

   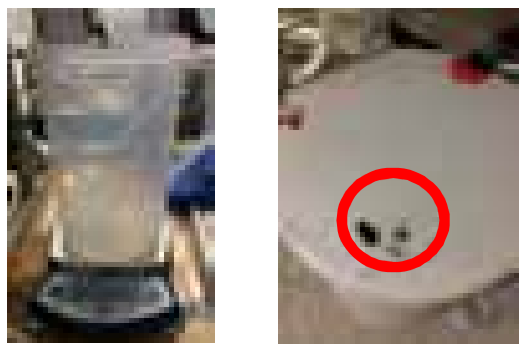

   Every ~8-9 hours, empty permeate tank into feed. Take samples or record salinity as needed. Check no issues have occurred (stable flow rates, temp, no leaks, water levels of tank & bag ok, etc)

------------------------------------------Phase 4 –Shut Down-----------------------------------------

9. **Stop Recording programs, Turn off Pumps & Electronics, Dry Membrane**

## 12.7 MD Experiment Recording

### 12.7.1 Permeate Mass Recording

----Phase 2 -- Start Experiment ---

Turn Scale on

Open RealTerm Program

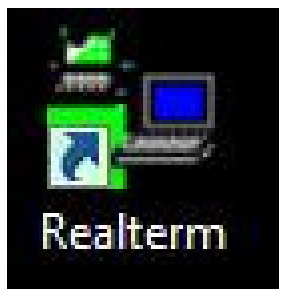

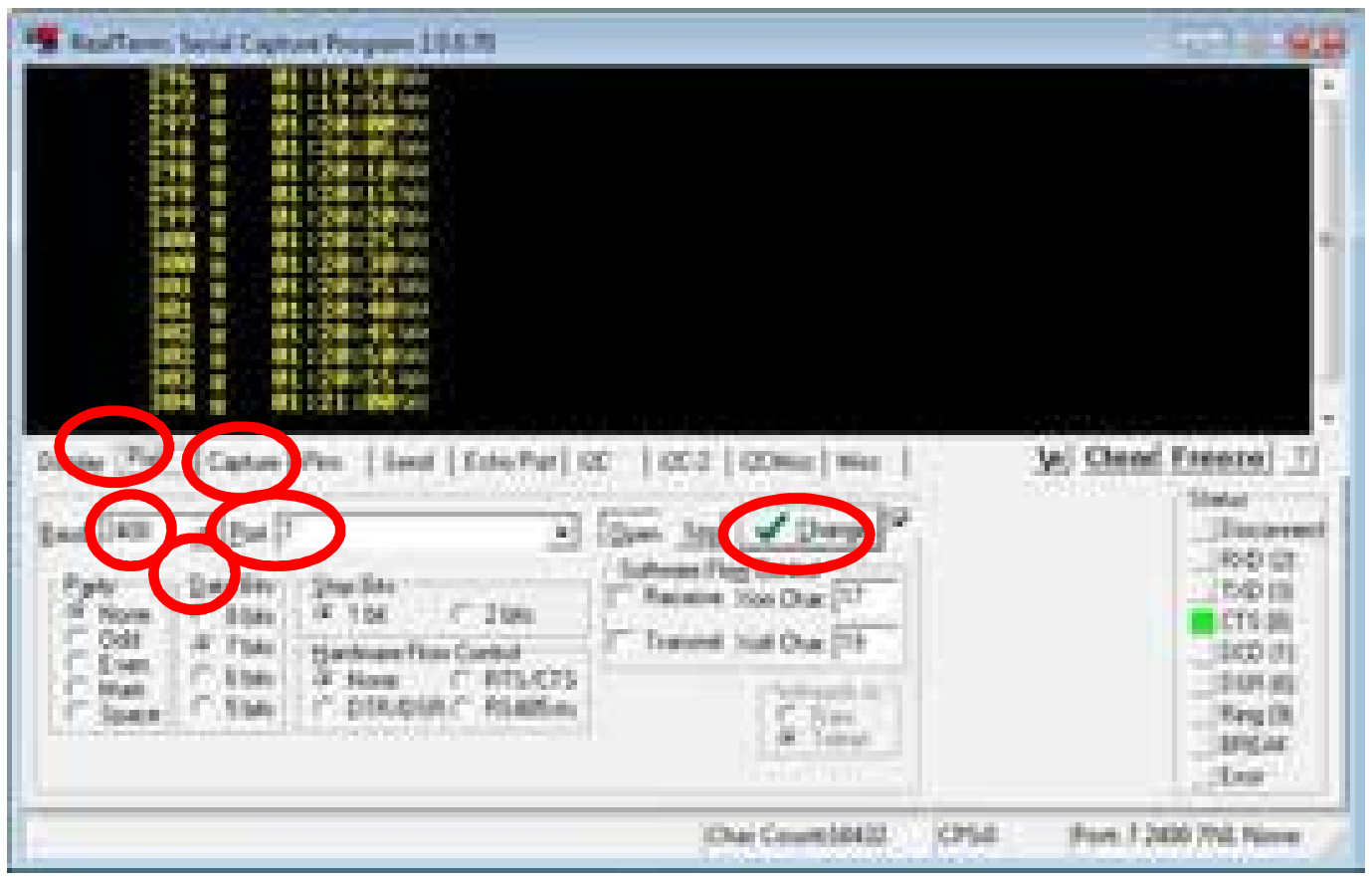

Port Tab

Set Baud field to 2400

Set Port # to 7

Set Data bits to 7

Press Change to confirm

(if yellow text doesn't update, restart computer)

Capture Tab

Click ... button

Go to desktop, Scroll down and click shortcut to Mass Scale Data

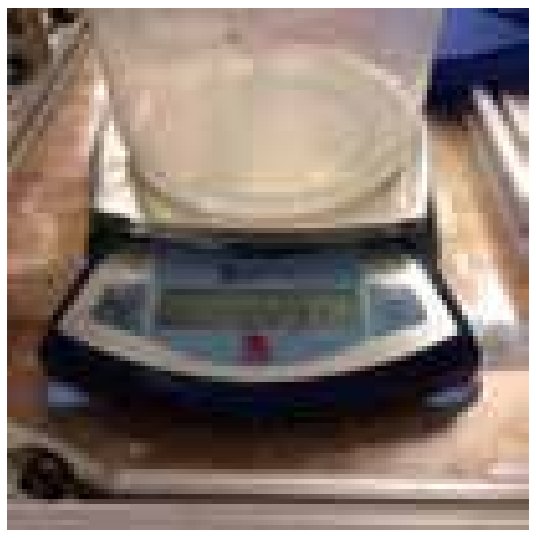

Goes to: Dropbox\Membrane Distillation\Data\Mass_Scale_Data )

Make a new folder for the experiment

Save the file according to the Naming Convention

eg Silica SI_3 3GPM 70C 50C 7_18_14

Set YMDHS under time stamp

Start recording by clicking Start Overwrite or Start Append

Start button becomes stop button

----Phase 4 –Shut Down----

Click "stop capture"

### 12.7.2 TEMPERATURE RECORDING

----Phase 2 -- Start Experiment ---

Open TracerDAQ

Strip Chart

Press play button, records for 4 hours

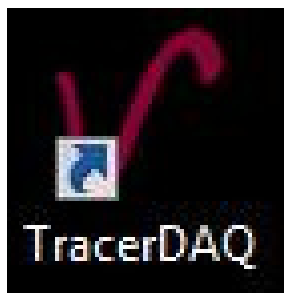

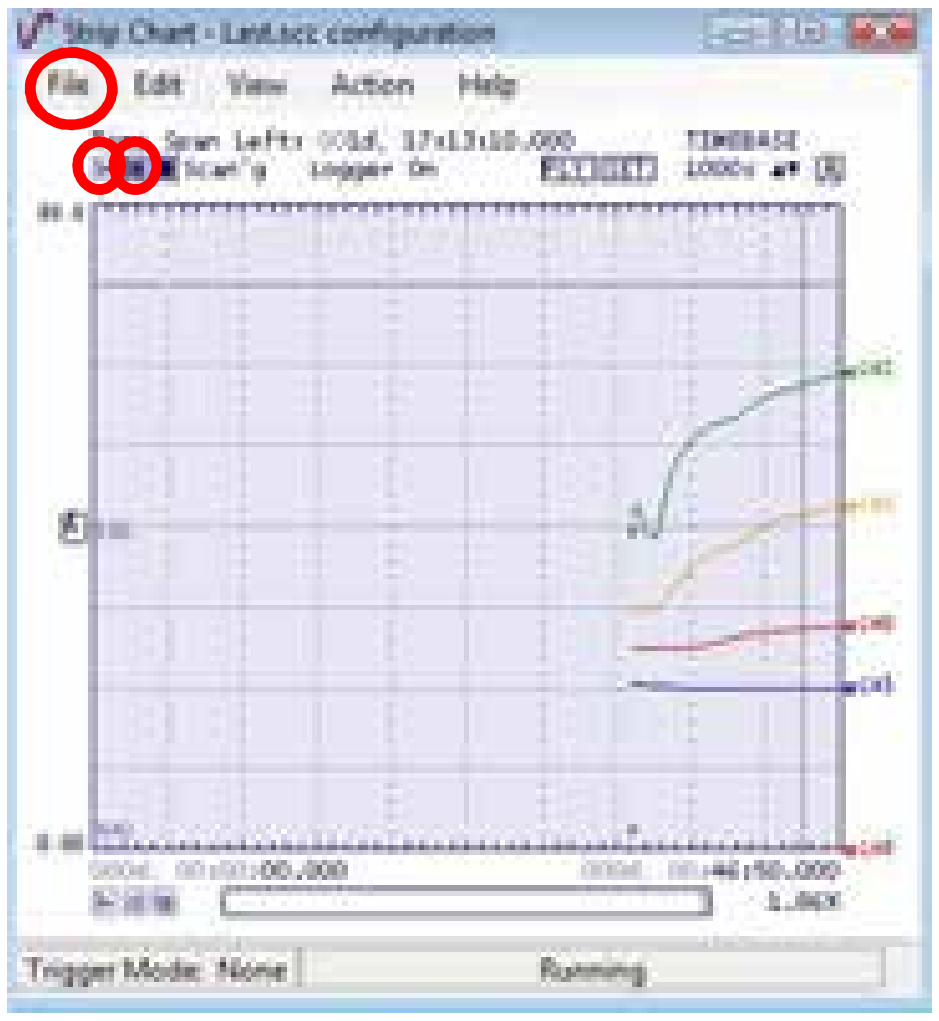

File, save, navigate to the same folder save as csv. Use naming convention like that for mass.

Note: for above 4 hours, go to Edit --> scan rate --> change form 0.2 to a smaller value & change Acquire data for 4 hours to a larger number

----Phase 4 –Shut Down----

Click the pause button

### 12.7.3 CONDUCTIVITY PROBE RECORDING

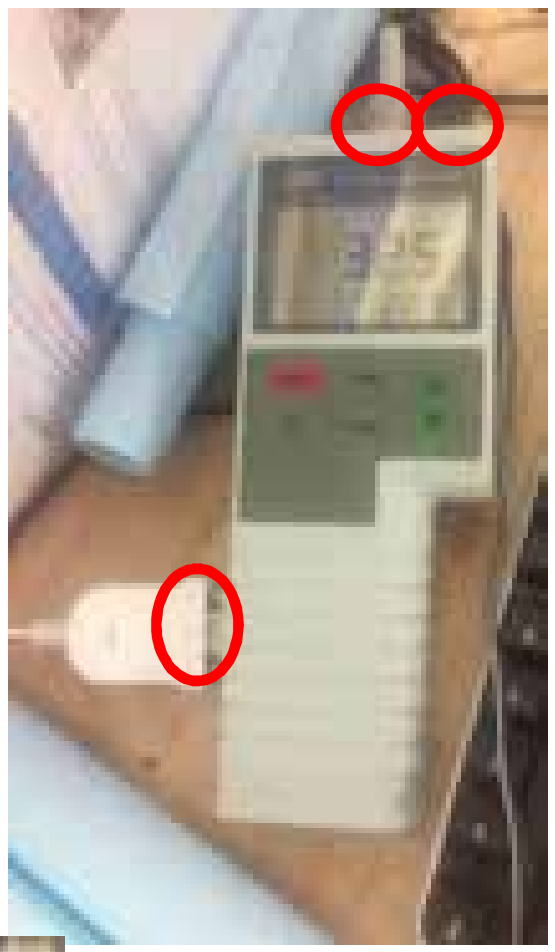

-Check all 3 ports are connected ( 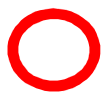 )

-turn meter on

-If probe is changed/uncalibrated:

-hit CAL 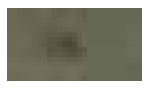 and press MODE 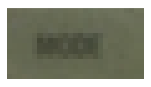 3 times

-Use up and down arrows 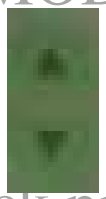 to pick the probe: 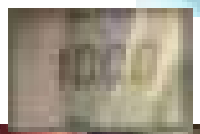

- Pure permeate: black probe, set to 0.1 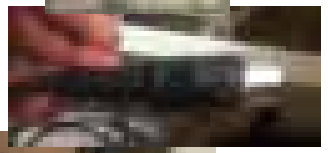

- Saline: white probe, set to 10.0 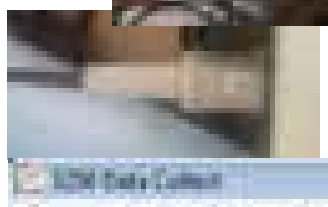

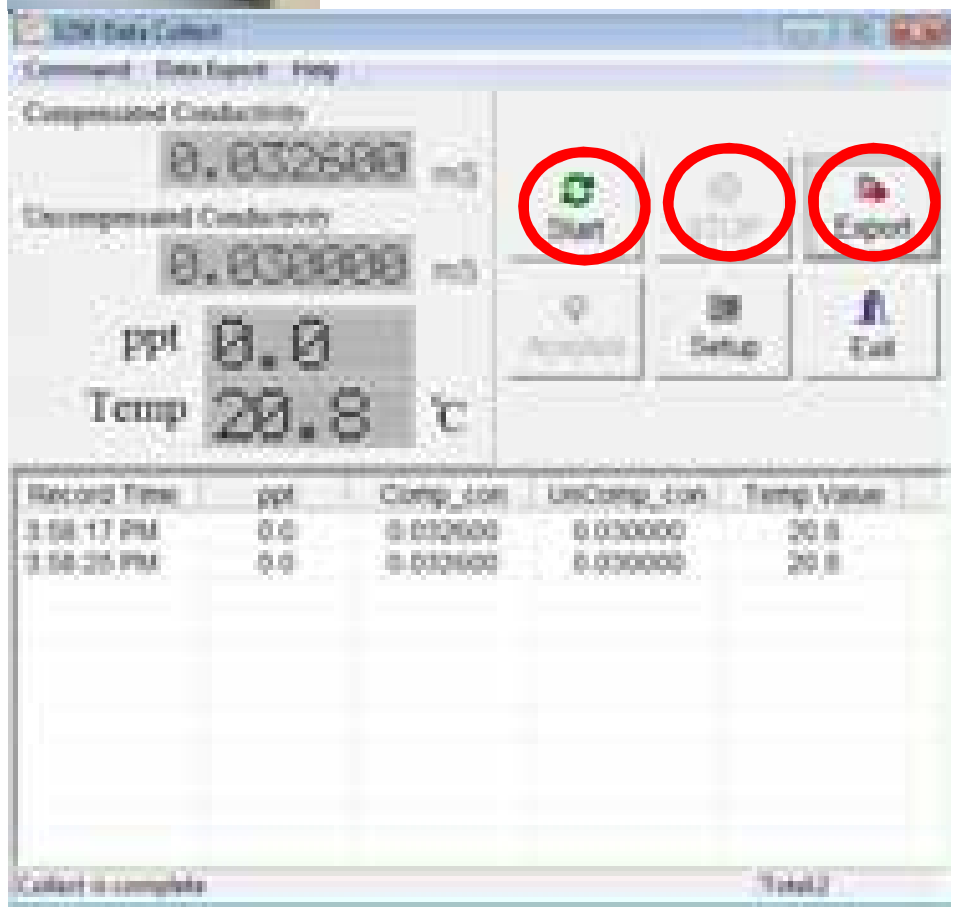

**Beginning Experiment:**

-Open Data Collection program 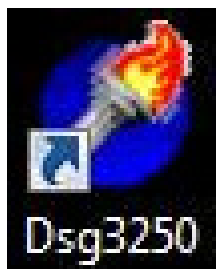

Press Start to record

**Ending Experiment**:

-Press STOP

-Press Export, save as txt into the directory with the other files

## 12.8 MD Controlling Operating Conditions

### 12.8.1 Temperature

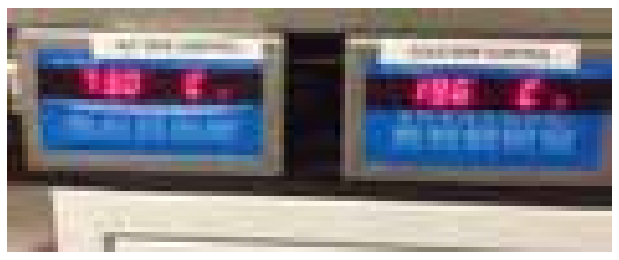

Long-press SP/PR to enter “Entr. Stpt” mode (A). Choose digit (B) and change (C). For 70C set 73.4C, for 60C set ~62.6C .. 40C set 42 C. Then hit SP/PR again (D)

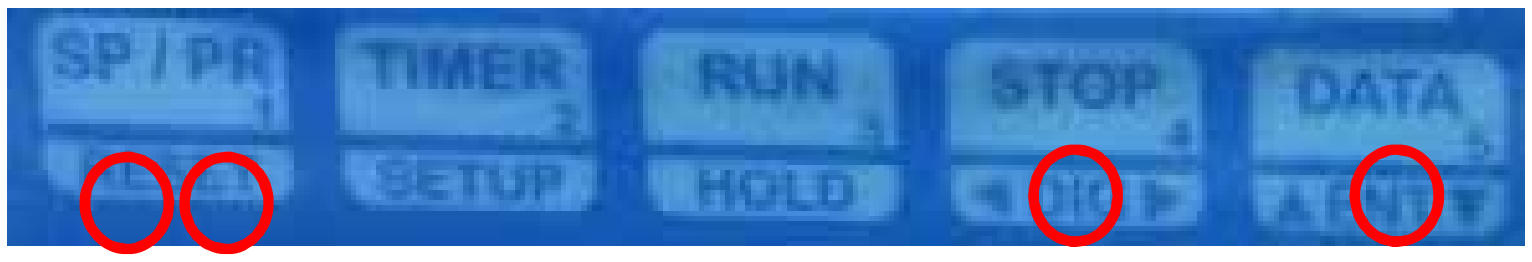

### 12.8.2 Flow Rate

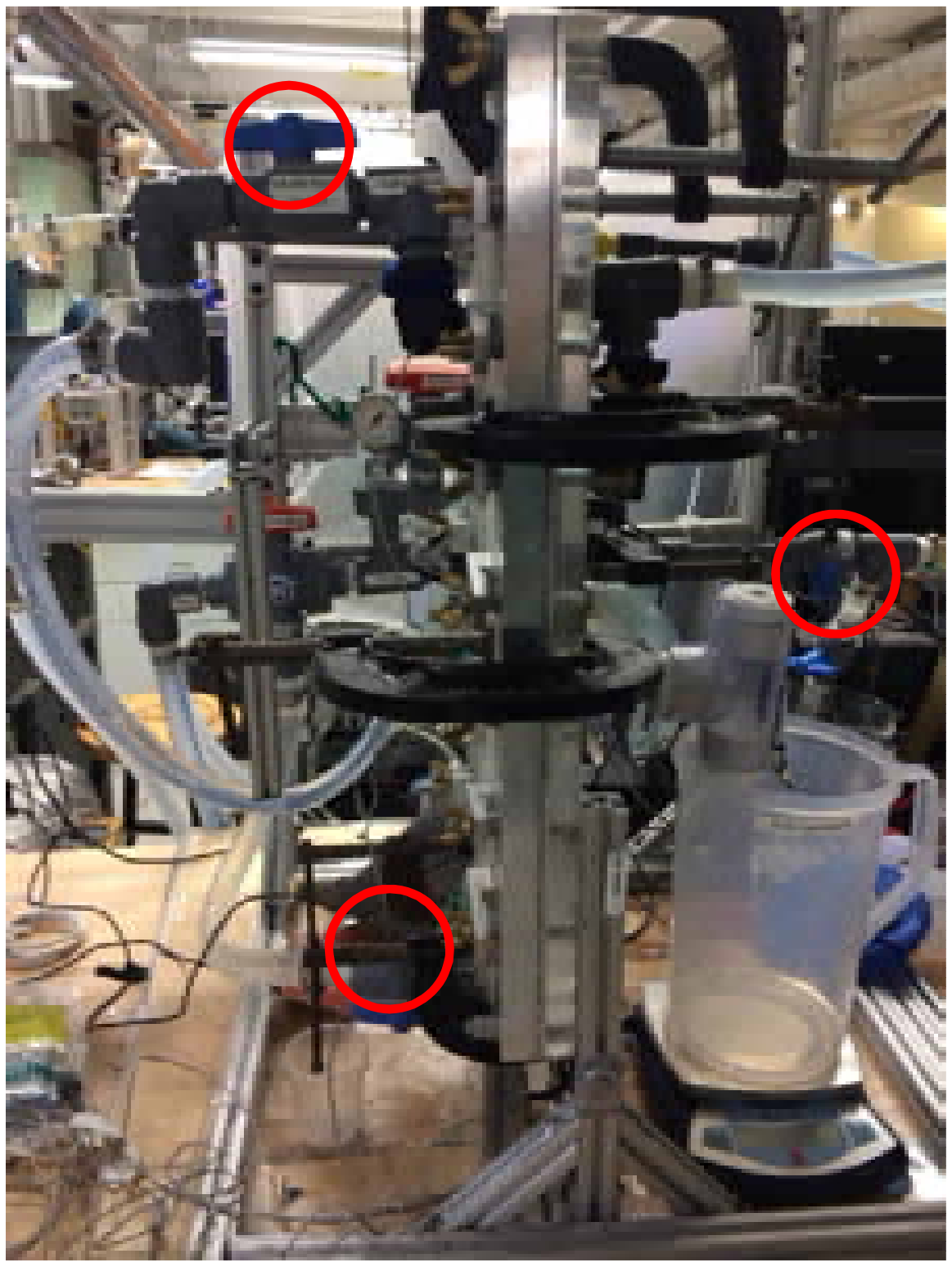

Flow rate is controlled with valves. Slightly close the valves before and after the module, and on the bypass loop.

Use the flow meter and pressure meter to adjust until conditions are met. (e.g. 3 PSI (default) and 3 GPM)

## 12.9 PhD Gantt Chart

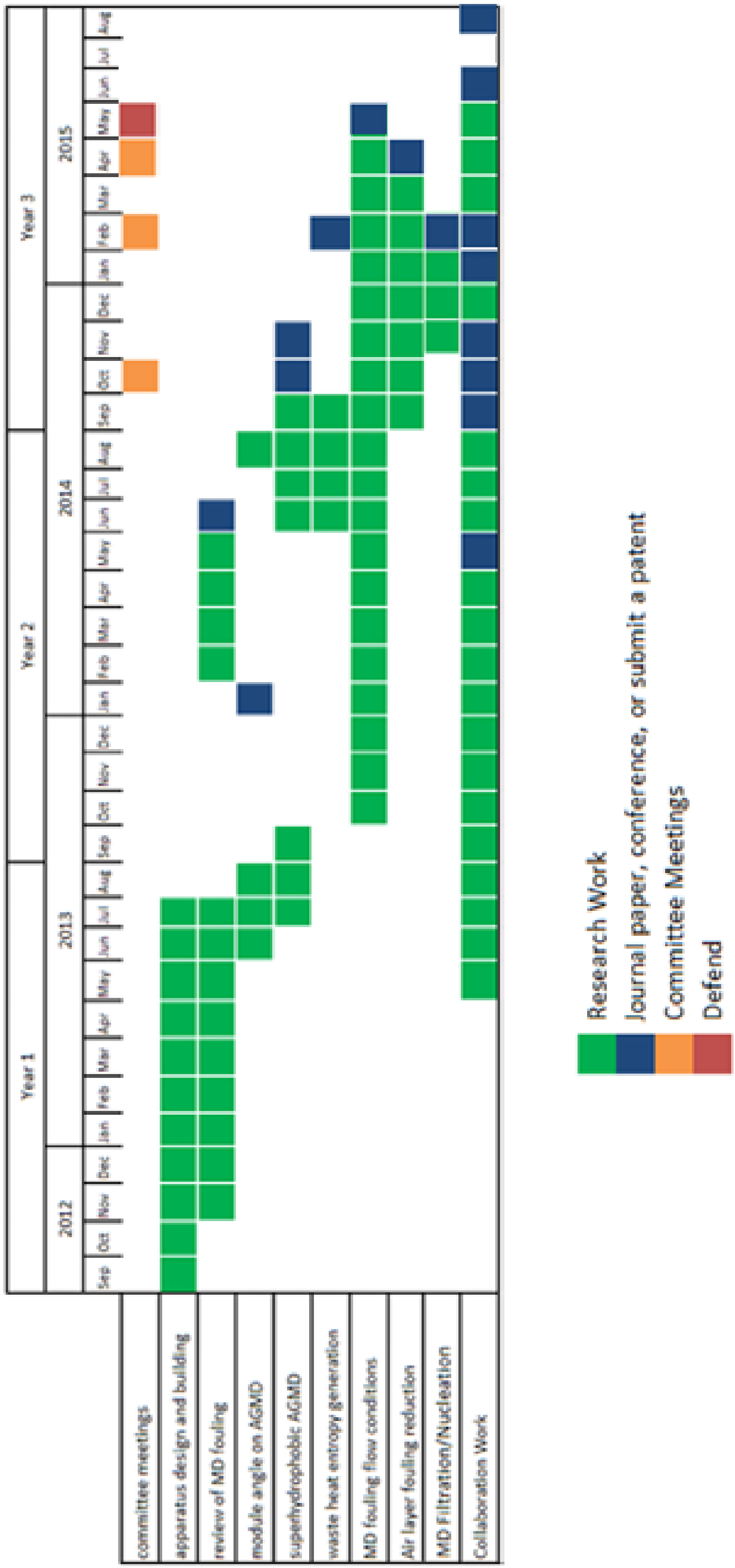

## 12.10 ENGINEERING EQUATION SOLVER CODE

```
"-----AGMD Modeling-------"
"Ed Summers, Jai Swaminithan, David Warsinger"
"June 17, 2011"
"Revised Oct 16, 2013"

"Modified for calculating dimensions of coupon sized scaling analysis apparatus -  with concentration polarization effect"
"Jaichander S."
"October 22, 2012"

"-----FUNCTIONS-----"
"-- Both Re and Sh based on Gnielinski's correlation --"
"Nu and Sh are needed in feed channel and cooling channel heat transfer calculations"

FUNCTION get_nusselt(Re, Pr, d_h, L)          "Nu"
{Returns nusselt number for appropriate range}
f := (0.79*ln(Re)-1.64)^(-2)
IF (Re<=2800) THEN get_nusselt := 7.54+((0.03*(d_h/L)*Re*Pr)/(1+0.016*((d_h/L)*Re*Pr)^(2/3)))
IF (Re>2800) THEN get_nusselt := ((f/8)*(Re-1000)*Pr)/(1+12.7*(f/8)^(1/2)*(Pr^(2/3)-1))

END

FUNCTION get_sherwood(Re, Sc, d_h, L)          "Sh"
{Returns Sherwood number for appropriate range}
f := (0.79*ln(Re)-1.64)^(-2)
IF (Re<=2800) THEN get_sherwood := 7.54+((0.03*(d_h/L)*Re*Sc)/(1+0.016*((d_h/L)*Re*Sc)^(2/3)))
IF (Re>2800) THEN get_sherwood := ((f/8)*(Re-1000)*Sc)/(1+12.7*(f/8)^(1/2)*(Sc^(2/3)-1))

END

FUNCTION  get_deltaP_long(m_dot,l,w,d,mu,rho_feed)     "Delta Pressure for any cell or for manifold"
{Returns deltaP across a long narrow channel}
d_h := 2*w*d/(w+d)
Re:= m_dot*d_h/(w*d*mu)
f := (0.79*ln(Re)-1.64)^(-2)
L_star := (2*(l)/d)/Re
f_app := ((3.44/(L_star)^(1/2))^2 + (f*Re)^2)^(1/2)/Re
K_channel := 4*f_app*(2*l/d_h)

get_deltaP_long := (K_channel)*(m_dot)^2/(2*rho_feed*(w*d)^2)
END

"-----DIRECTIVES------"
$UnitSystem SI C Pa J mass deg


"------------INPUTS--------------"
```

"--1. Numerical Parameters--"

n_cells = 120 "Number of disscrete cells that make up the length of the module"

"--2. Operational Parameters--"

m_dot_c_ex = 0.00[kg/s] {"Excess cooling water through condenser chamber, typically not used"}

"Yellow highlight indicates values to change to model experiments-"

m_dot_f_in = 3.453*0.063[kg/s] "Total Feed Mass Flow into Module" "(0.063 converts GPM to kg/s)"

m_dot_c_in = 3.675*0.063 [kg/s]

// T_f_in = 70[C] "Feedwater Inlet Temp"

T_c_in = 13[C] "Coolant stream inlet temperature"

S_in = 0.2 [g/kg] " Salinity of the feed (in ppt) "

P_hr_in = P_atm "Heat reservoir Pressure, for pressure drop calcs"

P_hr_out = P_atm

P_p = P_atm "Air gap pressure, typically same as feed water pressure (1 atm)"

P_c = P_atm "Coolant side pressure"

"--3. Membrane Properties--"

B = (16*10^(-7))[kg/m^2-Pa-s] "Membrane dist. coefficient"

delta_m = 0.0002[m] "Membrane Thickness"

xi = 0.8 "Membrane Porosity"

k_m = 0.2[W/m-K] "Membrane Polymer Thermal Conductivity (not including empty spaces)"

"--3. Module geometry --"

"--3a. Coupon sized membrane geometry--"

L = 0.16[m] "Module effective active length (in flow direction)"

w = 0.12[m] "Module/flow channel width"

d_ch = 0.004[m] "Flow channel depth"

d_gap = 0.00045 [m] "Air Gap depth" "May be altered to adjust mass transfer resistance"

d_cond = 0.01 [m] "Condenser channel depth" "This side can be bigger to reduce pressure drop required?"

n_sheets = 1 "Number of paralell membrane sheets, typically 1"

"-3b. Areas-"

dA = w*L/n_cells "differential membrane area of one cell"

```
A_m_tot = L*w*n_sheets                        "Total membrane area"


"-----QUANTATIES CONSTANT OVER MODULE LENGTH-----"

"---4. Fluid Properties---"

"Water - Feed"

h_fg = Enthalpy(Steam_IAPWS, P=P_atm, x=1)-Enthalpy(Steam_IAPWS, P=P_atm, x=0)
                                              "Enthalpy of vaporization." "To be used to
                                              calculate GOR only"

"Water - In the air gap region - subscript p for permeate"

h_f = Enthalpy(Steam_IAPWS, P=P_p, x=0)       "H2O enthalpy and density lookups"
h_g = Enthalpy(Steam_IAPWS, P=P_p, x=1)
rho_f = Density(Steam_IAPWS, P=P_p, x=0)
rho_g = Density(Steam_IAPWS, P=P_p, x=1)

"Air"
k_air = Conductivity(Air, T=(T_f_in+T_c_in)/2)     "Air Properties lookup"
M_H2O = MolarMass(Water)*convert(kg/kmol, kg/mol)
R = 8.3144[J/mol-C]     "Universal Gas constant -- Edited to make it J/mol-C to avoid unit error"
g = 0.8[m/s^2]                                "gravity"

D_wa = 1.97*10^(-5)*(P_f/P_p)*((273.15+40)/256)^1.685       "Diffusion of water in Air ins SI, fcn
                                                            of pressure"

" Cooling loop - Subscript c "

h_c_in = Enthalpy(Water, P=P_c, T=T_c_in)


{h_t_f = 2200[J/s-m^2-K]                      "Heat Xfer coefficient as input, typically not
used"}

"-----Calculated Heat Transfer coeff-----"
"Feed and Condenser Side                      Using T_h_in and T_c_in because variation in
temperatures along the channel length are small"

" Feed Side - calculations at T_f_in since deltaT across the membrane module is negligible "

mu_f = Viscosity(Water, P=P_f, T=T_f_in)      "viscosity, prandtly, conductivity, density lookups
for feed"
Pr_f = Prandtl(Water, P=P_f, T=T_f_in)
k_f = Conductivity(Water, P=P_f, T=T_f_in)
rho_feed = Density(Water, P=P_f, T=T_f_in)

d_h = 4*(w*d_ch)/(2*(w+d_ch))                 " feed hydraulic diameter, area, Re, friction
                                              factor, Nu"
A_f = w*d_ch
Re_f = (m_dot_f_in/n_sheets)*d_h/(A_f*mu_f)
f = (0.79*ln(Re_f)-1.64)^(-2)

Nusselt_f = get_nusselt(Re_f, Pr_f, d_h, L)
```

```
h_t_f = Nusselt_f*k_f/d_h

" Condenser side"

mu_c= Viscosity(Water, P=P_c, T=T_c_in)
Pr_c = Prandtl(Water, P=P_c, T=T_c_in)
k_c = Conductivity(Water, P=P_c, T=T_c_in)
rho_cond = Density(Water, P=P_c, T=T_c_in)

d_h_c = 4*(w*d_cond)/(2*(w+d_cond))
A_c = w*d_cond

Re_c = (m_dot_c_in/n_sheets)*d_h_c/(A_c*mu_c)
f_c = (0.79*ln(Re_c)-1.64)^(-2)

Nusselt_c = get_nusselt(Re_c, Pr_c, d_h_c, L)
h_t_c = Nusselt_c*k_c/d_h_c

" Average velocities in the streams "

v_f = m_dot_f_in/(rho_feed*A_f)                "Feed Velocity"

v_c= m_dot_c_in/(rho_cond*A_c)                 "Cooling Channel Velocity"

"----Calculated Mass transfer coefficients----"
" Only Feed side "

D_s_w = 1.5*10^(-9)      "Diffusivity of NaCl in water in m^2/s" "pink = check value"

Sc_f = mu_f/(rho_feed*D_s_w)                   "Schmidt & Sherwood numbers"
Sherwood_f = get_sherwood(Re_f,Sc_f,d_h,L)
k_mass = Sherwood_f*D_s_w/d_h                  "Mass transfer coefficient"

"---Conduction Through Membrane---"

K_cond = (k_m*(1-xi)+k_air*xi)/delta_m         "Membrane thermal conductivity"


" Others "

MW_solute = 60 [g/mol]                         "Molecular Weights"
MW_water = 18 [g/mol]

"-------MD Module Code----------"
"This part breaks the AGMD system into n cells and performs calculations in each cell.  This is a
1-D model, properties change as flow passes through the system."

"Initial inputs"

"Inlet"
m_dot_f[1] = m_dot_f_in/(1*n_sheets)           "Feed Mass flow rate"
T_f_b[1] = T_f_in                              "Feed temp"

delta[n_cells] = 0.00001[m]                    "Initial condensate film thickness." "small non-
                                               zero for computational reasons"
```

```
h_c_b[n_cells+1] = h_c_in                    "Coolant Inlet"

DUPLICATE i=1,n_cells

"Distance"
z[i] = i*(L/n_cells)

"Enthalpy-Temperature Relations"
h_f_b[i] = Enthalpy(Water, P=P_f, T=T_f_b[i])     "Feed Bulk"
h_f_m[i] = Enthalpy(Water, P=P_f, T=T_f_m[i])     "Feed Membrane"
h_p_m[i] = Enthalpy(Steam, P=P_p, T=T_a_m[i])"Air gap Memrbane"

h_c_b[i] = Enthalpy(Water, P=P_c, T=T_c_b[i])     "Coolant Bulk"
h_c_wall[i] = Enthalpy(Water, P=P_c, T=T_wall[i])"Collant channel wall"
h_c_i[i] = Enthalpy(Water, P=P_c, T=T_i[i])       "Condensate interface"

                                              "--Conservation Equations--"


                                              "Membrane"
                                              P_f_m[i] = P_SAT(Water, T=T_f_m[i])*(1 -
((x_f_m[i]/MW_solute)/((x_f_m[i]/MW_solute)+((1000[g/kg]-x_f_m[i])/MW_water))))
                                              "Raolt's law, vapor P, feed side"

                                              J[i] = B*(P_f_m[i]-P_p*x_a_m[i]) "Main Flux eqn.
   J = flux accross membrane. J = B*delta vapor pressure "

                                              "Feed Channel"


                                              x_f_b[i] =
S_in/(m_dot_f[i]/(m_dot_f_in/n_sheets))       "Concentration along flow direction"
                                              x_f_m[i] = x_f_b[i] * exp(J[i]/(k_mass*rho_feed))
                                              "Effect of concentration polorization. "

                                              h_fg_f[i] = Enthalpy(Steam, T=T_f_m[i], x=1)-
Enthalpy(Steam, T=T_f_m[i], x=0)              "Local h_fg at the membrane side where evap
                                              occurs"

q_out[i] = J[i]*(h_fg_f[i]) + q_m[i] - J[i]*(h_f_b[i]-h_f_m[i])
                                          "q_out = heat removed from feed side = Latent heat
                                          plus conduction losses and sensible cooling of feed"
T_f_m[i] = T_f_b[i] - (q_out[i]/h_t_f)            "BL Resistance"
q_m[i] = K_cond*(T_f_m[i]-T_a_m[i])               "Conductive loss through membrane"

m_dot_f[i+1] = m_dot_f[i]-J[i]*dA                 "conserve mass in bulk stream"
h_f_b[i+1] = h_f_b[i] - q_out[i]*dA/(m_dot_f[i])  "conserve energy in bulk stream (flow direction)"
                                                  "assumes average temp =

                                              "Air Gap"
q_conv[i] = q_m[i]                            "Only heat convected across gap is membrane
                                              conduction loss"
                                              "Sensible Cooling of vapor neglected"
                                              "Heat transfer from mass transfer considered
                                              separately"
```

```
(J[i]/M_H2O) = (c_a[i]*D_wa)/(d_gap-delta[i])*ln(1+((x_i[i]-x_a_m[i])/(x_a_m[i]-1)))
"Diffusion Equation"

c_a[i] = P_p/(R*(T_a_m[i]+T_i[i]+2*273.15[C])/2) "Changed unit of R to be C rather than K. Also changed 237.15 to 273.15 from Ed's original value"

T_a_m[i] - T_i[i] = (q_conv[i]/k_mix[i])*(alpha_mix[i]*rho_mix[i]/J[i])*(exp(J[i]/(rho_mix[i]*alpha_mix[i])*(d_gap-delta[i]))-1)

"Heat convection equation"

T_i[i] = T_SAT(Water, P=(P_p*x_i[i]))           "Condensate interface temp"
x_i[i] = ((w_i[i]/0.662)/(1+(w_i[i]/0.662)))     "Condensate mass ratio"

"-Average Quantaties in Gap-"
"evaluated at average temperature"
x_av[i] = (x_a_m[i] + x_i[i])/2     "Average vapor mole fraction, average humidity ratio, average temp,"

x_av[i] = ((w_av[i]/0.662)/(1+(w_av[i]/0.662)))     "Note: solving in terms of w_av may make code run faster"

T_av[i] = (T_i[i]+T_a_m[i])/2

"Conductivity, cp, density,"
k_mix[i] = Conductivity(AirH2O, P=P_p, T=T_av[i], w=w_av[i])

c_p_mix[i] = Cp(AirH2O, P=P_p, T=T_av[i], w=w_av[i])

rho_mix[i] = Density(AirH2O, P=P_p, T=T_av[i], w=w_av[i])

alpha_mix[i] = k_mix[i]/(c_p_mix[i]*rho_mix[i])

"-Condensate Stream-"
T_av_film[i] = (T_i[i]+T_wall[i])/2
nu_f[i] = Viscosity(Water, P=P_p, T=T_av_film[i])/rho_f

k_film[i] = Conductivity(Water, P=P_p, T=T_av_film[i])

J[i]*dA = g*(rho_f-rho_g)/(3*nu_f[i])*w*(delta[i-1]^3-delta[i]^3)

"Condesnation film thickness"

h_fg_c[i] = Enthalpy(Steam, T=T_i[i], x=1)-Enthalpy(Steam, T=T_i[i], x=0)

"h_fg at the condensate interface temp"

q_c[i] = J[i]*h_fg_c[i] + q_m[i]                 "heat absorbed by condensate"
q_c[i] = k_film[i]/delta[i]*(T_i[i]-T_wall[i])   "film conduction resistance"

"-Condenser Channel-"
T_wall[i] = T_c_b[i] + ((q_c[i] + J[i]*(h_c_i[i]-h_c_wall[i]))/h_t_c)

"BL resistance"
h_c_b[i+1] = h_c_b[i] - (q_c[i]+J[i]*(h_c_i[i]-h_c_wall[i]))*dA/(m_dot_f[1]+(m_dot_c_ex/1))
```

```
                                                "conserve energy in bulk stream"
END                                             "Ends iteration through n cells"

"---Post Processing Steps---"
" T out of cold fluid "
h_c_b[1] = Enthalpy(Water,P=P_c,T=T_c_out)

"Heat Input"
P_atm = 101320 [Pa]
0 = Q_dot_in - m_dot_f_in*(Enthalpy(Water, P=P_hr_out,T=T_hr_out)) +
m_dot_f_out*(Enthalpy(Water,P=P_hr_in,T=T_hr_in))                    "in heater"

" Pump "
s_p_in = Entropy(Water, P=P_hr_out, T=T_hr_out)
h_p_in = Enthalpy(Water,P=P_hr_out, T=T_hr_out)
h_p_out_ideal = Enthalpy(Water,s=s_p_in,P=P_p_out)
h_p_out = h_p_in + (1/eta_pump)*(h_p_out_ideal - h_p_in)
h_p_out = Enthalpy(Water, P=P_p_out, T=T_p_out)

power_pump = m_dot_f_in * (h_p_out - h_p_in)

" Heat loss in piping "
h_outer_pipe = 10 [W/m^2*K]
r_outer = 0.02 [m]
t_pipe = 0.002 [m]
r_inner = r_outer-t_pipe
k_pipe_material = 10^(-2)*0.14 [W/m*K]

L_pipe_in  = 2 [m]
L_pipe_out = 2 [m]

R_outside = 1/(h_outer_pipe*L_pipe_in*(2*pi*r_outer))
R_inside = 1/(h_t_f * L_pipe_in * (2* pi *r_inner))
R_pipe = ln(r_outer/r_inner)/(L_pipe_in*2*pi*k_pipe_material)

T_inf = 27 [C]

Q_dot_loss_f_inlet_pipe = (((T_f_in+T_p_out)/2)-T_inf)/(R_outside+R_inside+R_pipe)

        m_dot_f_in*(Enthalpy(Water,P=P_p_out,T=T_p_out)-
        Enthalpy(Water,P=P_f_in,T=T_f_in)) = Q_dot_loss_f_inlet_pipe

Q_dot_loss_f_outlet_pipe = (((T_f_out+T_hr_in)/2)-T_inf)/(R_outside+R_inside+R_pipe)

        m_dot_f_out*(Enthalpy(Water,P=P_f_out,T=T_f_out)-Enthalpy(Water,P=P_f,T=T_hr_in))
        = Q_dot_loss_f_outlet_pipe

        Q_dot_loss_f_module = m_dot_f_in*Enthalpy(Water,P=P_f_in,T=T_f_in) - m_dot_f_out *
        Enthalpy(Water,P=P_f_out,T=T_f_out)
" Permeate Rate "
m_dot_p = sum(J[1..n_cells])*dA*n_sheets        "total permeate mass flow in SI units"
J_tot = m_dot_tot/A_m_tot                       "average membrane flux in SI units"
m_dot_f_out = m_dot_f[n_cells+1]*n_sheets       "feed output mass flow rate"
h_f_b[n_cells+1] = Enthalpy(Water, P=P_f_out, T=T_f_out)"feed outlet temperature is the
                                                unknown"
GOR = m_dot_p*h_fg/Q_dot_in                     "GOR"
```

```
RR = m_dot_p/m_dot_f_in                          "Recovery Ratio"

"Non-SI quantaties"
m_dot_tot = m_dot_p*3600[s/hr]                   "Permeate Flow in kg/hr"
V_tot = m_dot_tot/rho_f*8[hr/day]                "Total Volume flowrate in m^3/day for 8HRDAY!"
{EC = 626.94/GOR                                 "Specific energy consumption in kWh/m^3"}"----"

"--------Pumping Power----------"
K_in = 1.2 "Manifold Losses"
K_out = 1.2
L_dev = 25*d_h                                   " Development length "

" Manifold pressure losses"
{
d_manifold_large = 0.02
w_manifold_large = 0.03
l_manifold_large = w

N_fissure = 8 [-]
w_fissure = w/(2*N_fissure+1)
m_dot_fissure = m_dot_f_in/N_fissure
d_fissure = 0.002 [m]
l_fissure = 0.03 [m]

DELTAP_fissures = get_deltaP_long(m_dot_fissure,l_fissure,w_fissure,d_fissure,mu_f,rho_feed)
DELTAP_main_channel =
get_deltaP_long(m_dot_f_in,l_manifold_large,w_manifold_large,d_manifold_large,mu_f,rho_feed)
}

"Long thin channels -- Pressure drop through the membrane"
DELTAP_chan = get_deltaP_long(m_dot_f_in,L,w,d_ch,mu_f,rho_feed)
DELTAP_dev = get_deltaP_long(m_dot_f_in,L_dev,w,d_ch,mu_f,rho_feed)
DELTAP_manifolds = (K_in+K_out)*(m_dot_f_in/n_sheets)^2/(2*rho_feed*A_f^2)
DELTAP_total_module = DELTAP_manifolds + DELTAP_dev + DELTAP_chan
P_f_in - P_f_out =  DELTAP_total_module
P_f = P_f_out + ((DELTAP_manifolds/2) +DELTAP_chan)/2

"Estimated Pipe Properties"
eta_pump = 0.7
d_pipe = 2*r_inner
Re_pipe_in = m_dot_f_in*1*d_pipe/(pi*d_pipe^2/4*mu_f)
f_pipe_in = (0.79*ln(Re_pipe_in)-1.64)^(-2)

DELTAP_pipe_inlet =
f_pipe_in*(L_pipe_in/d_pipe)*(m_dot_f_in*1)^2/(2*rho_feed*(pi*d_pipe^2/4)^2)

Re_pipe_out = m_dot_f_out*1*d_pipe/(pi*d_pipe^2/4*mu_f)
f_pipe_out = (0.79*ln(Re_pipe_out)-1.64)^(-2)

DELTAP_pipe_outlet= f_pipe_out*(L_pipe_out/d_pipe)*(m_dot_f_out*1)^2/(2*rho_feed*(pi*d_pipe^2/4)^2)
P_f_in = P_p_out - DELTAP_pipe_inlet
P_hr_in  = P_f_out - DELTAP_pipe_outlet
Pwr_pump=((DELTAP_pipe_outlet+DELTAP_pipe_inlet+DELTAP_total_module)*1)*(m_dot_f_in*1)/rho_f
eed/eta_pump
{Pwr_vac = 750[J/s] "Estimated"}
// Pwr_tot = Pwr_pump
```

## 12.11 NATIONAL DISSERTATION AWARD FOR THIS THESIS

**THE UNIVERSITIES COUNCIL ON WATER RESOURCES**

*universities and organizations leading in education, research, and public service in water resources*

April 6, 2016

Dr. David M. Warsinger
Massachusetts Institute of Technology
77 Massachusetts Ave, Room 7-034
Cambridge, MA 02139-4307

Dear Dr. Warsinger:

It is with great pleasure that we notify you that you have been selected as the first place recipient of the 2016 Ph.D. Dissertation Award in the category of Natural Science and Engineering.

The award consists of a certificate, a $750 check, reimbursement up to $1,000 for travel expenses, and waived conference registration for the 2016 UCOWR/NIWR Conference, June 21-23, at the Hilton Pensacola Beachfront Hotel in Pensacola Beach, Florida. Your award will be presented at the UCOWR Awards Banquet on June 22 at 6:00 p.m.

We would also like to invite you to prepare and deliver a 15-minute talk of your choice to the entire conference during our Wednesday morning plenary session. Please email your talk title and abstract to Melissa Pind at mpind@siu.edu. She will also need a 150-200 word bio and a high-resolution photo.

We are pleased to speak for the entire UCOWR delegation in congratulating you and have every confidence that your contribution to the field of water resources will be a significant one.

Sincerely,

Karl W.J. Williard

Karl W.J. Williard
Executive Director

KW:mp

Cc: John H. Lienhard V, Ph.D.

1231 Lincoln Drive, Room 118
Southern Illinois University – Mail Code 4526
Carbondale, IL 62901
Phone (618) 536-7571 • Fax (618) 453-2671 • E-Mail ucowr@siu.edu
www.ucowr.org

# Chapter 13. REFERENCES